\documentclass[12pt,index=totoc,openany]{book}
\usepackage{amsmath,amsfonts,amssymb,amsthm,a4wide,graphics}
\usepackage{makeidx}
\usepackage{nameref}
\usepackage[pdftex,bookmarks=true,bookmarksnumbered]{hyperref}
\hypersetup{colorlinks=true,%
            citecolor=black,%
            filecolor=black,%
            linkcolor=black,%
            urlcolor=black,%
         pdfsubject={},%
        pdfnewwindow=true,%
        pdffitwindow=true,%
        pdftoolbar=true,%
        pdfmenubar=true,%
        bookmarks=true,%
        unicode=false,%
                pdftex}
\numberwithin{equation}{section}

\makeglossary
\makeindex

\begin{document}

\title{{\bf\Huge Extended Quantum Mechanics}\footnote{This corrected and completed version of the publication: P.B., {\sl acta phys. slov.} {\bf 50} (2000) 1--198\newline\hspace*{.5cm}
 is now converted into book-format and endowed with functioning Index and \bf{hyper-references}.}}
 \author{Pavel B\'ona, \\ e-mail:\ bona@sophia.dtp.fmph.uniba.sk \\
Department of Theoretical Physics, Comenius University \\ SK-842 48
Bratislava, Slovakia}

\maketitle
\def\nl{\hfil\break}


\addcontentsline{toc}{section}{\protect\numberline{}{\noindent\it Preface}}
\centerline{\it\Large Preface}\vspace{1.2cm}
{
In this work, we are dealing with such formal and conceptual extensions of nonrelativistic
quantum mechanics (QM) which contain QM with its standard formalism and interpretation as
a subtheory.
The book can be considered as an essay on mathematical and conceptual
structure of QM which is
related here to some other (more
general, but also to more special -- ``approximative'') theories.
QM is here primarily equivalently reformulated in the form of
a Poisson system
on the phase space consisting of density matrices, where the
``observables'', as well as ``symmetry generators'' are represented by a
specific type of real valued (densely defined) functions, namely the usual
quantum expectations of corresponding selfadjoint operators.

It is shown that inclusion of additional
(``nonlinear'') symmetry generators (i.e. ``Hamiltonians'') into this
reformulation of (linear) QM leads to a considerable extension
of the theory: two kinds of quantum ``mixed states'' should be
distinguished, and operator -- valued functions of density matrices should
be used in the r\^ole of ``nonlinear observables''.

A general framework for
physical theories is obtained in this way: By different choices of the sets
of ``nonlinear observables'' we obtain, as special cases, e.g. classical
mechanics on homogeneous spaces of kinematical symmetry groups,
standard (linear) QM, or nonlinear extensions of QM. Time dependent Hartree-Fock
theory emerges  as
a specific application of our theoretical scheme.  Also various other
``approximations'' to (single- or many-particle) quantum-mechanical dynamics
 are subtheories of the presented
extension of QM - the {\em extended quantum mechanics} (EQM).

Interpretational questions of QM connected also with possible problems of its
extensions to NLQM are considered, discussed,  and resolved.
An interpretation scheme of EQM extending the usual statistical
interpretation of QM is also proposed. Eventually, EQM is shown to be
(included into) a $C^*$-algebraic (hence linear) quantum theory.

 Mathematical formulation of these theories is presented.
The presentation includes an analysis of problems connected with
differentiation of not everywhere defined functions on infinite -- dimensional
manifolds, e.g. a solution of some problems connected to work with
only
densely defined unbounded real--valued functions on the
(infinite dimensional) ``phase space'' corresponding to unbounded operators
(generators) and to their nonlinear generalizations. Also ``nonlinear
deformations'' of unitary representations of kinematical symmetry Lie groups
are introduced. Possible applications are briefly discussed, and some
specific examples are presented.

The text contains, in the introductory chapter, also brief reviews of
geometric form of Hamiltonian classical mechanics, as well as of QM.
Mathematical appendices make the work nearly selfcontained.}\nl\nl

\rightline{\it P.B.,  Bratislava, 1999 and 2012}


\pagebreak
\tableofcontents

\hyperlink{index}{{\bf Index}}{\hfill{\bf \pageref{index}}}

\hyperlink{symbols}{\bf List of Symbols}\hfill{\bf \pageref{symbols}}

\vspace*{\fill}

\newpage

\hyphenation{co-ad-joint re-pre-sen-ta-ti-on Bra-ti-sla-va}

\def\refer#1#2#3#4#5{#1:\ {\sl #2}\ {\bf #3}\ {(#4)}\ #5;\ }

\def\acc{$\!\!$\'{}$\!$}
\def\dti{\!\cdot\!}
\def\rref#1~{~(\ref{eq;#1})}
\def\dref#1~{~Definition~\ref{df;#1}}
\def\pgr#1~{~\pageref{eq;#1}}
\def\pgd#1~{~\pageref{df;#1}}
\def\pg#1~{~\pageref{#1}}
\def\bs#1{$\boldsymbol{#1}$}
\def\mbs#1{\boldsymbol{#1}}

\def\Ran{{\rm Ran}}
\def\Ker#1{${\rm Ker}(#1)$}
\def\mKer#1{{\rm Ker}(#1)}

\def\supp{{\rm supp}\ }

\def\deg#1{{\rm deg}($#1$)}
\def\mdeg#1{{\rm deg}(#1)}

\def\mT#1{\mcl T_{#1}}
\def\T#1{$\mcl T_{#1}$}
\def\cw{\curlywedge}

\def\LM#1{$\mbs{\Lambda^{#1}}(M)$}
\def\mLM#1{\mbs{\Lambda^{#1}}(M)}
\def\LN#1{$\mbs{\Lambda^{#1}}(N)$}
\def\mLN#1{\mbs{\Lambda^{#1}}(N)}

\def\mSs{{\cal S}_*}
\def\Ss{${\cal S}_*$}
\def\cS{${\cal S}$}
\def\mS{{\cal S}}
\def\Ej{$E_{\{j\}}$}
\def\mEj{E_{\{j\}}}
\def\mrj{\mrh_{\{j\}}}
\def\rj{$\mrh_{\{j\}}$}
\def\psitr{\psi_t^\mrh}

\def\mbf#1{{\mathbf #1}}

\def\pika{\ $\diamondsuit$}
\def\bpika{\ $\blacklozenge$}
\def\zel{\ $\spadesuit$}
\def\zal{\ $\clubsuit$}
\def\dovi{\ $\heartsuit$}

\def\pI{${\mbf p}_I$}
\def\pII{${\mbf p}_{II}$}
\def\mpI{{\mbf p}_I}
\def\mpII{{\mbf p}_{II}}

\def\bT{$\Bbb T$}
\def\mbT{{\Bbb T}}
\def\bA{$\Bbb A$}
\def\mbA{{\Bbb A}}
\def\bK{$\Bbb K$}
\def\mbK{{\Bbb K}}
\def\bC{$\Bbb C$}
\def\mbC{{\Bbb C}}
\def\bR{$\Bbb R$}
\def\mbR{{\Bbb R}}
\def\bN{$\Bbb N$}
\def\mbN{{\Bbb N}}
\def\bZ{$\Bbb Z$}
\def\mbZ{{\Bbb Z}}
\def\bI{$\Bbb I$}
\def\mbI{{\Bbb I}}
\def\bF{$\Bbb F$}
\def\bG{$\Bbb G$}
\def\mbF{{\Bbb F}}
\def\mbG{{\Bbb G}}
\def\bbf{${\bf f}$}
\def\mbbf{{\bf f}}
\def\cbF{${\bf F}$}
\def\mcbF{{\bf F}}
\def\bD{{\bf D}}
\def\bx{{\bf x}}
\def\by{{\bf y}}
\def\bz{{\bf z}}
\def\bv{{\bf v}}
\def\bw{{\bf w}}

\def\bbs#1{$\boldsymbol{#1}$}
\def\mbbs#1{\boldsymbol{#1}}
\def\bbr#1{$\boldsymbol{(#1)}$}
\def\mbbr#1{\boldsymbol{(#1)}}

\def\ia{{\it a}}
\def\ib{{\it b}}
\def\ix{{\it x}}
\def\iy{{\it y}}
\def\iz{{\it z}}
\def\iu{{\it u}}

\def\fpsxe{$\rf^\psi_{[\xi,\eta]}$}
\def\mfpsxe{\rf^\psi_{[\xi,\eta]}}
\def\fpsx{$\rf^\psi_\xi$}
\def\mfpsx{\rf^\psi_\xi}
\def\fpse{$\rf^\psi_\eta$}
\def\mfpse{\rf^\psi_\eta}

\def\rq{{\rm q}}
\def\rf{{\rm f}}
\def\rrh{{\rm h}}
\def\rx{{\rm x}}
\def\ry{{\rm y}}
\def\rz{{\rm z}}
\def\ra{{\rm a}}
\def\rrb#1{{\rm b}_{#1}}
\def\rc{{\rm c}}
\def\rd{{\rm d}}
\def\rQ{{\rm Q}}
\def\ru{{\rm u}}
\def\rv{{\rm v}}
\def\mrho{{\rm h}_0}
\def\mrhoo{{\rm h}^0}

\def\CH{${{\frak C}({\cal H})}$}
\def\mCH{{{\frak C}({\cal H})}}
\def\TH{${{\frak T}({\cal H})}$}
\def\UH{${\cal U(H)}$}
\def\mTH{{{\frak T}({\cal H})}}
\def\mUH{{\cal U(H)}}

\def\Lqn{$L^2(\mbR^n)$}
\def\mLqn{L^2(\mbR^n)}
\def\LHs{${\cal L(H)}_s$}
\def\mLHs{{\cal L(H)}_s}
\def\mH{{\cal H}}
\def\mK{{\cal K}}
\def\mLH{{\cal L(H)}}
\def\LH{${\cal L(H)}$}
\def\LHo{${\cal L(H}_\mome)$}
\def\mLHo{{\cal L(H}_\mome)}
\def\H{${\cal H}$}
\def\K{${\cal K}$}
\def\PH{$P({\cal H})$}
\def\mPH{P({\cal H})}
\def\A{${\cal A}$}
\def\mA{{\cal A}}
\def\Z{${\cal Z}$}
\def\mZ{{\cal Z}}
\def\C{${\cal C}$}
\def\Cc{${\cal C}_{cl}$}
\def\mCc{{\cal C}_{cl}}
\def\mC{{\cal C}}
\def\B{${\cal B}$}
\def\mB{{\cal B}}
\def\F{${\cal F}$}
\def\mF{{\cal F}}
\def\Fr{$\mF_{\mrh}$}
\def\mFr{\mF_{\mrh}}
\def\mFP{\mF_{\mPH}}
\def\FP{$\mF_{\mPH}$}

\def\PM{$\mcl P(\mfk M)$\ }
\def\mPM{\mcl P(\mfk M)}

\def\cl#1{${\cal #1}$}
\def\mcl#1{{\cal #1}}

\def\ben{$\beta_{\nu}$}
\def\mben{\beta_{\nu}}
\def\ber{$\beta_{\mrh}$}
\def\mber{\beta_{\mrh}}
\def\qr{$\rq_{\mrh}$}
\def\mqr{\rq_{\mrh}}
\def\pr{${\rm p}_{\mrh}$}
\def\mpr{{\rm p}_{\mrh}}
\def\qn{$\rq_{\nu}$}
\def\mqn{\rq_{\nu}}
\def\Tr{$T_{\mrh}$}
\def\mTr{T_{\mrh}}
\def\Trs{$T_{\mrh}^*$}
\def\mTrs{T_{\mrh}^*}
\def\Tn{$T_{\nu}$}
\def\mTn{T_{\nu}}
\def\Tns{$T_{\nu}^*$}
\def\mTns{T_{\nu}^*}

\def\Ty{$T_{\by}$}
\def\mTy{T_{\by}}

\def\P#1{$P_{#1}$}
\def\mP#1{P_{#1}}

\def\bcDp{${\boldsymbol{\mcl D}}_{r+}^1$}
\def\mbcDp{{\boldsymbol{\mcl D}}_{r+}^1}
\def\bcD{${\boldsymbol{\mcl D}}_r$}
\def\mbcD{{\boldsymbol{\mcl D}}_r}
\def\bcDF{${\boldsymbol{\mcl D}}(\mbF)$}
\def\mbcDF{{\boldsymbol{\mcl D}}(\mbF)}
\def\bcDFr{${\boldsymbol{\mcl D}_r}(\mbF)$}
\def\mbcDFr{{\boldsymbol{\mcl D}_r}(\mbF)}
\def\DdYa{$\mcl D_{ra}(\delta_Y)$}
\def\mDdYa{\mcl D_{ra}(\delta_Y)}
\def\DdX{$\mcl D_r(\delta_X)$}
\def\DdXa{$\mcl D_{ra}(\delta_X)$}
\def\DdXs{$\mcl D_{r*}(\delta_X)$}
\def\mDdX{\mcl D_r(\delta_X)}
\def\mDdXa{\mcl D_{ra}(\delta_X)}
\def\mDdXd{\mcl D_{rd}(\delta_X)}
\def\DdXd{$\mcl D_{rd}(\delta_X)$}
\def\mDdXs{\mcl D_{r*}(\delta_X)}
\def\DX{$\mcl D_r(X)$}
\def\DXa{$\mcl D_{ra}(X)$}
\def\DXd{$\mcl D_{rd}(X)$}
\def\DXs{$\mcl D_{r*}(X)$}
\def\mDX{\mcl D_r(X)}
\def\mDXa{\mcl D_{ra}(X)}
\def\mDXs{\mcl D_{r*}(X)}
\def\As{$A^*$}
\def\DAs{$D(A^*)$}
\def\DA{$D(A)$}
\def\HH{$\mH\oplus\mH$}
\def\mHH{\mH\oplus\mH}

\def\pEF{$\tilde{\mcl E_{\mbF}}$}
\def\mpEF{\tilde{\mcl E_{\mbF}}}
\def\EF{$\mcl E_{\mbF}$}
\def\mEF{\mcl E_{\mbF}}

\def\DGom{$\mcl D^{\mome}(G)$}
\def\mDGom{\mcl D^{\mome}(G)}
\def\DGomr{$\mcl D^{\mome}_r(G)$}
\def\mDGomr{\mcl D^{\mome}_r(G)}

\def\d#1,#2~{$d_{#2}{#1}$}
\def\md#1,#2~{d_{#2}{#1}}
\def\D#1,#2~{$D_{#2}{#1}$}
\def\mD#1,#2~{D_{#2}{#1}}
\def\Dfr{\D f,\mrh~}
\def\dfr{\d f,\mrh~}
\def\mDfr{\mD f,\mrh~}
\def\mdfr{\md f,\mrh~}
\def\Dfn{\D f,\nu~}
\def\dfn{\d f,\nu~}
\def\mDfn{\mD f,\nu~}
\def\mdfn{\md f,\nu~}
\def\Dhr{\D h,\mrh~}
\def\dhr{\d h,\mrh~}
\def\mDhr{\mD h,\mrh~}
\def\mdhr{\md h,\mrh~}
\def\mDhn{\mD h,\nu~}
\def\mdhn{\md h,\nu~}
\def\Dhn{\D h,\nu~}
\def\dhn{\d h,\nu~}

\def\mN#1,#2~{\|#1\|_{#2}}
\def\N#1,#2~{$\|#1\|_{#2}$}

\def\ad{{\rm ad}}
\def\madr{{\rm ad}_{\mrh}}
\def\madn{{\rm ad}_{\nu}}
\def\adr{${\rm ad}_{\mrh}$}
\def\adn{${\rm ad}_{\nu}$}
\def\madrs{{\rm ad}^*_{\mrh}}
\def\madns{{\rm ad}^*_{\nu}}
\def\adrs{${\rm ad}^*_{\mrh}$}
\def\adns{${\rm ad}^*_{\nu}$}

\def\OGF{$\mcl O_{F}(G)$}
\def\mOGF{\mcl O_{F}(G)}
\def\OWHr{$\mcl O_{\mrh}(\mGWH)$}
\def\mOWHr{\mcl O_{\mrh}(\mGWH)}
\def\OUr{$\mcl O_{\mrh}(\mfk U)$}
\def\OGr{$\mcl O_{\mrh}(G)$}
\def\OUn{$\mcl O_{\nu}(\mfk U)$}
\def\OGn{$\mcl O_{\nu}(G)$}
\def\OU{\cl O(\fk U)}
\def\OG{\cl O($G$)}
\def\mOUr{\mcl O_{\mrh}(\mfk U)}
\def\mOGr{\mcl O_{\mrh}(G)}
\def\mOUn{\mcl O_{\nu}(\mfk U)}
\def\mOGn{\mcl O_{\nu}(G)}
\def\mOU{\mcl O(\mfk U)}
\def\mOG{\mcl O(G)}
\def\Or{$\mcl O_{\mrh}$}
\def\mOr{\mcl O_{\mrh}}
\def\On{$\mcl O_{\nu}$}
\def\mOn{\mcl O_{\nu}}

\def\fk#1{$\frak #1$}
\def\mfk#1{\frak{#1}}

\def\fTs{$\mfk T_s$}
\def\mfTs{\mfk T_s}
\def\fNn{$\mfk N_{\nu}$}
\def\mfNn{\mfk N_{\nu}}
\def\fNr{$\mfk N_{\mrh}$}
\def\mfNr{\mfk N_{\mrh}}
\def\fMr{$\mfk M_{\mrh}$}
\def\mfMr{\mfk M_{\mrh}}

\def\eps{$\epsilon$\ }
\def\meps{\epsilon}
\def\veps{$\varepsilon$\ }
\def\mveps{\varepsilon}
\def\ome{$\omega$\ }
\def\mome{\omega}
\def\gam{$\gamma$\ }
\def\mgam{\gamma}
\def\alp{$\alpha$}
\def\malp{\alpha}
\def\mphi{{\varphi}}
\def\rh{$\varrho$}
\def\mrh{\varrho}
\def\sg{$\sigma$}
\def\sgo{$\sigma_0$}
\def\msg{\sigma}
\def\msgo{\sigma_0}
\def\mlam{\lambda}
\def\lam{$\lambda$}

\def\Xx{$X_{\xi}$}
\def\mXx{X_{\xi}}
\def\Xe{$X_{\eta}$}
\def\mXe{X_{\eta}}
\def\Xxe{$X_{[\xi,\eta]}$}
\def\mXxe{X_{[\xi,\eta]}}

\def\Ac{$\mcl A_{cl}$}
\def\mAc{\mcl A_{cl}}
\def\Cbs{$\mcl C_{bs}$}
\def\mCbs{\mcl C_{bs}}
\def\CG{${\mcl C}^G$}
\def\mCG{{\mcl C}^G}
\def\CGc{${\mcl C}^G_{cl}$}
\def\mCGc{{\mcl C}^G_{cl}}
\def\CGq{$\mcl C^G_{q}$}
\def\mCGq{{\mcl C}^G_{q}}
\def\GGc{${\mcl G}^G_{cl}$}
\def\mGGc{{\mcl G}^G_{cl}}
\def\GG{${\mcl G}^G$}
\def\mGG{{\mcl G}^G}
\def\nGcl{$\nu G$--classical\ }
\def\rGcl{$\mrh G$--classical\ }
\def\Gcl{$G$--classical\ }

\def\GWH{$G_{WH}$}
\def\mGWH{G_{WH}}

\def\hh#1,#2,#3~{$\hat {h_{\mfk#1}}(#2,#3)$}
\def\mhh#1,#2,#3~{\hat {h_{\mfk#1}}(#2,#3)}

\def\mh#1{{h}_{#1}}
\def\h#1{${h}_{#1}$}
\def\hSl{$\mh H^{Sl}$}
\def\mhSl{\mh H^{Sl}}

\def\ph#1,#2~{$\varphi_{#1}^{#2}$}
\def\mph#1,#2~{\varphi_{#1}^{#2}}
\def\cmrh#1~{{\varrho_{#1}}}
\def\crh#1~{$\varrho_{#1}$\ }

\def\pph#1,#2~{$\tilde{\varphi}_{#1}^{#2}$}
\def\mpph#1,#2~{\tilde{\varphi}_{#1}^{#2}}
\def\un#1,#2,#3~{${\rm u}_#1(#2,#3)$}
\def\mun#1,#2,#3~{{\rm u}_#1(#2,#3)}
\def\gQ#1,#2~{$g_\rQ(#1,#2)$}
\def\mgQ#1,#2~{g_\rQ(#1,#2)}
\def\taQ{$\tau^\rQ$}
\def\mtaQ{\tau^\rQ}
\def\mtQ#1,#2~{\tau^\rQ_{#1}#2}
\def\tQ#1,#2~{$\tau^\rQ_{#1}#2$}

\def\mv#1{{\bf v}_{#1}}
\def\w#1{${\bf w}_{#1}$}
\def\mw#1{{\bf w}_{#1}}
\def\vv#1,#2~{${\mathbf v}_{#1}(#2)$}
\def\mvv#1,#2~{{\mathbf v}_{#1}(#2)}
\def\vfn{${\mathbf v}_f(\nu)$}
\def\vvfn{${\mathbf{\check{v}}}_f(\nu)$}
\def\mvvfn{{\mathbf{\check{v}}}_f(\nu)}
\def\mvfn{{\mathbf v}_f(\nu)}
\def\vfr{${\mathbf v}_f(\mrh)$}
\def\vf{${\mathbf v}_f$}
\def\mvf{{\mathbf v}_f}
\def\mvfr{{\mathbf v}_f(\mrh)}
\def\vh{${\mathbf v}_h$}
\def\mvh{{\mathbf v}_h}
\def\Tpq{$\mcl T^p_q(M)$}
\def\Tpqx{$T^p_{qx}(M)$}
\def\mTpqx{T^p_{qx}(M)}
\def\mTpq{\mcl T^p_q(M)}
\def\mrTpq{T^p_q(M)}
\def\rTpq{$T^p_q(M)$}
\def\XN{$\mcl X(N)$}
\def\mXN{\mcl X(N)}
\def\XM{$\mcl X(M)$}
\def\mXM{\mcl X(M)}
\def\L#1~{$\pounds_{#1}$}
\def\mL#1~{\pounds_{#1}}
\def\dom{$\rd\mome$}
\def\mdom{\rd\mome}
\def\ip#1{$\boldsymbol{i_{#1}}$}
\def\mip#1{\boldsymbol{i_{#1}}}

\def\Lq{$L^2({\Bbb R},dq)$}
\def\mLq{L^2({\Bbb R},dq)}
\def\Wa{$W^*$-algebra}
\def\Wsa{$W^*$-subalgebra}
\def\Ca{$C^*$-algebra}
\def\Csa{$C^*$-subalgebra}
\def\rep{${}^*$--representation}
\def\Wrep{$W^*$--representation}
\def\autm{${}^*$-automorphism}
\def\aut#1{${}^*$-Aut\ #1}
\def\maut#1{{}^*$-Aut$\ #1}
\def\Aut#1{$Aut(#1)$}
\def\mAut#1{Aut(#1)}

\def\Psib{$\Psi^\flat$}
\def\mPsib{\Psi^\flat}

\def\lb{\langle}
\def\rb{\rangle}

\def\plr{$p_{\leftrightarrow}$}
\def\mplr{p_{\leftrightarrow}}

\def\eequiv{\Leftrightarrow}
\def\imply{\Rightarrow}

\def\wrt{with respect to\ }
\def\om#1,#2~{$\omega_{#1}^{#2}$}
\def\mom#1,#2~{\omega_{#1}^{#2}}
\def\ommr{$\mome_{\mu,\hat\mrh}$}
\def\mommr{\mome_{\mu,\hat\mrh}}

\def\prob{{\rm prob}}
\def\nbhd{neighbourhood\ }

\def\eq{{equation}}

\def\noidt{\noindent}

\def\emn#1~{{#1}\ind{#1}}
\def\emm#1~{{\bf #1}\ind{#1}}
\def\glss#1~{#1\glo{#1}}

\newcommand{\bequ}{\begin{equation}}
\newcommand{\barr}{\begin{eqnarray}}
\newcommand{\earr}{\end{eqnarray}}

\newcommand{\glo}{\glossary}
\newcommand{\ind}{\index}
\newcommand{\rarw}{\rightarrow}
\renewcommand{\Im}{{\rm Im}}
\renewcommand{\Re}{{\rm Re}}

\swapnumbers
\theoremstyle{plain}
\newtheorem{thm}{Theorem}[section] 
\newtheorem{defi}[thm]{Definition}    
\newtheorem{defs}[thm]{Definitions}   
\newtheorem{notat}[thm]{Notation}     
\newtheorem{prop}[thm]{Proposition}    
\newtheorem{lem}[thm]{Lemma}        
\newtheorem{lem*}[thm]{Lemma*}       
\newtheorem{pt}[thm]{}             
\theoremstyle{remark}
\newtheorem{intpn}[thm]{\bf Interpretation}  
\newtheorem{rem}[thm]{{\bf Remark}}         
\newtheorem{exmp}[thm]{{\bf Examples}}      
\newtheorem{exm}[thm]{{\bf Example}}         
\newtheorem{ill}[thm]{{\bf Illustration}}    
\newtheorem{note}[thm]{{\bf Notes}}       
\newtheorem{noti}[thm]{{\bf Note}}         


\chapter{Introduction}\label{I;int}
\def\autor{\ref{I;int}\quad Introduction}

We present in this work a straightforward, and  a ``very natural'' theoretical
extension of
traditional (linear) quantum mechanics (QM), providing a general framework
of several physical theories. It contains QM itself, its (almost all up to now
published) nonlinear modifications and extensions, and also
its ``semiclassical approximations'', together with the Hamiltonian
classical mechanics (CM). This is made formally by a
geometrical reformulation of QM and by its
subsequent nonlinear extension (containing the unchanged linear QM as a
subtheory); an interpretation scheme for this extended theory
is also proposed here.
Although rather ``trivial'' from a certain point of view, the obtained extended
quantum mechanics (EQM)\footnote{The obtained EQM provides rather
``metatheoretical framework'' for a broad class of physical theories than
a specific theory of a given class of physical systems.}
 seems to offer new insights into conceptual
foundations and also possible applications of quantum theory. It renders
also alternative views to
different approximations and modifications of QM like, e.g., the time
dependent Hartree Fock theory, WKB approximation, or the ``nonlinear
Schr\"odinger equation'', which are just subtheories of EQM.\footnote{Let us
note here that, for general dynamical systems (resp.\ systems of differential
equations), ``(non--)linearity'' is not an unambiguous specification: Any
linear equation can be transformed into a nonlinear form by a change of
variables and, conversely, many nonlinear equations can be rewritten into a
form of linear ones, cf.\ e.g.\ a Poincar\'e theorem~\cite[Chap. 5,\S 22]{arn3},
or the ``Koopmanism'' in e.g.,~\cite{koopman} and Remark~\ref{rem;koopm}.
Linearity in QM is
determined in our work with a help of structures on the projective Hilbert
space \PH.} The presented
theory provides also a global view onto solutions of dynamical equations of
many of its subtheories including a specification of ways to obtaining
their solutions. Having its origin in mathematically well
defined models of infinite quantum systems described by traditional (hence
linear!) nonrelativistic quantum field theory (QFT),
cf.\ \cite{hp+lie1,bon1,morch&stroc,stroc}, no mathematical
inconsistencies could be expected in the basic structure of EQM.

Next Section~\ref{motiv} contains a description of the present
author's motivation, including some of his presently accepted philosophical
ideas, and his mostly personal view on the history of this work.
The author is aware that motivation and history of writings can be
considered either from a subjective point of view of the author, or
from the point of view of more ``objective'' history based on a review
of existing published works
connected in some way with the contents of the presented work.
The second point of
view, if taken seriously, would need considerable historical effort of
experts in the related fields, and we shall not try to present it
in this work; we shall add, however, some comments
and references to compensate partially this gap, cf.\ also
Remark~\ref{rem;hist}.

Many important papers relevant to the contents of the present work became known
to the present author only after writing his own ``independent'' version of the
``story''.\footnote{This explains also some omissions of citations of some
relevant earlier published papers in the author's previous works:
The present author would like to apologize to the authors of those omitted
papers in this way.}
It is, however, important to have in sight also independently written works on
the considered subject, since alternative approaches to formulation of similar
theories might provide also some alternatives for
interpretation and/or application of the developed formal theory.
This is even more valid taken into account
that the author's formulation of the presented results
was rather  ``indirect'', obtained as a byproduct of other (a priori
unrelated) investigations.
We are trying to give here all the relevant citations and credits we are aware
of.\footnote{In spite of
this, the bibliography remains probably rather incomplete, and the present
author has to apologize repeatedly to authors of unnoticed relevant works.}

The Section~\ref{results} contains a heuristic description of the general
construction of main concepts, mathematical structures, as well as
interpretation problems, and possible applications  of the presented
nonlinear
extension of quantum mechanics (NLQM). The Section~\ref{notes} contains
some notes on organization of the presented work.
We include also into this chapter sections describing briefly
general structure of CM (cf.\ Section~\ref{I;clmech}), as well as of QM
(Section~\ref{I;qmech}), because it provides a starting framework for
forthcoming theoretical constructions.

\section{Notes on Motivation, Background Ideas, and History}
\label{motiv}
\def\autor{{
\ref{I;int}\quad Introduction}}
\def\nazov{{
\ref{motiv}\quad Notes on Motivation, Background Ideas,
and History}}

The present author is aware of problematic nature of claims about ``the
originality''
of ideas in Science, and of the corresponding ``priorities''.
Even if written in the author's relative isolation, the ideas might come
indirectly into the author's mind, through various cultural and social
manifestations, or simply by reading also scientific papers not manifestly
related to the considered problem.
The author will not try to do complicated introspective
psychological considerations on origins of his own ideas, what would be
necessary to give quite honest (but in any case subjective) answers
to questions on ``the originality'', or at least on ``the independence'', of
obtaining the presented results. We shall try, in the next paragraphs,
 to describe as honestly as possible in a brief exposition
the genesis and history of ideas resulting in this work. That might be
useful also for better understanding of place of the presented theory
in the framework of contemporary theoretical physics.

\begin{rem}[On contexts and contributions of this work]\label{rem;hist}
Let us mention here at least some references considered by the author as
important for a sight on the present work in the broader scientific context .
The presented work can be put in a connection with attempts at specific
nonlinear
generalizations of QM (NLQM) considered as a Hamiltonian field theory on the
projective Hilbert space as the ``phase space'' with a
specific (quantum) statistical interpretation; the present work generalizes
and unifies such theories. A pioneer work in this
direction was, perhaps, the short paper~\cite{kibble} by T.W.B. Kibble,
containing a sketch of nonlinear pure--state dynamics and also suggestive
motivation directed to applications and generalizations in
relativistic QFT and general relativity (GR). Trials (unsuccessful)
to formulate quantum statistical interpretation of such theories, as well as
some dynamics of mixed states contains the proposal~\cite{weinb} by S. Weinberg.
In the papers by R. Cirelli et al., eg. in~\cite{cir,cir4}, the authors
formulate in a
mathematically clear geometrical way standard QM, and they are looking for
general principles for possible generalizations of (pure state) quantum
kinematics. The papers by M. Czachor et al.~\cite{czachor} contain also
proposals for description of dynamics of density matrices in NLQM
(accepting essentially the author's proposal from~\cite{bon10}),
and also there are investigated methods to solutions
of dynamical equations for some classes of generators. The author's
paper~\cite{bon10} contains all the essentials of the here presented theory.
  Connections with older
formulations of NLQM and with semiclassical approximations, as well as some
proposals for a search for generalized (pure state) kinematics are contained
in~\cite{ashtekar}.

Any of the (to the present author) known published papers do not
contain consistent proposals of definitions and of quantum
statistical interpretation of \emm nonlinear
observables~;\footnote{This seems to be true also for the
papers~\cite{doebner,luecke} by Doebner, Goldin, L\"ucke et al.;
their ``Doebner--Goldin'' NLQM (DG) appears to be for some choice
of model-parameters non-Hamiltonian,
hence it does not fully ``fit'' into the kind of presently
analyzed theories: For testing the belonging of DG to the here
analyzed class of NLQM, one should, e.g.\ to check, whether the
r.h.s. of~\cite[Eq. (1.2)]{doebner} can be rewritten in the form
of the r.h.s. of\rref2.14~, resp.\ in a form
$\bigl(\hat{\malp_{P_\psi}}+{\bf f}^0(P_\psi)\bigr)\dti\psi$, with
\alp\ a closed one--form on \PH, and
$\hat{\malp}\in\mLHs=\mfTs^*$\ being its operator form (cf.\
page~\pageref{eq;2.rep}).} such a definition and interpretation of
observables is given in this work. It is given here also an
inclusion of the introduced (nonlinear) EQM  into a linear theory
of a bigger system described in framework of algebraic QT, cf.\
also~\cite{bon1,d+wer1}. Work with unbounded generators is
proposed here in a flexible way: One can restrict attention to a
certain set of submanifolds of the ``quantum phase space'' \Ss :=
the space of all density matrices, the union of which is not
necessarily dense in \Ss. Two kinds of ``mixed states'' are
introduced, what is a natural consequence of nonlinear dynamics,
cf.\ also~\cite{jordan,czachor}. A unitary representation of a Lie
group $G$ is chosen here as a ``parameter'' serving to specify all
the general elements of the theory: the domains of definitions (in
\Ss) of unbounded generators, the sets of generators, of
symmetries, of observables, and of states of the described system;
it specifies the $UG$--system $\Sigma_{UG}$. This allows us to
determine also the concept of a \emm \bs{UG_I}--subsystem~ of a
given $G$--system $\Sigma_G$; also a general definition of a
subsystem of a physical system in NLQM was not satisfactorily
established in the known literature. We shall not, however, look
here for generalized kinematics (i.e.\ alternatives to \Ss, cf.\
\cite{cir,cir4,ashtekar}),
 neither we shall try to
formulate here a solution of the ``problem of measurement in QM''
\label{1hist}
(understood, e.g.\ as a dynamical description of the ``reduction of wave
packet'').
\hfill\dovi\end{rem}
This work is a modified and completed version of the
preprint~\cite{bon10}.\footnote{The
author is deeply indebted to Vlado Bu\v zek for his strong
encouragement in the process of the author's decision to prepare and
publish this new version of~\cite{bon10}, as well as for
the kind support and also for the effective help he rendered
in the process of preparation of the publication of this work.}
The author
decided to publish it now also because of recently renewed interest in
nonlinear QM (NLQM) (see, e.g.
\cite{jordan,bug,jones,beltram,leifer,ashtekar,cir4,cant2},
or~\cite{doebner,czachor,goldin,luecke}),\footnote{The
author is indebted also to (that time) PhD students, esp.\ to  M. Gatti and
E. Gre\v s\'ak,
who helped him to make clear some technical features of the presented work,
cf.\ \cite{sikela,sladecek,gatti,gresak,polakovic}.}
as well as in foundational questions of connections of QM with CM, cf.\ e.g.
\cite{QTM1,penrose1,penrose2,zeil1,zeil2,zeil3,lands1,breuer,cir3,qm-hist}, or
also
\cite{peres2,f&lewis&c,gr&pr,tol&chadzi,fecko,schulman,streater}.\footnote{M.
Czachor and his coworkers are
 acknowledged for their repeated interest in the author's work, as well as
 for the kind submissions of information about the progress of their work.
The author expresses his dues
also to  S.T. Ali, P. Busch, V. Bu\v zek, G. Chadzitaskos, R. Cirelli,
V. \v Cern\'y, H.-D. Doebner, G.G. Emch, M. Fecko, G.A. Goldin,
K.R.W. Jones, N.P. Landsman, J.T. Lewis,
E. Lieb, W. L\"ucke, H. Narnhofer, P. Pre\v snajder, E. Prugove\v cki,
A. Rieckers, G.L. Sewell, R.F. Streater,
W. Thirring, J. Tolar, T. Unnerstall, R.F. Werner, A. Zeilinger, W.H. \.Zurek,
and other colleagues and friends for discussions, and/or for providing him
with their relevant papers, and/or for giving him moral support.}

 Moreover, it can be assumed that ideas contained in
this work will be useful for construction of some (not only physical) models.

\subsection{On initial ideas and constructions}\label{mot;modif}

The idea of a natural nonlinear generalization of QM
(leading to the paper \cite{bon10}) appeared to the present author
after an equivalent reformulation of QM in terms of CM on (infinite
dimensional) symplectic manifold \glss \PH\ ~\glossary{\PH - projective Hilbert
space} in the works~\cite{bon4,bon8}.
This was, in turn, a result of trials to understand
connections between
QM and CM more satisfactorily than via the
limits $\hbar\rightarrow 0$:\footnote{For a review and citations on various
approaches to
``quantization'' and ``dequantization'' with their rich history beginning
with the advent of QM see
e.g.~\cite{doeb&tol,tol,fronsdal,sniatycki,flato,isham,tolar,landsman};
some connections of CM and QM via $\hbar\rarw 0$ could be seen
from~\cite{hp3,hagedorn,kay}; for a recent trial to define
the limit $\hbar\rarw 0$\ in a mathematically correct way cf.
also~\cite{polakovic}.}  A part of the effort was a formalization of the
Bohr's beautiful argumentation, e.g.\ in~\cite{bohr1,bohr2}, on necessity of
using CM for a formulation of QM as a physical theory, combined with
the author's requirement on ``universality'' of quantum theory
(QT),\footnote{We distinguish here QM from QT, the later including
also mathematically well
defined parts or versions of QFT, e.g.\ the nonrelativistic \Ca ic theory of
systems ``with infinite number of degrees of freedom''. In this
understanding, QT can describe also macroscopic parameters of ``large''
quantal systems, composing their classical subsystems.} the effort possibly
hopeless if taken too literally.\footnote{\label{ft;1meas}The intention
of the author
was even to formulate a general model of the measurement in QM, being up to now
an unsolved fundamental problem of QM (if QM is considered as a
``universal theory''),~\cite{bon-m}. This author's effort started in 1961
at Charles University and/or Czech Technical University (\v CVUT) in Prague
(the Faculty of Technical and Nuclear Physics -- FTJF -- was
administratively moved between these two universities in those years), later
continued also in
 a small seminar formed by J. Jers\'ak, V. Petr\v z\'\i lka, J. Stern, and
the present author; in the framework of this seminar was formulated a simple
(unpublished) proof
of impossibility of information transmission by  ``reduction of wave
packets'' corresponding to the EPR--like quantum measurements according to the
 traditional (Copenhagen) formulation of QM, cf.\ Note~\ref{not;red-post} on
page~\pageref{not;red-post}.}

The papers \cite{bon4,bon8} resulted from the recognition
of quantum pure--state space \PH\ as a natural symplectic (even a K\"ahler)
manifold; this personal
``finding'' was gradually reached at studying of generalized coherent
states (GCS)\footnote{The author is indebted to P.Pre\v snajder for turning
his attention to GCS.} in QM,
\cite{klaud1,perel1,berez2,lieb1,sim1,davies,koh-st}, in
looking for their possible usage in describing connections between QM and
CM. We benefitted also from the description of symplectic
structure on (finite dimensional) complex projective spaces \cite{arn1}.
Works on their quantum mechanical connections/applications
\cite{berez1,cant,herman,rowe1,pr+v,rowe2} was encouraging in this effort.
As the author can judge today, many important results
have been obtained in the literature.
Unfortunately, not all of the details of the cited
works were clearly seen by him during the time when he formulated
his theory: There was a variability of languages and interpretations in
various papers, as well as a lack of sufficient mathematical rigor which
obviously was an obstacle for a better understanding. There were also important
unnoticed works containing some of the author's later results,
e.g.~\cite{strocchi,kibble,kra&sar}.\footnote{For the citation~\cite{kra&sar}, as
well as for some other useful notes made during the correspondence concerning
~\cite{bon8} the author is obliged to K.Hepp. The author obtained the
citation~\cite{cir5} from K.R.W.Jones. About the
citation~\cite{kibble} was the author informed by N.P.Landsman.}

Conceptually important in the search of QM\ $\leftrightarrow$\ CM
connections was appearance of symmetry groups $G$ allowing a unified
theoretical description of ``changes of objects with a specified
identity'', cf.\ mainly~\cite{curie,weyl,wigner2,HVW}, and givig a framework for
description of physical quantities;
we have restricted our attention to Lie groups, where distinguished
one--parameter subgroups correspond to specific physical quantities (cf.
Galileo, or Poincar\'e groups).
The cited papers using sets of GCS used them either as a tool for description
of some ``quasiclassical approximations'' to QM in various specific
situations, or as a formulation of a ``quantization'' procedure, cf.
also more recent literature, e.g.~\cite{ali2,ali&doeb,landsman}.

Generalized coherent states were usually considered as submanifolds of the
Hilbert space\ind{Hilbert space} determined either as some more or
less arbitrary
parametrically determined manifolds (usually finite dimensional), or as
orbits of continuous unitary representations of a Lie group $G$. An essential
r\^ole is ascribed to a symmetry Lie group $G$ also in the present work:
This corresponds to the accepted (hypothetical) point of view according to which
observables in physics are necessarily connected in some way with a group of
symmetries.\footnote{A possible generalization of this point of view might
lead to the assumption, that observables are determined by local
groups,~\cite{kiril},
or gruppoids,~\cite{landsman}; the Landsman's book~\cite{landsman}
contains also other relevant ideas and techniques, as well as citations.}

\begin{intpn}\label{intpn;G-quantit}
This ``philosophy'' can be substantiated by the following simple
intuitive consideration: Physical situations (e.g.\ different states of a
physical system) corresponding to different values of a ``physical quantity''
should be connected by some transformations which make possible to assure
that the
different values are really ``values of the same quantity''; the assumption
of transitivity and invertibility of these up to now unspecified
transformations seems to be natural for quantities without some exceptional
values in their range. This results in the hypothesis of presence of a group
defining physical observables (resp.\ quantities); some further ``physically
natural'' continuity requirements then end at a Lie group.\footnote{As
concerns a general gnoseological approach of the present author to Theoretical
knowledge, it is close in a certain feature to that of K. R. Popper,
\cite{popper,popper1}, cf.\ also~\cite{heisenberg}; we accept, e.g.\ that each
scientific assertion can be
considered just as a hypothesis: There is no ``final truth'' in our
Knowledge. Moreover, any ``meaningful'' assertion concerning possible
empirical situations should be falsifiable by some empirical tests.
 Let us add, however, that one should distinguish different
``degrees of certainty'' of various claims: Although mathematically
formulated, claims on empirical contents should undergo our identification
with specific ``extratheoretical'' situations, and this process cannot be
fully formalized.}\hfill\bpika
\end{intpn}

\begin{rem}\label{rem;G-unb}
The presence of a Lie group $G$ in the following considerations has,
however, also a technical function: it offers us an easy possibility to work
with specific unbounded observables described by not everywhere defined
functions
on the symplectic manifold \glss \PH~; such observables correspond (in the
linear case) to usual unbounded
operators describing physical quantities in QM. The corresponding technical
tool is the existence of the \emm \bs{C^\infty(G)}--domains~ (e.g.\ the
{\bf G\accent23 arding domains}\ind{G\aa rding domains}) of strongly continuous
unitary representations $U(G)$\glo{U(G)} of any Lie group $G$.\footnote{For
an application of this kind of ideas cf.\ also the theory of
``Op*-algebras'',~\cite{lassner}.}
\hfill\dovi
\end{rem}

The importance of Lie group representations for QM was stressed already by
founders of QM, let us mention especially Weyl and Wigner
\cite{wigner1,weyl,wigner2,wigner3}; applications of Lie groups in
 fundations of QM was afterwards elaborated by many
others, e.g., cf.\ \cite{ludwig,ludw1,kiril,varad,alonzo,gilmore,divak}.  Also
Prugove\v cki's and Twareque Ali's papers, e.g.
\cite{pru1,pru2,ali,pru3,prugovecki}, were stimulating for the present author's
work: Some intuitively convenient statistical interpretation of GCS in
QT was also looked there for.
The Weyl's book \cite{weyl} contains, in an implicit way (as it was
perceived by the present author), some of the main ideas concerning
connection of QM with CM formulated in the papers
\cite{bon4,bon8}.\footnote{The above mentioned inspiring ``ideas'',
``stimulations'',
etc. are difficult to specify and formulate clearly: They were often hidden
in the stylistic form of presentation of otherwise ``quite simple facts'' by
the cited authors; e.g.\ the Weyl's
considerations on ``Quantum Kinematics'' in~\cite[Chapter IV.D]{weyl},
presently known to every physics student as CCR, were perceived by the present
author as very stimulating -- much later than during his student's years.}

In our presentation, orbits\ind{orbits of coadjoint representation}
of coadjoint representations of $G$ play an
important r\^ole. They appear naturally in QM as orbits of expectation
functionals corresponding to GCS, which are calculated on generators of the
considered Lie group representation $U(G)$; these generators are usually
interpreted in QM as distinguished sets of quantummechanical
``observables''. The canonical symplectic structure on these
$Ad^*(G)$\glo{$Ad^*(G)$ -coadjoint representation}--orbits is
described, e.g., in the  monograph \cite{kiril}, cf.\ also Appendix~\ref{A;LieG}.
The general coordinate--free differential geometric
formalism of Ellie Cartan and its applications to CM is described, e.g.\ in
\cite{abr&mars,kob&nom,thirr1,3baby}, cf.\ Appendix~\ref{A;manif}.

   Generalized coherent states determined by  continuous  unitary
representations $U(G)$  of  finite  dimensional   Lie groups   $G$
provide a ``semiclassical background''  to
ap\-proximate descriptions of quantum theory. Points of the manifolds
of coherent states can be canonically parametrized in many  cases
by points of an orbit of the coadjoint representation $Ad^*(G)$.
In these ca\-ses, a canonical Poisson structure corresponding to that one
existing on the $Ad^*(G)$--orbit\label{AdGo}   can  be  defined  on  the
manifold of coherent states. It is possible  to  determine
cano\-nically  a specific ``projection'' of quantum mechanical (=:
\emm quantal~) dynamics to  such
a ``classical phase space'',~\cite{bon8}. Some satisfactory (unambiguous, and
general) interpretation of these canonical ``classical projections'' is,
however, still missing.\footnote{We have known the ``mean-field'' interpretation
of such quantum motions, cf.\ Subsection~\ref{mot;MF}, and
Section~\ref{sec;IIIB}; physical origin of such a classical ``background
field'' might be looked for in hypothetical, or sometimes even known, existence
of some ``long--range forces'', representing an influence of, e.g.\ (let us
allow some visions to ourselves) Coulomb forces with quantum
correlations of distant stars to the considered microsystem,
cf.\ \cite{zeh1}. }

Methods of the  ``time  dependent
Har\-tree--Fock description''   of fermionic systems, or more generally,
of the ``time dependent variational principle'' in QM,~\cite{kra&sar}, can be
redu\-ced in many cases to specifications of  the  general  procedure of
the mentioned ``classical projections''.
The ``classical projections'' of quantum dynamics to orbits of
co\-herent states can lead, in some formally chosen cases (i.e.\ chosen
regardless to existence of any possibility of a physical interpretation of the
considered dynamics), to such a  classical  dynamics  which
has little in common with the original quantum system, cf. Illustration \ref{ill;difrestr}. This left an open
question to us, in what cases ``classical projections'' are ``close'' to
the projected quantum dynamics,~\cite{bon8}.

The dynamics  of  an individual  subsystem  of the infinite quantum system in
mean--field theory (MFT) is described exactly  by  such  a  kind  of
``classical projections'',~\cite{bon3,bon1}.  In this is hidden a
connection of our EQM with a (linear) QT of infinite quantum systems, cf.
also Subsection~\ref{mot;MF}.

We shall show, in Section~\ref{sec;IIIF}, that the
dy\-namics in NLQM (modified with respect to that of Ref.~\cite{weinb} for the
cases of evolution of ``mixed states'') can also  be  described  in
this way. We obtain a mathematically correct and physically
con\-sistently interpreted standard type of  quantum  theory  (i.e.\  a
\Ca ic theory)      in the case of such a mean-field
rein\-terpretation of the ``classical projections of QM''.  We  shall
describe these theories in a form of a generalized quantum
mecha\-nics of autonomous physical systems. ``Observables'' in
the presen\-ted theory are expressible as operator--valued
functions ${\mfk f}:
{\bf F}\mapsto{\mfk f}({\bf F})$\ of a classical field with values ${\bf
F}$\
appearing in corresponding interpretation also in MFT. In  models  of
MFT the ``classical field'' ${\bf F}$\ can describe, e.g.\ collective
varia\-bles describing macroscopic quantum phenomena like
superconducti\-vity, or other  ``global  observables''  describing a  large
quantum system.  The classical field \bF\ (cf.\dref2.17~) aquiring values in
$Lie(G)^*\ni{\bf F}$\ is
here present in a r\^ole of a ``macroscopic  background''  of  the considered
quantum system. The (nonlinear) dynamics, as  well  as the probabilistic
interpretation of the theory can be  described, however, independently of
any use of ``background fields'': The in\-troduction of the field \bF\ (which
is a  function  of  the  quantum states $\mrh\in\mSs(\mLH)$\ appears like an
alternative description  (or an ``explanation'') of the dynamics which can
lead to simpler solu\-tions of problems. We have not specified
unambiguously a physical interpretation of dependence of the operators
${\mfk f}({\bf F})$\
on values {\bf F}\ of the macroscopic field \bF. It can be suggested, e.g.\ that
\bF\  takes part in determination of ``physical meaning'' of the quantum
obser\-vables: For each value {\bf F}\ of \bF, ``the same'' quantum
observable \fk f\
is described by a specific operator ${\mfk f}({\bf F})$. We have  introduced,
how\-ever, a
standard prescription for calculation of probabilities of measured results
of observables represented by the operator -- va\-lued functions $\mfk f:
{\bf{F}}\mapsto\mfk{f}({\bf F})$ which is consistent with the
tra\-di\-tio\-nal one, cf.\ formulas\rref2.42d~,\rref2.42c~, and\rref2.42a~.
We also expect that
traditional foundational problems in physics like the
``quantum measurement problem'',     or the question on ``origins of
irrever\-sibility'' might be fruitfully reformulated in the presented
framework.

\subsection{Relation to infinite systems}\label{mot;MF}
An important element in building the presented scheme of EQM
was construction of classical
quantities of an infinite quantal system. This was done in usual
\Ca ic\ind{\Ca} language~\cite{emch1,dix1,dix2,sak1,pedersen,bra&rob,takesI},
cf.\ also~\cite{bon8}. The author was especially inspired by the
papers~\cite{hp-meas,hp+lie1,berez1}, the monographs on quantum--theoretical
description of systems ``with infinite number of degrees of
freedom''~\cite{emch1,bra&rob}, some general ideas of  Einstein, Bohr,
Heisenberg and other thinkers expressed in many, nowadays difficult
identifiable, places (as introductions to books and papers, popular and
philosophical writings, quotations by other people, etc.), as well as by some
other, both ``technically $\&$\ ideologically'' composed papers, like the
review~\cite{yaffe} on ``large N limits'' in QM.

Let us describe briefly the obtained picture of kinematics of an infinite
quantum system\ind{infinite quantum system} in which a commutative
(``classical macroscopic'') subalgebra
\cl N\ of observables
is determined by a unitary representation $U(G)$ of a Lie group $G$.
Let the large quantum system consists of $N$ copies of equal systems
described in separable Hilbert spaces
\ind{Hilbert spaces - separable}
${\cal H}_m$ by algebras of observables ${\cal L}({\cal H}_m), m=1,2,\dots N$.
Then the algebra of observables ${\cal A}_N$ of the
composed system is isomorphic to ${\cal L}(\otimes_{m=1}^N{\cal H}_m)$, and
nothing essentially new is obtained: It has only one ``reasonable''
irreducible
representation (up to unitary equivalence). The so called $C^*$--inductive
limit for
$N\rightarrow \infty$ of ${\cal A}_N$, cf.\ \cite{sak1}, however, is an
algebra \cl A\ of a different type: It has uncountably many mutually
inequivalent faithful irreducible representations. Subsets of these
representations
could be parametrized by some ``classical quantities'', which can be
themselves realized as a (commutative) \Ca\ in the center \Z\ of the double dual
${\cal A}^{**}$ of \A. But the center \Z\ is an incredibly big algebra
which cannot be, probably, used as a whole to some useful description of
macroscopic properties of ``the system \A''. Here was used a Lie group $G$ for
obtaining a specification of a subalgebra of \Z\ of a ``reasonable size''.
The use of a Lie group $G$\ind{Lie group G} allowed also a natural
introduction of a Poisson structure~\cite{weinst,marle,arn1}\ind{Poisson
structure}, and consequently classical dynamics\ind{classical dynamics in
\Z} into the ``relevant part'' of \Z.\footnote{A Poisson structure is,
however, always present in any noncommutative \Ca\ in the form of the
commutator of any of its two elements, cf.\ also~\cite{d+wer1}. This can be
used to obtain, by a certain
limiting procedure, cf.\ \cite{bon1,bon2,d+wer1}, also a Poisson structure on
some subsets of the commutative
\Wa\ \Z. The Poisson structure obtained in this way is identical with that
one connected with a Lie group action. Lie groups are, however, useful
(besides for technicalities in dealing with unbounded generators) for
interpretation of abstract ``observables'', and for determination of
``proper subsets'' of the huge centre \Z.}

 These constructions were
motivated by some attempts to understand possible
quantummechanical basis of classical description of macroscopic bodies,
cf.\ \cite{hp-meas,bon8,bon-m}, as well as of interaction of that bodies
with microscopic
systems described by QM. This effort included trials to solve the old
problem of modeling the
``measu\-re\-ment process in QM'' \cite{hp-meas,bon-m}. Although this questions
were extensively studied during the whole history of existence of QM, cf.,
e.g.\ \cite{neum1,bell,w+z,ludw1,QTM}, no  approach to their solution, hence no
answers, are  generally accepted up to now. In the process of modeling of
interaction of microsystems with macroscopic bodies in QM framework,
a quantum description of macroscopic bodies was a necessary preliminary
step. The simplest possibility was a study of kinematics of an infinite set
of equal quantum systems in the framework of \Ca ic theory. This is formulated
in \cite{bon8}. One of the most important questions was a ``proper'' choice
of observable quantities of such a big system.\footnote{That a choice of
``observable observables'' is a nontrivial task also from a quite different
point of view is claimed in~\cite{nielsen}.} This was done by a choice of
the kinematical Lie group $G$ mimicking macroscopic motions of the large
(composed)
quantum system: The representation $U(G)$ acted equally on any ``elementary
subsystem'' described by ${\cal L}({\cal H}_m), m=1,2,\dots \infty$.

The resulting formulation of nonlinear quantum dynamics in the presented
extension of QM
can be connected with the specific form of the author's formulation of dynamics
of infinite quantum systems \cite{bon1,bon3,unner1,unner2,d+wer1}
with interactions of ``mean--field type'', having its roots in~\cite{hp+lie1}.
\footnote{From the personal author's point of view, it was obtained in a
sense ``occasionally''. The resulting dynamics of the infinite
mean-field\ind{mean field system} systems \cite{bon8,bon1} was a natural
answer
to a simple question: {\sl How to define a microscopic Hamiltonian dynamics on
the infinite quantum system leading to a given (arbitrarily chosen) classical
dynamics on the part of the centre \Z\ specified with a help of the mentioned
representation $U(G)$ ?}} Our citations of works relevant for the theory of
microscopic description of macroscopic phenomena in quantum systems
are incomplete; some other relevant citations can be found
in~\cite{bra&rob,bra&rob2,spohn,sewell}.

Many modifications and generalizations of the sketched description
of classical quantities of infinite quantum systems, including
their dynamics, are possible. Some of them will probably lead to
the same ``microscopic'' nonlinear dynamics, as it is in the case
of MFT. The presented results can be considered as just a first
step in investigation of macroscopic dynamics from
quantum--theoretical point of view. There were performed already
some works containing more sophisticated description (than ours)
and more detailed results in this direction, cf.\ e.g.\ Sewell's
papers~\cite{sewell1,sewell2}, or some works in algebraic quantum
field theory (QFT),\footnote{\label{AQFT} This is the theory
formulated by Araki, Haag, Kastler, and others, cf.\
\cite{haag4,borchers1,araki,haag&kast,dopl1,dopl2,dopl3,haag2,borchers}.}
e.g.\ in~\cite{fredenh}.

We shall briefly return to some technicalities of the description of
``macroscopic subsystems'' of large quantum systems in
Section~\ref{sec;IIIB}.

\subsection{Questionable ``subsystems''}\label{mot;subsys}

A general interpretation of EQM considered as a ``fundamental theory'' is
not formulated in this work. It can be, however, conjectured that a viable
possibility for its interpretation is (by admitting the linear QM as
``the fundamental theory of simple systems'') a description of ``relatively
isolated systems'', i.e.\ ``ordinary'' quantal systems moving in an external
field which is in turn influenced by (or correlated with) these quantal
systems. Let us give here some motivation and background to this rough idea.

One of the most basic concepts of contemporary physical theories, and, perhaps,
of the methodology of the whole Science, is the concept
of {\it isolated systems}\index{system isolated} the description of which is
especially ``simple'': It is supposed, that there are specific
``circumstances'' under which we can deal with phenomena independently
of the rest of the world. Examples are: idealized  bodies ``sufficiently
distant from all other bodies'' described in CM in framework of an
``inertial coordinate system'', realized, e.g., by atoms in a dilute gas during
a certain time intervals. More generally, we are used to think about any
specific
``object'' as determined ``relatively independently'' of other objects
(except of some generally accepted ``background'', e.g.\ inertial frame, or
vacuum).  Mere
possibility to formulate such concepts of various ``isolated systems''
which approximately describe some observed phenomena can be considered as one
of the miracles of human existence. More detailed
investigation (and specification) of any phenomenon usually shows that such
a simple description  is of a restricted use, and  better results might be
obtained by consideration of a ``larger piece of the world''; the identity
of the ``considered (sub-)system'' can be then, however, lost.\footnote{A
version of the concept of an ``isolated system'' necessarily appears in any
kind of reproducible reflection in human thinking. Its specification, however,
varies with accepted ``paradigms''~\cite{kuhn} (let us stay with a mere
intuition on these ambiguous philosophical concepts), e.g.\ the meaning of the
physical system representing a falling stone was different for Aristotle
from that of Galileo, and also it was different for Einstein from that
of Mach,~\cite{mach}.}

An often used ``first step'' to describe some ``influence of other
systems'' onto the ``considered one'' is an introduction of
an appropriate
(possibly time dependent) ``external field''. This procedure corresponds to
the formal construction (and logic) of nonrelativistic CM:
The motion of a body interacting with other ones can be expressed in CM as
its motion in a time dependent ``external field'' (determined by a known
motion of ``other bodies'' in the presence of ``the considered one'').
Subsystems in CM are, in this
sense, clearly definable (they are continually described by a point in
their phase sub-space), and we can consider them as {\it relatively
isolated}\index{subsystem relatively isolated}: They move according to certain
nonautonomous evolution laws (as if they have their own -- time dependent
-- Hamiltonians), what can be intuitively understood as ``just a (time
dependent) deformation'' of a background of
formerly isolated system, leaving the identity of the system ``essentially
untouched'',\footnote{It is an analogy to ``external'' gravitational
field in general relativity acting on a ``test body'': it is a ``deformed''
inertial frame corresponding
to the background determined by massive bodies (e.g.\ by distant stars) -- in a
sense similar to that of the
Mach's approach to CM \cite{mach,votruba}.}  and this has introduced a change
into the dynamical law of the system.

The determination of isolated systems, as well as of subsystems
in QM is much more problematic than in CM. Schr\"odinger equation
describes, in analogy with CM, dynamics of a physical system in a given
external field:  Systems  described in this way can be considered as
``relatively isolated''. This formulation was very successful in
description of scattering and motion in external (macroscopic)
fields, of dynamics of atoms and small molecules, as well as in approximate
descriptions of a lot of phenomena in
many--particle systems.
QM time evolution of mutually interacting systems
occurring initially in uncorrelated pure states (i.e.\ in a pure
``product--state'') leads
usually in later times to an ``entangled'' state of the composed
system.\footnote{Theoretical, as well as experimental investigation of
``entangled states'' in QM  is quite intensive in last years, cf.
e.g.~\cite{knight,lew&samp,3horod,buz3,gisin1,buz4,zeil3,zeil2}.}
The  states of constituent subsystems are described in such a state just by
density matrices (which are mathematical objects also used for description
of ``mixed states'' in a common sense interpretation, i.e.
in the ``ignorance'' interpretation which is common in classical statistical
physics), and time development of these (obtained
by taking the ``partial trace'' of the evolving pure state of the
whole ``isolated''
system) need not be Hamiltonian (e.g.~\cite{davies,buz3,gisin1}). Since
nontrivial interaction (and also entanglement) between states of charged
microscopic particles and quantum states of macroscopic bodies (if
considered as quantum many--particle systems) is present also in systems
whose constituent subsystems are separated by cosmic
distances, cf.\ \cite{zeh1}, an empirically realizable definition of
isolated systems in QM remains a problem. We assume that EQM provides also a
possibility of an approximative Hamiltonian evolution for some of such
``basically entangled'' situations.

Another problem of QM connected with the problem of determination of
subsystems is the classical ``problem of
measurement in QM'', cf.\ \cite{neum1,bell,w+z,QTM,bon-m}. It can be, perhaps,
considered as
an (up to now unknown) process of ``entanglement'' of the states of the measured
microsystem with macroscopically distinguishable states of the
apparatus.\footnote{\label{ft;2meas}Recently are quite popular ``solutions'' of the quantum
measurement problem via ``decoherence'', cf.\ \cite{zurek}, resp.\ via
``decoherent histories'' approach, cf.\ \cite{qm-hist,dowk&kent}; the
present author considers them at most as preliminary attempts to attack the
problem.}
A determination of a clear cut between
``microscopic'' and ``macroscopic'' is missing in both of these problems.
Perhaps, the only available, formally well defined formulation of the
``micro--macro difference'' can be found in the framework of the \Ca ic
formulation of QT,~\cite{haag2,emch1,bra&rob,sewell}. In this framework,
also some models for the measurement process in
QM were formulated,~\cite{hp-meas,bon-m,whit+emch}; the process needed there,
however, an infinitely long time interval for its completion. We expect that
EQM provides a way also for description of the mentioned micro--macro
``entanglement''.

\subsection{Some basic building blocks of EQM}\label{mot;basic}
Our Extended Quantum Mechanics contains many theories as exact (i.e.
obtained without any ``approximations'' in a usual sense) subtheories. They
are considered usually as
different (but inequivalent) possibilities of descriptions of the same
system, e.g.\ one of the theories is considered as an ``approximation'' of
another one. Examples are WKB, Hartree--Fock, or classical mechanical
approximations to descriptions of some problems in QM, or CM and QM
themselves. All these subtheories are obtained from the
general scheme of EQM by specifying three subsets (which are, however, mutually
consistently interconnected) of corresponding three general building sets
of theoretical objects.\footnote{Other
conventional relations
between CM and QM are ``quantizations'', and
``dequantizations'', the later understood usually as a limiting procedure
denoted by ``$\hbar\rarw 0$''.}

  In classical mechanics\ind{classical
mechanics} \cite{whitt,abr&mars,thirr1,arn1,mack1}
 (CM) as well  as  in  \emm quantum theories (QT)~
 \cite{mack1,dirac,neum1,ludw1},
three main (mutually  interconnected)  classes
of fundamental objects (corresponding to basic concepts of the theory) are
used: (i) \emm observables~,
(ii)  \emm states~,
and (iii) \emm symmetries~. A one parameter subgroup of symmetries specifies
a chosen \emm dynamics~ of the system, and the corresponding parameter is
called the \emm time~.\footnote{In the considered specific theories the time
parameter is in a sense ``global'', so that it is meaningful to speak about
states and observables of the (total) system {\em at a time} $t\in\mbR$.} The
mathematical
representation  of  these classes and formulation of their mutual connections
do not always use ``physically motivated''  properties only;  some
clarity in expression of connections between constructs of formalized theories
and empiri\-cal and conceptual analysis of phenomena is often  reached  by  a
subsequent specification and interpretation of the used
mathemat\-ical objects. Any fundamental theory of the process of
measure\-ment\ind{measurement} of an arbitrary
mathematically defined ``observable'',
consid\-ered as a dynamical process within QM is not known;
we  are  not
able generally decide which mathematically defined  ``observables''
are accessible to empirical identifications; similar  comment
applies to ``states'', and also to ``symmetries''. This lack of
``bi\-jective correspondence'' between classes of known  empirical  situations
and objects of a theory could make the theory, on the other  hand,  more
flexible.

   We shall reformulate and extend the formalism of QM so that it
will include QM and a class  of  its  (nonlinear) generalizations.
Such an EQM contains much larger variety of
``observa\-bles''\ind{observables},
``states''\ind{states}, and ``symmetries''\ind{symmetries}
than does the  traditional  QM. These extended sets of fundamental objects
contain different subsets representing different
``subtheories''\ind{subtheories} of the extended QM. Between these
subtheories we shall find, in addition to ordinary (linear) QM, also, e.g.
Hamiltonian CM\ind{Hamiltonian CM} with phase spaces being
homoge\-neous phase spaces\ind{phase spaces - homogeneous} of Lie
groups,\footnote{A homogeneity requirement on phase spaces \wrt some
topological group seems to us natural from an ``epistemological'' point of
view, cf.\ Subsection~\ref{mot;modif}, resp.
Interpretation~\ref{intpn;G-quantit}, and Remark~\ref{rem;G-unb}.
There would be no problem, however,
to find in EQM also Hamiltonian CM on a general, not necessarily specified by
 a group action, symplectic submanifold of \PH.}
several existing formulations of nonlinear quantum dynamics\ind{nonlinear
quantum dynamics}, cf.\ Subsection~\ref{IIIA1;NL-Sch}, and~\cite{ashtekar},
and also the frequently used  approximations
to quantum\ind{approximations to quantum dynamics} dynamics
consisting of its  specific  restrictions  to
manifolds of generalized coherent states\ind{generalized coherent states}
of the considered system, or also the WKB-approximation~\cite{ashtekar}
are in our extension obtained as ``subtheories'' (without  making  any
approximations).
The mentioned specifications are obtained by corresponding
choi\-ces of subsets of ``observables'', ``states'',  and  ``generators''  of
symmetry groups, and are usually mainly determined (cf.\ Section~\ref{IIB;gener})
by a choice of a
uni\-tary representation $U(G)$\glo{U(G) - unitary representation of G}
of a Lie group $G$ in the Hilbert space \H\
corresponding to the traditional quantummechanical description  of  the
considered
system:\footnote{The group $G$ {\em cannot be generally identified} with the
 group of
symmetries of the system!} E.g., QM corresponds to the choice $G
:= \{e\}$ (a  one-point set), cf.\ Section~\ref{sec;IIIA1} (this
does not exclude a use of  other  group representations $V(G_1)$
in the description of ``microscopic observa\-bles'', and
``symmetries'' in QM; $V(G_1)$ will play, however,  another r\^ole
than the picked out $U(G)$ in the theory!);  CM  of $N$ scalar
particles is specified by the Schr\"odinger  representation $U(G)$
of the $6N+1$ - dimensional Weyl--Heisenberg group $G$, and by
addi\-tional restrictions to the sets of  ``permitted''  (or
``physical'') states, generators and  observables, cf.\
point~\ref{pt;CM}. Other ap\-proaches (related to our one) to some
incorporation of classical observables into an exten\-ded quantum
formalism were published, e.g., in~\cite{piron}, cf. also our
Section~\ref{sec;IIIB}, Appendix~\ref{B;Ca},
and~\cite{bon1,bon2,bon-m}.

The dynamics (generally nonlinear) of EQM on the ``quantum phase space \Ss''
can
be recovered as a {\em
subdynamics of linear dynamics of a larger quantal system}. This can be seen
from Section~\ref{IIC;symm-obs}, where in\dref2.25b~(ii) a \Ca\
of observables \CG\ was introduced such, that our evolutions in EQM are
(linear) automorphism groups of this \Ca, cf.\ also Section~\ref{sec;IIIB}.
Looking on the obtained EQM ``from a side'', we could recover similarity
between our transition from QM (resp.\ NLQM) to EQM (and its linear
realization on the
\Ca\ \CG), and the ``Koopmanism'' in CM (cf.\ Remark~\ref{rem;koopm}):
While in the Koopman transition the CM was ``linearized'' by transferring
the phase space $(M;\Omega)$ as a sort of ``spectrum space'' (cf.
Appendix~\ref{B;Ca}, Example~\ref{ex;abel-C&W}) into the
infinite--dimensional Hilbert space $L^2(M,\mu_\Omega)$, and its
(nonlinear) dynamics into a (linear) unitary group, in our consideration of
EQM (leading to nonlinear evolution on the ``quantum phase space'' \Ss, i.e.
in a ``restricted Schr\"odinger picture'') as a
\Ca ic theory we obtain (in the corresponding ``Heisenberg picture'')
a linear quantum dynamics on a \Ca\ (namely \CG), cf.\ \cite{bon-tr}.
The state space of this
\Ca\ is, however, much larger than \Ss, or even than the space $\mcl
M_{+1}(\mSs)$ of probability measures on \Ss\ (of which is \Ss\ the subspace
of Dirac measures, in a canonical way).


\section{A Brief Description of the Contents}\label{results}

For better orientation of readers in the contents of the following text,
we shall give
here also a brief and heuristic explanation of some of the main points of the
contents of this paper, as well as some of their interconnections.
Some notes on the placing of different parts of the contents in the text can
be also found in Section~\ref{notes}.

   Let us introduce first some notation used in this paper:

\addtocounter{thm}{-1}\begin{notat} (i) We use usually different fonts
(e.g., f,  \bbf,
{\it f}, $\mfk f$, \~f, \^f, {\em f}) for different kinds of mathematical
objects.\footnote{Bold form of symbols will be used sometimes, mainly in their
definitions, however, also for the otherwise nonbold ones, which are of the
same typographic form.} By $\&$\ is denoted the logical conjunction ``and''.\nl
(ii)  The relation $ A(x)\equiv B(x)$ expresses (usually) {\it assertion},
that  values of  the  two  functions $A$  and $B$  are  mutually  equal  on
the intersection $D(A)\cap D(B)\ (\ni x)$ of the domains $D(A)$
(resp.\ $D(B)$)  of definition of the functions $A$ and $B$. The relation
$A:=B$\ defines a new symbol $A$\ by expressing it via an earlier introduced
 expression $B$.\nl
(iii) The symbol $f(\cdot,y)$ denotes the function $x\mapsto f(x,y)$. \hfill\pika
\end{notat}
\begin{pt}[QM, and NLQM]\label{pt;generQM}{\rm
 QM is traditionally formulated in terms of  selfadjoint
operators $X$ on a complex Hilbert space \H\ which play the double
r\^ole of the ``observables'', as well as of the ``generators'' of
sym\-metry groups in the theory. QM can be  equivalently  reformulated
in terms of (infinite dimensional) classical  Hamiltonian
mecha\-nics on the phase space \glss \PH\ ~ consisting of  one-dimensional
com\-plex subspaces ${\bf x, y},\dots $ of $\mH$.\footnote{Such a scheme
should be supplied by an interpretation scheme extending the probabilistic
interpretation of QM. Such an interpretation is given later, cf.
Interpretation~\ref{int;2.27}.}  Linear operators $X = X^*$  on \H\  then
correspond to  the  functions  $h_X: {\bf x}\mapsto h_X({\bf x})  :=
Tr(P_{\bf x} X)\equiv \lb x|X|x\rb/\lb x|x\rb$ on \PH, where
$P_{\bf x}:= P_{\it x}\quad (0\neq{\it x}\in{\bf x}\subset D(h_X)$, cf.
\rref2.18~) is the orthogonal projection onto \bx. The Poisson  bracket
is
\begin{equation}\label{(1.1)}
\{h_X,h_Y\}({\bf x}) := i\,Tr(P_{\it x}[X,Y]) = h_{i[X,Y]}({\bf x}),
\end{equation}
where $[X,Y] := XY - YX$ is the commutator. The  Schr\"odinger
equa\-tion is then equivalent to Hamiltonian  equations
corresponding to~\eqref{(1.1)}: If $H$ is the Hamiltonian operator
of a QM system, then  the evolution of the ``observables''
$h_X\equiv h_{X0}\mapsto h_{Xt},\ t\in\mbR,$ is described by the
Heisenberg-Hamilton (resp.\ von Neumann-Liouville) equations
\begin{equation}\label{1.2}
{\frac{d}{dt}}h_{Xt}({\bf
x})\equiv{\frac{d}{dt}}h_X(\varphi_t^H{\bf x}) =
\{h_H,h_X\}(\varphi_t^H{\bf x}), \ {\bf x}\in\mPH,\
t\in{\mbR};\quad h_{Xt}:=h_X\circ \varphi_t^H,
\end{equation}
where $\varphi_t^H $ is the ``Hamiltonian'' (resp.\ ``Poisson'') flow on \PH\
cor\-responding to the unitary evolution $t\mapsto exp(-itH){\it x}$ of vectors
${\it x}\in \mH$, i.e.\ a one-parameter group of transformations of \PH\
con\-serving Poisson brackets which can be determined from~\eqref{1.2}. This
immediate rewriting of QM differs from an  ``ordinary  Hamiltonian
CM'' on \PH\  by a specific restriction of the set \F(\PH) of
dif\-ferentiable real valued functions used as ``observables'' and
``generators'':  QM
uses only those ${\it f}\in\mcl{F}(\mPH)$ that have the form ${\it f}\equiv h_X
(X=X^*)$.
Let us call these $h_X$ \emm affine functions~ (or also ``K\"ahlerian
functi\-ons'',~\cite{cir}) on \PH: They can be considered as affine
functions defi\-ned on all convex combinations
$\mrh~:= \sum_j\lambda_j P_j\in\mSs$ of the pure states $P_j\in\mPH$; they
can be characterized, however, in a ``purely geometrical way'' in the
framework of \PH\ with a help of canonical metrics on it
(cf.\ \cite{cir,bon4,bon8}): affine functions
${\it f}\in\mcl{F}(\mPH)$
are exactly those $f$\ which generate Poisson flows conserving the metrics
(equivalently: conserving transition probabilities,~\eqref{1.4}),
and, in that case, they are expressible by linear operators $X$, i.e.\ $f=h_X$.
We shall sometimes call the affine functions $f$ also ``linear functions''.
Other ${\it f}\in\mcl{F}(\mPH)$ will be called  \emm nonlinear functions
on \bs{\mPH}~. The ``equation of motion''~\eqref{1.2} for general functions
${\it f,h}\in\mcl{F}(\mPH)$ has the form
\begin{equation}\label{1.3}
\frac{d}{dt}{\it f_t}({\bf x})\equiv\frac{d}{dt}{\it f}(\mph
t,h~{\bf x}) = \{{\it h,f}\}(\mph t,{\it h}~{\bf x}),\ {\bf
x}\in\mPH,\ t\in{\mbR},
\end{equation}
where the Poisson bracket is the unique extension of \eqref{(1.1)}
to mo\-re general real-valued functions $h,f,\dots$\ on \PH\  (cf.
Sec.~\ref{IIA;q-phsp}), and $f_t:=f\circ\mph t,{\it h}~,\ f:=f_0$.

The \emm formal transition from QM to NLQM~ consists (in our
``classical--like'' rewriting of QM)
in the addition to affine ``generators'' of QM of also the
nonli\-near ones. Such an infinite-dimensional  classical  mechanics  on
\PH\  is developed in Sections~\ref{sec;IIIA},~\ref{IIIA1;QM}.  Inclusion
of  these  nonlinear functions between generators implies, however, a
sequence of  problems for  quantum theory.}\hfill\zel \end{pt}
\def\nazov{{
\ref{results}\quad A Brief Description of the Contents}}
\begin{pt}[Evolutions and mixtures]\label{pt;evol-mixt}{\rm The basic
Wigner theorem (cf.\ Proposition~\ref{prop;3.4}) states that to any
bijective transformation $\mphi$ of \PH\  onto  itself  con\-serving the
``transition probabilities'', i.e.
\begin{equation}\label{1.4} Tr(P_{{\bf
x}}P_{{\bf y}})\equiv Tr(P_{\mphi {\bf x}}P_{\mphi{\bf y}}),
\end{equation}
there exists a unitary or antiunitary operator $U$ on \H\ such, that
\begin{equation}\label{1.5}
P_{\mphi{\bf x}}\equiv P_{U{\it x}}, {\rm
with}\ 0\neq x\in{\bf x},\ \forall\bx\in\mPH.
\end{equation}

The corresponding operators $U_t$
are unitary for  continuous  fami\-lies $t\mapsto \mphi_t, (\mphi_0 :=
id_{\mPH})$ of mappings $\mphi$ satisfying~\eqref{1.4}.  The unitary
operators in \eqref{1.5} are determined essentially uniquely by $\mphi$,
up to numerical factors. This means, that $\mphi$ from \eqref{1.4} with a
unitary $U$ uniquely determines a ${}^*$-automorphism $\alpha_{\mphi}$ of
the von Neumann algebra \LH\ of bounded operators on $\mH$.  Such an
automor\-phism, in turn, determines the dual mapping $\alpha_{\mphi}^*$
that affinely  and bijectively maps the space \Ss\  of all density matrices
onto itself and extends the mapping $\mphi: \alpha^*_{\mphi}{\bf x}\equiv
\mphi{\bf x}$. On the other  side, \cite[Theorem 3.2.8, Corolary 3.2.13,
Examples 3.2.14 and 3.2.35]{bra&rob}, if  a one parameter family $\mphi(t)$
of bijections of the pure  states  \PH\ onto itself can be extended by a
``sufficiently continuous'' family $t\mapsto \alpha^*_{\mphi(t)}$  of
affine bijections $\alpha^*_{\mphi(t)}$ of \Ss\  onto  itself,  then there
is a one-parameter family of \autm s $\alpha_{\mphi(t)}$  of \LH\
represented  by  unitary  operators $U(t)$  such  that $P_{\mphi(t){\bf x}}
\equiv U(t)P_{\bf x}U(t)^*$ . It can be shown \cite{cir} (cf.\ also
Proposition~\ref{prop;3.5}) that  the transformations $\mphi := \mphi_t^h
(t\in R)$ solving \eqref{1.3} satisfy \eqref{1.4} iff there is some $H =
H^*$ such that $h\equiv h_H$ . Hence, evolutions determi\-ned from
(\ref{1.3}) for nonlinear $h$ necessarily violate (\ref{1.4}), and \ph t,h~
cannot be (for all $t$) extended by affine mappings of \Ss\  onto it\-self.
This also means that $\mphi_t^h$ cannot be extended into a
transfor\-ma\-tion of density matrices $\mrh := \sum\lambda_jP_{{\bf
x}_j}=: \sum\lambda_j{\bf x}_j$, con\-ser\-ving affine combinations, i.e.
for any such extension ${\tilde{\mphi}}_t^h$ there is
\begin{equation}\label{time}
 \mpph t,h~\mrh\not\equiv\sum_j\lambda_j\mph t,h~{\bf x}_j.
\end{equation}
This has consequences described in Note~\ref{not;3.7}, and in
Interpretation~\ref{int;3.7a}, as well as in
Subsection~\ref{q-phsp;mixt}: An evolution $\tilde{\mphi}_t^h$ of
density matrices  cannot be expressed by ``the same'' evolution
\ph t,h~   of  pure components  of their decompositions. This has
several further consequences.}\hfill\zel \end{pt}
\begin{pt}[Emergence of nonlinear observables]\label{pt;why-obs}{\rm The
evolution \ph t,h~  (in the ``Heisenberg picture'') of affine ``observables''
$h_X$  does not lead
identically to affine observables, i.e.\ there are {\it no} such
one parameter sets of operators $X(t)^*\equiv X(t), X(0):=X$ that $h_X(\mph
t,h~{\bf y})\equiv h_{X(t)}({\bf y})$, for nonlinear $h$. Hence, inclusion
of nonlinear
gene\-rators implies necessity of inclusion of also ``nonlinear
observa\-bles'' into the theory. The probabilistic interpretation  of  such
observables is not possible in a traditional way, cf.
Interpretation~\ref{int;3.7a}. The interpretation inspired by ``mean--field
interpretation'' is  described in  Interpretation~\ref{int;2.27}, where the
expression of nonlinear functions $h\equiv h_{\frak f}, h_{\frak f}({\bf
x})  := Tr(P_{\bf x}{\frak f}(\mbF(\bf x)))$ with a help of conveniently
chosen operator-valued functions \fk f is used (see
Definition~\ref{df;2.17} for \bF). A restriction of  possible choices of
the functions $h$ , as well as of nonlinear  generators, can be determined
by a choice of the representation $U(G)$, cf.\ also
Definitions~\ref{df;2.19}, \ref{df;2.25a}--\ref{df;2.25d}.}\hfill\zel \end{pt}
\begin{pt}[Two kinds of mixtures]\label{pt;2mixt}{\rm Impossibility of a
unique extension of $\mph t,h~$ (determined by the function $h$ defined on
\PH\  only) to a mapping ${\bf \mphi}_t^h$ on \Ss\  leads to necessity of
investigation of a  natural  ``Poisson  structure'' and a consequent
definition of ${\bf \mphi}_t^h$  for ``Hamiltonian functions $h$'' defined
now on the whole \Ss , cf.\ Section~\ref{IIA;q-phsp}. This provides a
solution of problems arising in the earlier trials to formulate NLQM with
connection of evolution of mixed states, cf.\ also \cite{weinb,mobil}.
These facts lead also to  neces\-sity of distinction of two kinds of
``mixed states''  in  nonlinear extensions of QM. These are introduced in
Subsection~\ref{q-phsp;mixt}, and in Definition~\ref{df;2.25d}. The {\em
elementary mixtures} correspond to density matrices considered as points of
the {\em elementary phase space} \Ss; these elementary mixtures are
transformed by Poisson flows ${\bf \mphi}^h$ as points of \Ss,
independently of their possible convex decompositions.  Ano\-ther kind of
``mixed states'' is described by probability  measures $\mu$ on \Ss, which
are not
concentrated in one point: these are cal\-led the {\em genuine mixtures}
(corresponding to the term ``Gemenge'' used in~\cite{QTM}). Evolution of
states described by $\mu$'s is given by evolutions of points in the support
of $\mu$. This  offers, e.g., a possibility to distinguish between the
state described by an elementary mixture -- e.g.\ the density matrix $\mrh_I$ of
a subsystem I (obtained as the ``partial trace''~\cite{davies}) of a
composed system  I+II being as a whole in a pure state\footnote{Hamiltonian
evolutions of $\mrh_I$\ -- linear, or not -- are, however, rather rare
consequences of evolutions of the composed system $I+II$;
these evolutions should be rather specific in those cases.}
 belonging to the
manifold $P(\mcl{H}_I\otimes\mcl{H}_{II})$ on one side, and, on the other
side, a  state  with the same barycentre~\cite{bra&rob} $\mrh_I$ (expressed
now by a probability measure $\mu_I$\ on $\mS_{I*}$) obtained
after some ``reduction  of  the wave packet'', cf.\ \cite{neum1,w+z,bon-m}:
in the last case the different states occurring in the support of the measure
$\mu_I$\ of the microsystem $I$ are cor\-related with
macroscopically distinguishable states of the measu\-ring apparatus (usually
declared as ``pointer positions''); this correlation can be reflected
in  a de\-scription of states by genuine mixtures $\mu_I$.}\hfill\zel \end{pt}
\begin{pt}[Unbounded generators]\label{pt;unb-gener}{\rm Another (rather
``technical'', at first sight)
complication arising in our process of reformulation and extension of QM in
geometrical terms is connected with the neces\-sity of a use of unbounded
selfadjoint operators $X$ on the Hilbert space \cl H\ in QM. It is a generally
known mathematical theorem that such operators are defined on dense linear
subsets $D(X)$ of \cl H\ certainly different from the whole \cl H.
Hence, our  extension to nonlinear theory requires to use of also
(``linear'', or not) functions $f$ on \Ss\   in a r\^ole of generators
that are not defined everywhere on the corresponding manifold of quantum
states, and also are not locally bounded on \Ss ; such nonlinear $f$'s could
be  obtained,  e.g.\ as some nonlinear perturbations of the (only densely
defined, unbounded) function $h_X$ corresponding  to  unbounded $X$. The main
technical advantage of the use of the  representation $U(G)$ is that it
offers a possibility of definition of a class  of nonlinear unbounded
generators $h$ generating Poisson flows ${\bf\mphi}_t^h$ on \Ss\   that
extends the set of affine (unbounded) generators $h_X$ (the later generate
projections of the common unitary flows $U(t):=exp(-itX)$). This is done in
Section~\ref{IIB;gener} with a help of the ``macroscopic field \bF'', cf.
Definitions~\ref{df;2.17}~and~\ref{df;2.19}. The representation $U(G)$\
enters into the determination of the set of ``relevant generators $h$'';
taking part in determination of the ``considered physical system'' in this
way, the use of $U(G)$ has not only ``technical r\^ole'', but it has also a
``physical meaning''.}\hfill\zel \end{pt}
\begin{pt}[Structure of observables]\label{pt;str-obs}{\rm
Section~\ref{IIC;symm-obs} is devoted to definition of observables, to
investigation of their algebraic properties, and of their trans\-formation
groups. It is proposed, in the geometrical setting, to describe observables
by func\-tions ${\mfk{f}}: \mSs\times\mSs\rightarrow{\mbR},
(\mrh;\nu)\mapsto \hat{f}(\mrh,\nu)$ {\em of two variables}, the first one
is called {\em the quantum variable} and the function
${\hat{f}}(\cdot,\nu)$ is affine. The observables are related to the choice
of $U(G)$  that determines (cf.\ Definition~\ref{df;2.17},
Definition~\ref{df;2.19}, and Proposition~\ref{prop;2.24}) an affine
mapping $\mbF:\mcl D(\mbF) (\subset\mSs) \rightarrow\mcl E_{\mbF} (\subset
Lie(G)^*)$ descri\-bing a ``classical field''. The dependence of
observables \fk f on the second ``{\em macroscopic}'' (or ``classical'')
variable $\nu$ can be restricted to an ``indi\-rect dependence'',
i.e.\  ${\hat{f}}(\mrh,\nu)\equiv Tr(\mrh\mfk{f}(\mbF(\nu)))$ for some
operator-valued func\-tion \fk f on (a subset \cl E\ of) $Lie(G)^*$.
Restriction to such a type of dependece on the quantum states $\nu \in
\mSs$ provides a tool for dealing with the above mentioned (see
~\ref{pt;unb-gener}) unbounded functions. We see that a general type of
``quantum fields'' $\mfk{f}:\mcl{E}_{\mbF}\rarw\mLH$ enters naturally  into
the game, cf.\ Definitions~\ref{df;2.19}, \ref{df;2.25b}, as well as
Interpretation~\ref{int;2.27}.}\hfill\zel \end{pt}

\begin{pt}[Possible applications]\label{pt;appl}{\rm
The presented theory is still in a preliminary stage: Its mathematical form
is more elaborated than its possible physical interpretations. As a
consequence, we restrict our attention in this work to existing theories
and their incorporation into our conceptual scheme. We give here some general
technical procedures to approach solutions of nonlinear dynamical
(Schr\"odinger) equations (Section~\ref{sec;IIIE}). We propose also a general
mechanism for ``delinearizations'' of unitary group representations in
Proposition~\ref{prop;2.31}. A general interpretation scheme of EQM is
proposed, cf.\ e.g.\ Subsection~\ref{q-phsp;mixt},
Interpretation~\ref{int;obs},~\dref2.25d~, and
Interpretation~\ref{int;2.27}.

As concerns some proposals of {\em new applications} of the EQM
(in addition to all ones of QM), they could be found also {\em
without requiring a ``fundamental nonlinearity''} in laws of
Nature (i.e., now in QT). We consider here description of systems,
which can be considered as ``relatively closed'' subsystems of
larger (linear) QM systems. Such might be some ``mesoscopic
systems'' of large molecules, of ``trapped'' Bose--Einstein
condensates, etc. As concerns (non-)linearity of physical laws, it
can be suspected that pervasive scientific thinking is nowadays
``generally linear'': Even if dealing with nonlinear equations,
mappings or ``effects'', we express them eventually in terms of
linear spaces (real numbers, additive operations = commutative
groups, ``linearizations'' of different kinds, etc.). Linearity
seems to be one of the present time ``paradigms'' of our thinking.
As it is shown in several places of this work, any of considered
nonlinear theories can be extended to a linear theory ``of a
larger system'' (a generalized ``Koopmanism''). Hence, conversely,
we can expect nonlinear behaviour  by specific restrictions of
dynamics to subsystems. Possibilities of  various
interpreta\-tions of the presented general theoretical scheme of
EQM are left  open here for further development.}\hfill\zel\end{pt}

\begin{pt}[Notes on a Weinberg's proposal]\label{pt;weinb}{\rm
     In some papers,~\cite{weinb}, S. Weinberg posed a  question  on  a
possible nonlinear modification of QM (motivated by his aim to formulate a
way to testing fundamental principles of QM), and  sketched  a  specific
proposal of ``nonlinear quantum mechanics'' (NLQM). Trials  to
ob\-tain a consistent generalization of the  traditional
interpreta\-tion of QM to this theory led, however, to difficulties
connec\-ted mainly with the appearing lack of conservation of the
``tran\-sition probabilities''  under  nonlinear  transformations  in  QM.
There are difficulties with appearing possibility of superluminal
communication  via  Einsten-Podolsky-Rosen (EPR)-type
experi\-ments, difficulties with the statistical interpretation of  the
formalism (as will be shown in Subsection~\ref{IIIA1;QM}) and  also
difficulties
with description of composed systems.\footnote{There are, however, works
 devoted to search of some observable  deviations  from
the QM predicted by the Weinberg formulation of NLQM; in some of these
works also proposals for experimental tests of predictions of this formulation
of NLQM were given.} Weinberg's description of evolution of mixed states of
subsystems (it was basis dependent), as well as statistical interpretation
of predictions (it was based on an approximation motivated by KAM theory)
were even mathematically and conceptually ambiguous.

   We shall reformulate here NLQM in the  mathematically
unambi\-guous terms of symplectic reformulation of QM discovered some time
ago, cf.\ \cite{strocchi,bon4,PH,bon8,cir}  by extending it subsequently
by ``nonlinear
quanti\-ties''. This formulation admits the interpretation  suggested by a
specific formulation   of quantum mean-field  models: the given QM
system is considered as an individual subsystem  of  an  infinite
collection of equal quantum subsystems interacting mutually via a
very weak, long range, and permutation invariant interaction; its
dynamics can be described as a quantum dynamics of an  individual
subsystem moving in the time dependent ``external'' classical field
given by actual values of intensive quantities  of  the  infinite
system. Mathematical unambiguity of this MFT ensures such property
for our EQM. Since also more realistic interactions than that of MFT, e.g.
the Coulomb interaction, are ``of long range'' and lead in specific limits
to validity of a certain forms of MFT, cf.\ Thomas--Fermi
theory~\cite{thirr4}, one can expect existence of applications of the
presented theory in realistic situations.\nl
We shall return to a reformulation of a part of Weinberg's
theory in Section~\ref{sec;IIIF}.}\hfill\zel\end{pt}

   Let us note that we shall not present in this work any review of
   mean--field theory (MFT),
in spite of its (at least ``ideological'') importance for
understan\-ding of some constructions of the present paper,  as well  as  of
their proposed interpretation; for a brief review of MFT cf.,
e.g.,~\cite{bon3}, the
introductory sections of~\cite{bon2}, or
in~\cite{unner2,unner1,unner0}; cf.\ also Section~\ref{sec;IIIB}.\nl

\section{Remarks on the Text}\label{notes}

The text is divided into three Chapters, including this introductory one,
and of three appendices (numbered alphabetically) divided to (sub)sections.
The second chapter entitled ``Extended Quantum Mechanics'' contains the general
formal and interpretational scheme of the presented theory, the EQM. The last
one: ``Specifications and Applications'' contains a description of some more
specific theories which are included as subtheories into EQM.
Chapters are divided into sections, numbered separately in each chapter.
Subsections, formulas, and assertions (of all kinds, consecutively, including
Definitions, Theorems, Remarks, Interpretations, some unnamed paragraphs, etc.)
are numbered within each Section separately. For better orientation at
reading, the end of text of Theorems, Propositions and Lemmas is denoted by
$\clubsuit$, end of Definitions and Notations is denoted by $\diamondsuit$, and
that of
Interpretations by $\blacklozenge$; Notes, Remarks, Illustrations and Examples
are
finished by $\heartsuit$, and some other unnamed numbered paragraphs are ended
by the sign $\spadesuit$.

The appendices are written in a language, which is not always strictly
rigorous from mathematical point of view, what is due to the author's desire to
 make the mathematical text easier to read for more readers. The
contents of (sub)sections is briefly seen from the Table of contents.
Phrases and formulas typed {\bf boldface} are usually newly defined
expressions. The bibliography is far from complete; this is also due to
many sources and connections of EQM.

The text is written as a {\em physically motivated mathematical model}
intended, however, to provide a framework for solution of actual physical
problems.
Hence, it is not quite {\em physically neutral} as a purely mathematical
text, perhaps, should be. There are included paragraphs denoted by
``Interpretation'' containing some of these author's ideas and proposals,
but also some (perhaps) generally accepted parts of quantum theory (QT).

We did not try to use some ``up to date mathematics'', and
the level is ``slightly graduate''. Appendices might help readers to
refresh some mathematical concepts and facts.
They contain
some technical prerequisites on topology, differential calculus on Banach
spaces and differential geometry (also on Banach manifolds), on Lie groups,
basic facts on \Ca s, and \Wa s, and their representations and
automorphisms (i.e.\ symmetries), as well as a brief information
on unbounded symmetric
operators and their symmetric and selfadjoint extensions.
The appendices can serve, together with Sections~\ref{I;clmech},
and~\ref{I;qmech} presenting briefly general schemes of CM and QM, to fix
notation, and also to pedagogical purposes (independently of
Chapters~\ref{sec;II}, and~\ref{chap;III}).\nl

The given scheme of EQM contains also Hamiltonian
classical mechanics (CM) as a subtheory
in an obvious way, as it is mentioned in the paragraph~\ref{pt;CM}. In
Section~\ref{III;spec} also other subtheories and some invented
applications of EQM are listed, and an ``itinerary'' of the
Chapter~\ref{chap;III}\ is there given. It is not mentioned there a
possibility of
an application to a formulation of connections of the theory of general
relativity with QT, since the
present author is not acquainted with the actual status of these
problems.\footnote{It might be assumed that works by, e.g.\ C.J. Isham and/or A.
Ashtekar contain relevant attempts of this kind.}
Equally it is not discussed a hypothesis on possible application of methods
close to the
presented ones to the ``algebraic quantum field theory'' (QFT): Let us just
mention that a ``self-consistent approach'' could be, perhaps, useful in dealing
with such classical objects like ``domains in Minkowski space'' in a
framework of any quantum theory.

We were not intended to criticize here in details the Weinberg`s formulation
of a nonlinear modification of QM; some relevant criticism was presented in
published papers, e.g.\ in~\cite{noncausal,mobil}.
The Section~\ref{sec;IIIF} is devoted to just a reformulation of our NLQM on
\PH\ in terms close to those used in the Weinberg's paper~\cite{weinb}. This
allows us to compare in mathematically clear terms the two approaches to a
generalization of QM, which might be considered (up to the used
interpretations) practically identical on the set of vector states, resp.
on \PH. Some useful algorithms for solution of these nonlinear
Schr\"odinger equations might be found in Section~\ref{sec;IIIE}.
A reduction of solutions of a class of nonlinear
Schr\"odinger equations connected with a group action on \PH\
to two ``simpler'' problems: to solutions of
classical Hamilton's equations (possibly, finite dimensional), and to
solution of a linear time--dependent Schr\"odinger equation is described in
that Section~\ref{sec;IIIE}.

Other theories described here as subtheories of EQM entered to NLQM as
``approximate theories'' to problems of linear QM: It might be rather
interesting how nonlinearities enter into approximated linear theories of
QM. We shall present, e.g., (partly elaborated) cases of time--dependent
Hartree-Fock
theory in~\ref{IIIA1;H-F}, and a class of nonlinear Schr\"odinger equations
known also from
traditional attempts to formulate nonlinear modifications of QM, cf.
Subsection~\ref{IIIA1;NL-Sch}.

A connection of EQM with ``quantum theory of large systems'' (i.e.\ with a
class of nonrelativistic QFT) is sketched briefly in
Section~\ref{sec;IIIB}. This connection seems to us crucial from the
interpretational point of view, since the presented EQM appears (in a slightly
different form) as a well formulated {\em linear QT of large quantal
systems}. Such a linear QT contains also classical macroscopic observables in
a natural way, as a {\em consequence of local quantum kinematics}, where
a specific r\^ole of symmetry groups and a ``mean--field'' dynamics can be
introduced, to point out those of the obtained (unnecessary huge) set of
``observables''
which are interpretable, hence ``useful''.

At the beginning of the Chapter~\ref{sec;II}, in
Subsections~\ref{q-phsp;basic}, and~\ref{q-phsp;manif} of
Section~\ref{IIA;q-phsp}, the mainly ``kinematical structure'' of the
theory is described, whereas the next two Subsections~\ref{q-phsp;poiss},
and~\ref{q-phsp;flows} describe the way of constructing ``dynamics'', and
also more general one--parameter symmetry groups.
Only bounded and differentiable, hence ``nice'' objects are considered in
details in these subsections.
The following Sections~\ref{IIB;gener}, and~\ref{IIC;symm-obs} consist,
perhaps, the most technical parts of the paper containing also important
interpretation proposals. They contain both a
solution (and some hints for alternatives) of the technical problem of
dealing with unbounded nonlinear generators (``Hamiltonians''), as well as
definitions and interpretation proposals for ``observables''.
The Section~\ref{IIC;symm-obs} contains the basic definitions of a variety
of described (sub--)systems, and also a description of ``nonlinear
realizations'' of symmetry Lie groups.

Before starting with a description of tools for our generalization
of QM to EQM, let us, however, present in the next two sections
brief reviews of traditional CM, and also of QM, a knowledge and
understanding of which is a necessary prerequisite for successful
reading of Chapter~\ref{sec;II}.



\def\nazov{{
\ref{notes}\quad Remarks on the Text}}
\section{A General Scheme of Hamiltonian Classical Mechanics}
\label{I;clmech}

We present a brief review of geometric formulation of classical mechanics
in this section.
The presented scheme is standard~\cite{abr&mars} and represents an important
part of intuitive and technical background for our subsequent constructions.
The language used will be that of a simple version of global differential
geometry: We want to avoid as much as possible a use of coordinates for sake
of transparency and formal and conceptual simplicity; this will be our
``policy'' in all the following text. Some review of a necessary minimum of
mathematical background is presented in the Appendix~\ref{A;geom}.

\subsection{Classical phase space and dynamics}\label{clmech;phsp}
Let us first mention basic general concepts, and subsequently  some examples
will be given.

The space of \emm classical ``pure states''~ in a model of Hamiltonian
mechanics, i.e.
the \emm phase space \bs{(M;\Omega)}~, is a
differentiable manifold $M$ of finite (even) dimension endowed with a
symplectic (i.e.\ nondegenerate and closed) two-form $\Omega$. The
specification of the form $\Omega$ is equivalent to a specification of a
nondegenerate Poisson
structure on $M$, i.e.\ to definition of \emm Poisson brackets~ $\{f,h\}$ on
the set
$\mcl F(M)\ (\ni f,h)$ of infinitely differentiable real valued functions
on $M$.

The equivalence between Poisson and symplectic
structures on a (symplectic) manifold is only the case, however, of a
\emm nondegenerate Poisson structure~ , i.e.
that one satisfying all the five following defining properties:

\begin{defs}[\emm Poisson structure~]\label{df;P-strct}
Let $M$ be a differentiable manifold, and let a mapping $\{\cdot,\cdot\}:
\mF(M)\times\mF(M)\rarw\mF(M)$\ be given. Assume the following properties of
 $\{\cdot,\cdot\}$:
 \item{  (i)}  $\{\cdot,\cdot\}$ is \emm antisymmetric~: $\{f,h\}\equiv
-\{h,f\}$;
 \item{ (ii)}  $\{\cdot,\cdot\}$ is \emm bilinear~: $\{f,h_1+\lambda h_2\}
\equiv
  \{f,h_1\}+\lambda \{f,h_2\} $;
 \item{(iii)}  $\{f,\cdot\}$ is, for any fixed $f\in\mF(M)$, a
 \emm derivation~: $\{f,h_1h_2\}\equiv\{f,h_1\}h_2+h_1\{f,h_2\}$;
 \item{ (iv)} {\emm Jacobi identity~}:
 $\{h_1,\{h_2,h_3\}\} +\{h_3,\{h_1,h_2\}\} +\{h_2,\{h_3,h_1\}\}=0$
 is fulfilled;
 \item{  (v)}  $\{\cdot,\cdot\}$ is \emm nondegenerate~: If, for a fixed $f\in
 \mF(M)$, there is $\{f,h\}\equiv 0$ for all $h\in \mF(M)$, then $f\equiv
const.$ on
 each connected component of $M$.%

 If $\{\cdot,\cdot\}$ satisfies first four properties (i) - (iv), then it
 is called a \emm Poisson structure~ on $M$.

A manifold $M$ endowed with a Poisson structure is called a \emm Poisson
manifold~,~\cite{weinst}.

 Relation of a general
Poisson manifold $M$ to its canonically determined symplectic submanifolds is
such that $M$ decomposes
uniquely to union of disjoint manifolds $M_{\iota}$ each of them is endowed
with a
uniquely defined symplectic structure $\Omega^{(\iota)}$\ determined by the
Poisson
structure $\{\cdot,\cdot\}$, and canonically determining it on
corresponding $M_\iota$. The dimensions of the \emm symplectic leaves \bs
{M_{\iota}}~
might be mutually different. Any $h\in \mF(M)$ determines a unique
\emm Hamiltonian vector field~ $\bv_h$ on the whole $M$ by the formula
\begin{equation}\label{eq;ham-poiss}
df(\bv_h) \equiv \bv_h(f) := \{h,f\},\ \text{ for\ all}\ f\in\mF(M).
\end{equation}
\noidt The same formula can be obtained for a symplectic manifold by
combining~\eqref{eq;ham-poiss} with~\eqref{eq;ham-vect}.
Corresponding Hamiltonian flows leave each the symplectic leaf $M_\iota$\
invariant.
 \hfill\pika  \end{defs}

This allows us to ascribe to each function $h\in\mcl{F}(M)$ a
unique (local) flow \ph {},h~  on $M$ representing solutions of Hamilton's
dynamical equations
\begin{equation}\label{eq;ham-eq}
\frac{df_t}{dt}= \{h,f_t\},\ \ \text{ with}\ f_t(x):= f(\mph t,h~x),\ t\in\mbR,\
x\in M,
\end{equation}
\noindent with the Hamiltonian function $h$: for the initial state
$x(0):=x\in M$ the state in a time $t\in\mbR$ is expressed by
$x(t)=\mph t,h~x$. This is done in the following way:
The symplectic form $\Omega$ determines the {\em
Hamiltonian vector field}\ind{vector field, Hamiltonian} $\bv_h$ on the phase
space $M$ corresponding to an arbitrary differentiable function
$h\in\mF(M):=C^\infty(M,\mbR)$, by the formula
\begin{equation}\label{eq;ham-vect}
\Omega_x(\bv_h,\bw):= -d_xh(\bw), \text{ for\ all}\ x\in M, \bw\in T_xM.
\end{equation}
\noidt Then the {\emm Poisson bracket~} is defined by
\begin{equation}\label{eq;ham-Poiss}
  \{f,h\}:=\Omega(\bv_f,\bv_h),\ f,h,\in\mF(M),
\end{equation}
and the right hand side of the equation~\eqref{eq;ham-eq} is just $\bv_h(f_t)$.
The solutions $x(t)=\mph t,h~x$ of~\eqref{eq;ham-eq} needn't exist for
all times $t\in\mbR$ for any initial condition $x\in M$,\ and\ $\mph {},h~$
represents in general just a collection of {\em local flows}\ind{flow,
local}. If $\mph {},h~$ exists for all $t$ on $M$, it is called the (global
Hamiltonian) \emm flow of the vector field~ $\bv_h$.
A vector field with global flow is called \emm complete vector field~.
General criteria for deciding what Hamiltonian $h$ on a
given $(M;\Omega)$ has complete vector field $\bv_h$ are not known, although
some criteria are known for specific classes of (possibly symplectic)
manifolds; especially, on
compact $M$ all vector fields are complete. Let us note that
``completeness'' of a Hamiltonian vector field of $h$ on a dense invariant
subset
of $M$ is equivalent~\cite[2.6.14, and 2.6.15]{abr&mars} to essential
(anti-)selfadjointness of a densely defined linear operator on the complex
Hilbert space $\mH:=L^2(M,d\mu_{\Omega})(\ni f)$, cf.\ Appendix~\ref{C;oper}.
Here the measure $\mu_{\Omega}$ used in the definition of the square
integrability in the Hilbert space \H\ is the $n$-th power of $\Omega$, cf.
Appendix~\ref{A;manif}, if dim$(M)=2n$:
\begin{equation}\label{eq;ham-meas}
\mu_{\Omega}(\Lambda)=\int_{\Lambda}\wedge^n\Omega.
\end{equation}
\def\nazov{{
\ref{I;clmech}\quad A General Scheme of Hamiltonian Classical
Mechanics}}

The mentioned antiselfadjoint operator acts on differentiable functions
$f\in\mH$ as the differential operator determined by the vector field
$\bv_h$:
\[ f\mapsto \bv_h(f) :=df(\bv_h).\]

 A \emm symplectic transformation~ of $(M;\Omega)$ is a diffeomorphism $\mphi$
of $M$ onto itself conserving the form $\Omega$, i.e.:
$\mphi^*\Omega\equiv\Omega$. Hamiltonian flows are one-parameter groups of
symplectic transformations (hence, they conserve the measure~\eqref{eq;ham-meas}
- this is the \emm Liouville theorem~ used in classical statistical
mechanics).
Conversely, each one-parameter group of symplectic transformations defines its
(at least local -- in open neighbourhoods
of all points of $M$) Hamiltonian function generating the given
flow~\cite{abr&mars,arn1}. Any symplectic transformation can be considered
as a (kinematical) symmetry of the considered classical system. If the
dynamics is described by the Hamiltonian $h$ with the flow $\mph {},h~$,\ and a
symmetry one-parameter group is described by the flow \ph {},f~
 corresponding to its ``Hamiltonian'' $f$,\ and if, moreover, the Poisson
bracket of the corresponding Hamiltonians vanishes: $\{f,h\}=0$, then the
two flows mutually commute:
\[ \mph t,h~ \circ \mph s,f~ \equiv \mph s,f~ \circ \mph t,h~. \]
In this case, the function $f$ represents an
{\emm integral of motion~}, resp.\ a {\emm conserving quantity~}
of the system,
 cf.\ eq.~\eqref{eq;ham-eq}. If there is a Lie group $G$\ (cf.\ref{A;LieG})
 acting on $M$ \emm transitively~ (i.e.\ for any $x,y\in M
\ \text{ there is a}\ g\in G$ such, that
its action transforms $x$ to $y$) by symplectic transformations, the phase
space $M$ is called a {\emm symplectic homogeneous space of $G$~}.

Let us give now some simple examples:
\begin{exmp}\label{exmp;ham}
\item{(i)} The linear space $M:=\mbR^{2n}$ of $2n-$tuples of Cartesian
coordinates $(q_1,\dots q_n,p_1,\dots p_n)$ is endowed with the symplectic form
$\Omega := \sum_{j=1}^n dp_j\wedge dq_j$. The Poisson bracket is in the
given coordinates expressed in the standard form
\begin{equation}\label{eq;ham-PB}
\{f,h\}=\sum_{j=1}^n \left(\frac{\partial f}{\partial p_j} \frac{\partial
h}{\partial q_j} - \frac{\partial h}{\partial p_j} \frac{\partial
f}{\partial q_j}\right).
\end{equation}
\noidt Symmetries of this space contain linear symplectic transformations
described by $2n\times 2n$ matrices commuting with the matrix $S$ with
elements (in the considered ``canonical'' basis\ind{canonical basis})
$S_{j,k}\equiv 0$, except of $S_{j,j+n}\equiv -S_{j+n,j}=1\ (j=1,2,\dots n)$,
but also affine
transformations consisting of arbitrary parallel shifts $\mphi: x\mapsto
\mphi (x)\equiv x+a$, for any fixed $a\in \mbR^{2n}$. Symmetries are, of
course, all the symplectomorphism of the form \ph t,h~  (the above mentioned
linear
transformations, as well as affine ones, are also of this form; e.g.\ shifts
are generated by linear $h(q,p)\equiv \sum_{j=1}^n (c_j q_j + d_j p_j)$;
 quadratic $h$'s correspond to groups of linear transformations). Let us
 mention
 explicitly, that specific quadratic $h$'s describe the dynamics of
 ``harmonic oscillators''\ind{harmonic oscillators}, whereas those $h$'s
 which contain (in their Taylor expansion, e.g.) terms of higher than the
 second order in  the standard canonical coordinates $(p;q)$ lead to
 nonlinear canonical flows on $M$.
\item{(ii)} The complex projective space $CP^n := P(\mbC^{n+1})$ constructed
from the linear space $\mbC^{n+1}$ as the factor-space consisting of its
one-dimensional complex subspaces can be considered as $2n$-dimensional
real manifold endowed with a canonical symplectic structure~\cite[Appendix
3]{arn1}. This is a special case of complex projective Hilbert spaces \PH\
considered in Section~\ref{sec;IIIA}, and finite dimensional examples in
specific
charts can be straightforwardly constructed.
\item{(iii)} Cotangent bundles: Let $Q$ be any differentiable manifold and
$M:=T^*Q\equiv T^0_1Q$ be its cotangent bundle, cf.\ also
Appendix~\ref{A;manif}. Hence, points of $M$ are
linear functionals $p\in T_q^*Q:=(T_qQ)^*$ ``attached to points'' $q\in
Q$; the natural projection $\tau:T^*Q\rarw Q$\ maps $p\in T^*_qQ$\ to
$\tau(p)=q\in Q$. The derivative (i.e.\ the tangent mapping) of $\tau$\ is
\[ \tau_*:=T\tau:TM:=T(T^*Q)\rarw TQ.\]
The \emm canonical one form \bs{\vartheta}~ on the cotangent bundle $M=T^*Q$\
is defined by:
\bequ\label{eq;cot1}
\begin{split}
 \vartheta: p(\in M)\mapsto & \vartheta_p\in T_p^*M,\\
 \vartheta_p:\bv (\in T_pM)\mapsto & \vartheta_p(\bv):=p\circ\tau_*(\bv).
\end{split}
\end{equation}
Then $\Omega:= \rd\vartheta$\ is a symplectic form on $M$, the \emm
canonical symplectic form on \bs{T^*Q}~. If $\{q_1,q_2,\dots,q_n\}$\ are
local coordinates on $Q$, then $p\in T^*Q=M$\ is expressed (in the
corresponding chart on $M$) as $p\equiv\sum_{j=1}^np_j\rd q_j$. In this
coordinate \nbhd one has
\[ \vartheta = \sum_{j=1}^n p_j\rd q_j\circ\tau_*\equiv \sum_{j=1}^n
p_j\tau^*\rd q_j ,\]
and from commutativity of pull--backs with exterior differentiation \rd,
and from the basic property $\rd\circ\rd\equiv 0$, we
have the canonical expression for $\Omega$ in that \nbhd:
\[ \Omega:=\rd\vartheta= \sum_{j=1}^n \rd p_j\wedge\tau^*\rd q_j.\]
Hence, any cotangent bundle is a symplectic manifold in a canonical way.
Taking $Q:=\mbR^n$, we obtain the example (i), where the coordinates
$\{q_j,p_k\}\in\mbR^{2n}$\ can be chosen global (corresponding, e.g.\ to a
trivial coordinate (linear) chart on $Q=\mbR^n$).
\hfill\dovi
\end{exmp}
\begin{rem}[On the notion of ``chaos'']
 The Liouville theorem on noncontractibility of the phase
 volume, cf.\eqref{eq;ham-meas} and  the text following it,
 implies nonexistence of \emm attractors~,~\cite{abr&mars},
 of Hamiltonian flows. The attractors, especially so called \emm ``strange
 attractors''~,~\cite{str-atr}, are usually connected with the
 notion of \emm chaos~,~\cite{arn&avez,verh,gutzw}, in dynamical systems.
 This does not mean
 that in Hamiltonian systems does not occur a chaotic motion. The
 ``chaoticity'' of motion is characterized rather by its instability \wrt
 choices of initial conditions than by presence of some attractors. Such
 instabilities seem to occur generically in Hamiltonian systems. This fact
 remained hidden for most of physicists for several decades:
 Mainly so called \emm
 \bf{(completely) integrable systems}\index{completely integrable systems} were described in university
 textbook literature only:
 These are, roughly speaking, systems the dynamics of which can be fully
 described on
 surfaces of given values of integrals of motion, in conveniently chosen
 coordinates, as systems of independent linear harmonic oscillators;
 parameters of the oscillators might depend on values of the integrals of
 motion; the ``integrals-of-motion surfaces'' decompose the energy submanifolds
 and all they are
 diffeomorphic to tori $T^n$, or to cylinders,~\cite{arn1,arn2,abr&mars}.
 This was, perhaps, due to the fact that
  all known explicitly solved ($\equiv$ integrated) models were of this
 kind.\footnote{This seems to be generally believed, cf.\ also~\cite{arn1}.}
  It was proved~\cite{markus&meyer}, however, that the set of integrable
 systems is in a well defined sense rare in the set of all possible
 Hamiltonian systems. In the cited paper~\cite{markus&meyer} no
 restrictions to dynamics coming, e.g.\ from observed symmetries of physical
 systems were  considered; such restrictions could, perhaps, enlarge
 the ``relative
 size'' of integrable systems. But, on the other hand, some ``physically
 realistic'' systems in classical mechanics were proved to be {\emm
 nonintegrable~}, e.g.\ the three (and more) body problem
 in celestial mechanics (i.e.\ in the nonrelativistic model of planetary
 systems with point masses moving in $\mbR^3$\ and interacting via the
 Newton potential) is nonintegrable,~\cite{abr&mars}.\hfill\dovi
 \end{rem}

\subsection{Observables and states in classical
mechanics}\label{clmech;obs&st}

Also CM can be formulated in terms familiar from QM. This formal analogy is
useful for description of classical subsystems in the quantummechanical
framework. Concepts introduced in this subsection are useful also in
formulation of classical statistical
mechanics, see e.g.~\cite{landau5,uhl&ford,ruelle1,ruelle2}.

As a set of classical observables can be chosen, e.g.\ the \Ca\
(without unit, if $M$ is not compact) $C_0(M)$\ of all complex--valued bounded
continuous functions
on the phase space $M$\ tending to zero at infinity, cf.
Appendix~\ref{B;Ca}. This \Ca\ can be
completed by unit ($:=\mbI\equiv1=$identically unit function on $M$), and
this completion will be called the \emm \bs{C^*-}algebra of classical
observables~,
denoted by \glss \Ac~.\footnote{For the concepts and properties of
\Ca s and von Neumann (resp.\ $W^*-$)
algebras see the standard books~\cite{dix1,dix2,sak1,sak2,takesI,bra&rob},
and also our Appendix~\ref{B;Ca}.} The algebraic operations are defined
pointwise on $M$: for
$f,h\in\mAc$\ one has $(f\dti h)(m)\equiv f(m)h(m),\ (f+\mlam h)(m)\equiv
f(m)+\mlam h(m)$, $f^*(m)\equiv\overline{f(m)}$, and the norm is the
supremal one, i.e.
$\|f\|:=\sup\{|f(m)|:m\in M\}$. The spectrum space of \Ac\ is just the
one--point compactification of $M$. Further extensions of the algebra of
observables \Ac\ could lead us to abelian von Neumann algebras: Let, e.g.
the Borel measure $\mu_\Omega$ on $M$\ be given, and consider the Banach space
$L^1(M,\mu_\Omega)$ of integrable complex--valued Borel functions $f$\ on
$M$, with the norm $\mN f,1~:=\mu_\Omega(|f|)\equiv\int
|f(m)|\mu_\Omega(\rd m)$. Its topological dual, cf.\ \cite{R&S,bourb;vect},
$L^\infty(M,\mu_\Omega)$\ consisting of $\mu_\Omega$--essentially bounded
Borel functions on $M$ is a $W^*$--algebra containing \Ac. It can be
interpreted as a maximal commutative von Neumann algebra of bounded
operators in \LH, namely the operators of $M$--pointwise multiplication
by functions
$f\in L^\infty(M,\mu_\Omega)$\ of elements of the Hilbert space
$\mH:=L^2(M,\mu_\Omega)$. The mentioned duality is realized by the
sesquilinear relation
\bequ
\lb f;h\rb\equiv\int\overline{f(m)}h(m)\mu_\Omega(\rd m),\ \forall f\in
L^\infty(M,\mu_\Omega),h\in L^1(M,\mu_\Omega).
\end{equation}
This last definition of a (complexified, linear) set of ``classical
observables'' as a \emm {$W^*$--algebra of observables}~ has an
advantage that this \Ca\ contains also projections in \LH\ represented by
multiplication operators by characteristic functions of the Borel subsets
of $M$, by which is it generated.
Hence, (also unbounded) ``observables $f$'' could be defined
alternatively by \emm projection--valued measures~ $E_f$\ (with values in
multiplication projections in $\mcl L\bigl(L^2(M,\mu_\Omega)\bigr)$)
on Borel subsets $\mcl B(\mbR)$\ of \bR:
\[ E_f: B (\in\mcl B(\mbR))\mapsto E_f(B):=\chi_{f^{-1}(B)}\in\mLH, \]
with the characteristic function $\chi_\Lambda$\ of a Borel set
$\Lambda:=f^{-1}(B)\subset M$\ considered as an element of
$L^\infty(M,\mu_\Omega)\subset\mLH$.

The (mathematically defined,~\cite{dix2,sak1,bra&rob}) {\bf(classical) states}\index{classical states}
$\mcl S(\mAc)$\ on the \Ca\ \Ac\ are just the
probability measures $\mu\in\mcl M_{+1}(M)$\ on $M$, and the {\bf
(classical) pure states} \index{classical pure states} are all the the Dirac measures $\{\delta_m:m\in
M\cup\{\infty\}\}$, with $\delta_m(\Lambda)=1\eequiv m\in\Lambda$:
\[ \mu:f(\in\mAc)\mapsto \mu(f)(\in\mbC),\ \mu(f):=\int f(m)\mu(\rd m).\]
If one takes, on the other side, the \Wa\ $L^\infty(M,\mu_\Omega)$
as the \Ca\ of ob\-ser\-va\-bles , the set of all states on it will be ``much
larger'' than $\mcl S(\mAc)$ (which is included there as a proper subset),
but the normal states on $L^\infty(M,\mu_\Omega)$ restricted to the
subalgebra \Ac\ are just measures in $\mcl M_{+1}(M)$\ represented by
elements of $L^1(M,\mu_\Omega)$, i.e.\ just the \emm measures absolutely
continuous~ \wrt $\mu_\Omega$.

\begin{intpn}\label{intpn;cl-ensemb}
In any case, the Dirac measures $\delta_m, m\in M$, represent ``pure
states'', resp.\ in mathematical language, the extremal points of the convex
set of all Borel probability measures on $M$. Other probability measures of
this set have nontrivial, but {\bf unique decompositions} into the extremal
Dirac
measures. Their physical interpretation is probabilistic, in the sense of
statistical ensembles of Gibbsian statistical
mechanics,~\cite{hill,gugg,landau5}:
In the ensemble described
by a measure $\mu\in\mcl M_{+1}(M)$, the fraction of otherwise equal
physical systems having pure (=``microscopic'', but classical) states
represented by points in
the Borel subset $\Lambda$\ of the phase space $M$\ is equal to
$\mu(\Lambda)$. This interpretation is conceptually consistent, due to the
uniqueness of decomposition of $\mu$'s into the extremal points. This point
hides an essential difference between CM and QM: $\mcl M_{+1}(M)$ is a \emm
simplex~, what is not the case of the state space \Ss\ (or of \cl S)
of QM.\hfill\bpika
\end{intpn}

\subsection{Symplectic structure on coadjoint
orbits}\label{clmech;coorb}

We shall mainly restrict our attention to such classical phase spaces $M$ in
this work, which are homogeneous spaces of a connected, simply connected Lie
group $G$, on which the action
$g:m\mapsto g\dti m\ (g\in G, m\in M)$\ of $G$\ consists of symplectomorphisms:
\[ f_g(m):=f(g\dti m),\ \forall f,h\in C_\infty(M,\mbR):
\{f_g,h_g\}\equiv\{f,h\}_g.\]
In these cases, the phase space $(M;\Omega)$\ is (locally) symplectomorphic to
an orbit of the coadjoint representation (see Section~\ref{A;LieG}, and
below in this subsection) either of $G$, or of its central
extension by the additive Lie group \bR, cf.\
\cite[\S 15.2, Theorem 1]{kiril}.

Any (noncommutative) Lie group provides a canonical example of Poisson
manifold. Also the most common case of the $2n$-dimensional symplectic
linear space of the Example~\ref{exmp;ham}(i) can be considered as coming
in this way from the $2n+1$-dimensional Weyl-Heisenberg group. This will be
described in Subsection~\ref{IIIA1;CCR}.
Let us describe here the general construction.

Let $G$ be a finite dimensional connected (for simplicity) Lie group with
its Lie algebra $\mfk g := Lie(G)$, and with the exponential mapping
$\exp:\mfk g\rarw G, \xi\mapsto \exp(\xi)$. The canonical symplectic
manifolds will be found in the dual space $\mfk g^*$ of $\mfk g$. The
duality will be alternatively denoted by $F(\xi) \equiv <F;\xi>$\glo{$F(\xi)
\equiv <F;\xi>$},\ $F\in\mfk g^*$.\label{Fxi}
 The adjoint and the coadjoint representations of $G$ on its Lie algebra
 \fk g\ (resp.\ on its dual $\mfk g^*$) are defined in\dref2Ad~.\nl

 Let us fix any element $F\in\mfk g^*$. Then the subset (a
submanifold) \OGF\ of the linear space $\mfk g^*$\ defined by
\[ \mOGF:=\{F'\in\mfk g^*: \exists g'\in G, F'=Ad^*(g')F\} \]
is called the \emm {coadjoint orbit of $G$\ through F}~. The space $\mfk
g^*$\ is decomposed into coadjoint orbits of (in general) various
dimensions (as submanifolds).\nl

 Let us consider $\mfk g^*$\ as differentiable manifold in which,
as in any linear space, the tangent space $T_F\mfk g^*$\ in any of its
points $F$\ is canonically identified with the linear space $\mfk g^*$\
itself. The dual space $T^*_F\mfk g^*$\ then contains canonically (resp.\ for
finite
dimensional $G$: is identified with) the Lie algebra \fk g, which is
$w^*$--dense (i.e.\ $\msg(\mfk g^{**},\mfk g^*)$--dense) in the second dual
$\mfk g^{**}$\ of the Lie algebra $Lie(G)$, cf.\ \cite[Chap.IV,\S
5.1]{bourb;vect}. This allows us to define canonically a Lie algebra structure
on the second dual $\mfk g^{**}$. Let us denote this structure again by the
bracket $[\cdot,\cdot]$. Let $f,h\in C^\infty(\mfk g^*,\mbR)$. Then their
differentials $\rd_Ff,\dots$, are elements of $T^*_F\mfk g^*\sim\mfk g^{**}$,
and their commutator (i.e.\ the canonical Lie bracket) is defined. Then we
define the \emm {Poisson structure on $\mfk g^*$}~ by
\bequ\label{eq;coAd-poiss}
\{f,h\}(F):= - \lb F;[\rd_Ff,\rd_Fh]\rb,\ \forall F\in\mfk g^*,\ f,h\in
C^\infty(\mfk g^*,\mbR).
\end{equation}
The Hamiltonian vector fields $\bv f, \bv_h,\dots$, cf.\rref2.7~,
are then tangent to all the orbits
\OGF,~\cite{kiril,weinst}. \nl

The simplest examples of functions $f\in C^\infty(\mfk g^*,\mbR)$\ are
$f\equiv f_\xi, \xi\in\mfk g$, defined by $f_\xi(F):=F(\xi)\equiv\lb
F;\xi\rb$. Their Poisson brackets are trivially
\bequ\label{eq;poiss-basis}
\{f_\xi,f_\eta\}=-f_{[\xi,\eta]}.
\end{equation}
The functions $f_\xi$\ generate, if used as Hamiltonian functions, the
actions of one--dimen\-sio\-nal subgroups in the $Ad^*(G)$-representation, i.e.
the Hamiltonian flow of $f_\xi$\ on $\mfk g^*$\ is
\bequ\label{eq;basis-flow}
\mph t,{f_\xi}~F\equiv Ad^*(\exp(t\xi))F,\ \forall F\in\mfk g^*,\ \xi\in\mfk
g,\ t\in\mbR.
\end{equation}
\begin{exm}
Let us give a simple example of coadjoint orbits of a Lie group. Let
$G:=SU(2)$, the covering group of the rotation group $SO(3)$. These are
3--dimensional Lie groups with the Lie algebra generated by elements
$\xi_j, j=1,2,3$, corresponding to one parameter groups of rotations around
tree fixed mutually orthogonal axes, and satisfying the relations
(with the summation convention)
\[ [\xi_j,\xi_k]=\epsilon_{jkl}\xi_l, \meps_{jkl}\equiv
-\meps_{kjl}\equiv\meps_{klj},\meps_{123}:=1.\]
Then it is possible to show, that the coadjoint orbits (in the dual basis to
$\{\xi_j\}$) are just all the spheres centered at origin. Hence, in this
simple case, all the (symplectic) orbits \OGF\ are two--dimensional except
of their common centre, which is a unique zero--dimensional orbit. The
flows corresponding to the generators $f_{\xi_j}$ are just rotations around
the chosen axes in $\mfk {so}(3)^*$.\hfill\dovi
\end{exm}

\section{Basic Concepts of Quantum Mechanics}\label{I;qmech}
\def\nazov{{
\ref{I;qmech}\quad Basic Concepts of Quantum Mechanics}}

We shall give here a review of an abstract scheme of standard quantum
mechanics used for description of such systems, ``classical analogs'' (or
``classical limits'') of which
are described by CM with finite--dimensional phase spaces.

The basic intuition and terminology of QM comes from CM (supplemented with a
``nonclassical'' statistical interpretation). This is due to
the history of physics, but also, on  more fundamental level, due to the
intuitive
necessity to express empirical statements of QM (as well as of an arbitrary
theory) in terms describing macroscopic
bodies of everyday life, or in terms of (again macroscopic)
laboratory instruments. And states of macroscopic systems (resp.\ ``macroscopic
parameters'' of physical systems) are described by classical concepts.
Mathematical formalism of QM in its traditional form looks, however, rather
different from that of CM.
It will be shown in later sections of this work, in what aspects these two
formalisms can be made almost identical, and it can be also seen, where
differences are essential.

The presentation in this section will not be quite ``parallel'' to that of
CM in Section~\ref{I;clmech}, because we want to stress and to describe also
some technicalities specific to QM.

\subsection{Pure states and dynamics in QM}\label{qmech;dyn}

The r\^ole played in CM by a phase space plays in QM a {\em normed complete
(linear) space with norm determined by a scalar product} -- over
complex numbers, a separable Hilbert space \H. The correspondence to classical
phase
space is not, however, faithful enough, since there are classes of vectors
in \H\ corresponding to the same physical state: All vectors
$\{\mlam\psi;0\neq\mlam\in\mbC\}$\ with any chosen $0\neq\psi\in\mH$,
correspond to the same physical state. The space of these classes is
the \emm projective Hilbert space~ \bs{\mPH}; it is no more linear.
Linearity seemed to be, however, important in historical
development of QM,~\cite{deBrog1,deBrog2,schroed1,dirac,landau3}, and it is
still important in many
experimental projects due to its intuitively appealing content.
We shall return briefly to this point later.\footnote{It is still possible to
define a ``superposition of states'' also in this nonlinear setting, cf.\ e.g.
~\cite{pulmann,cant2,cir4}.}
The points of the projective Hilbert space \PH\ are faithfully represented
by one--dimensional projection operators $P_\psi, 0\neq\psi\in\mH,
P_\psi\psi\equiv\psi$.
As will be shown later, the space \PH\ is a symplectic manifold (of the real
dimension $\dim_{\mbR}\mPH=2\dim_{\mbC}\mH-2$) in a canonical way.
\begin{intpn}[QM--CM ``correspondence'']\label{int;qm-cm-corr}
In QM--description of many phenomena, it is customary to introduce into
theoretical, as well as into experimental considerations a vaguely defined
concept of a \emm classical analogue~ of the considered system described by
QM, i.e.\ a classical--mechanical system in some way ``corresponding'' to
the considered phenomena (resp.\ to QM--system). So, e.g., for a hydrogen atom
described by vectors in the
infinite--dimensional Hilbert space $\mH := L^2(\mbR^6,\rd^6q)$, the
corresponding ``classical analogue'' is the Hamiltonian system on the
(12--dimensional) phase--space $T^*\mbR^6$, with the canonical symplectic
structure (cf.\ Examples~\ref{exmp;ham}(i), and (iii)) the dynamics of which is
described by the Hamiltonian
\[ h(q,p;Q,P):=\frac{p^2}{2m}+\frac{P^2}{2M}-\frac{e^2}{|q-Q|};\ \quad
q,p,Q,P\in\mbR^3.\]
The classical observables $\{q_j, p_j, Q_j, P_j,\ j=1,2,3;\ h \in
C^\infty(\mbR^{12})\}$ help to
interpret the points $P_\psi$\ of infinite--dimensional symplectic ``phase
space'' \PH\ as states of the (``real'', or ``genuine'') QM hydrogen atom:

We associate with any of these classical functions on the phase space
$\mbR^{12}$\ a selfadjoint linear operator on \H\ in such a way, that the
``corresponding'' operators $\mfk X\in\{\mfk q_j, \mfk p_j, \mfk Q_j, \mfk
P_j,\ j=1,2,3\}$\
determine specific functions $h_{\mfk X}$ on (a dense subset of) the
phase--space \PH\ (in an analogy with the observables in CM):
\[ h_{\mfk X}(P_\psi):= Tr(P_\psi\mfk X),\quad\forall P_\psi\in\mPH.\]
These functions satisfy ``the same'' commutation relations (i.e.\ Poisson
brackets relations) as the
corresponding classical phase space variables $X\in\{q_j,p_j,Q_j,P_j,\
j=1,2,3\}$, as we
shall see later. They also form, surprisingly (cf., however,
Subsection~\ref{IIIA1;CCR}), an ``irreducible set of
variables'' on the infinite--dimensional manifold \PH\ (i.e., in some
sense, they generate a complete set of ``coordinate functions''), if a
noncommutative ``$*$-product'' between these functions (cf.
also~\cite{flato} for alternatives)
\[ h_{\mfk X_1}*h_{\mfk X_2}:=h_{\mfk X_1\mfk X_2}, \]
is defined.\footnote{That these functions on \PH\ are not
differentiable in the usual sense (they are not even everywhere
defined) is not important in the considered connections: they
could be replaced by some of their bounded ``versions''; we can
work, e.g.\ with bounded operators from the algebra generated by
projection measures (cf.\ Appendices~\ref{B;Ca}, and~\ref{C;oper})
of the (unbounded) operators \fk X.} In this way, the functions
$h_{\mfk Y}$\ (where \fk Y\ are algebraic expressions composed of
the above introduced operators \fk X) form a noncommutative
(infinite--dimensional) algebra.\footnote{For a possibility of
mathematical definition of such algebras of unbounded operators
see, e.g. ~\cite{lassner}.} Its elements are interpreted in such a
way, that a ``correspondence'' with finite dimensional phase space
$\mbR^{12}$\ remains valid as a ``many--to--one'' mapping
$\mbF:\mPH\rarw\mbR^{12}$, defined in coordinates by
\[ \mbF_X: P_\psi\mapsto h_{\mfk
X}(P_{\psi})\equiv Tr(P_\psi\mfk X)=:\mbF_X(P_\psi),\
X=q_j,p_j,Q_j,P_j,\ j=1,2,3.\]
This mapping is then \emm interpreted statistically~ as {\bf expectation of
``observables''} \bs{\mfk X}\ {\bf in the pure states} \bs{P_\psi}. Values
of higher
degrees (\wrt the $*$-product) of these functions are then interpreted as
higher momenta of statistical distributions of these ``observables $X$''.
Hence, different QM--states $P_\psi$\ with the same expectations
$\mbF_X(P_\psi)=
h_{\mfk X}(P_\psi)$\ (for all \fk X) differ mutually by probability
distributions of some of these observables $X$.

A specific feature of QM in description of such ``finite systems'' as
the hydrogen atom is that {\bf there are no pure states \bs{P_\psi\in\mPH}
with zero dispersion of all observables in an ``irreducible set''}, in our
case formed by $\{q_j,p_j,Q_j,P_j,\ j=1,2,3\}$. This means that for any
$P_\psi\in\mPH$\ there is at least one $X\in \{q_j,p_j,Q_j,P_j,\
j=1,2,3\}$\ such that for the corresponding quantum observable one has
nonzero dispersion, i.e.
\[ h_{\mfk X}*h_{\mfk X}(P_\psi)\neq h_{\mfk X}(P_\psi)^2 .\]

The statistical interpretation of (even pure) states in QM differs from
interpretation of states in classical statistical physics. This difference
can be
expressed roughly (cf.\ \cite{bell,holevo,mack1,mermin}) so that in QM there
is no
(in some sense ``natural'') ``phase space'' (resp.\ a ``space of elementary
events'' -- in terminology of Kolmogorovian probability theory)
consisting of points representing some (at least fictitious)
dispersion--free states, such that
probability measures on it would determine the quantum states. Pure states
are interpreted in QM as in a sense ``the most detailed possibility'' of
a description of states of
``quantum objects'' (resp.\ ``systems'').\footnote{Cf., e.g.~\cite{davies} for
comparison of dispersions of observables in ``mixed states'' with those in
their pure convex summands.} In what sense, in the case of the
absence of any dispersionfree states, these ``objects really exist'' is
still a discussed problem: ``Object'' is characterized by its state which
contains just statistical predictions on possible outcomes of its
interactions with other bodies at specified initial conditions, leading each
time to a stable trace (i.e.\ a reproducibly verifiable ``macroscopic
change of environment'' in each single case of the repeatedly obtained cases of
``events of detection'');
such a process, if it is correlated with values of a physical quantity, is
called a ``measurement in QM''. The formalism of QM does not contain
``single events''.
 \hfill\bpika\end{intpn}

The quantal time evolution of vectors in \H\ is supposed to be such, that
it transforms, by a family of transformations
\[ \phi_t\ (t\in\mbR): \mPH\rarw\mPH,\ P_\psi\mapsto \phi_t(P_\psi), \]
the classes of  the vectors in \H\ corresponding to the
same physical interpretation, i.e.\ the points of \PH, bijectively onto \PH.
Traditionally, there is another general requirement to these
transformations $\phi_t(P_\psi)$: They should conserve the \emm transition
probabilities~,
i.e.\ the values of the nonnegative function
\bequ\label{eq;trans-pr}
\Pr:\mPH\times\mPH\rarw\mbR_+, \Pr(P_\psi,P_\mphi):=Tr(P_\psi
P_\mphi)\equiv\frac{|(\mphi,\psi)|^2}{\|\mphi\|^2\|\psi\|^2}.
\end{equation}
It is required:
\bequ\label{eq;invar-pr}
 Tr(\phi_t(P_\mphi)\phi_t(P_\psi))\equiv Tr(P_\mphi P_\psi).
\end{equation}
Considerations on possible physical interpretation of this
requirement are postponed to later sections, cf.\
also~\cite{bon-tr}.\footnote{\label{ft;red-post}This requirement
can be connected with the \emm reduction postulate~ of Dirac and
von Neumann,~\cite{dirac,neum1}, stating that, by measuring a
quantity $X$ on a considered system, after obtaining a result $x'$
the system suddenly ``jumps'' into a dispersionfree state of the
quantity $X$ in which that quantity {\em has the value} $x'$;
or alternatively, that the statistical ensemble representing the
system in the initial state (i.e.\ all members of the ensemble are
initially in the same quantum state) jumps during the measurement
into the statistical ensemble consisting of  systems occurring in
such quantum states that are all dispersionfree of $X$ with values
equaling to the measurement results $x'$; these systems occur in
the ensemble with the frequencies equaling to the frequencies of
occurrence of the corresponding results $x'$\ obtained by
 the measurement.}
According to a Wigner's theorem (cf.\ Proposition~\ref{prop;3.4}), the
additional requirement of the group property of $t\mapsto\phi_t$, i.e.\
$\phi_{t_1+t_2}\equiv\phi_{t_1}\circ\phi_{t_2}$, and
of continuity of  the functions
\[ t\mapsto Tr(P_\mphi \phi_t(P_\psi)),\ \forall P_\psi,P_\mphi\in\mPH,\]
suffices to imply the existence of a strongly continuous one--parameter
unitary group $t\mapsto U(t)$\ on \H\ such, that it is
\[ \phi_t(P_\psi)\equiv U(t)P_\psi U(-t)\equiv P_{U(t)\psi}.\]
Then the Stone's theorem, cf.\ \cite{R&S,riesz&nagy}  and
Theorem~\ref{thm;Stone}, gives the existence of a (unique,
up to an additive constant multiple of identity $I_\mH$) selfadjoint
operator $H$\ such, that
\bequ\label{eq;1gr-H}
 U(t)\equiv \exp(-itH).
\end{equation}
This leads to the \emm Schr\"odinger equation~ for evolution of vectors
$\psi(t)\in\phi_t(P_\psi)\in\mPH$:
\bequ\label{eq;1gr-SH}
\psi(t):=U(t)\psi(0)\imply i\frac{\rd}{\rd t}\psi(t)=H\psi(t),\quad\psi(0)\in
D(H),
\end{equation}
with $D(H)$\ being the domain of the selfadjoint $H$, cf.\ Appendix~\ref{C;symm}.
Let us stress the trivial fact, that the
Schr\"odinger equation makes no sense for ``improperly chosen'' initial
conditions $\psi(0)\not\in D(H)$.

This is the general form of time evolutions in QM. The operator $H$\ is
called the Hamiltonian and it is interpreted (cf.\ next subsection) as an
operator
describing the energy observable. It should be stressed, that mere symmetry of
the operator
$H$\ (i.e.\ $(\mphi,H\psi)=(H\mphi,\psi),\ \forall\mphi,\psi\in D\subset D(H),
\overline D=\mH$) is not sufficient to define a one--parameter group
by\rref1gr-H~; $H$\ should be selfadjoint to generate a group,
Appendix~\ref{C;oper}. On the other hand, between selfadjoint $H$'s, and
strongly
continuous one--parameter unitary groups $U(t)$'s\ there is a canonical
bijection expressed by\rref1gr-H~, cf.\ Theorem~\ref{thm;Stone}.

\subsection{States and observables}\label{qmech;stsp}

States in QM (let us denote the whole set of them by \Ss) form a
convex set, with ``pure states'' described by one
dimensional projections $P_\psi$\ as its extremal (i.e.\ indecomposable into
nontrivial convex combinations) points. Convexity of the state space can be
traced back to the classical, essentially macroscopic notion of
{\em statistical
ensemble}, cf.\ Interpretation~\ref{intpn;cl-ensemb}, in which expectations of
all
observables are expressed by the same convex combination of their
expectations in \emm subensembles~, that intuitively correspond to
``maximally specified ensembles'' (in CM these ``pure ensembles'' are
dispersion--free for all observables) .\footnote{Importance of the convex
structure of state spaces, and its
relation to other theoretical concepts was stressed and analyzed, e.g.
in~\cite{ludw1,mielnik,gudder}.}
 It was pointed out above that in CM such a ``maximal decomposition''
is unique. This means, that the classical state space $\mcl S(\mAc)$\ forms
a \emm simplex~, cf.\
\cite{choquet,meyer,ruelle1,bra&rob}. {\em This is not the
case of QM}, what is one of its deepest differences from CM. The
``shape'' of \Ss\ is closely connected with the set of ``observables'',
cf.\ \cite{mielnik}. We shall not go into interesting details of these
connections, but we shall rather review the standard traditional setting.

The set of \emm bounded quantum observables~ is taken (in the theory
without superselection rules,~\cite{WWW,jauch}) to be the set of all
bounded selfadjoint operators on \H, i.e.\ \LHs, and as the \emm \Ca\ of
quantum observables~ will be taken \LH. The \emm set of quantum states~
will be for us here just a part of the set of all positive normalized
linear functionals on \LH, namely the normal states \Ss\ consisting of
functionals expressible in the (defining) faithful representation of \LH\
by \emm density matrices~ on \H, i.e.\ by positive operators on \H\ with unit
trace:
\bequ\label{eq;1dens}
 \mrh\in \mSs\imply\mrh=\sum_j\mlam_jP_{\psi_j},\
\mlam_j\geq0,\ \sum_j\mlam_j=1.
\end{equation}
The expression\rref1dens~ represents one of infinitely many different (if
$\mrh\neq P_\psi$, for any $\psi\in\mH\setminus\{0\}$) extremal decompositions
of $\mrh\in\mSs$. Hence, a ``statistical interpretation'', like
in Interpretation~\ref{intpn;cl-ensemb}, of density matrices is questionable,
cf.\ also~\cite{bon-dens} for a more detailed formulation.

Unbounded observables are usually given as unbounded selfadjoint operators
on (a dense domain of) \H.\footnote{The forthcoming technical concepts are
briefly described also in Appendices~\ref{B;Ca}, and~\ref{C;oper}.}
  They are faithfully expressible by \emm
projection valued measures~ (\glss PM~, cf.\ \cite[Ch.IX.4]{varad}) on the
real line \bR: To any
selfadjoint operator $A=A^*$\ corresponds a unique projection valued
mapping
\[ E_A: B(\in\mcl B(\mbR))\mapsto E_A(B)=E_A(B)^2=E_A(B)^*(\in\mLH) \]
such, that for countable number of pairwise disjoint Borel sets $B_j\in\mcl
B(\mbR)$\ is $E_A$\ additive (the sums converging in the strong topology of
\LH), and $E_A(\mbR):=I_{\mH}$, cf.\ Definitions~\ref{df;1p-measure}.
Such a correspondence between PM and selfadjoint operators is bijective,
hence we can (and we often shall) as an observable in QM consider a
PM. The standard useful formula connecting $A$ with $E_A$\ is expressed by
the strongly convergent integral, cf.\ Theorem~\ref{thm;b-spectral}, and
Proposition~\ref{prop;fA-dom}:
\[ A=\int_{\mbR}\mlam E_A(d\mlam).\]

\begin{noti}\label{PVM}
A generalization of PM leads to \emm positive operator valued measure~
POV (or POVM), which also represents a selfadjoint operator, but it is not
determined by that operator uniquely. It represents a generalization of
the concept
of observable given by PM. Any POVM on \bR\ is a positive--operator valued
function
\[ \Delta: B(\in\mcl B(\mbR)) \mapsto\Delta(B) (\in\mLHs),\quad
0\leq\Delta(B)\leq I_{\mH}, \]
which is also countably additive (in strong topology) \wrt the additions
of disjoint sets. In
this case, contrary to PM, different $\Delta(B)\ (B\in\mcl B(\mbR))$\ need
not mutually commute. POVM can be used to modeling of imperfect measurements,
reflecting nonideal sensitivity of measuring
apparatuses,~\cite{davies,bon7,buz1,buz2}. We shall not go into details of
this refinement of the concept of ``quantummechanical observable''; see also
\dref1p-measure~.\hfill\dovi
\end{noti}

Let us turn now our attention to time evolution of general states (the
``Schr\"odinger picture''), and also of observables (the ``Heisenberg
picture''). It is naturally defined from that of pure state space described
in Subsection~\ref{qmech;dyn}, due to linearity and/or affinity of all
relevant relations. Hence, for the one--parameter unitary group
$U(t):=\exp(-itH)$ describing the evolution of pure states, or also vectors
in \H, the corresponding evolution of density matrices from\rref1dens~ is
\bequ\label{eq;1t-dens}
t\mapsto\mrh_t\equiv\phi_t(\mrh):=U(t)\mrh U(-t)\equiv\sum_j \mlam_jU(t)
P_{\psi_j} U(-t),
\end{equation}
what is valid for all possible decompositions\rref1dens~ of the
density matrix \rh. This description of time evolution corresponds to the
\emm Schr\"odinger picture~.

The \emm Heisenberg picture~ of the time evolution in QM is the dual
(=transposed)
transformation group $\phi^*$\ to that of $\phi_t: \mSs\rarw \mSs$, since the
algebra of observables \LH\ is the (topological) dual space of the complex
linear space spanned by density matrices and completed in the trace norm
$\mN\mrh,1~:= Tr(|\mrh|)$. Since the duality is realized by the bilinear
form
\bequ\label{eq;expect}
 (\mrh;A)(\in \mSs\times\mLH)\mapsto \lb A;\mrh\rb \equiv Tr(\mrh A) =: \lb
A\rb_\mrh,
\end{equation}
the time evolution of the $A$'s in \LH,\ $(t;A)\mapsto A_t:=\phi^*_t(A)$\
is determined by the requirement
\[ \lb A_t;\mrh\rb \equiv\lb\phi^*_t(A);\mrh\rb:=\lb A;\phi_t(\mrh)\rb, \]
and we have
\bs{A_t \equiv \phi^*_t(A):= U(-t)AU(t)}.
Let us note that, according to the introduced definition of $\phi^*_t$,
one has the following invariance:
\bequ\label{eq;1phi-hat}
\lb \phi^*_{-t}(A);\phi_t(\mrh)\rb\equiv \lb A;\mrh\rb.
\end{equation}
Let us notice similarity of the equations\rref1phi-hat~, and\rref invar-pr~,
what will be of importance in the subsequent nonlinear extensions of QM,
cf.\ \cite{bon-tr}.

Interpretation of states {\em and} observables is given by determination of
a formula expressing the probability of obtaining
results $\mlam\in B$\ (:= a subset of the spectrum, i.e.\ of the set of
possible values of $A$) by measuring of an observable $A$ of a system
occurring in the state \rh. This probability will be denoted by\
$\prob(A\in B;\mrh)$.
It can be also useful to introduce the corresponding probability measure
$\mu^A_{\mrh}$\ on the real line \bR\ of values of the measured quantity
$A$:
\bequ
\label{eq;1prob}
\prob(A\in B;\mrh)\equiv \mu^A_{\mrh}(B):=Tr(E_A(B)\mrh).
\end{equation}
This formula allows us essentially to express all empirically verifiable
statements of
QM. The expectation (mean value) $\lb A\rb_\mrh$\  is given by\rref expect~.

The assertion on nonexistence of dispersion--free states for
all observables can be made precise in a form of general \emm Heisenberg
uncertainty relations~:
\begin{prop}\label{prop;HUR}
Let $A, B$\ be two bounded selfadjoint operators (representing two quantal
observables), and let $\lb A\rb_\mrh:=Tr(\mrh A)$\ be the expectation of
measured values of the arbitrary observable $A$\ in any state $\mrh,\
\mrh\in\mSs$. Let $\Delta_\mrh A:=\sqrt{Tr(\mrh(A-\lb A\rb_\mrh)^2)}$\ be
the dispersion of the measured values of $A$\ in the same state $\mrh$.
Then
\bequ\label{eq;1HUR}
\Delta_\mrh A\dti\Delta_\mrh B\geq\frac{1}{2}\bigl|Tr(\mrh(AB-BA))\bigr|
\equiv\frac{1}{2}\bigl|\lb i[A,B]\rb_\mrh\bigr|.\quad\clubsuit
\end{equation}
\end{prop}

This proposition can be generalized to unbounded operators, with
corresponding restrictions for the states $\mrh\in\mSs$.
This shows that noncommutativity of two observables leads to nonexistence of
their mutually sharp values in states with nonvanishing expectation of their
commutator. Remember that for the operators $\mfk Q_j, \mfk P_j$\
corresponding in QM to the classical $j$--th position and linear momenta
coordinates, one has $[\mfk Q_j,\mfk P_k]=i\hbar\mbI_\mH\delta_{jk}$\ (on a
corresponding dense domain in \H). Hence the ``observables $Q_j, P_j$''
cannot be both sharply determined in any state
$\mrh\in\mSs$.\footnote{For a discussion and citations on various
interpretations of\rref1HUR~ see e.g.~\cite{busch}.}
The formula\rref1prob~ leads also to convenient realizations of
(elements of) \H\ in terms of numerical functions.
\begin{rem}[``Representations'' in QM]\label{rem;QM-rep}
It might be useful to comment and formulate here, in some more general
terms than is it usually presented, what is traditionally
named ``the representation theory'' according to
Dirac,~\cite{dirac}. Physicists often work with specific realizations of
Hilbert space \H, according to specific physical systems to be described.
Elements of the ``Hilbert space of a given physical system'' are often
expressed as ``wave functions'', i.e.\ complex valued functions of
``configuration variables'' (e.g.\ positions of described particles).
Since all infinite--dimensional separable Hilbert spaces are isomorphic,
different realizations of \H\ can be specified only by an additional
mathematical structure. This is done by a choice of a ``complete set of
commuting observables'', i.e.\ by specifying a maximal commutative von Neumann
subalgebra~\cite[p.112]{sak1}\footnote{A commutative algebra of bounded
operators on \H\ is {\em maximal commutative} if its arbitrary nontrivial
extension by addition of an operator violates its commutativity. Such an
algebra is always weakly closed in \LH, i.e.\ it is a \Wa. cf.\ also
Appendix~\ref{B;Ca}.}
in \LH\ generated by (mutually commuting)
projection valued measures
$E_A,E_B,\dots$, of a set $A,B,\dots$, of mutually commuting selfadjoint
operators. These operators represent in QM some ``simultaneously measurable
observables''. The von Neumann algebra $\mcl R$\ generated by a set
\[ \mcl R_0:=
\{E_A(B_1),E_B(B_2),\dots; B_1,B_2,\dots\in\mcl B(\mbR)\} \]
of bounded operators (projections) in \H\ containing the unit operator
$I_{\mH}\in\mLH$\
is obtained by taking the double commutant, according to famous von
Neumann ``bicommutant theorem'',~\cite{najm,sak1,takesI,bra&rob},
$\mcl R=\mcl R_0''$, in \LH.\footnote{cf.\ also Appendix~\ref{B;Ca} for
technicalities.}
Here, the commutant $\mcl R_0'$\ of $\mcl R_0$\ is given by
\[ \mcl R_0':=\{B\in\mLH:[B,A]=0,\forall A\in\mcl R_0\},\]
and $\mcl R_0'':=(\mcl R_0')'$, for any subset $\mcl R_0\subset\mLH$. Any
commutant in \LH\ is a \Ca\ closed in weak--operator topology of \LH, and
such \Ca s are called \emm von Neumann algebras~, or \Wa s.
The \Wa\ \cl R\ is maximal commutative iff
$\mcl R=\mcl R'$, what is equivalent with the situation when the
commutative \Wa\ has a cyclic (then also separating) vector $\psi_0$\ in \H,
cf.\ \cite{R&S}.
Let $M_{\mcl R}$\ be the (compact Hausdorff) spectrum space (cf.
Example~\ref{ex;abel-C&W}) of \cl R, hence
the algebra of continuous complex valued functions $C(M_{\mcl R})$ is
isomorphic (denoted by $\sim$) to \cl R.
If $\psi_0\in\mH$\ is cyclic for $\mcl R\sim C(M_{\mcl R})$, then (denoting the
operators in \cl R\ by $\pi_0(f)$, for the corresponding functions
$f\in C(M_{\mcl R})$) the integral, i.e.\ the positive linear functional on
$C(M_{\mcl R})$  (according to the Riesz--Markov
theorem,~\cite{R&S})
\bequ\label{eq;meas-R}
 f\ (\in C(M_{\mcl R}))\mapsto \mu^{\mcl R}(f):=(\psi_0,\pi_0(f)\psi_0)
\end{equation}
determines (if $\psi_0$\ is normalized) a probability measure $\mu^{\mcl
R}$\ on $M_{\mcl R}$, and the mapping
\bequ\label{eq;LqM-W}
 U_{\mcl R}: \pi_0(f)\psi_0\ (\in\mH)\mapsto f\in C(M_{\mcl R})\subset
L^2(M_{\mcl R},\mu^{\mcl R})
\end{equation}
can be uniquely extended to an isomorphism of Hilbert spaces, \cite{sak1}.
Moreover (cf.\ \cite[Chap. I.9]{gamelin}), all the functions
$f\in C(M_{\mcl R})$\ are just all the (elements of equivalence classes of
$\mu^{\mcl R}$--essentially) bounded Borel functions on
$M_{\mcl R}$, i.e.\ $C(M_{\mcl R})=L^\infty(\mu^{\mcl R})$.\footnote{Let us
note, that these functions $f\in C(M_{\mcl R})$\ can be considered
either as elements of $\mcl
L(L^2(M_{\mcl R},\mu^{\mcl R}))$, or as elements of the Hilbert space
$L^2(M_{\mcl R},\mu^{\mcl R})$\ itself. Let us also note that the constant
unit function \bI\ is an element of this Hilbert space representing a
cyclic vector \wrt the maximal commutative algebra $C(M_{\mcl R})$\ of
operators.}

The spectrum space $M_{\mcl R}$ of an abelian \Wa\ \cl R\ has a rather
``wild'' topology, since
any \Wa\ is generated by its projections which are, in the commutative
case, continuous characteristic functions of clopen subsets of $M_{\mcl R}$,
which in turn form a basis of the Hausdorff topology of $M_{\mcl R}$,
cf.\ \cite{gamelin}.
As a
consequence of this \emm extremely disconnected~ topology (cf.\ \cite[Chap.
III.1]{takesI}), the function realization of \H\ in\rref LqM-W~ needn't seem
to be practically convenient. If, however, there is in \cl R\ a strongly dense
unital \Csa\ $\mcl A\equiv\mcl A_{\mcl R}$ with some ``nice'' spectrum space
$M$ (e.g., $M$ could be connected), then we can write for the corresponding
 isomorphism $U_{\mcl R}$, instead of\rref LqM-W~:
\bequ\label{eq;LqM-C}
 U_{\mcl R}: \pi_0(f)\psi_0\ (\in\mH)\mapsto f\in C(M)\subset
 L^2(M,\mu^{\mcl A}),
\end{equation}
where the measure $\mu^{\mcl A}$\ is defined, now from \cl A, by the same way
(i.e.\ via Riesz--Markov theorem) as it was done
in\rref meas-R~ from \cl R, since $\mcl A\psi_0 \bigl(\mcl A:=\pi_0(C(M))\bigr)$
\ is again
dense in \H.

Let \cl R\ be generated by $n$ projection measures
$\{E_{A_1},E_{A_2},\dots,E_{A_n}\}$, e.g.\ spectral measures of (possibly
unbounded) selfadjoint operators $\{A_j, j=1,\dots n\}$; i.e.\ \cl R\ is the
minimal \Wa\ containing these projections, and it is maximal commutative
with a cyclic vector $\psi_0\in\mH$.
Then the spectrum space $M=M_{\mcl A}$\ can be chosen
homeomorphic to a compact subset of a compactification of
$\mbR^n$, namely the support of the product--measure $E_{\mcl R}$\ (what is
again a projection measure) of the spectral measures
$E_{A_j},j=1,\dots,n$, cf.\ \cite[Chap. 5, \S 2, Theorem 6; Chap. 6, \S 5,
Theorem 1]{bir&sol}.
 We have then $L^2(M,\mu^{\mcl A})=L^2(\mbR^n,\mu^{\mcl A})$, and each
 operator $U_{\mcl R}A_jU_{\mcl R}^{-1}$\ acts on $L^2(\mbR^n,\mu^{\mcl A})$
 as ``multiplication by the $j$--th variable'':
 \[ U_{\mcl R}A_jU_{\mcl R}^{-1}\mphi(q)\equiv q_j\mphi(q),\ q\in\mbR^n,
 \mphi\in L^2(\mbR^n,\mu^{\mcl A}).\]

 We can speak now about the \emm \bs{\mcl A}--representation~ , resp.
\bs{\{A_j:j=1,2,\dots,n\}}{\bf--re\-pre\-sen\-ta\-tion}, of (Quantum
Mechanics represented in) the Hilbert space  \H.

Let us assume, that the product--measure $E_{\mcl R}$\ is absolutely
continuous
\wrt the Lebesgue measure $\rd^nq$\ on $\mbR^n$, i.e.\ absolutely continuous are
all the probability measures
\[ B(\in\mcl B(\mbR^n))\mapsto
 (\psi,E_{\mcl R}(B)\psi),\quad\forall\psi\in\mH,\quad\|\psi\|=1.
\]
Hence also all the probability measures
 \[ B(\in\mcl B(\mbR))\mapsto
 \mu_j^\psi(B):=(\psi,E_{A_j}(B)\psi),\quad\forall\psi\in\mH,\|\psi\|=1,\quad
j=1,\dots n, \]
 are absolutely continuous \wrt $\rd q$\ on \bR.\footnote{We do not formulate
here sufficient
 conditions for absolute continuity of $E_{\mcl R}$.}
Since the vector $\psi_0\in\mH$\ is cyclic and
separating for $\mcl A''=\mcl A'=\mcl R$, the measure $\mu^{\mcl A}$\ is
absolutely continuous \wrt the Lebesgue measure $\rd^nq$\ on $\mbR^n$. Let
\[ q\ (\in\mbR^n)\mapsto f_{\psi_0}(q):=\frac{\rd\mu^{\mcl
A}}{\rd^nq}(q),\quad \frac{\rd\mu^{\mcl A}}{\rd^nq}\in
L^1(\mbR^n,\rd^nq)\]
be a version of the Radon--Nikodym derivative (cf.\ \cite{najm,R&S}) of
$\mu^{\mcl A}$\ \wrt the Lebesgue measure. Then $L^2(M,\mu^{\mcl A})$\ can
be mapped onto a subspace of $L^2(\mbR^n,\rd^nq)$\ by the unitary mapping
\bequ\label{eq;LqM-R}
 \psi(q) \mapsto\psi(q)\sqrt{f_{\psi_0}(q)},\quad\forall \psi\in
L^2(M,\mu^{\mcl A}),\ q\in \mbR^n(\supset M).
\end{equation}
In this setting, on \H\ represented by (a subspace of) $L^2(\mbR^n,\rd^nq)$,
the operators $A_j,j=1,2,\dots,n$, are realized as multiplication operators
by the coordinates $q_j,j=1,2,\dots,n$, with
$\{q_1,q_2,\dots,q_n\}\equiv q\in\mbR^n$.\footnote{If the spectrum of some
$A_j$\ is not the whole \bR, then \H\ is represented by a proper subspace
 $L^2(\supp\!(E_{\mcl R}),\rd^nq)\subset L^2(\mbR^n,\rd^nq)$.}
The probabilities\rref1prob~ have now the form
\bequ\label{eq;2prob}
\prob(\{A_j\}\in B\subset\mbR^n;P_\psi)=Tr(E_{\mcl R}(B)P_\psi)
=\int_B|\psi(q)|^2\rd^nq,
\end{equation}
with $\|\psi\|=1$.
 A special case of this situation is the usually used ``position
 representation'' of the state vectors.
\hfill\dovi
\end{rem}
\begin{exm}[Position representation]\label{ex;C(M)}
Let \cl A\ be the subalgebra of \LH\ generated by the unit operator
$I_{\mH}$ and by the operators
$f(Q_1,Q_2,\dots,Q_{3N})$, with the functions $f$ from the Schwartz
space $S(\mbR^{3N})$, where the standard
position operators $Q_j, j=1,2,\dots,3N$, of the irreducible representation
of \GWH\ for an $N$--particle system (cf.\ Subsection~\ref{IIIA1;CCR}),
were introduced. Then the spectrum space $M$\ is the one--point
compactification of $\mbR^{3N}$\ with the ``usual'' topology.
The weak closure \cl R\ of \cl A\ in \LH\ is an abelian \Wa\ containing
also projection operators belonging to the spectral decompositions of
$Q_j$'s, i.e.\ elements $E_{Q_j}(B)$\ of their PM's.
If there is a cyclic vector $\psi_0$\ for \cl R\ in \H, then $\psi_0$\ is
cyclic also for \cl A. Then we can use the unitary transformation $U_{\mcl
R}:\mH\rarw L^2(\mbR^{3N},\rd^{3N}q)$ determined from\rref LqM-C~, and\rref
LqM-R~ by \cl A\ only. Hence
$\mH\sim L^2(\mbR^{3N},\rd^{3N}q)$. This is the usual ``position-coordinate
representation'' of \H. \hfill\dovi
\end{exm}
\subsection{Symmetries and projective representations in
QM}\label{qmech;symm}

The time evolution described in Subsection~\ref{qmech;dyn} was an example
of a continuous transformation group in QM. It can be considered as a
representation of a specific group ($G:=\mbR$) of symmetries of a physical
system, namely a representation of the observed (or postulated)
\emm homogeneity of time~: This symmetry, described by formulas expressing
fundamental laws of physics independent of the time variable,
can be considered as just an expression of possibility of formulation of
such laws. The invariance is encoded in the group property of the set of
time--evolution operators, what corresponds to time independence of its
generator (the Hamiltonian): ``Dynamics'' is time--independent, and differences
in various possible (or observed) evolutions of the system in its ``various
occurrences'' are ascribed to differences in ``initial
conditions'',~\cite{wigner2,HVW},
resp.\ in ``boundary conditions'' (including also ``external fields'').

The relevance of symmetries in physics was probably (at least) intuitively
clear since the advent of any considerations which now we call ``physical''.
Their formalization came, however, much later: Although importance of
symmetries for human activities was claimed already by Leonardo da Vinci
(according~\cite{wigner2,weyl}),
clear understanding of their importance for formulation of geometry and
laws of nature came only at about the beginning of 20th century, e.g.\ in
works of F. Klein~\cite{klein}, G. Hamel~\cite{hamel}, H.
Poincar\'e~\cite{poincare}, E. Mach~\cite{mach,mach1}, P.
Curie~\cite{curie}, A. Einstein~\cite{einst1,einst}, and others.\footnote{
Many historical notes on symmetries  can be found in~\cite{mars&rati}.}

Their
importance is clearly seen, e.g.\ in formulation of classical -- mechanical
problems on integrability (connected with the question of stability of
Solar system), in Einstein discovering of relativity theories, in Gibbs
formulation of statistical physics~\cite{landau5}, etc.
Clear mathematical connection of variational equations with symmetries and
with integrals of motion was formulated also due to the theorems by
Emmy Noether~\cite{noether}.
Nowadays is generally accepted the connection between Lie group invariance
of ``action integrals''
(or/and Lagrangians) of classical physics,
cf.\ ,e.g.,~\cite{landau1,landau2,abr&mars}, with conservation of some
nontrivial functions on phase space \wrt the time evolution determined by
the corresponding variational problem. These \emm integrals of motion~
determine submanifolds in phase space left by the time evolution invariant.
This leads to practical advantage of ``lowering dimensions'' of solved
problems.
Intuitively, this also allows a better specification of the (self)identity of
moving physical systems.

A ``quantum--field--rephrasing'' of the mentioned principles was one of the
leading tools in formulations of (heuristic, but successful) quantum
theories of elementary particles, with quantum electrodynamics as their
prototype. Also in foundations of mathematically clear (but, up to now not
very successful) ``axiomatic'' algebraic formulation of quantum field theory
(QFT), cf., e.g.~\cite{str&wight,haag2,borchers}, symmetry principles play a
key r\^ole.

\begin{noti}
We can suspect even more general meaning of
``invariances'' \wrt some group of transformations in physics: They help us to
determine physically (hence operationally) meaning of ``physical
quantities''; very pictorially expressed, symmetry means that {\bf some
mutually different things} (states, observed values of something,$\dots$)
{\bf are} in a
certain sense {\bf equal},~\cite{curie}, what might help us to specify
how to measure them.
A very fruitful principle in physics is, as is generally known, the requirement
of invariance \wrt Galileo, resp.\ Poincar\'e groups, cf.\ also
Interpretation~\ref{intpn;G-quantit}. \hfill\dovi
\end{noti}

The symmetry considerations in QM are even more important and useful than
in classical physics. This is, perhaps, due to the ``more mathematical'' and
less
intuitive nature of quantum theories. The a priori linear formulation of QM
offered a natural application of (linear) representation theory of groups
to solution of specific classes of problems in QM, esp.\ in classification
 of ``elementary systems'' (these might be ``elementary
particles'', but also molecules), of their spectra and interactions,
in scattering theory etc.,
cf.\ \cite{weyl,wigner1,wigner2,wigner3,str&wight,haag&kast}. One can say
that symmetry considerations are lying now somewhere ``in the heart'' of QM.
They belong, e.g., to the main tools in the search for new fundamental
interactions of elementary particles.

We shall restrict now our attention to a rather specific
technical question connected with
appearance of symmetry considerations in mathematical formulations in QM.
Symmetry groups are usually specified either from observations of specific
motions of macroscopic bodies (e.g.\ translations and rotations of ``rigid''
bodies), or by some theoretical hypotheses coming from an interplay of
presently accepted theoretical scheme and observations connected with it
(e.g.\ the isospin group, and other symmetries of elementary particle
theories). Groups appear then in formalisms of physical theories in a form
of their ``realizations'', cf.\ \cite{kiril}, i.e.\ in a form of their
actions on spaces of physically relevant theoretical objects like
``states'', ``observable quantities'', ``state vectors'', etc. In
traditional formulations of QM, symmetries are formalized as
transformations of Hilbert space vectors. It is important in some
considerations to understand connections of the transformations of vectors
in \H\ with corresponding transformations of quantal states.

The usually required general restrictions
to the set of symmetry transformations of the states of a QM--system
are the same as for $\phi_t$\ in Subsection~\ref{qmech;dyn}, esp.
in\rref invar-pr~.
There is, however, an additional complication for general (more than
one--dimensional) continuous groups
$G$ of transformations, $g(\in G)\mapsto \malp_g:\mSs\rarw\mSs$. Let us
assume that\rref invar-pr~ is again fulfilled:
\bequ\label{eq;inv-pr}
 Tr(\malp_g(P_\mphi)\malp_g(P_\psi))\equiv Tr(P_\mphi P_\psi).
\end{equation}
Then a trivial adaptation of arguments following\rref invar-pr~ (by
the assumptions of the group property and continuity, as above) leads to
the conclusion (cf.\ also~\cite{varad,kiril}) that a continuous family
\[ g(\in G)\mapsto U(g)(\in \mLH) \]
of unitary operators exists representing the mapping $g\mapsto\malp_g$ as
\bequ\label{eq;alph-U}
 \malp_g(P_\psi)\equiv U(g)P_\psi U(g)^* .
\end{equation}
This determines, however, the unitary operators up to phase factors, and we
obtain (for details see~\cite[Chaps.IX, and X, esp.\ Theorem 10.5]{varad})
\bequ\label{eq;proj-rep}
U(g_1\dti g_2)\equiv m(g_1,g_2)U(g_1)U(g_2),
\end{equation}
where $m:G\times G\rarw S^1\subset \mbC$ is a \emm multiplier~ for the group
$G$ satisfying the following identities implied by associativity of group
multiplication:
\bequ\label{eq;multiplier}
\begin{split}
{}&m(g_1,g_2\dti g_3)m(g_2,g_3)=m(g_1\dti g_2,g_3)m(g_1,g_2),\quad\forall
g_j\in G,\\ {}&m(g,e)=m(e,g)=1,\quad \forall g\in G,\quad e\dti g\equiv g.
\end{split}
\end{equation}
Multipliers for $G$ form a commutative group (by pointwise multiplication;
cf.\rref1pr-rep~ for additive notation)
with the unit element $\mbI(g,h)\equiv 1$.
 If the multiplier can be removed by multiplying
$U(g)\mapsto a(g)U(g)$ by some ``phase factors'' $a(g)\in S^1:=${\em the
complex numbers of unit modulus}, then it is {\em similar to $\mbI$},
or \emm exact~.
Two multipliers $m_1, m_2$\ are mutually similar, if the
multiplier $m_1\dti m_2^{-1}$\ is similar to $\mbI$.
The unitary family satisfying\rref proj-rep~ is called a \emm projective
representation~ of $G$\ with the multiplier $m$. All projective
representations of $G$\ obtained from the same $\malp_G$\ have mutually
similar multipliers, and, to any projective representation $U$\ with
a multiplier $m$, and to each multiplier $m'$\  similar to $m$, there is a
projective representation $U'$\ with the multiplier $m'$\ leading to the same
 $\malp_G$\ according to\rref alph-U~ as $U$.

Hence, if the multiplier in\rref proj-rep~ is exact, it is possible to
choose a unitary representation (i.e.\ with $m\equiv 1$) corresponding to
the $\malp_G$. Otherwise, it is possible to find another group $G_m$\
containing $G$\ as a normal subgroup, a \emm
central extension~ of $G$\ by the commutative group $S^1$\ corresponding to
the multiplier $m$, and such that the formula\rref proj-rep~ determines its
unitary representation: Elements of $G_m$\ are couples $(g;\mlam)\in G\times
S^1$\ with the group multiplication
\bequ\label{eq;G-m}
(g_1;\mlam_1)\dti(g_2;\mlam_2)=(g_1\dti g_2;m(g_1,g_2)\mlam_1\mlam_2).
\end{equation}
The corresponding unitary representation $\tilde U(G_m)$\ is
\[ \tilde U(g;\mlam) := \mlam^{-1}U(g),\quad\forall g\in G,\mlam\in S^1.\]
The check that $\tilde U(g;\mlam)$\ leads (for all $\mlam\in S^1$) to the same
symmetry transformation $\malp_g$\ of the states than the element $U(g)$\
of the projective representation $U(G)$\ is straightforward.
\begin{exmp}
(i) Only projective representations of the inhomogeneous Galileo group with
nontrivial
multipliers can be interpreted,~\cite{varad}, in the usual interpretation
schemes of QM,
 as transformations of states of systems in QM
representing the corresponding (relative) motions of macroscopic
background. The unitary representations of this group are all
``unphysical''.\nl
(ii) The most basic application of group representations in QM
is, perhaps, the case of canonical commutation relations (CCR). These
relations determine a Lie algebra structure in a set of basis
elements (i.e.\ of ``elementary observables'' completed by the ``trivial
element'') in
a ``Hamiltonian system on $\mbR^{2n}$''-- both quantum and classical. These
relations are expressed in CM by Poisson brackets between
canonical position and momenta coordinates, and in QM they are commutators
between ``corresponding'' selfadjoint operators (representing in some way
also physical position and momenta observables). The connection with group
representations is, that these operators are generators of a
projective representation of the {\em commutative group of translations
in the classical flat phase space $\mbR^{2n}$}, or they are generators
(together with a unit operator) of a {\em unitary representation} of a
one--dimensional central extension of this commutative group, i.e.
of the $2n+1$--dimensional
(noncommutative) Weyl--Heisenberg group \GWH.\footnote{Remember that
commutative
groups have only one--dimensional irreducible {\em unitary}
re\-pre\-sen\-ta\-tions.}
  All such (nontrivial, i.e.\ more
than one--dimensional) irreducible projective representations are parametrized
(up to unitary equivalence) by all nonzero reals,~\cite{zelob&stern,kiril}.
Remarkable \emm physical
feature of CCR~ is, that they can correspond, to reach agreement of
theoretical predictions with experiment, just to one of the infinite
number of mutually inequivalent representations of classical shifts in
phase space, and namely the ``correct'' choice of the representation {\bf fixes
the} \emm value of Planck constant \bs{\hbar}~; cf.\ also
Section~\ref{IIIA1;CCR} for corresponding technicalities. \nl
(iii)  The (covering group of the) connected component of the
Poincar\'e group is \emm rigid~, i.e.\ it has no
nontrivial multipliers,~\cite{varad,woron,kiril}. It follows that any
projective
representation of the (connected) Poincar\'e group can be obtained from
the corresponding unitary representation of its covering group.\hfill\dovi
\end{exmp}

\subsection{On the causality problem in QM}\label{qmech;causal}

With discussions on Einstein causality in NLQM, cf.\ \cite{noncausal,luecke},
or also Interpretation~\ref{int;3.7a},
it is interesting to pose such a question
also in frameworks of {\em linear} QT. In a renewal of such a
discussion~\cite{hegerfeldt,buch&yng}
(initiated probably by Fermi in 1932~\cite{fermi}),
there was discussed a simple
mathematical theorem with impressive consequences for possibilities on
``instantaneous spreading of wave packets'' in QM. It can be formulated as
follows:
\begin{thm}[Long distance action in QM]\label{thm;acausal}
Let a selfadjoint lower bounded operator $H=H^*$\ on a Hilbert space \H\ be
given: $H-\mveps_0 \mbI_\mH\geq 0$. Assume that, for a bounded operator
$B$: $0\leq B\in\mLH$, and for a vector $0\neq\psi\in\mH$, there is:
$\quad Tr(P_\psi B)=0$.
Let us define $\psi(t):=\exp(-itH)\psi,\forall t\in\mbR$. Then either\nl
\quad(i)\quad $Tr(P_{\psi(t)}B)\equiv 0$\ for all $t\in\mbR$, or\nl
\quad(ii)\quad $Tr(P_{\psi(t)}B)\neq 0$\ for all $t\in\mcl T_{\psi,B}\subset
\mbR$, with $\mcl T_{\psi,B}$\ open and dense in \bR, and of the total
Lebesgue measure $m$: $m(\mbR\setminus\mcl T_{\psi,B})=0$.
\hfill\zal\end{thm}

Let, e.g.\ $P_\psi$\ be a state of a composed system $I+II$\ (say, consisting
of two mutually spatially distant atoms $I$\ and $II$), and let
$B=B^*=B^2\neq0$\ be a projection on a subspace of \H. Assume that the
vectors of $B\mH$\  correspond to those states of $I+II$\ in which the atom
$II$\ is in its excited state (we assume a possibility of determination of
such states of $I+II$). Then $Tr(P_{\psi(t)}B)\neq0$\ might be interpreted as
excitation in time $t>0$\ of the formerly not exited atom $II$\ ``due to an
influence of the atom $I$''. The theorem could be tried then
interpreted so that if
there will be some influence at all (sometimes, in the mentioned sense),
then it  is {\it always immediate}, i.e.\ there is nonzero probability
that it is realized instantaneously!

Above considerations seem to show that in QM, in the described sense, the
 Einstein
causality is never fulfilled.
This result is a general consequence of the assumptions
of the positivity of the generator $H$,
of the interpretation of projection operators $B$\ as
observables measuring of arbitrary ``properties'' of
described systems in QM, as well as due to occurring of arbitrary projections
between the observables;
all these assumptions might seem to belong to general assumptions of an
arbitrary quantum theory (QT). But even in this framework, it is impossible
to draw any {\it physical consequences} from the above mentioned result: The
considered initial condition  $Tr(P_{\psi(t_0)}B)=0$, if not fulfilled
identically for all $t_0\in\mbR$, cannot be fulfilled by all $t_0\in I$ for
any interval $I\subset\mbR$ of nonzero lenght, and a physical preparation of
a state needs nozero time interval. This fact is automatically
encompassed in definition of algebras of observables in
``algebraic forms of relativistic QFT''.

One can now ask whether Einstein causality is
fulfilled in relativistic QFT. \label{R-causal}
It is argued in~\cite{buch&yng} that it is so in
the algebraic formulation of QT (e.g.~\cite{haag&kast,haag2,borchers,horuz}),
as a consequence of the relativistic
covariance of/and local structure of algebras of observables. This can be
seen, roughly, due to consequent specific structure of algebras of
localized observables (cf.\ Note~\ref{note;types}), as
well as due to the Reeh--Schlieder theorem, cf.\ \cite[Theorem 3.1]{horuz}.
This theorem implies that in ``most of
interesting states'' of ``sufficiently'' localized subsystems (e.g.\ in the
states extendable to states of the total system with
restricted total energy, if the space--time region of the localization has the
space--like complement with nonvoid interior) any localized positive observable
has nonzero expectation, cf.\ also~\cite{schlieder,haag4,haag2}. Hence
the above assumption $Tr(P_\psi B)=0$\ cannot be fulfilled for such systems,
states,
and observables. Moreover, the assumed locality together with Einstein
covariance lead to positive result on Einstein causality,~\cite{buch&yng}.

\begin{noti}[Impossible signals due to measurements]\label{not;red-post}
It might be useful to recall here that it is impossible in QM to send
signals by
the process of quantum measurements even if one accepts the instantaneous
``reduction of wave packets'', cf.\ Footnote~\ref{ft;red-post}:

Let two (mutually spatially distant) quantal subsystems $I$ and $II$ be
``EPR--like'' correlated (cf.\ Interpretation~\ref{int;3.7a}) in a given
state $\Psi$ of the composed system $I+II$. The only
available
``information'' which could be transferred (=signalled) from $II$ to $I$, as a
result of the mere measurement of a quantity $A$ of the subsystem $II$, might
be the choice (and its possible changes) of the quantity $A$, resp.\ of its
eigenbasis (i.e.\ its PM $E_A$) $\{\Phi_k\}\subset\mH_{II}$.  The only
way, on the other side, of
perceiving of the signal by $I$ might be the measurement of the state \rh\ of
$I$, what is, however, independent on the choice of $A$.
The point is, that QM is a
statistical theory not containing in its formalism any objects corresponding
to our intuition on a ``single system'' (possibly, as an element of some
``ensemble of equally prepared systems'') resulting in a ``single event''
at a measurement; cf.\ also~\cite{sch&busch}.
\hfill\dovi\end{noti}


\vspace*{\fill}
\newpage


\chapter{Extended Quantum Mechanics}
\label{sec;II}
\def\autor{\ref{sec;II}\quad Extended Quantum Mechanics}

This chapter contains a description of technical features, as well as of
the proposed interpretation of the theoretical construction called here
extended quantum mechanics (EQM). We can emphasize here several types of
problems posed and solved in this chapter; let us call them:
(i) kinematical, (ii) dynamical, (iii) analytical, and (iv)
interpretational.\nl
Questions in (i) include topics that could be named
``the geometry of phase space'',
into (ii) can be included questions connected with dynamics, as well as
with continuous actions of symmetry groups on the ``phase space'', under
(iii) we shall understand mainly technical problems connected with infinite
dimensionality of the ``phase space'', with unboundedness of generators of
the group actions etc.; interpretation in (iv) is understood as a series
of notes and proposals concerning a {\em general scheme for interpretation} of
the theory; however, many questions on possible specific empirically
verifiable applications of EQM are left open here.

These sets of questions are mutually interconnected, e.g.\ in dealing with
``geometry of phase space'' we cannot avoid some technical problems
connected with its infinite--dimensionality, including different
topological and differential--geometrical technicalities. Similarly by
dealing with ``symmetry group actions'' one has to deal simultaneously also
with some ``algebraic'', or ``structural'' questions, and with problems
connected with the (only)
densely defined generators of these actions and their domains of definition.
Hence, it is impossible to distinguish clearly the forthcoming sections
according
to the sort of problems solved in their scope. Keeping this in mind, we
shall try to characterize at least roughly the contents of the sections in
the present Chapter.

 Section~\ref{IIA;q-phsp} is mainly devoted to a description of the
 ``geometrical features'' (i), describing the canonical manifold and Poisson
 structures on the space  \Ss\ of  all density matrices of conventional QM.
 Also a preliminary description of Hamiltonian vector fields and
 corresponding induced dynamics is included into that section.
 Also in this case, as in finite dimensional ones, the Poisson ``manifold''
 \Ss\ decomposes into ``symplectic leaves'' left by all Hamiltonian flows
 invariant. All these leaves are homogeneous spaces (i.e.\ orbits) of the
 unitary group \fk U\ of the Hilbert space \H\ \wrt its natural
 coadjoint action. There are, however two kinds of these orbits (leaves):
 The ``finite dimensional'' ones consist of density matrices of finite range
 (i.e.\ only finite number of their eigenvalues are positive), and
 the induced symplectic structure is ``strongly nondegenerate'', the tangent
 spaces having a canonical Hilbert space structure; these properties make these
 symplectic leaves in some sense similar to finite dimensional symplectic
 manifolds. The ``infinite dimensional'' leaves consisting of density
 matrices with infinite numbers of nonzero eigenvalues are only ``weakly
 symplectic'', and the naturally defined ``tangent spaces'' are not closed
 in their (again  ``naturally chosen'') topology. The set of ``finite
 dimensional'' leaves
 is, fortunately, dense in the whole \Ss, so that we can restrict, for many
 purposes, our attention onto them.

 Section~\ref{IIB;gener} contains an analysis of questions connected with
 unboundedness of generators (i.e.\ ``Hamiltonians'') of group actions
 corresponding to linear, as well as to nonlinear cases. The unbounded
 generators always appear in any description of ``nontrivial'' actions of
 noncompact Lie
 groups, and cannot be avoided in the considered framework. The domain
 problems and description of the induced dynamics (flows) are solved in
 the cases when the (nonlinear) generators are
 constructed in a certain way from a continuous unitary representation of a
 Lie group $G$. For more general cases, we formulate at least some proposals.

 In the last Section~\ref{IIC;symm-obs}, interconnections between all the
 sets (i) -- (iv) of problems are especially obvious. Introduction of
 ``nonlinear observables'' is a consequence of the nonlinear dynamics.
The interpretation of such observables extending the usual one leads (in
the scheme proposed in this work) to introduction of observables as
numerical functions of {\em two
 variables from \Ss}. It is also presented a (preliminary) classification
 of theories
 according to the choice of a Lie group $G$ determining (sub)sets of {\em
 observables, generators}, and {\em states} of the considered (abstract)
 ``physical system''. For a given $G$, a further classification of
 generators, observables, and states is proposed. A general scheme of
 constructions of nonlinear generators and a description of their flows is
 given. The chapter ends with a description of ``nonlinear'' actions of Lie
groups obtained from linear ones by (mathematically perhaps trivial)
``symplectic deformations'', with inclusion of EQM into a (linear) \Ca ic
scheme and with a description of its general symmetries.
The section contains also a description of some interpretation proposals, cf.
 Interpretations~\ref{int;obs},~\ref{int;2.27},~\ref{int;tr-prob},
 and~\ref{intpn;macro-ex}.

\def\autor{{
\ref{sec;II}\quad Extended Quantum Mechanics}}
\section{Elementary Quantum Phase Space}
\label{IIA;q-phsp}
\def\nazov{{
\ref{IIA;q-phsp}\quad Elementary Quantum Phase Space}}
States of a ``considered physical system'' in QM are described (under a
natural continuity requirement) by density matrices $\mrh\in\mSs:=\mfk
T_{+1}\subset\mfk T$\ on the corresponding Hilbert space \H\
(cf.\ next Subsection~\ref{q-phsp;basic}), but
linearity of QM allows often to reduce the theoretical work to the work
with vector states
described by one--dimensional density matrices $\mrh=P_\psi,\
0\neq\psi\in\mH$\ (i.e.\ pure states, if superselections are missing). Any
state of that quantum-mechanically described system is expressible with a
help of these ``elementary vector states''. In nonlinear versions of QM,
operations like symmetry transformations, and specifically time evolutions,
are nonlinear, resp.\ nonaffine; hence, if they are
performed on states described by density matrices, these operations are not
reducible to those on vector states. In EQM, the whole set \Ss\ of density
matrices will play a r\^ole of the set of ``elementary states'' in such an
intuitive sense,
where each density matrix $\mrh\in\mSs$\ is considered as an analogue of a
point of phase space of CM, irrespective of dimension of the range of the
operator \rh. This implies, e.g., that the time evolution of density matrix
states \rh\ of a ``relatively isolated system'' in EQM can be determined only
with a help of determination of corresponding Hamiltonian flow in a \nbhd
of \rh, that can be independent of determination of the flow in
neighbourhoods
of the vector states into which \rh\ can be formally decomposed.

\subsection{Basic mathematical concepts and notation}
\label{q-phsp;basic}
Let \H\  be a separable complex Hilbert space with scalar product ${\it (x,y)}
\ ({\it x,y}\in\mH$) chosen linear in the second factor {\it y}. Let $\mfk{F}
\subset\mfk{T}\subset\mfk{H}\subset\mfk{C}\subset\mLH$\ be the
\glo{$\mfk{F},\mfk{T},\mfk{H},\mfk{C},\mLH$}
subsets of linear operators in \H\
consisting of the all finite--rank, trace--class, Hilbert--Schmidt, compact,
and bounded operators respectively. All these subsets are considered as
complex associative ${}^*$-subalgebras (in fact ideals) of the algebra \LH,
i.e.\ they are also invariant \wrt the involution $a\mapsto a^*$\ defined as
the operator adjoint mapping. The algebras \fk C and \LH\   are \Ca s (\LH\ is
in fact a $W^*$-algebra), and all of them except of \fk F  are Banach spaces
($\equiv$ B--spaces) if endowed with proper norms: \fk T is endowed by the
trace--norm $\|a\|_1:= Tr|a|$\ with $|a|:=(a^*a)^{\frac 12}$, \fk H\ is endowed
by the Hilbert--Schmidt norm $\|a\|_2:=\sqrt{(a,a)_2}$\ corresponding to the
Hilbert--Schmidt scalar product $(a,b)_2:=Tr(a^*b)$\ of operators
$a,b\in\mfk H$, whereas \fk C and \LH\   are endowed with the usual operator
norm (which is equal to the spectral radius for selfadjoint operators)
denoted by $\|a\|$\glo{$\|a\|$}. Here $Tr$\ denotes the trace of the operators
in $\mfk T$.
Note also that \fk T contains all products of
at-least-two Hilbert--Schmidt operators, and each element of \fk T is of
this form; the last statement follows from the polar decomposition of
closed densely defined (hence also bounded) operators in \H~\cite{najm}.

The linear space \fk F  is dense in the Banach spaces \fk T, \fk H,
and \fk C,
and it is dense also in \LH\   in its \sg-strong operator topology.
The Banach space \fk T will be considered also as the topological du\-al
space to the $C^*$-subalgebra \fk C of $\mLH$, the duality being  given
by the bilinear form $(\mrh;\ra)\mapsto \mrh(\ra) := Tr(\mrh\ra)\equiv
\lb\mrh;\ra\rb$\ on
$\mfk C\times\mLH$\ ($\supset\mfk T\times\mLH$);  the
same bilinear form describes the duality between \fk T and\ $\mLH\equiv\mfk
T^*$.

Let us introduce also the\label{PH}
\emm projective Hilbert space~
\PH\    which is
obtained from \H\  as the factor-space consisting of  all  its
one-dimensional complex subspaces $\bx := \{{\it y}\in\mH:{\it y}
=\lambda{\it x},
\lambda\in\mbC\}\ 0\neq{\it x}\in\mH$, with the factor-topology
induced by the norm of $\mH$; it can be
identified with the subset of \fk T consisting of all one-dimensional
orthogonal projections $P_{{\it x}}$\ (projecting \H\  onto $\bx,\ 0\neq{\it
x}\in\mH$)  endo\-wed with the relative topology of the trace-norm
topology, or
with $\msg(\mfk T,\mfk C)$-topology (these topologies are equivalent on
$\mPH$,~\cite{bra&rob}).

We shall also use some elementary concepts of differential ge\-ometry
on (also infinite-dimensional)
manifolds,~\cite{abr&mars,bourb;Lie,3baby,kob&nom,bourb;manif}, see also
our Appendix~\ref{A;geom}, as  well as some concepts of the theory
of \Ca s~\cite{pedersen,bra&rob,dix1,dix2,takesI}, cf.\ also our
Appendix~\ref{B;Ca}, in this paper.

Let us denote by\glo{\Ss}  $\mSs\subset\mfk T$\ the state space of
$\mfk C$; it can be canon\-ically identified with the convex set
of  all  normalized  normal positive linear functionals on
$\mLH:\mSs := \mSs(\mLH) = \mS(\mfk C)$. The general (not
necessarily normal) states $\mS(\mLH)$\ of \LH\   will be denoted
by $\mS$\glo{\cS}. Let \fk U\glo{\fk U} denote the unitary group
\glss \UH~\ of $\mLH:\ru\in\mfk U \eequiv \{\ru\in\mLH\ \&\
\ru\ru^* = \ru^*\ru = I\}\eequiv \ru\in\mUH$, where $I\in\mLH$\ is
the identity operator. Let $\tilde{\mfk A} := \mfk U/\mfk J$\ be
the fac\-tor-group of \fk U with respect to the central subgroup
$\mfk J := \{\ru := \lambda I: |\lambda|=1,\lambda\in\mbC\}$.
Since all \autm s of \LH\   are inner~\cite{pedersen},
$\tilde{\mfk A}$ is isomorphic to the group of all  \autm s (cf.
also~\cite[Vol.I]{varad}): $\alpha\in \maut{\mLH}\imply\exists
\ru\in\mfk U:\alpha({\rm b}) = \ru{\rm b}\ru^* (\forall {\rm
b}\in\mLH)$, and if also $\rv\in\mfk U$\ represents \alp\  in this
sense, then $\ru^*\rv\in\mfk J$. Let $\mgam:\ru\mapsto \mgam_{\ru}
\in\tilde{\mfk A} = \maut{\mLH}$\glo{$\mgam_{\ru}$- inner
automorphisms of \LH } be the corresponding representation of
$\mfk U,\ \mgam_{\ru}({\rm b}) := \ru{\rm b}\ru^*$; the kernel of
\gam is $\mfk J$. Moreover, \fk U (and $\tilde{\mfk A}$) is a
(infinite--dimensional for $\dim\mH=\infty$) Lie group; the Lie
algebra $Lie(\mfk U)$\ of \fk U is the real subspace $\mLH_a
:=\{\rx\in\mLH: \rx^* = -\rx\}$\ of antihermitean elements of
$\mLH$~\cite{bourb;Lie}. Let $[\ra,{\rm b}]\ (:={\rm ab - ba})$\
be the commutator in $\mLH$. We shall use the selfadjoint
generators $\rx^* = \rx\in\mLH_s := i\mLH_a$\ to represent the Lie
algebra elements $i\rx\in \mLH_a$. The Lie bracket will be defined
on \LHs\   as $(\rx;\ry) \mapsto i[\rx,\ry],\ \rx,\ry\in \mLHs$\
what corresponds to the commutator $[i\rx,i\ry]$\ in $\mLH_a :
[i\rx,i\ry] =: i\rz \imply \rz = i[\rx,\ry]$. The Lie algebra of
$\tilde{\mfk A}$\ is $Lie(\mfk U/\mfk J) = \mLH_a/\{\mbR I\}$\ -
the factoralgebra by the central ideal of real multiples of
identity. Let $Ad(\mfk U)$\ be the adjoint  representation  of \fk
U  on $Lie(\mfk U) = i\mLHs$, i.e.\ $Ad(\ru)$\glo{$Ad(\ru)$} is
the restriction of $\mgam_{\ru}$\ to \LHs:
\begin{equation}\label{def;Ad}
           Ad(\ru){\rm b}\equiv Ad(\ru)({\rm b}) := {\rm ubu}^*,\ {\rm
           b}\in\mLHs,\ \ru\in\mfk U.
\end{equation}

The (topological) dual of \LHs\   is the  real  subspace $\mLHs^*$
of $\mLH^*$\ consisting of symmetric bounded linear functionals on
\LH, i.e.\ $\nu\in \mLHs^*\imply \nu({\rm b^*b})\in\mbR\ (\forall
{\rm b}\in \mLH)$, where
$\lb\nu;\ry\rb\equiv\nu(\ry)$\glo{\nu(\ry)} denotes  the value of
$\nu\in \mLH^*$\  on the element $\ry\in \mLH$. The state space
$\mS := \mS(\mLH)$\ is a compact convex subset of $\mLHs^*$, if it
is  endowed with the $w^*$-topology, i.e.\ with the
$\msg(\mLH^*,\mLH)$-topology~\cite[Theorem IV.21]{R&S}. Let
$Ad^*(\mfk U)$\ be the coadjoint representation of $\mfk U\ (\ni
\ru)$\ on $\mLHs^*$:
\begin{equation}\label{eq;2.1}
[Ad^*(\ru)\nu]({\rm b}):=\nu(Ad(\ru^{-1}){\rm b}),\ \nu\in\mLHs^*,\ {\rm
b}\in \mLHs.
\end{equation}

   It is clear~\cite[Chap. 3.2]{bra&rob} that the state-spaces \cS\  and
   \Ss\  are
   both  $Ad^*(\mfk U)$-invariant subsets of $\mLHs^*$.

   Let $\mcl O_{\nu}(\mfk U) := Ad^*(\mfk U)\nu:=\{\mome\in\mLHs^*:\mome=
   Ad^*(\ru)\nu, \ru\in\mfk U\}$\ be the
 $Ad^*(\mfk U)$-orbit of $\nu$. The state space \cS\ decomposes into union of
$Ad^*(\mfk U)$-orbits. Let $\mrh\in\mfk T_s :=\mfk T\cap\mLHs^*$\ be a
density matrix (i.e.\ $\mrh\geq 0,\ Tr\ \mrh=1$)
describing equally denoted state $\mrh\in \mSs:\mrh({\rm b}) :=Tr(\mrh{\rm
b})\equiv\lb\mrh;{\rm b}\rb,\
{\rm b}\in\mLH$\glo{$\mrh({\rm b})$}.

We shall use for the density matrices spectral decomposition in the form
\begin{equation}\label{eq;2.2}
\mrh=\sum_{j\geq 1}\mlam_jE_j,
\end{equation}
\noidt where we choose the ordering of the eigenvalues $\mlam_j>\mlam_{j+1}>0$,
and the spectral projections $E_j$\ are all finite dimensional.
Let us denote $E_0:= I-\sum_{j\geq1}E_j$\glo{$E_j$- spectral
projection of a density matrix}.

\begin{lem}\label{lem;2.1}
 Each orbit $\mcl O_{\mrh}:= Ad^*(\mfk U)\mrh\ (\mrh\in\mSs)$\glo{orbit
 $\mcl O_{\mrh}$} consists of all the
 density matrices which have the same set of eigenvalues  (including
multiplicities). Hence, the state space \Ss\  is $Ad^*(\mfk
U)$-invariant.\hfill\zal
\end{lem}
\begin{proof}
The $Ad^*(\ru)$-mapping is a unitary mapping conserving spectral invariants,
i.e.\ spectrum and spectral multiplicities, cf.\ \cite{halmos}. Hence all the
elements of the orbit $\mcl O_{\mrh}$\ are density matrices with the same
spectra and multiplicities.
The spectral resolution of any density matrix $\mrh'$\ of the same spectral
invariants as \rh\   in~\eqref{eq;2.2} has the form:
\begin{equation*}
\mrh'=\sum_{j\geq 1}\mlam_jE'_j,
\end{equation*}
with equal dimensions of $E'_j$\ and $E_j, \forall j$. Then there is a
unitary operator $\ru$ mapping all the $E_j$'s onto the corresponding
$E'_j$'s for all $j\geq 0$. E.g., one can choose orthonormal bases
$\{{\it x}_m\}$, resp.\ $\{{\it y}_n\}$\ in \H\  containing
subbases, for all $j\geq 0$,\ of $E_j\mH$, and $E'_j\mH$, respectively,
to order them in
accordance with orderings of $E_j$'s, i.e.\ so that $E_j{\it x}_k={\it
x}_k\eequiv E'_j{\it y}_k={\it y}_k$\ and define
$\ru$ by the formula
\begin{equation*}
\ru{\it x}_k := {\it y}_k,\ \text{ for\ all\ } k.
\end{equation*}
Then $\mrh' = Ad^*(\ru)\mrh$, what proves the lemma.
\end{proof}

   Hence, the projective space \PH\    coincides with the orbit
$\mcl O_{\mrh}(\mfk U)$\ with $\mrh^2 =\mrh$, what characterizes
one-dimensional projections \rh\   in $\mH$.


\subsection{The manifold structure of \Ss}
\label{q-phsp;manif}
   We shall now introduce a natural manifold structure on the or\-bit
$\mOr\ (\mrh\in \mSs)$. Let $\mfk U_{\mrh}\subset\mfk U$\ be the
stability  subgroup  for  the
point $\mrh\in\mSs$\ at $Ad^*(\mfk U)$-representation. Let us note that
$\mfk U_{\mrh}$\glo{$\mfk U_{\mrh}$-stability subgroup of \fk U} is the
unitary group of the $W^*$ -algebra \bs{\{\mrh\}'}\ := the (\emm commutant
of the density matrix \bs{\mrh}\ in \bs{\mLH}~), hence it is a Lie group,
and  its  Lie algebra $Lie(\mfk U_{\mrh}) =: i\mfk M_{\mrh}$\glo{$\mfk M_{\mrh}$}
consists~\cite[Chap.3,
\S 3.10]{bourb;Lie} of antisymmetric elements
of the commutant $\{\mrh\}'$. The proof of the following simple lemma
exemplifies methods used here in dealing with infinite-dimensional spaces.
\begin{lem}~\label{lem;2.2}
The stability subgroup $\mfk U_{\mrh}$\  is a Lie subgroup of $\mfk U$,
\cite{bourb;Lie,bourb;manif}.\hfill\zal
\end{lem}
\begin{proof}
 We shall prove that the Banach subspace $\mfk M_{\mrh}\subset\mLHs$
 has  a topological complement~\cite{bourb;vect} in \LHs, i.e.\ $\mLHs=
 \mfk M_{\mrh}\oplus \mfk N_{\mrh}\ \equiv$ the  topological direct sum
 with a Banach subspace $\mfk N_{\mrh}$\glo{$\mfk N_{\mrh}$} of $\mLHs$.
 Let \rh\   be
expressed in the form~\eqref{eq;2.2}.
We shall use also the projection $E_0$\ corresponding to the eigenvalue \lam
 = 0, hence always $\sum_{j\geq 0} E_j = I$.
Let \bs{{\rm p}_{\mrh}: \ry\mapsto {\rm p}_{\mrh}(\ry) :=
\sum_{j\geq 0} E_j\ry E_j}\glo{${\rm p}_{\mrh}$- a
projection onto $\mfk M_{\mrh}$} be a
projection of $\mLHs$\ onto $\mfk M_{\mrh}$\ defined by the strongly
convergent series.
One has
\begin{equation*}
\|{\rm p}_{\mrh}(\ry)\|\leq \sup_j\|E_j\ry E_j\| \leq
\|\ry\|,\quad\forall\ry\in\mLHs,
\end{equation*}
hence the projection ${\rm p}_{\mrh}$\ is continuous, what implies~\cite[Chap.I.
\S 1.8 Proposition 12]{bourb;vect} the
complementability of $\mfk M_{\mrh}$. The Lie group \fk U can be modeled (as a
 mani\-fold) by its Lie algebra $i\mLHs$\  via the inverse of the exponential
mapping~\cite[Chap. III.6.4. Theorem 4]{bourb;Lie}, and the subgroup
$\mfk U_{\mrh}$\  is modeled via the  same  mapping
by  the  complementable  subspace $i\mfk M_{\mrh}\subset \mLH_a$.
This   gives   the result~(\cite[Chap.III.\S 1.3]{bourb;Lie},
\cite[5.8.3]{bourb;manif}).
\end{proof}

\begin{defs}\label{df;2.3}
Let $\mrh = \sum \mlam_jE_j$\ be a density matrix, $\sum_{j\geq 0} E_j = I$,
as above.
\item{(i)} Let \bs{\mqr :\mLHs\rarw \mfk N_{\mrh}}\glo{\qr - a projection onto
$\mfk N_{\mrh}$} be the \emm complementary projection~
to ${\rm p}_{\mrh}$, $\mqr({\rm b})={\rm b}-{\rm p}_{\mrh}({\rm b})$\
(cf.\ proof of
Lemma~\ref{lem;2.2}):
\begin{equation}\label{eq;2.3a}
   \mqr({\rm b}) := \sum_{j\neq k} E_j {\rm b}E_k\ \ \text{ for}\
{\rm b}\in\mLHs,
\end{equation}
\noidt which leaves $\mfk T_s$\ invariant; the sum is here strongly (resp., in
$\mfk T\ni {\rm b}$, trace-norm-) convergent. We shall define $\mqr(X)$\
also for unbounded $X=X^*$\ by the formula\rref2.3a~\ with ${\rm b} := X$\
for those \rh\ for which it is unambiguously defined (i.e.\ the expressions
in the sum and its strong limit exist).
\item{(ii)} Let $\ad^*:\mLHs\rarw\mcl L(\mLHs^*)$\glo{$\ad^*$} be defined by
\bs{\ad^*(\ry): \nu\mapsto \glss\ad^*(\ry)~\nu}:
\begin{equation}\label{eq;2.3b}
  [\ad^*(\ry)\nu](\rz) := i\nu([\ry,\rz]),\ (\forall \ry,\rz\in\mLHs,
\nu\in\mLHs^*).
\end{equation}
\noidt We can see that the space $\mfk T_s$\ is invariant with respect to all
operators
$\ad^*(\ry),\ \ry\in\mLHs$, \cite[Proposition 3.6.2]{pedersen}.
\item{(iii)} Let $\mrh\in\mOn(\mfk U)$. Let us denote \bs{\mTr\mOU :=
\mTr\mcl O_{\nu}(\mfk U):=}$\{\rc\in\mfk T_s:\rc = i[\mrh,{\rm b}],
{\rm b}\in\mLHs
\}$\glo{$\mTr\mOU$}
the set of vectors in $\mfk T_s$\  tangent to the
curves $c_{{\rm b}}: t\mapsto c_{{\rm b}}(t) := Ad^*(\exp(-it{\rm b}))\mrh$\
at $\mrh,\ {\rm b}\in\mLHs$;
these curves cover a neighbourhood of \rh\   on the orbit \OUr=$\mcl
O_{\nu}(\mfk U)\subset\mfk T_s$. We shall also denote $\ad^*_\mrh:\mLHs\rarw
\mTr\mOU,\ {\rm b}\mapsto \mbs{\ad^*_{\mrh}({\rm b}) :=
\ad^*({\rm b})\mrh}\equiv
i[\mrh,{\rm b}]\in\mfk
T_s$\glo{\adrs } for $\mrh\in\mfk T_s$. One can easily  check  that \qr\
leaves
$\mTr\mOU$\ pointwise invariant, i.e.\ $\mTr\mOU\subset\mfk N_{\mrh}$.

\item{(iv)} For an arbitrary $\rc\in\mLHs$, and $n\in
{\Bbb Z}_+\setminus\{0\}$,  let
\begin{equation}\label{eq;2.3c}
 \mber^{(n)}(\rc):=i\sum_{j\neq k}^{\max(j;k)\leq
 n}E_j\rc E_k(\mlam_k-\mlam_j)^{-1},
\end{equation}
\noidt where in the summation are included also the values $j=0,k=0$\ of the
indices.

 Let $\beta_{\mrh}\glo{\ber}: \mTr\mOU\rarw\mfk N_{\mrh}$\  be the mapping
\begin{equation}\label{eq;2.3d}
 \beta_{\mrh}(\rc):= i\sum_{j\neq k}E_j\rc E_k(\mlam_k -\mlam_j)^{-1}.
\end{equation}
\noidt
The {\bf limits \bs{\mber(c)} of the strongly
convergent sequences} \bs{\left\{\mber^{(n)}(\rc): n\geq 1\right\}} define the
mapping \ber. We shall define \ber(\rc)
in this way also for those $\rc\in\mLH$,  as  well  as  for those unbounded
operators $\rc$, for which this  sequence  is  defined  and  converges
strongly in $\mLH$.
\item{(v)} Let $\|\rc\|_{\mrh}:=\|\mber(\rc)\|$,\glo{\N\cdot,\mrh~} where
$\|{\rm b}\|$\ denotes the operator
norm of ${\rm b}\in\mfk N_{\mrh}$\ in $\mLHs$.
\item{(vi)} Let \OUr\ be endowed with the
canonical~\cite[Chap.III,\S 1.6]{bourb;Lie}  analytic  manifold structure
of the homogeneous space $\mfk U/\mfk U_{\mrh}$. We shall call  this
structure the \emm canonical manifold structure~
on \OUr, and the notion of the \emm manifold \bs{\mOUr}~
will mean namely the set \OUr\ endowed with this structure.\hfill\pika
\end{defs}

\begin{note}\label{note;2.3d}
\item{(i)}
A direct inspection shows that \ber\  is a linear bijection of $\mTr\mOU$
onto $\mfk N_{\mrh}=\mqr(\mLHs)$: if $\rc:=i[\mrh,{\rm b}]$,
with ${\rm b}\in\mfk
N_{\mrh}$, then $\mber(\rc) ={\rm b}$.
It is the inverse mapping to the mapping $\ad^*_{\mrh}:\mfk N_{\mrh}\rarw
\mTr\mOU$;
let us note that $\ad^*_{\mrh}$\ is $\|\cdot\|\mapsto\|\cdot\|_1$- continuous,
hence also $\|\cdot\|\mapsto\|\cdot\|$-continuous:
$\|[\mrh,{\rm b}]\|\leq\|[\mrh,{\rm b}]\|_1\leq 2\|\mrh\|_1\|{\rm b}\|$.

\item{(ii)}
It is clear that $\rc\mapsto\|\rc\|_{\mrh}$\ is a norm on
\Tr\OU. The mapping $\ad^*_{\mrh}$\ is $\|\cdot\|\mapsto\|\cdot\|_{\mrh}$\
isometric;  the
corresponding ``\N\cdot,\mrh~-topology'' of \OUr\ is finer than  the
``\N\cdot,1~-topology''
induced by the trace-norm topology of $\mfk T_s$.\hfill\dovi
\end{note}

   The following proposition specifies the manifold properties of
the orbits \OUr\ in $\mfk T_s$\ (endowed with its \N\cdot,1~-topology).
\begin{prop}\label{prop;2.4}
Let us consider \Tr\OU\  as the normed space  with
the norm $\mN\cdot,\mrh~$. Then \Tr\OU\  is a B-space, and \ber\  is a Banach
space
isomorphism. This B-space structure on \Tr\OU\  coincides with  the
one induced by the canonical manifold structure of \OU\  on its
(equally denoted) tangent spaces $\mTr\mOU$. Furthermore, the
following four statements (i) -- (iv) are then equivalent:
\item{(i)} $\mrh\in\mSs$\ is finite-dimensional, i.e.\ $\mrh\in\mfk F$; we shall
write also $\dim(\mrh)<\infty$\ in this case.
\item{(ii)} The range \fNr\  of the mapping \qr\   coincides with \Tr\OU\
(considered now as a linear subspace of \fTs).
\item{(iii)} The set \Tr\OU\  is a closed subspace of \fTs.
\item{(iv)} \OUr\ is a regularly embedded~\cite[p.550]{3baby} submanifold of \fTs.
\item{Moreover, one has}:
\item{(v)} For $\mrh\in\mfk F$, the subspace \Tr\OU\ of \fTs\ is reflexive.
\item{(vi)} For any $\mrh\in\mSs$, \Tr\OU\  is dense (in the strong topology of
\LHs) in $\mfk N_{\mrh}:=\mqr(\mLHs)$.\hfill\zal
\end{prop}
\begin{proof}
\fNr\   is a B-subspace of \LHs, and \ber\  is a linear isometry
(hence homeomorphism) of \Tr\OU\  (with the  norm \N\cdot,\mrh~)  onto
$\mfNr$,
what follows directly from definitions, cf.\ Notes~\ref{note;2.3d}.
This gives the first
assertion. The second one follows  because  of complementability
of the space $\mfMr = i\dti Lie(\mfk U_\mrh)$,
$\mfMr\subset\mLHs=\mfMr\oplus\mfNr$,
and the inverse  mapping of  the  mapping $Ad^*(\exp(-i(\cdot)))\mrh:
\mfNr\rarw \mOUr,\ \ra\mapsto Ad^*(\exp(-i\ra))\mrh$, if restricted to an open
neighbourhood of the zero point of $\mfNr$, can be chosen as a chart of the
manifold \OUr.

\item{(i)$\imply$(ii):}
If $\mrh\in\mfk F$, then~\eqref{eq;2.3a} shows that also $\mqr(\ra)\in\mfk F$
for  any  $\ra\in\mLHs$, since \fk F  is an ideal in \LH. The application of the
formula~\eqref{eq;2.3d} to $\rc := \mqr(\ra)$\ shows  that $\mqr(\ra) =
i[\mrh,\mber(\mqr(\ra))]\in \mTr\mOU$. Also, \qr\   leaves \Tr\OU$\subset\mLHs$
pointwise invariant. Hence $\mfNr = \mTr\mOU$.

\item{(i)$\imply$(iii):}
It follows now that for $\mrh\in\mfk F$\ the set \fNr\   is a subset of
\fk T. Since \fNr\ is
closed in the norm-topology of \LHs\   and on the subset $\mfk T_s\subset
\mLHs$
the trace-topology determined by \N\cdot,1~ is finer than the topology of
\LHs\   determined by \N\cdot,{}~, $\mN \rx,{}~\leq \mN \rx,1~\ (\forall
\rx\in\mfk
T)$, it follows that $\mfNr$\ = \Tr\OU\  is closed also in trace-topology,
i.e.\ (iii).

\item{(ii)$\imply$(i):}
Let $\mrh\not\in\mfk F$. Let $e_j\equiv |e_j\rb (j\geq 1)$\ be an infinite
orthonormal  set in \H\  such that $E_je_j = e_j (\forall j)$,
cf.\eqref{eq;2.2}.  Let  us  define $\ra\in\mLHs$\ by the formula
(in the Dirac notation,~\cite{dirac})
\begin{equation}
  \label{eq;2.3e}
  \ra :=\sum_{j\geq 1}\alpha_j(|e_{2j}\rb\lb e_{2j+1}|+|e_{2j+1}\rb\lb
  e_{2j}|),\quad   \|\ra\| < M < \infty.
\end{equation}
We have $\ra = \mqr(\ra)\in\mfNr$\ for any bounded real sequence
$\{\alpha_j\}$, but for some choices of
$\{\alpha_j\}$\ (e.g. $\alpha_j\equiv 1$) one has
$\ra\not\in\mfTs\supset\mTr\mOU$. This proves that $\mfNr\neq\mTr\mOU$.

Let us make now a technical remark providing an alternative proof of the
last statement, as well as a device to further work:
\begin{rem}\label{rem;incomplet}
Let us chose in\rref2.3e~ $\malp_j:=\gamma_j(\mlam_{2j}-\mlam_{2j+1})$,
where $0<\gamma_j\rarw\infty$\ for $j\rarw\infty$, but still $\sum_{j\geq1}
\gamma_j(\mlam_{2j}-\mlam_{2j+1})<\infty$. Such a choice of strictly
positive divergent sequence $\{\gamma_j\}$, for any given
$\mlam_j>0,\sum_{j\geq 1}\mlam_j=1$, is always possible. Then $\ra\in\mfTs$.
Let us now calculate $\mber^{(n)}(\ra)$\ according to\rref2.3c~:
\bequ\label{eq;diverg}
\mber^{(2n+1)}(\ra)=i\sum_{j=1}^{n}\gamma_j \bigl(|e_{2j+1}\rb\lb
e_{2j}|-|e_{2j}\rb\lb e_{2j+1}|\bigr).
\end{equation}
Due to divergence of $\{\gamma_j\}$, it is clear that the result
``$\mber(\ra)$''\ diverges for $n\rarw\infty$, i.e.\ we can obtain
in this way at the best an unbounded operator. This shows that our
$\ra\not\in\mTr\mOU$, although it is still in \fTs. This is another proof of
the inequality $\mfNr\neq\mTr\mOU$, because $\mber: \mTr\mOU\rarw\mfNr$\ is
$\|\dti\|_1\rarw\|\dti\|$--continuous.
\hfill\dovi\end{rem}

\item{(vi):} Since the sequence $\{i[\mrh,\mber^{(n)}(\ra)]: n\geq 1\}\subset
\mTr\mOU$
converges strongly to $\ra\equiv\mqr(\ra)\ (\forall\ra\in\mfNr:=\mqr(\mLHs))$,
it is
seen that \Tr\OU\  (considered as a subspace  of $\mfTs\subset\mLHs$) is
strongly
dense in $\mfNr$. This proves (vi).

\item{(iii)$\imply$(i):} Let us chose
$\ra\in\mfNr\cap\mfTs\setminus\mTr\mOU$.
The preceding considerations also  show that \Tr\OU\
is not closed in \fTs\ if $\mrh\not\in\mfk F$; namely, according to the
Remark~\ref{rem;incomplet}, and the formula~\eqref{eq;diverg}, one can
choose $\ra\in \mfTs\cap\mfNr$\ such that the sequence
$\{\mN{i[\mrh,\mber^{(n)}(\ra)]-\ra},1~: n\geq 1\}$\ converges to zero. This
means that the sequence $i[\mrh,\mber^{(n)}(\ra)]\in\mTr\mOU$\ converges to
$\ra\not\in\mTr\mOU$.

\item{(iv)$\eequiv$(iii):}
   The restriction of the projection $\mqr:\mfTs\rarw\mfTs$\ is  continuous
    also  in  the trace-norm topology, what follows from continuity of
${\rm p}_{\mrh}$\ in that topology: For positive operators $\rc\in\mfk T$, all
    $E_j\rc E_j \geq 0$,  hence
\begin{equation*}
\mN {\rm p}_{\mrh}(\rc),1~=Tr(\sum_jE_j\rc E_j)=\sum_j Tr(E_j\rc E_j)=\sum_j
Tr(E_j\rc)=
Tr(\rc) = \mN \rc,1~,
    \end{equation*}

\noidt and the continuity of ${\rm p}_{\mrh}$\ follows. The equivalence
of  the  norms
\N \cdot,\mrh~  and \N \cdot,1~  on \Tr\OU\  in the case of $\mrh\in\mfk F$\
can
be shown  as follows: Let $\rc:=i[\mrh,{\rm b}]\in\mTr\mOU\imply\mqr({\rm
b})=\mber(\rc)$,  and
from~\eqref{eq;2.3d} and the definition of the norm \N\cdot,\mrh~ we obtain
\begin{equation}\label{eq;2.3f}
               \mN \rc,\mrh~\leq\left(\sum_{j\neq k}|\mlam_j-\mlam_k|^{-1}
\right)\mN \rc,1~,
\end{equation}

\noidt where the sum is taken over a finite index set. The opposite  inequality
is obtained by the known  property  of  the  trace-norm:
\begin{equation*}
\mN \rc,1~\equiv\mN{[\mrh,\mqr({\rm b})]},1~\leq 2\mN\mqr({\rm b}),{}~
\mN\mrh,1~=2\mN \rc,\mrh~,
\end{equation*}

\noidt since $\mqr({\rm b})=\mber(\rc)$, and \N \mrh,1~  = 1. This
fact, and the derived implications  of  finite dimensionality of
$\mrh\in\mfk F$\ give  the validity
 of  the  assertion (iv).\footnote{The {\em regularity} of the embedding
 is not proved here in detail; it is not necessary for validity of
 this proposition (if modified by simple scratching out the word
 `regularly' in the item (iv)), and it is not used in the
 following text of this work. A completion of the proof of the
 full text of this proposition can be found in \cite{bon-orbit}, or on in:
 \url{arXiv.org/math-ph/0301007.}}
  It is clear that (iv) cannot be  true  if
 (iii) were not valid.

 \item{(v)} If $\mrh\in\mfk F$, then the B-space \Tr\OU\ is a
 Hilbert space, cf.\ Theorem~\ref{thm;2.10}, hence \Tr\OU\ is reflexive.
\end{proof}

The proved Proposition~\ref{prop;2.4} shows, that only finite-dimensional
density matrices \rh's generate Ad$^*(\mfk U)$\ orbits with mathematically
convenient properties: Their tangent spaces are in the \fTs-induced
topology closed and reflexive. This has important consequences for the
following theoretical implications. Hence, we ask the question, whether it
would be possible to restrict our attention, in some appropriate sense,
to these ``finite dimensional orbits'', and simultaneously not to loose the
control on the whole
space \Ss. The next lemma indicates, that it might be possible.

\begin{lem}\label{lem;dens}
The set-union of the orbits $\{\mOUr:\mrh\in\mfk F\}$\ is a dense subset of
\Ss\  $\subset \mfTs$, in the norm-topology of \fTs.\hfill\zal
\end{lem}
\begin{proof}
Any density matrix $\mrh\in\mSs$\ is approximated in \N \cdot,1~  by finite
dimensional ones, what is seen, e.g.\ from its spectral resolution:
\begin{equation*}
\mrh=\sum_j \mlam_j E_j = \mN \cdot,1~-\lim_{n\rarw\infty}\kappa_n\sum_{j=1}^n
\mlam_j E_j,\ {\rm with}\ \kappa_n:=\left(\sum_{k=1}^n\mlam_k
\dim(E_k)\right)^{-1}.
\end{equation*}
\end{proof}


\subsection{Poisson structure on quantum state-space}
\label{q-phsp;poiss}

We shall consider the set \Ss\  ($\subset
\mLH^*$) as the set of relevant physical states in the following
considerations, i.e.\ the \emm quantum phase space~ will mean for us the set
of \emm normal states~.\footnote{It might be mathematically
interesting, and, perhaps, also physically useful, to formulate analogies
of the following constructions on the
space \cS\  of all positive normalized functionals on \LH. This leads to
technical complications and, for purposes of our physical interpretations,
it would be unnecessary. cf.\ also~\cite{bon10}, where a
(heuristic) trial for such a formulation was presented. A
nice and useful property of \cS\ is its compactness in the $w^*$-topology,
what is not the case of \Ss.}

Let us now introduce a Poisson structure~\cite{marle,weinst},
\cite[Appendix 13]{arn1}
on the linear space \fTs\ containing \Ss\   as a bounded convex subset. The
Poisson structure will allow us to ascribe (Poisson-) Hamiltonian
vector fields (on \fk F, at least) with the corresponding flows leaving the
state space \Ss\ invariant.

It will be useful to use, in the following mathematical formulations, the
standard differential calculus on Banach
manifolds~\cite{jt-schw,bourb;manif,3baby}
based on the Fr\'echet differential calculus of mappings between (linear)
Banach spaces~\cite{h-cartan,l-schw,3baby}.\footnote{Let us note, for a
preliminary information, that in this infinite--dimensional differential
calculus
``most'' of the usual differential operations in finite--dimensional
spaces remain formally, under certain conditions, unchanged: the
differential is the ``linear part of difference'', where should be used the
Banach-norm limit for its definition. The rules for writing the Taylor
expansion, differential of composed maps, for calculation of derivatives of
``products'' etc. have the same formal expressions as in finite--dimensional
case, see also the Appendix~\ref{A;diff}.}

\noidt If the Fr\'echet derivative of a function $f: \mfk T\rarw \mbR$\ exists,
then there exists also directional
(so called Gateaux) derivative:
\begin{equation}\label{eq;2.4}
\mD f,\nu~(\mome)=\lim_{t\rarw
0}\frac{1}{t}\left[f(\nu+t\mome)-f(\nu)\right],\ \forall \mome\in\mfk T.
 \end{equation}

 Conversely, if the Gateaux derivative~\eqref{eq;2.4} exists in a \nbhd $U$\ of
 a point $\nu\in\mfk T$, and {\em if it is continuous linear, continuously
 depending on} $\nu\in U$,\ $\mD f,\cdot~:$\
 $ U \rarw {\cal L}(\mfk T,\mbR)$, then also the Fr\'echet
 derivative~\eqref{eq;frechet} exists~\cite{h-cartan}.

   We shall be mainly interested, in the following text, in the
$Ad^*(\mcl U)$-invariant  subset
\Ss\     consisting of normal states on \LH. Let $\mcl F := \mcl F(\mfTs)$
 be the
set of infinitely norm-differentiable real functions on \fTs, with its
trace-norm \N \cdot,1~.
Let us denote  by \glss \F(\B)~
the set of the restrictions of $f\in\mF$\ to some subset \B\ of
\fTs.
\begin{rem}
Noncompactnes of \Ss\   allows, e.g.\ that $\mF(\mSs)$\ contains
also unbounded functions on \Ss, e.g.\ any $f\in\mF$\ with the restriction
$f:\mrh\mapsto f(\mrh) := \ln(Tr(\mrh^2))$\ for $\mrh\in\mSs\subset\mfTs$\ is
unbounded. Put, e.g., with orthonormal $e_j$'s,
$\mrh_N:=\sum_{j=1}^N\frac{1}{N}|e_j\rb\lb e_j|$, whence
$Tr\mrh_N^2=\frac{1}{N},\ \lim_{N\rarw\infty}\ln
Tr(\mrh^2_N)=-\infty$.\hfill\dovi
\end{rem}
 The definition of the
F-derivative and its expression~\eqref{eq;2.4} also apply to $f\in\mF$,  and
the notation \D f,\nu~ will not lead to any ambiguity for $f\in\mF$.

We shall often work with infinite--dimensional manifolds modelled by Banach
(specifically, e.g., in the case of pure state manifold \PH, or of any
\OUr\ with $\dim\mrh<\infty$, by Hilbert) spaces, cf.
Appendix~\ref{A;manif}. The main ideas, and many of general constructions
and theorems work in that cases similarly as in the case of more common finite
dimensional
manifolds. We shall point out differences in specific cases, if it will be
needed.
In the case of the linear manifold \fTs, and for a differentiable
function $f\in\mF$, the derivative \D f,\nu~ belongs to the cotangent space
\Tns(\fTs)$=\mLH$, and we shall deal with it also as with an operator in the
sense of this canonical isomorphism.

\begin{defs}\label{df;2.5}
\item{(i)} Let \Fr\ denote the algebra $C^{\infty}(\mOr(\mfk U),\mbR)$\ of
functions  on  the
manifold \OUr. The restrictions of functions from \F\ to \OUr\ belong to \Fr,
because the topology on the manifold \OUr\ is  finer
than the relative topology coming from \fTs\ (cf.\ proof of  Proposi-
tion~\ref{prop;2.4}).
\item{(ii)} The mapping from $\mF\times\mF$\ to $\mF: (f;h)\mapsto
 \{f,h\}$, where\glo{$\{f,h\}$ - Poisson bracket}
 \begin{equation}\label{eq;2.5b}
                \{f,h\}(\nu) := \nu(i[\mD f,\nu~,\mD h,\nu~]),\quad
                \nu\in\mfTs,
\end{equation}
will be called the \emm Poisson structure~ on \fTs. The function $\{f,h\}\in
\mF$\ is the \emm Poisson bracket~ of the functions $f$ and $h$ from \F.
\item{(iii)} The functions $h_{\rm y}\in\mF\ ({\rm y}\in\mLHs)$\ are defined
by $\glss h_{\rm y}(\nu)~ :=\nu({\rm y})\equiv Tr(\nu{\rm y}),\
\forall\nu\in\mfTs$.
Then $\mD{h_{\rm y}},\nu~={\rm y}$, the second derivative $D^2_{\nu}h_{\rm
y}=0$,
and the Poisson bracket of two such functions is
\begin{equation}\label{eq;2.5c}
        \{h_{\rm x},h_{\rm y}\}(\nu)=i\,\nu([{\rm x,y}])\equiv h_{i[{\rm
        x,y}]}(\nu).
   \end{equation}
From this we obtain Poisson brackets for polynomials in functions \h{\rm
x}\ (${\rm x}\in\mLHs$) with a help of derivation property (cf.
Proposition~\ref{prop;2.6}), in accordance
with~\eqref{eq;2.5b}.       \hfill\pika
\end{defs}

   The space \fTs\ can  be  considered  as  an infinite--dimensional manifold
with the atlas consisting of one chart  determined  by the identity mapping
on \fTs. Then  the  tangent space \Tn\fTs\ to \fTs\ at each point $\nu$\
will be canonically identified with the vector  space \fTs\
itself. The space \LHs\   is then  canonically identified with \Tns\fTs.
In this interpretation, we can also consider the derivative (cf.
Appendix~\ref{A;diff})  \D f,\nu~$\in$\Tns\fTs\ as differential of $f\in\mF$
on the manifold \fTs, as  it  is   used   in   differential
geometry.  The usual symbol \d f,\nu~ will be used, however,
to  stress  the  restriction  of  the  differentiation  to   some
``smaller'' manifold in \fTs. For a real  function  $f$  continuously
differentiable as a function on  the  manifold \OUr\  we  shall
denote by \d f,\nu~ the differential of $f$ in the point $\nu$\ on  the  orbit
\OUr\ $(\ni\nu)$. We shall also identify $\md f,\mrh~:=\mqr(\mDfr)\in \mfNr
\subset \mLHs$\  considered as an element of the
cotangent space $\mTrs\mOU\ := (\mTr\mOUn)^*
:= \mTrs\mOUn$; this identification (resp.\ representation) of the cotangent
space is possible due to the identities:
\begin{equation}\label{eq;2.rep}
\begin{split}
{} &\mdfr(c) :=  Tr(c\mqr(\mDfr)) =
i\,Tr([\mrh,\mber(c)]\mqr(\mDfr))  =              \\
&i\,Tr(\mber(c)[\mqr(\mDfr),\mrh]) = i\,Tr(\mber(c)[\mDfr,\mrh]) =   \\
&i\,Tr([\mrh,\mber(c)]\mDfr)  =
Tr(c\mDfr),\ \text{for all}\ c\in\mTr\mOU.
\end{split}
\end{equation}
The  operator \dfr\ represents  the
pull-back of $\mDfr \in \mTrs\mfTs$\ with respect to the  embedding  of \OUr\
into \fTs, if $\mrh\in\mfk F$. Now we can write the Poisson  bracket  in  the
form:
\begin{equation}\label{eq;2.6}
   \{f,h\}(\nu) = i\nu([\mdfn,\mdhn]).
\end{equation}
The form~\eqref{eq;2.6} shows, that the value of the Poisson
bracket~\eqref{eq;2.5b} in a point $\nu\in \mfTs$\ depends on the restrictions
of the functions $f,h\in
\mF$\ onto the orbit \OUn\ only. This is due to the fact, that the orbits
\OUr\ are the ``symplectic leaves'' of the Poisson manifold
\fTs,~\cite{weinst}, as will be seen from the following. The orbits are \emm
Poisson submanifolds~~\cite{weinst} of the Poisson manifold \fTs. We shall
now prove that~\eqref{eq;2.5b} really determines a structure of a \emm Poisson
manifold~ on the Banach manifold \fTs:

\begin{prop}\label{prop;2.6}
The Poisson bracket from \eqref{eq;2.5b} has all the general properties of
the Poisson structure~\cite{weinst,arn1} (coinciding with that of
Hamiltonian classical mechanics, except of nondegeneracy), i.e.\ for all
$f,h,k\in\mF$, and all $\mlam\in \mbR$\ one has:
\[ \begin{array}{cll}
(i)   &\{f,h\}=-\{h,f\};  & \text{(antisymmetry)} \\
(ii)  &\{f,h+\mlam
k\}=\{f,h\}+\mlam\{f,k\};&((ii)\&(i)\imply\text{bilinearity})\\
(iii) &\{f,hk\}=\{f,h\}k+h\{f,k\}; &\text{(derivation property)}\\
(iv)  &\{f,\{h,k\}\}+\{h,\{k,f\}\}+\{k,\{f,h\}\}=0;&\text{(Jacobi
identity)}
\end{array}  \]\hfill \zal

 \end{prop}
\begin{proof}
The first three properties are immediate consequences of
Definitions~\ref{df;2.5}, cf.\ also \eqref{eq;2.7}. The validity of (iv)
follows immediately from \eqref{eq;2.5c} and from the properties (i) -
(iii) for such functions $f,h,k$\ which have form of polynomials in the specific
type of functions $h_{\rm a}\in\mF,\ \ra\in\mLH$,~\eqref{eq;2.5c}. For general
$f,h,k$\ one can prove (iv) directly as follows:

\noidt Let us first express $\mD{\{h,k\}},\nu~\in \mLHs$\ according to
~\eqref{eq;2.4},~\eqref{eq;2.5b},
\begin{eqnarray*}
\mome(\mD{\{h,k\}},\nu~)&=&\left.\frac{d}{dt}\right|_{t=0}(\nu+t\mome)(i\,[\mD
h,\nu+t\mome~,\mD k,\nu+t\mome~])\\
&=& \mome(i\,[\mD h,\nu~,\mD
k,\nu~])+\nu(i\,[D^2_{\nu}h(\mome,\cdot),\mD k,\nu~])+\nu(i\,[\mD
h,\nu~,D^2_{\nu}k(\mome,\cdot)]),
\end{eqnarray*}
where the second derivatives in any point $\nu$\ are symmetric bilinear \N
\cdot,1~-continuous functions on \fTs. Hence, the linear mapping
$D^2_{\nu}k(\mome,\cdot):\mrh\mapsto
D^2_{\nu}k(\mrh,\mome)=D^2_{\nu}k(\mome,\mrh)\equiv
\mrh(D^2_{\nu}k(\mome,\cdot))$
can be (and is here) considered as an element of \LHs. We need to calculate
$\{f,\{h,k\}\}(\nu) := i\,\nu([\mD f,\nu~,\mD{\{h,k\}},\nu~]).$\ With a help
of the notation~\eqref{eq;2.3b} and of the above derived formula for
\D{\{h,k\}},\nu~ we obtain
\begin{eqnarray*}
\{f,\{h,k\}\}(\nu)&=&[\ad^*(\mD f,\nu~)\nu](\mD{\{h,k\}},\nu~) \\
&=&-\nu([\mD f,\nu~,[\mD h,\nu~,\mD k,\nu~]])-[\ad^*(\mD
k,\nu~)\nu](D^2_{\nu}h(\ad^*(\mD f,\nu~)\nu,\cdot))+\\
&&[\ad^*(\mD h,\nu~)\nu](D^2_{\nu}k(\ad^*(\mD f,\nu~)\nu,\cdot)) \\
&=&-\nu([\mD f,\nu~,[\mD h,\nu~,\mD k,\nu~]])-D^2_{\nu}h(\ad^*(\mD
f,\nu~)\nu,\ad^*(\mD k,\nu~)\nu)+\\
&&D^2_{\nu}k(\ad^*(\mD f,\nu~)\nu,\ad^*(\mD h,\nu~)\nu).
\end{eqnarray*}
From the symmetry of second derivatives, and from validity of Jacobi
identity for commutators of operators in \LH, we obtain the result.
\end{proof}


\subsection{Hamiltonian vector fields and flows}\label{q-phsp;flows}

In the case of a finite--dimensional Poisson manifold M, the Poisson structure
determines a vector field \vf\ to each differentiable function $f$ on M,
so called \emm Hamiltonian vector field~ corresponding to the \emm Hamiltonian
function~ $f$:
\begin{equation}\label{eq;2.7}
\mL \mvf~(h)\equiv dh(\mvf):=\{f,h\},
\end{equation}
where \L{\mathbf v}~ denotes the \emm Lie derivative~ (uniquely extendable to a
derivation of any tensor field on M,~\cite{kob&nom,bourb;manif})
\wrt the vector field {\bf v}: The Poisson bracket $\{f,h\}(\nu)$\ at fixed
$f$ and $\nu$\ is a (first order)
differential operator on real valued functions differentiable at $\nu$, which
determines unique - in the case of finite--dimensional M - vector
$\mvf\in\mTn$M.

In the case of infinite--dimensional manifolds, the relation between
(first order) differential operators and tangent vectors is not always an
isomorphism of normed spaces,~\cite{3baby}. The following lemma is,
however, valid,~\cite[Chapter VII.A.1]{3baby}:
\begin{lem}\label{lem;reflex}
Let M be a manifold modeled by a Banach space E, hence the tangent spaces
$T_mM,\ m\in M$, are isomorphic to E. Let us assume that E is reflexive: E =
E$^{**}$\ (:= the double topological dual of E). Let a differential
operator satisfying the Leibniz rule (i.e. a derivation)
$\Delta: \mF(U)\rarw\mF(U),\
f\mapsto\Delta f,\ U\subset M$\ (with
domain $U$ of a chart $(U;\mphi;E)$\ containing $m\in M$), satisfy the
following inequality for a $K>0$:
\begin{equation}\label{eq;reflex}
|\Delta f(m)|\leq K\,\mN{\mD{(f\circ\mphi^{-1})},\mphi(m)~},E^*~.
\end{equation}
Then the operator $\hat{\Delta}_m: f\mapsto \Delta f(m)$\ can be
identified with the vector $\Delta_m\in T_mM \cong E^{**}:
\Delta_m(d_mf):=\Delta f(m)$.\hfill\zal
\end{lem}
\begin{proof}
The equation~\eqref{eq;reflex} shows that the kernel of the
operator $\hat{\Delta}_m$\ contains the kernel of \d f,m~, and
also is bounded. Hence, it is defined as bounded linear functional
on $T_m^*M\ (\ni\md f,m~\ \forall f\in\mcl F(U)$), i.e.\ as an
element of $(T_m^*M)^*\cong E^{**}=E$.
\end{proof}

Let us check validity of \eqref{eq;reflex} for the Poisson bracket
$\hat{\Delta}_{\nu}(\cdot):= \{h,\cdot\}(\nu)$:
\begin{equation*}
 |\{h,f\}(\nu)|\leq
 2 \mN \nu,1~\,\mN{\mD h,\mrh~},\mLH~\,\mN{\mD f,\mrh~},\mLH~.
\end{equation*}
Reflexivity of the tangent spaces \Tn\OU\ is the case for
``finite--dimensional'' orbits \OUn, cf.\ Proposition~\ref{prop;2.4}(v).

\begin{rem}\label{rem;pois-embed}
The considerations preceding \eqref{eq;2.6}
show, that the Poisson bracket $\{h,f\}$\ in a point $\nu\in\mfk F\cap\mSs$\
can be
calculated with a help of restrictions $h\rceil_{\mOUn},
f\rceil_{\mOUn}$\ only,
cf.\
\eqref{eq;2.6}, i.e.\ the orbits \OUr\ are themselves Poisson manifolds
regularly embedded into \fTs, and this embedding is a Poisson
morphism~\cite{weinst}.\hfill\dovi
\end{rem}

Let us restrict our attention, for a while, to ``finite--dimensional'' orbits
\OUn.
From the Lemma~\ref{lem;reflex} and the above mentioned
facts we can
see that to any $f\in\mF$\ there is associated, for any $\nu\in\mfk F$, the
\emm Hamiltonian vector field~ \vf\ on \OUn, $\mvf(\mrh)\in \mTr\mOUn$,
expressed by
\begin{equation}\label{eq;2.8a}
\mvf(\mrh)=\madrs(\mdfr)=\madrs(\mDfr).
\end{equation}
\noidt Note that \dfr = \qr(\Dfr) $\in$ \fk F\ for all $\mrh\in\mOUn$, and
that \Dfn $\in$ \LHs, hence the unitary group
\begin{equation*}
\ru_{f,\nu}: t\mapsto \ru_{f,\nu}(t) := \exp(-it\mDfn)
\end{equation*}
generates a curve on \OUn\ = \OUr\ determining \vf($\nu$):
\begin{equation}\label{eq;2.8b}
\mdhn(\mvf)=\left.\frac{d}{dt}\right|_{t=0}h(Ad^*(\ru_{f,\nu}(t))\nu).
\end{equation}
This again indicates the ``usual'' (i.e.\ as in finite--dimensions) connection
between differentiable curves and tangent vectors $\mvf(\nu)\in\mTn\mOUr$.

\begin{note}\label{note;eachv}
\item{(i)} Each element of \LHs\   can be written in the form \Dfn\ for some
smooth real-valued  function $f\in\mF$: For a given b $\in\mLHs$\ one can
chose $f(\nu):=Tr({\rm b}\nu)$; then \Dfn = b.
\item{(ii)} The reflexivity of \Tr\OU, for $\mrh\in\mfk F$,\ implies that each
vector
${\mathbf v}\in \mTr\mOU$\ is of the form\rref2.8a~ for some \Dfr $\in \mLHs$.
\item{(iii)} Although the Hamiltonian vector fields were defined on
orbits \OUn\ for $\nu\in\mfk F$\ only, they are extendable by~\eqref{eq;2.8a}
to the whole space \fTs:
\begin{equation}\label{eq;extend}
\mvf: \mfTs\rarw\mfTs,\ \nu\mapsto \mvf(\nu):= \madns(\mDfn).
\end{equation}
Since
\begin{equation*}
\mN{\mvf(\nu)},1~=\mN{[\nu,\mDfn]},1~\leq 2\,\mN \nu,1~\,\mN{\mDfn},{}~,
\end{equation*}
and the function $\nu\mapsto \mDfn$\ is infinitely (continuously)
differentiable, the uniqueness of the extension of \vf\ to \fTs\ follows from
the density of \fk F\ in \fk T. \hfill\dovi
\end{note}

\begin{defi}\label{df;ham-vf}
Let $f\in\mF,\ \nu\in\mfTs$, and let $\mvfn\in\mTn\mOUn\subset\mTn\mfTs$\ be
determined by
equation~\eqref{eq;2.8a}. The smooth vector field $\nu\mapsto\mvfn$\ is
called the \emm Hamiltonian vector field~ on \fTs.\hfill\pika
\end{defi}

Now we could proceed further also with the Hamiltonian vector fields \vf\
restricted to ``finite--dimensional''
orbits \OUr\ being the Hamiltonian vector fields on Poisson manifolds \OUr,
$\mrh\in\mfk F$.

Each \vf\  from~\eqref{eq;2.8a} ($f\in\mF$) determines a differential
equation~\cite{bourb;manif} on the infinite--dimensional Banach manifold
\fTs\  with a \emm maximal solution \bs{\mpph {},f~}~,
$\mpph {},f~(t,\mrh)\in\mfTs$,
defined on an open domain in $\mbR\times\mfTs\ni(t;\mrh)$\ containing
$\{0\}\times\mfTs$. For values of $t_j$'s for which the objects
entering into~\eqref{eq;2.9} are
defined,  the formula
\begin{equation}\label{eq;2.9}
\mpph {},f~(t_1+t_2,\mrh)\equiv \mpph{},f~(t_2,\mpph{},f~(t_1,\mrh))
\end{equation}
is satisfied.
If the domain is the whole $\mbR\times\mfTs$, what means that the \emm
vector field \vf\ on \bs{\mfTs} is complete~, we obtain a one-parameter
group of
diffeomorphisms \pph t,f~ ($t\in \mbR$) of \fTs:
\begin{equation}\label{eq;2.10}
\mpph t,f~(\mrh) := \mpph{},f~(t,\mrh)\ \text{for all}\
t\in\mbR,\mrh\in\mfTs.
\end{equation}

\noidt We shall now express the (local) flow \pph t,f~, i.e.\ the solution
of Hamilton's equations (obtained by combining~\eqref{eq;2.7}
and~\eqref{eq;2.8b}, or\rref2.5b~), in a form of Schr\"odinger (resp.\ Dyson) equation.

\begin{prop}\label{prop;2.7}
Let $\nu\in\mfTs,\ f\in\mF, \nu(t):=\mpph t,f~(\nu)~$\ for $t$ in an open
interval $J_{\nu}\subset\mbR$\ containing zero. Let we represent the
differentials \Dfn, resp.\ \dfn\ of $f\in\mF$\ by operators in \LH,
e.g., as above: $\mdfn:=\mqn(\mDfn)$. Let us consider the equation
\begin{equation}\label{eq;2.11}
i\,\frac{d}{dt}\mun f,t,\nu~=\md f,\nu(t)~\cdot\mun f,t,\nu~,
\end{equation}
where $\md f,\nu(t)~\cdot$\ denotes the (left) multiplication in the
algebra \LHs.
The equation~\eqref{eq;2.11}, with the initial condition \un
f,0,\nu~$\equiv I_{\mH}$, has a
unique (unitary) solution $t\mapsto\mun f,t,\nu~\in\mLH,\ t\in J_{\nu},\
\nu\in\mfTs$. This solution satisfies the ``cocycle identity''
\begin{equation}\label{eq;2.12}
\mun f,t+s,\nu~=\mun f,s,{\mpph t,f~\nu}~\mun f,t,\nu~
\end{equation}
for those $t,s\in J_{\nu}$, for which both sides of~\eqref{eq;2.12} are
defined. One has, moreover,
\begin{equation}\label{eq;2.13}
\mpph t,f~\nu:=\mpph t,f~(\nu)=Ad^*(\mun f,t,\nu~)\nu,
\end{equation}
and this, together with~\eqref{eq;2.12} shows fulfillment
of~\eqref{eq;2.9}.\hfill\zal
\end{prop}
\begin{proof}
Unique solvability of~\eqref{eq;2.11} on each interval $J'_{\nu}\subset
J_{\nu}$\ on which the function $t\rarw \mN{\md f,\nu(t)~},{}~$\ is uniformly
bounded follows
from  general  theory  of  differential   equations   in   Banach
spaces, cf.\ \cite[Chap.V.\S 2.Theorem 4]{l-schw}.  Unitarity and the
property~\eqref{eq;2.12} can be
proved,  e.g.\ by the method of the proof of~\cite[Theorem X.69]{R&S} using
the Dyson expansion, since $t\mapsto \md f,\nu(t)~$\ is norm-continuous.
Finally,~\eqref{eq;2.13} can be verified by differentiation and by the
uniqueness of the local flow \pph {},f~  of the vector field \vf.
\end{proof}

\begin{note}\label{note;dyson}
\item{(i)} The equation~\eqref{eq;2.11} is a generalized form of the Dyson
equation known from QM, which in turn is a time-dependent generalization of
Schr\"odinger equation. For $f(\nu)\equiv h_{{\rm H}}(\nu):= Tr(\nu{\rm
H})$, with H $\in\mLHs$, and with $\nu\in\mPH$, the equation reduces to
the Schr\"odinger equation with the Hamiltonian H.
\item{(ii)} The substitution $\nu(t):=Ad^*(\mun f,t,\nu~)\nu$\ into\rref2.11~
makes that equation for $\mun f,t,\nu~$\  manifestly nonlinear.
 We shall see in Section~\ref{sec;IIIF} that the
equation~\eqref{eq;2.11}
can be equivalently rewritten, in the case $\nu\in\mPH$, into the form of
the nonlinear version of QM proposed in~\cite{weinb}, and also into the more
traditional versions of ``nonlinear Schr\"odinger equations'', cf.
Subsection~\ref{IIIA1;NL-Sch}.
\item{(iii)} The equation~\eqref{eq;2.13} shows, that the obtained form of
Hamiltonian flows on ``quantum phase space'' \fTs\  can be expressed with a
help of coadjoint action of the unitary group \fk U\ of the algebra \LH,
hence it leaves invariant the orbits \OUn. This gives the invariance of the
quantum state space \Ss, as it is formulated in the following theorem.\hfill\dovi
\end{note}

\begin{thm}\label{thm;2.8}
Let $f\in\mF,\ \mrh\in\mSs$. Then \OUr\ is \pph{},f~-invariant. Hence
also \Ss\   is \pph{},f~-invariant.\hfill\zal
\end{thm}
\begin{proof}
 The result follows from the relation~\eqref{eq;2.13} showing  that
\pph t,f~ can be realized by the $Ad^*(\mfk U)$-action, and \Ss\   consists
of  the $Ad^*(\mfk U)$-orbits \OUr, $\mrh\in\mSs$.
\end{proof}

Let us specify non-uniqueness of cocycles~\eqref{eq;2.12}
satisfying~\eqref{eq;2.13}. We obtain ``physically equivalent'' evolution
equations connected by a ``gauge transformation'', cf.\ also
Section~\ref{sec;IIIF}, Remark~\ref{rem;2.9}, and
Proposition~\ref{prop;AxC-symm}.

\begin{rem}\label{rem;2.9}
 The cocycle $u_f$\ satisfying~\eqref{eq;2.13} is nonunique.
The same evolution
\pph {},f~ is obtained also from the solutions $u'_f$
of the equations resulting after the insertion \dfn\ +\ ${\mathbf
f}^0(\nu)$\ in the place of
\dfn\ into~\eqref{eq;2.11}, where
${\mathbf f}^0 :\nu\mapsto {\mathbf f}^0(\nu)$\ is a norm-continuous function
from \Ss\   (or from the whole \fTs)
 to \LHs\   with values in $\mfk M_{\nu} = i\dti Lie(\mfk U_{\nu})$, i.e., as
 an operator in \LH, the value ${\mathbf f}^0(\nu)$\ commutes with the
 operator $\nu$\ for any $\nu$:
\begin{equation}\label{eq;2.14}
         i\,\frac{d}{dt}\mun f,t,\nu~= \left[\md f,\nu(t)~+{\mathbf
         f}^0(\nu(t))\right]\cdot\mun f,t,\nu~.
 \end{equation}

\noidt Specifically, one can use \Dfn = p$_{\nu}$(\Dfn) + \qn(\Dfn) instead of
\dfn := \qn(\Dfn)  in~\eqref{eq;2.11}. Let us mention, moreover, that the
continuity
requirement to the function $t\mapsto \md f,\nu(t)~ +{\mathbf f}^0(\nu(t))$\ in
the assumptions of the Proposition~\ref{prop;2.7} can be weakened: For validity
of the  conclusions as well as of the proof of the proposition it suffices to
assume strong-operator continuity of this  ``time-dependent  Hamiltonian''
together  with  its  locally  uniform  (in   the   parameter   t)
boundedness.\hfill\dovi
\end{rem}

Now we shall investigate the geometry of manifolds \OUr\ for
``finite--dimensional'' $\mrh\in\mfk F$, especially a naturally determined
metric and
symplectic structures on them. It will be seen in the Section~\ref{sec;IIIA}
that the obtained structure leads to the standard symplectic, and also
metric (known as the ``Fubini-Study metric'') structures on the space of pure
quantum states \PH, this both structures connected by complex structure
coming from that in the underlying Hilbert space \H\ (this is called a
K\"ahlerian structure):

\begin{thm}\label{thm;2.10}
 Let $\dim \mrh < \infty$. Let  us  define  a  complex-valued
tensor field $\Psi:\mrh\mapsto
\Psi_{\mrh}\equiv\Gamma_{\mrh}-i\,\Omega_{\mrh}$\  on the manifold \OUr,
where $\Gamma_{\mrh}$\ and $\Omega_{\mrh}$\ are real two-covariant tensors on
\Tr\OU\ ($\ni\mbf v,\mbf w$):
\begin{subequations}
\label{eq;2.15a}
\begin{equation}
\Psi_{\mrh}(\mbf{v,w}):=\Gamma_{\mrh}(\mbf{v,w})-i\,\Omega_{\mrh}(\mbf{v,w}):=
2Tr\left(\mrh\,\mber(\mbf v)\mber(\mbf w)\right).
\end{equation}
Then the B-space \Tr\OU\ is a real Hilbert space with scalar  product
$\Gamma_{\mrh}$\  endowed
 also with the two-form $\Omega_{\mrh}$\ (here $[\cdot,\cdot]_-$\ is the
 commutator, and $[\cdot,\cdot]_+$\ is the anticommutator in \LH, and \ber\
is as in\rref2.3d~):
\begin{equation}
\Gamma_{\mrh}(\mbf{v,w})\equiv Tr(\mrh[\mber(\mbf v),\mber(\mbf w)]_+),\ \
\Omega_{\mrh}(\mbf{v,w})\equiv i\,Tr(\mrh[\mber(\mbf v),\mber(\mbf w)]_-).
\end{equation}
\end{subequations}
 $\Gamma$\ is a Riemannian metrics, and
 $\Omega$\ is  a symplectic form on
\OUr, both are strongly  nondegenerate,~\cite{3baby}. The symplectic form $\Omega$\ ascribes to each
$f\in\mFr$,~\ref{df;2.5}, the vector field \vf:
\begin{equation}\label{eq;2.15b}
\Omega_{\nu}(\mvf,\mbf w)\equiv -\mdfn(\mbf w),
\end{equation}
 coinciding with \vf\ from~\eqref{eq;2.8a} for $f\in\mF(\mOUr)$, and the
 corresponding Poisson bracket
 \begin{equation}\label{eq;2.15c}
                     \{f,h\}\equiv \Omega(\mvf,\mvh)
\end{equation}
coincides with the one defined in~\eqref{eq;2.5b} and~\eqref{eq;2.6}.

   Moreover, the following norms are all mutually  equivalent  on
\Tr\OU: \N\cdot,{}~, \N\cdot,1~, \N\cdot,2~, \N\cdot,\mrh~, and
\N\cdot,\Gamma~:= $\Gamma(\cdot,\cdot)^{\frac{1}{2}}$.\hfill\zal
\end{thm}
\begin{proof}
 The equivalence of the norms \N\cdot,1~, and \N\cdot,\mrh~,  as  well  as
the completeness of \Tr\OU\ was proved in Proposition~\ref{prop;2.4}.
To prove equivalence of norms \N\cdot,{}~, and \N\cdot,1~, let us write
$\mrh=\sum_{j\geq 1}\mlam_jE_j$, with $\sum_{j\geq 0} E_j =I,\
\mlam_1>\mlam_2>\dots>\mlam_N,\ \mlam_0:=0$, as before. Let
\rc = \qr(\rc) := $i\,[\mrh,\ra] = i\,[\mrh,\mqr(\ra)]\in\mTr\mOU,\
\forall\ra\in\mLHs$.  Then $\mN
\rc,{}~\leq \mN \rc,1~\leq\sum_{j\neq k}\mN E_j\rc E_k,1~=2\sum_{j>k}\mN
E_j\rc E_k,1~\leq2\sum_{j>k}\mN E_j,1~\mN \rc E_k,{}~\leq 2\mN
\rc,{}~\sum_{k\geq
0}\sum_{j(>k)}\mN E_j,1~$,
where the degeneracy of $\mlam_j$\  equals $\mN E_j,1~ <\infty$\ for $j\neq 0$,
and  the number $N + 1$\ of mutually different eigenvalues $\mlam_j$\  of
\rh\ is
finite. This proves equivalence of the norm \N\cdot,1~ with \N\cdot,{}~, hence
 also their equivalence with \N\cdot,2~, since always $\mN \rc,{}~\leq
 \mN \rc,2~ \leq\mN \rc,1~$.  We have further $\frac{1}{2}\mN \rc,\Gamma~^2 =
 Tr(\mrh\mber(\rc)^2) = Tr(\mrh\,\mqr(\ra)^2) = \mN \mrh\,\mqr(\ra)^2,1~\leq
 \mN\mrh,1~\mN \mqr(\ra)^2,{}~ = \mN \mqr(\ra)^2,{}~\equiv\mN \rc,\mrh~^2$.
 On the other hand, since $0\leq\mlam_j\leq 1$, one has
 \begin{eqnarray*}
\mrh(\mber(\rc)^2)&=&\sum_{k\neq j}\mlam_j Tr(E_j\ra E_k\ra E_j)\geq\sum_{k\neq
 j}\mlam_j^2 Tr(E_j\ra E_k\ra E_j) \\
&\geq & \sum_{j>0}\sum_{k(\neq j)}\mlam_j^2 Tr(E_j\ra E_k\ra) -\sum_{k\neq
 j}\mlam_j\mlam_k Tr(E_j\ra E_k\ra) \\
&=&\frac{1}{2}Tr([\mrh,\mqr(\ra)][\mqr(\ra),\mrh]) = \frac{1}{2}\mN \rc,2~^2.
 \end{eqnarray*}
 These inequalities together with
the  previously  proved  equivalences  show  also   the   desired
equivalence of \N\cdot,\Gamma~. This proves also  nondegeneracy of $\Gamma$;
 its analytic dependence on the point \rh\ of the orbit \OUn\ can  be  proved
from its dependence on elements of the group \fk U acting  on \OUn.
The explicit form of $\Omega$
\begin{equation}\label{eq;2.15d}
    \Omega_{\mrh}(\mbf{v},\mbf{w})\equiv i\,\mrh([\mber(\mbf v),\mber(\mbf w)])
\end{equation}
shows, after inserting into it $\mbf v:=\madrs(\mdfr)$, and $\mbf w
:=\madrs(\mdhr)$,
that it can be expressed by our  Poisson  bracket~\eqref{eq;2.5b}: we obtain
\eqref{eq;2.15c}, according to \eqref{eq;2.8a}. The closedness $d\Omega = 0$
 follows  from
the proved Jacobi identity for the Poisson brackets
(Proposition~\ref{prop;2.6}). The mapping \dfr
$(\in\mTrs\mOU)\mapsto\mvf(\mrh) :=\madrs(\mdfr)\in\mTr\mOU\
(f\in\mFr)$\ is an isomorphism, what is a consequence of  the  proved
equivalence of topologies on \Tr\OU, of the  surjective  property
of the mapping $\madrs :\mfNr\rarw \mTr\mOU,\ \mdfr\mapsto \madrs(\mdfr)$, as
well as of the reflexivity of the Hilbert space (\Tr\OU;\ \N\cdot,\Gamma~).
This  proves that $\Omega$\ is strongly nondegenerate.
\end{proof}

\begin{noti}[\emm Symplectic and Poisson structures~]\label{note;sympl-poiss}
Existence of symplectic form $\Omega$\ is useful to easy introduction of
a canonical (induced) Poisson structure on submanifolds of $M$\ = \OUr\
determined, e.g.\ by actions of symmetry groups: The pull back by embeddings
is well defined for covariant tensor fields (i.e.\ for elements of $\mcl
T^0_n(M)$, whereby $\mcl T^0_1(M)$\ are one-forms on $M$), what is not the case
of Poisson bracket (remember that the Poisson structure is determined by a
two-contravariant antisymmetric tensor field, i.e.\ the element of $\mcl
T^2_0(M)$, cf.\ also~\eqref{eq;2.6},~\cite{marle,weinst}).

One could try to introduce a symplectic form $\tilde\Omega$\  on the whole
space \fTs\ in such a way, that the forms $\Omega_\mrh$ on \OUr's
($\mrh\in\mSs$) are its
restrictions by embeddings $\iota_\mrh:\mOUr\rarw\mfTs$, i.e.\ $\Omega_\mrh
\equiv\iota^*_\mrh\tilde\Omega$. This cannot be done by a naive
``extension'' of the formula\rref2.15d~ to the whole \fTs; e.g., for
$\dim\mrh=\infty$, the mapping \ber\ has not a ``natural'' extension to \fTs,
cf.\rref diverg~. We shall not investigate this possibility here (it can be
connected with considerations in Remarks~\ref{rem;mixinPH}).\hfill\dovi
\end{noti}

Let us note, that $\mvf \equiv 0$\ for a Hamiltonian vector field \vf\  does not
mean $f(\cdot)\equiv const.$\ on connected components of a considered
Poisson manifold $M$, as it is valid for a nondegenerate Poisson structure
(of Hamiltonian classical mechanics, e.g.),
cf.\ Definition~\ref{df;P-strct}. The vanishing of \vf\ only implies
constancy of restrictions of f to connected components of symplectic leaves
of $M$, e.g.\ the leaves \OUr\ of \Ss, resp.\ of \fTs.




\subsection{On interpretation: Subsystems and two types of mixed
states}\label{q-phsp;mixt}

The space \Ss\    with  the  introduced
Poisson structure will play in EQM a r\^ole  similar  to  the  phase
space of classical mechanics. It contains pure states of standard
QM described by points $\nu$ of the orbit \OUr\ = \PH\ with $\mrh=\mrh^2$,
i.e.\ consisting of one-dimensional orthogonal projections  on \H,
as well as the states  described  by  density  matrices $\mrh\neq\mrh^2$
traditionally called ``mixtures''. This type of mixture can  always
be obtained (cf., e.g.~\cite{gisin,davies,bon-dens}) by the  restriction
\begin{equation*}
\mpI:\mSs^{I+II}\rarw \mSs:=\mSs^I,
\end{equation*}
(the ``partial trace'',\cite{davies,QTM}, i.e.\ $\mpI\equiv Tr_{II})$ of a pure
state $\mrh_{I+II}=(\mrh_{I+II})^2\in\mSs^{I+II}$
of a composed system ``$I+II$'' (described with a  help  of  the  Hilbert
space $\mH_{I+II}:=\mH_I\otimes\mH_{II}$, with $\mH_I:=\mH$) to a
given state $\mrh_I := \mrh\in\mSs$  of
the considered subsystem, $\mrh = \mpI(\mrh_{I+II})$.\footnote{A more general
definition of ``subsystems'' can be found in~\dref G-subsyst~.}

Work with EQM requires introduction of two different types of
``mixed states'', cf. also~\cite{davies2}:\footnote{The concept of
``states'' will be reconsidered and generalized after introduction
of ``the observables'' of the considered systems in
Section~\ref{IIC;symm-obs}.}
\begin{defi}\label{df;men-mix}
Let the states
described by density matrices be  called \emm elementary  states~
(also \emm elementary mixtures~ to stress possibility of $\mrh\neq\mrh^2$).
The topological space \Ss\   endowed with the Poisson structure  will  be
then called the \emm elementary phase space~ for QM.

   Another type of states (let us call them \emm genuine mixtures~) are
described by probability measures $\mu$ on the set \Ss\   of normal states on
\LH\ endowed
with a Borel structure, cf.\ also~\cite{bon-dens}.
The set of elementary mixtures can be considered as the subset of the set of
genuine mixtures consisting of the Dirac measures (each concentrated on its
own one--point subset of \Ss).
\hfill\pika\end{defi}
\begin{rem}\label{rem;2.11}
We shall not investigate in details, in this paper, various
possible convenient Borel structures on \Ss\ , i.e.\  \sg-algebras
of subsets of \Ss\   generated by open subsets in a topology; we
shall not need it in our general considerations. From the point of
view of measure theory, cf.\ \cite{choquet,bra&rob}, it is
convenient to work on locally compact spaces. There are two ways
how to introduce a ``relatively compact'' topology on \Ss, coming
as the relative topology from its compactification in a natural
way:
\item(i) The space \Ss\   is a subset of \cS\ -- the set of all states on \LH\
which is compact in $\msg(\mLH^*,\mLH)$ topology. The induced topology from
this $w^*$-topology coincides on \Ss\   with the (topology induced from the)
natural norm topology on $\mLH^*$,~\cite[Proposition 2.6.15]{bra&rob}.
Observe that the restriction of the norm of $\mLH^*$ to \Ss\ coincides with
the trace-norm \N\cdot,1~ of \fTs. Moreover, \Ss\ is $w^*$-dense in
\cS,~\cite[Example 4.1.35]{bra&rob}. Hence, \cS\ is a natural compactification
of \Ss.
\item(ii) Another way of introduction of a ``relatively compact'' topology
in \Ss\
is (a priori different than that in (i)) $w^*$--topology coming from the duality
$\mfk C(\mcl H)^* = \mfk T(\mcl H)$, i.e.\ the \sg(\fk T,\fk C)-topology,
where the duality is expressed by the formula $\lb\mrh;{\rm \rc}\rb \equiv
Tr(\mrh{\rm \rc})$. By the same argument as in (i),~\cite[Proposition
2.6.15]{bra&rob}, the $w^*$--topology on \Ss\ coincides with the
norm-topology of $\mfk C^*=\mfk T$, hence again with the trace-norm
topology.
\hfill\dovi\end{rem}
These ways of introduction of (the same, as we see)
topology on \Ss\ leads us to another compact set (let us denote it
$\lb\mSs\rb$), a subset of which is \Ss:

\begin{defi}
The set $\lb\mSs\rb$ is the ($w^*$-compact) convex span of \Ss\ and of the
zero element of $\mfk C^*$. The compact $\lb\mSs\rb$ is sometimes
called~\cite{pedersen} the \emm quasi state space~ of the \Ca\ \fk C.\hfill\pika
\end{defi}

Let us return now to description of the genuine mixtures.
Let $f\in\mF$ be an ``observable'' (cf., however,
Definitions~\ref{df;2.25b}, and Interpretation~\ref{int;2.27}, for more
elaborated
concepts). If we interpret, in accordance with the standard  interpretation
of formalism of QM, its value $f(\mrh)$ as ``the expectation value $\lb
f\rb_{\mrh}$ of $f$ in the state \rh'', then the expectation in a (genuine
mixture--) state
$\mu$  would be naturally determined by the formula
\begin{equation}\label{eq;2.16}
             \mu(f):=\lb f \rb_{\mu}:=\int f(\mrh)\mu(d\,\mrh).
\end{equation}
If $f$ is an \emm affine function~, i.e.\ $f:= h_{\ra}$ for some $\ra\in\mLH$
(later the denotation ``affine'' will be used also for functions $f$
which  are  not  everywhere  defined  and  which  correspond   to
unbounded operators $X$, $f\equiv h_{X}$, cf.\ Sec.~\ref{IIB;gener}), and if
$\mfk b(\mu)\in \mSs$ is the \emm barycentre~ (also\label{resultant}
\emm resultant~, resp.
intuitively the ``center of mass'') of $\mu$~\cite{bra&rob}, then
\begin{equation}\label{eq;2.17}
 \mu(h_{\rm a})=h_{\rm a}(\mfk b(\mu))=\mfk b(\mu)({\rm a}),\ \forall
 \ra\in\mLH.
\end{equation}
This shows, that there is no observable difference between the genuine mixture
$\mu$ and the corresponding elementary mixture $\mfk b(\mu)\in\mSs$\ in the
case, if only affine functions can be observed. For  other
continuous $f$ (i.e.\ for $f\not\equiv h_{\ra}\ \text{for any a} \in\mLH$,
let us call such  functions
(bounded) \emm nonlinear functions~; they will appear as a new kind  of
\emm observables~, resp.\ \emm generators~, cf.\ Definitions~\ref{df;2.25a},
~\ref{df;2.25b}) one
has $\mu(f)\neq f(\mfk b(\mu))$ for a general $\mu$ (identity $\mu(f)\equiv
f(\mfk b(\mu))$ for all $\mu$
would lead to $f = h_{\rm a}$ for some $\ra\in\mLH$). Moreover, if  the  time
evolution \pph{},f~ is generated by  the  Hamiltonian  vector  field \vf\
corresponding to a nonlinear $f$, then, even for affine $h_{\rm a}$, one  has,
contrary to the case of affine generators $f,\ \mu_t(h_{\rm a})\not\equiv
h_{\rm a}(\mpph t,f~\mfk b(\mu))$,  where $\mu_t:= \mu\circ\mpph -t,f~$ (cf.
also Note~\ref{not;3.7}). This shows some reasons for making
distinctions between two kinds of ``mixtures'' in the  presence  of
nonlinear observables (and nonlinear evolution generators). If we
accept a sufficiently large class of nonlinear ``observables'' $f$,
e.g.\ $f\in\mF_b(\mSs)\equiv$ \emm uniformly bounded
infinitely differentiable functions~ on \Ss, then a  genuine  mixture
$\mu$ coincides  with  an
elementary mixture \rh\  iff $\mu = \delta_{\mrh}$ := {\em the Dirac measure
concentrated on the one-point set}\ $\{\mrh\}\subset\mSs$.

   Mutually different genuine mixtures $\mu\neq\mu'$ ``corresponding'' to
a given elementary state $\mrh = \mfk b(\mu) = \mfk b(\mu')$\ can be
interpreted as different extensions of
a given state of the ``considered microsystem'' (the observables of
which are described in the traditional way - exclusively  by  the
affine observables) to mutually different states of a larger system (say, a
macrosystem, cf.\ Section~\ref{sec;IIIB}, and also~\cite[Section II.C]{bon1})
described by a larger set of
observables, see Definition~\ref{df;2.25b}. In this sense, the
formalism described in this work, and describing (many -- also ``most'' of the
earlier known -- versions of) nonlinear dynamics in QM can be shown as a
restriction to a subsystem of a linear evolution of some larger (say
\emm macroscopic~) quantal system, cf.\ also~\cite{bon-tr}.

\begin{intpn}\label{int;3.7a}
\item{(i)} Let us consider a density matrix $\mrh\in\mfk T_{+1}(\mH_I)$\ of
a ``system I'', and a normalized vector $\Psi\in\mH_I\otimes\mH_{II}$\ of a
``composed system I+II'' such, that its restriction to the ``subsystem I''
(i.e.\ the partial trace \wrt the ``system II'') gives the density
matrix \rh:
\[ Tr\bigl(\mpI(P_\Psi)\ra\bigr) := Tr\bigl((\ra\otimes\mbI_{II})\dti
P_\Psi\bigr) = Tr(\mrh\dti \ra),\quad\forall \ra\in\mcl{L(H}_I).\]
Such a ``system II'', and a vector--state $P_\Psi$\ (resp.\ vector $\Psi$),
always exist for any given \rh. Let $\{\Phi_k:k\in K\}$ be an orthonormal
basis in $\mH_{II}:\sum_{k\in K}P_{\Phi_k}=\mbI_{II}$,
and let $\{\mphi_j:j\in J\}$\ be an arbitrary basis in the Hilbert space of
the ``considered system'' $\mH_I$. Then the set of vectors
$\{\mphi_j\otimes\Phi_k:j\in J,k\in K\}\subset\mH_I\otimes\mH_{II}$\ forms
a basis in the Hilbert space of the composed system, and there is
a unique decomposition
\[ \Psi=\sum_{j\in J,k\in K}c_{jk}\mphi_j\otimes\Phi_k= \sum_{k\in
K}\left(\sum_{j\in J}c_{jk}\mphi_j\right)\otimes\Phi_k.\]
Let us define the nonnegative numbers
\[ \mlam_k:=\|\sum_{j\in J}c_{jk}\mphi_j\|^2,   \]
for which the normalization property of $\Psi$\ gives $\sum_{k\in
K}\mlam_k=1$, and let us define also the unit vectors (in general mutually
{\em nonorthogonal}) for $\mlam_k\neq 0$:
\[ \psi_k:=\frac{1}{\sqrt{\mlam_k}}\sum_{j\in J}c_{jk}\mphi_j \]
in the Hilbert space $\mH_I$. Then we obtain for the given density matrix
\rh\ the expression:\footnote{It might happen also $P_{\psi_k}=P_{\psi_m}$\
for some $k\neq m$.}
\begin{subequations}\label{eq;1red}
\bequ\label{eq;red}
 \mrh=\sum_{k\in K}\mlam_k P_{\psi_k}.
 \end{equation}
 This decomposition does not depend on a choice of the basis
 $\{\mphi_j:j\in J\}\subset \mcl H_I$.
We see that an arbitrary orthonormal basis $\{\Phi_k:k\in K\}$\ in
$\mH_{II}$\ determines a unique decomposition of \rh. The vector $\Psi$ is
here considered fixed, and it is written in the form:

\[\Psi=\sum_{k\in K}\sqrt{\mlam_k}\psi_k\otimes\Phi_k. \]

Let us assume now, that an observable is ``measured'' on the composed
system I+II such, that it just \emm performs a filtering~ of the subsystem
II according to the chosen basis $\{\Phi_k:k\in K\}$, corresponding (in a
sense of the classical ``reduction postulate''~\cite{neum1}, cf.\
Footnote~\ref{ft;red-post}) to the
\emm measurement of the quantity \bs{A:=\sum_{k\in K}\malp_k P_{\Phi_k}}~,
where
$\malp_j (j\in K)$\ are arbitrary, mutually distinct real numbers. One can
imagine a situation similar to that in the Bohm version of the
Einstein-Podolsky-Rosen (EPR) ``gedanken experiment'',~\cite{EPR,bell,w+z},
that
the systems I and II are in the instant of measurement (being in the state
$P_\Psi$\ in that time) mutually very distant and noninteracting, so that
the measurement of the quantity $A\in\mcl L_s(\mH_{II})$\ (or, what is the
same in QM, of the quantity $\mbI_I\otimes A\in\mcl
L_s(\mH_I\otimes\mH_{II})$\ of the composed system) ``does not affect'' the
state of the subsystem I. After the measurement, according to the ``reduction
postulate'', the state of the composed system is
\[ \mrh_\Psi^A=\sum_{k\in K}\mlam_k P_{\psi_k}\otimes P_{\Phi_k},\]
and the reduced density matrix, if calculated {\em after the measurement},
is again the same \rh, as in\rref red~. Hence, the state (i.e.\ the reduced
density matrix) of the subsystem I does not depend on a choice of the
measured quantity $A$ of the subsystem II, but its decomposition\rref red~
is dependent on this choice.\footnote{The state $P_{\psi_k}$\ can be called,
in accordance with~\cite{ever&whee}, the {\em relative state} of $I$ \wrt
the state $P_{\Phi_k}$\ of $II$, if the state of $I+II$ is $\Psi$. We shall
not discuss here, however, consequences of EQM on possible mutual influence
of different ``branches'' in the {\em many world interpretation} of QM,
cf.\ \cite{noncausal}.}

The situation can be generalized to the measurement of a quantity $A$ with
degenerate discrete spectrum: \bs{A=\sum_l\malp_l E_l}, where $E_l$\ are
orthogonal projections in $\mH_{II}$\ commuting with all $P_{\Phi_k}$, and
$\sum_l E_l=\mbI_{II}$. Then the state of the composed system {\em after the
reduction} is
\[\mrh^A_\Psi:=\sum_l(\mbI_I\otimes E_l)\dti P_\Psi\dti(\mbI_I\otimes E_l)=
\sum_l \kappa_l\, P_{\Psi_l},\]
with $\Psi_l\propto(\mbI_I\otimes E_l)\Psi,\ \kappa_l:=\sum_{k\in(l)}\mlam_k$,
and
$(l):=\{k\in K: P_{\Phi_k}E_l=P_{\Phi_k}\}$. The above decomposition of the
reduced density matrix \rh\ corresponding to this alternative measurement
situation is
\bequ\label{eq;2red}
\mrh=\sum_l\kappa_l\,\mrh_l,\quad Tr(\mrh_l\dti\ra):=
Tr\bigl((\ra\otimes\mbI_{II})\dti P_{\Psi_l}\bigr)\ (\forall\ra\in\mcl
L(\mH_I)).
\end{equation}
Here it is \bequ\label{eq;2ared}
\kappa_l\,\mrh_l:=\sum_{k\in(l)}\mlam_k P_{\psi_k}, \quad
\kappa_l:=Tr[(\mbI_I\otimes E_l)\dti P_\Psi].
\end{equation}
\end{subequations}

Now we can try, however, to interpret
the density matrix \rh\ obtained by the restriction to the subsystem $I$\
{\em after the measurement} of the quantity $A$ of the subsystem $II$\
{\em not as an indecomposable entity}, i.e.\ as an
elementary state, but {\bf we are going to interpret its different
decompositions {\rref1red~} as} \emm different genuine mixtures~.
Hence we shall assume that the process of measurement of $A$ on the
correlated subsystem $II$\ {\bf transforms} the elementary mixture \rh\
(what is an empirically indecomposable quantity) into the corresponding
genuine mixture determined by the (empirically identifiable)
decomposition~\rref red~ into the elementary components $P_{\psi_k}$, resp.\ by
the decomposition~\rref2red~ into the elementary components $\mrh_l$, with the
same barycentre \rh.
This is an important difference in interpretations for nonlinear
dynamics: If the evolution of the subsystem I
after the measurement on the subsystem II is nonlinear, its state \rh\ will
evolve, generally, in course of some time after the measurement,
into different
states, in dependence of what quantity was measured on the distant (and
noninteracting) but
correlated system II.
 We see now that {\bf if we accept instantaneous ``reduction of the wave
packet'' of the composed system}, and, moreover, {\bf if we qualify the
obtained decomposition\rref red~ as the genuine mixture of the components
\bs{P_{\psi_k}}\ (resp.\ the decomposition\rref2red~ as the genuine mixture
of the components \bs{\mrh_l})}, then the subsequent different evolutions
of the mixtures
with the same initial barycentre (obtained at different choices of the measured
quantity $A$) can lead to distinguishable states before a light signal
coming from the distant system II can bring any information about the
quantities $A$ measured on the system II, cf.\
also~\cite{noncausal}.\footnote{\label{causal}This is, perhaps, a different
situation from that one discussed in~\cite{luecke}, where a sudden
``localized'' change of a nonlinear evolution generator led to instantaneous
change of time evolution ``at distant places''.}

\item{(ii)}\label{3meas}
Let us try to give at least a vague, intuitive formulation of an (hypothetical)
alternative for the above described {\em transformation of an elementary
mixture into a genuine one}, by which the ``action at a
distance'' is avoided:

In our understanding, a quantum measurement is a physical process
by which a quantum interaction of the micro-object with an
``apparatus'' leads to specific macroscopic changes of the
apparatus states, by which statistical distribution of eigenstates
of the micro-object corresponding to its measured observable in
its given quantal state is ``copied'' into a corresponding
classical statistical distribution of mutually classically
different (i.e.\ mutually noninterfering) ``pointer positions''. A
generally accepted description of such a dynamical process is
still missing,~\cite{w+z,bell,bon-m}. Let us assume, however, that
we have some description of this process in a framework of QT. Let
us consider the combined quantal system $I+II+III$, where we added
to our combined system $I+II$\ also an apparatus $III$. Then,
during the (many times repeated) measurement of $A$\ on $II$, the
apparatus states (let us denote them $\tilde\Psi_k$) corresponding
to the classes of states of $II$\ with sharp values $\malp_k$\ of
$A$\ (with their vectors lying in the subspaces
$E_k\mH_{II}\subset\mH_{II}$) become (in some short but nonzero
time) eventually correlated with the states $\mrh_k$\ of the
``distant'' subsystem $I$\ (that was left undisturbed during the
measurement). This correlation with ``pointer positions''
$\tilde{\Psi}_k$\ correspods to the ``reduction of the wave
packet'' and it has no observable influence on the system $I$.
According to our present (rather provisional) hypothesis, the
presumed process of transformation of \rh\ into the genuine
mixture $\sum\kappa_k\,\mrh_k$\ begins either {\em after the
measurement}, or already {\em at installing (and activating) of
the apparatus}. This corresponds to two eventualities:

(first) Since different pointer positions $\tilde\Psi_k$\ represent
different macroscopic states of ``the environment'' for the system $I$\
(we need
not be any more interested in the future fate of the measured subsystem $II$),
these macroscopic states might have different ``influences'' (as different
values of an external potential, or a ``field'') on the correlated states
$\mrh_k$. These ``influences'' might be very weak, just to provide a
possibility to distinguish between different states $\mrh_k$\ in the mixed
state~\rref2red~.

(second) The environment of the system $I$\ was changed by installation of
 an apparatus for measurement of $A$\ on $II$, and this change (providing
information about the set $\{E_k\}$\ of projections characterizing $A$)
performs an ``influence'' on $I$ transforming \rh\ into the status of the
genuine mixture from\rref2red~.

We expect, however, that {\bf this ``influences'' will be spread in both the cases
{\em with finite velocity}}. Hence, in a presently badly understood way, the
elementary mixture \rh\ changes into the corresponding genuine mixture (given
by the decomposition of \rh\ specified by the measured quantity $A$ of $II$)
in a finite time, avoiding the above described ``nonlinearity reason'' for a
superluminal
communication between $I$\ and apparatuses measuring different
observables $A$\ of $II$; other possible ``sources of noncausality''
mentioned in Subsection~\ref{qmech;causal}, or in the Footnote~\ref{causal},
needn't be improved by such a ``mechanism''.
Let us note finally that these considerations, to lead to a consistently
formulated part of QT, should be reconsidered in frameworks of Einstein
relativistic theories, cf.\ remarks and citations on page~\pageref{R-causal}.
\hfill\bpika\end{intpn}

   Let us note that earlier attempts~\cite{deBrog} to   introduce
nonlinearities into QM were connected with trials to make drastic
changes in interpretation of the formalism of QM and, contrary to
the here presented theory, they did not  include  the  traditional ``linear''
theory as a specific ``subtheory''.

\section{Unbounded Generators}
\label{IIB;gener}
\def\nazov{{
\ref{IIB;gener}\quad Unbounded Generators}}

   We have introduced the Poisson  structure  on  the  elementary
state space of QM with the help of the group \fk U  (resp.\
$\tilde{\mfk A}$) which can be considered as  a  ``maximal
possible  symmetry  group''  of described systems in the sense,
that each orbit \OUr\ is a homogeneous space of its action in
\fTs, whereas any ``physically acceptable'' (unitary) operation
leaves all the orbits \OUr\ $(\mrh\in\mfTs)$ invariant. In the
setting of preceding sections, generators of all there described
transformations (including time evolution) were functions
$f\in\mF$; the corresponding ``linear'' generators $f\:=\mh{\ra}$
correspond to {\em bounded} selfadjoint operators ${\rm
a}\in\mLH$\ only. The ``realistic models'' describing particles
and fields which are, e.g.\ invariant \wrt Poincar\'e, or Galilei
symmetries have, however unbounded Hamiltonians, and the
generators of many symmetry subgroups are also unbounded. These
symmetry groups are usually finite--dimensional noncompact Lee
groups, hence there are no ``interesting'' unitary representations
with all the generators bounded. Such a ``more realistic''
situation cannot be described by the formalism developed up to
now: To keep general ideas of our (nonlinear) extension of quantum
theory untouched, mathematically correct description requires more
sophisticated considerations: It leads to ``Hamiltonian
functions'' only defined on dense subsets of \fTs, and these
Hamiltonians  are not even locally bounded.\footnote{The
difference from the infinite dimensional Lie group \fk U\ of all
unitaries in \LH\ consists in discontinuity of the relevant
unitary representations $U(G)$ of noncompact finite--dimensional
Lie groups $G$: The one--parameter subgroups $t\mapsto
U(\exp(t\xi))\in\mfk U,\ \xi\in Lie(G),$\ of $U(G)\subset\mfk U$\
representing Lie subgroups of $G$\ are not all {\em Lie subgroups}
of \fk U: Some of them are discontinuous in norm--topology of \LH,
what is the topology \wrt which \fk U\ is endowed with a Lie group
structure.} We shall proceed  stepwise, starting with the linear
theory.


\subsection{Some probabilistic aspects of selfadjoint operators}
\label{gener;lin1}

To  obtain  structures  useful  to  effective
description of measurable quantities  of  a  specific  considered
system, as well as to obtain their empirical interpretation,  one
has to specify symmetry  groups $G$ ``smaller'' than \fk U.  These
groups  are related to the quantal system by their continuous (in some
topologies) representations
in \fk A, resp.\ by their projective representations $U(G)$ in \fk U,
cf.\ \cite{varad}.  Such realizations of $G$ leave the structure of the
elementary
(quantum) phase space invariant. These representations may not be analytic,
and  their  weaker  continuity  properties  are  connected   with
existence of unbounded generators.
Then we  are  faced  with the problem of description of locally unbounded
functions on \Ss, playing the r\^ole of
``observables'' or  ``generators'' $f\not\in\mF$,  corresponding  to  the
unbounded operators. These  functions  are  not  defined  on  any
nonempty open subset of \Ss, nevertheless they could generate  (in
a specific way) one parameter subgroups of transformations of \Ss.
This functions appear usually in the form $f :=\mh X$, where $X$ is  an
unbounded selfadjoint operator generating the unitary group $U^{X}: t\mapsto
\exp(-it{X})$, and
\begin{equation}\label{eq;2.18}
    \mh X(\mrh) := i\,\left.\frac{d}{dt}\right|_{t=0}\mrh(\exp(-itX))
\end{equation}
for such \rh, for which the derivative exists; this set of $\mrh\in\mSs :
=\mSs(\mLH)$ {\bf will be denoted by} \glss \bs{\mcl D(\mh X)}~. Let \glss
$D(X)$~$\subset\mH$
be the domain of $X$, and let \bs{D(\mh X) :=} $\{\ix\in\mH:
P_{\ix}\in\mcl D(\mh X)\}$. Clearly, $D(X)\subset D(\mh X)$,
and $D(\mh X)$  is \glss $U^{X}$~ --invariant.

   One of the main problems considered in this  section  will  be
the question of possibility of generalization  of  the  developed
Poisson formalism to locally unbounded (not everywhere  defined)
nonlinear generators  of  transformation  groups,  e.g.\  to  some
nonlinear perturbations  of  unbounded  affine  generators \h X. A
partial solution of this problem will be reached with a  help  of
group representations.

\begin{subequations}\label{eq;2.12p}
\begin{pt}\label{note;2.12}
 Some other characterizations of \cl D(\h X)   are  relevant
also from the point of view of possible  interpretations  of  the
presented formalism.
   Let\ $\mbf E_{X}$  denote the projection--valued (spectral)  measure of
a selfadjoint operator $X$. Let $\mu_{\mrh}^X$ be the probability measure on
the spectrum of $X$: sp($X$) $\subset\mbR,\ \mu^{X}_{\mrh}(B):=Tr(\mrh\mbf
E_{X}(B))$,  corresponding  to  any $\mrh\in\mSs$. The
characteristic function of $\mu_{\mrh}^{X}$ is $t\mapsto
Tr(\mrh\,\exp(it{X}))$. The domain
\cl D(\h X)  consists of all such points $\mrh\in\mSs$, for which the following
limit exists and is continuous in the real parameter t,~\cite{feller}:
\begin{equation}
\exp(it\mh
X(\mrh)):=\lim_{n\rarw\infty}\left(Tr(\mrh\,\exp\left(i\frac{t}{n}
X\right))\right)^n.
\end{equation}
The  probability  measure  corresponding  to  the  characteristic
function $t\mapsto \exp(it\mh X(\mrh))$ is  the  Dirac  measure
$\delta_{\mlam}$ on \bR\ concentrated at \lam\ = \h X(\rh). [It can be
shown, that this \lam\ can
be interpreted as ``a sharp value of a macroscopic observable $X_\Pi$'' in
a quantum theory of infinitely large systems,
cf.\ \cite{bon1,bon10}, cf.\ also Section~\ref{sec;IIIB}.]\hfill\dovi
\end{pt}
\begin{pt}\label{pt;2.12a}{\rm
  Let us mention still another (probabilistic)  characterization
of the domain \cl D(\h X), \cite[Chap.XVII,\S2.a, and Chap.XV,\S4]{feller}:
Let $\chi_n$  be the characteristic function (indicator)  of
the interval $(-n;n)\subset\mbR$, let $id_{\mbR}$  denote the identity function
$\mlam\mapsto\mlam$ on \bR, and let $\Bbb I$ denote the function identically
equal to 1 on \bR. Let
$\mu(f)$ denote the value of the integral  of  the  function  $f$  with
respect to a measure $\mu$. Then \cl D(\h X) consists of those
$\mrh\in\mSs(\mLH)$
for which the sequence of integrals $\mu^{X}_{\mrh}(\chi_n\,id_{\mbR})$\
(cf.\ref{note;2.12})
has a finite  limit
for $n\rarw\infty$, and for which simultaneously
\begin{equation}
\lim_{n\rarw\infty}\mu^{X}_{\mrh}(n({\Bbb I}-\chi_n)) = 0.
\end{equation}
We have in that  case
\begin{equation}
\lim_{n\rarw\infty}\mu^{X}_{\mrh}(\chi_nid_{\mbR}) = \mh X(\mrh).
\end{equation}
This shows that the existence of the first
momentum  $\mu^{X}_{\mrh}(id_{\mbR})$  of  the  probability  measure
$\mu^{X}_{\mrh}:\ id_{\mbR}\in L^1(\mbR,\mu^X_\mrh)$\ (i.e.\ the
existence of the expectation of $X$ in the state \rh, i.e.\ the
integrability of the absolute value $|id_\mbR|$)  implies $\mrh\in\mcl
D(\mh X)$.}\hfill\dovi
\end{pt}
\begin{pt}\label{pt;2.12b}
Similar considerations show, that $\mrh\in\mDX$\ (where \DX\ is
specified in Definition~\ref{df;domains} below) is equivalent to the existence
of the second momentum: $\mu^X_{\mrh}((id_{\mbR})^2)<\infty$\ for
$\mrh\in\mfk F_s\cap\mSs$.  Since the existence of
second momentum of a probability measure on \bR\   implies the existence of
the first one, we have $\mDX\subset\mcl D(\mh X)$.\hfill\dovi
\end{pt}
\end{subequations}
   Defined according to\rref2.18~, \h X  uniquely determines $X$, which in turn
uniquely determines the one parameter  unitary  group $U^{X}(t)$.  We
intend to determine the \glss{flow \pph t,{X}~\rh}~  := $Ad^*(\exp(-itX))\mrh$
from  the (densely defined) generator \h X, or rather from its ``differential''
$d\mh X$, as  a  Poisson  flow  corresponding  unambiguously  to ``the
Hamiltonian \h X'', and we shall generalize such a determination  of
flows to nonlinear unbounded generators.


\subsection{Unbounded ``linear'' generators}\label{gener;lin2}

Let us now start an investigation of possible generating of Hamiltonian
flows by real--valued functions defined on a dense set of \Ss, and locally
unbounded. It is clear that this will be only possible for a restricted
class of
functions, especially if chosen from the ``nonlinear'' ones. We shall
consider now the
most simple and basic case of a ``linear'' function, namely the function \h
X corresponding to an unbounded selfadjoint operator $X$ defined in the
subsection~\ref{gener;lin1}. We shall need to choose some subsets of the
domain \cl D(\h X) where \h X  will be in a convenient sense
``differentiable'', so that we shall be  able to define on sufficiently
large subset of \Ss\ the corresponding vector field, and subsequently its
flow, so that this flow will be coincident with the canonical unitary flow
generated by $X$.

Let us restrict our attention to subsets of ``finite dimensional'' density
matrices $\mrh\in\mfk F$ only, what is motivated by technical consequences of
Proposition~\ref{prop;2.4}.
\begin{defs}[{\bf Domains}\index{domains}]\label{df;domains}
\item{(i)} The \emm domain of the selfadjoint operator $X$~ on the Hilbert
space
\H\ will be denoted $\glss D(X)\subset\mH~$; the \emm subdomain of its analytic
vectors~ is denoted by \glss $D_a(X):=D^{\omega}(X)\subset D(X)$~. The space
of \emm infinitely differentiable vectors~
\begin{subequations}\label{eq;difdom}
\begin{equation}
 D^{\infty}(X):=
 \left\{\ix\in\mH:\left.\frac{d^n}{dt^n}\right|_{t=0}\exp(itX)\ix
\in\mH,\quad
\forall  n\in\mbN\right\}
\end{equation}
will also be denoted by \glss $D_d(X)\subset D(X)$~. Clearly $D_a(X)\subset
D_d(X)\subset D(X)$.
\item{(ii)} The \emm domain of the generator $\delta_{X}$~ of the
group
$t\ (\in\mbR)\mapsto Ad^*(\exp(-it{X}))\mrh$, $\forall\mrh \in \mfTs$, of
the B-space
automorphisms of \fTs\  will be denoted by \glss {$\mcl D(\delta_{X})$}~:
\begin{equation}
\mrh\in \mcl D(\delta_X)
\eequiv\left.\frac{d}{dt}\right|_{t=0}\bigl(\exp(-itX)\mrh\exp(itX)\bigr)\in
\mfTs,\ \forall\mrh\in\mfTs.
\end{equation}
\item{(iii)} The \emm restricted domain~ of the generator $\delta_X$ is
\bequ
 \glss{\mDdX\ :=\mcl D(\delta_X)\cap\mfk F_s\cap\mSs}~.
\end{equation}
\item{(iv)} \glss $\mcl D_r(X)$~ will denote the set of all finite real--linear
combinations of one--dimensional projections $P_{\ix},\ \ix\in D(X)$, i.e
the set of all selfadjoint finite rank operators with range in $D(X)$.
\bs{\mDX}\ will be called the \emm restricted domain of $X$~.
\item{(v)} The subset of \DX\ consisting of operators with their range
in the set of
analytic vectors of $X$ will be denoted by \glss \DXa~, and called the
\emm restricted analytic domain of $X$~. The operators in
\DX\ with range in $D_d(X)$ will be denoted by \glss \DXd~.
\item{(vi)} Let
\[\glss \mDdXa\ := \mDdX\cap\mDXa~.\]
This is the \emm restricted analytic domain of $\delta_X$~.\hfill\pika
\end{subequations}
\end{defs}
The following lemma expresses some important properties
of the domain \DdX.
\begin{lem}\label{lem;subdom}
 For any selfadjoint operator $X$ on \H\ one has:
\item{(i)} The domain \cl D($\delta_X$) of the generator $\delta_X$ contains
exactly those $\mrh\in\mfTs$ for which the following two conditions are
fulfilled:
\subitem{a.} The operator $\mrh\in\mLH$ leaves the domain $D(X)\subset\mH$
of $X$ invariant.
\subitem{b.} The operator $i\,[\mrh,X]$ (a priori defined, in the case of
validity of (a), on the domain $D(X)$) is uniquely extendable to an operator
lying in \fTs\  $\subset\mLH$.
\item{(ii)} The inclusion $\mDdX\subset\mcl D(\mh X)$ is valid.
\item{(iii)} For all $\mrh\in\mDdX$\ it is $\mrh X\in\mDX\subset\mfk F\ \&\
X\mrh\in\mfk F$\ (the products are considered here
\noidt as unique continuous
extensions of the operators initially defined on $D(X)$).
\item{(iv)} For $\mrh\in\mDdXa$ we have also $X\mrh\in\mDXa$.\hfill\zal
\end{lem}
\begin{proof}
(i) is proved in~\cite[Lemma 5.1 of Chap.5]{davies}. It implies, that
$\mrh\in\mDdX\imply\mrh X\in\mfk F\ \&\ X\mrh\in\mfk F$, where the products
with $X$ are considered as the corresponding (unique) bounded extensions
in \LH. From these facts we see, that, for the considered \rh, the expectation
$\mu^X_{\mrh}(id_{\mbR})=h_X(\mrh)$ exists, cf.\eqref{eq;2.12p},
what in turn implies $\mrh\in\mcl D(\mh X)$, i.e.\ (ii). With \rh\ as in
(iii),  $X$ and \rh\ are defined on the domain $D(X)$, and the range of
\rh\ is in $D(X)$; hence, both products are densely defined finite--range
operators, the first one in \DX. The last statement (iv) is valid due the
fact that the set of analytic vectors of $X$ is invariant also \wrt the action
of the operator $X$.
\end{proof}
It will be useful to introduce the following
\begin{notat}
Let us denote \DdXs, resp.\ \DXs, resp.\ $D_*(X)$\ the variable symbols with
possibilities
$*\in\{\circ,d,a\}$, where $D_{\circ}(X):=D(X)$, e.g.. An assertion
containing the symbol $*$\ (in the
described contexts) will be valid for all choices of the alternatives
(with the same value chosen in all places of the assertion simultaneously),
if something else will not be
 specified for it; the assertion might be expressed by a sequence of sentences.
That assertion might be
also numbered by attached $*$ corresponding to any of the choices.\hfill\pika
\end{notat}
Let us formulate  several useful simple implications of these
facts in the following
\begin{lem*}\label{lem;qqq}
\item{(*i)} The domain \DdXs\   consists of all finite convex combinations of
one--dimensional projections $P_{\ix},\ \ix \in D_*(X)\subset\mH$, i.e.
$\mDdXs\subset\mDXs$. All domains \DdXs\ (for $*=\circ,d,a$) are dense in
\Ss, resp.\ the domains \DXs\ are
dense in \fTs, in the topology induced by \N\cdot,1~\ of \fTs.
\item{(*ii)} For $\mrh\in\mDdXs$, one has
\begin{subequations}\label{eq;qqq}
  \begin{equation}
  \delta_X(\mrh)=i\,[\mrh,X]=i\,[\mrh,\mqr(X)]\in\mTr\mOU\subset\mfTs(\mH);
  \end{equation}
  \begin{equation}
  \mber(\delta_X(\mrh))=\mqr(X)\in\mfNr\subset\mfk F\subset\mLH.
  \end{equation}
\end{subequations}
\item{(*iii)} The sets of vectors $\{i\,[\mrh,{\rm b}]:\mrh\in\mDdXs,\
{\rm b}\in\mDXs\}$
are all dense in $\mTr\mOUr,\ \forall\mrh\in\mDdXs$\ in its topology given
by any of the equivalent
norms mentioned in Theorem~\ref{thm;2.10}.\hfill\zal
\end{lem*}
\begin{proof}
\ (*i)\  From Lemma~\ref{lem;subdom}(i), and the definition
in~\ref{df;domains}(iii), as well as from the corresponding definitions
of $\mDdXs:=\mDdX\cap\mDXs$, with the help of spectral decompositions of
$\mrh\in\mDdXs$, the first assertion of (*i) follows immediately.
It is sufficient to prove the density for $*=a$. Density of the set \DXa\ in
\fTs\ will be proved from its density in \fk F\ in \N\cdot,1~--topology,
because \fk F\ is dense in \fTs\ in this topology. But it suffices to prove
arbitrary close approximatebility of one--dimensional projections by such
projections from \DXa, i.e.\ by $\{P_{\ix}:\ix\in D_a(X)\}$. Since $D_a(X)$
is linear and dense in \H, unit vectors in $D_a(X)$ are dense in unit
sphere of \H\ (by triangle inequality). Then, for two unit vectors
$\ix,\iy\in\mH$, we can use:
\[ \mN \ix-\iy,{}~^2=2(1-\Re(\ix,\iy))>1-|(\ix,\iy)|^2=\frac{1}{4}\mN
P_{\ix}-P_{\iy},1~^2 , \]
where the second equation is proved by calculation of eigenvalues of
$\rd_{xy} :=
P_{\ix}-P_{\iy}; \rd_{xy}$\ is selfadjoint with trace zero, and range
two--dimensional, hence its two eigenvalues are opposite reals $\pm\mlam$;
then, by calculating $Tr({\rm d}^2_{xy})=2(1-Tr(P_{\ix}P_{\iy}))=2\mlam^2$ one
obtains the desired equation. This easily leads to a proof of density of \DXa\
in \fTs. The density of \DdXa\ in \Ss\ follows then by a use of
convexity of both sets.
\item{(*ii)} This is a consequence of Lemma~\ref{lem;subdom}(i), as well as of
 our constructions in Section~\ref{q-phsp;manif}, see esp.
 Definitions~\ref{df;2.3}.
\item{(*iii)} For any $\mrh\in\mDdXs$, it is $\{i\,[\mrh,{\rm b}]:\
 {\rm b}\in\mDXs\}\subset\mTr\mOUr$. Due to inequality
 \[ \mN{[\mrh,{\rm b}]},1~\leq 2 \mN \mrh,1~\mN {\rm b},{}~,\
 \forall\mrh\in\mfTs,{\rm b}\in\mLHs, \]
we know, that the linear mapping ${\rm b}\mapsto i\,[\mrh,{\rm b}]$\ is
continuous and can {\rm b}e uniquely extended to the whole \LHs\
($\ni {\rm b}$), the range of the extended mapping being the whole \Tr\OUr.
This leads eventually to validity of the statement.
 \end{proof}

The following assertion is important for our subsequent
constructions.

\begin{prop}\label{prop;r-orbits}
 Let $\mrh\in\mDdXs,\ {\rm b}\in\mDXs$. Then
$Ad^*(\exp(-it{\rm b}))(\mrh)\nl (\equiv \exp(-it{\rm b})\mrh\exp(it{\rm b}))\
\in\mDdXs$, i.e.
\DdXs\ is invariant \wrt the unitary flows generated by ${\rm b}\in\mDXs$.\hfill\zal
\end{prop}
\begin{proof}
There is a projection $P_{\rm b} \in\mDXs$ such that ${\rm b}={\rm b}P_{\rm b}$
($P_{\rm b}$ might be
chosen to be the range projection of ${\rm b}$). Then $\exp(i{\rm b})=
\exp(i{\rm b})P_{\rm b} + I-P_{\rm b}$, hence
\begin{eqnarray*}
&& Ad^*(\exp(-it{\rm b}))(\mrh) = \\
&& P_{\rm b}\exp(-it{\rm b})\mrh\exp(it{\rm b})P_{\rm b} + \mrh -
\mrh P_{\rm b} - P_{\rm b}\mrh + P_{\rm b}\mrh P_{\rm b} - \\
&& P_{\rm b}\exp(-it{\rm b})\mrh + \mrh\exp(it{\rm b}) P_{\rm b}
 - P_{\rm b}\mrh\exp(it{\rm b})P_{\rm b} -
P_{\rm b}\exp(-it{\rm b})\mrh P_{\rm b} .
\end{eqnarray*}
The expression consists of a sum of elements of \DXs\   with ranges contained
in the Hilbert subspace determined by the orthogonal projection
$(\sum_{j\geq 1}E_j)\vee P_{\rm b} \in \mDXs$, where we used the spectral
projections $E_j$ of \rh. Hence $Ad^*(\exp(-it{\rm b}))\mrh\in\mDXs$. Due to
unitarity of the transformation of \rh, we have also
$Ad^*(\exp(-it{\rm b}))\mrh\in\mDdXs$. This proves the assertion.
\end{proof}

Let us now define \glss ${\md \mh X,\mrh~}\in\mTrs\mOUn$~ for $\mrh\in
\mDdX\subset\mcl
D(\mh X)$. For these \rh's, we can write $\mh X(\mrh)= Tr(\mrh X)$. According
to the Proposition~\ref{prop;r-orbits}, we can write for $b\in\mDX$:

\begin{eqnarray}\label{eq;dhX}
&d_{\mrh}h_X(i[\mrh,{\rm b}])=\left.\frac{d}{dt}\right|_{t=0}
h_X(\exp(-it{\rm b})\mrh
\exp(it{\rm b}))= \nonumber \\
&Tr(i\,[\mrh,{\rm b}]X) = i\,Tr({\rm b}[X,\mrh]) =
i\,Tr([\mrh,{\rm b}]\mqr(X)),
\end{eqnarray}
so that \d h_X,\mrh~ is represented by the operator $\mqr(X)$. In the
calculations in~\eqref{eq;dhX}, there was used (iii) and (iv) of
Lemma~\ref{lem;subdom}, as well as Lemma~\ref{lem;qqq}.
In this way, we arrived to the
\begin{defi}\label{df;dhX}
Let $\mrh\in\mDdXs$. Then the \emm generalized differential of \bs{h_X}~, \
\bs{\md h_X,\mrh~}, is the element of \Trs\OUr\ represented by
$\mqr(X)\in\mfNr$, according to the correspondence
\[ i\dti[\mrh,{\rm b}]\ (\in\mTr\mOUr)\mapsto i\,Tr([\mrh,{\rm b}]\mqr(X)),\
{\rm b}\in\mLHs, \]
as explained above, cf.\eqref{eq;dhX}.\hfill\pika
\end{defi}
The definition can be abbreviated as
\begin{equation}\label{eq;2.19}
\md h_X,\mrh~(i\,[\mrh,{\rm b}])=\ad^*_{\mrh}({\rm b})(\md \mh
X,\mrh~)=\madrs({\rm b})(\mqr(X)).
\end{equation}

Such a ``differential'' \d \mh X,\mrh~ is defined till now in points
$\mrh\in\mDdXs$ as a linear functional on vectors $i\,[\mrh,{\rm
b}]\in\mTr\mOUr$
for ${\rm b}\in\mDXs$\ only. But these vectors are dense in \Tr\OUr\ (in
any of the
equivalent norms mentioned in Theorem~\ref{thm;2.10}), because \DX\   is dense
in \LHs, and \fk F   is dense in \fTs, cf.\ Lemma~\ref{lem;qqq}. Consequently,
we can uniquely extend
\d \mh X,\mrh~ to a bounded linear functional, \d \mh X,\mrh~=
$\mqr(X)\in\mTrs\mOUr\subset\mLH$.

We shall turn now to the question, whether and how the ``differential'' \d
\mh X,{}~ defined just on a subset $\mrh\in\mDdXs$ of \Ss\ can determine the
``unitary flow'' $Ad^*(\exp(-itX))$\ on the whole state space \Ss\ in a
``geometric way''. We define the ``Hamiltonian vector field'' \vv X,\mrh~
 corresponding
to the function \h X  via its ``differential'' \d \mh X,\mrh~  in the point
$\mrh\in\mDdXs$ with a help of Poisson brackets according to~\eqref{eq;2.6}
and~\eqref{eq;2.7}, i.e.\ in the representation of tangent vectors in
\Tr\OUr\   used above, we have
\begin{equation}\label{eq;2.24}
\mvv X,\mrh~=i\,[\mrh,\mqr(X)]\equiv\madrs(\md \mh X,\mrh~),\
\forall\mrh\in\mDdXs,
\end{equation}
in accordance with equation~\eqref{eq;2.8a}. It is clear, that vectors \vv
X,\mrh~ are tangent to curves $t\mapsto Ad^*(\exp(-itX))(\mrh)$ in each
point $\mrh\in\mDdXs$ of their definition. These curves are all lying in the
domain \DdXs, since the unitary flow $Ad^*(\exp(-itX))$ leaves \DdXs\
invariant. But the closure of \DdXs\    in \N \cdot,1~ --topology is the whole
\Ss. Moreover, the functions $\mrh\mapsto Ad^*(\exp(-itX))(\mrh),\ \forall
t\in\mbR$ are continuous in \N \cdot,1~, hence could be uniquely extended
by continuity from \DdX\  on the whole \Ss. In this way, we have seen that
a complete flow on \Ss\ is uniquely determined by the ``Hamiltonian vector
field''~\eqref{eq;2.24} defined on a dense subset \DdXs\ of \Ss\ only.
It remained, however, partially open the question here, how to determine
the flow ``from
the function \h X  alone'', i.e.\ without an explicit use of the operator
$X$, with having given the function \h X  and its ``directional
(Gateaux,~\eqref{eq;2.4}), or partial derivatives'' on the corresponding
domains only. The known
properties of the linear operator $X^*=X$\ might serve to us as a hint
to look for relevant properties of \h X only. A description of the
resulting dynamics might be given as follows:

\begin{pt}\label{pt;2.15}
The flow $\mrh\mapsto\mpph t,{X}~\mrh$ on $\mrh\in\mDXa$ corresponding to
the vector field~\eqref{eq;2.24} can be described
by unitary cocycles (what are just unitary groups in these cases),
 according to eq.~\eqref{eq;2.14} (with interchanged
$\nu\leftrightarrow\mrh,\ f\leftrightarrow h$).\hfill\zal
\end{pt}

We want to generalize the described situation to ``Hamiltonian
functions'' generating Poisson (or Hamiltonian) flows,
also not being of the form \h X  for any selfadjoint $X$ and, moreover,
are also
only densely definable in \Ss. The most simple generalization is, probably,
the generator $h(\mrh):=f(\mh X(\mrh))$, where $f$ is a sufficiently
differentiable real valued function on \bR. We shall go further: We shall
generalize and investigate the preceding constructions to functions
$h(\mrh):=f(\mh{X_1}(\mrh),\mh{X_2}(\mrh),\dots\mh{X_k}(\mrh))$, $f\in
C_{\mbR}^{\infty}(\mbR^k)$, for
``conveniently chosen'' sets of (in general noncommuting) selfadjoint
operators $X_j\ (j=1,2,\dots k)$ on \H. Before that, however, a more
general framework will be sketched.


\subsection{On unbounded nonlinear generators}\label{gener;nonlin}

As we saw in the example of selfadjoint operators and the corresponding
``linear'' generators -- locally unbounded Hamiltonian functions \h X, the
definition of a (Poisson) flow from such a function \h X  might be
possible, if we determine from it a densely (in \Ss) defined vector field
\vv f,\cdot~ having integral curves (lying, of course, in its domain), in
an agreement with~\eqref{eq;2.8a}.
Hence, the domain $\{\nu\in\mSs: \mvv f,\nu~\ \text{exists}\}$ should consist
of (at least) one dimensional
differentiable ($C^1$\ --) submanifolds of (sufficiently many of)
\OUr's (we shall again consider $\mrh\in\mfk F_s\cap\mSs$ only).

\begin{rem}[Speculating on ``integral'' submanifolds]
\label{rem;submanif} To make possible a use of the Poisson
structure at construction of smooth vector fields on {\em smooth
manifolds}, as well as their integral curves from only densely
defined functions on $\mOUr,\ \mrh\in\mfk F_s$, and also to have
possibility to define Poisson brackets for several such densely
defined functions, we would need algorithms to construct some
``convenient'' more than one--dimensional smooth submanifolds in
domain of definition of our densely defined objects, and this
seems to be a nontrivial question in a general case. A solution
will be found in subsequent sections for a specific class of
densely defined generators and vector--fields determined by Lie
group representations: A given continuous unitary representation
of a Lie group determines in the state space \Ss\ smooth
submanifolds (orbits of GCS). Hamiltonian vector fields on these
submanifolds can be defined from given ``nonlinear'' real--valued
functions with a help of the existing ``K\"ahlerian'' structure
$\Psi$\ $\bigl($cf.\rref2.15a~; let us note that this structure is
K\"ahlerian only if restricted to \PH$\bigr)$. Existence of such
apriori defined domains of definition is typical also for some
standard approaches to not--everywhere defined vector fields
and/or Hamiltonian functions, cf.\ \cite{chern&mars,mars}. Let us
speculate a little now on alternative possibilities for
construction of some smooth submanifolds of \OUr's, determined by
some apriori given objects, e.g.\ by an (only densely defined)
vector field \vv f,\nu~.\footnote{The vectors \vfn\ needn't belong
to a (possibly Hamiltonian, in some sense) vector field determined
by a function $f$; the letter ``$f$'' might be here just a label.}
The rough idea consists in looking for possibility of construction
of some submanifolds in \Ss\ of more than one dimension from such
a ``relatively poor'' object as a vector field. These submanifolds
might become a ``playground'' for definition of other vector
fields and their integral curves.

Let us formulate here just some ``toy examples'' how
to define, to a given (possibly not everywhere defined) vector field
\vv f,\mrh~, other vector fields such that they both together (perhaps) span a
symplectic submanifold
of \OUr. Our proposals might be useful as hints for a search of alternatives
to cases described in literature, if the
 assumptions required there are not fulfilled.   This new vector field
will be constructed via the symplectic and metric structures
on \OUr\ given by~\eqref{eq;2.15a}, i.e.\ by $\Omega_{\mrh}$ and
$\Gamma_{\mrh}$ respectively. Note, that these structures are invariant
\wrt ``unitary automorphisms'' of \OUr, i.e.\ for a given unitary operator
$\ru\in\mfk U:=\mcl U(\mH)$ the corresponding mapping $Ad^*(\ru):\mOUr\rarw
\mOUr$ leaves invariant not only the symplectic form, but also the metric;
the \emm push--forward~ $(Ad^*(\ru))_*\mbf v$ of a vector field
$\mrh\mapsto\mvv{},\mrh~\equiv i\,[\mrh,{\rm b}_{\mbf v}(\mrh)]\in
\mTr\mOUr$ is
$\left.\frac{d}{dt}\right|_{t=0} \ru\exp(-it{\rm b}_{\mbf v}(\mrh))\mrh
\exp(it{\rm b}_{\mbf v}(\mrh))\ru^* =
i\,[\ru\mrh \ru^*,\ru{\rm b}_{\mbf v}(\mrh)\ru^*]$, hence the \emm pull--back~
of the bilinear form
$\Psi_{\mrh}$ by the same mapping is
\begin{eqnarray*}
\bigl((Ad^*(\ru))^*\Psi\bigr)_{\nu}(\mbf v(\nu),\mbf w(\nu))&=&\Psi_{\ru\nu
\ru^*}(\bigl(Ad^*(\ru)\bigr)_*\mbf v(\nu),\bigl(Ad^*(\ru)\bigr)_*\mbf w(\nu))
\\ &=&2 (\ru\nu \ru^*)\bigl(\beta_{\ru\nu \ru^*}((Ad^*(\ru))_*\mbf v)
\beta_{\ru\nu
\ru^*}((Ad^*(\ru))_*\mbf w)\bigr) \\
&=&2\,Tr\bigl(\ru\nu {\rm b}_{\mbf v}{\rm b}_{\mbf w}\ru^*\bigr) =
2\,Tr\bigl(\nu
{\rm b}_{\mbf v}{\rm b}_{\mbf w}\bigr)=\Psi_{\nu}(\mbf v(\nu),\mbf w(\nu)).
\end{eqnarray*}
We shall present here two possibilities of construction of linear
independent vector fields from a given one. We do not, however, even
formulate precisely a question of their ``integrability'' to some integral
submanifolds containing these vector fields as sections of their tangent
bundles, e.g.\ in a sense of the Frobenius theorem,
cf., e.g.~\cite{abr&mars,3baby,arn3}. The integrability questions would need
more specific assumptions on the (domain of the) vector field ${\bf v}_f$.
\item{(i):}
Let us fix a point $\nu\in\mOUr$, and a vector
$\mvfn\in\mTn\mOUr$.
 We shall construct another vector \glss \vvfn~\  forming with it a
 ``canonical pair''
(\wrt the form $\Omega$). Let, for any subset $N\subset\mTn\mOUr$, its
orthogonal complement (in sense of the real Hilbert space structure
given by $\Gamma$) be denoted by $N^{\perp}$, and the skew--orthogonal
complement by $N^{\angle}:=\{\mbf v\in\mTn\mOUr: \Omega_{\nu}(\mbf v,\mbf
w)=0\ \forall \mbf w\in N\}$. It is clear that $N^{\angle}$ is a closed
linear subspace of \Tn\OUr, and that $N^{\angle\angle\angle}=N^{\angle}$,
resp.\ also $N^{\angle\angle}=N$ for a closed linear subspace $N$, similarly
as it is valid for orthogonal complements. For any nonzero $\mbf
v\in\mTn\mOUr$
the space $[\mbf v]^{\angle}$ is of codimension one. Hence
$\left[[\mbf v]^{\angle}\right]^{\perp}$ is one--dimensional, the nonzero
vectors of which
have nonzero ``skew-product'' with $\mbf v$, and are orthogonal to it. Let
us choose for any $\nu\in\mcl D(\mvf)\equiv$\ the domain of \vf:
\begin{subequations}\label{eq;1vf}
\begin{equation}
\mvvfn\in\left[[\mvfn]^{\angle}\right]^{\perp},\ \
\Omega_{\nu}(\mvfn,\mvvfn):=1,\ \ \Gamma_{\nu}(\mvfn,\mvvfn)=0.
\end{equation}
We can ascribe, in this way, to any vector field \vfn\ a ``canonically
conjugated'' vector field \vvfn.

\item{(ii):}
An alternative way to construct another vector field $\nu\mapsto
\mvvfn$\ to a given $\nu\mapsto\mvfn$\ might be as follows:
\begin{equation}
 \mvvfn\propto [\nu,[\nu,\mben(\mvfn)]],\ \mvfn:=i\,[\nu,{\rm b}(\nu)].
\end{equation}
This proposal allows us to construct also more than two--dimensional subspaces
of
\Tn\OUn\ ($\nu\in\mDdXs,\ {\rm b}(\nu)\in\mDXs$) containing a given field
$\mvf^{(1)}(\nu):=\mvfn$ together with the
vector field $\mvf^{(2)}(\nu)\propto\mvvfn$. In terms of our operator
representations of \Tn\OUn\ we can construct a sequence of (a finite number
of linearly independent) vector fields by the formula:
\begin{equation}
\glss \mvf^{(n)}(\nu)~:=i\dti[\nu,\mvf^{(n-1)}(\nu)]:=
i^n\dti[\nu,{\rm b}(\nu)]^{(n)},
 \end{equation}
where $[\nu,{\rm b}]^{(n+1)}:=[\nu,[\nu,{\rm b}]^{(n)}],\ [\nu,{\rm b}]^{(1)}
:=[\nu,{\rm b}]:=\nu
{\rm b}-{\rm b}\nu.$ Let us mention some properties of these vector--fields
\wrt the bilinear form $\Psi_{\nu}$, cf.\ Theorem~\ref{thm;2.10}; they are
derivable from simple properties of the commutators and traces:
\begin{eqnarray}\label{eq;1vvv}
 \Psi_{\nu}(\mvf^{(n)},\mvf^{(m)})&=&(-1)^k\dti\Psi_{\nu}
(\mvf^{(n-k)},\mvf^{(m+k)}) \\
&=&(-1)^{n-m}\dti\overline{\Psi_{\nu}(\mvf^{(n)},\mvf^{(m)})}.
 \end{eqnarray}
Since the symplectic form $-\Omega_{\nu}$ is the imaginary part of
$\Psi_{\nu}$, and the metric $\Gamma_{\nu}$ is the real part, we see that
the fields $\mvf^{(n)}$ and $\mvf^{(n+1)}$ are pointwise mutually
orthogonal, whereas $\mvf^{(n)}$ and $\mvf^{(n+2)}$
are mutually skew--orthogonal $(\forall n\in\mbf N)$. Observe also, that all
these fields have, in a given point $\nu$, nonzero values simultaneously:
this is due to the fact, that for $\nu\in\mDdXs$ the mapping \ben\ is
an isomorphism (resp.\ it can be considered as an automorphism, after a
natural identification, cf.\ Notes~\ref{note;2.3d}, and
Proposition~\ref{prop;2.4}) of \fNn\ and \Tn\OUn:
\begin{equation}
\mben(i\,[\nu,\mqn({\rm b})])=\mqn({\rm b})=\mqn(\mqn({\rm b})).
\end{equation}
This allows us to extend the sequence of vector fields $\mvf^{(n)},\
(n=1,2,\dots)$\ to all integers $n\in\mbZ$. We shall assume here that
${\rm b}(\nu)\in\mfNn\ (\forall\nu\in \mcl D(\mvf))$. We define:
\begin{equation}
\mvf^{(0)}(\nu):=\mben(\mvf^{(1)}(\nu))\equiv {\rm b}(\nu),\
\mvf^{(-n)}(\nu) :=\mben^{n}\bigl({\rm b}(\nu)\bigr),\ \forall n\in\mbZ.
\end{equation}
Since the ranges of $\nu$\ and ${\rm b}(\nu)$ are finite--dimensional, only a
finite number of elements of $\{\mvf^{(n)}:n\in\mbZ\}$ are linearly
independent.
It is also easily seen that the bilinear form $\Psi_\mrh$\ is nonzero on any
pair of these vectors, what follows from\rref1vvv~ and from:
\barr
\Psi_\mrh(\mvf^{(n+1)},\mvf^{(n+2)}) &=& Tr\bigl(\mrh\,
[\mber(\mvf^{(n+1)}),\mber(\mvf^{(n+2)})]\bigr)
 = Tr\bigl(\mrh\,[\mvf^{(n)}(\mrh),\mvf^{(n+1)}(\mrh)]\bigr)\nonumber\\
 &=& i\dti Tr\bigl(\mrh\,[\mvf^{(n)}(\mrh),[\mrh\,,\mvf^{(n)}(\mrh)]]\bigr)
\nonumber\\
&=& i\dti Tr\bigl([\mrh\,,\mvf^{(n)}(\mrh)]^2\bigr)\neq 0,
\end{eqnarray}
since all the $\mvf^{(n)}(\mrh)$'s are represented by selfadjoint trace
class operators on \H.
\hfill\dovi
\end{subequations}
\end{rem}

We shall proceed, also in nonlinear generalizations, in the framework of
Hilbert space \H, since this allows us to use some usual techniques with
linear mappings and scalar product, as well as intuition and/or
interpretation from the standard QM. We believe, however, that the
developed ideas can be used also in a ``purely geometrical'' transcription
(and possible modifications),~\cite{cir4,ashtekar}, of the theory developed
in this paper.

\begin{notat}[Domains]\label{notat;domains}
Let us assume, that a norm--dense linear subset $\mbf D$  of \H\ is given. This
means also, that any finite linear combination $\sum_{\malp=1}^k
c_{\malp}\ix_{\malp}$ of vectors $\ix_{\malp}\in\mbf D$  also belongs to
$\mbf D$, hence finite--dimensional subspaces generated by such
vectors are subspaces of $\mbf D$.
Let us denote by \glss \bcD~\ the set of all finite real--linear
combinations of finite dimensional projections to subspaces of \bD,
$\mbcD\subset\mfk F_s$. In the
general scheme constructed here in an analogy with preceding subsection, the
set \bcD\ is  here the object corresponding to \DXs\ in
Subsection~\ref{gener;lin2}. Let $\mbcDp:=\mbcD\cap\mSs$\ be the object
corresponding to \DdXs\ in Subsection~\ref{gener;lin2}. \bcDp\ is dense in
\Ss, in the $\|\cdot\|_1$--norm topology.\hfill\pika
\end{notat}
\begin{defs}[{\bf Generalized fields and
integrability}\index{generalized fields}]\label{df;g-dif}
\item{(i)} Let $h:\mbcDp\rarw\mbR$ be such that there exist
\begin{subequations}\label{eq;g-dif}
\begin{equation}\label{eq;1g-dif}
\glss {\md h,\mrh~}(i[\mrh,{\rm b}])~:=
\left.\frac{d}{dt}\right|_{t=0}h(\exp(-it{\rm b})\mrh\exp(it{\rm b})),\
\forall \mrh\in\mbcDp, {\rm b} \in\mbcD,
\end{equation}
\noidt and that it is bounded linear in the variable\ $i\,[\mrh,{\rm b}],\
{\rm b}\in\mbcD$; let its unique bounded linear extension is expressed by the
operator $\glss \mqr(\mbcD h)~:=\md h,\mrh~\in\mfNr\subset\mbcD\subset\mLHs$:
\begin{equation}\label{eq;2g-dif}
\md h,\mrh~(i[\mrh,{\rm b}])=
i\,Tr\left(\mqr(\mbcD h)\!\cdot\![\mrh,{\rm b}]\right),\
\forall {\rm b}\in\mLHs.
\end{equation}
This densely defined function \d h,\cdot~: \glss $\mrh\mapsto \mqr(\mbcD
h)\in\mfNr,\ \mrh\in\mbcDp$~, will be called
the \emm \bcD--generalized differential~ of $h$.
\item{(ii)}
The corresponding (densely defined in \Ss) \bf{(generalized)\
\bcD--Hamiltonian vector field} \index{generalized\
\bcD--Hamiltonian vector field} is:
\begin{equation}\label{eq;3g-dif}
\glss{\mvv h,\mrh~}~:=\madrs(\mqr(\mbcD h))\in\mTr\mOUr,\ \mrh\in\mbcDp.
\end{equation}
\noidt Let us stress that values of this vector field also belong to
$\mbcD\subset\mfTs$.
\item{(iii)}
\label{cl V}
Let us assume that \bcD\ contains the set \cl V of mutually disjoint
submanifolds $\mcl V_{\iota}$,
$\mcl V:=\{\mcl V_{\iota}:\ \iota\in\Upsilon:= \text{an index set}\}$, such
that their union $\cup\mcl V$ :=
$\cup_{\iota\in\Upsilon}\mcl V_{\iota}$ is dense in \bcD. Further assume
that for a given $h:\mbcD\rarw\mbR$\ with \bcD--generalized differential its
\bcD--Hamiltonian vector field is tangent to $\mcl V_{\iota}$ in any point
$\nu\in\mcl V_{\iota},\ \forall\iota\in\Upsilon$, so that the restrictions
of \vv h,\nu~ to $\mcl V_{\iota}\ni\nu$ are smooth vector fields on the all
$\mcl V_{\iota}$'s. Then we call the \bcD--generalized differential of $h$ to
be \emm \cl V--integrable~.
\item{(iv)}
Consider the situation from (iii) above, and let the differential
$\mqn(\mbcD h)$ be \cl V--integrable. Let us assume that the local flows \pph
t,h~ of these vector fields on \cl V\ depend continuously on initial
conditions, i.e.\ the functions
\begin{equation}\label{eq;4g-dif}
 (\nu;t)\mapsto\mpph t,h~(\nu),\ \forall (\nu;t)\in\mcl
 D_{\Upsilon}\subset\cup\mcl V\times\mbR\ (\mcl D_{\Upsilon}
 \supset\cup\mcl V\times\{0\}),
\end{equation}
are all continuous on the union $\cup\mcl V$ in the topology induced from
\N \cdot,1~.
Here $\mcl D_{\Upsilon}$ is the domain of the definition of the local flows,
and it is $\mcl D_{\Upsilon}=\cup\mcl V\times\mbR$ if the flows are
complete (i.e.\ defined for all $t\in\mbR$).
In this case the flows on leaves of \cl V\ can be uniquely
extended to a flow on \Ss. Then  we call the \bcD--generalized
differential to be \emm \Ss--integrable~.\footnote{Some variations on these
definitions allowing more refined classification of flows, what are extendable
to submanifolds of \Ss\ only, are sketched in~\cite{bon10}.}\hfill\pika
\end{subequations}
\end{defs}

We shall look now, for a moment, back to the ``linear cases'' to show that
they are contained in our present generalized scheme:

\begin{prop}[Differentials for ``linear'' generators]\label{prop;lin-gen}
Let $X$ be a selfadjoint operator on \H, let \bcDp:= \DdXs,\ \bcD:=\DXs.
Then the \bcD--generalized differential $dh_X$ of \h X, $\mh X(\mrh):=
Tr(\mrh X)$, exists. The differential $dh_X$ is \cl V--integrable for $\mcl
V:=\bigl\{\mcl V_\nu:\quad \mcl
V_\nu:=\{\exp(-itX)\nu\exp(itX):t\in\mbR\},\nu\in\mDdXa\bigr\}$.\hfill\zal
\end{prop}
\begin{proof}
The proof is contained in the text following the
Definition~\ref{df;dhX}.
\end{proof}
\begin{note}
We could choose in the Proposition~\ref{prop;lin-gen} more than one--dimensional
$\mcl V_\nu$ as submanifolds with smooth Hamiltonian vector
field~\eqref{eq;2.24} constructed with a help of
Proposition~\ref{prop;r-orbits}. Our simplest choice was, however, enough
to demonstrate a consistency feature of the theory.\hfill\dovi
\end{note}

The ``Schr\"odinger equation'' for the unitary cocycles describing the
Hamiltonian flow of the \bcD--Hamiltonian vector field ${\bf v}_h$\
can be written as in~\eqref{eq;2.11},
resp.~\eqref{eq;2.14}:
\begin{equation}\label{eq;nlinSch}
i\,\frac{d}{dt}
\ru_h(t,\mrh(0))=\left[q_{\mrh(t)}(\mbcD h)+\mrhoo(\mrh(t))\right]
\ru_h(t,\mrh(0)),\quad\ru_h(0,\mrh(0)):=I_{\mH},
\end{equation}
where $\mrh(t)\equiv \ru_h(t,\mrh(0))\mrh(0)\ru^{-1}_h(t,\mrh(0)),\
\forall\mrh(0)\in\mbcD$. {\em The equation~\eqref{eq;nlinSch} is an expression
of general form of dynamical (nonlinear Schr\"odinger) equations.}
We intend to discuss various specifications of this equation in subsequent
parts of this work.
If the function $\mrhoo$ on \bcD\ is chosen ``sufficiently nice'' (e.g.
sufficiently continuous, with
values in $\mfMr\cap\mbcD$), the objects in this equation are well defined
on the dense domain \bcD. In specific cases, the
equation~\eqref{eq;nlinSch} can be considered as a nonautonomous (i.e.\ time
dependent) linear Schr\"odinger--Dyson equation provided
the dependence $t\mapsto q_{\mrh(t)}(\mbcD h)$ is known; this
``time--dependence of Hamiltonian'' can be sometimes obtained in an
independent way, without solving this nonlinear equation.  Such a
possibility of ``elimination of nonlinearity'' will arise in specific
applications investigated in Section~\ref{sec;IIIE}.


\subsection{Nonlinear generators from group
representations}\label{gener;nl-grp}

We have sketched in Subsection~\ref{gener;nonlin} a formulation of the problem
of construction of some ``convenient''
submanifolds in \OUr, with $\mrh\in\mDdXs$, on which some (on \OUr\ only)
densely defined vector
 fields could be determined as smooth vector fields in the
 corresponding tangent subbundles. This was the case, e.g., of
 densely defined ``nonlinear'' Hamiltonian vector fields from
 Definitions~\ref{df;g-dif}, but
 also the case of the ``linear'' Hamiltonian function \h X, if we wanted to
 proceed in the determination of the corresponding Hamiltonian flow in a
 geometric way (i.e.\ without a return to the functional analysis connected
 with the selfadjoint operator $X$\ on \H).
 The proposals outlined
 in Remark~\ref{rem;submanif} were left in a very preliminary form.
 Analogical theory of that one for generators in ``linear case'' would be,
 e.g.\  some hypothetical nonlinear
 generalization of the von Neumann theory of symmetric and selfadjoint
 operators (\emm ``deficiency--indices'' theory~,
 cf.\ \cite{R&S}, and also Appendix~\ref{C;symm});\footnote{It is known that,
e.g.\ completeness of locally
 Hamiltonian vector fields is (up to subsets of measure zero) equivalent to
 essential selfadjointness of their generators in the ``Koopman version''
 of CM; this follows from a Povzner theorem, cf.\ \cite{povzner},
~\cite[Theorem 2.6.15 and Proposition 2.6.14]{abr&mars}.}
 we are  not aware
 of existence of such a theory.\footnote{An exception might be a theory of
unbounded derivations on \Ca s, cf.\
\cite{sak2}; this could be used in our case
 after an ``embedding'' of our nonlinear system into a larger linear one,
 cf.\ also~\cite{bon8,bon1}.}
We have worked above with a ``large'' domain \DdXs, containing one--dimensional
solutions of the equation~\eqref{eq;nlinSch}. Rigorous and systematic methods
for solving that equation were, however, missing.\footnote{Cf.,
however,~\cite[\S 4.1]{chern&mars}, where the concept of ``manifold domain''
was introduced; this can be applied, in the case of single selfadjoint
generator $X$, to its domain $D(X)\subset\mH$\ endowed with the
graph--norm, cf.\ also\rref1GrA~.}
Now we shall use Lie group
representations to allow us rigorous work with nonlinear unbounded
generators of specific kind; its specification to solution of~\rref
nlinSch~ is described in Section~\ref{sec;IIIE}.

Let $G$ be a real Lie group~\cite{bourb;Lie}, and let \glss $U(G)$~ be its \emm
strongly continuous unitary or projective representation~ in \H, hence $U:g(\in
G)\mapsto U(g)(\in\mfk U),\ g\mapsto Tr(\mrh U(g))$
being continuous on $G$ for all $\mrh\in\mSs$.
Assume that $U(G)$ has a $U(G)$--invariant dense set \glss
$D^{\mome}(G)~\subset\mH$
of analytic vectors, i.e.\ $\ix\in D^{\mome}(G)\eequiv$ the function
$g\mapsto U(g)\ix$ is
real analytic in a neighbourhood of the identity $e\in G$. This  is
the case~\cite{bar&racz} of  each  strongly  continuous $U(G)$ of  any  finite
dimensional Lie group $G$, as well as of an analytic representation
$U$ of an arbitrary  Lie  group, e.g.\ the defining representation of the
unitary group $\mfk U:=\mcl{U(H)}$\ in\H.  Let \bs{\mcl D^{\mome}(G)}\ be the
(norm--dense)
$Ad^*(U(G))$--invariant set of analytic elements $\nu\in\mSs$, i.e.\ the
functions $g\mapsto Ad^*(U(g))\nu$ from $G$ to \fTs\ are real analytic  around
$e\in G$. Let us write also $g\cdot\nu := Ad^*(U(g))\nu$. Let \glss
$Lie(G)\equiv\mfk g$~
denote the \emm Lie algebra of \bs G~, and let \glss $\exp: Lie(G)\rarw G$~ be
the
exponential mapping.  Then we have $U(\exp(t\xi)) =: \exp(-itX_{\xi}), \xi\in
Lie(G)$,  for  a
selfadjoint (in general unbounded) operators \Xx\ on \H.
The mapping \bs{\xi\mapsto X(\xi):=\mXx}\ is a Lie  algebra  morphism: It is
linear, and on a dense ($U(G)$--,  and  also $X(Lie(G))$--)invariant
domain (common for all \Xx, $\xi\in Lie(G)$), e.g.\ on $D^{\mome}(G)$,
 satisfies
the relation,~\cite{bar&racz}:
\begin{equation}\label{eq;2.25}
         [\mXx,\mXe] := \mXx\mXe-\mXe\mXx =i\,\mXxe.
\end{equation}
Here $[\xi,\eta]\in Lie(G)$ denotes the Lie bracket. Let
$\mOGr\subset\mOUr\cap \mcl D^{\mome}(G)$ be the $Ad^*(U(g))$--orbit of the
$G$--action on \fTs\ through \rh, $\mOGr:=\{U(g)\mrh U(g)^*: g\in G\}$. Let
\begin{equation}\label{eq;2.26}
       \mh{X(\xi)}(\nu):=\nu(\mXx):=i\,\left.\frac{d}{dt}\right|_{t=0}
       \nu(\exp(-it\mXx)),
\end{equation}
for $\nu\in\mcl D(\mh{X(\xi)})$, cf.\eqref{eq;2.18}, and
~\eqref{eq;2.12p} with $X := X(\xi)$. Let
us denote \bs{G_{\nu}:=} $\{g\in G: U(g)\in\mfk U_{\nu}\}$\ the stability
subgroup of
$G$\ at $\nu\in\mSs$ with respect to the action $Ad^*(U(\cdot)):(g;\nu)\mapsto
g\cdot\nu$.
The following lemma shows that the set of nice (i.e.\ ``analytic finite
dimensional'') orbits of the action of $G$ on \Ss\ satisfy not only conditions
on \bcD\ stated in Definition~\ref{df;g-dif}, but these orbits also can be
used in the r\^ole of the submanifolds mentioned in the
Remark~\ref{rem;submanif}. Let us first introduce notation
\bequ\label{eq;1Dr}
 \glss \mDGomr~:=\cap\{\mcl D_{ra}(X_{\xi});\xi\in Lie(G)\},\quad \dim
G<\infty,
\end{equation}
i.e.\ the $Ad^*(U(G))$--invariant set \bs{\mDGomr\subset\mDGom}\
consists of finite dimensional density matrices with ranges in
$D^{\mome}(G)$.

\begin{lem}\label{lem;2.16}
 Let $G$ be a finite--dimensional Lie group, and let $\nu\in\mDGom$. Then
 \OGn\  is an embedded submanifold~\cite{3baby} of \fTs\ lying in \Ss.
If $\mrh\in\mDGomr$, then $\mrh\in\mcl D_{ra}(\mXx)$, and
$\md{\mh{X(\xi)}},\mrh~\in\mfNr$, for all $\xi\in\mfk g$. The
vectors $\mvv{X(\xi)},\mrh~:= \madrs(\md{\mh{X(\xi)}},\mrh~) (\xi\in\mfk
g\equiv Lie(G))$  form  the  linear
space \Tr\OGr.  The  union  of  the  submanifolds \OGn\
($\nu\in\mDGomr$) composes a norm--dense subset of \Ss.
The  vectors $\mvv{X(\xi)},\mrh~,\ \mrh\in \mDGomr$, compose
 generalized  vector  fields
\vv{X(\xi)},\cdot~\ ($\xi\in Lie(G)$) on \Ss\  generating the flows
$(t;\mrh)\mapsto\mpph t,\xi~(\mrh) := Ad^*(U(\exp(t\xi)))\mrh$.\hfill\zal
\end{lem}
\begin{proof}
 Due to  the  continuity  of $U(G)$,  and  because \fk T\ is  a
Hausdorff space, $G_{\nu}$ is a closed (hence Lie) subgroup of $G$. This
implies that $Ad^*(U(\cdot))\nu$ can be considered as a bijective  mapping
of the analytic manifold $G/G_{\nu}$ onto the orbit \OGn. This  mapping
is analytic, and its differential (i.e.\ the tangent map) maps the tangent
space $T_e(G/G_{\nu})$
onto  a   finite--dimensional   subspace   of \Tn\fTs,  which   is
complementable. This fact together with the $Ad^*(U(\cdot))$--invariance
of \DGom\ implies,~\cite{bourb;manif}, that \OGn\ is an embedded submanifold of
\fTs. The second, and the third assertions are implied by the  considerations
developed in the Subsection~\ref{gener;lin2}, since the vector--fields
\vv{X(\xi)},\nu~:= \adrs(\d{\mh{X(\xi)}},\mrh~) generate the flows
\pph{},\xi~  which  were used to formation of the orbit \OGr.
The existence of a dense subset of \Ss\   of  analytic
elements lying in \DGomr\ with respect to the norm--topology of \fTs in \Ss\
implies  the  fourth  assertion. Differentiation  of  these
flows demonstrates also validity of the last statement.
\end{proof}

   Let us extend now our definition of Poisson brackets~\eqref{eq;2.5b}  to
densely defined functions \h{X(\xi)} ($\xi\in Lie(G)$) defined on  a  dense
subset of \Ss\ consisting of orbits \OGn. According to the construction of
orbits \OGn\  from the ``flows of \vv X(\xi),\cdot~\ generated by \h{X(\xi)}'',
it is clear that the vector fields \vv X(\xi),\cdot~\ are tangent to those
orbits everywhere where they are defined. Let $\nu\in\mDGomr$.
Since $\mqn(\mXx)=\md{\mh{X(\xi)}},\nu~\in\mfNn\ (\xi\in Lie(G))$, and also
$\nu\mXx\in\mfk F$,  we
can define the commutator
$i\,[\md{\mh{X(\xi)}},\nu~,\md{\mh{X(\eta)}},\nu~] \in\mLHs$,  and  the
Poisson bracket according to the relation~\eqref{eq;2.6}, cf.\ also
Definitions~\ref{df;2.3}:
\begin{subequations}
\label{eq;2.27}
\begin{equation}\label{eq;2.a27}
\{\mh\mXx,\mh\mXe\}(\nu):=i\,\nu([\md\mh\mXx,\nu~,\md\mh\mXe,\nu~])=
\madns(\mqn(\mXx))(\mqn(\mXe)).
\end{equation}
On the other hand, according to~\eqref{eq;2.25}, one also has
\begin{equation}\label{eq;2.b27}
\mh\mXxe(\nu)=Tr(\nu\mXxe)=-i\,Tr(\nu[\mXx,\mXe])=
             -i\,\nu([\mqn(\mXx),\mqn(\mXe)]),
\end{equation}
what gives the result:
\begin{equation}\label{eq;2.c27}
      \{\mh\mXx,\mh\mXe\}(\nu)=-\mh\mXxe(\nu).
\end{equation}
\end{subequations}

We shall consider this relation as the definition of the Poisson
bracket in the Lie algebra  of  functions \h{X(\xi)}\ ($\xi\in Lie(G)$)
defined on their common domain
\begin{subequations}\label{eq;2.27a}
\begin{equation}\label{eq;2.a27a}
 \mbcDF:= \{\nu\in\mSs: \text{the Fr\'echet differential of}\
g\mapsto\nu(U(g))\ \text{exists}\},
\end{equation}
what implies\footnote{Here the Fr\'echet
differential can be understood as the differential of a mapping defined on
the Banach manifold $G$, cf.\ \cite{bourb;manif,jt-schw,3baby}.}
that\footnote{For explanation of the notation \bcDF\
see Definition~\ref{df;2.17} below.}
\begin{equation}
\mbcDF\subset \cap\{\mcl D(\mh{X(\xi)}):\xi\in Lie(G)\}.
\end{equation}
The intersection $\cap\{\mcl D(\mh{X(\xi)}):\xi\in Lie(G)\}$ is the domain
consisting of those $\nu\in\mSs$ for which the function $g\mapsto\nu(U(g))$
is Gateaux differentiable. If $\dim G<\infty$, then the
(continuous) Gateaux differentiability implies Fr\'echet differentiability,
cf.\ \cite[Lemma 1.15]{jt-schw}, hence
\begin{equation}
\mbcDF=\cap\{\mcl D(\mh{X(\xi)}):\xi\in Lie(G)\},\ {\rm for}\ \dim G<\infty.
\end{equation}
\end{subequations}
The derivation property of  Poisson  brackets  (Proposition~\ref{prop;2.6})
allows us to extend  definition  of  this  Poisson  bracket  to
polynomials in variables \h\mXx\ $(\xi\in Lie(G))$ on the domain \bcDF.
The derivation property for the Poisson bracket of our not everywhere
defined functions follows from the derivation property of commutators (also
of unbounded operators on common invariant domains) via the
equations\rref2.27~ valid on \bcDF.
   If we want to  use  polynomials  in  the  variables \h\mXx\ as
generators of  evolution  of  our  generalized  quantummechanical
system determined by the described Poisson structure  on \Ss,  we
have to define also Poisson brackets of  these  polynomials  with
differentiable (locally bounded)  functions $f\in\mcl F$.  These  are
naturally determined  for $\mrh\in\mDGomr\subset\mbcDF$\
by  the formula:
\begin{equation}\label{eq;2.27b}
          \{\mh\mXx,f\}(\mrh):= i\,\mrh([\mqr(\mXx),\md f,\mrh~]).
\end{equation}

This relation determines the vector fields \vv\mXx,\cdot~ on \OGr\ in accordance
with\nl
Lemma~\ref{lem;2.16}.

Now we shall define the mapping \bF, what appears to be one of the most useful
objects for our subsequent considerations:

\begin{subequations}
\label{eq;2.28}
\begin{defs}[{\bf Domains and momentum mapping \bF}\index{momentum mapping \bF}]\label{df;2.17}
Let \glss $Lie(G)^*\equiv\mfk g^*$~ denote the dual space to  the  Lie
algebra of $G$  (recall  that  $Lie(G)$  is  a  normable  topological
algebra also for infinite--dimensional $G$). Define also the \emm
restricted domain~
\glss $\mbcDFr:=\mDGomr\subset\mbcDF$~, cf.\rref1Dr~, and\rref2.27a~,
of the mapping
\glss \bF~\ (the \emm Momentum
mapping~), cf.\ \cite{arn1,abr&mars}, which is defined on the domain \bcDF\
as follows:
\begin{eqnarray}
&&\glss \mbF:\ \mbcDF\rarw Lie(G)^*~,\ \glss
\mrh\mapsto\mbF(\mrh):=F_{\mrh}~,\\
&& \text{with}\ \ F_\xi(\mrh)\equiv F_{\mrh}(\xi):=\mh{X(\xi)}(\mrh). \nonumber
\end{eqnarray}

 Let us denote also by  \glss $f_{\xi}:Lie(G)^*\rarw\mbR$~ the functions
$f_{\xi}(F)
 := F(\xi) := \left(\text{the value  of}\right.$\ $F \in Lie(G)^* \text{on the
 vector}\ \left.\xi\in Lie(G)\right)$. The \emm domain of \bF~, i.e.\ the set
 $\mbcDF := \cap\{\mcl
 D(\mh{X(\xi)}): \xi\in Lie(G)\}\subset\mSs(\mLHs)$\ is $Ad^*(U(G))$
 --invariant.\footnote{For a general definition, and also for various
applications of momentum mappings cf., e.g.~\cite{abr&mars,mars&rati}.}

   One can prove immediately validity of the following equivariance
   property:
\begin{equation}
F_{g\cdot\mrh}:=\mbF(Ad^*(U(g))\mrh)=Ad^*(g)\circ\mbF(\mrh),\ \text{for all}\
\mrh\in\mbcDF,\ \text{and all}\ g\in G,
\end{equation}

\noidt since $U(g)\mXx U(g)^*= X_{Ad(g)\xi}$ for all $\xi\in Lie(G)$; here
$Ad^*(G)$  is the \emm coadjoint representation~ of $G$ in $Lie(G)^*$, i.e.\
the dual representation to the adjoint  representation $Ad(G)$, cf.\dref2Ad~,

 \begin{equation}\label{AdG}
 Ad(g)\xi:= \left.\frac{d}{dt}\right|_{t=0} g\cdot\exp(t\xi)\cdot g^{-1}.
\end{equation}

  Let \bF(\rh) be  called  the  (value  of  the) \emm U(G)-field \bs{\mbF}~\
  corresponding to the microscopic state \rh.\hfill\pika
  \end{defs}
\end{subequations}

\begin{rem}\label{rem;contF}
  The continuity of of the mapping
$\mbF(\mrh):\xi\mapsto\mbF(\mrh)(\xi)$ for $\mrh\in\mbcDF$ is trivial for
finite dimensional $G$, since each finite dimensional linear function is
continuous (in the unique l.c.--topology); in the case of a general Lie group
representation (we restrict our attention to the representations with
a dense analytic domain $\mDGomr\subset\mbcDF$) the continuity
for $\mrh\in\mbcDF$ is implied by the definition of points
$\mrh\in\mbcDF$: Fr\'echet differentiability means linearity and continuity of
the obtained mapping
\[\xi\mapsto \mh\mXx (\mrh)\equiv Tr(\mrh\mXx)=i\,\md{[\mrh(U(g))](\xi)},g=e~.
\]
We shall usually consider in the following, however, finite--dimensional
Lie groups $G$.\hfill\dovi
\end{rem}

\begin{rem}\label{rem;macro}
 \begin{subequations}
  Let us note that the states $\mrh\in\mbcDF$ are exactly those  normal
states of a constituent microsystem of a macroscopic one (in  the
description of infinite quantal systems composed of equal ``microscopic
constituents'', cf.\ Section~\ref{sec;IIIB})  in   infinite (symmetric)
tensor products \om\mrh,{}~
of which the ``macroscopic observables'' $X_{\xi\Pi}\ (\xi\in Lie(G))$ are
defined:
\begin{eqnarray}
 & & \mom\mrh,{}~ := \bigotimes_{p\in\Pi}\mrh_p\in\mSs(\mcl A^{**})\
  (\mrh_p\equiv\mrh), \\
 & & \mom\mrh,{}~(X_{\xi\Pi})=\mbF(\mrh)(\xi)\equiv
  F_{\mrh}(\xi):=\mh\mXx(\mrh), \\
 & & X_{\xi\Pi} := ("w")-\lim_{|\Lambda|\rarw\infty}\frac{1}{|\Lambda|}
\sum_{p\in\Lambda} X_p,
\end{eqnarray}

\noidt where $p\in\Pi$ distinguishes copies of the ``microscopic constituents'',
$\Lambda$ is a finite subset of these copies, and $X_p$ are ``equal
observables'' for distinguished copies $p\in\Pi$. The limit in the formula
above is taken in a specific weak ($"w"$) topology (we shall not
specify it here,
see, e.g.~\cite{bon1}). In this connection, the
introduced function $\mbF(\mrh)$\ is called also the
\emm \bs{U(G)}--macroscopic field~
corresponding to the ``microscopic state'' \rh.

\noidt  Observe also, that the value of the \emm \bs{\mfk U}--macroscopic field~
corresponding to $\mrh\in\mSs$ (for the defining representation $\mfk U\rarw
\mfk U$\ of the unitary group of \H)\ is \rh\ itself: The dual space to the
$Lie(\mfk U) := i\,\mLHs$ can be identified with $\mLHs^*$ containing the
 (normal) state space \Ss\ as an $Ad^*(\mfk U)$--invariant subset. This is in a
  sense maximal ``classical macroscopic phase space'' $\mSs\ :
\mbF(\nu)\equiv\mbF_{\mfk U}(\nu)=\nu\ (\forall\nu\in\mSs)$.
Such a  ``macroscopic  field'' separates points of the
elementary quantum phase space, i.e.\ the macroscopic field $\mbF_{\mfk U}$\
determines corresponding microscopic states. {\bf This can be considered as a
formalization
of the conventional belief of QM that a macroscopically determined
``preparation procedure'' determines the corresponding microscopic state
of a considered quantummechanical system uniquely.}\hfill\dovi
\end{subequations}
\end{rem}

We could temporarily take the point of view that only
``macroscopic properties'' of the
system (in the sense of Remark~\ref{rem;macro}) described by the values of
\bF\  are interesting for us.
 Then it would be interesting to know in what extent the values \bF($\nu$)
separate the points $\nu$ of an orbit \OGr.

\begin{lem}\label{lem;2.18}
 Let $\mrh\in\mbcDFr,\ \xi,\eta\in Lie(G)$. Then
             \begin{equation}\label{eq;2.29}
 \frac{d}{dt} F_{\exp(t\eta)\cdot\mrh}(\xi)=
 F_{\mrh}([Ad(\exp(-t\eta))\xi,\eta]),
 \end{equation}
 for all $t\in\mbR$. In particular, if we have a fixed $\eta\in Lie(G)$ such
that the derivative in~\eqref{eq;2.29} vanishes for all $\xi\in Lie(G)$ at
one value of $t\in\mbR$, then it vanishes for all $\xi$ at all values of
$t\in \mbR$.\hfill\zal
\end{lem}
\begin{proof}
By a use of the identity
\begin{equation*}
U(g)X_{\xi}U(g^{-1})=X_{Ad(g)\xi},
\end{equation*}
as well as of the relation
\begin{equation}\label{eq;dcom}
\left.\frac{d}{dt}\right|_{t=0}F_{\exp(t\eta)\cdot\mrh}(\xi)=
F_{\mrh}([\xi,\eta]),\
\forall \xi,\eta\in\mfk g,
\end{equation}
cf.\eqref{eq;2.25}, and~\rref2.28~, we obtain
\begin{subequations}
\begin{eqnarray}
F_{\exp(t\eta)\cdot\mrh}([\xi,\eta])&=&Tr\Bigl(U(\exp(t\eta))\mrh
U(\exp(-t\eta))\mXxe\Bigr) \\
&=&-i\,Tr\Bigl(\mrh\, U(\exp(-t\eta))[\mXx,\mXe]U(\exp(t\eta))\Bigr) \\
&=&-i\,Tr\Bigl(\mrh\, [U(\exp(-t\eta))\mXx U(\exp(t\eta)),\mXe]\Bigr) \\
&=&-i\,Tr\Bigl(\mrh\, [X_{Ad(\exp(-t\eta))\xi},\mXe]\Bigr) \\
&=&\ F_{\mrh}([Ad(\exp(-t\eta))\xi,\eta]).
\end{eqnarray}
\end{subequations}
After a subsequent application of~\eqref{eq;dcom} with $\mrh\mapsto
\exp(t\eta)\cdot\mrh$, the preceding calculation gives the result.
\end{proof}
 This lemma gives an answer to the question on separation properties of
 \bF\ on \OGr:

   Let $\eta\in Lie(G)$ be such that $F_{\mrh}([\xi,\eta]) = 0, \forall
   \xi\in Lie(G)$. Then $\mbF(\exp(t\eta)\cdot\mrh) = \mbF(\mrh)$ for  all
   $t\in\mbR$,  hence  the  points
   $ exp(t\eta)\cdot\mrh\in\mOGr$ for different values of $t$ cannot be
   distinguished by the values of the field \bF. The  vectors $\eta$
 form the Lie algebra of the \emm stability subgroup
 of $G$ at the point \bF(\rh)~  with  respect to  the action of the
 $Ad^*(G)$--representation denoted by
 \glss \bs{G_{\mbF(\mrh)}}~.
 Clearly, it is valid
 \begin{lem}\label{lem;stabG}
 Let $G_{\mrh}\subset G$\ be the stability subgroup of the
 $Ad^*(U(G))$--action of G on \Ss, at the point $\mrh\in\mSs$. Then
 $G_{\mrh}\subset G_{\mbF(\mrh)}$, and the equality
 $G_{\mrh}=G_{\mbF(\mrh)}$ is valid iff the restriction of the mapping \bF\ to
 \OGr\ is a bijection onto an $Ad^*(G)$--orbit in $Lie(G)^*$.\hfill\zal
 \end{lem}

 \begin{rem}\label{rem;Om-restr}
A definition of Poisson bracket on \OGr, with $\mrh\in\mDGomr$, equivalent
to that in~\eqref{eq;2.27},
can be given with a help of the (strongly) symplectic structure
$\Omega_{\mrh}$ by definition of a closed
two--form $\boldsymbol{\iota}_{\mrh}^*\Omega_{\mrh}$ -- the pull back of
the ``overlying'' form $\Omega$ by the embedding $\boldsymbol{\iota}$ of
the manifold \OGr\ into \OUr, in the case if the obtained two--form on the
submanifold \OGr\ is nondegenerate. If the restricted symplectic structure
$\boldsymbol{\iota}_{\mrh}^* \Omega_{\mrh}$
is degenerate, we can obtain a symplectic manifold by factorization of
\OGr\ according to the orbits of stability subgroups $G_{\mbF(\nu)}$ leaving
the values $\mbF(\nu)\in\mfk g^*,\ \nu\in\mOGr$
invariant,~\cite{bon4,bon8}.\hfill\dovi
\end{rem}

 One can
 construct examples of representations $U(G)$ with both even-- and
odd--dimensional
 orbits \OGr\ $(\mrh\in\mbcDFr)$\ (for  finite--dimensional $G$~\cite{bon8},
 cf.\ also our Subsection~\ref{IIIA1;restr}).
 Orbits of the $Ad^*(G)$--representation are always
 ``even--dimensional'': They  are  endowed  with  a  canonical
Kirillov--Kostant
 symplectic structure corresponding to the standard  Poisson structure (called
 also {\em Berezin brackets}) on $Lie(G)^*\equiv\mfk
 g^*$:\footnote{These considerations might also be valid for
 infinite--dimensional Lie groups, cf.\ \cite[Appendix 13]{arn1}.}
 \begin{equation}\label{eq;2.30}
            \{f_{\xi},f_{\eta}\}(F)= -F([\xi,\eta]):= -f_{[\xi,\eta]}(F).
\end{equation}

If $\mvv\xi,F~\in T_F(\mfk g^*)\ (\xi\in Lie(G))$ are the vectors tangent at
$F\in\mfk g^*\equiv Lie(G)^*$ to  the  flows $(t;F)\mapsto
Ad^*(\exp(t\xi))F$, then the Kirillov--Kostant symplectic form $\Omega^K$ can be
 expressed as
 \begin{equation}\label{eq;2.31}
         \Omega^K_F(\mbf v_{\xi},\mbf v_{\eta}) = -F([\xi,\eta]).
 \end{equation}

Comparison of the relation~\eqref{eq;2.30} with~\eqref{eq;2.27} shows,  that
the mapping \bF\ is a \emm Poisson morphism~,~\cite{weinst}: The
functions
\[\mbF^*f_{\xi} := f_{\xi}\circ\mbF = \mh{X(\xi)}=:F_{\xi}\equiv\rf_{\xi}\]
 on \bcDFr\ satisfy~\eqref{eq;2.27},
what leads to  a  definition of Poisson brackets for all functions f on
$\mOGr\ (\mrh\in\mbcDF)$ which are expressible in the form
\footnote{We shall usually distinguish typographically, in the following
text, real valued
functions $f,h$\ defined on the dual of the Lie algebra, $\mfk g^*\equiv
Lie(G)^*$,
from the ``corresponding'' functions $\rf:=\mbF^*f,{\rm h}:=\mbF^*h$\ defined
on domains lying in \Ss. To stress the difference of domains, we shall write
also \rf, e.g.\ for arbitrary functions $\rf\in\mcl F(\mSs)$.}
\begin{equation}\label{eq;2.32}
 \rf :=\mbF^*f := f\circ\mbF,\ \ f\in C^{\infty}(\mfk g^*,\mbR).
\end{equation}

\begin{rem}\label{rem;infdG}
In the case of infinite--dimensional groups, we cannot expect reflexivity of
\fk g: For \fk g := \LHs = $Lie(\mfk U) = \mfTs^*$ and infinite--dimensional
Hilbert space \H\ one has $\mfk g^*=\mLHs^*\neq \mfTs$, and $\mfk g^{**}$
is strictly larger than \fk g. Then we have to be careful in
reading~\eqref{eq;2.32}: If the differentiation of $f\in C^{\infty}$ is taken
in the canonical norm--topology of $\mfk g^*$, then the first differentials
of $f$'s belong generally to  $\mLH^{**}$, and needn't be expressible as
bounded operators on \H. The space $\mLH^{**}$ is, however a von Neumann
algebra in a canonical way,~\cite{sak1,takesI,dix1,dix2,bra&rob}, hence
also endowed with a canonical Poisson--commutator structure, which is unique
extension of that of \LH. Another possibility would be to take derivatives
on $\mfk g^*$ in the weak$^*$--topology (in some sense,
cf.\ \cite{kriegl;michor} for a theory of differentiation on locally convex
spaces), in which case we could stay in \fk g ($\ni df$); in this case we
would work with a restricted set of functions $f$ differentiable in a weaker
than norm--sense. We shall consider norm differentiability, if another
possibility is not mentioned explicitly. Most of formulas can be
considered, however, also in another interpretation.\hfill\dovi
\end{rem}

The functions $\rf:\mOGr\rarw\mbR$ of the form~\eqref{eq;2.32} will play a
r\^ole  of (nonlinear, unbounded -- in general) generators of transformation
groups (e.g.\ of time evolution) in our  theory, cf.
Proposition~\ref{prop;2.31}.  Their mutual Poisson
brackets  are defined in accordance with~\eqref{eq;2.27} in the following way:
\begin{subequations}\label{eq;2.33}
\begin{equation}\label{eq;2.a33}
\{\mbF^*f,\mbF^*h\}(\nu) :=\mbF^*\{f,h\}(\nu)\ \forall\nu\in\mbcDF,\forall
f,h\in C^{\infty}(Lie(G)^*,\mbR),
\end{equation}

\noidt where the bracket on the right side of the relation is the  Berezin
bracket.
The equation~\eqref{eq;2.a33} shows that the mapping \bF\ of \bcDF\ onto
its image in $\mfk g^*$ is a \emm Poisson
morphism ({\rm resp.} mapping)~, cf.\ \cite{weinst}. It
follows, that trajectories of the Hamiltonian flow corresponding to
Hamiltonian function h := {\sl h}$\circ\mbF$ on \bcDF\ are projected onto
trajectories of the Hamiltonian flow corresponding to the Hamiltonian
function {\sl h} on coadjoint
orbits of $G$. We shall find later also a possibility of determination of
flows on \bcDF\ from given Hamiltonian flows on $\mfk g^*$.
For $\nu\in\mbcDFr, \rf\in\mcl F$,  and $h\in C^{\infty}(Lie(G)^*,\mbR)$, we
shall extend our definitions of the Poisson brackets as follows:
\begin{equation}\label{eq;2.b33}
  \{\mbF^*h,\rf\}(\nu) :=\md h,\mbF(\nu)~\circ\{\mbF,\rf\}(\nu),
  \end{equation}

\noidt where $\md h,\mbF(\nu)~\in\mcl L(Lie(G)^*,\mbR)\ (= Lie(G),$ in the
case of
weak differentiability, cf.\ e.g.\ Remark~\ref{rem;infdG}) is the differential of
 $h$ in the point $\mbF(\nu)\in Lie(G)^* , \{\mbF,\rm f\}(\nu)\in Lie(G)^*$ is
 defined by its values $\{\mbF_{\xi},\rf\}(\nu):=
 \{\mh{X(\xi)},\rf\}(\nu)\in\mbR$ on the elements $\xi\in Lie(G)$,
 and $\{h_{X(\xi)},\rf\}$ is defined in~\eqref{eq;2.27b}.

 Let us note also, that
 \[ d_{\mrh}(\mbF^*f)=\mqr(X(\md f,\mbF(\mrh)~)),\ for\ \mrh\in\mbcDFr.\]

Let $\{\xi_j: j= 1,2,\dots \dim(G)<\infty\}$ be a basis of $\mfk g=Lie(G)$
and let
$F_j:=F(\xi_j)$ be coordinates of $F\in\mfk g^*$ in the dual basis. Then
the Poisson bracket~\eqref{eq;2.b33} can be expressed as
\bequ\label{eq;2.c33}
 \{\mbF^*h,\rf\}(\nu)=\sum_j\partial_jh(\mbF(\nu))\{{\rm
h}_{X(\xi_j)},\rf\}(\nu),
\end{equation}
and the Poisson bracket ~\eqref{eq;2.a33} can also be written in the form:
\begin{equation}\label{eq;2.d33}
\{\mbF^*f,\mbF^*h\}(\nu)=\sum_{j,k}\partial_jf(\mbF(\nu))\partial_kh(\mbF(\nu))
\{F_j,F_k\}(\mbF(\nu)).
\end{equation}
Observe that (cf.\ Theorem~\ref{thm;2.10}) the restriction to the
submanifold \OGr\ of the symplectic form $\Omega$ defined
in~\eqref{eq;2.15a}
on \OUr\ (i.e.\ the pull--back of $\Omega$ by the embedding of \OGr\ into \OUr)
coincides with the pull--back of the Kirillov--Kostant form $\Omega^K$ by the
mapping \bF:
\begin{equation}\label{eq;2.e33}
\left(\mbF^*\Omega^K\right)_{\nu}(\mbf v,\mbf w)=\Omega_{\nu}(\mbf v,\mbf w),
\end{equation}
for $\nu\in\mOGr,\ \mbf v,\mbf w\in\mTn\mOGr$, and $\mrh\in\mbcDFr$.
\end{subequations}

The formulas~\eqref{eq;2.b33},\rref2.c33~ show that the function
Q := $\mbF^*Q,\ Q\in
C^{\infty}(\mfk g^*,\mbR)$, generates a generalized (densely defined) vector
field $\mbf v_{\rm Q}$ on \Ss\ with values (for $\dim G<\infty$):
\begin{subequations}
\label{eq;2.34}
\begin{equation}
\label{eq;2.34a}
\mbf v_{\rm Q}(\nu)=\sum_j\partial_jQ(\mbF(\nu))\ \mvv{X(\xi_j)},\nu~.
\end{equation}
For an arbitrary $G$, and such $Q$ that $\md Q,\mbF(\nu)~\in\mfk
g\subset\mfk g^{**}$ one has:
\begin{equation}
\label{eq;2.34b}
\mvv{\rm Q},\nu~= \madns(\mqn(X(\md Q,{\mbF(\nu)}~))).
\end{equation}
\end{subequations}
This describes a class of Hamiltonian (resp.\ Poisson) generalized
vector fields generating the flows \pph t,Q~ leaving the
corresponding $U(G)$--orbits in the state space \Ss\ invariant.
One can see that the generating Hamiltonian functions Q are
constant on the orbits of the action $Ad^*(U(G_{\mbF(\nu)}))$,
i.e.
\[ {\rm Q}(Ad^*(U(\exp(t\eta))\nu))\equiv{\rm Q}(\nu), \ \ \eta\in
Lie(G_{\mbF(\nu)}).\]
This suggests an idea how to restrict the Poisson actions of other
generators to the orbits \OGr, cf.\ref{df;2.19}. We shall also introduce

\begin{defi}[\emm Poisson structure on submanifolds of \OUr~]
Let \cl N\ be a submanifold of \OUr, and $\Omega_{\nu},\ \nu\in\mOUr$ be
the symplectic form from~\eqref{eq;2.15a}. Let the restriction of $\Omega$
to \cl N, i.e.\ the pull back \wrt embedding $\iota_{\mcl N}:\mcl
N\rarw\mOUr$, $\Omega^{\mcl N}:=\iota^*_{\mcl N}\Omega$ be nondegenerate.
Then the symplectic structure $\Omega^{\mcl N}$ on \cl N\ will be also
called the \emm restriction of the Poisson structure on \Ss\ to \cl N~.\hfill\pika
\end{defi}

Let us formulate now a theorem containing some results and consequences of
the preceding considerations:
\begin{thm}\label{thm;G-flow}
Let $Q\in C^{\infty}(\mfk g^*,\mbR),\ {\rm Q}:=Q\circ\mbF$, hence {\rm Q} $\in
C^{\infty}(\mOGn,\mbR),\ \forall\nu\in\mbcDFr$. Assume that $\md
Q,\mbF(\mrh)~\in\mfk g$ for some $\mrh\in\mbcDFr$ (this assumption might be
nontrivial for infinite--dimensional $G$). Then \vv Q,\nu~
from~\eqref{eq;2.34b} is a Hamiltonian vector field on \OGr\  (hence, it is
tangent to \OGr, everywhere on \OGr) corresponding to the
Poisson structure on \OUr\ determined by the pull--back of
\bF,~\eqref{eq;2.a33}, or, equivalently, to the ``original'' Poisson
structure
on \Ss\ restricted to the (\N\cdot,1~--dense) collection of orbits \OGn\ lying
in \bcDFr. Then the (local) flow \pph Q,t~ leaves the orbits \OGn\
invariant.\hfill\zal
\end{thm}
We shall now formulate concepts describing Hamiltonian dynamics and
symmetries on ``allowed'' submanifolds of \OUr.

\begin{defs}[{\bf Classical and restricted G--dynamics}\index{G--dynamics}]\label{df;2.19}
\item{(i)}
Let \glss {\rm Ran}(\bF)~$\subset\mfk g^*$ denote the image of \bcDF\ under
\bF. We shall consider $\mfk g^*$ either with its canonical (coming from that
of \fk
g) norm--topology, or with its $w^*$--topology (again \wrt the
canonical norm--topology of \fk g,~\cite{bourb;Lie}; this will be different
from the norm--topology for
infinite--dimensional $G$). Let \glss \bs{\mcl E_{\mbF}}~
denote the closure of \Ran(\bF) in that topology. The space $\mcl E_{\mbF}$
will be also called the {\emm \bs G--classical\ ({\em alternat.}:
\bs G--macroscopic) phase
space~}
\ind{\bs G--classical\ ({\em alternatively}:\bs G--macroscopic) phase space}
of the system. Let by $C^{\infty}(\mcl M,\mbR),\ \mcl M\subset\mfk
g^*,$ be denoted the set of all infinitely differentiable functions on
an (arbitrary) open \nbhd of
\cl M\ in the corresponding topology (we shall not specify here the way of
differentiation on nonnormable lc--spaces, cf.
however~\cite{kriegl;michor}).
\item{(ii)}
If the \bcD--generalized differential of \rf\ is \Ss--integrable
we say that \emm {\rf\ generates the Poisson flow \pph{},\rf~ on \Ss}~.\index{Poisson flow}
\item{(iii)}
Let a densely defined real function $\rf:\mbcD\rarw\mbR$\ generate
a Poisson flow on \Ss, and let there is a differentiable function f on (an
open -- in the corresponding topology -- \nbhd of) \EF,
$f\in C^{\infty}(\mEF,\mbR)$\ such, that $\rf\equiv\mbF^*f:=
f\circ\mbF\ {\text on}\ \mbcD$. Then {\bf \rf\ is a \Gcl generator}\index{\Gcl generator}.
\footnote{More sophisticated and more distinctive (and also more
complicated) work with domains was presented
in~\cite{bon10}; the corresponding modifications of concepts connected with
domains
presented in this paper can be, however, seen without being explicitly
formulated here.}
\item{(iv)}
Let \rf\ generate a Poisson flow on \OUr\ (the submanifold
$\mOUr\subset\mSs$\ can be substituted for \Ss\ in obvious
modifications of preceding definitions). Let $\nu$ be such that
$\mOGn\subset\mbcDF\cap\mOUr\cap\mbcD$, and let the restriction $\rf^{\nu}$\
of \rf\ to \OGn\ can be expressed in the form
 \bequ
 f^{\nu}(\mbF(\nu'))\equiv\rf^{\nu}(\nu'):=\rf(\nu'),\ \text{for}\ \nu'\in\mOGn,
 \end{equation}
with some $f^{\nu}\in C^{\infty}(Ad^*(G)\mbF(\nu),\mbR)$, hence
\glss $\rf^{\nu}=\mbF^*f^{\nu}$~. Then the function \rf\
 will be called a \emm \nGcl generator~. (Hence, the same
 function \rf\ can be a \nGcl generator for several different
 orbits \OGn.)
 \item{(v)}
 Let \rf\ be a \nGcl generator. Its flow \pph{},\rf~\ needn't
 leave the orbit \OGn\ ($\subset\mOUr$) invariant.\footnote{This is a
 difference \wrt \Gcl generators, cf.\ also Proposition~\ref{prop;2.26}.}
  Let \pph{},{\nu,\rf}~\
 be the (Poisson) flow on the orbit \OGn\ corresponding to the vector field
 on \OGn\ generated by $\rf^{\nu}$\ according to~\eqref{eq;2.b33}
 and~\eqref{eq;2.34} (with h, resp.\ Q replaced by $f^{\nu}$). The flow
 \pph{},{\nu,\rf}~\ will be called the \emm {$\nu G$--restriction of the flow
 \pph{},\rf~}~ to the orbit \OGn.\hfill\pika
\end{defs}

Let us present now, without detailed explanation (hence without an analysis
and proofs), some
examples of \nGcl generators.
\begin{exmp}\label{exmp;nGcl}
Let a representation $U(G)$\ be given as above, and let \bF\ be the
corresponding momentum mapping. Let Y be a selfadjoint
operator on \H, and let \h Y\ be the corresponding
(densely defined, generalized) generator. Let
$\mOGn\subset\mbcDFr\cap\mDdYa$, with \DdYa\ denoting the set of analytic
elements of $\delta_Y$ belonging to $\mfk F_s$.
Then \h Y\ is a \nGcl generator, e.g., in any of the following cases:
\item{(i)} $Y:=X_{\xi}$\ for some $\xi\in\mfk g$.
\item{(ii)} $Y:=i^N[X(\xi_1),[X(\xi_2),[\dots[X(\xi_N),A]\dots]]]$, where
$\xi_j\in\mfk g,\ (j=1,2,\dots N)$, and $A$\ is such a selfadjoint
operator on \H\ that \h A\ is a \nGcl generator. The commutators
can be considered here in a generalized sense,~\cite{bon8}, so
that it ensures existence of \h Y\ in the points $\mrh\in\mOGn$ in
the sense of~\eqref{eq;2.12p}. This can lead to \nGcl generator \h
Y\ even in some cases, when the above expression does not
determine a well defined linear operator $Y$.
\item{(iii)} All stability subgroups $G_{\mbF(\mome)}\ (\mome\in\mOGn)$
of points $\mbF(\mome)\in\mbF(\mOGn)\equiv Ad^*(G)\mbF(\nu)$ are symmetry
groups of the operator $Y$:
\[ \ U(g)YU(g^{-1})=Y,\ \forall g\in\cup\{G_{\mbF(\mome)}:\mome\in\mOGn\}, \]
and, moreover, $h_Y^{\nu}\in C^{\infty}(\mOGn,\mbR)$.
\item{(iv)} The orbit \OGn\ is such, that $G_{\mbF(\mome)}=G_{\mome},\
\forall \mome\in\mOGn$. The subgroups $G_{\mome}\subset G$ are stability
subgroups of the points \ome\ of the orbit \OGn\ for the considered
action:
$g\mapsto Ad^*(U(g))\mome = \mome\eequiv g\in G_{\mome}$.\hfill\dovi
\end{exmp}

Restrictions of ``true quantum--mechanical dynamics'' to various
submanifolds of ``coherent states'' (i.e.\ to orbits \OGn) are often
considered~\cite{klaud1,rowe1,rowe2} as approximations (sometimes called
``quasiclassical'') to the ``true dynamics''.\footnote{\label{1cl-proj}These
``restrictions'' were called in~\cite{bon8} {\bf ``classical projections''}
of quantummechanical evolutions.}
   This is not, however, a ``good approximation''
for a general (linear) quantum dynamics, and what are conditions for
well controlled validity (i.e.\ a relevance) of such approximations is not
yet, as far as the present author knows, generally established.

\begin{rem}\label{rem;2.21}
The $U(G)$--restriction\ $\mpph{},{\nu,\rf}~$\ of the quantal flow
\pph{},\rf~\ needn't be ``close'' to \pph{},\rf~\ for a general \nGcl
generator \rf, not even for ``classical'' (or ``macroscopic'') quantities
described by expectations of a distinguished subset of selfadjoint
operators. One can compare, e.g., the evolution of the expectations
\h{X(\xi)}\ of quantum ``observables'' \Xx\ under \pph{},\rf~, i.e.\ the
function
\begin{equation}\label{eq;classrestr}
(t;\xi)\mapsto \mh{X(\xi)}(\mpph t,\rf~\nu)\equiv \mbF_{\xi}(\mpph
t,\rf~\nu)\in\mfk g^*,
\end{equation}
with the restricted evolution $\mbF_{\xi}(\mpph t,{\nu,\rf}~(\nu))$, for the
same initial conditions.\hfill\dovi
\end{rem}
Let us illustrate this remark by a simple example,
cf.\ also~\cite[4.1.10]{bon8}.

\begin{ill}[Restricted and ``global'' flows might be ``very'' different]
\label{ill;difrestr}\nl
Let us take $\mH := \mLq$, and let
\[ \psi\in\mH,\
\psi(q):=\pi^{-\frac{1}{4}}\exp\left(-\frac{1}{2}q^2\right);\]
let us set $\psi_z:=U_z\psi$, with $z:=q-ip\in\mbC$, and
$U_z:=\exp(i(pQ-qP)).$
Here $Q$\ and $P$ are the Schr\"odinger operators of position and linear
momentum in QM:
\label{eq;1GWH}
\[ Q\chi(q)\equiv q\chi(q),\ P\chi(q)\equiv
-i\frac{\partial}{\partial q}\chi(q),\quad \chi\in\mH.\]
Let the (artificial) ``generator of time evolution'' be
$H:=\alpha\cdot P_{\psi},\ \malp\in\mbR$, i.e.\ it is proportional to a
one--dimensional
projection. We shall consider the restriction of the corresponding flow to
the orbit $\mcl O_{\boldsymbol{\psi}}(G_{WH})$ of the 3--dimensional
 Weyl--Heisenberg group \GWH\ (cf.\ also Subsection~\ref{IIIA1;CCR})
defined by the injective mapping of
the ``classical phase space'' $\mbC\ni z$\ into the projective Hilbert space:
$z\ (\in\mbC)\mapsto P^z_{\psi}:= U_zP_{\psi}U^*_z\in\mPH$. If we parameterize
points of the orbit by $z\in\mbC$, then the restriction ${\rm h}^{\psi}_H$
of the corresponding Poisson generator \h H\ to the orbit is:
\begin{subequations}
 \label{eq;clHam1}
 \begin{equation}\label{eq;1clHam}
 {\rm h}^{\psi}_H(z)\equiv Tr(P^z_{\psi}H) =
\alpha\exp\left(-\frac{1}{2}\overline{z}z\right),
\end{equation}
with $z\mapsto\overline{z}$ being the complex conjugation. The restricted
flow is identical (by the identification $z\longleftrightarrow
P^z_{\psi}$) to the Hamiltonian flow
 \begin{equation}\label{eq;2clHam}
 \mpph t,{\psi,H}~z\equiv \exp(-it{\rm h}^{\psi}_H(z))z,
 \end{equation}
generated by the Hamiltonian
function~\eqref{eq;1clHam} on the classical phase space $\mbR^2$ with the
symplectic form $\Omega\equiv dp\wedge dq$.
The ``true quantal flow'' with the same initial condition $z=q-ip$ is
\bequ\label{eq;3clHam}
 \begin{split}
 \mpph t,H~z&:=Tr(\exp(-itH)P^{\psi}_z\exp(itH)(Q-iP)) \\
 &\ \equiv
(1-\alpha^{-1}{\rm h}^{\psi}_H(z))z+\alpha^{-1}{\rm
h}^{\psi}_H(z)\exp(-it\alpha)z.
\end{split}
\end{equation}
\end{subequations}
By comparing these two evolutions of ``the same classical quantities'',
i.e.\ the two motions in \bC, we see two uniform motions on mutually tangent
circles
with different radii and different frequencies. This shows that, for
general Hamiltonians, the ``classical projections'' needn't be any
approximations to the ``true quantum dynamics''.\hfill\dovi
\end{ill}
The next assertion shows in what sense the restricted generators also are
of relevance for the (unrestricted) quantum theory.
\begin{prop}\label{prop;2.22}
Let \rf\ be a \nGcl generator and let $\rf^{\nu}$\ be its restriction to
\OGn. Then, by considering the definitions~\eqref{eq;2.33} of Poisson
brackets on \OGn, for all $\nu'\in\mOGn$, and for any $Q\in C^{\infty}(\mfk
g^*,\mbR)$, the following relations are valid:
\begin{equation}\label{eq;2.35}
\{{\rm Q},\rf\}(\nu')=\{{\rm
Q},\rf^{\nu}\}(\nu')=\mbF^*\{Q,f^{\nu}\}(\nu'),
\end{equation}
where ${\rm Q}:=\mbF^*Q$, and $\rf^{\nu}=:\mbF^*f^{\nu}$, and where the
first bracket is defined according to~\eqref{eq;2.b33} (or, equivalently,
by the formula~\eqref{eq;2.6} with a help of generalized differentials of
{\rm Q}\ and \rf).\hfill\zal
\end{prop}
\begin{proof}
The second equation in~\eqref{eq;2.35} is just the first equation
of~\eqref{eq;2.33}. The unrestricted Poisson bracket on \Ss\ occurring on
the left side of~\eqref{eq;2.35} is equal, according to~\eqref{eq;2.b27},
to the derivative of \rf\ along to the vector field~\eqref{eq;2.34} at each
point $\mrh\in\mOGn$. This implies, that the derivative $\md{\rf({\bf v}_{\rm
Q})},\nu'~=\{{\rm Q},\rf\}(\nu')$ in any point $\nu'\in\mOGn$ of an
arbitrary function only depends on its restriction $\rf^{\nu}$\ to \OGn.
One has $\md{\rf({\bf v}_{\rm Q})},\nu'~ \equiv\md{\rf^{\nu}({\bf v}_{\rm
Q})},\nu'~$ on \OGn, and the last derivative is expressed by the Poisson
bracket~\eqref{eq;2.33} on \OGn. This proves the first equation.
\end{proof}
\begin{rem}
The definitions~\eqref{eq;2.27} of the Poisson bracket {\em on an
$Ad^*(U(G))$--orbit} \OGn\ were formulated with a help of
selfadjoint operators on (dense domains of) \H, so that our construction of
the Poisson structure on \OGn\ is not an ``intrinsic construction'' on the
orbit alone: It uses the values of the differentials of the functions
\h{X(\xi)}\ and \rf\ as elements of infinite--dimensional spaces \Tr\OUn\
for points of a $G$--orbit, $\mrh\in\mOGn\subset\mOUr$ (which is
finite--dimensional in the case $\dim G < \infty$). The differential
\d\rf,\mrh~\ cannot be calculated in general cases from the restriction of
$\rf\in\mcl F$\ to the orbit \OGn\ only. If \OGn\ is a symplectic manifold
with the symplectic structure obtained by pull--back of the
Kirillov--Kostant form $\Omega^K$ on $\mbF(\mOGn)=Ad^*(G)\mbF(\nu)$, or
equivalently, if the restriction of the bilinear forms $\Omega_{\mrh},\
\mrh\in\mOGn$\ to \OGn\ (i.e.\ to $\mTr\mOGn\times\mTr\mOGn,\
\forall\mrh\in\mOGn$) is nondegenerate, then we have defined on \OGn\ the
necessary isomorphism (at least for $\dim G<\infty$) between \Tr\OGn\ and
\Trs\OGn\ ($\mrh\in\mOGn$). In this special case, we can calculate
restrictions \pph{},{\nu,\rf}~\ of the flows \pph{},\rf~ to the orbit \OGn\
with a help of the restrictions $\rf^{\nu}$, cf.
Definition~\ref{df;2.19}(iv), only.\hfill\dovi
\end{rem}
Let us look now on some properties of the ``classical phase space''
$\mEF\subset Lie(G)^*$. Let \glss conv$_0(B)$~ be the convex hull of a
subset $B$
of some locally convex space, and let \glss conv$(B)$~ be its closure.
Let \glss $\mEF^0:=\mPH\cap\mbF(\mDGom)$~. Then we have:
\begin{prop}\label{prop;2.24}
The range of \bF, \Ran(\bF), is a convex, $Ad^*(G)$--invariant subset of
$Lie(G)^*$\ containing
conv$_0(\mEF^0)=\mbF(\mbcDFr)$. If $\dim(G)<\infty$, then \Ran$(\mbF)=\mEF$,
i.e.\ it is a closed subset of $Lie(G)^*$.\hfill\zal
\end{prop}
\begin{proof}
The mapping $\mbF:\mbcDF\rarw Lie(G)^*$ is affine, and \bcDF\ is convex,
since \cl D(\h{X(\xi)}) is convex and \h{X(\xi)} is affine. Hence \Ran(\bF)
is convex, and conv$_0(\mEF^0)\subset{\rm Ran}(\mbF)$. One can see from the
definitions that conv$_0=\mbF(\mbcDFr)$, and that \bcDFr\ is norm--dense in
$\mbcDF\subset\mSs$. The $Ad^*(G)$--invariance follows
from~\eqref{eq;2.28}, and from the $Ad^*(U(G))$--invariance of \bcDF.

Let $\dim G<\infty$. The closedness of \Ran(\bF) can be proved by
construction of a projection--valued measure on $\mfk g^*$ representing the
commutative group (linear space) \fk g, resp.\ the commutative algebra of
``classical observables'' $C^\infty(\mfk g^*,\mbR)$ (generated by the
functions $f_{\xi}(F):=F(\xi),\ \xi\in\mfk g, F\in\mfk g^*$),
\cite{bon8,bon1}. The support of this measure is identical with \Ran(\bF),
hence \Ran(\bF)\ is closed.
\end{proof}



\section{Symmetries, Dynamics and Observables}\label{IIC;symm-obs}

It was shown in Sections~\ref{IIA;q-phsp}, and~\ref{IIB;gener}, how
real--valued functions $\rf:\ \nu\mapsto\rf(\nu),\ (\nu\in\mSs)$
 can be used  in  the  r\^ole  of  generators  of  the
one--parameter  families \pph{},\rf~\ of  transformations  of   elementary
states. Differentiable functions on phase spaces of CM  are  used
in double r\^ole: as the generators, as well as ``observables'', i.e.
as a certain objects ascribing  (numerical)  values  of  possible
results of  specific  measurements  to  states $\nu$  to  which  the
measurements are applied. Selfadjoint operators $X$ represent  both
these objects in QM: they are generators of  the  unitary  groups
$\exp(-itX)$  on \H,   and   also   observables   with   probability
distributions $\mu^X_\nu$, cf.\rref1prob~, and point~\ref{note;2.12},
of their (real) values measured in the state $\nu$.
We include into the presented scheme also such a double r\^ole  and
the standard interpretation for the functions \h X: Besides of being
generators  according  to  assertion~\ref{pt;2.15}, they also could be
considered as observable quantities with $n$--th momenta $\nu(X^n)$  (if
they exist) of the probability measures $\mu^X_\nu$   calculated  directly
from \h X, as it is indicated in formula~\eqref{eq;3.14} of
Subsection~\ref{IIIA1;QM},  cf.
also~\cite{cir}.  Difficulties  arise,  however,  in   trials   to
interpret a nonlinear function \rf\ defined on (a subset of) \Ss\ in
a r\^ole of an observable in the traditional way,  as  it  will  be
shown in Interpretation~\ref{int;tr-prob}, in Note~\ref{not;3.7}, as well as
in Interpretation~\ref{int;expect}.

   Now we shall show that, on the other hand, the use of nonlinear
   generators of transformation groups in QM implies also necessity of
introduction of some nonlinear ``observables''  together  with
the affine ones.

   Let us assume that we have a flow \pph{},f~\  generated by a  nonlinear
generator $f$\ according to Section~\ref{IIA;q-phsp}, and let u$_f$  be  the
 corresponding solution of~\eqref{eq;2.11}. For any ``observable'' $\mh \ra\
 (\ra \in\mLHs)$, one has ``a natural time--evolved form'':
 \[ \mh \ra^t(\mrh):=\mh \ra(\mpph
 t,\rf~\mrh)\equiv Tr(\mrh\mun\rf,t,\mrh~^*\,\ra\,\mun\rf,t,\mrh~), \]

\noidt and the functions $\mh \ra^t$ are not generally of the form \h{\ra(t)},
i.e.
\[ \mrh\mapsto Tr(\mrh\mun \rf,t,\mrh~^*\,\ra\,\mun\rf,t,\mrh~) \]
are not affine functions of \rh\ for all $t\in\mbR$;
this can be seen, e.g., from~\cite[Proposition 4.3]{bon1}.

\begin{intpn}\label{int;obs}
   We propose an interpretation scheme, in which a numerical--valued
function \rf\ on \Ss\ can
have several different interpretations as  ``observables''  in
EQM.  The  ``appropriate  choice''  of  the  class   of
observables of the system depends also  on  the  chosen  symmetry
group $G$ entering into the description of the  considered  system.
From our point of view, the specified symmetry group $G$  could  be
interpreted as a group of motions of (a  relevant  part  of)  the
macroscopic  background  determining  physical  meaning  of   the
``observables'', i.e.\ quantities used for description of  empirical
specification of states  of  a  given  physical  system.  We  can
interpret  the  genuine  mixtures  (cf.\ Subsection~\ref{q-phsp;mixt}) as
describing  states  of a ``microscopic subsystem'' of a composed system
consisting of the
``microscopic  subsystem''  (i.e.\  the  considered   one)   and   a
``macroscopic background''.  This ``background'' can interact with the
considered
quantum  system also without  being  influenced  by  it;  it  can  be
represented, e.g.\ by an infinite number of copies  of  the  ``considered
quantum  system''  interacting  mutually  by  a  type  of  quantum
mean--field
interaction,~\cite{hp+lie1,bon1,bon2,bon3,unner0,unner1,unner2,unner3}.
The genuine mixture of the ``microsystem'' corresponds to  a  nontrivial
statistical
distribution of values of macroscopic observables $X_\Pi(\xi)$): The values
of some ``macroscopic observables'' of this ``macrosystem''
(describable in classical terms) are
 correlated with the states of the ``microsystem'' entering into the
support of the measure determining the genuine mixture, cf.\ also
Remark~\ref{rem;macro}. \hfill\bpika
\end{intpn}

\def\nazov{{
\ref{IIC;symm-obs}\quad Symmetries, Dynamics and Observables}}

   We shall introduce now a (in  a  certain  sense  minimal) set of
nonlinear functions representing observables and  containing  all
the usually used ``linear observables'' of QM  which  is  invariant
with respect to a sufficiently large class of  (nonlinear)  dynamics
and also with respect to the symmetry group specified by the
representation $U(G)$, as it was introduced in Section~\ref{gener;nl-grp}.
We shall introduce also other concepts (generators of different kinds, e.g.)
forming with the chosen set of observables a consistent closed theory.
This set of concepts specifies a method of determination a subtheory from
the overwhelmingly large set of possible (mathematically admissible)
``generators'', and ``observables'' of possible formally extended quantum
theories. The usefulness of the (representation of the) group $G$ is here
(at least) twofold, interpretational, and technical:
\begin{description}
\item[(interpretation)]
The group $G$, if interpreted a priori in terms of some
``macroscopic variables'', cf.\ Remark~\ref{rem;macro}, can serve as a
theoretical tool for specification of interpretation of mathematically
specified
``observables'', as well as symmetry transformations generated by a
distinguished class of ``generators''.
\item[(technicality)]
The strongly continuous unitary representation $U(G)$ is an
effective device to select the dense set \bcDF\ of points, as well as
of submanifolds \OGr, where the
differential--geometrical objects as ``differentials'', or ``vector
fields'' can be defined from a specified (by the same representation) set
of generators, which are locally unbounded for many physically relevant
cases (``generically'' for physically relevant noncompact group
representations).
\end{description}
\begin{defs}[\emm G--generators~]\label{df;2.25a}
\nopagebreak
\item{(i)}
Let \glss \GGc~\ denote the \emm Poisson algebra of G--classical
generators~: $\rf\in \mGGc\eequiv \rf= \mbF^*f:= f\circ\mbF$ for some $f\in
C^\infty(\mEF,\mbR)$.
Let $\mpEF(f)\subset Lie(G)^*$  be
some  (for  each  f  separately
chosen) open neighbourhood of \EF\ in $Lie(G)^*$ endowed with one of the
canonical topologies, cf.\ Definitions~\ref{df;2.19}.
The Poisson structure on \GGc\ is expressed by~\eqref{eq;2.33}.
\item{(ii)}
Let \rf\ be densely defined real--valued function on \Ss\ such that its
\bcD--generalized differential exists and it is \Ss--integrable, cf.
Definition~\ref{df;g-dif}. Let $\nu\mapsto\mvv\rf,\nu~$ be the corresponding
\bcD--Hamiltonian vector field and assume, that its flow \pph {},\rf~ is
complete, and leaving \bcDF\ invariant. Let, moreover, the flow can be
described by \glss ${\mun\rf,\cdot,\cdot~}:\mbR\times\mSs\mapsto \mfk U$~
satisfying~\eqref{eq;2.12},~\eqref{eq;2.13}, and also~\eqref{eq;2.11} on a
``sufficiently large'' subdomain of \bcDFr\ (cf.\ Definition~\ref{df;g-dif}).
Then \rf\ will be called a {\bf(quantum) G--generator}\index{quantum G--generator}.
\item{(iii)}
Let, for the quantum G--generator \rf\  of the above definition (ii),
$\mbF(\mpph
t,\rf~\nu)=\mbF(\mpph t,\rf~\nu')$\ for all $\nu'\in\mbF^{-1}[\mbF(\nu)]$,
for any $(t;\nu)\in\mbR\times\mbcDF$; the G--generator \rf\ is called then
a \emm
G--(classically) deterministic generator~. In this case, we shall denote
\[{\mph t,\rf~}[\mbF(\nu)]\equiv\mbF({\mpph t,\rf~}\nu);\]
this relation determines a flow \ph{},\rf~ on \EF.
\item{(iv)}
A quantum G--generator which is not G--(classically)
deterministic is called a \emm G--(classically) stochastic generator~: The
quantum flow \pph{},\rf~ does not determine a classical flow, and a
``corresponding'' classical evolution might be considered (?) as a stochastic
process.
\item{(v)}
Let a G--(classically) deterministic generator \rf\ be such that one can
choose
\begin{equation}\label{eq;sym-gen}
\mun\rf,t,\nu~\equiv\mun\rf,t,\nu'~,\ \forall\nu'\in\mbF^{-1}[\mbF(\nu)],\
(\forall (t;\nu)\in\mbR\times\mbcDF);
\end{equation}
then we can (and we shall) write \glss ${\mun \rf,t,F~:=\mun\rf,t,\nu'~}$~ for
$\nu'\in\mbF^{-1}[F],\ F\in\mEF$. Let the mappings (cf.
Definitions~\ref{df;2.25b} for \CG)
\begin{equation}
\glss \tau^{\rf}_t~:\mCG\rarw\mCG,\ \mfk h\mapsto\mfk h_t,\ \mfk
h_t(F):=\mun\rf,t,F~^{-1}\mfk h(\mph t,\rf~F)\mun \rf,t,F~
\end{equation}
be C\autm s of \CG for all $t\in\mbR$. Then \rf\  is called a \emm G--symmetry
generator~. The set of all G--symmetry generators will be denoted \GG. It
is $\mGGc\subset\mGG$, as will be shown in Proposition~\ref{prop;2.26}, and
Theorem~\ref{thm;2.29}.\hfill\pika
\end{defs}
These definitions of different types of generators (of evolutions, or
symmetry groups) specify also their relations to the corresponding
transformations induced in the set of ``classical variables'' determined by
the chosen (unitary representation $U(G)$ of the) group $G$.
``Observables'' in EQM are not sufficiently determined by real--valued
functions on \Ss; the quantummechanical interpretation needs possibility of
determination of probability distributions in any point $\mrh\in\mSs$\ for
general observable quantities. The following definitions of observables
respect also the requirement of their invariance \wrt
``Heisenberg--picture--transformations'', into which nonlinearities bring
modifications \wrt the linear case: One has to distinguish between
transformations of elementary states (described by density matrices) and
corresponding transformations of observables (described, e.g., by operator
valued functions of density
matrices). This distinction ensures ``conservation of transition
probabilities'' also in nonlinear QM.

\begin{defs}[\emm G--observables~]\label{df;2.25b}
\item{(i)}
Let the \glss $s^*(\mLH,\mbcDF)$--topology~ on \LH\  be  given  by  the
family of seminorms \glss $p_\nu, p^*_\nu\ (\nu\in\mbcDF)$~ determined by
their  values
$p_\nu(x) := \nu(x^*x)^{1/2}$, and $p_\nu^*(x):=\nu(xx^*)^{1/2}$\ on $x\in\mLH$.
Let \fk f\ and \fk h\ be uniformly bounded operator--valued functions on
\EF, \fk f: $\mEF\rarw \mLH,\ F\mapsto\mfk f(F)$, $\mN \mfk f,{}~:= \sup\{\mN
\mfk f(F),{}~: F\in\mEF\} < \infty$, which are $s^*(\mLH,\mbcDF)$ --
continuous.\footnote{Remember that a topology on $\mfk g^*$ is here
understood to be one of
the two canonical topologies, which are mutually equal for finite--dimensional
group $G$, cf.\ definition (i) in~\ref{df;2.19}.}
Let \glss \Cbs~\ be  the  set  of  all  such
functions endowed with (pointwise) operations: $(\mfk f+\mlam\mfk
h)(F):=\mfk f(F)+\mlam\mfk h(F)$,\ $(\mfk f\mfk h)(F):=\mfk f(F)\mfk h(F)$,
and $\mfk f^*(F):=\mfk f(F)^*,\ \mlam\in\mbC$.  It  can
be shown~\cite{bon8} that \Cbs\  with these algebraic operations and the  norm
is a \Ca. The elements of \Cbs\ are \emm unrestricted bounded G--observables~.
\item{(ii)}
Let $\mfk B_U:=U(G)''$\ be the von Neumann subalgebra of \LH\ generated by
U(G).
Let \glss\bs{\mCG}~\ (resp.~\glss\bs{\mCG_U}~)
be  the  \Csa,~\cite{bon8}, of \Cbs\ generated  by  the
uniformly bounded operator--valued functions
\[ \mfk h_{\rx,\gamma,f}: F (\in\mEF)\mapsto\mfk
h_{\rx,\gamma,f}(F):=U(\gamma(F))^*\,\rx\, U(\gamma(F))f(F),\] for
all $\rx\in\mLH$\ (resp.\ $\forall\rx\in\mfk B_U$), $\gamma\in
C(\mEF,G),\ f\in C_b(\mEF,\mbR)$; elements of \CG\ will be
considered also as operator--valued functions on \bcDF\  obtained
by pull--back by \bF: $$\mfk f\in{\cal {C}}^G\imply\mfk
f:\mrh(\in\mbcDF)\mapsto\mfk f(\mbF(\mrh)).$$

The set \CG\ (resp.\ $\mCG_U$)
is called the {\emm \bs{C^*}--algebra of
G--observables~}
(resp.\ the \emm \bs{C^*}--algebra of UG--observables~) of
the system. Any $\mfk f=\mfk f^*\in\mCG$ will be called a \emm
G--observable~. Elements $\rx\in\mLH$ are considered as elements of \CG\ for
any $U(G)$: They are identified with the constant functions $\mfk
h_\rx:F\mapsto\mfk h_\rx(F):=\rx$ on \EF. Elements $\mfk h_\rx\
(\rx\in\mLHs)$ generate a (complex) \emm subalgebra of elementary quantum
observables~ denoted by \glss\bs{\mCG_q}~ which is a subset of \CG\
isomorphic to \LH\ (for any choice of $U(G)$). Any uniformly bounded
element $\rf=\mbF^*f\in\mGGc$\ will be considered also as the element of
$\mCG_U\ (\subset\mCG$) described by the scalar--valued function $\mfk h_f:
F\mapsto\mfk
h_f(F):=\mbI\!\cdot\! f(F)$\ on \EF. The G--observables of this form will be
called the (bounded) \emm G--classical observables~. They belong to \glss
$\mCGc:= \mbI\!\cdot\!C(\mEF,\mbC)\subset\mCG_U$~, $\mbI:=I_\mH$.
\item{(iii)}
The \emm unbounded G--observables~ (resp.\ {\bf UG--observables}) are functions
$Y:F\mapsto Y(F)$ on \EF\
with values in unbounded selfadjoint operators \glss $Y(F)$\ ~on \H, with the
spectral measures \glss $E_{Y(F)}$~ such that the functions
$E_{Y(\cdot)}(B):F\mapsto E_{Y(F)}(B)\in\mfk P(\mLH)$\ (cf.
Note~\ref{note;types}) belong to \CG\
(resp.\ to $\mCG_U$)\
for any Borel set
$B\subset\mbR$. Note that we needn't specify the domains of the operators
$Y(F)$\ here. \hfill\pika
\end{defs}

\begin{defs}[\emm Function representation of observables~]\label{df;2.25c}
\item{(i)} Let us denote $h_{\mfk f}:\nu\mapsto h_{\mfk f}(\nu):=\nu(\mfk
f[\mbF(\nu)])$. The mapping $\mfk f(\in\mCG)\mapsto h_{\mfk f}$ is
not injective.  Let us introduce the functions \hh f,\cdot,\cdot~
of two variables $(\mrh;\nu)\in\mSs\times\mbcDF$,
$(\mrh;\nu)\mapsto\glss{\mhh f,\mrh,\nu~}~:=\mrh(\mfk
f[\mbF(\nu)])$. Then $ \mh{\mfk f}\equiv\mhh f,\nu,\nu~,\
\nu\in\mbcDF$. The mapping $\mfk f\mapsto\mhh f,\cdot,\cdot~$ is
an injection into the set \glss $\hat{\mCG}$~ of real-valued
 functions $\hat{\rf}$  defined
on the product $\mSs\times\mbcDF$\ such that the dependence
$\mrh\mapsto\hat{\rf}(\mrh,\nu)$\ is affine bounded continuous for
each fixed $\nu$, and
$\hat{\rf}(\mrh,\nu)\equiv\hat{\rf}(\mrh,\nu')$ for all
$\nu'\in\mbF^{-1}[\mbF(\nu)]=$\ (a \emm level set of the mapping
\bF~), for each fixed $\mrh\in\mSs$. Continuity properties of the
functions $\nu\mapsto\mhh f,\mrh,\nu~$ are determined by
properties of \bF\ and by the continuity of $F\mapsto\mfk f(F)$.
The element $\hat{\mh{\mfk f}}\in\hat{\mCG}$\ will be called the
\emm function representative of the (bounded) G--observable \fk f~
of the system; elements of $\hat{\mCG}$ will also be called the
\emm G--observables~. The first variable $\mrh\in\mSs$ in
$\hat{\rf}(\mrh,\nu)$\ will be called the \emm quantum variable~,
and the second one, $\nu\in\mbcDF$, will be called the \emm
G--classical variable~ (cf. Section~\ref{sec;IIIB} for motivation
of such terminology) of the (function representative of the)
observable $\hat{\rf}\in\hat\mCG$. The function $h_{\mfk
f}:\nu\mapsto h_{\mfk f}(\nu):=\nu(\mfk f[\mbF(\nu)])$ will be
called the \emm reduced function representative~ of $\mfk
f\in\mCG$.
\item{(ii)} Functions \h Y, and $\hat{\mh Y}$, for unbounded observables
$Y$, can be introduced as (not everywhere defined) \emm function
representatives of unbounded observables~, in analogy with the
case (i) of bounded observables, i.e.\ $\hat{\mh
Y}(\mrh,\nu)\equiv Tr\bigl(\mrh Y\bigl(\mbF(\nu)\bigr)\bigr)$ on a
corresponding domain in $\mSs\times\mbcDF$\ (the domain
specification would be here, generally, difficult).\hfill\pika
\end{defs}

 We shall next introduce states (as linear functionals on an algebra of
``observables'') corresponding to the general concept of ``genuine
mixtures'' introduced in the Subsection~\ref{q-phsp;mixt}. They will be
``suited'' also to the just introduced constructions determined by the
representation $U(G)$.
\begin{defs}[\emm G--states~]\label{df;2.25d}
\item{(i)}
Let \glss $\mcl M^G$~ be the set of regular Borel probability measures on \bcDF\
(with its Borel structure coming from the metric topology of \Ss). The
genuine mixtures $\mu\in\mcl M^G$\ determine the set \glss $\mS^{cl}_G$\ ~of the
\emm G--classical states \bs{\mome_\mu}~ of the considered system: The
elements $\mome_\mu\in\mS(\mCG):=(\mCG)^*_{+1}$\ (:= the state space of the
\Ca\ \CG) are determined by their values $\mome_\mu(\mfk f),\ \mfk
f\in\mCG$\ expressed by the integrals:
\begin{equation}\label{eq;2.37}
\mome_\mu(\mfk f):=\mu(\mh{\mfk
f})\equiv\int\nu(\mfk f\bigl(\mbF(\nu)\bigr))\,\mu(d\nu),\ \forall \mfk
f\in\mCG.
\end{equation}
\noidt Elementary states $\mrh\in\mbcDF$\ are represented by Dirac measures
$\delta_\mrh$\ concentrated at \rh.

{\rm In these states, the values of the quantum variable \rh\ of \hh f,\mrh,\nu~
copy those of the classical variable. If the ``microscopic state''
described by the quantum variable is not connected with the classical
variable in this way, one arrives at definition of more general states:}
\item{(ii)}
Let $\hat\mrh:\mbcDF\rarw\mSs,\ \nu\mapsto\hat\mrh(\nu)$ be a Borel
function. Let the state \glss $\mome_{\mu,\hat\mrh}\in\mS(\mCG)$~ be defined by
\begin{equation}\label{eq;2.42d}
\mome_{\mu,\hat\mrh}(\mfk f):=
\int\hat\mrh(\nu)(\mfk f\bigl(\mbF(\nu)\bigr))\,\mu(d\nu)\equiv
\int Tr\Bigl(\hat\mrh(\nu)\mfk f\bigl(\mbF(\nu)\bigr)\Bigr)\,\mu(d\nu).
\end{equation}
The set of all such states $\mome_{\mu,\hat\mrh}\in\mS(\mCG)$ will be
denoted by \glss $\mS_G$~. The elements \bs{\mome_{\mu,\hat\mrh}} of $\mS_G$\
will be called \emm G--states~. Clearly $\mS_G^{cl}\subset \mS_G$. The
functions $\hat\mrh$ playing the
described r\^ole will be called here \emm quantum deviation functions~. For
$\mome_{\mu,\hat\mrh}\in\mS_G^{cl}$\ one has $\hat\mrh(\nu)\equiv\nu$.
\hfill\pika
\end{defs}

\begin{defi}[\emm G--systems~]\label{df;G-syst}
Let a unitary continuous representation $U(G)$ of a Lie group $G$ be given.
The model of a (quantummechanical) physical system of EQM in
which the sets of its (``system determining'') generators, states, and
(bounded) observables
coincide with the sets of the G--symmetry generators \GG, G--classical
states $\mS^{cl}_G$, and G--observables \CG\ (resp.\ UG--observables
$\mCG_U$) respectively is called the
\emm G--classical ({\it resp.} UG--classical) quantum system~, or just the
\emm G--system ({\it resp.} UG--system)~, based on the
representation $U(G)$. The G--system (resp.\ UG--system) will be also denoted
by $\Sigma_G$ (resp.\ by $\Sigma_{UG}$). One has $\Sigma_G=\Sigma_{UG}$\ for
irreducible $U(G)$.\hfill\pika
\end{defi}
\begin{rem}\label{rem;G-syst}
This (basic) definition will need, probably, further elaboration. The
bracketed expressions ``system determining'', and ``bounded'' has to
indicate, that also other generators etc. are possibly acceptable in the
theory. Similar remarks might be, probably, added to several other parts of
the here presented (working) version of the theory, called here ``EQM''.
\hfill\dovi\end{rem}
The definition of ``G--systems'' leads to a formally (and, perhaps, also
intuitively) natural,
and also ``operationally'' transparent, definition of ``subsystems'':
\begin{defi}[\emm \bs{G_I}--subsystems~]\label{df;G-subsyst}\nl
Let a G--system be given by $U(G)$, and let $G_I\subset G$\ be a Lie subgroup
of the Lie group $G$. The restriction $U(G_I)$\ of $U(G)$\ to $G_I$\ is a
continuous unitary  representation of $G_I$. The $G_I$--system
$\Sigma_{G_I}$\ (resp.
$UG_I$--system $\Sigma_{UG_I}$) determined by this
restriction is the \emm \bs{G_I}--subsystem~ (resp.
\emm\bs{UG_I}--subsystem~) of $\Sigma_G$.
\hfill\pika\end{defi}
Let us note that the definition of states of a subsystem given in
Subsection~\ref{q-phsp;mixt} with a help of the ``partial trace'' fits into a
special case of the presently introduced definition of the $UG_I$--subsystems:
It should be chosen $G:=\mcl U(\mcl H_{I+II})$--the unitary
group of the set of all bounded operators on $\mcl H_{I+II}=\mcl
H_I\otimes\mcl H_{II}$, and as the Lie subgroup we choose
$G_I:=\mcl U(\mcl H_I)\sim\mcl U(\mcl H_I)\otimes\mbI_{\mcl H_{II}}$, with
$U(\cdot)$\ being their defining (identical) representation(s).
The linear QM can be considered here as described by the subalgebra of
$\mCG_U$\ consisting of constant functions only (what is an alternative to
the choice $G:=\{e\}$, cf.\ point~\ref{pt;QM}).

\begin{pt}\label{pt;PB}
Let us express now the Poisson bracket between the reduced function
representatives of two observables \fk f,\ \fk l in \CG. This is done by a
repeated use of the composite--mapping theorem, \cite{abr&mars,3baby}. For
the case $n:=\dim G<\infty$, from~\eqref{eq;2.27}, and~\eqref{eq;2.33}
we have:
\begin{eqnarray}\label{eq;2.38}
\{\mh{\mfk f},\mh{\mfk l}\}(\nu)&\equiv&i\,\nu\bigl([\mfk f(\mbF(\nu)),\mfk
l(\mbF(\nu))]\bigr)+i\,\sum_{j=1}^n\nu\left(\frac{\partial\mfk
f(\mbF(\nu))}{\partial
F_j}\right)\,\nu\bigl([X(\xi_j),\mfk
l(\mbF(\nu))]\bigr)\nonumber\\
&&+i\,\sum_{j=1}^n \nu\left(\frac{\partial \mfk l(\mbF(\nu))}{\partial
F_j}\right)\,\nu\bigl([\mfk f(\mbF(\nu)),X(\xi_j)]\bigr)\nonumber\\
&&+
\sum_{j,k}\nu\left(\frac{\partial\mfk f(\mbF(\nu))}{\partial
F_j}\right)\,\nu\left(\frac{\partial \mfk l(\mbF(\nu))}{\partial
F_k}\right)\,\mbF^*\{F_j,F_k\}(\nu).
\end{eqnarray}
One can immediately deduce from this expression also expressions for Poisson
brackets of specific cases of elementary quantum and G--classical
observables.\hfill\dovi
\end{pt}
We shall formulate now the solution of a quantummechanical dynamical
equation of a G--system in terms of a classical equation on the group
manifold $G$. The solution will also show us that G--systems are
``self-consistent'' in the sense that the G--(classical) generators generate
flows leaving the sets of G--observables, G--generators, as well as the
G--classical states together with their algebraic and topological structures
invariant.

Let us assume $\dim G<\infty$.
Let $\gamma:G\rarw G$\ be a differentiable mapping, let $e\in G$\ be the unit
element, $\mfk g\equiv T_eG$. The tangent mapping $T_e\gamma:T_eG\rarw
T_{\gamma(e)}G$ is defined by
\[T_e\gamma(\xi):=\left.\frac{d}{dt}\right|_{t=0}\gamma(\exp(t\xi))
\equiv T_{t=0}\gamma(\exp(\cdot\xi)),\ \xi\in\mfk g.\]
Let $R_g:g'\mapsto R_gg':=g'g\ (g,g'\in G)$ be the right action of $G$ onto
itself. Let us identify the tangent space $T_F\mfk g^*$ in any point
$F\in\mfk g^*$ with $\mfk g^*$\ itself in the canonical way (as any tangent
space to a
linear space), and let its dual $T^*_F\mfk g^*$\ be identified with $\mfk
g^{**}=\mfk g$\ (canonical identification for reflexive spaces). Then, for
any $Q\in C^\infty(\mEF,\mbR)$, and any $F\in\mEF$, we have $\md
Q,F~\in\mfk g$. The set \EF\ is $Ad^*(G)$--invariant, cf.
Proposition~\ref{prop;2.24}.

\begin{prop}\label{prop;2.26}
Let $U(G)$ be as above, and $Q\in C^\infty(\mEF,\mbR)$, with complete
Poisson flow \ph{},Q~\ on \EF. Let $\rQ:=\mbF^*Q\in\mGGc$, i.e.\ \rQ\ is a
\Gcl\ generator, cf.\dref2.19~(iii).  Then there is a unique infinitely
differentiable solution $g_Q:\mbR\times\mEF\rarw G,\ (t;F)\mapsto g_Q(t,F)$
of the differential equation on the group manifold:
\begin{subequations}
\label{eq;2.39}
\begin{equation}\label{eq;2.39a}
\frac{d}{dt}g_Q(t,F)=T_eR_{g_Q(t,F)}(d_{F_t}Q)\quad\in T_{g_Q(t,F)}G,
\quad g_Q(0,F)\equiv e,
\end{equation}
with $F_t:=\mph t,Q~F$, for all $F\in\mEF$. The function $g_Q$ satisfies the
cocycle identity:
\begin{equation}
g_Q(s,\mph t,Q~F)g_Q(t,F)\equiv g_Q(s+t,F),
\end{equation}
\end{subequations}
and it determines the flow \ph{},Q~ according to the following relation:
\begin{equation}\label{eq;2.40}
\mph t,Q~F\equiv Ad^*(g_Q(t,F))F.
\end{equation}
The flow \pph{},\rQ~ generated by the Hamiltonian vector field
\vv\rQ,{\cdot}~\
from~\eqref{eq;2.34} is then given on \bcDF\  by
\begin{equation}\label{eq;2.41}
\mpph t,\rQ~\mrh\equiv Ad^*\bigl(U[g_Q(t,\mbF(\mrh))]\bigr)\mrh,\
\mrh\in\mbcDF,
\end{equation}
with $g\mapsto U(g)$ being the given unitary representation of $G$. Hence,
\pph{\cdot},\rQ~\ leaves all the orbits \OGr\ invariant.
\hfill\zal
\end{prop}
\begin{proof}
The flow \ph{},Q~\ leaves the $Ad^*(G)$--orbits invariant, since it is a
Poisson flow and the $Ad^*(G)$--orbits are symplectic leaves of the Poisson
(--Berezin) structure on $\mfk g^*$,~\cite{marle,arn1,weinst}. Hence
$F_t\in\mEF\ (F\in\mEF)$ for all $t\in\mbR$, and $d_{F_t}Q\in T^*_{F_t}\mfk
g^*\equiv\mfk
g:=Lie(G)$. The vectors $T_eR_g(d_FQ)\in T_gG\ (g\in G)$ form a
right--invariant vector field on $G$ for each $F\in\mEF$, and $\mvv
Q,{g;t;F}~:=T_eR_g(d_{F_t}Q)\ (t\in\mbR,g\in G)$\ are values of
$t$--dependent vector fields (for any $F\in\mEF$) on G. Their infinite
differentiability follows from the properties of $Q$. The existence and
uniqueness of the solution $g_Q$ of~\eqref{eq;2.39}
fulfilling~\eqref{eq;2.39} are then consequences of the theory of ordinary
differential equations on manifolds, cf.\ \cite{bourb;manif}.

Let $\xi\in\mfk g$. The derivative of $[Ad^*(g_Q(t,F))F](\xi)\equiv
F(Ad(g_Q(t,F)^{-1})\xi)$ at $t=0$ equals, according to~\eqref{eq;2.39}, to
$F([\xi,d_FQ])$, what can be rewritten in the form of Berezin bracket for
$\xi:=d_Fh,\ h\in C^\infty(\mfk g^*,\mbR)$:
\[ \left.\frac{d}{dt}\right|_{t=0}h\bigl(Ad^*(g_Q(t,F))F\bigr)=
d_Fh\left(\left.\frac{d}{dt}\right|_{t=0}Ad^*(g_Q(t,F))F\right)=
\{Q,h\}(F).\]
This, together with~\eqref{eq;2.39}, proves~\eqref{eq;2.40}.

The generator $\rQ\in\mGGc$ generates, on the other hand, a Poisson flow
\pph{},Q~\ on \Ss. Since $\rQ=\mbF^*Q$,~\eqref{eq;2.41} is proved
by~\eqref{eq;2.33},~\eqref{eq;2.39}, and~\eqref{eq;2.40}.
\end{proof}
\begin{intpn}\label{int;2.27}
\item{(i)} Let us assume that a standard measuring procedure can be
associated with a given mathematical quantity $\mfk f\in\mCG$\ (or with a
quantity that can be described by an unbounded selfadjoint operator--valued
function $F\mapsto Y(F)$) which leads
to a numerical result \lam\ at each individual repetition of the measuring
performed on the system--object. We understand here that with each such
individual measuring act there is necessarily accompanied a \emm registration
$\equiv$ detection~ of
a copy of considered system--object. This means that, contrary to often
accepted definition of ``measurement process'' in QM, performing a statistical
empirical test measuring the
 (average/per time) number of
incoming systems in a beam (leavig a preparation apparatus), as well as of
the (average/per time) number of systems approaching (entering)
the apparatus, a knowledge of efficiency parameters
 of the apparatus, and also exact knowledge of (calculated) final state
of measured objects
``entered into the apparatus'' (i.e.\ the state just before being detected
by a ``counter''), {\bf all of this together is not
sufficient for presence of a measuring act}. Or, in other words, the result
\lam\  of each individual measuring act should be represented by a
(macroscopic) change of initial state of measuring device which is observable
as a stable mark (i.e.\ a ``trace'' repeatably testable
 by different, namely by any ``correspondingly educated'', human observers
 with the same result of the tests with, possibly,
standard statistical deviations), e.g.\ a ``new pointer position
\lam\ of the measuring apparatus''.\footnote{According to this
understanding of the content of the
``process of measurement in QM'', the measurement of a spin--coordinate of
a $1/2$--spin particle by a Stern--Gerlach apparatus is not realized after
 passage of the particle across the inhomogeneous magnetic field, in spite
 of the fact that the wave function of the state of such a particle is
 splitted into two  ``macroscopically separated'' beams: QM does not
 exclude a possibility of rejoining and interference of the two beams,
 hence they are not yet ``macroscopically distinguished''. The
 spin--component is measured only after detection of the particle described by
 the two--beam state vector, i.e.\ only after the ``in which
 beam--question'' is practically resolved (by an appearance of a ``macroscopic
 trace'' corresponding to just one of the eventualities).}
(It might be useful to stress also here that such a measurement process is not
yet satisfactorily formalized in QT.)

We assume that this \lam\ belongs to the union of the spectra sp[\fk f(F)]
of selfadjoint operators \fk f(F)\ (resp.\ spectra of generally unbounded
$Y(F)\mapsto\mfk f(F)$):
\[ \mlam\in\cup\{{\rm sp}[\mfk f(F)]:F\in\mEF\}\subset\mbR. \]
\item{(ii)}
We propose the following interpretation of the introduced observables $\mfk
f\in\mCG$, or, more generally, of any (``sufficiently measurable'', so that
the integrals in\rref2.42c~ can be defined, cf.
Definition~\ref{df;2.25b}(iii)) selfadjoint operator--valued function
$Y:F(\in\mEF)\mapsto Y(F)=\int_\mbR\mlam E_{Y(F)}(d\mlam)$, cf.
also~\cite{bon8,bon3,bon1,unner1,unner2,unner3}:

Let $\mu\in\mcl M^G$\ be a genuine mixture, and let $\hat\mrh$\ be a
quantum deviation function, both together defining the corresponding state
\ommr, cf.\ Definition~\ref{df;2.25d}.
Let $B\subset\mbR$ be a Borel set. The probability of realization of the
detected values $\mlam\in B$ at repeated measurements of the observable
$Y:F\mapsto Y(F)$ in the (repeatably ``identically'' prepared) state
\ommr\ is expressed by:
\begin{equation}\label{eq;2.42c}
\prob(Y;\mu;\hat\mrh)(B)\equiv\int_{\mbcDF}\hat\mrh(\nu)
\bigl(E_{Y(F(\nu))}(B)\bigr)
\mu(d\nu),\ {\rm with}\ \hat\mrh(\nu)(E):=Tr(\hat\mrh(\nu)E),
\end{equation}
if the integral exists.\hfill\bpika
\end{intpn}
Let us illustrate this general interpretation scheme by more specific
examples:
\begin{exmp}\label{exmp;2.42}
\item{(i)} Let $\mu\in\mcl M^G$ be a genuine mixture describing the state
$\mom\mu,{}~\in\mS_G^{cl}$ of a system, let $B\subset\mbR$ be a Borel set,
and let $E_{Y(F)}$ be a projection (spectral) measure of the selfadjoint
operator $Y(F),\ F\in\mEF$. Then probability of finding in $B$ the obtained
value (i.e.\ the result) of a measurement of the observable: $\nu\mapsto
Y(\mbF(\nu))$ in the state \om\mu,{}~\ is
\begin{subequations}
\label{eq;2.42a}
\begin{equation}
\prob(Y\in B;\mu)\equiv\prob(Y;\mu)(B)=\int_{\mbcDF}
\nu\bigl(E_{Y(\mbF(\nu))}(B)\bigr)\,\mu(d\nu).
\end{equation}
For the specific choice of the measure $\mu:=\delta_\nu$, we have then
\begin{equation}
{\rm prob}(Y;\delta_\nu)(B)=\nu\bigl(E_{Y(\mbF(\nu))}(B)\bigr)\equiv
Tr\bigl(\nu\dti E_{Y(\mbF(\nu))}(B)\bigr),
\end{equation}
what is the usual probability distribution of the measuring results in QM of the
observable described by the operator $Y(\mbF(\nu))$ performed on the system
prepared in the (elementary) state $\nu\in\mbcDF$. The expectation (if it
exists) of an observable $\mfk f\in\mCG$\ in any state \om\mu,{}~\ is
expressed by~\eqref{eq;2.37}.
\item{(ii)} Let us choose in the above formulas $Y(F):=\mfk h_\xi(F):=
f_\xi(F)\mbI\equiv F(\xi)\mbI$\ ($\xi\in Lie(G)$); then
\begin{equation}
 E_{\mfk h_\xi(F)}(B)=\delta_{F(\xi)}(B)\mbI=\chi_B(F(\xi))\mbI,
 \end{equation}
where $\chi_B$\ is the characteristic function (= indicator) of the set B.
Hence $\mfk h_\xi\in\mCGc$\ is a classical observable. Let us denote \glss
\bs{\mbF_\xi(\nu):=\mbF(\nu)(\xi)}$\in \mbR,\ \xi\in\mfk g,\ \nu\in\mbcDF$~.
In the considered case we have
 \begin{equation}
 \prob(\mfk h_\xi,\mu)(B)=\int_{\mbcDF}\chi_B(\mbF_\xi(\nu))\,\mu(d\nu)=
\mu\bigl(\mbF_\xi^{-1}[B]\bigr)=:\glss \mu_\xi(B)~,
\end{equation}
where the measure $\mu_\xi\equiv\mu\circ\mbF_\xi^{-1}$\ on the real line
\bR\ was introduced.
\end{subequations}
\hfill\dovi
\end{exmp}

We shall now define transformation laws $\tau^\rQ$\ for observables,
\[\mfk f\mapsto\mfk f_t:=\tau^\rQ_t(\mfk f),\ \mfk f\in\mCG,\]
corresponding to the actions of the flows \pph{},\rQ~\ on \Ss\ described in
the Proposition~\ref{prop;2.26}. We shall assume that
\begin{equation}
\mbs{\mh{\mfk f}(\mpph t,\rQ~(\nu))\equiv\mh{\mfk f_t}(\nu)},
\end{equation}
what corresponds to
the transition from the Schr\"odinger to the Heisenberg picture in QM. This
assumption is reflected in the following definitions.
\begin{subequations}\label{eq;2.43}
\begin{defs}[\emm G--transformations~]\label{df;2.28}
\item{(i)} Let us consider a G--system. Let us choose some $\rQ\in\mGGc,\
\rQ=\mbF^*Q$, with complete flow \pph{},\rQ~\ on \Ss. Then \ph{},\rQ~\
determined by $\mph t,\rQ~\mbF(\nu)\equiv\mbF(\mpph t,\rQ~\nu)$\ is the flow
with Hamiltonian $Q$\ on \EF. Let $\ru_\rQ$\ be the solution
of~\eqref{eq;2.11} (with $f$\ replaced by \rQ), cf.\ also
Definition~\ref{df;2.25a}(v). Then, for an arbitrary G--observable $\mfk
f\in\mCG$, we set:
\begin{equation}
\mfk f_t(F):=\tau_t^\rQ(\mfk f)(F):=\mun\rQ,t,F~^{-1}\mfk f(\mph
t,\rQ~F)\mun\rQ,t,F~.
\end{equation}
In terms of~\eqref{eq;2.40} and~\eqref{eq;2.41}, we can write also
$\mun\rQ,t,F~=U(g_\rQ(t,F))$, hence:
\begin{equation}
\glss \tau^\rQ_t~(\mfk f)(F)\equiv U(\mgQ t,F~^{-1})\,\mfk
f(\mph t,\rQ~F)\,U(\mgQ
t,F~)\equiv Ad\bigl(U(\mgQ t,F~^{-1})\bigr)\,\mfk f(\mph t,\rQ~F).
\end{equation}
We shall call \bs{\tau^\rQ}\ the one--parameter \emm G--symmetry group
generated by \rQ~.
\item{(ii)} Let a Lie group continuous unitary representation $U(G)$ be
given. Elements of the Lie algebra \fk g\ of $G$\ are represented by affine
functions $\mh{X(\xi)}\in\mGGc,\ \xi\in\mfk g$, which are generators of
one--parameter groups of symplectic isometries of our elementary phase
space \Ss. Let a subgroup $\msg(G)\subset$\,\mbox{\aut\CG}\ of \autm s of
the \Ca\ of observables \CG\ be determined by:
\begin{equation}
[\msg(g)\mfk f](F):=U(g)\,\mfk f(Ad^*(g^{-1})F)\,U(g^{-1}),\ \forall\quad\mfk
f\in\mCG,\ g\in G,\ F\in\mEF.
\end{equation}
The function $\mh{X(\xi)}\ (\xi\in\mfk g)$\ generates the flow \pph{},\xi~\ on
\Ss, and for $\mfk f\in\mCG$ one has:
\begin{equation}
\mh{\mfk f}(\mpph t,\xi~\nu)\equiv\nu\bigl((\msg[\exp(t\xi)]\mfk
f)(\mbF(\nu))\bigr)=\mh{\mfk f}(Ad^*\bigl(U(\exp(t\xi))\bigr)\nu).
\end{equation}
The automorphism group \glss $\msg(G)$~ is \emm induced by the unitary
representation $U(G)$~. We also
have the expression of an arbitrary one--parameter G--symmetry group
$\tau^\rQ\subset$
\mbox{\aut\CG}\ in terms of $\msg(G)$, cf.
Theorem~\ref{thm;2.29}:
\begin{equation}
(\tau_t^\rQ\mfk f)(F)\equiv[\msg(\mgQ t,F~^{-1})\mfk f](F).
\end{equation}
The group $\msg(G)$ is called the \emm G--automorphism group of \CG~.

\item{(iii)}
Let $\hat{\rf}$\ be a function--representative of an observable.
Its evolution \glss $\hat\tau_t^\rQ~:\hat{\rf}\mapsto\hat{\rf}_t$\
under the G--symmetry group $\tau^\rQ$\ is expressed with a help
of the function \gQ\cdot,\cdot~ from~\eqref{eq;2.39} as
\begin{equation}
\hat{\rf}_t(\mrh,\nu)\equiv\hat\tau^\rQ_t(\hat{\rf})(\mrh,\nu):=
\hat{\rf}(Ad^*\bigl(U(g_\rQ\bigl(t,\mbF(\nu)\bigr))\bigr)\mrh,
\mpph t,\rQ~\nu).
\end{equation}
The transformation group \bs{\hat\tau_t^\rQ}\ is the one--parameter
\emm G--symmetry
group of the function representatives generated by \bbs Q~.\hfill\pika
\end{defs}
\end{subequations}

\begin{subequations}\label{eq;tr-prob}
\begin{rem}[Transition probabilities~]\label{rem;tr-prob}
Let us stress here that in the general case
\begin{equation}
Ad^*\bigl(U(g_\rQ\bigl(t,\mbF(\nu)\bigr))\bigr)\mrh\not\equiv\mpph
t,\rQ~\mrh,\quad{\rm for}\ \mbF(\mrh)\neq\mbF(\nu).
\end{equation}
The transformation law for observables described in Definitions~\ref{df;2.28}
leads to a natural nonlinear generalization of
the usual (``linear'') transformation of ``transition probabilities''.
\item{{\bf(linear case):}}
In the linear case, time evolution is described in QM by a
strongly--continuous one--parameter group $U(t)$ of unitary
transformations, i.e.\ $U(t)\equiv \exp(-itX)$ for a selfadjoint
Hamiltonian operator $X$. Expectation values of an arbitrary (``linear'')
observable $Y=Y^*$ in time evolved states $\mrh_t\equiv Ad^*(U(t))\mrh$ are
 \begin{equation}
 Tr(\mrh_t Y)=Tr(U(t)\mrh U(-t) Y)=Tr(\mrh U(-t)YU(t))=:Tr(\mrh Y_t)
 \end{equation}
where the ``\emm Heisenberg picture~'' of the time evolution $t\mapsto
Y_t:=U(-t)YU(t)$\ (expressed in terms of observables, instead of the
evolution of states) was
introduced. It is now trivial to see that the expression
\begin{equation}\label{eq;trans}
Tr(\mrh_t Y_{-t}) \equiv Tr(\mrh Y)
\end{equation}
remains constant in $t\in\mbR$\ for any selfadjoint
``observable $Y$''.

If one inserts now into $Tr(\mrh Y)$ for the observable $Y^*=Y$\  a
one--dimensional projection \P y, and for the density matrix another
projection \P x, then one obtains the well known ``\emm conservation of
transition probabilities~''\footnote{This interpretation of ``\emm transition
probabilities~'', by which one of the vectors represents state preparation
(``source''), and the another corresponds to a detector, connected with
 their invariance at symmetry transformations, is also in accordance
 with~\cite[I.3.1]{haag2}.}
 \begin{equation}
 Tr(\mP x\mP y)\equiv |\lb U(t)x|U(t)y\rb|^2 = |\lb x|y\rb|^2.
 \end{equation}
This seems to be usually interpreted as a trivial consequence of equal
unitary transformation of the two vectors $x,y\in\mH$\ entering into the
scalar product. Hence it is usually interpreted as an expression of the
fact that ``{\bf the \emm transition amplitude~ between two state vectors} $x,
y\in\mH$'' does not depend on time, if both states are evolved by the same
time transformation $U(t)$.

 This (mis--)interpretation is repeatedly presented in
connections with definitions of ``symmetries'' in QM,~\cite{wigner1},
and with the
celebrated Wigner's theorem, which can be formulated in the following way:
\item{{\bf(Wigner's theorem):}}
Let $\phi:\mPH\rarw\mPH$ be a bijection conserving ``transition
probabilities'', i.e.
\begin{equation}\label{eq;WT}
 Tr(\mP x\mP y)\equiv Tr(\phi(\mP x)\phi(\mP y)),\ \ \forall x,y\in\mH,
 \end{equation}
then there is either unitary or antiunitary bijection $U_\phi:\mH\rarw\mH$\
such that $\phi(\mP x)\equiv \mP{U_\phi x},\ \forall x\in\mH$.

Symmetries in QM are then defined as transformations $\phi$, resp.
$U_\phi$, satisfying conditions of the Wigner's theorem.

After reformulating the two mentioned interpretations of the
``transformations of probability amplitudes'' in the nonlinear case, we shall
return to the problem of a choice between these two interpretations
in Interpretation~\ref{int;tr-prob}.

\item{{\bf(nonlinear case 1):}}
Extending the above last mentioned (mis--)interpretation mechanically
to nonlinear
case, one obtains \emm ``non-conservation of transition
probabilities''~:\footnote{We write here $U^*(g)\equiv U(g)^*$.}
 \begin{eqnarray}
& Tr(\mpph t,\rQ~(\mP x)\mpph t,\rQ~(\mP y))= \\
&  Tr\bigl(U(\mgQ t,\mbF(\mP
x)~)\mP x U^*(\mgQ t,\mbF(\mP x)~)U(\mgQ t,\mbF(\mP y)~)\mP y U^*(\mgQ
t,\mbF(\mP y)~)\bigr)\nonumber,
\end{eqnarray}
what {\em need not be (and often isn't) constant in time $t\in\mbR$} since for given \P x~,\P
y~ the unitary-valued function $W^Q_{x,y}(t)$ need not belong to the commutant of $\mP x$ or $\mP y$:
\[W^Q_{x,y}(t):= U^*(\mgQ t,\mbF(\mP x)~)U(\mgQ t,\mbF(\mP
y)~)\not\in \{\mP x\}'\cup\{\mP y\}',\]
and is (`generically') time dependent together with the functions $t\mapsto W^{Q*}_{x,y}(t)\mP x W^Q_{x,y}(t).$
 Hence, if we calculate the ``transition probabilities'' according to the
 algorithm taken from the linear QM in the case of nonlinear evolutions, we
 obtain ``generically'' their dependence on the parameter of transformations
 (on the time). This seems to be in contradiction with the usual meaning of
 ``transformation groups'' in quantum theory.

\item{{\bf(nonlinear case 2):}}
 Let us now, however, accept the first mentioned interpretation of the
 ``\emm transition probabilities~'', i.e.\ that $|\lb x|y\rb|^2$\  is the expectation
 value of the ``observable \P y'' in the ``state \P x'' (or vice versa),
 according to equation~\eqref{eq;trans}.
 ``Observables'' in our generalized (nonlinear) quantum mechanics are
 represented by operator valued functions of ``elementary states''
 $\mrh\in\mSs$, possibly via the momentum mapping \bF, or by the corresponding
 function representatives. The transformation groups act on them in
 accordance with the equations~\eqref{eq;2.43}, hence the transformations
 depend (generally) on points \bF(\rh)\ of $Lie(G)^*$, hence on the states
 \rh. Expectation of an observable $\mrh\mapsto\mfk f(\mbF(\mrh))$\ in the
 elementary state \rh\ equals $Tr(\mrh\mfk f\bigl(\mbF(\mrh)\bigr))$,
 cf.\eqref{eq;2.37}. If
 we transform \rh\ as \pph t,\rQ~(\rh), and the observable \fk f\ is
 transformed simultaneously by
 the automorphism group transformation $\tau^\rQ_{-t}\mfk f$,~\rref2.43~,
  and we calculate
 then the expectation of the transformed observable in the transformed
 state, we obtain in accordance with\rref WT~ and\rref trans~
\begin{eqnarray}
&Tr\bigl(\mpph
t,\rQ~\mrh\!\cdot\!(\tau_{-t}^\rQ\mfk f)(\mbF(\mpph t,\rQ~\mrh))\bigr)\\
&= Tr\bigl(U(\mgQ t,\mbF(\mrh)~)\mrh U^*(\mgQ t,\mbF(\mrh)~)U(\mgQ
t,\mbF(\mrh)~)\mfk f(\mbF(\mrh))U^*(\mgQ t,\mbF(\mrh)~\bigr) \nonumber \\
&
=Tr\bigl(\mrh \mfk f(\mbF(\mrh))\bigr),\nonumber
\end{eqnarray}
 i.e.\ the result independent of $t\in\mbR$, as it is usually
 required. If the observable is, e.g.\ $\mfk f(\mbF(\mrh))\equiv\mP y$\ ,
 i.e.\ it is
 independent of \rh, then again it should be transformed by the same way, so
 that the transformed observable becomes, in general case, a function of
 \rh. Hence, for $\mrh:=P_x$, one has
 \begin{equation}
Tr\bigl(U(\mgQ t,\mbF(\mP x)~)\mP x U^*(\mgQ t,\mbF(\mP x)~)U(\mgQ
t,\mbF(\mP x)~)\mP y U^*(\mgQ t,\mbF(\mP x)~\bigr)\equiv Tr(\mP x\mP y),
 \end{equation}
 and the time invariance of transition probabilities is, trivially,
  again obtained.
 \hfill\dovi
\end{rem}
\end{subequations}

We shall now return to the interpretation question of the ``transition
probability'' $Tr(\mP x\mP y)$.\nl

\begin{intpn}[{\bf Probabilities and measurements}]\label{int;tr-prob}
If we use the concept ``probability'' in connection with our empirical
experience, it is always (perhaps) connected with a quantification of
``observed phenomena'', or of ``occurred events''. A meaning of sentences
like: ``The probability of the chosen {\em value of possible
eventuality} is $\malp>0$'' appears to us (in {\em empirical} sciences)
unspecified without the
``eventuality'' being in some sense ``realizable''. After an
experience with QM, we know that ``an event'' is always correlated with a
change of some {\em macroscopically observable} (hence classical, in a
general sense) parameter value. We conclude from this that probabilities
ascribed
to states in QM should be connected with the quantummechanical ``process of
measurement'': They express some ``weights'' connected with (macroscopic)
results of measurement; these weights are usually interpreted as
``frequencies of occurrence'' of specific results at repeated preparations
of ``the same microscopic state'' and consecutive measurements of ``the
same physical observable'' (let's note that this time--ordering corresponds
to our, perhaps a priori, demand of causality).

 All empirically interpretable (and verifiable) assertions of QM are
formulated in terms of ``probabilities'', expressed usually by squares of
moduli of ``\emm probability amplitudes~''. These probabilities are often called,
cf. e.g.\ \cite{peres}, the {\bf ``\emm transition probabilities~''}.
Let us ask now, what
``transitions'', or/and transitions between what things are meant in this
formulation? The mentioned probabilities are of the form $Tr(\mP
x\mP y)\equiv |\lb x|y\rb|^2$ for normalized vectors $|\cdot\rb\in\mH$\
corresponding to pure states of the considered microscopic system. The
standard interpretation scheme of QM (cf.\ \cite{dirac,peres}) tells us that
if a system is prepared in the state $|x\rb$ and the measured observable
$Y$ has nondegenerate pure point spectrum (i.e.\ a complete orthonormal set of
eigenvectors
$|y_j\rb,\ j\in J\equiv$\ an index set, $Y|y_j\rb=\mlam_j|y_j\rb,\
\mlam_j\in\mbR, \forall j\in J,\ \mlam_j\neq\mlam_k$\ for $j\neq k$), hence
 if it
is possible to write
\[ |x\rb=\sum_{j\in J} \lb y_j|x\rb |y_j\rb,\quad\forall x\in\mH, \]
then only possible results of the measurement of the quantity $Y$\ are the
numbers $\mlam_j,\ j\in J$, and the probability of obtaining the result
$\mlam_j$ in a vector state $|x\rb$ at measuring of $Y$\ equals to
\[ \prob(Y=\mlam_j;x)=|\lb y_j|x\rb|^2. \]

This interpretation is the generally accepted one (according to the present
author's knowledge). The denotation of this probability as ``\emn transition
probability~'' can be understood in connection with the Dirac--von Neumann
``projection (resp.\ reduction) postulate'',~\cite{dirac,neum1}\index{reduction ($\equiv$ projection) postulate},
stating that
after obtaining the result $\mlam_j$ the measured microsystem changes
abruptly its initial state $|x\rb$\ into the eigenstate of the measured
quantity $Y$\ corresponding to the obtained result $\mlam_j$. Hence,
there is assumed a ``transition $x\mapsto y_j$'' of the
microsystem.\footnote{This postulate, however, needn't be accepted: It
cannot be usually (or even always?) verified if the measured system is really
detected. As an exception might be considered the ``indirect'' measurement,
 when a correlated system is detected, what is the case of EPR--like processes.
We prefer not to formulate any assumptions on the form of states of
measured systems arising after measurements of a general type.
}

A remarkable (in the presented formulation mathematically trivial) fact is
the symmetry of $\prob(Y=\mlam_j;x)$\ \wrt interchange of the vectors $x$\ and
$y_j$. This formal mathematical symmetry (although not being without some
deep physical content) might (mis--)lead us to consider occurrence of the
vectors $x$\ and $y_j$ in the ``transition probability'' also as physically
symmetric. We have to keep in mind, however, that the eigenvectors $y_j$\
are here in the r\^ole of labels of macroscopic ``pointer positions'',
whereas the vector $x$\ represents a preparation procedure for the
microsystem. This can be expressed with a help of the spectral measure
$E_Y$\ of $Y$:
\[ \prob(Y\in B;x)=Tr(P_xE_Y(B)),\quad B\in\mcl B(\mbR),\]
where we have $E_Y(\{\mlam_j\})\equiv P_{y_j}$, in the considered specific case.
This physical asymmetry remains valid irrespective of
(non-)acceptance of the ``\emn projection postulate~'' of Dirac and von Neumann.

To conclude, we hope that it is seen from the above considerations that in the
(mathematically symmetric) expression $Tr(\mP x\mP y)$ for probability of a
certain measurable (i.e.\ observable) phenomenon described in QM, the
interpretation of the two vectors $x,\ y$\ should be mutually different: One
of the vectors represents a given
(prepared) state of the micro-object, and the second represents a measured
observable.  This leads also to formulation of the symmetry transformation
rule for these
expressions generalized to our nonlinear EQM. Those symmetry
transformations leave the ``transition probabilities'' invariant also for
nonlinear generators. An a priori requirement for such an invariance
is, however, of little
determinative power, from the point of view of our presently defended
interpretation, cf.\ also~\cite{bon-tr}. \hfill\bpika
\end{intpn}

\begin{thm}\label{thm;2.29}
Any G--symmetry group $\tau^\rQ$\ of a G--system (resp.\ UG--system) is a
$\msg(\mCG,\mS_G^{cl})$--con\-ti\-nu\-ous one--parameter group\
of $^*$-automorphisms $\tau^\rQ\subset\ \maut\mCG$\ (resp.\ $\subset\
\maut\mCG_U$).
 The relation
\begin{equation}\label{eq;2.44}
\mh{\mfk f}(\mpph t,\rQ~\nu)=\nu(\tau_t^\rQ\mfk f(\mbF(\nu))),\ \forall\mfk
f\in\mCG,\ \forall\nu\in\mbcDF,\ \forall t\in\mbR.
\end{equation}
is satisfied for this group of automorphisms of the \Ca\
of G--observables \CG.\hfill\zal
\end{thm}
\begin{proof}
The algebraic properties of \taQ, and also the \taQ--invariance of
$\mCG_U$\ are consequences of~\eqref{eq;2.43}, and of the cocycle
identities~\eqref{eq;2.39},~\eqref{eq;2.12}.
Relation~\eqref{eq;2.44} is a consequence of~\eqref{eq;2.40},
\eqref{eq;2.43}, \eqref{eq;2.28}, and of the relation
~\eqref{eq;2.41}. The $\msg(\mCG,\mS_G^{cl})$--continuity, i.e.\
that for all $\mu\in\mcl M^G$, $\mfk f\in\mCG$ the functions
$t\mapsto\mom\mu,{}~(\mtQ t,{\mfk f}~)$ are continuous, and
$\mom\mu,{}~\circ\mtQ t,{}~\in\mS_G^{cl}\ (\forall t\in\mbR)$,
follows from~\eqref{eq;2.43},~\eqref{eq;2.37}, the continuity
properties of \fk f, $g_\rQ$, and $U$, as well as from the
Lebesgue dominated convergence theorem.
\end{proof}
\begin{rem}\label{rem;2.30}
The flow \pph{},\rQ~\ is determined by the automorphism group \taQ\
uniquely. This association needn't be, however, injective: Different
automorphism groups of \CG\ can, for a general $U(G)$, lead to the same flow
\pph{},\rQ~\ on the elementary state space \Ss. This possible ambiguity can
be seen from~\eqref{eq;2.14}, where different
operator--valued functions $\nu\mapsto\mbf f^0(\nu)$\ with values in the
commutant $\{\nu\}'$ can be chosen, cf.\ also~\cite[eqs. (2.29),
(2.30)]{bon1}. The
whole state space--transformation groups of $\mS(\mCG)$\ defined as the dual
mappings to the one--parameter groups \taQ\ are, of course, different for
the different \taQ. We could try, e.g., to transform by them general
 states from $\mS_G$.\hfill\dovi
\end{rem}
\begin{intpn}\label{intpn;macro-ex}
The theorem~\ref{thm;2.29} shows, that our nonlinear dynamics can be
described with a help of a \autm\ group of our algebra of observables \CG,
resp.\ of $\mCG_U$,
which is a \Ca, hence it corresponds to standard {\em linear}
descriptions of quantum systems, cf.\ \cite{haag&kast,bra&rob,emch1,haag2}.
Since our \Ca\ \CG\ is essentially (a weak completion of) the
tensor--product algebra $\mLH\otimes C(\mEF,\mbC)$\ (let us ignore here
some topological aspects of definitions), it corresponds intuitively
to a quantummechanical system composed of the ``traditional'' one, described
by observables in \LH, and of a ``classical subsystem'' with the
``generalized phase--space'' \EF. Hence, our nonlinear quantum dynamics can
be considered as a specific restricted description of dynamics (in
Schr\"odinger picture) of a
general quantum (``linear'') system obtained by expressing just the
evolution of
``microscopic elementary states (resp.\ mixtures) $\in\mSs$''
(as states on the algebra of ``microscopic
observables'' in \LH) only, and leaving the evolution of other degrees of
freedom of the composed system explicitly unnoticed.
For some further comments of this point cf.\ Section~\ref{sec;IIIB}.
\hfill\bpika
\end{intpn}

\begin{rem}\label{rem;nlin-sym}
We shall be interested now in the possibility to represent the Lie algebra
elements $\xi\in\mfk g$ by some nonlinear generators ${\rm h}_\xi\in\mGGc$,
and, correspondingly, to represent the group G by continuous ``nonunitary''
Poisson automorphisms of \Ss. We shall formulate here one of such
possibilities obtained ``trivially'' by a ``nonlinear'' Poisson morphism
from the linear representation $U(G)$. This possibility was in~\cite{weinb}
classified as ``equivalent'' to
the linear representation. This equivalence is, of course present from the
abstract mathematical point of view of theory of Poisson systems. But the
quantummechanical interpretation depends on the metric structure
$\Gamma_\nu$ on \Ss, which is not invariant \wrt such Poisson morphisms.
Hence, the {\bf physics obtained by such a ``\emm trivial delinearization~'' of}
\bs{U(G)}, as well as of other ``G--structures'' based on $U(G)$
 {\bf might be
quite different from physics coming by traditional way from the linear
representation \bs{U(G)}}.\hfill\dovi
\end{rem}

The following proposition describes an example of mechanism of  the mentioned
``delinearization'' (cf.\ Remark~\ref{rem;nlin-sym}) of the
{\em G--structures} based on $U(G)$.

\begin{prop}[Nonlinear G--realizations]\label{prop;2.31}\index{nonlinear G--realizations}
Let the G--system based on a unitary continuous representation U(G) be
given. Let $\psi$ be a Poisson automorphism of \EF\ (specified, e.g.\ with a
help of an open \nbhd of \EF) leaving each symplectic leaf invariant:
\begin{equation}\label{eq;2.45}
\psi^*\{f,h\}=\{\psi^*f,\psi^*h\}\ \text{for}\ f,h\in C^\infty(\mEF,\mbR).
\end{equation}
Let ${\rm h}_\xi:=\glss \mfpsx~ :=\mbF^*\circ\psi^*f_\xi,\ \xi\in\mfk g$.
Then $\mfpsx\in\mGGc$, and
\begin{equation}\label{eq;2.46}
\{\mfpsx,\mfpse\}(\nu)=-\mfpsxe(\nu)\ \text{for}\ \nu\in\mbcDF,\
\xi,\eta\in\mfk g,
\end{equation}
and the association $\mh{X(\xi)}\mapsto\mfpsx\ (\xi\in\mfk g)$ is a Poisson
Lie algebra isomorphism.

Let $\Phi^\psi(g):=\psi^{-1}\circ Ad^*(g)\circ\psi:\mEF\rarw\mEF$; the
mappings $\Phi^\psi(g)$ form a group of Poisson automorphisms of \EF\ such,
that its one--parameter subgroups
\[ \Phi^\psi_\xi: t\mapsto\Phi^\psi_\xi(t):=\Phi^\psi(\exp(t\xi)) \]
are the flows generated by $f^\psi_\xi:=\psi^*f_\xi\ (\xi\in\mfk g)$. Then
\fpsx\ are generators of their ``lifts'' $\tilde\Phi^\psi_\xi$\ to the
Poisson automorphism groups of \Ss\ determined by the G--symmetry groups
$\tau^{\xi,\psi}:=\mtaQ$\ with $\rQ:=\mfpsx$\ according to the
equations~\eqref{eq;2.43}, hence also
\[ \mh{\mfk
k}(\tilde\Phi^\psi_\xi(t)\nu)\equiv\nu\bigl((\tau^{\xi,\psi}_t(\mfk
k))(\mbF(\nu))\bigr),\quad\forall \mfk k\in\mCG.\ \]\hfill$\clubsuit$
\end{prop}
\begin{proof}
Recall that (cf.\ Definition~\ref{df;2.17})
\[ \mh{X(\xi)}(\nu)\equiv f_\xi(\mbF(\nu))=\mbF^*f_\xi(\nu),\ \nu\in\mbcDF,
\]
and the pull--back has trivial kernel in $C(\mEF,\mbR)$. Since Ran(\bF)
consisting of $Ad^*(G)$--orbits is dense in \EF, a continuous function
$f_\eta$
identically vanishes on each orbit lying in \EF, hence vanishes on \EF, iff
there vanishes $\psi^*f_\eta$. It follows that the association
$\mh{X(\xi)}\mapsto\mfpsx\ (\forall\xi\in\mfk g)$\ is a bijection. It is
linear in $\xi$, and the formulas~\eqref{eq;2.33},\ \eqref{eq;2.45}, and\
\eqref{eq;2.30} show the conservation of the Poisson brackets, hence the
validity of ~\eqref{eq;2.46}.

It remains to prove, that the ``deformed'' flows $\Phi^\psi_\xi$ are
generated by $f^\psi_\xi$. Let $h\in C^\infty(\mEF,\mbR)$. Then
\begin{multline}\label{eq;poiss-morph}
\left.\frac{d}{dt}\right|_{t=0}h\bigl(\Phi^\psi_\xi(t)\mbF\bigr) =
\md{\left((\psi^*)^{-1}h\right)},{\psi F}~\circ{\rm ad}^*_\xi(\psi F)
=(\psi F)([\md{((\psi^*)^{-1}h)},{\psi F}~,\xi]) \\
=\{f_\xi,(\psi^*)^{-1}h\}(\psi F) = \psi^*\{f_\xi,(\psi^*)^{-1}h\}(F)=
\{f^\psi_\xi,h\}(F),
\end{multline}
where we define $(-{\rm ad}^*_\xi):=(ad_\xi)^*$, the dual mapping of
the inner differentiation of the
Lie algebra, $ad_\xi:\eta\mapsto[\xi,\eta]$. This proves the
proposition.
\end{proof}
\begin{exmp}
As a large class of examples of mappings $\psi$ occurring in the
Proposition~\ref{prop;2.31}, we can choose $\psi:=\mph t,Q~$\ for any
nonlinear $Q\in C^\infty(\mEF,\mbR)$\ with complete Hamiltonian vector
field (hence flow) on \EF, with a fixed value of $t\in\mbR$. The question
of a physical
interpretation of such ``nonlinear deformations'' of the ``linear''\ $U(G)$\ is
left open here.\hfill\dovi
\end{exmp}

Let us consider now the specific case of a physical system described (in the
sense of EQM) by a \Ca\ $\mcl C:=C(\mcl E,\mfk A)$, with \cl E\ an
Hausdorff compact, and \fk A\ a simple unital \Ca, cf.\dref repres~; the
continuity of $\mfk f(\in\mcl C):F(\in\mcl E)\mapsto\mfk f(F)(\in\mfk A)$\
is here uniform in the norm of \fk A.
E.g., we can use $\dim\mH<\infty$\
and $U(G)$\ irreducible
in our previous costructions, and then we shall have $\mfk A:=\mLH$, and
$\mcl E(\subset \mfk g^*)$\ some compact convex $Ad^*$--invariant set.
In this case $\mcl C\sim\mfk A\otimes
C(\mcl E)$,~\cite{sak1}, and the structure of such systems can be
described now with some additional
details. Let us mention first,~\cite[Proposition 2.6]{bon1}:
\begin{lem}\label{lem;AxC}
The pure states $\mome\in\mS(\mcl C)$\ (i.e.\ extremal points of the
$\msg(\mcl C^*,\mcl C)$--compact \cS(\cl C)$=\mcl C^*_{+1}$) are of the form
\bequ\label{eq;p-states}
 \mome(\mfk f)=\mome_{\mfk A}\bigl(\mfk f(F_{\mome})\bigr),\ \forall\mfk
f\in\mcl C,
\end{equation}
where $\mome_{\mfk A}\in\mS(\mfk A)$\ are pure states on \fk A, and
$F_\mome\in\mcl E$\ is fixed.
\hfill\zal\end{lem}
It could be useful to compare this assertion with\dref2.25d~\ to see
what states of the \Ca\ \cl C\ are not contained in the set of states
determined by that definition.

Let us now describe
the general form of symmetry--transformations (i.e.\ the automorphisms of \cl
C) of such a system, cf.\ \cite{bon-sym}, and~\cite[Remarks 3.15]{bon2} for
more complete (but there unproved) formulations:
\begin{prop}\label{prop;AxC-symm}
Let a \Ca\ $\mcl C:=C(\mcl E,\mfk A)$\ be given as above. Then there is a
canonical bijection between $\gamma\in\maut\mcl C$, and couples
$\{\mphi_\gamma;\hat\gamma\}$, where $\mphi_\gamma$\ is an arbitrary
 homeomorphism of \cl E, and $\hat\gamma$\ is an arbitrary mapping
 $\hat\gamma:\mcl E\rarw\maut\mfk A,\
F\mapsto\hat\gamma_F$, with the functions $F\mapsto\hat\gamma_F(\rx)\
(\forall\rx\in\mfk A)$\ being all norm--continuous. The bijection is
determined by the identity
\bequ\label{eq;3autom}
 \gamma(\mfk f)(F)\equiv\hat\gamma_F\bigl(\mfk f(\mphi_\gamma F)\bigr),
\end{equation}
valid for all $\mfk f\in\mcl C$.
\hfill\zal\end{prop}
\begin{proof}
Due to the simplicity of \fk A, the abelian subalgebra $C(\mcl E)$\ of \cl C\
coincides with its center $\mcl Z:=\mcl Z(\mcl C)$. The center \cl Z\ is
invariant \wrt any  \autm\ of \cl C, hence the restriction of $\gamma$\ to
$C(\mcl E)$\ is also an automorphism. The Gel\acc fand--Najmark theory of
commutative \Ca s (cf.\ \cite{najm,GRS,gamelin}, and also
Example~\ref{ex;Ca&Wa}(iii)) implies that the $^*$--automorphisms of $C(\mcl
E)$\ are in a bijective correspondence with homeomorphisms $\mphi$\
of \cl E\ onto itself.
\item{(i)}\ Let $\gamma\in\maut\mcl C$. Then the corresponding
homeomorphism $\mphi_\gamma$\ is defined by:
\[ (\gamma f)(F)=:f(\mphi_\gamma F),\quad\forall f\in C(\mcl E)\subset\mcl
C,\ \forall F\in\mcl E.\]
Let an arbitrary $\rx\in\mfk A$\ be considered as a constant function -- an
element $\hat\rx\in\mcl C:=C(\mcl E,\mfk A),\ \hat\rx(F)\equiv\rx\equiv
\rx\dti\mbI(F)$, where $\mbI(F)=1,\ \forall F\in\mcl E$. Then the value
$\gamma(\hat\rx)(F),\ F\in\mcl E,$\ of $\gamma(\hat\rx)\in\mcl C$\ will be
denoted by
\[ \hat\gamma_F(\rx):=\gamma(\hat\rx)(F),\quad\forall F\in\mcl
E,\quad\forall \rx\in\mfk A.\]
The pointwise character of algebraic operations in $C(\mcl E,\mfk A)$\
implies that in this way defined $\hat\gamma: F\mapsto\hat\gamma_F$\
is a mapping to $^*$--morphisms of \fk A\ into itself.

We shall show that $\hat\gamma_F$\ is a nonzero morphism (hence an
isomorphism, due to simplicity of \fk A) for any $F\in\mcl E$. A general
element $\mfk f$\ of \cl C\ is uniformly approximated by elements of the form
\bequ\label{eq;1sum-j}
 \mfk f':=\sum_j\rx_j\dti f_j,\ \mfk f'(F)\equiv\sum_j\hat\rx_j(F)\dti
f_j(F),\quad \rx_j\in\mfk A,\ f_j\in C(\mcl E)\subset\mcl C,
\end{equation}
hence also by the elements of the form $\gamma(\mfk f')$, since $\gamma$\ is
a \autm\ of \cl C. For the elements of the form\rref1sum-j~\ one has
\bequ\label{eq;2sum-j}
\gamma(\mfk f')(F)\equiv\sum_j\hat\gamma_F(\rx_j)\dti f_j(\mphi_\gamma F).
\end{equation}
For a zero morphism $\hat\gamma_{F_0}$\ it would be $\gamma(\mfk f)(F_0)=0$\
for all $\mfk f\in\mcl C$, what cannot happen, since both \fk A, and $C(\mcl
E)$\ are unital. It follows that $\hat\gamma_F\in\maut\mfk A,\ \forall
F\in\mcl E$.

The formula\rref2sum-j~ implies then\rref3autom~ due to continuity of all
the $\hat\gamma_F$, as well as of $\gamma$:
\[ \gamma(\mfk f')(F)\equiv\hat\gamma_F\left(\sum_j\rx_j\dti
f_j(\mphi_\gamma F)\right).\]
The continuity of $\hat\gamma:F\mapsto\hat\gamma_F$\ follows from the
continuity of each function $F\mapsto\gamma(\mfk f)(F),\ \mfk f\in\mcl C$.
\item{(ii)}\ Let us now have given any homeomorphism $\mphi_\gamma$\ of the
Hausdorff compact \cl E\ onto itself, as well as an arbitrary strongly
continuous family $\hat\gamma:\mcl E\rarw\maut\mfk A,\
F\mapsto\hat\gamma_F$. Let us define the mappings $\mphi:\mcl C\rarw\mcl C$,
and $\gamma_0:\mcl C\rarw\mcl C$\ as follows:
\bequ\label{eq;2-morphC}
\mphi(\mfk f)(F):=\mfk f(\mphi_\gamma F),\qquad\gamma_0(\mfk
f)(F):=\hat\gamma_F\bigl(\mfk f(F)\bigr),\qquad\forall\mfk f\in\mcl C,\
F\in\mcl E.
\end{equation}
The continuity and the morphism properties of the given
$\hat\gamma_{\cdot}$,\
and $\mphi_\gamma$ show that both the mappings $\mphi$, and
$\gamma_0$\ introduced in\rref2-morphC~ are $^*$--automorphisms of \cl C.
Hence the formula\rref3autom~ determines an automorphism $\gamma:\mcl
C\rarw\mcl C$\ as the composition of these two automorphisms:
$\gamma:=\gamma_0\circ\mphi\in\maut \mcl C$.
\end{proof}

This proposition allows us to view also a degree of generality of
the before introduced
``$G$-- transformations'', cf.\dref2.28~, at least in the present
simple case: Without further symmetry requirements, the general automorphism
group contains much more continuous subgroups than we have introduced
in\dref2.28~. It might be worth mentioning that, under some continuity
requirements onto $\gamma\in\maut\mcl C$\
($\pi_{\mfk g}$--normality,~\cite{bon-sym,bon2}),
the homeomorphism $\mphi_\gamma$\ can be, in specific models (mean--field,
cf.\ also Subsection~\ref{mot;MF}, and Section~\ref{sec;IIIB}), uniquely
determined by the set of automorphisms $\{\hat\gamma_F:F\in\mcl E\}$.

\vspace*{\fill}
\newpage



\chapter{Specifications and Applications}
\label{chap;III}
\def\autor{{
\ref{chap;III}\quad Specifications and Applications}}

It will be shown in this chapter that the general scheme for dynamical
theories developed in Chapter~\ref{sec;II} applies to a wide scale of
existing physical theories. Different specifications of the formal scheme
mainly consist in choices of classes of ``generators'',
``observables'',
and ``states'', cf.\ Definitions~\ref{df;2.25a},~\ref{df;2.25b},
and~\ref{df;2.25d}.

A choice of the representation $U(G)$ belongs to important general tools for
such a specification, i.e.\ a determination of a G-system, cf.
Definition~\ref{df;G-syst}. There are, however, also other possible ways for
determination of a specification; some of them are connected with models
already published in
literature. Let us give first a review of some of these specifications.

\section{A Review of Considered Specifications}\label{III;spec}
\def\nazov{{
\ref{III;spec}\quad A Review of Considered Specifications}}

There is a whole array of general physical theories and/or their
``caricatures'',
resp.\ ``approximations'' covered by the general model of our \emm ``Extended
Quantum Mechanics'' (EQM)~ described in Chapter~\ref{sec;II}.
Let us mention here briefly some typical of them, resp.\ some of
possible applications of EQM; most of these will be described in some details
later in this chapter:
\begin{pt}[{\bf Quantum Mechanics}]\label{pt;QM}
\end{pt}
Traditional (linear) quantum mechanics (QM) is obtained as the
specification corresponding to the G--system with trivial group
$G\equiv\{e\}\equiv\{${\it one--point set consisting of the unit element}\
$e\in G\}$, cf.\ Section~\ref{sec;IIIA1}, esp.\ Subsection~\ref{IIIA1;CCR}.
Another possibility of obtaining QM from EQM is the restriction of any
G--system to the subalgebra $\mCG_q\subset\mCG$, and accepting only
such generators the flows of which leave $\mCG_q$\ invariant.\hfill\zel

\begin{pt}[{\bf Nonlinear Quantum Mechanics}]
\end{pt}
Nonlinear extensions of QM ``living'' on the projective Hilbert space \PH,
and containing in their sets \GG\ all ``relevant'' generators is called here
the nonlinear QM (NLQM). A specific choice of ``observables'' also depends on
the accepted interpretation scheme; the same concerns the set of states of a
chosen theory. Note here that, in the framework of this ``specification'',
it is possible to describe also the ``general theory'', because,
e.g.\ all density
matrices could be expressed by unit rays in the Hilbert space \fk H\ of
Hilbert--Schmidt operators, cf.\ Remark~\ref{rem;mixinPH}; such an approach
seems, however, in a certain sense ``unnatural'', because it needs some
additional restrictions. NLQM will be shortly discussed in
Subsections~\ref{IIIA1;QM}, and~\ref{IIIA1;NL-Sch}.\hfill\zel
\begin{pt}[{\bf Subsystems in Macroscopic Environment}]
\end{pt}
The ideas leading to the theory presented in this work are closely connected
 with models of
infinite quantum systems with specific dynamics ``of mean--field (MF)
type'', cf.\ \cite{hp+lie1,bon1,bon2,bon3,unner1,morch&stroc,duf&rieck}. We
shall not go into details on this point in this paper. It was shown,
however, in Theorem~\ref{thm;2.29}, that all the nonlinear evolutions
generated by
G--(classical) generators $Q\in\mGGc$\ can be described as one--parameter
groups of \autm s of a \Ca, namely the \Ca\ \CG. Let us note here
that this \Ca ic description allows us, in this formal framework, to
{\bf distinguish elementary mixtures from genuine ones}: In the case if the
domain $\mEF\subset\mfk g^*$ can be identified with the whole \Ss\ (what is
possible in the case of the choice $G:=\mfk U:=\mcl U(\mH)$), elementary
mixtures are just the pure (i.e.\ extremal) states of the abelian
\Csa\ $\mCGc\subset\mCG$. For some further comments see
Section~\ref{sec;IIIB}.\hfill\zel

\begin{pt}[{\bf Classical Mechanics}]\label{pt;CM}
\end{pt}
Classical mechanics (CM) of a system with a symmetry group $G$ is also
contained in EQM:

Let a ``kinematical'' symmetry group $G$ of a classical Hamiltonian system
be given; we shall assume for simplicity that it is a connected and simply
connected Lie group.
Let the phase space of this system be a homogeneous space of $G$,
the action of $G$ being there symplectic. This phase space can be identified
with a coadjoint orbit of the
group $G$, or of its one--dimensional central extension,~\cite{kiril}. Natural
generalizations of these systems are Poisson systems ``living'' on an
$Ad^*(G)$--invariant part of $Lie(G)^*\equiv\mfk g^*$, cf.
Section~\ref{I;clmech}, and also Appendix~\ref{A;LieG} (for literature on
classical mechanics see
also~\cite{whitt,abr&mars,arn1,thirr1,mack1,arn2,mars&rati}).
We can see from Section~\ref{IIC;symm-obs} that our scheme of EQM
restricted to the algebra \CGc\ leads to description of any such
``$G$--symmetric'' CM--system as a subsystem of our (quantal) $G$--system.\hfill\zel
\def\autor{{
\ref{chap;III}\quad Specifications and Applications}}

\begin{pt}[{\bf Hartree--Fock Theory}]\label{2HF}
\end{pt}
Specific ``quasiclassical'' and/or ``selfconsistent'' approximations for QM
described as dynamics on manifolds of generalized coherent
states,~\cite{rowe1,klaud1,perel1,koh-st}, i.e.\ the ``classical projections
of QM'',~\cite{bon8}, are contained in EQM as well; these specifications
include the systems obtained by
the ``time--dependent variational principle'',~\cite{koh-st,kra&sar}. An
important special case of these is the time--dependent Hartree--Fock
approximation; the corresponding (infinite dimensional) set of generalized
coherent states consists now from
all Slater determinants of an $N$--fermion system, and the group $G$ is the
whole unitary group of one--particle Hilbert space,
cf.\ Subsection~\ref{IIIA1;H-F}.\hfill\zel

\begin{pt}[{\bf Specific Time--Dependent QM}]
\end{pt}
A class of quantummechanical systems with time--dependent Hamiltonians can
be found as a subtheory of EQM; it
appears to be identical with the corresponding (time independent) NLQM. This
class includes,
as a special case, the nonlinear dynamics (for pure states) proposed by
Weinberg in~\cite{weinb}.\footnote{There is no need of any restriction
by $U(G)$ in finite dimensional Hilbert spaces.} The ``integrability'' of such
systems is
determined by integrability of corresponding classical Hamiltonian systems;
cf.\ Sections~\ref{sec;IIIE},~\ref{sec;IIIF}.\hfill\zel

\begin{pt}[{\bf Aspects of Quantum Measurement}]\label{4meas}
\end{pt}
The developed theory provides a possible (?) framework for dealing with the
old fundamental question of QM -- the measurement problem, resp.\ the problem
of ``collaps of wave packets''. Such a possibility is here, however,
except of several included remarks and notes at various places of the text,
left just on the level of this unspecified hypothesis. Some models of ``effective'' measurement
in QM were developed (inspired by Hepp's work \cite{hp-meas}) and published
in \cite{bon6,bon7,bon-sir}.\hfill\zel
\nl\nl
Since the large part of this chapter will consist of formal constructions
on a unique K\"ahlerian orbit \OUr, i.e.\ the projective Hilbert space \PH\
consisting of one--dimensio\-nal projections $\mrh\equiv\mrh^2$, it would be
useful {\em first to examine the manifold structure of \PH} in some details.\footnote{For some additional
mathematical aspects see also \cite{bon-orbit}.\label{orbit}} Our
analysis is mainly based on the earlier author's works~\cite{bon4,bon8};
the obtained structures, as well as used mathematical devices are essentially
identical with independently composed papers by Cirelli et al., cf.
mainly~\cite{cir,PH}.

\section{Structure of Projective Hilbert Space}
\label{sec;IIIA}

Let \H\ be a complex Hilbert space. Elements \ix\ of \H\ will be naturally
associated with the corresponding elements $\ix^*$ of the topological dual
space $\mH^*$ of \H\ via the Riesz lemma, i.e.\ $\ix^*(\iy):=(\ix,\iy)\
(\forall \iy\in\mH)$; the mapping $\ix\mapsto\ix^*$\ is an antilinear
isometry of \H\ onto $\mH^*$. The space of Hilbert--Schmidt operators \fk
H\ will be (linearly and isometrically) identified with the tensor product
$\mH\otimes\mH^*$ in such a way, that the operator (in the Dirac
notation,~\cite{dirac}) $|\ix\rb\lb\iy|\equiv\ix\otimes\iy^*\in\mLH$ acts as
follows:
\[ |\ix\rb\lb\iy|\iz:=|\ix\rb\lb\iy|\iz\rb:=(\iy,\iz)\ix,\quad
\forall\iz\in\mH. \]
The scalar product in \fk H\ is then
\[
\bigl(\ix\otimes\iy^*,\iz\otimes\iu^*\bigr)_2=Tr\bigl(|\iy\rb\lb\ix|\!
\cdot\!|\iz\rb\lb\iu|\bigr)=(\ix,\iz)(\iu,\iy)\equiv
\lb\ix|\iz\rb\lb\iu|\iy\rb.\]

\begin{rem}\label{rem;mixinPH}
There is a natural question whether the dynamics (in general --
nonlinear) of density matrices described in Chapter~\ref{sec;II}
as dynamics on the orbits \OUr\ with arbitrary $\dim(\mrh)$ can be
described equivalently as a ``corresponding'' dynamics on the
projective Hilbert space $P(\tilde{\mfk H})\ni\mrh$ of some
``another'' Hilbert space $\tilde{\mfk H}$, i.e.\ a dynamics on
the ``one--dimensional'' orbit $\mcl O_\mrh(\tilde{\mfk U})$\ with
$\dim(\mrh)=1,\ \tilde{\mfk U}:=\mcl U(\tilde{\mfk H})\equiv$\
(the unitary group of $\mcl L(\tilde{\mfk H})$). This question can
be motivated, e.g., by the fact, that any density matrix \rh\ in a
separable Hilbert space \H\ can be considered as the ``partial
trace'',~\cite[Section 10.1]{davies}, of a one--dimensional
projection \P x\ on a tensor--product space $\mH\otimes\mK$
interpreted usually as the Hilbert space of a composed QM--system
containing the considered system (occurring in the state \rh) as a
subsystem described in \H\ (this assertion is almost trivial: it
can be proved by an explicit construction of \P x\ from \rh); cf.\
also (iii) below. Let us mention here three possibilities of
description of ``mixed states dynamics'' by a dynamics of
vector--states projected to some $P(\tilde{\mfk H})$:
\item{(i)}
Let us recall that the trace--class operators are $\mfk T\subset\mfk
H\subset\mLH$, where the space \fk H\ is the space of Hilbert--Schmidt
operators endowed with a canonical Hilbert space structure. This shows
that one could formulate all the
theory of Chapter~\ref{sec;II} ``in principle'' on the orbit of the unitary
group ${\cal U}(\mfk H)$ of \fk H\ consisting of one--dimensional
projections of \fk H, i.e.\ the projective Hilbert space $P(\mfk H)$\ of \fk
H.
This would need, however, an
additional work to distinguish what elements of $P(\mfk H)$\ are relevant in
what physical situations and, moreover, what unitary transformations in \fk
H\ correspond to those used in Chapter~\ref{sec;II}.

\item{(ii)}
Another possibility to describe also density matrices and their
(possibly nonlinear) dynamics in framework of a projective Hilbert
space comes from elements of the Tomita--Takesaki theory of
modular Hilbert algebras, cf.\ \cite{tomita,bra&rob}: Let
$\tilde{\mfk H}$\ be Hilbert space of a faithful
weakly${}^*$--continuous representation of the considered von
Neumann algebra of observables (in our case it is \LH) with a
cyclic and separating vector; that part of $P(\tilde{\mfk H})$
which is the image by the canonical projection $\tilde{\mfk
H}\rarw P(\tilde{\mfk H})$\ of the natural positive cone $\mcl P$\
in $\tilde{\mfk H}$, cf.\ \cite[Section 2.5.4]{bra&rob}, describes
the whole $\mSs(\mLH)$.

\item{(iii)}
The last possibility to be mentioned here is that one considering the
density matrix $\mrh\in\mTH$\ as the partial trace of some $\mP
x\in P(\mH\otimes\mK)$. Let us assume that both the Hilbert spaces \H\ and
\K\ are infinite--dimensional separable. If $\{\mph k,(1)~; k\in\mbZ_+\}
\subset\mH$\ is
an orthonormal basis of \H\ such that the given density matrix \rh\ is
\[ \mrh=\sum_{j=1}^\infty \mlam_j |\mph j,(1)~\rb\lb\mph j,(1)~|, \]
then, for any orthonormal basis $\{\psi_k^{(2)}, k\in\mbZ_+\}$\ of $\mK$, the
vector $\ix\in\mH\otimes\mK$\ defined by
\[ \ix:=\sum_{j\in\mbZ_+} \sqrt{\mlam_j}\mph j,(1)~\otimes\psi_j^{(2)} \]
has the desired property: For any $A\in\mLH$, with $I_\mK$\ the identity in
${\cal L(K)}$, one has
\[ Tr(\mrh A):=Tr_\mH(\mrh\!\cdot\!A)\equiv Tr_{\mH\otimes\mK}(\mP
x\!\cdot\!A\otimes I_\mK). \]
This formula defines the mapping $Tr_\mK:\mP x\mapsto\mrh$ called the \emm
partial trace~, cf.\ \cite{davies}. The mapping $Tr_\mK$ can be extended by
linearity to whole
space $\mfk T(\mH\otimes\mK)$ of trace--class operators on the Hilbert
space of the ``composed system''. Let, e.g., \un f,t,\mrh~ be a unitary
cocycle describing (nonlinear) evolution of $\mrh\in\mTH$\ according to
Proposition~\ref{prop;2.7}. Then
\[ \mrh(t)\equiv\mun f,t,\mrh~\mrh\mun f,t,\mrh~^{-1}, \]
and the corresponding evolution of $\ix\in\mH\otimes\mK$ can be chosen as
\[ \ix(t)\equiv \sum_{j=1}^\infty\sqrt{\mlam_j}\bigl(\mun f,t,\mrh~\mph
j,(1)~\bigr)\otimes\psi_j^{(2)}.\] Now one has to solve the
problem whether and how this evolution can be described by a
unitary cocycle $\tilde{\mun h,t,\ix~}$\ acting on
$P(\mH\otimes\mK)$.

We do not intend to elaborate further these remarks in this work. They were
mainly mentioned here to stress importance
of the special orbit of \UH: the projective Hilbert space \PH. Cf. also Footnote \ref{orbit}.
\hfill\dovi
\end{rem}

The projective Hilbert space \PH\ will be considered as a complex--analytic
manifold, the structure of which will be presently described.
\def\nazov{{
\ref{sec;IIIA}\quad Structure of Projective Hilbert Space}}

\begin{notat}\label{notat;bfPH}
The elements of \PH\ will be identified with one--dimensional projections
and denoted also by boldface lowercase letters: $\by\equiv P_\iy\in\mPH,\
\iy\in\by$, i.e.\ we shall consider elements of \PH\ interchangeably as
equivalence classes in \H: $\by:=\{\ix\in\mH: \exists \mlam\in\mbC,
\ix=\mlam\iy\}$, and as one--dimensional projections $P_\iy\equiv P_\by$.
In the case if $(0\neq)\iy\in\mH$\ is expressed by a {\em formula} written
in any type of letters, then we shall use the boldface expression in
boldface brackets to write down the corresponding symbol for the class
$\by\in\mPH$, $\iy\in\by$\ := \bbr{formula}.\hfill\dovi
\end{notat}

 Let us define now an atlas on the manifold \PH:
 \begin{defs}[{\bf Atlas on \bs{\mPH}}\index{atlas on \bs{\mPH}}]\label{df;atlas}
 \item{(i)}
 The \emm topology of \PH\ ~ will be defined as the factor--topology coming
 from the Hilbert--space norm--topology of \H. It can be
 shown,~\cite{bon8}, that this topology is equivalent to several other
 natural topologies induced on \PH\ by its embedding to the Banach spaces
 \LH, \fk H, \fTs, or also to several weak topologies coming from the
 duality relation $(\mP x;C)\mapsto Tr(C\mP x)\equiv\lb C;\mP x\rb$.

 \item{(ii)}
 The \emm charts on \PH~\ consist of neighbourhoods
 \begin{subequations}\label{eq;3.1}
 \begin{equation}
\glss \mcl V_\by~:=\{P_\ix\in\mPH:Tr(P_\ix P_\iy)\neq 0\}
 \end{equation}
 of the points $\by\in\mPH$, and their (\iy--dependent) mappings
 \begin{equation}
 \theta_\iy: \mcl V_\by~\rarw [\iy]^\perp, P_\ix\mapsto\theta_\iy(\bx):=
 \|\iy\|^2(\iy,\ix)^{-1}(I-P_\iy)\ix
 \end{equation}
 onto the complex orthogonal complements $[\iy]^\perp$ (considered as complex
 Hilbert subspaces of \H) of nonzero $\iy\in\mH,\ \iy\in\by$.
 \end{subequations}
 \end{defs}

\begin{subequations}
 \begin{prop}\label{prop;2.1.6}
The mapping $\theta_x$ is a homeomorphism of $\mcl V_\bx$ onto $[x]^\perp$\
(with the norm--topology of \H). The set
\begin{equation}
 \{ (\mcl V_\bx;\theta_x): 0\neq x\in\mH\}
\end{equation}
is an atlas on \PH\ defining a complex--analytic manifold structure
consistent with the topology of \PH.\hfill\zal
 \end{prop}
 \begin{proof}
For any $\by_j\in\mcl V_\bx$, and any $y_j\in\by_j\ (j=1,2)$, it is
$\by_1\neq\by_2$\ iff $(x,y_2)y_1\neq(x,y_1)y_2$, hence according
to~\eqref{eq;3.1}, $\theta_x$\ is injective.

For any $z\in[x]^\perp$ and $y:=z+x$, we have $\by\in\mcl V_\bx$\ (since
$x\neq0$), and $\theta_x(\by)=z$, hence $\theta_x$\ is bijective. Let
$\|x\|:=1$. For $z_j\in[x]^\perp,\ y_j:=z_j+x\ (j=1,2)$ the identity
\begin{equation}
 1-Tr(\mP{y_1}\mP{y_2})= \frac{1}{(\|z_1\|^2+1)(\|z_2\|^2+1)}
\left(\|z_1-z_2\|^2+\|z_2\|^2\!\cdot\!\|(I-\mP{z_2})(z_1-z_2)\|^2\right)
\end{equation}

\noidt implies the bicontinuity of $\theta_x$. For any $0\neq x_j\in\mH,\
j=1,2$,
and for $z\in\theta_{x_1}(\mcl V_{\bx_1}\cap\mcl V_{\bx_2})$, we have
\begin{equation}
\theta_{x_2}\circ\theta^{-1}_{x_1}(z)=\|x_2\|^2\frac{x_1+z}{(x_2,x_1+z)}-x_2,
\end{equation}
and we can see, cf.\ \cite{bourb;manif,h-cartan}, that the mapping
\begin{equation}
\theta_{x_2}\circ\theta^{-1}_{x_1}:\theta_{x_1}\left(\mcl V_{\bx_1}\cap\mcl
V_{\bx_2}\right) \rarw \theta_{x_2}\left(\mcl V_{\bx_1}\cap\mcl
V_{\bx_2}\right)
\end{equation}
is a complex analytic function.
 \end{proof}
 \end{subequations}

 The tangent space $T_\by\mPH$ of \PH\ at $\by\in\mPH$\ will be identified
 with the linear space of classes of mutually tangent differentiable curves
 at \by\ as in the finite dimensional case,~\cite{abr&mars,3baby,kob&nom};
 this is in accordance with our results from
 Subsection~\ref{q-phsp;manif}, cf.\ Definitions~\ref{df;2.3}, and
 Proposition~\ref{prop;2.4}. For any differentiable mapping $\theta$\ of a
 \nbhd of \by\ onto a \nbhd of $\theta(\by)$\ in another differentiable
 manifold, the corresponding tangent mapping $T_\by\theta$\ maps the vector
 $\bv\in T_\by\mPH$\ represented by a curve $t\mapsto c_\bv(t)\
 (c_\bv(0)=\by)$\ onto the vector tangent at $\theta(\by)$\ represented
 by the curve $t\mapsto\theta(c_\bv(t))$. If $\bx\in\mcl V_\by$ (hence
 $\by\in \mcl V_\bx$), $T_\by\theta_\ix$\ maps $T_\by\mPH$\ onto
 $[\ix]^\perp$, and the choice of $\ix:=\iy\in\by$\ (in the index of
 $\theta_\ix$) leads to a natural (\iy--dependent) identification of
 \Ty\PH\ with $[y]^\perp$\ (and also with $\mcl V_\by$). The vector ${\bf
 v}\in\mTy\mPH$\ (let \by\ be fixed) is mapped onto $\mv x$\ :=\
 \Ty$\theta_\ix(\bv)\in[\ix]^\perp$,
 \begin{subequations}\label{eq;3.2}
 \begin{equation}
 \mv x:=\left.\frac{d}{dt}\right|_{t=0}\theta_\ix(c_\bv(t)).
 \end{equation}
 One can choose, e.g.,
  \begin{equation}
  c_\bv(t):=\mbbr{\exp(-itB(\bv))y}=Ad^*\bigl(\exp(-itB({\bf
 v}))\bigr)\mP y,
 \end{equation}
 where $B(\bv)$ is a selfadjoint element of \LH\ representing an
 arbitrarily chosen vector $\bv\in\mTy\mPH$\ in this way. With such a
 choice of $B(\bv)$, one has the expression
 \begin{equation}
 \mv x=-i\,(\ix,\iy)^{-1}\|\ix\|^2\left(I-\frac{\mP y\mP x}{Tr(\mP y\mP
 x)}\right)B(\bv)\iy.
 \end{equation}
 For $\ix=\iy$, this leads to
 \begin{equation}
 \mv y=-i\,(I-\mP y)B(\bv)\iy.
 \end{equation}
 Specifying \bv\ by the choice of any ${\it v}\in[\iy]^\perp$, and by
 the choice
 \begin{equation}
 B(\bv):=i\,\|\iy\|^{-2}(|{\it v}\rb\lb\iy|-|\iy\rb\lb{\it v}|),
 \end{equation}
 \end{subequations}
 one obtains $\mv y={\it v}$. With a chosen $\iy\in\by$, and the
 corresponding ``identification $\theta_\iy$'' of \Ty\PH\ with
 $[\iy]^\perp$, one can identify $\bv\equiv{\it v}\equiv\mv y$. For
 different choices of \ix\ in~\eqref{eq;3.2}, on the other hand, one
 obtains expressions $\mv x$\ and $\mv z$\ of \bv\ in different charts
 $\theta_\ix$\ and $\theta_\iz$ related mutually by
 \begin{equation}\label{eq;3.4}
 \mv z=\|\ix\|^{-2}(\iz,\iy)^{-1}\|\iz\|^2(\ix,\iy)\left(
 I- \frac{|\iy\rb\lb\iz|}{(\iz,\iy)}\right)\mv x.
 \end{equation}
 Let us note that $(\iz,\iy)^{-1}|\iy\rb\lb\iz|=\mP y\mP z/Tr(\mP y\mP z)$.
 One can now also check validity of the following two mutually
 inverse relations:
 \begin{subequations}\label{eq;3.4a}
\begin{equation}
\mv x =(\ix,\iy)^{-1}\|\ix\|^2\left(I-\frac{\mP y\mP x}{Tr(\mP y\mP
x)}\right)\mv y,
\end{equation}
and
\begin{equation}
\mv y=\|\ix\|^{-2}(\ix,\iy)(I-\mP y)\mv x.
\end{equation}
\end{subequations}
We shall consider \PH\ as a real analytic manifold endowed with
(integrable,~\cite{cir}) complex structure
\[ J\in\mcl T_1^1(\mPH)\]
($\mcl T^r_s(M)$\ denotes the vector space of all $r$--times contravariant and
$s$--times covariant smooth tensor fields on a manifold $M$) defined as the
section $\by\mapsto J_\by\in\mcl L([\iy]^\perp)$ with $(J_\by{\bf
v})_\iy:=i\,\mv y$, i.e.\ the complex structure is determined by the given
multiplication by the imaginary unit ``$i$'' in \H. The K\"ahler metrics
$\Gamma\in\mcl T^0_2(\mPH)$ on \PH, cf.\ \cite{bon4,bon8,cir}, called also
the {\em Fubini--Study metrics}, can be expressed in the following form:
\begin{equation}\label{eq;3.5}
\Gamma_\by(\bv,\bw):= 2\|y\|^{-2}\Re(\mv y,\mw y),\ \bv,\bw\in\mTy\mPH.
\end{equation}
The corresponding symplectic form $\Omega\in\mcl T_2^0(\mPH)$ is then
expressed by:
\begin{equation}\label{eq;3.6}
\Omega_\by(\bv,\bw):=\Gamma_\by(\bv,J\bw)=-2\|y\|^{-2}\Im(\mv y,\mw
y).
\end{equation}
These structures coincide on \PH\ with those coming from the tensor field
$\Psi$, cf.\rref2.15a~.
\begin{lem}\label{lem;3.3}
The two--form $\Omega$\ in~\eqref{eq;3.6} coincides with the restriction to
the orbit \PH\ of the form $\Omega$\ from~\eqref{eq;2.15a}.\hfill\zal
\end{lem}
\begin{proof}
\begin{subequations}\label{eq;3.6a}
The mapping \ben\ defined in~\eqref{eq;2.3d} for $\nu:=\by\in\mPH$\ has the
form $\beta_\by(\rc)=i[\rc,\mP y]$, where $\rc\in\mTy\mcl O_\by(\mfk
U)\subset\mfTs$\ is represented (cf.\ Definitions~\ref{df;2.3}(iii)) by a
bounded operator. If a vector $\bv\in\mTy\mPH$\ corresponds to the curve
\begin{equation}
 t\mapsto\rc_\bv(t):=\mbbr{\exp(-itB(\bv))y}= {\rm
Ad}^*\bigl(\exp(-itB(\bv))\bigr)\mP y,
\end{equation}
then the corresponding operator is
\begin{equation}
\rc=\dot\rc_\bv:=\left.\frac{d}{dt}\right|_{t=0}\rc_\bv(t)=i[\mP
y,B(\bv)].
\end{equation}
By a use of Definitions~\ref{df;2.3}(iv), one obtains
$\beta_\by(\dot\rc_\bv)=q_\by(B(\bv))$. Inserting these expressions
to~\eqref{eq;2.15a}, we obtain the relation
\begin{equation}
\Omega_\by(\bv,\bw)=i\,Tr\bigl(\mP y[q_\by(B(\bv)),q_\by(B(\bw))]\bigr),
\end{equation}
what is identical with the result of the corresponding insertions
from equations~\eqref{eq;3.2} into~\eqref{eq;3.6}.
\end{subequations}
\end{proof}
\noidt Expressed in the chart $\theta_x$, the K\"ahler structure $\Psi$ on
\PH\ has the form:
\begin{equation}\label{eq;3.7}
\Gamma_\by(\bv,\bw)-i\Omega_\by(\bv,\bw)=2\|y\|^{-2}Tr(\mP x\mP y)(\mv
x,(I-\mP y)\mw x).
\end{equation}

Inserting from~\eqref{eq;3.2} into~\eqref{eq;3.5} and~\eqref{eq;3.6}, one
obtains an expression of the K\"ahler structure in terms of the selfadjoint
operators $B(\bv(\bx))$\ and $B(\bw(\bx))$ representing the vector fields
\bv\ and \bw\ in any point $\bx\in\mPH$, cf.\ also~\eqref{eq;2.15a}:
\begin{equation}\label{eq;3.8}
\begin{split}
&\Psi_\by(\bv,\bw)=2\|y\|^{-2}(\mv y,\mw y)= \\ & 2Tr\bigl(\mP
y B(\bv(\by))B(\bw(\by))\bigr)-2Tr\bigl(\mP y B(\bv(\by))\bigr)Tr\bigl(\mP
y B(\bw(\by))\bigr).
\end{split}
\end{equation}

It can be shown,~\cite{PH}, that the distance function d(\bx,\by) on \PH\
corresponding to the Riemannian metrics $\Gamma$\ is expressed
by\footnote{The
derivation of the distance d(\bx,\by) is easy after accepting the
(plausible looking) assumption, that any
geodesic is contained in the submanifold of \PH\ homeomorphic to a real
two--dimensional sphere representing the projective Hilbert space of the
two--dimensional complex subspace of \H\ spanned by $\{x,y\}$.
The nontrivial part of the proof
consists in justification of this assumption,~\cite{bon-d}.}
\begin{equation}\label{eq;3.9}
\rd(\bx,\by)=\sqrt{2}\arccos\sqrt{Tr(\mP x\mP y)},
\end{equation}
with values in the interval $\left[0;\frac{\pi}{\sqrt{2}}\right]$.

The linearity of conventional quantummechanical time--evolutions, as well
as other symmetry transformations is closely connected with the
metrics~\eqref{eq;3.9}. The corresponding mathematical formulation is in
fact a rephrasing of the very well known Wigner theorem, cf.
Remark~\ref{rem;tr-prob},~\cite{wigner1,woron}:

\begin{prop}\label{prop;3.4}
Let $\Phi$ be any bijection of \PH\ onto itself conserving the distance
function
\rd\ from~\eqref{eq;3.9}. Then there is a linear, or antilinear
isometry $\ru_\Phi$\ of \H\ onto itself representing $\Phi$ in the sense
that $\Phi(\by)=\mbbr{\ru_\Phi y}$\ for all $\by\in\mPH\ (0\neq\iy\in\by)$.
If $\ru_\Phi$\ is linear, then $\Phi$\ conserves also the symplectic form
$\Omega$:
\begin{equation}\label{eq;3.10}
\left(\Phi^*\Omega\right)_\by(\bv,\bw):=\Omega_{\Phi(\by)}
\left(\Phi_*\bv,\Phi_*\bw\right)=\Omega_\by(\bv,\bw),
\end{equation}
i.e.\ $\Phi$ is an isometric symplectomorphism of \PH. The mapping $\Phi$
changes the sign at the symplectic form $\Omega$ in the case of antilinear
 $\ru_\Phi$: $\Phi^*\Omega\equiv -\Omega$. \hfill\zal
\end{prop}
\begin{proof}
Conservation of \rd\ means conservation of the ``transition probabilities''
$Tr(\mP x\mP y)$, $\forall\ix,\iy\in\mH\setminus\{0\}$; this means also
conservation of the metric tensor $\Gamma$. According to the Wigner theorem
there is unitary or antiunitary bijection $\ru_\Phi:\mH\rarw\mH$, as stated
in the proposition. But the symplectic form is invariant \wrt unitary
transformation, as was shown in the Remark~\ref{rem;submanif}.
The last part of the proposition is a consequence of the fact that
antiunitary mappings \ru\ change the value of the scalar product in \H\ to
its complex conjugate:
$(\ru x,\ru y)=(y,x)$. For more details cf.\ \cite{PH,bon8,cir}.
\end{proof}
\begin{rem}\label{rem;3.4a}
A general (``nonlinear'') symplectomorphism of \PH\ does not conserve
$\Gamma$ (equivalently: the distance function \rd). This might be
considered as a strong argument for linearity of QM, since, as we shall see
soon in Section~\ref{sec;IIIA1}, the metric tensor leading to this distance
function is a tool for geometric reformulation of
the probability interpretation of QM. By introducing the ``nonlinear
observables'' and their nonlinear transformations, and also the
corresponding interpretation based on the ``two point function
representatives'' of observables, cf.\ Definitions~\ref{df;2.25b},
and~\ref{df;2.25c}, we have overcame the difficulty with noninvariance of
this ``interpretational device'' \wrt general symplectomorphisms.\hfill\dovi
\end{rem}


\section{Symplectic Form of QM and NLQM; \\ Restrictions
of QM}\label{sec;IIIA1}

The traditional (linear) quantum mechanics (QM) is completely described by
kinematics and dynamics on \PH, i.e.\ the effects connected with other parts
of the ``elementary quantum phase space'' \Ss\ containing density matrices
$\mrh\neq \mrh^2$ which are described by the formalism of
Chapter~\ref{sec;II} can be reproduced by the restriction of that formalism
to the ``one--dimensional'' orbit \PH\ only, and by ``dynamics
independent'' manipulations with objects defined on it. This is due to
linearity, since the
used transformations (time evolutions, symmetries) of \Ss\ are then affine
mappings, and expectations also affinely depend on $\mrh\in\mSs$.

In the terminology of Chapter~\ref{sec;II}, QM can be obtained as
the $G$--system
on an infinite--dimensional separable Hilbert space \H\ with the trivial group
$G:=\{e\}$. In this case, the set \GGc\ of $G$--classical generators consists
of constants. The set \GG\ of $G$--symmetry generators, on the other hand,
contains (densely defined) functions \h Y\ corresponding to all selfadjoint
operators $Y$. Observables can be represented by affine functions \h{\mfk
f}\ only, since their function representatives $\hat h_{\mfk f}$ do not
depend on the $G$--classical variable $\nu$ in the ascription
$(\mrh;\nu)\mapsto\mhh f,\mrh,\nu~$. The ``genuine mixtures''
$\mu,\mu'\in\mcl M^G$ corresponding to the same barycentres $\mfk
b(\mu)=\mfk b(\mu')$\ are not mutually distinguishable by measurements of
the $G$--observables, neither they could be distinguished after a use of
symmetry
transformations (resp.\ evolutions) in the framework of this $G$--system. The
``permitted'' (possibly unbounded) generators and observables include
(densely defined) affine functions $\nu\mapsto\mh X(\nu):=\nu(X)$\
corresponding to selfadjoint operators $X$.

We shall consider ``nonlinear extensions'' of this QM--system (i.e.\ of the
$G$--system with trivial $G:=\{e\}$) by allowing {\em evolutions of states by
nonlinear generators}.\footnote{Such an extension of QM can be obtained {\em by
restriction} of a $G$--system with nontrivial $G$ in the way, that we shall
admit {\em linear observables only}, i.e.\ the observables represented by
nonconstant operator--valued functions on \EF\ will be ignored (cf.
Definitions~\ref{df;2.25b}).} To be able to deal also with
the questions of ``integrability'' of also nonlinear functions of these \h
X's, see Definitions~\ref{df;g-dif},
it is useful to (choose and to) consider these $X$'s as selfadjoint
generators of some unitary representation $V(S)$ of a ``symmetry group
$S$'' associated with the considered system (e.g.\ $S$ could be the
$2n+1$--dimensional Weyl--Heisenberg group, i.e.\ the standard
one--dimensional central extension,~\cite{kiril,varad}, of the commutative
$2n$--dimensional
group of translations in classical linear $2n$--dimensional phase space,
see below in this section). These versions of nonlinear quantum
mechanics (NLQM) are not symmetric \wrt transitions between {\em
Schr\"odinger and  Heisenberg pictures}: They can be used in Schr\"odinger
picture only, since a nonlinear (i.e.\ nonaffine) transformation of \Ss\
cannot be expressed by some ``transition to adjoints'',~\cite{bourb;vect},
as a transformation of the {\em algebra of linear observables}: this
algebra could not stay invariant \wrt such a transformation.

Another way of ``transitions to nonlinearity'' in QM
consists in restrictions of (linear) dynamics of QM to submanifolds of \PH\
(or also of \fTs), e.g.\ to some orbits $\mcl O_{\mrh}(S)$\ of a
representation $V(S)$. We can obtain in that way also usual ``quasiclassical'',
or ``self-consistent'' approximations, e.g.\ WKB, or Hartree--Fock
approximations as versions of NLQM, cf.\ also our
Subsections~\ref{IIIA1;restr} and~\ref{IIIA1;NL-Sch}.
The group $S$ needn't be interpreted, however, as a
group of transformations of a ``classical background'' (cf.
Section~\ref{sec;IIIB}) being {\em
dynamically connected} with the system, as it is in the case of $G$--systems
with nontrivial $G$ and general (nonlinear) $G$--generators. Only affine
functions
$\nu\mapsto\rf(\nu)$\ (and their restrictions)\footnote{Restrictions of
affine functions to submanifolds $\mcl O_{\mrh}(S)$\ considered as
Hamiltonians on the phase spaces $\mcl O_{\mrh}(S)$\ lead generally,
however, to nonlinear dynamics on these submanifolds.}
 defined on dense sub--domains of \Ss\ are used
here in the r\^oles of the generators as well as observables.
All the ``traditional'' quantities are ``essentially contained'' in the
sets of corresponding quantities of {\em any $G$--system}: \bcDF\ is dense in
\Ss, and for calculation of any bounded (hence continuous) observable
$\mfk f^*=\mfk f\in\mCGq$\ ($\equiv\mLH$, cf.\ Definition~\ref{df;2.25b})
one can use
values $\mh{\mfk f}(\nu)$\ for $\nu\in\mbcDF$\ (cf.\eqref{eq;2.37}, and
Interpretation~\ref{int;2.27}). The general observables \h X\ used in the
r\^ole of generators could, however, violate the relation $\mpph
t,X~\mbcDF\subset\mbcDF$\ for some $G$--systems.

\def\nazov{{
\ref{sec;IIIA1}\quad Symplectic Form of QM and NLQM; Restrictions
of QM}}

We shall describe in this section the symplectic reformulation
(equivalent to the usual Hilbert space formulation) of
traditional (linear) QM, as well some of its restrictions to submanifolds
of \PH\ leading to nonlinear dynamics (corresponding, e.g., to some
``quasiclassical approximations'') the general form of which was described
in Chapter~\ref{sec;II}.
Let us first, however, formulate briefly a general nonlinear quantum
mechanics (NLQM) on the projective Hilbert space \PH\ to point out
some differences between QM and NLQM.

\subsection{Generalized quantum mechanics on\ \PH}\label{IIIA1;QM}

We shall consider here a general (nonlinear) EQM, but we shall restrict our
attention to
dynamics and kinematics restricted to \PH\ only. Let us  choose also a
Lie group $G$
and its unitary representation $U(G)$\ such, that the space of generators
\GG\ includes all the (nonlinear) generators we want to use in the theory.
Let us consider, however, only elementary quantum observables \CGq,
cf.\dref2.25b~(ii), in the r\^ole of bounded observables we intend to
interpret in the considered model. Hence, for nonlinear evolutions, the
Heisenberg picture will not be used.
We shall call the chosen system a \emm restricted $G$--system~ (i.e.\
restricted to the ``restricted quantum phase space'' $\mPH\subset\mSs$, with
the restricted set of observables \CGq).

If $X_j=X_j^*$ are elements of the representation $dU(\mfk g)$\ of the Lie
algebra \fk g\ of $G$, then the typical form of the (``restricted'') generators
 $\rQ\in\mGG$ will be
\[ \rQ:\nu (\in\mPH)\mapsto \rQ(\nu)\equiv Q(\nu(X_1),\nu(X_2),\dots),\]
with $Q\in C^\infty(\mfk g^*,\mbR)$. The corresponding nonlinear
Schr\"odinger equations are discussed in
Sections~\ref{sec;IIIE},~\ref{sec;IIIF}, and also in
Subsection~\ref{IIIA1;NL-Sch}.

Let us denote $\mFP:=C^\infty(\mPH,\mbR)$\ the differentiable functions on
the Banach manifold \PH. The differential $df\in\mcl T^0_1(\mPH)$\ of
$f\in\mFP$ can be determined by the formula
\bequ\label{eq;rdf}
d_{\bx}f(\bw(\bx)):=\left.\frac{d}{d
t}\right|_{t=0}f\bigl(\mbs{(\exp(-itB(\bw(\bx)))x)}\bigr),
\end{equation}
with $B(\bw)$\ specified in\rref3.2~, for any vector field
$\bw\in\mcl T_0^1(\mPH)$. The symplectic form $\Omega$\ is strongly
nondegenerate~\cite{3baby} on \PH\ (cf.\ Theorem~\ref{thm;2.10}), hence it
associates with each $f\in\mFP$\ a unique Hamiltonian vector field \vf\
on \PH\ such that
\bequ\label{eq;3.11}
\Omega(\mvf,\bw)=-df(\bw), \forall\bw\in\mcl T_0^1(\mPH).
\end{equation}
The (local) flow \pph{},f~\ of \vf\ leaves $\Omega$ invariant,
hence for the Lie derivative \L{\mvf}~\ we have:
\[ \mL\mvf~\Omega=0.\]
The Poisson bracket $\{f,h\}:=\Omega(\mvf,\mvh)\in\mFP$\ determines the
differential equation (equivalent to the Schr\"odinger equation for affine
f) for the Hamiltonian flow \pph{},f~. Also the following formula (well
known from CM,~\cite{arn1,abr&mars}) is valid here:
\[ dh(\mvf)=\{f,\mh{}\}\quad(\forall h\in\mFP). \]
We shall formulate now a necessary and sufficient condition under which a
function $f\in\mFP$\ is affine, i.e.\ is expressed by a linear operator:

\begin{prop}\label{prop;3.5}
Let $f\in\mFP$, and let \vf\ is the corresponding Hamiltonian vector
field on \PH. Let $\Gamma$\ be the canonical (K\"ahlerian) metrics on \PH.
Then $\mL\mvf~\Gamma\equiv0$\ iff there is a bounded selfadjoint operator
$\ra=\ra^*\in\mLH$\ such that:
\begin{subequations}
\label{eq;3.12}
\bequ
f(\bx)\equiv\mh{\ra}(\bx):= Tr(P_x\ra),\quad 0\neq x\in\bx.
\end{equation}
In the case of $f=\mh{\ra}$, \vf\ is complete, and one has
\bequ
\mpph t,f~(\bx)=\mbs{(\exp(-it\ra)x)},\ t\in\mbR,\ \bx\in\mPH,\ 0\neq
x\in\bx.
\end{equation}
Hence the flows of those Hamiltonian vectors fields \vf\ which conserve the
metrics $\Gamma$\ on \PH\ cor\-res\-pond to  norm continuous one--parameter
unitary groups on \H.\hfill\zal
\end{subequations}
\end{prop}
\noidt A proof is contained in~\cite[Propositions 3.4, and 3.5]{cir}, resp.
in~\cite{bon8}.

Let us introduce also the \emm Riemann bracket \bs{[[\cdot,\cdot]]}~ in
accordance with~\cite{cir}:
\bequ\label{eq;3.13}
[[f,\mh{}]]:=\Gamma(\mvf,\mvh).
\end{equation}
An immediate consequence of\rref3.8~ and of the Proposition~\ref{prop;3.5}
is the following lemma, cf.\ also~\cite{cir}:
\begin{lem}\label{lem;3.6}
Let $\mh{\ra}\in\mFP$\ be defined for any $\ra\in\mLH$\ by\rref3.12~.
Then for any selfadjoint $\ra,{\rm b}\in\mLH$, the following formula holds:
\bequ\label{eq;3.14}
2\mh{\ra}*\mh{{\rm b}}:=2\mh{\ra\dti{\rm b}}=
[[\mh{\ra},\mh{{\rm b}}]]-i\{\mh{\ra},\mh{{\rm
b}}\}+ \mh{\ra}\mh{{\rm b}}.
\end{equation}
The mappings $\ra\mapsto\mh{\ra}(\bx)\ (\bx\in\mPH)$\ are continuous in
the weak operator topology.\hfill\zal
\end{lem}

Due to\rref3.14~, we can calculate $Tr(P_x\ra^n)\ (\forall n\in\mbN)$\ in
terms of the function $\mh{\ra}$, what allows us to express the
probability interpretation of QM in differential geometrical terms on \PH:
The formula\rref3.14~ leads us (via the functional calculus) to a rule for
calculation of $\mh{f(\ra)}$\ for an arbitrary real bounded Borel
function $f$ defined on the spectrum \sg(\ra)\ of $\ra=\ra^*\in\mLHs$. Then
the number $\mh{f(\ra)}(\bx)\equiv Tr(P_x f(\ra))$\ can be interpreted as the
expectation value of the ``observable'' (represented by the operator)
$f(\ra)$\ obtained by
averaging of repeated measurements of $f(\ra)$\ in the (repeatedly
prepared) pure quantum state $\bx\in\mSs$.
 The probability of finding measured values of a selfadjoint $\ra\in\mLHs$\ in an
 interval $J\subset\mbR$\ is then expressed by taking for $f$\ the
 characteristic function $\chi_J$\ of that interval:
\bequ\label{eq;1probJ}
  \prob(\bx,\ra\in J)=Tr(P_x\chi_J(\ra)).
  \end{equation}

 Calculating expectations of
arbitrary selfadjoint \ra\ in \LH\ for any (elementary) mixture $\mrh\in\mS$\
in the standard way from the expectations in pure states (by corresponding
convex combinations), we obtain the result $Tr(\mrh\ra)$, in accordance
with QM.

Let us stress also here that each \rh\ such that $\mrh^2\neq\mrh\in\mSs$\
can be decomposed in uncountably many different ways into (not necessarily
orthogonal) convex combinations
\[ \mrh=\sum_j\mlam_j P_{x(j)},\quad \mlam_j\geq0,\ \sum_j\mlam_j=1, \]
of one--dimensional projections $P_{x(j)}, x(j)\in\mH$\ (representing pure
states).\footnote{This is an essential difference of QM from CM.}
Different decompositions $(P_{x(j)};\mlam_j;j\in
J)$, and $(P_{x'(j')};\mlam'_{j'};j'\in J')$\ of a given \rh\ can be
represented, in another language, by probability measures $\mu_{\mrh},\
\mu'_\mrh$\ on the state space \Ss\ with the same \emm barycentre \bs{\mfk
b(\mu_\mrh)=\mfk b(\mu'_\mrh)=\mrh}~,\ the measures being concentrated on
at most countable sets of points
(i.e.\ on the sets $\{P_{x(j)}:j\in J\}$):
\[\mu_\mrh(\{P_{x(j)}\})=\mlam_j,\ \forall j\in J,\]
hence the states $\mome_\mu\ (\mu=\mu_\mrh,\mu'_\mrh,\dots)$, all
representing the same \rh, give the following expressions for expectation
values of $\ra\in\mLHs$:
\bequ\label{eq;3.15}
\begin{split}
\mome_\mu(a):= &\int_{\mSs}\nu(a)\mu(d\nu)\equiv \int_{\mSs}Tr(a\nu)
\mu(d\nu) \\
=& \int_{\mPH}Tr(a\nu)\mu(d\nu)= \int_{\mPH}\mh{\ra}(P_x)\mu(d P_x) \\
=& \sum_{j\in J}\mh{\ra}(P_{x(j)})\mu(\{P_{x(j)}\})=:Tr(\mfk
b(\mu)\ra)=Tr(\mrh\ra).
\end{split}
\end{equation}
In ``orthodox'' linear QM the states corresponding to measures on \Ss\ with
the same resultant (cf. page \pageref{resultant}) are indistinguishable.
This is one of the important
differences of QM from NLQM (also in the framework of our restricted
model of EQM).
\begin{noti}\label{not;3.7}
Let \rQ\ be a nonlinear generator of time evolution in our theory. Then,
according to
Proposition~\ref{prop;3.5}, its flow \pph{},\rQ~\ does not conserve the
canonical metrics $\Gamma$, hence it does not conserve the distance
function $\rd:\mPH\times\mPH\rarw\mbR_+$.
From the expression\rref3.9~ of the distance
function $\rd(P_x,P_y)$\
we see that, in turn, it does not conserve the ``transition
probabilities'' $Tr(P_xP_y)=|\lb x|y\rb|^2$ between the states
$\bx,\by\in\mPH$. This shows, however, that different measures $\mu\neq\mu'$\
with the same barycentres $\mfk b(\mu)=\mfk b(\mu')$\ can have different
barycentres after some time $t\neq0: \mfk b(\mu\circ\mpph-t,\rQ~)\neq\mfk
b(\mu'\circ\mpph-t,\rQ~)$, and validity of some of the equalities in\rref3.15~
will depend on time (cf.\ also Subsection~\ref{q-phsp;mixt}).
This might lead to prediction of superluminal
communication (for a specific, but rather conventional, interpretation of
the process of measurement in QM), as is pointed out in the
Interpretation~\ref{int;3.7a}.
\hfill\dovi
\end{noti}

\begin{intpn}\label{int;expect}
In the traditional interpretation of QM, the expectation value of the
numerical results of measurement of an ``observable f'' (i.e.\ a
scalar--valued function f of quantum states \by, in this case an ``affine''
one, resp.\ K\"ahler function in the terminology of~\cite{cir})
in an arbitrary
(pure) state $\by\in\mPH$\ equals to its value f(\by), i.e.\ for f:=\h
X, the expectation is $\mh X(\by)=Tr(P_yX)=\lb y|X|y\rb,\ \text{if}\
\|y\|=1$. The calculations of these expectations are closely connected, in
the orthodox QM, with eigenstates of the operators $X$ (assume now, that
$X$ has pure point spectrum). In terms of the
presented ``geometric formulation'', the eigenvectors
$x(k)\equiv|x(k)\rb\in\mH$,
\[ X|x(k)\rb=\kappa_k|x(k)\rb,\ k\in K,\quad \sum_{k\in K} P_{x(k)}=I_{\mH},\]

\noidt resp.\ the one--dimensional eigenprojections $P_{x(k)}\equiv\bx(k)
\in\mPH$, are exactly the ``stationary points'' of the generators \h X,
cf.\ref{lem;df=0}.

Stationarity of the points $\bx(k)$\ is rather a ``dynamical property''. The
observable probabilities can be expressed with a help of the projection measure
$E_X$ of $X:\quad E_X(J):=\chi_J(X)$, as above, see\rref1probJ~.

Let us denote the \emm eigenprojections of \bs X~ corresponding to single
eigenvalues $\kappa_k$\ by $E_k$,
\[ E_k:=E_X(\{\kappa_k\}):= \sum_{j\in K:\kappa_j=\kappa_k}P_{x(j)}.\]
Then the probability of obtaining the result $\kappa_k$, if $X$ is
measured on the system prepared in the state $\by\in\mPH$, is
\[ \prob(\by;X=\kappa_k)=Tr(E_kP_y).\]
The values of these probabilities, for $\dim E_k=1$, i.e.\ $E_k=P_{x(k)}$, are
the above discussed (cf.\ Remark~\ref{rem;tr-prob})
 ``\emm transition probabilities~'', and the values of
the function \h X\ in these points $P_{x(k)}$\ are just the measured
eigenvalues, $\mh
X(\bx(k))=\kappa_k$. Hence the expectation value of $X$ with pure point
spectrum in an arbitrary $\by\in\mPH$\ is
\begin{equation}\label{eq;2expect}
\begin{split}
 \lb X\rb_{\by}:= & \mh X(\by)=\sum_{k\in K}Tr(P_{x(k)}P_y)\mh
X(\bx(k))  \equiv
Tr\left(\sum_{k\in K}\mh X\bigl(\bx(k)\bigr)P_{x(k)}P_y\right)\\
= & Tr\left(\sum_{k\in K}\kappa_k P_{x(k)}P_y\right)  \equiv  Tr(XP_y).
\end{split}
\end{equation}

The first sum is often interpreted in the sense of classical
probability,~\cite{feller}, by considering occurrences of different
$\kappa_k$ (better: of different orthogonal eigenstates) as independent
``events'', and the function $\bx(k)\mapsto Tr(P_yP_{x(k)})$\ is a measure
on the space of these ``events'' determined by the state $\by\in\mPH$, and
consisting of the ``transition probabilities''. If the concept of the
``transition probabilities'' (which is coming from an interpretation of
quantum measurement) were taken seriously also for NLQM, and the r\^ole
accepted for the stationary points $\{\bx(k)\in\mPH:\md f,\bx(k)~=0\}=:
\mbs{S(f)}$\ of a
``nonlinear observable'' $f$\ (cf.\ \cite{weinb}) were formulated as above,
in
the case of linear observables,\footnote{This means that we would work with
such an ``observable'' as with a random variable in the sense of Kolmogorov
formulation of probability theory, by which the stationary states form the
the whole space of ``elementary events''.}
 with keeping unchanged the above formula\rref2expect~ for
calculation of expectations, i.e.\ if we postulated something
like
\[ \lb f\rb_{\by}:=\sum_{\bx\in S(f)}Tr(P_xP_y)f(\bx),\
\forall\by\in\mPH,\]
then we would come to a contradiction: The ``nonlinear'' function $f$\
would be affine:
\[ f\equiv h_Y,\quad Y:= \sum_{\bx\in S(f)}f(\bx)P_x.\]
This consideration indicates that a ``traditional--like'' interpretation
of observables expressed as numerical functions on \PH\ (our ``reduced function
representations'', cf.\ Definitions~\ref{df;2.25c}) cannot be used in NLQM.
\hfill\bpika\end{intpn}

\begin{lem}\label{lem;df=0}
Let $X=X^*$\ be any selfadjoint operator on \H\ with corresponding (densely
defined) function \h X\ on \PH. Let $\bx\mapsto\md\mh X,\bx~$\ be its
generalized differential (cf.\dref dhX~) defined on the domain $\mcl D(\mh
X)$, cf.\ also Lemma~\ref{lem;qqq}, and Proposition~\ref{prop;r-orbits}.
Then the points $\bx(k)$\ lying in the  domain of \d\mh X,{}~ in which the
differential vanishes, satisfy the relation
\[ \md\mh X,\bx(k)~=0\eequiv X|x(k)\rb=\kappa_k|x(k)\rb,\quad
(x(k)\in\bx(k)\in\mPH),\]
i.e.\ they are exactly the one--dimensional eigenspaces of $X$.\hfill\zal
\end{lem}
\begin{proof}
The differential \d\mh X,\bx~\ can be represented, according to considerations
in Subsection~\ref{gener;lin2}, on its domain by the bounded operator
\[ \md\mh X,\bx~=\rq_{\bx}(X)\equiv P_xX(I_{\mH}-P_x)+
(I_{\mH}-P_x)XP_x=P_xX+XP_x-2P_xXP_x,\]
and its vanishing implies commutativity of $P_x$\ with $X$, i.e.
invariance of the one--dimensional subspace \bx\ \wrt the action of $X$. For
proof of the converse, the arguments  go in the  reversed order.
\end{proof}

Let us stop here with general considerations, and we shall turn to
more specific cases now.
\subsection{The Weyl--Heisenberg group and CCR}\label{IIIA1;CCR}
The $2n+1$--dimensional \emm Weyl--Heisenberg group~ \glss \GWH\ ~(it is
also called
the {\em Heisenberg group}) can be chosen in our theory
either in the r\^ole of the group $G$ defining a $G$--system, cf.
Definition~\ref{df;G-syst}, or in the r\^ole of the above mentioned Lie
group $S$ determining domains for generalized fields (cf.
Definition~\ref{df;g-dif}). We shall investigate here the
action of the standard irreducible Schr\"odinger representation
$U(\mGWH)$\ of \GWH\ on \H\ in some details, as well as the quantum kinematics
and dynamics constructed with a help of it. As an expression of the
corresponding Lie algebra relations between generators we obtain the usual
definitions of \emm canonical commutation relations~ (CCR).

Let us recall that the $2n+1$--dimensional group \GWH\ can be defined as the
group of square $(n+2)\times(n+2)$--matrices,~\cite{kiril,zelob&stern}:
\begin{subequations}\label{eq;GWH-m}
\begin{equation}
g(q,p,s):=\begin{pmatrix} 1&-q&s\\ 0&I_n&p^T\\ 0&0&1\end{pmatrix},
\end{equation}
where $q:=\{q_1,q_2,\dots q_n\}\in\mbR^n$, $p:=\{p_1,p_2,\dots
p_n\}\in\mbR^n$, $s\in\mbR$, $I_n$\ is the unit $n\times n$--matrix, $p^T$
is the transposed row $p$\ (i.e.\ the column vector), and $0$'s have an
appropriate meaning of zero submatrices according to their place in the matrix.
The group
multiplication is represented by the matrix multiplication:
\begin{equation}
g(q,p,s)g(q',p',s')=g(q+q',p+p',s+s'-q\!\cdot\!p'),
\end{equation}
with $q\!\cdot\!p':= \sum_j q_jp'_j$. Let's note that $p^T$\ (or $p$) can be
considered as an element of the dual $(\mbR^n)^*$, hence its value on
$q\in\mbR^n$ is $\lb p^T;q\rb:=q\!\cdot\!p:= \sum_j q_jp_j$.
\end{subequations}

The group \GWH\ is a central extension,~\cite{kiril,varad}, of the commutative
group $\mbR^{2n}\ni(q;p)\equiv x$\ (\wrt the addition $x+x'$) by the additive
group \bR, corresponding
to the \emm multiplier~ (in additive notation)~\cite[Chap. X]{varad} $\tilde
m(x,x')\equiv -p'\!\cdot\!q$.

\begin{noti}[Multipliers and quantization]
\label{not;cl-quanti}
The commutative group $\mbR^{2n}$\ is naturally identified with a classical
phase space, or with the group of its translations. As any commutative
group, it has only one--dimensional linear (unitary) irreducible
representations. It has, however, many (mutually inequivalent)
infinite--dimensional \emm projective representations~, i.e.
``unitary representations up to a phase factor''. Namely multipliers
$m(x,y),\ x,y\in\mbR^{2n}$, i.e.\ real--valued functions on the direct
product of two copies of the group, $\mbR^{2n}\times\mbR^{2n}$, satisfying
\bequ
\label{eq;1pr-rep}
 m(x+y,z)+m(x,y)\equiv m(x,y+z)+m(y,z),\ m(x,0)\equiv m(0,x)\equiv0,
\end{equation}
are the (logarithms/$i$ of the) phase factors of the (noncommutative)
projective representations. In a more general setting, let $G$ be a Lie
group, and say $V(G)$ be its continuous projective representation with a
multiplier $m$:
Let $g_1\dti g_2\in G$\ denotes the multiplication in $G$\ (e.g.\ addition
in $\mbR^{2n}$). Let $m$ be a multiplier of $G$, i.e.\ $m:\ G\times
G\rarw\mbR$\ satisfying the
relations in\rref1pr-rep~, with, e.g., $m(g_1\dti g_2,g_3)\mapsto m(x+y,z)$,
etc. The projective \emm \bs m--representation~ $V(G)$ is characterized by
unitarity of the all $V(g)$'s, and by the relation:
\begin{subequations}\label{eq;GWH-multip}
\bequ
V(g_1\dti g_2)\equiv \exp(i\dti m(g_1,g_2))V(g_1)V(g_2).
\end{equation}
One can make from these ``unitary up to factors''
representations $V(G)$ genuine unitary representations of larger noncommutative
groups $G_m$\ constructed from the original group $G$
(e.g.\ from our $G:=\mbR^{2n}$) with
a help of the corresponding multipliers $m$. These \emm central extensions
\bs{G_m}~ of a Lie group $G$\ are constructed as follows:

 Let $(g;\mlam)\in G\times S^1$, with $S^1:=\{\mlam\in\mbC: |\mlam|=1\}$.
Then the central extension $G_m$\ of the group $G$\ by the commutative
group $S^1$\ (resp.\ by \bR, if the ``corresponding logarithms'' are taken)
corresponding to the multiplier $m$ consists of
the couples $(g;\mlam)$, and the group multiplication is defined by
\bequ
(g_1;\mlam_1)\dti(g_2;\mlam_2):=\Bigl(g_1\dti
g_2;\exp\bigl(i\dti m(g_1,g_2)\bigr)\mlam_1\mlam_2\Bigr) .
\end{equation}
This simple procedure makes from a (say, commutative) group $G$ another
(noncommutative) group $G_m$, provided $m$ is not \emm exact~; exactness of
$m$ means the existence of a real function $a:G\rarw\mbR$\ such, that
 \bequ
 m(g_1,g_2)\equiv a(g_1\dti g_2)-a(g_1)-a(g_2).
\end{equation}
If the difference of two multipliers $m_1-m_2$ (what is always again a
multiplier) is exact, $m_1$\ and $m_2$\ are (mutually) similar, or
\emm cohomologous~.

Let us take now the $m$--representation $V(G)$. It can be ``translated'' into
a unitary representation $V(G_m)$ of $G_m$ as follows:
\bequ
V\bigl((g;\mlam)\bigr):=\mlam^{-1}V(g).
\end{equation}
\end{subequations}

There is a certain ``both-sided'' correspondence between projective
``$m$--re\-pre\-sen\-ta\-tions'' of $G$, and a class of unitary
representations of $G_m$. For details cf.\ \cite[Theorem 10.16]{varad}.
As we shall see in a while, traditional ``quantization'' of classical flat
phase spaces corresponds to specific choice of a multiplier of
$G:=\mbR^{2n}$, determined by the experimental value of the Planck constant
$\hbar$.\hfill\dovi
\end{noti}

 \begin{subequations}\label{eq;2param}
To any \emm similar multiplier~ $\tilde m'$\ related to $\tilde m$\ by a
real--valued function $a: (G\ni)x\mapsto a(x),\ a(0)=0$:
\bequ
\tilde m'(x,x')\equiv\tilde m(x,x')+a(x+x')-a(x)-a(x')
\end{equation}
corresponds the central extension isomorphic to \GWH. The choice
$a(q;p):=\frac{1}{2}p\!\cdot\!q$ gives the following group--multiplication
in \GWH\ (corresponding to a reparametrization of the abstract group \GWH)
\begin{equation}
\tilde g(q,p,s)\tilde g(q',p',s')=\tilde
g(q+q',p+p',s+s'+\frac{1}{2}(q'\!\cdot\!p-p'\!\cdot\!q)),
\end{equation}
and the corresponding matrix representation is
\begin{equation}
\tilde g(q,p,s):=\begin{pmatrix} 1&-q&s-p\!\cdot\!q/2\\ 0&I_n&p^T\\
0&0&1\end{pmatrix},
\end{equation}
\end{subequations}
what corresponds to the form usually used in QM, as will be clear soon.

\begin{subequations}\label{GWH*-orbits}

The Lie algebra of \GWH\ can be described as the matrix algebra consisting of
derivatives of matrices $g(q,p,s)$\ \wrt the parameters.
Let the basis $\{\xi_j,\xi_0; j=1,2,\dots 2n\}$ in the Lie algebra
$Lie(\mGWH)$\ be chosen such, that an arbitrary element $\xi\in Lie(\mGWH)$\ is
of the form
\bequ
\xi(\malp,\gamma,\beta)\equiv\sum_{j=1}^n \bigl( \malp_j\xi_{n+j}+\gamma_j\xi_j
\bigr)+\beta\xi_0\equiv
\begin{pmatrix} 0&-\malp&\beta\\ 0&0_n&\gamma^T\\ 0&0&0\end{pmatrix},
\end{equation}
where $\malp_j,\gamma_j,\beta\in\mbR,\ j=1,2,\dots n$. We shall use this
parametrization here.

The commutation relations on $Lie(\mGWH)$\ are expressed by this basis as
\bequ
\begin{aligned}
   & [\xi_{j+n},\xi_{k+n}]=0,\quad [\xi_j,\xi_{k}]=0,\\
   & [\xi_j,\xi_{k+n}]=\delta_{jk}\xi_0,\quad j,k=1,2,\dots n.
\end{aligned}
\end{equation}
Those elements $F$ of the
dual $Lie(\mGWH)^*$ for which $F(\xi_0)\neq 0$\ can be parametrized in the
basis dual to the chosen one in $Lie(\mGWH)$
by parameters $q_0,p_0\in\mbR^n,\ s_0\in\mbR\setminus\{0\}$ in such a way, that
they can be conveniently described by the matrix
\bequ
F(q_0,p_0,s_0)\equiv\begin{pmatrix} 0&0&0\\ s_0p_0^T&0_n&0\\
s_0&s_0q_0&0\end{pmatrix}.
\end{equation}
The value of the linear functional $F$\ with $F(\xi_0)\neq 0$\ on the element
$\xi$\ is then
\bequ
F(\xi):=\lb F;\xi\rb\equiv Tr\bigl[F(q_0,p_0,s_0)\xi(\malp,\gamma,\beta)\bigr]=
(q_0\!\cdot\!\gamma-p_0\!\cdot\!\malp+\beta)s_0.
\end{equation}
For $F$'s with $F(\xi_0)=0$, we have
\bequ
F(\xi)\equiv\sum_{j=1}^n \bigl( \malp_jF(\xi_{n+j})+\gamma_jF(\xi_j)
\bigr)+\beta F(\xi_0)=\sum_{j=1}^n \bigl( \malp_jF(\xi_{n+j})+\gamma_jF(\xi_j)
\bigr).
\end{equation}
 The coadjoint action on elements with $F(\xi_0)\neq 0$\ is expressed then by
 \bequ
Ad^*\bigl(g(q,p,s)\bigr)\begin{pmatrix} 0&0&0\\ s_0p_0^T&0_n&0\\ s_0&s_0q_0&0
\end{pmatrix}=\begin{pmatrix}0&0&0\\ s_0\bigl(p_0+p\bigr)^T&0_n&0\\
s_0&s_0\bigl(q_0+q\bigr)&0\end{pmatrix}.
 \end{equation}
 It is easy to see that the points $F\in Lie(\mGWH)^*$\ with $F(\xi_0)=0$\
 are all stable \wrt $Ad^*(\mGWH)$--action.
 Hence, the $Ad^*(\mGWH)$--orbits are either single points $F$\ with
 $F(\xi_0)=0$\
 covering a $2n$--dimensional hyperplane, or the whole hyperplanes
 with fixed $s_0\neq 0$.
 Let us note that the action of $Ad^*\bigl(g(q,p,s)\bigr)$\ does not
 depend on the parameter $s\in\mbR$, hence its above expression is independent
also on the
 considered reparametrization of \GWH.
\end{subequations}

All irreducible unitary representations $\pi_\mlam$ of \GWH\ (which are
more than one--dimensional) can be parametrized by a real parameter
$\mlam\neq0$, and they are realized in \H:=\Lqn\ as follows (in the
parametrization of~\eqref{eq;GWH-m},~\cite{zelob&stern}):
\bequ\label{eq;GWH-rep1}
[\pi_\mlam(g(q,p,s))\psi](q')=e^{i\mlam(s+p\cdot q')}\psi(q'-q),\
\forall\ q,q',p\in\mbR^n, s\in\mbR,\psi\in\mLqn.
\end{equation}
The generators of $\pi_\mlam$\ corresponding to the chosen parameters are
\begin{subequations}
\label{eq;GWH-gener}
\begin{eqnarray}
-i\,\mlam P_j:=&\left. {\frac{\partial}{\partial q_j}} \right|_0
\pi_\mlam =&-\frac{\partial}{\partial q'_j}\quad =\ -i\,X(\xi_{j+n}),\\
i\,\mlam Q_j:=&\left. \frac{\partial}{\partial p_j} \right|_0
\pi_\mlam=&i\,\mlam q'_j\dti\ (\text{i.e.\ multiplication by the variable}\
q'_j) \nonumber\\
&&\qquad\qquad=-i\,X(\xi_j),\ j=1,\dots n,\\
i\,(\mlam)^2X_0:=&\left. \frac{\partial}{\partial s} \right|_0
\pi_\mlam =& i\,\mlam I\qquad =\ -i\,X(\xi_0),
\end{eqnarray}
\end{subequations}
where the labels ``$0$'' at the derivatives denote differentiations in
the unit element of \GWH.

The Schr\"odinger representation of CCR can be considered as that one given by
the generators of $\pi_\mlam$\ with $\mlam:=\frac{1}{\hbar}$. We shall
need, however, the ``corresponding'' group representation of \GWH\ expressing
the Weyl form of CCR. If we use the parametrization of \GWH\
from~\eqref{eq;2param},
we obtain the rewriting of the representation $\pi_\mlam$\
from~\eqref{eq;GWH-rep1} in the
form $W_\mlam$\  (the Weyl form):
\bequ\label{eq;GWH-rep2}
[W_\mlam(q,p,s)\psi](q')=e^{i\mlam(s+p\cdot q'-\frac{1}{2}p\cdot q)}
\psi(q'-q),\ \forall
q,q',p\in\mbR^n, s\in\mbR,\psi\in\mLqn.
\end{equation}
\begin{notat}\label{notat;CCR}
Let us denote
the \emm projective representation~ of the commutative group $\mbR^{2n}$\
usually
referred to as ``the Weyl form of (the representation of) CCR'',
by $W_\mlam(x):=W_\mlam(q,p,0)$\,
with $\{q;p\}=:x\in\mbR^{2n}$.
Note that the projective representation $W_\mlam$\ of $\mbR^{2n}$\ differs from
the ``corresponding'' unitary representation of \GWH\ just by a ``phase
factor'':
\[ W_\mlam(q,p,s)\equiv W_\mlam(x)\dti e^{i\mlam s}, \]
hence the $Ad^*(\cdot)$--action of both representations on \fk T(\H)\ is
identical -- it depends on elements of the factorgroup
$\mGWH/\mbR=\mbR^{2n}$\ only.

Let $X_j:=Q_j,\ X_{j+n}:=P_j,\
j=1,2,\dots n,\ X_0:=\frac{1}{\mlam}I$, cf.\eqref{eq;GWH-gener}, be
selfadjoint generators of
$W_\mlam$. Let $S^T=-S=S^{-1}$ be the $2n\times 2n$--symplectic matrix with
elements
\[S_{j\ j+n}=-S_{j+n\ j}:=1,\ \  j=1,2,\dots n;\ \ S_{j k}=0\
\text{otherwise}. \]
For selfadjoint operators $X,Y$\ on \H, let $[X,Y]\equiv XY-YX$\ denote the
commutator on its domain, and $[X,Y]=i\,Z$, for selfadjoint $X,Y,Z$\ on \H\
will mean equality of operators restricted to their common domain.\hfill\pika
\end{notat}
The operators $X_j, j=0,1,2,\dots 2n $\ satisfy the (Heisenberg form of)
the \emm canonical commutation relations~ (CCR) on a common dense domain:
\begin{subequations}\label{eq;H-CCR}
\bequ
 [X_j,X_k] = i\,S_{jk}X_0,\ \ j,k\neq 0,
\end{equation}
 \bequ
 [X_j,X_0] = 0,\ \ j=1,2,\dots 2n.
\end{equation}
\end{subequations}
The \emm Weyl form of CCR~ is
\bequ\label{eq;Weyl}
W_\mlam(x+x')=\exp\left(\frac{i\mlam}{2}x\!\cdot\!S\!\cdot\!x'\right)
W_\mlam(x)W_\mlam(x'),
\end{equation}
and the unitary operators $W_\mlam$\ are expressed by
\bequ\label{eq;W-op}
W_\mlam(x)\equiv\exp\left(i\,\mlam X\dti S\dti
x\right):=\exp\left(i\,\mlam \sum_{j,k=1}^{2n}X_jS_{jk}x_k\right).
\end{equation}
The following useful relation is then valid:
\bequ\label{eq;W-shift}
W_\mlam(x)^{-1}X_jW_\mlam(x)\equiv X_j+x_jI,\quad\forall
j\in\{1,2,\dots2n\},\ x_j\in\mbR.
\end{equation}
\begin{notat}\label{notat;W(x)}
Let us write
$W(x)\equiv W_{1/\hbar}(x)$. Let us define $\mcl
V_\nu:=\{Ad^*(W(x))\nu:x\in\mbR^{2n}\}=\mOGn$\ with $G:=\mGWH$, and
$\mcl V:=\{\mcl
V_\nu:\nu\in\mbcDFr\}$, with \bF\ given by $W_{1/\hbar}(\mGWH)$,
cf.\dref2.17~
. Let $X(x):= X\dti S\dti x$\ be a selfadjoint
generator of the (projective) representation $W(\mbR^{2n})$, i.e.\ the
generator of the one--parameter unitary group $t\mapsto\exp(-itX(x))$.
The densely defined Hamiltonian function generating the corresponding flow
on the Poisson manifold \fTs\ is \h{X(x)}. Let us denote also
\glss $x\dti\nu:=W(x)\nu W(x)^* \equiv Ad^*(W(x))\nu$~.
\hfill\pika
\end{notat}
\begin{prop}\label{prop;GWH-orbits}
(i) With the notation from~\ref{notat;W(x)}, the (densely defined) function
 \h{X(x)}\
has \bcDFr--generalized differential, which is \cl V--integrable.
\item{(ii)} The orbits $\mcl V_\nu$ are embedded submanifolds of \fTs, each
diffeomorphic to the ``flat phase space'' $\mbR^{2n}$.
\item{(iii)} The restrictions of the symplectic forms $\Omega_\nu$\
introduced on
\OUn\ to the orbits $\mcl V_\nu$ are nondegenerate, and the restrictions
of the Momentum mapping \bF\ to these orbits are symplectomorphisms onto the
coadjoint orbit of \GWH\ ``corresponding'' to the choice of
$s_0=-\mlam:=-\frac{1}{\hbar}$, cf.\eqref{eq;GWH-orb}. \hfill\zal
\end{prop}
\begin{proof}
(i) The proof of integrability trivially follows from Lemma~\ref{lem;2.16}\
and Proposition~\ref{prop;lin-gen}
, since the integral
curves of the Hamiltonian vector field corresponding to $dh_{X(x)}$\ leave
all $\mcl V_\nu,\ (\nu\in\mbcDFr)$\ invariant.
\item{(ii)} The second assertion follows from the Lemma~\ref{lem;2.16}, and
from its proof. A more intuitive argument is seen from the
Momentum mapping \bF\ restricted to any $\mcl V_\nu$\ with a help
of~\eqref{eq;W-shift}:\nl
 In view of~\eqref{eq;2.28}, the $j$--th
component of $\mbF(x\dti\nu)$ can be expressed as:
\bequ\label{eq;GWH-orb}
 F_{x\cdot\nu}(\xi_j)=Tr\bigl(\nu W(x)^*X(\xi_j)W(x)\bigr)=
 \begin{cases}
F_\nu(\xi_j)-\mlam q_j,&\text{for}\ j=1,2,\dots n,\\
F_\nu(\xi_j)+\mlam p_{j-n}, &\text{for}\ j=n+1,\dots2n.
\end{cases}
\end{equation}
what proves bijection of $\mcl V_\nu$\ onto $\mbR^{2n}$.
\item{(iii)}
The (densely defined) vector fields $\mv j:=\mv{\xi_j}$\ corresponding to the
basis $\{\xi_j,\ j=0,1,\dots 2n\}$ of \GWH\ form a basis of
$\mTr(\mcl V_\nu)$\ for any $\mrh\in\mcl V_\nu$\
for all $\mcl V_\nu$. These vector fields are proportional to the Hamiltonian
vector fields
corresponding to \h{X_j}\ for the selfadjoint generators $X_j$ of the
representation $W_{1/\hbar}(\mGWH)$. The vector field $\mv 0$\ is
identical zero. According to~\eqref{eq;2.15d},~\eqref{eq;2.24},
and~\eqref{eq;qqq}, one has
\begin{equation}\label{eq;restr-CCR}
 \Omega_\mrh(\mv\xi,\mv\eta)=i\,Tr\bigl(\mrh[X(\xi),X(\eta)]\bigr),
\end{equation}
resp.\ from~\eqref{eq;2.33} one has
\[\mbF^*\{\mh\xi,\mh\eta\}=\{\mbF^*\mh\xi,\mbF^*\mh\eta\}. \]
The Kirillov--Kostant symplectic form on an $Ad^*(\mGWH)$\ orbits through
$F$ has the form~\eqref{eq;2.31}:
\[ \Omega^K_F(\mbf v_{\xi},\mbf v_{\eta}) = -F([\xi,\eta]).\]
From the CCR~\eqref{eq;H-CCR} one has
\barr\label{eq;Om-PH}
 \Omega_\mrh(\mv j,\mv{j+n})=-\Omega_\mrh(\mv{j+n},\mv j) &=&
 i\,Tr\bigl(\mrh[X(\xi_j),X(\xi_{j+n})]\bigr) \\
 &=&-i\,\mlam^2
 Tr\bigl(\mrh[Q_j,P_j]\bigr)\\ &=&-Tr(\mrh X(\xi_0))=-s_0=\mlam,
\earr
and for the remaining indices $j,k:\ \Omega_\mrh(\mv j,\mv{k})=0$.
For the Kirillov--Kostant form we have
\bequ\label{eq;Om-K}
 \Omega^K_F(\mbf v_{j},\mbf v_{j+n}) = -F([\xi_j,\xi_{j+n}])=-F(\xi_0),
 \end{equation}
 what corresponds to~\eqref{eq;Om-PH} in accordance with the
 equation~\eqref{eq;2.d33}; this proves the symplectomorphism property of \bF.
The commutation relations~\eqref{eq;H-CCR} show nondegeneracy of the
restricted form \glss $\Omega^{\mcl V}_\mrh$~\ :=\
$\iota^*_{\mcl V}\Omega_\mrh$\
for all relevant \rh.

 It remains to prove that \bF\ maps all $\mcl
V_\nu$\ onto a unique orbit. The basis $\{\xi_j,\ j=0,1,\dots2n\}\subset
Lie(\mGWH)$\ determines global coordinates on the dual $Lie(\mGWH)^*$,\
$F(\xi_j)$\ being the coordinates of $F\in Lie(\mGWH)^*$\ in the dual basis.
It is clear from~\eqref{eq;GWH-orb}\  that on any $\mcl V_\nu$\ there is a
point \glss $\mrh_0$~\label{rho-0} such that $\mbF(\mrh_0)(\xi_j)=0,\
j=1,2,\dots2n$. The
coordinates of other points on those orbits are then
$\mbF(x\dti\mrh_0)(\xi_j)=(-1)^{\left[\frac{n+1+j}{n+1}\right]}\mlam x_j,\
j=1,2,\dots2n$, and the remaining
coordinate $\mbF(x\dti\mrh_0)(\xi_0)\equiv Tr(\nu
X(\xi_0))$ $=-\mlam=-\hbar^{-1}$,~\eqref{eq;GWH-gener}, hence it is constant on
the orbit and of equal value on all orbits, i.e.\ on
all the images $\mbF(\mcl V_\nu)\subset Lie(\mGWH)^*$. This
proves the last statement.
\end{proof}
\begin{rem}\label{rem;h-bar}
These considerations showed that the choice of a specific value of Planck
constant in QM corresponds mathematically to the choice of the coadjoint
orbit of
\GWH\ labelled by $s_0=-\mlam=-\frac{1}{\hbar}$\ determining a unitary
representation of this group and, in this way, also determining Heisenberg
uncertainty
relations and many physical effects connected with them. Since unitary
equivalent representations lead, as a rule, to indistinguishable physics,
validity of the mathematical theorem about unitary inequivalence of
representations~\eqref{eq;GWH-rep1}, or~\eqref{eq;GWH-rep2}, for different real
values of \lam, ``can be seen'' also from the known physical measurability
of the Planck constant $\hbar$.
\hfill\dovi
\end{rem}




\subsection{Restricted flows with linear generators on \protect{$\mcl
O_\mrh(\mGWH)$}}\label{IIIA1;restr}

Let $X$ be a selfadjoint operator on a Hilbert space \H, and let $U(G)$\ be a
continuous unitary representation of a connected Lie group $G$. Assume that
the orbit of $Ad^*(U(G))$\ through $\mrh\in\mbcDFr$, cf.\
Definition~\ref{df;2.17}, belongs to the domain of $X$,
$\mOGr\subset\mDdXd$. {\em Let us assume further in this subsection
 that the function $\rrh_X$\ is constant on
the submanifolds $Ad^*\bigl(U(G_{\mbF(\nu)})\bigr)\nu$\ for all
$\nu\in\mOGr$}, i.e.\  that it is a $\mrh G$--classical generator,
cf.\dref2.19~(iv). This would be trivially the case, if the momentum
mapping \bF\
is injective on the orbit \OGr, i.e.\ if the orbit \OGr\ is diffeomorphic
to the coadjoint orbit $Ad^*(G)\mbF(\mrh)$. For further examples of \rGcl
generators cf.\ref{exmp;nGcl}.
We can now define the corresponding classical Hamiltonian $h_X^\mrh$\
 \bequ\label{eq;hXcl}
 h_X^\mrh: Ad^*(G)\mbF(\mrh)\rarw \mbR,\ h_X^\mrh(F_\nu):=\rrh_X(\nu)
\equiv Tr(\nu
X), \forall \nu\in\mOGr
\end{equation}
what is an infinitely differentiable function on the coadjoint orbit through
$F_\mrh$. The function \glss $h_X^\mrh$~\ will be also called \emm the
(classical) Hamiltonian induced by $(X;U(G))$~\ on the orbit
$Ad^*(G)\mbF(\mrh)$.
The restriction of $\rrh_X$:
\bequ\label{eq;hX}
 \rrh_X^\mrh: \mOGr\rarw \mbR,\ \nu\mapsto \rrh_X^\mrh(\nu):=Tr(\nu X)
 \end{equation}
generates the restricted flow of $X$, to the orbit \OGr.

Let us choose now $G:=\mGWH$, and $X:=H$, with
\bequ\label{eq;H}
H:=\frac{1}{2}\sum_{j=1}^n\frac{1}{m_j}P_j^2+V(Q_1,Q_2,\dots Q_n),
\end{equation}
(cf.\ref{notat;CCR}) with some ``convenient'' real function
$V:\mbR^n\rarw\mbR$.
The ``correct quantum evolution'' given by $z(\in\mH)\mapsto
z_t:=\exp(-itH)z$\ (we set here $\mlam=\hbar=1$) leads to the \emm
Ehrenfest's relations~ for expectations $\lb X_j\rb_t :=\lb z_t|X_j|z_t\rb,\
(j=1,2,\dots,n)$:
\begin{subequations}\label{eq;0Ehren}
\bequ\label{eq;1Ehren}
\frac{\rd}{\rd t}\lb Q_j\rb_t=\frac{1}{m_j}\lb P_j\rb_t,
\end{equation}
\bequ\label{eq;2Ehren}
\frac{\rd}{\rd t}\lb P_j\rb_t=-\lb\partial_jV(Q_1,\dots,Q_n)\rb_t.
\end{equation}
\end{subequations}
These relations {\em are not differential equations} for the functions
$t\mapsto \lb X_j\rb_t$, if the potential energy $V$ is not at most
quadratic polynomial in $Q$'s, cf.\ the text following Eq.~\rref A~.
In the case of quadratic $H:=A$\ from~\rref A~, the Hamiltonian evolutions
given by the Hamiltonians\rref hHcl~ lead to the results identical with
those of QM (hence satisfying also\rref0Ehren~\ with $x_j(t)\equiv\lb
X_j\rb_t$).

Let us express $h_H^\mrh$\ corresponding to $H$\ from~\eqref{eq;H} according
to~\eqref{eq;hXcl}. Let $\mrh:=\mrh_0$\ (cf.\ the text on page~\pageref{rho-0}
following~\eqref{eq;Om-K}) with
$Tr(\mrh X_j)=0,\ \forall j=1,2,\dots 2n$, with the notation of
Subsection~\ref{IIIA1;CCR}. We write $(q,p)$\ instead of
$\mbF(\nu)$\ with components $\mp\mlam x_j$, cf.\rref GWH-orb~:
\bequ\label{eq;hHcl}
h_H^\mrh(q,p)\equiv \frac{1}{2}\sum_{j=1}^n \frac{1}{m_j}p_j^2
+V_\mrh(q_1,q_2,\dots q_n),
\end{equation}
with
\[\glss V_\mrh(q)~:=Tr\bigl(\mrh V(Q+q)\bigr)+\sum_j\frac{1}{2m_j}Tr\bigl(\mrh
P_j^2\bigr).\]
The last sum in this expression is a constant term on the orbit \OGr, hence
the flow generated by $\rrh_H^\mrh$\ on \OGr\ is independent of this
constant. This flow (the restricted ``linear'' flow of $H$) is projected
by the momentum mapping \bF\ onto the flow on the coadjoint
orbit of \GWH\ with $s_0:=s_0(\mbF(\mrh))=-\frac{1}{\hbar}$ generated by the
Hamiltonian $h_H^\mrh(q,p)$\ via the standard symplectic form
$\rd p\wedge\rd q$.

\begin{exm}\label{exmp;T+V}
Let us take, e.g.\ $\mrh:=P_z$\ with $0\neq z\in\mLqn:\lb
z|X_j|z\rb=0,\forall j$.
Let \glss$\tilde z(q):=z(-q)$~. Then
\[ V_\mrh(q)=\text{const.}+V_z(q),\quad {\rm with}\ V_z(q):=|\tilde z|^2*V(q),\]
the symbol $a*b(q):=\int a(q-q')b(q')\rd^nq'$\ denoting convolution of two
complex--valued functions $a,\ b$\ on $\mbR^n$.
Let, e.g., $n=3$, and $V(q):=\frac{\malp}{|q|}$\ be the Coulomb potential.
Let the above $z\in L^2(\mbR^3)$\ be rotationally (i.e.\ $O(3)$)\ symmetric
normalized function with support ``concentrated'' near $q=0$.
Then $q\mapsto V_z(q)$ is again, for large values of $|q|$, approximately
(resp.\ exactly, for compact support of $z$)
of the Coulomb form.\hfill\dovi
\end{exm}
We see that the \rh\GWH--restrictions of the flow \pph{},H~ are identical
to the flows of classical Hamiltonian mechanics on $\mbR^{2n}$\ with the
Hamiltonian function $h_H^{\mrh}$\ from~\eqref{eq;hHcl} differing from the
usually considered ``classical limit'' of the quantum flow \pph{},H~ by the
``\rh--smearing'' of the potential V only.

 A specific interesting choice of $V$\ in~\eqref{eq;H} is a
quadratic function, describing, e.g.\ harmonic oscillators. This case can be
generalized to any quadratic operator $X:=A$:
\bequ\label{eq;A}
A:=\frac{1}{2}\sum_{j,k=1}^{2n}a_{jk}X_jX_k,
\end{equation}
with real constants $a_{jk}\equiv a_{kj}$, and with $X_j,\ j=1,2,\dots 2n$\
defined in Notation~\ref{notat;CCR}. This case is specific in that the
operator $A$\ is essentially selfadjoint on a common domain of all $X_j$'s,
and it generates, together with the $X_j$'s, a unitary representation of a
$2n+2$--dimensional Lie group containing \GWH\ as a subgroup.
This follows from the following considerations.

It is clear that the operators
$\{A;X_j,\ j=0,1,\dots 2n\}$\ form a basis of a Lie algebra of (unbounded)
operators in \H=\Lqn.
It is also easily seen that an arbitrary number of quadratic symmetric
operators of the form~\eqref{eq;A} together with all $X_j$'s can be
included into a {\em finite dimensional} Lie algebra $X(\mfk g)$\ of operators
(\wrt the operator commutation $i\,[A,B]$) containing operators at most
quadratic in
$X_j$'s. Less trivial is the assertion, that these Lie algebras of
operators are composed of essentially selfadjoint operators on the domain
$D^\mome(\mGWH)$\ of essential selfadjointness of all $X_j$'s, and
{\em that they are integrable into continuous unitary representations of some
Lie groups}. The maximal Lie algebra obtained in this way is (isomorphic to)
the algebra called
$\mfk{st}(n,\mbR)$\ (see also~\cite[\S 15.3, and \S 18.4]{kiril}).
Let us formulate and prove the mentioned facts for the
algebra $\mfk{st}(n,\mbR)$:
\begin{prop}\label{prop;Stn}
Let $X(\mfk{st}(n,\mbR))$ denote the above mentioned ``maximal'' Lie
algebra of ``at most quadratic'' symmetric operators acting on the
Hilbert space \H\
of representation $W_\mlam(\mGWH)$. Let $\tilde{St}(n,\mbR)$\ be
the corresponding connected, simply connected Lie group with the Lie
algebra $\mfk{st}(n,\mbR)$. Then the representation $W_\mlam(\mGWH)$ has a
unique extension to the continuous unitary representation
$\tilde W_\mlam(\tilde{St}(n,\mbR))$\
in \H\ such that all the closures of the operators from
$X(\mfk{st}(n,\mbR))$ are exactly all of its selfadjoint generators.\hfill\zal
\end{prop}
\begin{proof}
The analytic domain $D^\mome(\mGWH)$\ is a common invariant domain for all
operators from $X(\mfk{st}(n,\mbR))$. According to Nelson's
theorem (cf.\ \cite[Theorem 11.5.2]{bar&racz}) it suffices to prove
essential selfadjointness of the operator $\Delta$, what is sum of squares
of a basis of $X(\mfk{st}(n,\mbR))$. We chose here the basis consisting of the
generators $X_j$\ of $W_\mlam(\mGWH)$, and of all their
symmetrized products $\frac{1}{2}(X_jX_k+X_kX_j)$. Then
\begin{eqnarray*}
\Delta&:=&\sum_{j=1}^{2n}X_j^2+\frac{1}{4}\sum_{j,k=1}^{2n}(X_jX_k+X_kX_j)^2\\
      & =&\frac{3}{2}nI+\sum_{j=1}^n(P_j^2+Q_j^2)(I+\sum_{k=1}^n(P_k^2+Q_k^2)),
\end{eqnarray*}
where we used notation from~\ref{notat;CCR}, and the CCR~\eqref{eq;H-CCR}.
From the known properties of the Hamiltonians $P_j^2+Q_j^2$\ of independent
linear oscillators, we conclude (with a help of, e.g.~\cite[Theorem
VIII.33]{R&S} on operators on tensor products of Hilbert spaces) that
$\Delta$\ is essentially selfadjoint. The Nelson's
theorem states now integrability of $X(\mfk{st}(n,\mbR))$ onto a unitary
representation. Selfadjointness and uniqueness now easily follow.
\end{proof}
Hence, also any Lie subalgebra of $X(\mfk{st}(n,\mbR))$ integrates onto a
continuous group representation. Let us denote by
$A\mGWH$\ the simply connected Lie
group represented by this unitary representation with the basis of
generators $\{A;\ X_j,\ j=0,1,2,\dots 2n\}$, with $A$ from \rref A~.
The $2n+2$--dimensional group $A\mGWH$\ contains \GWH\ as its
normal subgroup.

In the ``quadratic case''~\eqref{eq;A} the expression~\eqref{eq;hHcl} has
the form
\bequ\label{eq;hAcl}
h_A^\mrh(x)\equiv \frac{1}{2}\sum_{j,k=1}^{2n}a_{jk}x_jx_k + const.,
\end{equation}
with the $const.$ depending on the choice of orbit only (we always assume
$\mrh:=\mrh_0$, according to the definition of $\mrh_0$ in the
notes on page~\pageref{rho-0}). This is
valid regardless the orbit $\mcl O_\mrh(A\mGWH)$ is $2n$--, or
$2n+1$--dimensional. Hence for the (at most) quadratic Hamiltonian $A$, the
projected quantal evolutions $\mpph t,A~(\nu)$ to the orbits and the
``corresponding'' classical
evolution ``coincide'' in the sense that the $2n$\ coordinates $Tr(\mpph
t,A~(\nu)X_j),\ j=1,\dots 2n$\ (of the possible total $2n+1$) satisfy
classical equations with the Hamiltonian $h_A^\mrh$\ from~\eqref{eq;hAcl}
corresponding to the canonical symplectic form $\rd p\wedge\rd q$.

\begin{rem}\label{rem;dimOz}
Let us stay on \PH, and let $\mrh_0=P_z$. Let $A$ be quadratic as in\rref A~.
Then a general assertion tells us
that the dimension of the $A\mGWH$--orbit through $P_z$ is $2n$\ iff it
contains an eigenstate of $A$,~\cite{bon8}. In that case, each point
$W(x)P_zW(-x)$\ of the orbit is (i.e.\ represents) an eigenstate of some
selfadjoint operator
of the form $\sum_jc_jX_j+A$. Other orbits are $2n+1$--dimensional.
This assertion is a consequence of the fact that the dimension of any
connected finite dimensional manifold is constant in all of its points and
equals to the dimension of the tangent spaces which are in turn generated by
vector fields corresponding to the flows \pph t,X~\
($X=A,X_1,\dots,X_{2n}$).
These $2n+1$ vectors in any point of the orbit are mutually linearly
independent except in a stationary point $P_z$\ where $\mpph t,X~(P_z)\equiv
P_z$, for some $X:=\sum_jc_jX_j+A$, i.e.\ $z\in \mH\ (z\neq 0)$\ is an
eigenvector of $X$\ (the linear
independence of the $2n$\ vectors determined by the $X_j$'s is a consequence
of CCR). \hfill\dovi
\end{rem}


\subsection{Time dependent Hartree--Fock theory}\label{IIIA1;H-F}

We shall consider here the ``approximation to QM'' which is very well known in
nonrelativistic quantummechanical many--particle theory as
the {\em Hartree--Fock theory}\index{Hartree--Fock theory}. It consists,
expressed briefly in our terminology, in the restriction of a given (linear)
QM problem to a
manifold (a G-orbit) of quantum states $\Psi$, and, in its stationary
setting,
in looking for the points $\Psi_0$\ of the manifold that minimize the
expectation value
$\lb\Psi|H|\Psi\rb$\ of a given Hamiltonian $H$. In the Hartree--Fock theory
of systems consisting of $N$ interacting fermions in an external potential
the manifold consists of all ``Slater determinants'' for the considered
$N$ fermions. The point $\Psi_0$\ then satisfies the Hartree--Fock
equation~\eqref{eq;HF-eq}, what is a condition for the zero value of the
 derivative
$\rq_{P_{\Psi}}(\mD{\mh H},P_{\Psi}~)$\ of the corresponding restricted
generator. It is assumed that in many interesting cases stationary points
$\Psi_0$\ (resp., more correctly: $P_{\Psi_0}$) of the orbit approximate
the ground state (states) of the unrestricted system. This theory is
expressible in terms of ``one--particle states'', due to a natural
bijection (see e.g., also~\cite{rowe1}) between the set of all Slater
determinants and an orbit in the
one--particle state space $\mSs(\mLH)$\ of unitary group $\mfk U:=\lb$
all unitary operators on the one--particle Hilbert space $\mH\rb$.

Let us consider a system of $N$\ identical fermions described in the
Hilbert space $\mH_N:=\otimes^N\mH$, where $\mH:=\mH_1$\ is ``the one--particle
Hilbert space''. The vectors in $\mH_N$\ are expressed by linear
combinations of ``product--vectors'' $\Phi:=
\phi_1\otimes\phi_2\otimes\dots\otimes\phi_N,\ \phi_k\in\mH$,
and the scalar product linear in the second factor is determined by
\[ \lb\Phi'|\Phi\rb:=\prod_{k=1}^{N}(\phi'_k,\phi_k).\]
Let $\psi_j\in\mH, j\in\mbN$\ be an orthonormal basis in
\H. Then an orthonormal basis in $\mH_N$\ consists of vectors labelled by
ordered $N$--tuples $(j):=(j_1;j_2;\dots j_N)\in\mbN^N$:
\bequ\label{eq;N-base}
\Psi_{(j)}\equiv|(j_1;j_2;\dots j_N)\rb:=\psi_{j_1}\otimes\psi_{j_2}\otimes
\dots\otimes \psi_{j_N},\ \forall j_1,j_2,\dots j_N\in\mbN.
\end{equation}
Let $\Sigma(N)$\ be the permutation group (=``symmetric group'') of $N$\
elements, and, for any $\sigma\in\Sigma(N)$, let its action on ordered
$N$--tuples of integers be denoted by
\[ \sigma(1;2;\dots N)\equiv(\sigma(1);\sigma(2);\dots\sigma(N)).\]
Then a unitary representation $\sigma\mapsto p_\sigma$\ of $\Sigma(N)$\ in
$\mH_N$\ is determined by
\[p_\sigma\Psi_{(j)}\equiv\Psi_{\sigma(j)}:=|(j_{\sigma(1)};j_{\sigma(2)};
\dots j_{\sigma(N)})\rb.\]
Let the \emm Fermionic subspace~ $\mH^F_N$\ of $\mH_N$\ consists of all
 vectors $\Psi\in\mH_N$\ satisfying
 \[ p_\sigma\Psi=\mveps_\msg\Psi,\quad
 \mveps_\msg\mveps_{\msg'}\equiv\mveps_{\msg\dti\msg'}\in\{-1;1\},
 \forall\msg,\msg'\in\Sigma(N),\]
 with $\mveps_{\msg}:=-1$\ for \sg\ corresponding to a mere interchange of
 two elements.
 The orthogonal projection $P^F_N$\ onto $\mH^F_N$ is the maximal of all
 projections $P$\ satisfying: $p_\msg P\equiv\mveps_\msg P$. It can be
 expressed:
 \[ P^F_N=\frac{1}{N!}\sum_{\msg}\mveps_\msg p_\msg.\]
 An orthonormal basis in $\mH^F_N$\ is given by {\em ``Slater
 determinants''} $\Psi_{\{j\}}$\ labelled by  the ordered $N$--tuples
 $(j):=(j_1;j_2;\dots j_N)$\ with $j_1<j_2<\dots<j_N$, and defined by
 \bequ\label{eq;slater}
 \Psi_{\{j\}}:=\frac{1}{\sqrt{N!}}\sum_\msg\mveps_\msg\Psi_{\msg(j)}.
 \end{equation}
Let \fk U\ be the unitary group of \LH, and let its unitary representation in
$\mH_N$\ be given by its action on the product vectors
\barr\label{eq;unit-N}
 \ru\mapsto\ru^{\otimes N},\quad \ru^{\otimes N}\Phi&\equiv&
\ru^{\otimes N}\left(\phi_1\otimes\phi_2\otimes\dots\otimes\phi_N\right)\\
&:=&\ru\phi_1\otimes\ru\phi_2\otimes\dots\otimes\ru\phi_N,\ \forall\ru\in\mfk U.
\earr
One orbit of this representation in $\mH^F_N$\ consists of all the Slater
determinants, and its canonical projection to the projective Hilbert space
$P(\mcl H_N)$\ is homeomorphic, cf.\ also~\eqref{eq;1-N}, to the coadjoint
orbit of \fk U\ in $\mcl T_{1+}(\mH)=:\mSs$\ (cf.\ page~\pageref{eq;2.1})
consisting of all the ``$N$--dimensional'' density
matrices $\ru\mrh_{\{j\}}\ru^{-1}$\ with maximally degenerate spectrum, cf.
also~\cite{rowe1}. These density
matrices are obtained as ``partial traces'' from Slater determinants by
restriction to ``one particle observables'', i.e.\ for all $\ra\in\mLH$ one
has
\bequ\label{eq;r-slater}
 Tr(\mrh_{\{j\}}\dti\ra):=\lb\Psi_{\{j\}}|\ra\otimes
 \mbI_{\mH}^{\otimes(N-1)}|\Psi_{\{j\}}\rb,
\end{equation}
and the resulting density matrix has an explicit expression of the form
\bequ\label{eq;dens-slater}
\mrh_{\{j\}}=\frac{1}{N}\sum_{k=1}^N |\psi_{j_k}\rb\lb\psi_{j_k}|=:
\frac{1}{N}E_{\{j\}},
\end{equation}
where the projector $E_{\{j\}}$\ onto the subspace of \H\ spanned by the
$N$\ orthonormal vectors $\{\psi_k:k\in(j)=(j_1;j_2;\dots j_N)\}$ was
introduced. Conversely, as can be proved by elementary techniques, any
$N$--dimensional subspace of \H\ determines a Slater
determinant (uniquely up to a numerical factor),
namely the space spanned by $N$\ one--particle orthonormal vectors $\psi_j$\
determines the Slater determinant constructed by the same vectors,
cf.\eqref{eq;slater},~\eqref{eq;N-base}, and different orthonormal bases
of that subspace give (up to a phase factor) the same Slater determinant.
Hence the mapping
\bequ\label{eq;1-N}
\mbG:\quad \mrh:=\mrh_{\{j\}}(\in Ad^*(\mfk U)\mrh_{\{j'\}})\mapsto\mbG(\mrh):=
P_{\Psi_{\{j\}}},\ (\Psi_{\{j\}}\in\{\ru^{\otimes N}\Psi_{\{j'\}}:\ru\in\mfk
U\})
\end{equation}
is a bijection (here $\Psi_{\{j'\}}$\ is an arbitrary Slater determinant).

Reduced one particle density matrices corresponding to arbitrary state-vectors
$\Phi\in\mH^F_N$\ cannot be expressed in such a simple way.

Let $H:=H_N$\ be a selfadjoint Hamiltonian on $\mH_N$, and let us assume
that it is permutation symmetric, i.e.\ that
\[ p_\msg H_N\equiv H_N p_\msg,\quad\forall\msg\in\Sigma(N).\]
The corresponding generator, as a function on (a dense subset of)
$P(\mH_N)$, is $\mh H^N(P_\Phi):=Tr(P_\Phi\dti H_N)$. Its
restriction to the \fk U--orbit of Slater determinants is
\bequ\label{eq;hH(r)} \tilde{\mh
H}^{Sl}(P_{\Psi_{\{j\}}}):=\lb\Psi_{\{j\}}|H_N|\Psi_{\{j\}}\rb=
\sum_\msg\mveps_\msg\lb
p_\msg\Psi_{(j)}|H_N|\Psi_{(j)}\rb=N!\lb\Psi_{(j)}|P_N^FH_N|\Psi_{(j)}\rb.
\end{equation}
With a help of the bijection \bG\ from~\eqref{eq;1-N}, we can
write the restricted function $\tilde{\mh H}^{Sl}$\ as a function
on the orbit $Ad^*(\mfk U)\mrh_{\{j\}}\subset\mSs$, i.e.\ as a
generator on one--particle states. We shall write the
corresponding ``one--particle energy'' as $\mh H^{Sl}:=
N^{-1}\tilde{\mh H}^{Sl}$: \bequ N\dti\mh
H^{Sl}(\mrh_{\{j\}}):=\lb\Psi_{\{j\}}|H_N|\Psi_{\{j\}}\rb=
Tr\bigl(\mbG(\mrh_{\{j\}})H_N\bigr).
\end{equation}
This relation can be made more explicit for specific choices of $H_N$.

Let us take for $H_N$\ a nonrelativistic spin--independent Hamiltonian of
$N$ point particles with symmetric pair potential interaction, i.e.
\bequ\label{eq;2-part}
H_N:= \sum_{j=1}^N{\rm h}_{0j}+\frac{1}{2}\sum_{j\neq k}\rv_{jk},
\end{equation}
where the indices $j,k=1,\dots N$ specify ``one--, resp.\ two--particle
factors''
\H\ in the tensor product $\mH_N$\ on which the corresponding operators
 act; the ${\rm h}_{0j}$'s are
copies of the same one particle Hamiltonian $\mrho$\ (kinetic energy plus
external
fields) ``acting at the $j$--th factor'' in the tensor product $\mH_N$,
and $\rv_{jk}\equiv \rv_{kj}$\ are also copies of a two--particle operator
$\rv\in\mcl L(\mH\otimes\mH)$ ``acting on the $j$--th and $k$--th factor''
in $\mH_N$.
Let us introduce the linear \emm exchange operator~ \glss$\mplr$~\ on the
two--particle spaces (commuting with \rv) by
\[ \mplr(\phi\otimes\psi):=\psi\otimes\phi,\ \forall\phi,\psi\in\mH. \]
We can calculate now~\eqref{eq;hH(r)} with $H:=H_N$\ from~\eqref{eq;2-part}:
\barr\label{eq;H-F-hamilt}
N\dti\mh H^{Sl}(\mrh_{\{j\}})&:=&\lb\Psi_{\{j\}}|H_N|\Psi_{\{j\}}\rb =
N!\lb\Psi_{(j)}|P^F_NH_N|\Psi_{(j)}\rb\nonumber \\
&\equiv &\sum_\msg\mveps_\msg
\lb\psi_{\msg(j_1)}\otimes\dots\otimes
\psi_{\msg(j_N)}|\left[\sum_{k=1}^N{\rm h}_{0k}+
\frac{1}{2}\sum_{k\neq
l}\rv_{kl}\right]|\psi_{j_1}\otimes\dots\otimes\psi_{j_N}\rb\nonumber \\
&=&\sum_{k=1}^N(\psi_{j_k},\mrho\psi_{j_k})+\frac{1}{2}\sum_{k\neq l}
(\psi_{j_k}\otimes\psi_{j_l}-\psi_{j_l}\otimes\psi_{j_k}|\rv_{kl}|
\psi_{j_k}\otimes\psi_{j_l})\nonumber \\
&=&N\dti Tr(\mrh_{\{j\}}\mrho)+\frac{1}{2}\sum_{k\neq l}
(\psi_{j_k}\otimes\psi_{j_l}|\rv\dti(\mbI_{\mH}\otimes\mbI_{\mH}-\mplr)|
\psi_{j_k}\otimes\psi_{j_l}) \nonumber \\
&=&N\dti Tr(\mrh_{\{j\}}\mrho)
+\frac{N^2}{2}Tr(\mrh_{\{j\}}\otimes\mrh_{\{j\}}\dti
\rv\dti(\mbI_{\mH}\otimes\mbI_{\mH}-\mplr)).
\earr
This function $\mh H^{Sl}$\ can be used as a generator of the (quantum
nonlinear) motion on the \fk U--orbit of $N$--dimensional projections
(which, divided by $N$, are density matrices of the domain of
\bG,~\eqref{eq;1-N}) in the one--particle state space \Ss.
The resulting dynamical equation describes the \emm time dependent
Hartree--Fock theory~, cf.\ \cite{rowe1,rowe2,kra&sar}, and the equations
describing its stationary points are just the \emm Hartree--Fock
equations~, cf.\ \cite{hartree,march,lieb2}. Let us show how it looks in our
formalism.

We shall need an expression for the derivative \D{\mhSl},\mrh~\ to be able
to write a dynamical equation for \rh, e.g.\ the ``Schr\"odinger
equation''~\eqref{eq;2.11}, resp.~\eqref{eq;2.14}.\footnote{Our
considerations will be a little ``heuristic'' from mathematical point of
view in this Subsection: we shall not consider here the domain
questions, hence we shall not be able to
write equations as~\eqref{eq;nlinSch} with precisely defined
generalized differentials.}
The differential will be calculated according to~\eqref{eq;2.4}, and with a
help of~\eqref{eq;2.rep}. We shall need, however, derivatives along the
curves $t\mapsto\exp(-itb)\mrh\exp(itb)$, corresponding to tangent vectors
$i[\mrh,b]$, cf.\ Notes~\ref{note;2.3d}:
\bequ\label{eq;DhSl}
\mD{\mhSl},\mrh~(i[\mrh,b])=i\,Tr(\mrho[\mrh,b])+i\frac{N}{2}Tr\left(
[\mrh,b]\otimes\mrh\dti (\mbI_{\mH}^{\otimes2}-\mplr)\rv+\mrh\otimes[\mrh,b]
\dti(\mbI_{\mH}^{\otimes2}-\mplr)\rv\right).
\end{equation}

This equation can be rewritten by inserting the unit operators
$\mbI_{\mH}$\ expressed with a help of convenient complete systems
$\{\mphi_k\}\subset\mH$\  of
orthonormal vectors into several places in between of the multiplied operators
in the above formula. E.g., from the
trace in $\mLH\otimes\mLH$\ of the product $(A\otimes\mrh)\dti B$\
($B\in\mLH\otimes\mLH,\forall A\in\mLH$)\ one can find an operator
$D\in\mLH$\ defined by
\[ Tr(A\dti D):=Tr(A\otimes\mrh\dti B) \]
as follows:
\bequ
\begin{split}
Tr(A\dti
D)&:=\sum_{j,k}\sum_{l,m}(\mphi_j\otimes\mphi_k|A\otimes\mrh|\mphi_l\otimes
\mphi_m)(\mphi_l\otimes\mphi_m|B|\mphi_j\otimes\mphi_k) \\
=&\sum_{j,l}(\mphi_j|A|\mphi_l)\underbrace{\sum_{k,m}(\mphi_k|\mrh|\mphi_m)
(\mphi_l\otimes\mphi_m|B|\mphi_j\otimes\mphi_k)}_{(\mphi_l|D|\mphi_j)}.
\end{split}
\end{equation}
Since
\[ (\mphi_m\otimes\mphi_l|(\mbI_{\mH}^{\otimes2}-\mplr)\rv|
\mphi_k\otimes\mphi_j)=(\mphi_m\otimes\mphi_l|\rv|\mphi_k\otimes\mphi_j)
-(\mphi_m\otimes\mphi_l|\rv|\mphi_j\otimes\mphi_k), \]
we can write for the operator representation of $\mD{\mhSl},\mrh~\in
T^*_\mrh\mfTs\sim\mLHs$:
\bequ\label{eq;(DhSl)}
\begin{split}
(\mphi|\mD{\mhSl},\mrh~|\psi)&=(\mphi|\mrho|\psi)+
\frac{N}{2}\sum_{k,m}(\mphi_k|\mrh|\mphi_m)\bigl[(\mphi\otimes\mphi_m|\rv|
\psi\otimes\mphi_k)-\\ &(\mphi\otimes\mphi_m|\rv|\mphi_k\otimes\psi)+
(\mphi_m\otimes\mphi|\rv|\mphi_k\otimes\psi)-(\mphi_m\otimes\mphi|\rv|
\psi\otimes\mphi_k)\bigr].
\end{split}
\end{equation}
Let us consider now that, on the chosen orbit \OUr, one has
\[\mrh_t\equiv
\mrh_{\{j\}}^t:=\mun H,t,\mrh~\mrh_{\{j\}}\mun H,t,\mrh~^*:=
N^{-1}\sum_{k=1}^N|\psi_{k}^t\rb\lb\psi_{k}^t|.\]
Let us denote by \bs{\mEj^t:=N\mrj^t}\ the $N$--dimensional projection
corresponding to
the Slater vector--determinant $\Psi_{\{j\}}^t$,
according to the mapping \bG\ from~\eqref{eq;1-N}.
For each such $\mrh=\mrj^t$\ occurring in~\eqref{eq;(DhSl)},
let us choose in
the r\^ole of the complete
orthonormal system $\{\mphi_k\}$\ such a complete orthonormal system
$\{\psi_j^t:j=1,2,\dots\}$\
that contains the $\mrj^t$--defining one--particle vectors
$\{\psi_j^t:j=1,\dots,N\}$ numbered as the initial segment. Then we
have for the obtained orthonormal bases $\{\{\psi_j^t:j\in\mbN\}:\
t\in\mbR\}$:

\[  (\mphi_k|\mrh_t|\mphi_m)\equiv (\psi_k^t|\mrh_t|\psi_m^t)=\begin{cases}
\frac{1}{N}\delta_{k,m}&{\rm for}\ k,m\leq N, \\ 0&{\rm for}\ \max(k;m)>N;
\end{cases}
\]
 these should be inserted into the formulas
like~\eqref{eq;(DhSl)}.
 Matrix elements of the Schr\"o\-din\-ger equation~\eqref{eq;2.11} for
 arbitrary $\mphi,\psi\in\mH$\ are then of the form
 \bequ\label{eq;HF-Sch}
\begin{split}
 i\frac{d}{dt}(\mphi|&\mun H,t,\mrh~|\psi)=(\mphi|\mD{\mhSl},\mrh_t~\dti\mun
 H,t,\mrh~|\psi)=(\mphi|\mrho\dti\mun H,t,\mrh~|\psi)+\\
 &\frac{1}{2}\sum_{k=1}^N\sum_j
 \bigl[(\mphi\otimes\psi_k^t|\rv|\psi_j^t\otimes\psi_k^t)-
(\mphi\otimes\psi_k^t|\rv|
 \psi_k^t\otimes\psi_j^t)+\\ &(\psi_k^t\otimes\mphi|\rv|\psi_k^t\otimes
\psi_j^t)-
 (\psi_k^t\otimes\mphi|\rv|\psi_j^t\otimes\psi_k^t)\bigr]
 (\psi_j^t|\mun H,t,\mrh~|\psi).
\end{split}
\end{equation}
Let us rewrite this equation in ``configuration representation'', if
$\mH:=\mLqn$, and operators $A$ are (formally) expressed by their
 ``kernels'':
\[ A(x,y):= \sum_{j,k}\psi_j(x)(\psi_j|A|\psi_k)\overline{\psi_k(y)}.\]
Let us, moreover, consider that $\rv_{12}=\rv_{21}$, hence for the
multiplication operator \rv\ (in this representation) one has:
$\rv(x,y)\equiv\rv(y,x)$. The projections $\mEj^t$\ have now the kernels
\[ \mEj^t(x,y):=\sum_{j=1}^N\psi_j^t(x)\overline{\psi_j^t(y)},\quad
\|\psi_j^t\|\equiv 1.\]

We obtain then, with $\psitr:=\mun H,t,\mrh~\psi$,
 the usual \emm time--dependent Hartree--Fock equation~:
\bequ\label{eq;TDHF}
i\frac{d}{dt}\psitr(x)=\left[\mrho+\int\rd y\,\mEj^t(y,y)\rv(y,x)\right]
\psitr(x)-\int\rd y\,\rv(x,y)\mEj^t(x,y)\psitr(y).
\end{equation}
We can insert into\rref TDHF~ $\psitr:=\psi_j^t,\ j=1,2,\dots N$, to obtain
coupled nonlinear equations for $\psi_j^t$'s.
Evolution of the whole density matrices on \OUr\ is expressed by
\bequ\label{eq;TDHF-r}
i\frac{d}{dt}\mrh_t=[\mD{\mhSl},{\mrh_t}~,\mrh_t].
\end{equation}
Its stationary solutions $\mrh_t\equiv\mrh=N^{-1}\mEj$ commute with
\D{\mhSl},\mrh~, hence the selfadjoint operator \D{\mhSl},\mrh~ leaves the
subspace \Ej\H\ of the Hilbert space \H\
invariant: $\mD{\mhSl},\mrh~\mEj\mH \subset\mEj\mH$. This means that its
restriction to the subspace \Ej\H\ can be diagonalized, and the basis
$\{\psi_j:j\in\mbN\}$\ can be chosen (in the point \rh: the bases are
point--dependent, according to their definition) such that the vectors
$\psi_1,\psi_2,\dots,\psi_N$\ are eigenvectors of \D{\mhSl},\mrh~. Hence we
have from~\eqref{eq;TDHF} the corresponding eigenvalue equation, what is
the (time independent) \emm Hartree--Fock equation~, cf.\ \cite[\S 10]{march}:
\bequ\label{eq;HF-eq}
\left[\mrho+\int\rd y\mEj(y,y)\rv(y,x)\right]\psi_k(x)-
\int\rd y\mEj(x,y)\rv(x,y)\psi_k(y)=\epsilon_k\psi_k(x).
\end{equation}

We have shown how the time dependent, as well as the
stationary Hartree--Fock theory is described in the framework of our formalism.


\subsection{Nonlinear Schr\"odinger equation and mixed states}
\label{IIIA1;NL-Sch}

Let us give here another example of description of ``a system'' in the
framework of NLQM.
We shall show here that a traditional ``nonlinear Schr\"odinger
equation''~\cite{deBrog,ashtekar} can be included in the scheme of EQM.
We shall partly proceed, in the following example (taken
from~\cite{bon-dens}), in a
heuristic way, by ``plausible'' formal manipulations; the necessary
mathematical comments will be omitted here. This example will be also used
to show, in a nontrivial concrete case, that the barycentre of a genuine
mixture $\mu\in\mcl M(\mSs)$ evolves
under nonlinear evolution differently from the evolution of the elementary
mixture \rh\
being its initial barycentre,\rref 2.17~: $Tr(\mrh \ra):=
\int_{\mSs}Tr(\nu\ra)\mu(\rd\nu),\ \forall\ra\in\mLHs$.

Let us first recall that, for a given generator $Q\in
C^\infty(\mSs,\mbR)$, the Schr\"odinger equation (resp.\ the Liouville--von
Neumann equation) for the flow $\mphi^Q$\
corresponding to the Poisson structure on \Ss\
(cf.\ Subsection~\ref{q-phsp;poiss}) can be written on \Ss\ in the form
(cf.\rref 2.11~ and\rref g-dif~)
\bequ\label{eq;rho-t}
i\,\frac{d}{dt}\mrh(t)=[\mD Q,\mrh(t)~,\mrh(t)],
\end{equation}
where
\bequ\label{eq;evol}
 \mrh(t)\equiv\mph t,Q~(\mrh):=\mun Q,t,\mrh~\mrh\mun
 Q,t,\mrh~^*,\quad\mrh(0):=\mrh,
\end{equation}
and \un Q,t,\mrh~\ satisfies the equation (one can use alternatively, in
the r\^ole of the ``Hamiltonian operator'' in the following equation, an
operator of the form
$D_{\mrh(t)}Q+{\bf f}^0(\mrh(t))$, with ${\bf f}^0(\mrh)\in\{\mrh\}'$,
cf.\ Remark~\ref{rem;2.9})
\bequ\label{eq;nl-Sch-Q}
i\,\frac{d}{dt}\mun Q,t,\mrh~=D_{\mrh(t)}Q\dti\mun Q,t,\mrh~,\quad
\mun Q,0,\mrh~\equiv0.
\end{equation}
The equation\rref rho-t~ can be rewritten for wave functions $\psi\in\mH$,
$\psi(t):=\mun Q,t,P_\psi~\psi\in\mH$\ (we set $D_\psi:=
D_{P_\psi}$):
\bequ\label{eq;psi-t}
i\,\frac{d}{dt}\psi(t)=D_{\psi(t)}Q\dti\psi(t).
\end{equation}

Let us take now \H=\Lqn\ with
$\lb\psi|\mphi\rb:=\int\overline\psi(x)\mphi(x)\,d^nx$. Let us write
density matrices \rh\ ``in the $x$--representation'' with a help of their
operator kernels $\mrh(x,y)$:
\bequ\label{eq;kern}
 [\mrh\psi](x)\equiv\int\mrh(x,y)\psi(y)\,d^ny,\quad\psi\in\mH.
 \end{equation}
 Projection operators $P_\psi$ have their kernels
 $P_\psi(x,y)\equiv\|\psi\|^{-2}\psi(x)\overline\psi(y)$.
Let the Hamiltonian function $Q$\ will be taken as the (unbounded) functional
 \bequ\label{eq;QP}
 Q(P_\psi):= Tr(P_\psi\dti
H_0)+\frac{\mveps}{\malp+1}\int P_\psi(x,x)^{\malp+1}d^nx,
\end{equation}
with $H_0$\ some selfadjoint (linear) operator on \Lqn, and
$\malp>0$.
Let $t\mapsto P_{\psi(t)}, \psi(0):=\psi$\ be any differentiable curve
through $P_\psi\in\mPH$, and let $\dot{P}_\psi\in T_{P_\psi}\mPH$\ be its
tangent vector expressed by an operator according to equations\rref 2.rep~.
Then the (unbounded, nonlinear) Hamiltonian \D Q,\psi~ can be expressed by:
\bequ
Tr(\mD
Q,\psi~\dti\dot{P}_\psi):=\left.\frac{d}{dt}\right|_{t=0}Q(P_{\psi(t)}),
\end{equation}
what leads to the corresponding form of
``nonlinear Schr\"odinger wave--equation'' for $\psi_t:=\psi(t)$:
\bequ\label{eq;nlin-weq}
i\,\left[\frac{d}{dt}\psi_t\right](x)=[H_0\psi_t](x)+\mveps
|\psi_t(x)|^{2\malp}\psi_t(x),\quad\|\psi_t\|\equiv 1.
\end{equation}
One possible extension of this nonlinear dynamics from \PH\ to the
whole space \Ss\ is obtained by ``the substitution $\mrh\mapsto P_\psi$'',
i.e.\ by the choice of the Hamiltonian
\bequ\label{eq;exten}
Q(\mrh):= Tr(\mrh\dti H_0)+\frac{\mveps}{\malp+1}\int\mrh(x,x)^{\malp+1}d^nx,
\end{equation}
and the corresponding dynamics is then described by~\eqref{eq;rho-t}
with\footnote{\label{ft;20} The notation \D Q,\mrh~\ represents here the
linear functional according to the standard definition of the Fr\'echet
derivative, as well as its operator representative, cf.\rref2.rep~, and
Definitions~\ref{df;g-dif}.}
\bequ
\mD Q,\mrh~(\nu)\equiv Tr(\mD Q,\mrh~\dti\nu)\equiv
Tr(\nu\dti H_0)+\mveps\int\mrh(x,x)^\malp\nu(x,x)\,d^nx.
\end{equation}

We shall compare the evolutions of the mixed states
described by the same initial barycentre $\equiv$\ density matrix \rh\
\bequ\label{eq;mrh}
\mrh:=\sum_j\mlam_jP_{\psi_j},\quad\sum_j\mlam_j=1,\ \mlam_j\geq 0.
\end{equation}
for the two distinguished interpretations.
The evolution of the elementary mixture \rh\ is described by \rref rho-t~,
while the evolution of the ``corresponding'' genuine mixture $\mu$,
\bequ\label{eq;mu-r}
\mu:=\sum_j\mlam_j\delta_{\psi_j},
\end{equation}
(where $\delta_{\psi}\equiv\delta_{P_\psi}$\ is the Dirac measure
concentrated on
$P_\psi\in\mPH\subset\mSs$)
 is described by
\bequ
   (t;\mu)\mapsto \mu_t\equiv\mu\circ\mph -t,Q~.
\end{equation}
This corresponds to such an evolution of the measure $\mu$\ representing a
state, when each of the vectors $\psi_j$\ entering into\rref mu-r~ evolves
according to the equation\rref psi-t~.

Let us illustrate, by explicit calculation, the difference between time
evolutions of the same initial
density matrix considered in its two different interpretations.

Let us take the system with its above determined ``extended'' dynamics,
and let us fix a nontrivial
(i.e.\ $\mlam_j<1,\ \forall j$)
mixture \rh\ of several vector states $P_{\psi_j}$, as in\rref mrh~.
Let us calculate the difference between the
derivatives \wrt the time in
$t=0$\ of the two evolutions: (i) of the {\em barycentre of the time evolved
genuine mixture}
$\sum_j\mlam_j\mph t,Q~(P_{\psi_j})$, and (ii) of the {\em elementary mixture
evolution} $\mph t,Q~(\mrh)$.  We shall calculate the right hand side of\rref
rho-t~ for the two cases and take their difference. Let us write the
kernel ``in $x$--representation'' of \rh\ as the convex
combination of the vector--state kernels:
\bequ
\mrh(x,y)\equiv\sum_j\mlam_j\|\psi_j\|^{-2}\psi_j(x)\overline\psi_j(y).
\end{equation}
The (symbolic) ``kernel'' of the Hamiltonian \D Q,\mrh~\ can be written:
\[ \mD Q,\mrh~(x,y)=H_0(x,y)+\mveps\delta(x-y)\mrh(x,x)^\malp. \]
Here, $\delta(\cdot)$\ is the Dirac distribution on $\mbR^n$. We have to
express the difference $\Delta^{\{\mrh\}}_t(x,y)$\
between the kernels (in $x$--representation) of the operators
\[ \sum_j\mlam_j[\mD Q,\psi_j(t)~,P_{\psi_j(t)}],\ {\rm and}\ [\mD
Q,\mrh(t)~,\mrh(t)], \]
what expresses the difference between time derivatives of ``the
same density matrix'' $\mrh=\sum\mlam_jP_{\psi_j}$ in the two
interpretations. The linear operator $H_0$\ does not contribute into this
difference. The kernels of commutators entering into the calculation are
(for all $\nu\in\mSs$) of the form
\[ [\mD Q,\nu~,\nu](x,y)=[H_0,\nu](x,y)+\mveps\nu(x,y)(\nu(x,x)^\malp
-\nu(y,y)^\malp).\]
We can (and we shall) take all $\|\psi_j\|\equiv1$.
Let us denote
\[\chi_j^{\{\mrh\}}(x):=|\psi_j(x)|^{2\malp}-\left(\sum_k\mlam_k
|\psi_k(x)|^2\right)^\malp.\]
Then the wanted difference at $t=0$\ is
\bequ
\Delta_{\{\mrh\}}(x,y):=\Delta^{\{\mrh\}}_0(x,y)=\mveps\sum_j\mlam_j
\psi_j(x)\overline\psi_j(y)(\chi^{\{\mrh\}}_j(x)-\chi^{\{\mrh\}}_j(y)).
\end{equation}
By proving that the operator $\Delta_{\{\mrh\}}$\ is not identical zero for all
$\{\mrh\}$,
we can prove
 nontrivial difference of the two time evolutions explicitly. This can be
proved easily for $\mlam_1:=1-\mlam_2$, and for $\psi_1,\psi_2$\ chosen to
be specific
three--valued (i.e.\ with two mutually distinct nonzero values) functions
concentrated on disjoint compact subsets of $\mbR^n$:
Each $\psi_j,\ (j=1,2)$\ has its nonzero constant values on domains with
different nonzero Lebesgue ($\rd^nx$) measure.

Analogical examples could be constructed for, e.g.\ unbounded functions $Q$
on dense domains of \Ss\ expressed by the formula
\bequ\label{eq;Q=Kr}
Q(\mrh):=Tr(\mrh\dti H_0)+\mveps\int_{\mbR^n}\mcl K(\mrh(x,x))\rd^nx,
\end{equation}
where $\mcl K\in C^\infty(\mbR,\mbR)$\ can be chosen (in this abstract
approach) rather arbitrarily. Such possibilities were mentioned (in a
framework of Schr\"odinger equations\rref psi-t~ for wave functions) also
in~\cite{ashtekar}; they include, e.g.\ the equations proposed
in~\cite{bial&myc}, and also WKB--equations.

Differentiation of\rref Q=Kr~ gives a formula for the corresponding
``Hamiltonian'' \D Q,\mrh~\ (cf.\ Footnote~\ref{ft;20}):
\[ \mD Q,\mrh~(\nu)\equiv Tr(\mD Q,\mrh~\dti\nu)= Tr(H_0\dti\nu)+
\mveps\int\mcl K'(\mrh(x,x))\dti\nu(x,x)\rd^nx, \]
or in terms of formal ``operator kernels'' (with $\mcl K'(s):=\frac{\rd\mcl
K(s)}{\rd s},\ s\in\mbR$):
\[ \mD Q,\mrh~(x,y) = H_0(x,y)+\mveps\dti\delta(x-y)\dti\mcl K'(\mrh(x,x)),
\]

We did not consider any domain questions here: that would need more time
and space. It seems, however, that the above formally given operators
\D Q,\mrh~\ could be correctly defined on (a dense subset of)
\H, at least as symmetric operators.

\begin{rem}[Koopmanism]\label{rem;koopm}
We have restricted, in the above considerations, our attention
to the ``Schr\"odinger picture'', hence the algebra
of observables \cl A\ was not investigated: It could be chosen \cl A\ =
\LH, as in linear QM. Its completion to an algebra of operator--valued
functions could give a ``linear extension'' of the system. Let us note,
however, that such extensions might be considered as a version of
``noncommutative Koopmanism'', cf.\ a Koopman formalism in CM (i.e.\ the
``commutative'' one), e.g.\ in~\cite[Chap.X.14]{R&S},
resp.~\cite{emch1,koopman}. This can be expressed schematically as follows.

In Hamiltonian classical mechanics, the system is described by a
$2n$--dimensional symplectic manifold $(M;\Omega)$, where the symplectic
form $\Omega$\ provides for ascription to Hamiltonian functions
(i.e.\ generators)
$h\in C^\infty(M,\mbR)$\ the symplectic flows \ph t,h~\ on $M$. Since the
$n$--th exterior power of $\Omega$, i.e.\ the Liouville volume
$\wedge^n\Omega$\ (corresponding to a measure $\mu_\Omega$\ on $M$)
is conserved by $\mphi^h$, it is possible to introduce the Hilbert space
$L^2(M,\mu_\Omega)$, where the  flow $\mphi^h$\ (let us assume, that it is
complete) is described by the
continuous unitary group $U^h(t)$ defined by:
\[ f_t(m):=[U^h(t)f](m):= f(\mph t,h~(m)),\quad\text{for all $f$\ in
classes}\ f\in L^2(M,\mu_\Omega).\]
The selfadjoint (cf.\ \cite[Proposition
2.6.14]{abr&mars}) generator $L_h$\ of $U^h(t)\equiv\exp(-itL_h)$\ is called
the Liouville operator of the CM system. It is in fact  the
differential operator given by the Poisson bracket (up to domain
questions concerning possible choice of $f$):
\[ L_hf\equiv i\dti\{h,f\},\quad f\in D(L_h)\subset L^2(M,\mu_\Omega).\]
In this way, the nonlinear finite--dimensional Hamilton's equations are
transformed formally into a linear Schr\"odinger--like ``Liouville equation''
\[ i\dti\frac{d}{dt}f= L_hf \]
on infinite--dimensional Hilbert space. This Hilbert space description of
CM is called the \emm Koopman formalism~.
Let us note that here, in the ``commutative case'' of CM, the
transformation $U^h(t)f$\ of elements $f\in L^2(M,\mu_\Omega)$\ of this
``extended phase space'' is uniquely given by the transformation \ph t,h~\
of the phase space $M$, i.e.\ of the space of arguments of scalar--valued
functions (the real values of $f(m)$\ should stay real also for $f_t(m)$,
due to their physical interpretation).

The situation in EQM can be considered in analogy with the preceding
Koopman transition in CM, cf.\ also~\cite{bon-tr}: The (nonlinear)
transformations \pph t,Q~\
of \Ss\ are extended to one--parameter \autm\ group \taQ\ of a \Ca\ \Cbs\
(cf.\dref2.28~ and Theorem~\ref{thm;2.29}),
or some of its subalgebras, what is a standard picture of {\em linear quantum
theories}.
The difference from the ``commutative case'' is, that \Cbs\ is generated by
functions on the ``quantum phase space'' \Ss\ {\em with values in
noncommutative \Ca}\ \LH. Hence, to obtain the automorphism group \taQ\
corresponding to nontrivial \pph t,Q~, we
have to introduce in a consistent way {\em also transformations of values}
of these functions. These are, however, nonunique, and the nonuniqueness is
pointed out, e.g., in Remark~\ref{rem;2.9}.\hfill\dovi
\end{rem}



\section{``Macroscopic'' Reinterpretation of EQM}
\label{sec;IIIB}

It might be interesting from technical, as well as from physically intuitive
point of view to show a simple way how our {\em nonlinear
quantum--mechanical dynamical systems} considered in this paper
(in the framework of EQM) can be considered as subsystems of infinite
physical systems
described in a framework of traditional (linear) quantum theory. A Hilbert
space description of such a ``large'' system would necessitate, however,
also usage of a nonseparable Hilbert space, e.g.\ the space of universal
representation of a certain algebra \cl A\ having uncountably many mutually
inequivalent faithful representations (each one corresponding to a specific
value of macroscopic variables), cf., e.g.~\cite{sak1,pedersen,sewell}.
It is not very comfortable to
have all these representations simultaneously as subrepresentations in one
nonseparable Hilbert space. A way to describe this situation
in a more transparent way can be found in the framework of
formalisms used in quantum
field theory (QFT), or in theories of systems ``with infinite number of
degrees of freedom'',
cf.\ \cite{haag&kast,haag1,emch1,bra&rob,bra&rob2,haag2}. The main
mathematical tool of these theories are \Ca s and their automorphism
groups, cf.\ also~\cite{sak2,pedersen}. These theories are usually used to
describe ``thermodynamic systems'' considered as infinitely large in the
sense of intuitive notion of spatial extension, and also containing an
infinite number of particles. Such an {\em infinite approximation to
finite, but large systems} is conceptually acceptable and technically
useful: It allows clear mathematical description of \emm macroscopic
subsystems~ of physical systems consisting of very large number of
microscopic constituents -- so large that any detailed practical (e.g.
numerical) description and measurement of their states, taken even in any
nontrivial a priori restricted precision for individual subsystems,
is hopeless;\label{5meas}
their macroscopic subsystems consist, on the other side, of
manageable sets of
classically described parameters. Mathematically clear description of
states and dynamics of such sets of classical parameters of quantal
systems is
up to now possible, however, only in the framework of ``infinitely
extended'' systems. Its possibilities include, e.g., a description of phase
transitions, cf.\ \cite{bra&rob2,sewell}.

We shall sketch briefly in this section a possibility, how to introduce a \Ca\
\cl C\ describing a ``large'' QM--system, ``containing'' in a certain
sense the traditional observable algebra \LH\ of a finite quantum
system, as well as a commutative subalgebra of continuous complex--valued
functions $C(M)$. This subalgebra $C(M)$\ is interpreted as the \Ca\ of a
classical subsystem in such a way
that \cl C\ {\em is determined by these two subalgebras}, together with
an infinite index set $\Pi$\ containing labels of the ``elementary''
(mutually equal) finite subsystems.
This algebra \cl C\ can be chosen so that it ``contains'' the \Ca\ \CG\
(cf.\dref2.25b~) describing any of the infinite number of
equal ``microscopic'' subsystems composing the large system,
as well as its
extension by a classical system (= a ``mean--field''); the later can
describe collective influence of all the other subsystems onto the specified
one,~\cite{hp+lie1,bon1,bon3,unner0,unner1,unner2}.
In such systems,
the dynamics can be determined by a sequence of local Hamiltonians. If a
 Lie group $G$ is given so that it determines (via
its unitary continuous representation) selfadjoint generators
entering into the expressions of the local Hamiltonians of the
(arbitrarily large but) finite subsystems, the spectrum space $M$
of the classical subalgebra is the range \EF=\bF(\bcDF)\ of
$\mbF$\ in the dual $\mfk g^*$\ of the Lie algebra of $G$,
cf.\dref2.17~; we can even have $\mEF=\mSs$, for $G:=\mfk U:=\mcl
U(\mH)$. This approach can be considered either as a
``phenomenological'' introduction of a formal classical system to
``complete'' a given nonlinear quantum system to a linear one, or
as a dynamical theory of a large system with a long range
interaction. The dynamics is then a \autm\  subgroup of the \autm\
group of \cl C.

Since the \Ca\ \CG\ is ``essentially'' (i.e.\ up to its completions in
weaker--than--norm topologies) the tensor product $\mLH\otimes\mCGc$, it
corresponds (in the sense of usual QM constructions -- again
``essentially'') to a compound system consisting of a ``standard (with
finite number of degrees of freedom) QM--system'' described by the algebra
of observables \LH, and of a ``classical subsystem'' described by the
commutative \Ca\ \CGc\ which is isomorphic to the space of all complex valued
continuous functions on the quantum phase space of elementary mixtures \Ss,
$C(\mSs,\mbC)$. Hence, the ``nonlinear'' EQM can be
embedded as a subsystem theory to a linear quantum theory, cf. also
Theorem~\ref{thm;2.29}. This linear
theory can be considered in turn as a subtheory of a (nonrelativistic)
quantum field theory (QFT), i.e.\ a theory of an infinite number of
``standard'' QM systems,~\cite{haag&kast,emch1,haag1,bra&rob,bon1}; this
can be done not only kinematically, i.e.\ by construction of the sets of
observables, but also by postulating a ``microscopic'' evolution in local
subalgebras (given by local -- linear -- Hamiltonians {\em of mean--field
type},~\cite{hp+lie1}, depending on size of the local
subsystems) and taking the thermodynamic limit,~\cite{bon1,bon2}. As a
result of such a limiting procedure, it is obtained, besides the
\emm quasilocal
algebra~ \cl A\ of observables of arbitrarily large, but finite subsystems,
also the \emm algebra of
classical quantities~ \CGc\ (belonging to the ``algebra of observables at
infinity'',~\cite{ruelle,hp-meas,bra&rob,sewell}), without which the time
evolution cannot be defined as a (semi--) group of transformations of an
algebra of observables,~\cite{morch&stroc,bon1}. A ``simplest'' and  a
``most natural'' quasilocal algebra of an infinite system in
nonrelativistic QFT is $\mcl A:=\otimes_{p\in\Pi}\mcl L(\mcl H_p)$, as it
is introduced below. If the
``standard'' QM system under consideration (the extension of which is the
``considered'' system described by EQM)
is described in finite--dimensional Hilbert space, then we have, as the
algebra of observables of the corresponding infinite system, directly the
tensor product \Ca\ $\boldsymbol{\mcl
C=\mcl A\otimes\mCGc}$. $\bigl[$In the case of infinitedimensional
 Hilbert spaces
\H, the algebra ${\mcl C}$\ contains $\mcl A\otimes\mCGc$\ as a
(possibly proper) subalgebra,~\cite{bon8}, (the fact, that $\mcl C\neq
\mcl A\otimes\mCGc$\ in this case is a consequence of weak, but not norm,
continuity of
corresponding unitary groups, resp.\ of unboundedness of generators,
cf.\ also~\cite{bra&rob,sak2} for some mathematical refinements).$\bigr]$
 In these cases of
infinite systems, the elements of the classical subalgebra \CGc\
are naturally interpreted as {\bf (global) intensive quantities}\index{global intensive quantities}
of the infinite system; hence, they correspond to \emm macroscopic
variables~ of this large quantal system, cf. also Remark
\ref{rem;macro}.
\def\nazov{{
\ref{sec;IIIB}\quad ``Macroscopic'' Reinterpretation of EQM}}

Let us introduce an example of such macroscopic algebraic elements.
The description is realized on, e.g., infinite tensor product
$\mH_\Pi:=\otimes_{p\in\Pi}\mH_p$\ (with $\Pi:=$\ an infinite index set
labelling the
``constituent microsystems''),~\cite{neum2}, of equal copies of the Hilbert
space
$\mH\equiv\mH_p$, and the quasilocal algebra \cl A\ is generated (via algebraic
operations and norm limits) by the subalgebras (isomorphic to)
\begin{subequations}
\label{eq;ql-alg}
\bequ
\mcl A_\Lambda:=\otimes_{p\in\Lambda}\mcl L(\mcl H_p),\quad
\Lambda\subset\Pi,\ |\Lambda|<\infty,
\end{equation}
where $|\Lambda|:=$\ {\em the number of elements in} $\Lambda\subset\Pi$,
with the natural inclusions
\bequ
\Lambda\subset\Lambda'\imply\mcl A_\Lambda\subset\mcl A_{\Lambda'},
\end{equation}
acting on
$\mH_\Pi\equiv\mH_\Lambda\otimes\mH_{\Pi\setminus\Lambda}$\ (the
tensor product of Hilbert spaces is, for finite number of factors,
associative,~\cite{neum2}) in the obvious way: Let us define an
``identification'' of the Hilbert spaces $\mH_p,\ p\in\Pi$\ with
\H\ by defining unitary mappings $\ru_p,\ p\in\Pi$\ of $\mH_p$\
onto \H. $\mcl A_\Lambda$ is generated by elements $\ra_p$, where
$\ru_p\ra_p\ru_p^{-1}\in\mLH$, with $\ra_p$'s ``acting on the
$p$-th factor'' $\mH_p,\ \forall p\in\Lambda$, i.e.\ for a vector
$\Phi\in\mH_\Pi$, one has \bequ
\Phi:=\otimes_{p\in\Pi}\phi_p,\quad
\ra_q\Phi:=(\ra_q\phi_q)\otimes
\left(\otimes_{p\in\Pi\setminus\{q\}}\phi_p\right).
\end{equation}
Now we choose any $X\equiv\ru_pX_p\ru_p^{-1}\in\mLHs$,\footnote{We can
choose also unbounded $X$'s here, but in that case nontrivial domain questions
should be considered,~\cite{bon8}.} corresponding to
``an observable of individual subsystems'', and define
\bequ
X_\Lambda:=\frac{1}{|\Lambda|}\sum_{p\in\Lambda} X_p \in\mcl A_\Lambda.
\end{equation}
\end{subequations}

Let a Lie group $G$ be unitarily and continuously represented in \H\ by
$U(G)$, and let $U_p(G):=\ru_p^{-1}U(G)\ru_p$\  be the corresponding action
on each $\mH_p$. Let the $X=X^*$\ above be an arbitrary generator of
$U(G)$.
The set of ``intensive observables'' in \CGc\ is generated by {\em limits
$X_\Pi$ in some topology}\footnote{The topology, in which this limit exists is
rather special: It cannot be norm--topology of $\mcl A$ for $X$ such, that
their spectrum contains at least two points,~\cite{bon1}. The
set $\{X_\Lambda:\Lambda\subset\Pi,|\Lambda|<\infty\}$ considered as a
net in the von Neumann algebra $\mcl A^{**}$ (:= the second topological
dual of \cl A) has more than one cluster points,~\cite{bon1}. Hence, the
topology on $\mcl A^{**}$\ has to be chosen weaker than its
$w^*$--topology,~\cite{bon1}.}
 of $X_\Lambda$, for $\Lambda\nearrow\Pi$, for all selfadjoint generators
 $X:=X_\xi,\ \xi\in\mfk g$\ of $U(G)$. These limits do not belong to
 $\mcl A$; their introduction to the algebra of
 observables is, however,  necessary for the ``standard--type''
 description of dynamics with  long--range interactions,~\cite{morch&stroc},
 e.g.\ for MF--type evolutions,~\cite{bon1}. The quasilocal algebra \cl A\
 is canonically included into its second topological dual $\mcl A^{**}$
 which is in turn a \Wa\ in a canonical way~\cite{sak1,pedersen};  the
 limits $X_\Pi$\ are then associated,~\cite[Definition 2.7.13]{sak1}, with
 a certain \Wsa s of $\mcl A^{**}$.
 On the algebra of functions  \CGc, Poisson brackets can be naturally defined.
 Then the mentioned
 mean--field type dynamics defined with a help of local Hamiltonians on
 $\mcl A_\Lambda$'s does not leave, in the thermodynamic limit, the \Ca\
 \cl A\ invariant (invariant \wrt\ such evolutions is, however, the
 ``classical subalgebra''  \CGc).\footnote{The r\^ole of the group
 $G$ is here
 in choice of the topology mentioned in the preceding footnote, as well as
 in determination of dynamics: It is very useful especially in the presence
 of unbounded generators (i.e.\ local Hamiltonians).}
 The dynamics of the classical algebra \CGc\ is Hamiltonian (\wrt the
 mentioned Poisson structure), and {\em the dynamics of any local subsystem
 (described by $\mcl L(\mH_p)$) ``essentially coincides'' with some of the
 nonlinear dynamics
 described by Theorem~\ref{thm;2.29}, and by Definitions~\ref{df;2.28}}:
The considered dynamics of the infinite system is constructed as follows:
Let us consider a function $Q\in C^\infty(\mfk g^*,\mbR)$\ as a Hamiltonian
for the dynamics (\wrt\ the natural Poisson structure on $\mfk g^*$)
described by the Poisson diffeomorphisms $\mphi^Q_t:\mfk g^*\rarw\mfk g^*$.
The local Hamiltonians of the infinite quantum system are (consider, in the
following formulas, $Q$ as a polynomial in $F_j:=F(\xi_j),\ F\in\mfk g^*$,
for simplicity, otherwise cf.\ \cite{d+wer1})
\begin{subequations}\label{eq;0Hl}
\bequ\label{eq;1Hl}
H_\Lambda:=|\Lambda|\dti Q(X_{\xi_1\Lambda},\dots,X_{\xi_n\Lambda}),
\end{equation}
with $\{\xi_j,j=1,\dots,n\}$\ a basis in \fk g, and the ``ordering'' of
operators is such, that all $H_\Lambda$'s are selfadjoint.\footnote{The
ordering and symmetry of the operators is not here very important, since
in the limit $\Lambda\nearrow\Pi$\ the elements $X_{\xi_j\Pi}$\ commute with
all observables: They belong to (a subspace of) the centre of $\mcl
A^{**}$.}
Then the limiting dynamics
\bequ\label{eq;2Hl}
\tau^Q_t(\rx)=
(?)-\lim_{\Lambda\nearrow\Pi}\exp(-itH_{\Lambda})\rx\exp(itH_\Lambda),
\quad\rx\in\mcl A
\end{equation}
can be defined, but only (for nonlinear $Q$) as a dynamics of the
``extended'' algebra containing also the classical (macroscopic)
quantities, cf.\ \cite{bon1,bon3,d+wer1}. If, in an initial state
$\mome=\mome_0\in\mcl{S(A)}$, the values $s-\lim_{\Lambda\nearrow\Pi}
\pi_{\mome}(X_{\xi\Lambda})$\ exist, then in its time evolved states
$\mome_t:=\mome\circ\tau^Q_t$\ (if canonically extended to the states on
$\mcl A^{**}$) we have
\bequ\label{eq;3Hl}
\mome_t(X_{\xi\Pi})\equiv Ad^*(g_Q(t,F_0))F_0(\xi),\qquad
F_0(\xi):=\mome_0(X_{\xi\Pi}),\forall \xi\in\mfk g,
\end{equation}
where the cocycle $g_Q(t,F)$\ is as in Proposition~\ref{prop;2.26}.
\end{subequations}

The subdynamics of the system ``living'' on any one of Hilbert
 spaces $\mH_p,\ p\in\Pi$, say on \H\ (if the index $p$ is skipped), is
given then by  a unitary cocycle $\{U(g_Q(t,F)): t\in\mbR,F\in\mfk
 g^*\}$\footnote{These cocycles are nonunique, cf.\ Remark~\ref{rem;2.9}.}, cf.
 Proposition~\ref{prop;2.26}:
\begin{subequations}\label{eq;0MFd}
\bequ\label{eq;1MFd}
 \psi(t)\equiv U(g_Q(t,F))\psi(0),\ \psi(0)\in\mH,
\end{equation}
 {\em if the initial condition $\psi(0)$\ is chosen such, that for the
initial value of ``the macroscopic field $F=F_0$'',
$F_0(\xi):=\mome_0(X_{\xi\Pi})$ at $t=0$, it is  fulfilled}
\bequ\label{eq;2MFd}
 F(\xi):=\lb F;\xi\rb \equiv
 (\psi(0),X_\xi\psi(0)),\quad\forall\xi\in\mfk g.
\end{equation}
 We can see that for
\bequ\label{eq;3MFd}
 F_t:=Ad^*(g_Q(t,F_0))F_0\equiv\mphi_t^Q(F_0),\quad t\in\mbR,
\end{equation}
 the following relation is valid:
 \bequ\label{eq;4MFd}
F_t(\xi)\equiv (\psi(t),X_\xi\psi(t)),\ \text{for}\
F_0(\xi)\equiv(\psi(0),X_\xi\psi(0)).
 \end{equation}
\end{subequations}
After insertion of\rref4MFd~ for $F_t$\ into the time dependent Schr\"odinger
equation for $\psi(t)$\ obtained from\rref1MFd~ by differentiation, we obtain
a nonlinear Schr\"odinger equation of EQM, describing now the evolution of a
``small'' subsystem of an infinite (linear) system of traditional QT, cf.\
\rref3.17~. Hence, the subdynamics of an infinite quantum system with
an automorphic (hence ``linear'') time evolution appears as nonlinear
evolution in a NLQM.

 We shall return to the relations\rref0MFd~ in Section~\ref{sec;IIIE},
where we
 shall rewrite nonlinear QM--equations as a couple of equations: one
 nonlinear classical Hamilton's equation, and a linear time--dependent
 Schr\"odinger one.



\section{Solution of Some Nonlinear Schr\"odinger
Equations}\label{sec;IIIE}

Let $\dim G=n<\infty$, for a Lie group $G$. Let an arbitrary $U(G)$--system
be given, with $\{\xi_j:j=1,2,\dots,n\}$\ a basis of $\mfk g := Lie(G)$, and
$F_j:=F(\xi_j),\ j=1,2,\dots,n,\ \forall F\in\mfk g^*$. Let $Q\in
C^\infty(\mEF,\mbR)$ be chosen. The selfadjoint generators of unitary
one--parameter subgroups $U(\exp(t\xi_j))$\ in the Hilbert space \H\
are $X_j\equiv X(\xi_j)$. Let
us consider the function $Q$\ as a function of $n$\ real variables
$\{F_j\}$, i.e.\ $Q(F)\equiv Q(F_1,F_2,\dots,F_n)$\ is expressed by vector
coordinates of the linear space $\mfk g^*$. Let there exists
complete classical flow \ph t,Q~\ on \EF, and let $F(t)\equiv\mph
t,Q~(F(0))$, with $F(0):=\mbF(P_{x_0})\equiv\mbF(\rx_0)$, where
$x_0\in\mDGom$.

We intend to look for
continuously differentiable curves $t\mapsto x_t\in\mDGom\subset\mH,\
x_{t=0}$ $:= x_0$\
satisfying the following nonlinear Schr\"odinger equation:
\bequ\label{eq;3.17}
i\frac{d}{dt}|x_t\rb=\sum_{j=1}^n\frac{\partial}{\partial F_j}
Q(\lb x_t|X_1|x_t\rb,\lb x_t|X_2|x_t\rb,\dots,\lb x_t|X_n|x_t\rb)\dti
X_j|x_t\rb,
\end{equation}
where the quantities $\lb x_t|X_j|x_t\rb$\ are inserted for the components
$F_j$\ of $F\in\mfk g^*$\ into
\[ \frac{\partial Q(F)}{\partial F_j},\quad F\in\mfk g^*.\]
It depends on the choice of the group $G$, and of its
representation $U(G)$, and also on the choice of realization of
the Hilbert space what a specific form this abstract differential
equation will attain: It can be partial differential equation, and
possibly also an integro--differential equation, and for nonlinear
(in variables $F_j$) function $Q$ it is always nonlinear. We shall
show, however, that in all of these cases the equation\rref3.17~
can be equivalently rewritten (for solutions lying in \DGom) in a
more transparent form of two connected problems:

(i) The problem of finding solutions of CM--problem for Hamilton's equations
on the (generalized) classical phase space $\mfk g^*$\ with its canonical
Poisson structure, and with the Hamiltonian $Q$, leading to the Poisson
flow \ph t,Q~\ on $\mfk g^*$.

(ii) Then, after insertion into the expression of \d Q,F~\ in the
equation\rref3.17~ for the argument $F$\
the appropriate solution (specified by the initial conditions)
$F(t):=\mph t,Q~(F(0))$, solving the obtained time dependent
linear Schr\"odinger equation
(resp.\ equivalently: solving\rref2.39~ to find the cocycle \gQ t,F~).

\def\nazov{{
\ref{sec;IIIE}\quad Solution of Some Nonlinear Schr\"odinger
Equations}}

Let us formulate and prove this result:
\begin{thm}\label{thm;3.8}
Let the conditions imposed above on the objects entering into the
equation\rref3.17~ are fulfilled. Then, for any $x_0\in\mDGom$, there is
a solution $\{x_t: t\in\mbR\}$\ of\rref3.17~ lying in \DGom. It can be
obtained as a solution of the time dependent linear equation
\bequ\label{eq;3.18}
i\,\frac{\rd|x_t\rb}{\rd t}=\sum_{j=1}^n\frac{\partial Q(F(t))}{\partial
F_j} X_j|x_t\rb,
\end{equation}
where $F(t)$\ is the solution of the classical Hamilton's equations
corresponding to the symplectic flow $\mphi^Q$\ on that $Ad^*(G)$--orbit
which contains the initial classical state $F(0):=\mbF(P_{x_0})$. If \gQ
t,F(0)~\ is the solution of the equations\rref2.39~, then a solution
$|x_t\rb$\ can be expressed by the relation:
\bequ\label{eq;3.19}
|x_t\rb\equiv U\bigl(\mgQ t,F(0)~\bigr)|x_0\rb.
\end{equation}
Each (global) solution of\rref3.17~ satisfies also\rref3.18~, with
$F_t := F(t),\ \lb x_t|X(\xi)|x_t\rb \equiv F_t(\xi)$\ satisfying the classical
equations: $F_t\equiv\mph t,Q~((F(0))$.\hfill\zal
\end{thm}
\begin{proof}
\DGom\ is $U(G)$--invariant, $x_0\in\mDGom$, hence also $U\bigl(\mgQ
t,F(0)~\bigr)\in\mDGom$\ for all $t\in\mbR$. The function $|x_t\rb$\
from\rref3.19~ leads to the identity
\bequ\label{eq;3.20}
\lb x_t|X(\xi)|x_t\rb\equiv\mph t,Q~F_\xi(P_{x_0}),
\end{equation}
what is a consequence of\rref2.41~,\rref2.40~, and
Definitions~\ref{df;2.17}. Hence we have $F_j(t) \equiv \lb x_t|X_j|x_t\rb$.
Differentiation of \rref3.19~ with a help of\rref2.38~,\rref2.39~, and of
the group--representation property of $U$\ gives:
\bequ\label{eq;3.dU}
\begin{split}
\frac{d}{dt}U\bigl(\mgQ t,F(0)~\bigr)|x_0\rb &=
\left.\frac{d}{ds}\right|_{s=0}U\bigl(\mgQ s,F(t)~\bigr) U\bigl(\mgQ
t,F(0)~\bigr)|x_0\rb \\ &=-i\,X(\md Q,F(t)~)U\bigl(\mgQ
t,F(0)~\bigr)|x_0\rb,
\end{split}
\end{equation}
what is the relation\rref3.18~ with $|x_t\rb$\ from\rref3.19~. Insertion
of\rref3.20~ into\rref3.18~ gives\rref3.17~, what proves that the function
$|x_t\rb$\ from\rref3.19~ solves the equation\rref3.17~.

Let $|x_t\rb$\ be some global solution of\rref3.18~ with $\|x_0\|=1$, and
fulfilling $F_j(0)\equiv\lb x_0|X_j|x_0\rb$. Then it satisfies\rref3.20~,
what follows from the differentiation of $\lb x_t|X_j|x_t\rb$\ with a help
of\rref3.18~,\rref2.27~, and\rref2.28~. Consequently, this $|x_t\rb$\
satisfies also\rref3.17~.

Conversely, let $|x_t\rb$\ be some global solution of\rref3.17~. Again by
differentiation of $F_j(t):=\lb x_t|X_j|x_t\rb$, one obtains, as above, the
identity (cf.\ also Notation~\ref{notat;bfPH})
\[ \frac{d}{dt}F_k(t)\equiv\sum_{j=1}^n\frac{\partial Q(F(t))}{\partial
F_j} \{F_j,F_k\}(\mbF(\mbf x_t)),\]
with $F(t)\equiv\mbF(\mbf x_t)$. This means that each global solution
of\rref3.17~ fulfills also the equations\rref3.18~ and\rref3.20~.
\end{proof}

We can see that all the norm--differentiable solutions of\rref3.17~
conserve their norms: $\lb x_t|x_t\rb \equiv\lb x_0|x_0\rb$, since
the generator on the right side of\rref3.18~ is selfadjoint for all
$t\in\mbR$\ (because it belongs to the generators of $U(G)$). It follows
also that, for $\|x_0\|=1$, one has $F_j(t):=Tr(P_{x_t}X_j) \equiv\lb
x_t|X_j|x_t\rb$.



\section{On an Alternative Formulation of NLQM}
\label{sec;IIIF}

It might be fair and also useful to look onto another, a rather popular
formulation of general NLQM (on the set of pure states \PH) published
by Weinberg in~\cite{weinb}. His proposal contained some ambiguities, and
also it had some physically unacceptable consequences discussed
already in literature,
cf.\ e.g.~\cite{noncausal,mobil}.\footnote{In the presented formulation of
EQM, some of these ``unacceptable consequences'' remain valid, as it was
discussed, e.g.\ , in Subsection~\ref{q-phsp;mixt}. We have overcome here, as
the present author believes, at
least the difficulties connected with the inappropriate work with mixed
states and subsystems (resp. composed systems; these we do not try to
introduce here as a general concept) in the
Weinberg's papers. We also proposed a consistent interpretation scheme, in
which possible ambiguities are well understood.}
   Its mathematical framework can be,
however, consistently presented if it is restricted to \PH. In that case,
it is in fact equivalent to our formulation of NLQM on \PH.

Weinberg mostly worked with finite dimensional Hilbert spaces, and he used
formalism depending on
components in a chosen basis of Hilbert space. We shall try to reformulate
the Weinberg's theory~\cite{weinb} in a coordinatefree way, but simultaneously
preserving, as far as possible, the main ideas\footnote{We mean here mainly
the {\em formalism determining --
mathematical ideas}, as they were understood by the present author.}
of the original formulation.

Let the {\em nonlinear observables (and generators)} be
differentiable functions $a,b,\dots$, of two variables $x\in\mH$,
and $y^*\in\mH^*$\ from the Hilbert space \H\ and its dual (here
$y^*$ corresponds to $y\in\mH$ via the Riesz lemma,
$y^*(x)\equiv(y,x)$): $(x;y^*)\mapsto a(x,y^*)\in\mbC$. It is
assumed that the functions $a,\dots$, are homogeneous of the first
degree in each of the variables, i.e. \bequ\label{eq;homog}
a(\mlam x,y^*)\equiv a(x,\mlam y^*)\equiv \mlam a(x,y^*),\
\forall\mlam\in\mbC\setminus\{0\}.
\end{equation}
Another requirement is the ``reality condition'':
$a(x,x^*)\in\mbR,\forall x\in\mH$.
A specific ``observable'' is $n(x,y^*):=y^*(x)\equiv(y,x)$; the observables
$a,b,\dots$, corresponding to traditional ``observables'' of QM determined
by selfadjoint operators $A,B,\dots$, are of the form
$a(x,y^*)\equiv(y,Ax),\dots$, where $\mlam(y^*)\equiv(\overline\mlam y)*$\
(since the bijective mapping $x\mapsto x^*$\ of \H\ onto $\mH^*$\ according
to the Riesz lemma is antilinear). In Ref.~\cite{weinb}, only values
$a(x,x^*),\dots$, of observables $\ia,b,\dots$, in ``diagonal'' points
$(x;x^*)\in\mH\times\mH^*$\ corresponding to a specific vector state
$\ix\in\mH$\ are used, except of the instants when $a,b.\dots$, are
differentiated according to only one of the variables $\ix,\iy^*$\ (in
points with \iy\ = \ix). The Fr\'echet differentials are
\bequ\label{eq;3.21a}
\begin{split}
\mD a,x~:=&\left.\frac{\partial a(x,y^*)}{\partial x}\right|_{y=x}\in\mH^*,
\\ D_x^*a:=&\left.\frac{\partial a(y,x^*)}{\partial x^*}\right|_{y=x}\in
\mcl L(\mH^*,\mbC)=\mH.
\end{split}
\end{equation}
Then we can write the nonlinear Schr\"odinger
equation,~\cite[(b):Eq.(2.12)]{weinb}, with a generator $h\equiv h(x,y^*)$\
in the form:
\bequ\label{eq;3.21}
i\,\frac{\partial x(t)}{\partial t} =D_{x(t)}^*h.
\end{equation}
As concerns the interpretation, let us only mention that the expectation
value of an observable \ia\ in the state described by a vector $x\neq 0$\
is expressed by the number
\bequ\label{eq;3.22}
\ra(\bx):=\frac{a(x,x^*)}{n(x,x^*)}.
\end{equation}

\def\nazov{{
\ref{sec;IIIF}\quad On an Alternative Formulation of NLQM}}

\noidt This is in accordance with our interpretation from\rref2.42d~,
and\rref2.42c~: The function \ra\ in\rref 3.22~ depends on elements
$\bx\in\mPH$\ only; it can be identified with one of our observables and/or
generators restricted to \PH. In the case of finite--dimensional \H, any
\ra\ in\rref 3.22~ can be written as a function
$\tilde a(\rf_1,\rf_2,\dots,\rf_n)$\ of a finite number of quantities
$\rf_j(\bx)$
given by an equation:
\bequ\label{eq;3.23}
\rf_j(\bx):=\rf_j(x,x^*),\ {\rm with}\ \rf_j(x,y^*):=\frac{(y,{\rm
X}_jx)}{(y,x)},
\end{equation}
with ${\rm X}_j\in\mLHs$. In the finite dimensional case, we can insert
into the nonlinear Schr\"odinger equation\rref3.21~ the function
\[
h(x,y^*):=n(x,y^*)Q(\mbbf(x,y^*)),\quad\mbbf:=(\rf_1;\rf_2;\dots;\rf_n),\]
where we write $Q$\ instead of $\tilde h$\ from the text above\rref3.23~.
Let also $\mbF(\bx):=\bbf(x,x^*)$, with components $F_j(\bx)$.
An easy computation then gives:
\bequ\label{eq;3.24}
\begin{split}
D^*_{x(t)}h&\equiv\sum_{j=1}^n\frac{\partial
Q\bigl(\mbF(\bx(t))\bigr)}{\partial F_j}X_j|x(t)\rb \\
&+\left(Q\bigl(\mbF(\bx(t))\bigr)-\sum_{j=1}^n \frac{\partial
Q\bigl(\mbF(\bx(t))\bigr)}{\partial F_j}F_j(\bx(t))\right)|x(t)\rb,
\end{split}
\end{equation}
what expresses the right hand side of the nonlinear Schr\"odinger equation
written in the form\rref3.21~ in accordance with Ref.~\cite{weinb}.
The notation in\rref3.24~ literally corresponds to that introduced in
Section~\ref{IIC;symm-obs}, because the selfadjoint operators
$\{X_j:j=1,\dots,n\}$\ in finite dimensional Hilbert space \H\ generate a Lie
algebra of operators $dU(\mfk g)$\ of a (finite--dimensional) simply connected
Lie group $G$\
with representation $U(G)$\ in \H\ generated by (integration of) $dU(\mfk
g)\ni X_j$, and \bF\ is then the corresponding momentum mapping, cf.
also~\cite[Sec.IV]{bon2}. Direct inspection shows that if
$t\mapsto\bx(t):=P_{x(t)}$\ corresponds to a solution of\rref3.21~, then
the function $t\mapsto\mbF(\bx(t))$\ is solution of Hamilton's equations
on $\mfk g^*$\ with Poisson brackets
\[ \{F_j,F_k\}(\mbF(\bx)):=i\,Tr(P_x[X_j,X_k]), \]
with the Hamiltonian
function $\mcbF\mapsto Q(\mcbF)$, in correspondence with the canonical
Poisson structures on \PH\ and on $\mfk g^*:= Lie(G)^*$. Let us denote
\[ \malp(\bx(t)):=Q\bigl(\mbF(\bx(t))\bigr) -\sum_{j=1}^n \frac{\partial
Q\bigl(\mbF(\bx(t))\bigr)}{\partial F_j}\,F_j(\bx(t)). \]

Since this is a real numerical function of time (for a given solution
$x(t),\ t\in\mbR$), any solution $|x(t)\rb$\ of\rref3.21~ can be
transformed into a solution $|x_t\rb$\ of the corresponding equation of the
form\rref3.17~ by a gauge--transformation, namely by multiplication of the
vectors $|x(t)\rb$\ by a phase factor $\exp(i\,\beta(t,x_0))$:
\[  |x_t\rb\equiv \exp(i\,\beta(t,x_0))|x(t)\rb,\]
where the phase $\beta(t,x_0)$\ is a solution of the equation
\bequ\label{eq;3.25}
\frac{\rd\beta}{\rd t}=\malp(\bx(t)),
\end{equation}
corresponding to the initial condition $x(0):=x_0$. The two solutions,
$|x_t\rb$\ of\rref3.17~, and $|x(t)\rb$\ of\rref3.21~, corresponding to the
same initial condition $x(0)=x_0$\ are mutually physically indistinguishable.
A comparison of the ``Weinberg type'' nonlinear Schr\"odinger
equations with that of geometric formulation of QM (essentially identical
with the ours one, as appeared already in~\cite{bon10})
was presented in~\cite{ashtekar}.


\vspace*{\fill}
\newpage

\appendix

\chapter{Selected Topics of Differential Geometry}\label{A;geom}
\def\autor{\ref{A;geom}\quad Selected Topics of Differential Geometry}

 We shall give in this appendix a brief review of
some basic definitions, illustrative examples, and some facts
(theorems) concerning the elements of differential geometry and some of related
topics.

\section{Introduction to topology}\label{A;topol}
\def\nazov{{
\ref{A;topol}\quad Introduction to topology}}
The general concept of topology is basic for mathematical description of
continuity, stability, connectedness, compactness, etc. This concept is
useful for clear understanding of several issues of this paper.

Let, for a given set \cl X the collection of all
its subsets (i.e.\ the \emm power set of \bs{\mcl X}~) be denoted by \glss
\cl{P(X)}~.
\begin{defs}[{\bf Topology}\index{topology}]\label{Adf;top}
\item{(i)} A \emm topology~ on the set \cl X is a collection $\mcl T\subset
\mcl{P(X)}$ of
subsets $\mcl U,\mcl V,\dots \subset\mcl X$ satisfying:
\item{\ \ \ t1.} Union of an arbitrary set of members of \cl T also belongs
to \cl T.
\item{\ \ \ t2.} Intersection of an arbitrary finite set of members of \cl
T is a
member of \cl T.
\item{\ \ \ t3.} The empty set $\emptyset$, as well as the whole \cl X, are
 members of \cl T.\vspace{3pt}\nl
\noidt The elements $\mcl U\in\mcl T$\ of the given topology \cl T\ are the
\emm open sets~ (in this
specific topology!). The complements $\mcl X\setminus\mcl U$ are called the
\emm closed sets~.   Topologies $\{\mcl T_{\gamma}\}$
are naturally ordered by inclusion: $\mcl T_1\prec \mcl T_2$ iff
$\mcl T_1\subset \mcl T_2$, and, in this case, \emm $\mcl T_2$ is stronger
($\equiv$ finer) than $\mcl T_1$~\ (also: \emm $\mcl T_1$ is weaker
($\equiv$ coarser) than $\mcl T_2$~). The set of all possible topologies on
\cl X is a
directed set; it is, moreover, a complete lattice (i.e.\ each subset has
supremum and infimum). The strongest of all topologies is
the \emm discrete topology~ for which each subset of \cl X is both open and
closed (=:\ \emm clopen sets~). The weakest topology is the trivial one:
only open
(and closed) subsets of \cl X are the empty set $\emptyset$, and the whole
space \cl X. For any subset $\mcl M\subset\mcl X$, for a given topology,
there is a unique minimal
(\wrt the set inclusion) closed subset $\overline{\mcl M}$ of \cl X\
containing \cl M, called the
\emm closure of \bs{\mcl M}:\ \bs{\overline{\mcl M}}~; as well as there is
a unique maximal open subset of \cl X\ contained in \cl M, called the \emm
interior of \bs{\mcl M}~, denoted by \bs{\mcl M^\circ}. If the closure of
\cl M\ is the whole space \cl X, then \emm \cl M\ is dense in \cl X~.
Given an arbitrary subsystem $S\subset\mcl{P(X)}$, there
is a minimal topology on \cl X containing $S$; it is the \emm topology
generated by $S$~.
The couple (\cl X;\cl T) is a \emm topological space~, or also the {\bf
topological space \bs{\mcl X}}. If
cardinality of a dense subset of \cl X\ is at most countable, then the
\emm topological space \bs{\mcl X}\ is separable~.
Any subset \cl M\ of \cl X, such that $x\in\mcl M^\circ$, is a \emm\nbhd of
 \bs{x\in\mcl X}~.
\item{(ii)}  Any subset $\mcl Y\subset\mcl X$ of the topological space (\cl
 X;\cl T)
is endowed with the \emm relative ({\rm or} induced) topology~ $\mcl
T_{\mcl Y}:=\{\mcl Y\cap\mcl V: \mcl V\in\mcl T\}$. With this topology, the
 subset
\cl Y is a \emm topological subspace~ of \cl X.
\item{(iii)} A topological space is \emm disconnected~ iff it is union of
two nonempty disjoint open (equivalently: closed) subsets. In the opposite case
it is \emm connected~. The union of all connected topological subspaces
each of which contains the point \rh\ is the \emm connected component~ of the
point $\mrh\in\mcl X$.
\item{(iv)} A topological space (\cl X;\cl T) is \emm compact~ iff {\em for
any}
collection $\{\mcl V_j: j\in J\}\subset\mcl T$ (with $J$ an arbitrary index
set) covering \cl X: $\cup_{j\in J}\mcl V_j = \mcl X$, there exists a
\emm finite subcovering~, i.e.\ there is a finite subset $K\subset J$ such,
that $\cup_{j\in K}\mcl V_j = \mcl X$. A subset $\mcl Y\subset\mcl X$ of
any topological space (\cl X;\cl T) is
compact, if it is compact in the relative topology.
\item{(v)} Topologies used usually in analysis are \emm Hausdorff~, i.e.\ for
any two distinct points \rh, \sg\  of the considered topological space there
are disjoint open sets $\mcl V_{\mrh}, \mcl V_{\msg}$ each containing one
of the
chosen points. This is one of the types of possible topologies which \emm
separate points~ of topological spaces, cf., e.g.~\cite{kiril}.
In Hausdorff spaces, each one--point set is closed, and any compact subset
is also closed. A Hausdorff space \cl X\ is \emm locally compact~ iff each
point
$x\in\mcl X$\ has a compact neighbourhood.

\item{(vi)} Let a topological space \cl X be decomposed into a collection of
its (mutually disjoint nonempty) subsets: \cl X = $\cup_j N_j: j\in J$, the
decomposition being denoted by $N$. Let us form the \emm factor--space $\mcl
X/N$~ (resp.\ also the \emm quotient--space~) the points of which are the
subsets $N_j$ (it is essentially equivalent
to the index set $J$ -- as a set).
Let $p_N$ be the natural projection of \cl X onto \cl X/$N$, $x\in
N_j\eequiv p_N(x)=N_j$.
The natural topology on \cl X/$N$, the \emm factor--topology~ (resp.\ \emm
quotient--topology~), is the
strongest topology for which $p_N$ is continuous, cf.\ the
Definitions~\ref{Adf;cont}.
\item{(vii)} Let \cl X, \cl Y be topological spaces, \cl{X\times Y} be their
Cartesian product, i.e.\ the set of ordered couples $(x;y), x\in\mcl X, y\in
 \mcl Y$.
The
\emm product topology~ is generated on this space by Cartesian products of
 all the couples of their open subsets $\mcl U\times\mcl V, \mcl U\in\mcl
{T_X}, \mcl V\in\mcl{T_Y}$. This concept is uniquely extended to products
of any finite numbers of topological spaces (by associativity of the
Cartesian product).\hfill\pika
\end{defs}

\def\autor{{
\ref{A;geom}\quad Selected Topics of Differential
Geometry}}

The perhaps most important ``topological'' concept is that of continuity.
\begin{defs}[{\bf Continuity}\index{continuity}]\label{Adf;cont}
\item{(i)} Let \rf\ be a mapping (i.e.\ a function) from a topological space
$(\mcl X, \mcl{T_X})$, into $(\mcl Y, \mcl{T_Y})$, $\rf:\mcl X\mapsto \mcl
Y$. Then
\rf\ is \emm continuous~
iff $\rf^{-1}(\mcl U)\in\mcl{T_X}, \forall \mcl U\in\mcl {T_Y}$.
\item{(ii)} The mapping $\rf:\mcl X\rarw\mcl Y$ is continuous in the point
$x\in\mcl X$\ iff for any open \nbhd \cl U of $\rf(x)\in\mcl Y$,
$\rf(x)\in\mcl U$, there is an open \nbhd \cl V of $x,\ \mcl V\ni x$ such,
that its image under \rf\ is contained in \cl U:\ $\rf(\mcl V)\subset\mcl
U$.
\item{(iii)} Any continuous bijection f of a topological space \cl X onto
another
topological space \cl Y such, that its inverse f$^{-1}$ is also continuous
is a \emm homeomorphism~ of the spaces \cl X and \cl Y. Spaces mutually
homeomorphic are indistinguishable from the topological point of view - they
are \emm topologically isomorphic~.
\item{(iv)} Any given set of functions $\{f_j: \mcl X\rarw \mcl Y_j, j\in J\}$,
where $\mcl Y_j$ are arbitrary topological spaces, determines a unique
topology on \cl X such, that it is the weakest topology for which all the
functions $\{f_j, j\in J\}$ are continuous. This topology on \cl X is the \emm
topology determined by the functions~ $\{f_j, j\in J\}$.\hfill\pika
\end{defs}

\begin{exmp}[Various topologies]\label{Aex;top}
We shall introduce here some examples of to\-po\-lo\-gies.
\item{(i)} The topology on a metric space (\cl X,$d$) generated by the
\emm open balls~ $B_{\mveps,x}:=\{y\in \mcl X: d(x,y)<\mveps\}\ (x\in\mcl
X, \mveps>0)$\ is the \emm metric
topology~. For \cl X := \bR, the metric topology given by the \emm distance
function~ $d(x,y)\equiv |x-y|$ is the ``usual topology''. The metric
topology is always Hausdorff. The ``usual'' topology on $\mbR^n$ is the
product topology of $n$ copies of the spaces \bR\ with their ``usual
topologies''. The complex line \bC\ is considered as topologically
equivalent (i.e.\ homeomorphic) to $\mbR^2$.
\item{(ii)} Very different kind of topology on \bR\ is the following one: Let
the open sets on \bR\  be $\{x\in\mbR: x<a\},\ \forall
a\in\mbR$. The obtained topology is not Hausdorff. Observe that there are
no nonempty mutually disjoint open subsets now. This implies that the only
continuous
real-valued functions (where the image-space \bR\  is endowed with the ``usual
topology'') on \bR\ endowed with this topology are constants.
\item{(iii)} Consider the identity mapping $id_{\mbR}:x\mapsto x, \forall
x\in \mbR$; it is discontinuous if the image-topology is finer than the
``domain-topology''. E.g., the identity mapping on an arbitrary set \cl X from
discrete to arbitrary
topology is continuous, and the inverse mapping is also continuous only in the
case, if both copies of the mapped set are endowed by the same (now
discrete) topology.
\item{(iv)} Let $\mcl B(I)$\ be any set of real--valued functions (i.e.\ the
image--space \bR\ is endowed with the ``usual'' topology) on the unit
interval $I:=\{r\in\mbR:0\leq r\leq 1\}$. It generates the weakest topology on
$I$ such, that all functions $\rf\in\mcl B(I)$ are continuous. Consider, e.g.
the cases, where $\mcl B(I)$\ contains also some characteristic functions of
subintervals of $I$: such a topology makes the interval $I$ disconnected.
\hfill\dovi\end{exmp}
We are often working with linear spaces endowed with some topologies.
Finite--dimensional spaces $\mbR^{3N}$ of particle configurations, as well as
infinite--dimensional spaces of functions with values in linear spaces (with
pointwise additions),
are linear spaces in a natural way . To be useful in dealing with linear
mappings, topologies introduced on such spaces should be in a
``correspondence'' with the existing linear structures on them.\pagebreak
\begin{defs}[{\bf Topological linear spaces}~\index{topological linear spaces}]\label{df;top-ls}
\item{(i)} Let \cl L\ be a linear space over ${\Bbb K}\in\{\mbR;\mbC\}$,
where \bK\ is considered with its canonical (:=``usual'') topology. Let a
topology \cl T\ on \cl L\ be given.
Let us consider the multiplication of elements $\ix\in\mcl L$ by scalars
$\mlam\in\mbK$ as mapping from the topological product--space
$\mbK\times\mcl L$ into $\mcl L:\ (\mlam;\ix)\mapsto \mlam\ix\in\mcl L$, and
the addition: $(\ix;\iy)(\in\mcl L\times\mcl L)\mapsto\ix+\iy(\in\mcl L)$,
also with the product--topology of $\mcl L\times\mcl L$.
Then the topological space $(\mcl
L;\mcl T)$ is a \emm topological linear space (=t.l.s.)~ iff the addition and
multiplication by scalars are (everywhere) continuous functions. This
allows us to define any topology of a topological linear space
on \cl L\
by giving just all the open sets containing an arbitrarily chosen point
(e.g.\ \ix = 0).
\item{(ii)} Most often used in applications are such t.l.s. which are
Hausdorff, and their topology is determined by \emm seminorms~: T.l.s. \cl L\ is
\emm locally convex space (= l.c.s.)~ iff its topology is determined by a set
$\{p_j:j\in J\}$ of mappings (=seminorms) $p_j:\mcl L\rarw\mbR_+,\ \ix\mapsto
p_j(\ix)\geq 0$\ such that
\[ p_j(\mlam\ix)\equiv|\mlam|p_j(\ix),\ p_j(\ix+\iy)\leq
p_j(\ix)+p_j(\iy),\ \forall \ix,\iy\in\mcl L,\ \forall j\in J.\]
It is supposed (to be the topology Hausdorff) that the set of seminorms is
``sufficient'', resp.\ that it \emm separates points~:
\[ \forall\ix\in\mcl L,(\ix\neq 0)\,\exists j\in J: p_j(\ix)> 0.\]
The topology is the weakest one for which all the seminorms are continuous.
On finite--dimensional linear spaces there is just one such a \emm
l.c. (locally convex)--topology~.
\item{(iii)} If the topology of l.c.s. \cl L\ is determined by just one
seminorm $p_\malp$, it is necessarily a \emm norm~ (i.e.
$p_\malp(\ix)=0\imply\ix=0$). A norm topology is naturally metric topology
with the distance function $d(x,y):=p_\malp(x-y)$. If
the space is complete as the metric space, \cl L\ is called a
\emm Banach space~, simply \emm B--space~. The norm of $\ix\in\mcl L,\
p_\malp(\ix)$ will be usually denoted by
\N\ix,\malp~, where the index \alp\ can distinguish different norms on \cl L.
\item{(iv)} Let \cl L\ be a B--space, its norm being denoted \N\cdot,{}~.
A linear mapping $\mrh:\ x(\in\mcl L)\mapsto\mrh(x)\equiv\lb\mrh;x\rb\in\mbK$\
is a \emm linear functional~ on \cl L. On general
(infinite--dimensional) B--spaces, there are also discontinuous linear
functionals. The set of all \emm continuous linear functionals on \cl L~ is
denoted by $\mcl L^*$, and it is called the \emm topological dual (space) of
\cl L~. In $\mcl L^*$, there is a canonical norm--topology determined by that
 of \cl L:
\[ \mN\mrh,{}~\equiv\sup\left\{\frac{|\mrh(\ix)|}{\mN\ix,{}~}:0\neq\ix\in\mcl
L\right\},\quad \mrh\in\mcl L^*.\]
With this norm, $\mcl L^*$\ is a B--space. Its dual space $\mcl L^{**}$\
contains, as a canonically isometrically embedded subspace, the original
B--space \cl L: $\ix\in\mcl L$ is interpreted as the mapping
$\mrh\mapsto\mrh(\ix)\equiv\lb\mrh;x\rb$, i.e.\ an element of $\mcl L^{**}$.
\item{(v)} Let \cl M\ be a linear set of linear functionals on a linear
space \cl L. Assume, that \cl M\ separates points of \cl L, i.e.
$\mrh(\ix)=0, \forall\mrh\in\mcl M\imply\ix=0$. The topology on \cl L\
determined by all $\mrh\in\mcl M$\ is called the \emm \cl M--weak topology
on \cl L~, or the \glss \sg(\cl L,\cl M)~--topology. If we consider \cl L\
as linear functionals on \cl M, and if \cl L\ separates points of \cl M,
then we have also the \sg(\cl M,\cl L)--topology on \cl M. If \cl L\ is a
B--space, the $\msg(\mcl L,\mcl L^*)$--topology is the \emm weak topology on
\cl L~. \label{sLM}
The $\msg(\mcl L^*,\mcl L)$--topology on the dual space $\mcl L^*$\ is
called the \emm $w^*$--topology on $\mcl L^*$~.The closed unit ball
$B_1:=\{\mrh\in\mcl L^*:\mN\mrh,{}~\leq 1\}$ of the dual to a B--space \cl
L\ is compact in the $w^*$--topology (=Banach-Alaoglu theorem).\hfill\pika
\end{defs}

\section{Elements of differentiation on Banach
spaces}\label{A;diff}

The differential calculus of mappings $f:\mfk T\rarw\mfk R$ between two Banach
spaces \fk T, and \fk R\ is largely similar to calculus in finite dimensional
spaces. A formal difference appears because of coordinate free notation, what
is useful also in the case, when the B-spaces \fk{T,R} are finite-dimensional.
We shall need mainly the case of an infinite dimensional \fk T (e.g.
$\mfk{T=T}_s$) and of the one dimensional \fk R = \bR.
Let us define the {\emm Fr\'echet differential~}
$D_{\nu}f$ at the point $\nu \in\mfk T$
of an \fk R-valued function $f:\mfk T\rarw \mfk R$:
\begin{defs}\label{def;frechet}
\item{(i)}
 Let \fk{T,R} be Banach spaces with (arbitrary) norms \N\cdot,{}~
(equally
denoted for both spaces) leading to
their Banach-space topologies. Let $U\subset \mfk T$ be an open subset
containing $\nu$. The {\em Fr\'echet differential}, resp.\ the \emm Fr\'echet
derivative~, of $f$ at the point
$\nu\in\mfk T$ is the unique (if it exists) continuous linear mapping
$D_{\nu}f:\mfk T\rarw\mfk R,\ \eta\mapsto D_{\nu}f(\eta)\ (\forall
\eta\in\mfk T)$
satisfying
\begin{subequations}\label{eq;derivative}
\begin{equation}\label{eq;frechet}
\lim_{\eta\rarw 0}\mN{\eta},{}~^{-1}\,\|f(\nu+\eta)-f(\nu)-D_{\nu}f(\eta)\|=0.
\end{equation}
If the derivative $D_{\nu}f$ exists, the function $f$ is
\emm differentiable at the point~ $\nu$. If the \emm derivative~\glo{\D f,{}~}
$Df:\nu\mapsto
D_{\nu}f \in \mcl L(\mfk T,\mfk R)$\glo{\D f,{\nu}~} exists in $U$, $f$ is
{\bf differentiable on \bs U};
if, in that case, $U=\mfk T$, then $f$ is called {\em Fr\'echet
differentiable function}, or just: $f$ is \emm \bs F--differentiable~.
\item{(ii)}
Let, by the above notation, the derivative of $t\mapsto f(\nu+t\eta)$:
\bequ\label{eq;gateaux}
D_{t=0}f(\nu+\cdot\eta)(1)\equiv
\left.\frac{\rd f(\nu+t\eta)}{\rd t}\right|_{t=0} =: Df(\nu,\eta),\quad
\forall\eta\in\mfk T,
\end{equation}
\end{subequations}
exists for all $\nu\in U$. Then $f$ is \emm G--differentiable~,
$Df(\nu,\cdot)$\ is \emm Gateaux derivative~
of $f$ at $\nu\in U$, and its value $Df(\nu,\eta)$ is the \emm derivative of
\bs f\ at \bs\nu\ in the direction \bs\eta~.

F--differentiability implies G--differentiability, and then it is
$Df(\nu,\eta)\equiv D_\nu f(\eta)$. Conversely, if $f$ is
G--differentiable in $U=U^\circ\subset\mfk T$, if the
 G--derivative
$\eta\mapsto Df(\nu,\eta)$\ is bounded linear\footnote{Let us stress, that
the linearity of G--derivative is a nontrivial requirement in general
B--spaces.} for all $\nu\in U$, and if the
function $\nu\ (\in U)\mapsto Df(\nu,\cdot)\ (\in\mcl L(\mfk T,\mfk R))$\ is
continuous, then $ Df(\nu,\eta)\equiv D_{\nu}f(\eta)$, cf.\ \cite[Lemmas
1.13--1.15]{jt-schw}.
\hfill\pika\end{defs}

In formulation of this definition, we have included also important
assertions on
uniqueness, and on the relation of the two concepts,~\cite{h-cartan}.

\def\nazov{{
\ref{A;diff} Elements of differentiation on Banach spaces}}
It is seen that the derivative $f\mapsto\mD f,\cdot~$\ is a linear
operation (functions $f$
with values in a linear space \cl R form naturally a linear space).

\begin{note}\label{note;frechet}
\item{(i)} In finite--dimensional case, i.e.\ for $\dim\mfk T<\infty$,
and also $\dim\mfk R<\infty$,
$D_{\nu}f$\  (expressed in some bases of \fk T, and of \fk R) is just
 the Jacobi matrix of $f$ at $\nu$. For \fk R := \bR, the function
$\nu\mapsto D_{\nu}f$\  is the ordinary first differential of $f$ (understood
as a linear functional on \fk T: the ``differentials of coordinates
$d\nu_j,\ j=1,2,\dots \dim\mfk T$'' are coordinates of vectors in \fk T);
in the case \fk T := \bR, $D_{\nu}f(1)\in\mfk R$\
is just the derivative of the (\fk R -- valued) function $f$ according to
the parameter $\nu\in\mbR$\ (here $1\in\mbR=\mfk T$\ is the ``number $1$''
considered as a vector from \fk T).
\item{(ii)} If \fk T is a function space, the derivative \D f,{\nu}~\ is the
\emm functional derivative~, cf.\ \cite{DNF,3baby}. If $f$ is
expressed in a form of integral over the space $M$ of arguments of the
functions $\eta\in\mfk T,\ \eta:x(\in M)\mapsto\eta(x)(\in\mbR) $,
then \D f,{\nu}~\ is usually expressed as an integral kernel:
\[ \mD{f_{(x)}},{\nu}~\equiv\frac{\delta f(\nu)}{\delta\nu(x)},\quad
 \nu\in\mfk T ,\]
and \D f,{\nu}~$(\eta)$\ is the integral containing in its integrand the
function $\eta$\ (and its derivatives \wrt its arguments, denoted together
by a vector--symbol $\tilde\eta$) linearly, e.g.
\[\mD f,{\nu}~(\eta)=\int_M\mD{f_{(x)}},{\nu}~\dti\tilde\eta(x)\mu(dx).\]
To be more specific, $f$ can be here, e.g., an ``action integral'' of
the  classical field theory,~\cite{noether,landau2},
$f(\nu):=\int_ML(x,\tilde\nu(x))\rd^4x$, the functions $\nu:x\mapsto
\nu(x)\in\mbR^K$\ are  finite collections ($K<\infty$) of classical fields
on the Minkowski space $M$, and the function $L$\ is a {\em Lagrangian
density}, i.e.\ it is a numerical differentiable function of a finite number
of $rK+4$\ real variables, $x$ and $\tilde\nu(x)\in\mbR^{rK}$, attaining
values $x\in M$, resp.\ equal
to values of components of $\nu(x)$, and
of (a finite number of) their derivatives taken simultaneously in the same
point $x\in M$\ (locality): $\tilde\nu:=\{\nu,\partial^{\malp_0}_0
\partial^{\malp_1}_1\dots\partial^{\malp_3}_3\nu:1\leq\sum\malp_j\leq r\}$.
Then the derivative of $f:\ \mD f,\nu~(\eta)$,
is expressed by an integral over $M$\ of the integrand
$D_{\nu}L_{(x)}\dti\tilde\eta(x)$, what can be considered as an application
of the chain rule for composed mappings,~\eqref{eq;composed}: $\nu\mapsto
L(\cdot,\tilde\nu(\cdot))\in C(M,\mbR)$\ might be considered as a mapping
between the Banach spaces $\mfk T:=C^r(M,\mbR^K)\ni\nu$, and $\mfk B:=
C(M,\mbR)$\ (endowed with some ``appropriate'' norms,
if, e.g., the domain of integration in $M$\ is bounded), with its derivative
$D_\nu L_{(\cdot)}\ (=\text{\sl a ``multiplication operator''}\in\mcl L(\mfk T,
\mfk B)): \eta\mapsto D_\nu L_{(\cdot)}\dti\tilde\eta(\cdot)$, and the integral
is the next (linear) mapping in the chain.

\item{(iii)} The derivative~\eqref{eq;frechet} of $f:\mfk T\rarw\mbR$ in
any point $\nu$ belongs to ${\cal L}(\mfk T,\mbR)=\mfk T^*_{\mbR}$. Hence,
for $\mfk T:=\mLHs^* \supset \mS$ the derivative would be in the double
dual $\mLHs^{**}$, what is strictly larger than $\mLHs=\mfTs^*$, whereas the
space \fTs\ is the
(\bR-linear envelope of the) normal state space, i.e.\ the ``density matrices
space'', which is, in turn, the space of {\em all symmetric linear
functionals}, i.e.\ the state space of the \Ca\ \fk C of compact operators
on \H.
\item{(iv)} It might be useful to stress here a (rather trivial) fact, that
the derivative of a linear function $f\in\mcl L(\mfk T,\mfk R)$\ equals, in
any point $\nu\in\mfk T$, to the element
$f\equiv\mD f,\nu~\in\mcl L(\mfk T,\mfk R)$\ itself. \hfill\dovi
\end{note}
An important formula can be proved for differentiation of composed
mappings~\cite[Ch.1.\S 2.2]{h-cartan}. Let \fk{T,R,L}\ be three B-spaces,
and let $f:\mfk T\rarw\mfk R,\ g:\mfk R\rarw\mfk L$\ be differentiable
mappings. Then the composed mapping $h:=g\circ f:\mfk T\rarw\mfk L$\ is
differentiable, and
\begin{subequations}
\label{eq;composed}
\bequ
\mD h,\nu~\equiv \mD g,f(\nu)~\circ\mD f,\nu~.
\end{equation}
Since \D f,\nu~\ is a linear mapping from \fk T\ into \fk R, and \D
g,f(\nu)~\ is a linear mapping from \fk R\ into \fk L, we have for all
$\eta\in\mfk T$:
\bequ
\mD h,\nu~(\eta)=\mD g,f(\nu)~\bigl(\mD f,\nu~(\eta)\bigr)\equiv
\bigl(\mD g,f(\nu)~\circ\mD f,\nu~\bigr)(\eta).
\end{equation}
\end{subequations}
Specifications of these concepts lead to infinite dimensional analogs of
\emm partial derivatives~, cf.\ \cite[Chap.1,\S 5.2]{h-cartan}.

\begin{defs}\label{def;Dn}
\item{(i)} The \emm second derivative~ $D^2_{\nu}f(\cdot,\cdot)$
of the differentiable function $f:\mfk T\rarw\mfk R$,
i.e.\ the first derivative of the function $\mD f,\cdot~:\mfk
T\rarw\mcl L(\mfk T,\mfk R)$, $\eta\mapsto\mD f,\eta~$,
at a point $\nu\in\mfk T$ belongs to a subspace $\mcl L_s^{(2)}(\mfk T,\mfk
R)$\ of the space $\mcl
L(\mfk T,\mcl L(\mfk T,\mfk R))$, what is canonically isomorphic to the space
$\mcl L(\mfk T\times\mfk T,\mfk R)$ of bilinear continuous functionals on $\mfk
T$. The subspace
$\mcl L_s^{(2)}(\mfk T,\mfk R):=\mcl L_s(\mfk T\times\mfk T,\mfk R)$
is the space of {\em symmetric} bilinear functionals: $D^2_{\nu}f(\mrh,\mome)
\equiv D^2_{\nu}f(\mome,\mrh)$.
\item{(ii)} Similarly as above, the $\boldsymbol{n-}${\bf th derivative}
\ind{Fr\'echet derivatives, higher}
$D_{\nu}^{(n)}f(\underbrace{\cdot,\cdot,\dots,\cdot}_{n-
\text{times}})$ is a symmetric continuous
$n-$linear functional on $\mfk T$, an element of the canonically defined Banach
space $\mcl L_s^{(n)}(\mfk T,\mfk R):=\mcl L_s(\times^n\mfk T,\mfk R)$.
\item{(iii)} The space of $k-$times continuously differentiable functions
$f$\ on
the B-space \fk T with values in \fk R will be denoted by $C^{k}(\mfk T,\mfk
R)$. The space of all
infinitely differentiable functions on \fk T will be denoted by
$C^{\infty}(\mfk T,\mfk R)$\ ($\equiv\mcl F(\mfk T)$, if $\mfk R:=\mbR$).\hfill\pika
\end{defs}
Also the notion of the Taylor expansion can be introduced similarly as in
finite--dimensional case,~\cite{h-cartan,jt-schw,3baby}.
It is clear from the point (iv) in Notes~\ref{note;frechet} that the second
derivative of any linear function (\wrt to the same argument) equals to
zero.

To deal with differential equations, resp.\ dynamical systems in different
conditions, it is useful to generalize the
differential calculus to more general spaces $M$ replacing the linear space
\fk T. Such convenient spaces $M$ are for us topological spaces endowed
with the structures called ``manifold structures'', and these $M$'s are
called ``manifolds''.



\section{Basic structures on manifolds}\label{A;manif}
\def\nazov{{
\ref{A;manif}\quad Basic structures on manifolds}}
We shall start with the concept of differentiable manifold as a basis of
further geometrical constructions, cf.\ \cite{deRham,bourb;manif}.
Intuitively, a manifold is a set ``piecewise similar'' to a t.l.s.\pagebreak
\begin{defs}\label{df;1manif}
\item{(i)} A \emm chart~ on a topological space $M$ is a triple
\glss $c:=(U;\mphi;\mcl L)$~, where $U^\circ=U\subset M$, $\mphi$\ is a
homeomorphism of $U$
onto an open subset of a Banach space \cl L. We shall often take
$\mcl L:=\mbR^n$ for some natural number $n<\infty$; in this case, the
existence of such a
chart means possibility of introducing $n$ continuous (local) coordinates on the
open subset $U$ of $M$. $U$ is the {\emm domain~} of $c$, resp.\ (also)
of $\mphi$: We shall call the ``chart $c$'' alternatively also
``the chart $\mphi$''.
\item{(ii)} A {\emm topological (Banach) manifold~} $M$ (simply: {\em a
manifold}\ind{manifold - topological}) is a
Hausdorff
topological space $M$ every point of which has an open \nbhd\ homeomorphic to
some open subset of a Banach space \cl L. This means that $M$ can be
covered by domains of charts defined on it.
\item{(iii)} A ${\boldsymbol{C^m}}$-{\emm atlas~} on a manifold $M$ is a
collection of charts
$\{c_j:=(U_j;\mphi_j;\mcl L_j): j\in J\}$ such, that the open subsets
$\{U_j: j\in J\}$ (J is an index set) cover $M$: $\bigcup_{j\in J} U_j = M$,
satisfying simultaneously the condition that for the set of homeomorphisms
\ph j,{}~  the mappings
$\mph j,{}~\circ\mph k,{-1}~:\mph k,{}~(U_j\cap U_k)\rightarrow
\mph j,{}~(U_j\cap U_k)$ are, for all $j,k \in J,\ C^m-$diffeomorphisms, i.e.
the mappings together with their inverses are m-times continuously
differentiable in all local coordinates. Two $C^m$-atlases are {\bf
equivalent}\ind{atlas - equivalent} if their union is again a $C^m$-atlas.
All the equivalent atlases compose the \emm maximal atlas~.
If all the B-spaces \cl L\ of the charts of the atlases are finite
dimensional $\mbR$-spaces, and an atlas is
$\{c_j:=(U_j;\mphi_j;\mbR^{n(j)}):\ j\in J\}$,
the numbers $n(j)$ occurring in the
specifications of charts are {\emm local dimensions~}
of $M$. For a
connected $M$ it follows that $n(j)\equiv n$ in which case $n$ is the {\emm
dimension of \bs M~}, \glss $n=\dim(M)$~.\label{dimM}
In the case of a manifold $M$\ with the
image-spaces \cl L\ being infinite--dimensional B-spaces, $M$\ is a \emm
manifold of infinite dimension~. The manifold $M$ endowed with a
$C^m$-atlas (equivalently:
with an equivalence class of $C^m$-atlases) is called a $C^m$-{\em
manifold}: The atlas(-es) defines a \emm structure of
($\boldsymbol{C^m-}$)differentiable manifold~ on $M$.
Equivalent atlases determine \emm equivalent manifold structures~ on $M$.\hfill\pika
\end{defs}
It is a theorem,~\cite{hirsch}, that on any finite dimensional
$C^m$-manifold with $m\geq 1$
there is also a $C^{\infty}$-atlas in the equivalence class defining the
manifold structure. Hence, on differentiable manifolds of finite dimension
we can always introduce local coordinates the transformations of which on the
intersections of their domains are all infinitely differentiable.
In the following, any manifold will be a $C^{\infty}$-manifold. Let us note
also that on a given (topological) manifold it might be possible to introduce
many nonequivalent differentiable structures; e.g., on the sphere $S^n$, for
$n\leq 6$, it can be introduced exactly one differentiable structure, but
for $n\geq
7$ there are several dozens of nonequivalent differentiable structures,
cf.\ \cite{kiril}.

Let us introduce now some examples.

\begin{exmp}\label{exmp;1exA}
\item{(i)} Let $M=\mbR^n$ considered with the (unique) locally convex
topology of $\mbR^n$. Let an atlas consisting of a unique chart with
domain $M$ and $\mphi$\  being the identity map be given. This atlas defines a
$C^{\infty}$-manifold structure on $M$.
\item{(ii)} Let $M:=S^n\subset \mbR^{n+1}$ be the n-dimensional unit sphere. We
can construct charts of an atlas on $M$ by stereographic projections onto
hyperplains $\mbR^n\subset\mbR^{n+1}$ orthogonal to coordinate axes: If $M$
is described by the equation $\sum_{k=1}^{n+1} x_k^2=1$, then, for the j-th
projection \ph j,{}~,  the point with coordinates $\{x_k: k=1,\dots n+1\}$
is mapped into the point $\{y_l:=2x_l/(1-x_j), l\neq j\}$, for all the points
in $\{x\in S^n: x_j\neq 1\} =: U_j$ composing the domain of \ph j,{}~.
As a simplest case of these manifolds, the circle $S^1$ needs at least two
charts to compose an atlas.
\item{(iii)} The torus $T^n=(S^1)^n$ is an example for multiply--connected
(cf.\ below)
manifold. Its charts are constructed, e.g., as Cartesian products of the
charts of circles.
\item{(iv)} Let a set $N$ be homeomorphic to the subset of $\mbR^2$
consisting of several mutually different straight lines intersecting in
some
points, with the induced topology. Then $N$ cannot be endowed with a
structure of manifold, since any point of intersection has not a \nbhd
homeomorphic to \bR\ (or to $\mbR^n$, for any $n\geq0$).\hfill\dovi
\end{exmp}
The real line~\bR\  will be always (if not mentioned contrary) considered with
its usual topology generated by open intervals. Similarly, the complex plane
\bC\ is considered with the usual product topology of $\mbR^2$. The manifold
structures of these spaces are given as in Example~\ref{exmp;1exA}(i).
We shall define now important subsets of a manifold, that are endowed with
canonically induced manifold structures.
\begin{defi}\label{df;1subm}
\item{(i)} A subset $N\subset M$ is a {\emm submanifold of \bs M~},
$dim(M)=n$, if every point $x\in N$
is in the domain $U$ of such a chart $(U;\mphi)$, that for all $x\in U\cap
N$ one has $\mphi(x) = \{x^1,x^2,\dots x^k,a^1,a^2,\dots a^{n-k}\}$, where
$\{a^1,\dots a^{n-k}\}$ is a constant in $\mbR^{n-k}$. The obvious manifold
structure on $N$\ determined by these charts is the \emm induced manifold
structure~ from the manifold $M$. Dimension of the manifold $N$\ is $\dim
N=k$.\hfill\pika
\end{defi}
The usual model of a submanifold $N$ in $M:=\mbR^n$ is realized as the
a ``surface'' in $\mbR^n$, i.e.\ as the inverse
image $f^{-1}(\{a\})=:N$ of a point $a\in \mbR^{n-k}$ by a differentiable
function $f: \mbR^n\rarw \mbR^{n-k}$ (i.e.\ $n-k$\ real differentiable
functions of $n$\ real variables) with its Jacobi matrix of constant maximal
rank on $N$; this means, that $N$ (with $dim(N)=k$) consists of roots
$x\in\mbR^n$\ of the equation
\begin{equation}\label{eq;subm}
f(x)-a=0\ (\in\mbR^{n-k}).
\end{equation}
Hypersurfaces of the
dimension $n-1$ are determined by real-valued functions $f$ on $M$ with
nonvanishing differential $df$ at
points $x$ satisfying~\eqref{eq;subm} .

Let $M, N$ be two manifolds, and let a function
$f: M\rarw N, x \mapsto f(x)$\ be given.
\begin{defi}\label{df;1difmap}
A function (resp.\ {\em mapping}) $f: M\rarw N, x \mapsto f(x) $ is
{\bf differentiable}\ind{function - differentiable}\ind{mapping -
differentiable}
in $x\in M$ iff there are charts $(U;\mphi), (V;\psi)$ on
$M, N$, respectively, with $x\in U, f(x)\in V$ such that the function
$\psi\circ f\circ\mphi^{-1} : \mphi(U) \rarw \psi(V)$ is differentiable in
$\mphi(x)$. That \emm \bs f\ is differentiable~  means differentiability
in each point $x\in M$. If $f:M\rarw N$\ is a bijection and both $f$ and
$f^{-1}$\ are differentiable, then f is a \emm diffeomorphism~ of the
manifolds $M$ and $N$.

Let $I\subset\mbR$\ be an open interval containing $0$. A differentiable
function $c: I \rarw M$ is a \emm differentiable curve
on \bs M~.\hfill\pika
\end{defi}
These concepts do not depend on a specific choice of charts in an
equivalence class of atlases. We shall mean in the following  by
``differentiability'' the infinite differentiability, if not stated
otherwise.
Differentiable mappings $f: M\rarw\mbR$
compose the space \glss $\mF(M)$~\ind{space=algebra $\mF(M)$} of infinite
differentiable real-valued functions
on $M$. The real linear space $\mF(M)$\glo{$\mF(M)$} is
also an \emm associative
algebra~ \wrt pointwise multiplication: $(fh)(x)\equiv f(x)h(x)$.

These concepts allow us to introduce an intrinsic definition of tangent
space to $M$ at a point $x\in M$. This has an advantage \wrt intuitive
notions of tangent spaces as a certain ``plains'' in some higher
dimensional linear space
containing our manifold $M$ as a submanifold: Such intuitive notions
needn't be
invariant \wrt diffeomorphisms, since after a diffeomorphic deformation of
$M$ the ``tangent plain'' might become ``tangent'' in more than one points of
$M$, or even intersect $M$ if this is not embedded in an ``appropriate
way''. Our definition is, however, physically intuitive, since it directly
defines tangent vectors as invariantly specified ``instantaneous
velocities'' of motions along curves lying on the manifold.

\begin{defi}\label{df;1tangent}
Let $c_j,j\in J$\ be differentiable curves on a manifold $M$\
through a point $x\in M: c_j(0)=x, \forall j\in J$. Let $(U;\mphi;\mcl E)$\
be a chart on $M$\ at $x\in U$. Then the derivatives
\[ \bv^\mphi_j:=\left.\frac{\rd \mphi(c_j(t))}{\rd t}\right|_{t=0}:=
D_{t=0}(\mphi\circ c_j)(1)\in\mcl E \]
exist. If they are equal for different $j\in J$, as vectors $\bv^\mphi_j$\ in
the B-space \cl E, this mutual equality is independent of a chosen chart
$\mphi$. We shall call $c_j$ and $c_k$, with $\bv^\mphi_j=\bv^\mphi_k$,
\emm equivalent curves at $\boldsymbol{x\in M}$~. Hence, the differentiable
curves at $x\in M$\ are distributed into \emm equivalence classes
$\boldsymbol{[c]_x}$~\ of curves $c$ at $x\in M$.\glo{$[c]_x$} \hfill\pika
\end{defi}

Let $\mphi$\ be a chart on $M$\ as above in\dref1tangent~, and let \cl E\
be considered as a manifold with the atlas consisting of single chart
given by the identity mapping $id_{\mcl E}$\ on \cl E. Then the equivalence
classes $[d]_\eta$\ of all curves $d_j:I_j\rarw \mcl E, d_j(0)\equiv
\eta\in\mcl E$
through $\eta$\ are in canonical bijection with vectors in \cl E\ given by
$[d]_\eta\leftrightarrow D_0d_j(1)\in\mcl E,\quad d_j\in[d]_\eta$. Any curve
$d_j$\ through $\eta:=\mphi(x)\in \mcl E$\ gives a curve $t\mapsto c_j(t)
:=\mphi^{-1}(d_j(t))$\ through $x\in M$. This helps us to see that there is a
bijection between the above defined equivalence classes $[c]_x$\ of curves
on $M$, and vectors in \cl E. Now it is possible to introduce linear
operations into the set $\{[c]_x:c\ \text{is a differentiable curve on $M$\
through}\ x\}$\ of equivalence classes of the curves, by extending the
above bijection to a linear mapping. It is important that {\em the linear
structure on the set of classes $[c]_x$\ does not depend on a chosen
chart}. This leads us to important

\begin{defs}\label{df;2tangent}
\item{(i)} Let $M$ be a differentiable manifold, $x\in M$. The above introduced
linear space of equivalence classes $[c]_x$\ of differentiable curves through
$x$ is called the \emm tangent space to $M$ at $x$~, and will be denoted by
\glss $T_xM\equiv T_x(M)$~. An element $\bv_x^c:=[c]_x\in T_xM$\ is a \emm
tangent vector~ at $x$ to $M$. If $U\subset M$ is an open subset (considered
as a
submanifold of $M$) containing $x$, we shall identify the tangent spaces
$T_xU\equiv T_xM$, since $T_xM$ is determined by ``the local structure'' of $M$.

\item{(ii)} Let $f: M\rarw N$\ be a differentiable mapping
(cf.\dref1difmap~) of manifolds. Let $c'\in [c]_x\in T_xM$. Then the
equivalence class of the curves $t\mapsto f(c'(t))\in N$\ through $f(x)$ is
independent of a representative $c'\in [c]_x$, hence the mapping $f$ induces
a well defined mapping \glss $T_xf$~\ of
classes $[c]_x$ into classes $[f\circ c]_{f(x)}\in T_{f(x)}N$:
\bequ\label{eq;1Tf}
T_xf\equiv T_x(f): T_xM\rarw T_xN,\quad \bv^c_x:=[c]_x\mapsto
T_xf(\bv^c_x):=[f\circ c]_{f(x)}.
\end{equation}
The mapping $T_xf$ is called the \emm tangent of \bs f\ at \bs x~.

\item{(iii)}
Let a manifold $M$ with an atlas $\{(U_j;\mphi_j;\mcl E_j):j\in J\}$\ be
given. Let \glss $TM$~\ be the manifold determined as the set
\[ \{[c]_x\in T_xM:x\in M\} \]
of all tangent vectors in all points of the manifold $M$, endowed by
the atlas consisting of charts
\[ \bigl(\cup\{T_xM:x\in U_j\};\Phi_j;\mcl E_j\times\mcl E_j\bigr), \]
where the mapping $\Phi_j$\ is defined:
\[ \Phi_j([c]_x):= \bigl(\mphi_j(x);T_x\mphi_j([c]_x)\bigr)\in
\mphi(U_j)\times\mcl E_j\subset\mcl E_j\times\mcl E_j. \]
In the last relation, the image of the tangent of $\mphi_j$\ on the vector
$[c]_x$ at $x\in U_j\subset M$\ equals to the derivative of the curve
$t\mapsto\mphi_j\circ c(t)\in\mcl E_j$\ in point $t=0$\ taken at ``the vector''
$1\in T_0\mbR\cong\mbR$, cf.\ Definition~\ref{def;frechet}, and
$T_{\mphi_j(x)}\mcl E_j$\ is identified with $\mcl E_j$. Moreover, let
the \emm projection \bs{\pi_M}~ be defined on the manifold $TM$ by
\bequ\label{eq;1projTM}
\pi_M:TM\rarw M;\quad [c]_x\mapsto x.
\end{equation}
The differentiable manifold $TM$\ endowed with the projection\rref1projTM~ is
the \emm tangent bundle of \bs M~. The projection $\pi_M$\ is the \emm
tangent bundle projection~ of $M$.

\item{(iv)} The tangent bundle is an example of a \emm vector bundle
\bs{(P,\pi_M,E)}~ , i.e.\ of a manifold $P$ with a differentiable mapping
$\pi_M:P\rarw M$\ onto another manifold $M$ with a given open covering
$\mcl U_M:=\{U_j:j\in J\}$ by
domains $U_j$\ of its charts, and a topological vector
(let it be Banach) space $E$ (considered with its natural manifold
structure) such that $\pi_M^{-1}(\{x\})=:E_x\subset P$\ is homeomorphic to
$E$, the homeomorphism being the restriction of a diffeomorphism of
$\pi_M^{-1}(U_j)$\ onto $U_j\times E$, and the homeomorphisms corresponding
to $j\neq k$\ and to points $x\in U_j\cap U_k$\ induce a group of linear
transformations on $E$ in a natural
way,~\cite{3baby,kob&nom,abr&mars,bourb;manif}, called the \emm structural
group~ of the bundle. Such homeomorphisms $E_x\leftrightarrow E$ allow us to
introduce a natural linear structure on all $E_x,\ x\in M$, by transferring
it from that on $E$.

\item{(v)} Let $T^*_x(M):=(T_xM)^*$\ be the topological dual of $T_xM$.
This space is called the  \emm cotangent space to M at x~.
Let us take $p$ copies of $T_x(M)$, and $q$ copies of $T_x^*(M)$, and let
us form the tensor product spaces
\bequ\label{eq;2TrM}
T^p_{qx}M:=\otimes_{j=1}^pT_x(M)\bigotimes\otimes_{k=1}^qT^*_x(M),\ x\in M.
\end{equation}
Let us denote \glss $T^p_qM\equiv \mrTpq$~ the set theoretic union of these
linear sets.
With a use of the manifold structure on $M$, they can be ``sewed
together'', i.e.\ there can be introduced a
manifold structure on the set \rTpq\ in an obvious analogy with that of
$TM$. The resulting manifold will be denoted by the same symbol \rTpq;
it will be called the (vector) \emm bundle of tensors of type
\bs{\binom{p}{q}}~, or of the tensors \emm contravariant of order \bs p~,
and \emm covariant of order \bs q~. The manifold $T^0_1(M)=:T^*(M)$\ is the
\emm cotangent bundle~ of the manifold $M$, and $T^1_0(M)=TM$.

\item{(vi)} Let a bundle $(P,\pi_M,E)$\ be given, and let $\bv:M\rarw
P,x\mapsto \bv(x)$\ be a differentiable mapping such, that
\bequ
\pi_M(\bv(x))\equiv x.
\end{equation}
Such mappings are called \emm sections of the (vector) bundle~. A section
of the tangent bundle $TM:=(TM,\pi_M,\mcl E)$\ is a \emm vector field on
\bs M~. Sections of the tensor bundle \rTpq\ are \emm tensor fields of type
\bs{\binom{p}{q}}~. The tensor fields of the type $\binom{p}{q}$\ form an
infinite dimensional vector space $\mcl T^p_q(M)$. The space of vector
fields is $\mcl T^1_0(M)$, the space $\mcl T^0_0(M)$\ is identified with
$\mcl F(M)$. The direct sum $\mcl T(M):=\oplus_{p\geq0,q\geq0}\mTpq$\ is
the \emm algebra of tensor fields~ on $M$, the algebraic operation being
the pointwise tensor product.

\item{(vii)} Let $f: M\rarw N$\ be as in (ii). The \emm tangent mapping of
\bs f~ is the mapping $Tf: TM\rarw TN$\ defined by (cf.\ eq.~\rref1Tf~)
\bequ\label{eq;Tf}
Tf: \mv x\ (\in T_xM)\mapsto T_xf(\mv x)\equiv T_xf\cdot\mv x\ (\in T_{f(x)}N).
\end{equation}
The tangent mapping is also denoted by \glss$f_*:=Tf$~. If $f$\ is a
diffeomorphism, then we denote by $f_*$\
also the {\em unique natural extension} of this mapping to the whole algebra of
tensor fields, $\mbs{f_*: \mcl T(M)\rarw \mcl T(N)}$, determined by its
``commuting with contraction'', and conserving the type
$\binom{p}{q}$,~\cite[Chap.I, Propositions 2.12 and 3.2]{kob&nom}.
\hfill\pika
\end{defs}
Any vector field \bv\ on $M$ uniquely determines a differentiation \L{\bv}~
(i.e.\ a linear mapping satisfying the Leibniz rule for its action on
products) of the associative algebra $\mcl F(M)$. Let $\bv(x)$ corresponds
to the class $[c^{\bv}]_x$\ of curves through $x\in M$, and let $c^{\bv}$\ be in
this class. Then \L{\bv}~ is defined  by the formula
\bequ\label{eq;1lie}
 \mL{\bv}~f(x):=\bv_x(f):=\left.\frac{\rd}{\rd t}\right|_{t=0}f(c^{\bv}(t)).
\end{equation}
Let us stress that this definition depends on vectors $\bv_x\in T_xM$\
only, independently of their possible inclusions as values of some
vector fields: The mapping $\bv(\in T_xM)\mapsto\mL\bv~$\ is well defined
for any fixed $x\in M$.
On finite dimensional manifolds, any differentiation on the algebra $\mcl
F(M)$\ is given by a vector field \bv\ according to\rref1lie~;
cf.\ \cite{3baby} for comments on infinite dimensional cases (cf. also
Lemma~\ref{lem;reflex}).
Hence, each vector field \bv\ determines a differential operator \L\bv~,
and the mapping $\bv_x(\in T_xM)\mapsto \mL\bv~$ is a linear injection into
the set of differential operators on the ``algebra of germs of functions
$\mcl F(M)$\ in the point $x\in M$''; this injection is also onto (i.e.
surjective) for $\dim M<\infty$. We shall often identify \L\bv~\ with
$\bv\in TM$.
The derivation \L\bv~\ can be naturally (under the requirement of
``commutativity with contractions'',~\cite{kob&nom,abr&mars}, and
of satisfaction of the Leibniz rule) uniquely extended to a derivation on all
spaces \Tpq. It acts on the vector fields as
\bequ\label{eq;2lie}
\mL\bv~\bw=[\mL\bv~,\mL\bw~]:=\mL\bv~\mL\bw~-\mL\bw~\mL\bv~\equiv\mL[\bv,\bw]~,
\end{equation}
and, for given vector fields \bv\ and \bw, it represents a vector
field,~\cite{bourb;manif}, denoted by $[\bv,\bw]$.
\begin{defi}\label{df;1lie}
The above determined mapping $\mL\bv~:\mcl T(M)\rarw\mcl T(M)$ (leaving
each \Tpq\ invariant) is the \emm Lie derivative of tensor fields \wrt
\bs{\bv\in\mcl T^1_0(M)}~. The result of its action on a vector field $\bw:
\mL\bv~(\bw)\ =\ [\bv,\bw]$ is the \emm commutator (or Lie bracket) of the
vector
fields \bv\ and \bw~. This Lie bracket satisfies the \emm Jacobi identity~;
\[ [\mL{\bf u}~,[\mL\bv~,\mL\bw~]]+[\mL\bw~,[\mL{\bf
u}~,\mL\bv~]]+[\mL\bv~,[\mL\bw~,\mL{\bf u}~]]\equiv0,\]
what is a consequence of the definition.
\hfill\pika
\end{defi}
Let us note that the mapping
\bequ\label{eq;1df}
d_xf: T_xM\rarw\mbR, \bv\mapsto d_xf(\bv):=\mL\bv~(f)\equiv\bv(f), \forall
\bv\in T_xM,
\end{equation}
is a bounded linear functional on $T_xM$  (this is a consequence of
definition of Fre'chet
differentiability of $f\in\mcl F(M)$; $d_xf$\ equals to $T_xf$, if
$T_{f(x)}\mbR\equiv\mbR$ is the canonical identification): $d_xf\in T^*_xM$.
Each element of $T^*_xM$\ has the form \glss $ d_xf$~\ of \emm differential of
the function \bs{f}~ for some $f\in\mcl F(M)$. Hence, each tensor in \Tpqx\
can be expressed as a linear combination of tensor products of the form
$\otimes_{j=1}^p\bv_j\bigotimes\otimes_{k=1}^q d_xf_k, \bv_j\in T_xM,
f_k\in\mcl F(M)$.

Any vector field $\bv\in\mcl T^1_0(M)=:\mXM$\ determines a \emm differential
equation~ on the manifold $M$, written symbolically for an \emm initial
condition~ $x(0)=x$:
\bequ\label{eq;1eqv}
 \dot x(t)=\bv(x(t)),\quad x(0):=x\in M.
\end{equation}
Its solutions are \emm integral curves of the vector field \bv~, i.e.
curves $t(\in I_x=I^\circ_x\subset\mbR)\mapsto x(t)$\ through
$x$ such that for any $t_0\in I_x$, the curve $\{t\mapsto x(t+t_0)\}\in$
``{\em the class of curves determined by} $\bv(x(t_0))$''. The open
interval $I_x$\ can be (and is supposed to be) chosen maximal. Let us
define the set $\mbs{\mcl D_{\bv}}:=\{(t;x):t\in I_x,x\in M\}$, called the
\emm domain of the (local) flow of \bv\ \bs{\in\mXM}~. There is defined on
it the mapping
\bequ\label{eq;1flow}
\mph\cdot,\bv~:(t,x)(\in\mbs{\mcl D_{\bv}})\mapsto\mph t,\bv~(x):=x(t), x(0)=x,
\end{equation}
where $x(t)$\ is the solution of\rref1eqv~; the mapping \ph\cdot,\bv~\ is
called the {\bf(local) flow of \bv}\index{local flow of \bv}. The locality means, that there might
be for some $x\in M:I_x\neq\mbR$. If for all the intervals one has:
$I_x\equiv\mbR,\forall x\in M$,
the vector field \bv\ as well as its flow are called \emm complete~. On
an arbitrary compact manifold $M$, any vector field is complete. Any (local)
flow satisfies {\em on its domain} the \emm group property~:
\bequ\label{eq;2flow}
\mph{t_1+t_2},\bv~=\mph{t_1},\bv~\circ\mph{t_2},\bv~.
\end{equation}

Vector fields are typically used to determine flows on manifolds as
solutions of the corresponding differential equations. There
are, on the other hand, other kinds of (covariant) tensor fields typically
used for integration on
manifolds. We shall not review here the integration theory on (finite
dimensional)
manifolds leading to the general Stokes' theorem generalizing the particular
Gauss', Green's, and Stokes' theorems connecting some integrals
on {\em manifolds $N$ with boundary}\footnote{A \emm manifold with
boundary~ has, besides the usual manifold charts, also charts
$\mphi_\malp$\ whose ranges
are intersections of open subsets of linear spaces $\mcl E_j$\ with their
closed ``halfspaces'',~\cite[\S 11.1]{bourb;manif}. The boundary of the
manifold consists of its points lying in inverse images of the boundaries
of the halfspaces \wrt the chart-mappings $\mphi_\malp$, cf.
also~\cite[p. 137]{abr&mars}.}
$\partial N$ with corresponding integrals on the boundary $\partial N$.
 The formal expression of the general Stokes' theorem is the following
\emm Stokes' formula~:
 \bequ\label{eq;Stokes}
\int_N d\omega=\int_{\partial N}\omega.
\end{equation}
If $N\subset M,\dim N=n$, and $M$ is a manifold ($\dim M>\dim N:=$\ the
dimension of the submanifold $N^\circ$), inducing on $N$\ its
structure of a {\em submanifold with boundary}, the objects entering into
the Stokes' formula are tensor fields $\omega\in\mcl T^0_{n-1}(M)$,\
$ d\omega\in\mcl T^0_n$\ of special kind called {\em differential forms}.
Another usage of differential forms is in formulation of some partial
differential equations on manifolds with a help of {\em exterior differential
systems},~\cite{3baby}.
 We need such tensor fields, in the present work, in
connection with Hamilton's formulation of mechanics on ``nondegenerate''
phase spaces (i.e.\ on {\em symplectic manifolds}), and also in some modified
situations (e.g.\ on {\em Poisson manifolds}).

Let us consider the
elements of $T^0_{px}(M)$\ as $p$--linear forms on $T_x(M)$, e.g.
$ d_x f_1\otimes d_x f_2\otimes\dots\otimes d_x f_p\in T^0_{px}(M)$ is
determined by specification of the mapping
\bequ\label{eq;1multil}
(\bv_1;\bv_2;\dots;\bv_p)\ \bigl(\in \times^pT_x(M)\bigr)\ \mapsto
\prod_{j=1}^p\bv_j(f_j);
\end{equation}
the space of bounded $p$--linear forms $\mcl L_p(T_xM,\mbR)$\ can be
identified with $T^0_{px}(M)$\ by the linear extension of this correspondence.
Let us introduce the \emm alternation mapping~ \bA\ of this space into
itself. Let for $\msg\in\Sigma(p):=$\ the permutation group of $p$
elements, and let $\meps_\msg=\pm 1$\ be the ``parity'' of \sg, i.e.\ the
nontrivial one--dimensional representation of $\Sigma(p)$. Let now \bA\
be the linear mapping determined by
\bequ\label{eq;2multil}
\mbA\mbf t(\bv_1,\bv_2,\dots,\bv_p):=\frac{1}{p!}\sum_{\msg\in\Sigma(p)}
\meps_\msg\mbf
t(\bv_{\msg(1)},\bv_{\msg(2)},\dots,\bv_{\msg(p)}),\forall\mbf t\in\mcl
L_p(T_xM,\mbR).
\end{equation}
One can see that this mapping is idempotent: $\mbA\circ\mbA=\mbA$.
Let us define now the subspace \bs{\Lambda^p_x}(M)\ of $T^0_{px}(M)$\ by
\bequ\label{eq;3multil}
\mbs{\Lambda^p_x}(M):=\mbA T^0_{px}(M).
\end{equation}
Let us denote \LM p\ the space of tensor fields $\mome:x\mapsto\mome_x$\
on $M$ with values
 $\mome_x\in\mbs{\Lambda^p_x}(M)$, for any integer $0\leq p<\dim M+1\leq\infty$.
Such \ome are called \emm \bs p--forms on \bs M~. We identify $0$--forms
with differentiable functions, i.e.\ $\mLM 0:=\mcl F(M)$. A useful associative
algebraic structure on the space $\mLM{}:=\oplus_{p=0}^{\dim M}\mLM p$ can
be introduced: The \emm wedge--product~ $\wedge: \mLM p\times\mLM
q\rarw\mLM{p+q}, (\mome_1;\mome_2)\mapsto\mome_1\wedge\mome_2$, where $\mLM
p:=\{0\}$, if $p>\dim M$.\label{1wedg} For arbitrary
$f_j\in\mcl F(M)$\ we define the wedge--product of their differentials (for
the consistency of various definitions of $\wedge$\ cf.\ \cite{arn1}):
\bequ\label{eq;1w-prod}
 d f_1\wedge d f_2\wedge\dots\wedge d f_p:=p!\dti\mbA ( d f_1\otimes d
f_2\otimes\dots\otimes d f_p),\ p=2,3,\dots,\dim M,
\end{equation}
where the alternation mapping \bA\ acts pointwise on $M$.
More general formula for an arbitrary  wedge--product of a
$p_1$--form $\mome_1$, and a $p_2$--form $\mome_2$ reads:
\begin{subequations}
\bequ\label{eq;2w-prod}
\mome_1\wedge\mome_2=\frac{(p_1+p_2)!}{p_1!p_2!}\mbA(\mome_1\otimes\mome_2).
\end{equation}
Then we have
\bequ\label{eq;3w-prod}
\mome_1\wedge\mome_2=(-1)^{p_1p_2}\mome_2\wedge\mome_1,
\end{equation}
\bequ
f\wedge\mome=\mome\wedge f\equiv f\cdot\mome,\ \forall f\in\mLM
0,\mome\in\mLM p\ (p=0,\dots,\dim M).
\end{equation}
\end{subequations}
\begin{defi}\label{df;p-forms}
The linear space $\mLM{}:=\oplus_{p=0}^{\dim M}\mLM p$\ endowed with the
above introduced wedge--product $\wedge$\ is called the \emm algebra of exterior
differential forms~ on $M$. Its elements lying in the subspace \LM p\ are
called \emm \bs p--forms on \bs M~; specifically, the elements of $\mLM 1=\mcl
X^*(M)$\ are \emm one--forms~, and the elements of $\mLM 0=\mcl F(M)$\ are
\emm zero--forms~.\hfill\pika
\end{defi}
Let us introduce now some operations on the algebra \LM{}, i.e.\ some linear
mappings of \LM{}\ into itself.
Let us first note that the {\bf Lie derivative} \L\bv~, as it was extended to
the
whole tensor algebra $\mcl T(M)$, leaves its linear subspace \LM{}\ invariant,
and the Leibniz rule \wrt the wedge--product is fulfilled:
\bequ\label{eq;w-lie}
\mL\bv~(\mome_1\wedge\mome_2)=(\mL\bv~\mome_1)\wedge\mome_2+
\mome_1\wedge(\mL\bv~\mome_2).
\end{equation}
Another important linear mapping $\rd:\mLM{}\rarw\mLM{}$ called the \emm
exterior differential~ is uniquely determined by the below listed
properties,~\cite[Theorem 2.4.5]{abr&mars}:
\begin{thm}\label{thm;ext-d}
The following properties determine a unique linear mapping \rd\ on \LM{}
(called the exterior differential on $M$):
\item{(i)} $\rd \mLM p\subset\mLM{p+1}$;
\item{(ii)} $\rd(\mome_1\wedge\mome_2)=(\rd\mome_1)\wedge\mome_2+
(-1)^{p_1}\mome_1\wedge(\rd\mome_2),\ \forall\mome_j\in\mLM{p_j}$;
\item{(iii)} $\rd\circ\rd\equiv0$;
\item{(iv)} For any $f\in\mcl F(M), \rd f\in\mLM 1=\mcl X^*(M)$:
$\rd f(\bv)\equiv \bv(f):=\mL\bv~(f),\ \forall\bv\in\mXM$. This means, that
the exterior differential of a function $f$ coincides with the differential
 df\ introduced above, in~\rref1df~.\hfill\zal
\end{thm}
Explicit expression of the differential \dom\ of an $\mome\in\mLM p$ given
by
\begin{subequations}
\bequ
\mome:=\sum_{j_1<j_2<\dots<j_p}h_{j_1j_2\dots j_p}\rd f_{j_1}\wedge\rd
f_{j_2}\wedge\dots\wedge\rd f_{j_p},
\end{equation}
with $h_{j_1j_2\dots j_p},f_j\in \mcl F(M)$,
is easily obtained by linearity and by the (modified) ``Leibniz rule'', as
well as by the property $\rd\circ\rd\equiv0$:
\bequ
 \mdom =\sum_{j_1<j_2<\dots<j_p}\rd h_{j_1j_2\dots j_p}\wedge
\rd f_{j_1}\wedge\rd
f_{j_2}\wedge\dots\wedge\rd f_{j_p}.
\end{equation}
\end{subequations}
\begin{defi}\label{df;i-prod}
Let a vector field \bv\ on $M$ be given, $\bv\in\mXM$. Then the linear
mapping $\mip\bv:\mLM{}\rarw\mLM{}, \mLM p\rarw\mLM{p-1}$, determined by
\[ (\mip\bv\mome)(\bv_1,\bv_2,\dots,\bv_{p-1}):=
\mome(\bv,\bv_1,\bv_2,\dots,\bv_{p-1}),\quad\mip\bv f:=0\ (\forall
f\in\mLM0) \]
is the \emm inner product of \bv\ and \bs\mome~.\hfill\pika
\end{defi}

One of the main statements of this section will be a list of mutual
relations between introduced operations on exterior differential forms.
Before quoting it, let us introduce still one transformation which allows
us to ``transfer'' differential forms (and other tensor fields) from a
manifold to another one.
\begin{defi}\label{df;1pullb}
Let $\beta:N\rarw M$\ be a differentiable mapping of a (differentiable)
manifold $N$ into a manifold $M$, and let $T\beta:TN\rarw TM$\ be its
tangent mapping. For any $p$--form on $M:\ \mome\in\mLM p$, let us define a
$p$--form $\beta^*\mome\in\mbs{\Lambda^p}(N)$ on the manifold $N$ by the
formula:
\[ (\beta^*\mome)_y(\bw_1,\dots,\bw_p)\equiv
\mome_{\beta(y)}(T_y\beta\dti\bw_1,\dots,T_y\beta\dti\bw_p),\
\forall y\in N,\bw_j\in T_yN.\]
The mapping $\beta^*:\mLM{}\rarw\mbs{\Lambda}(N)$ is called the \emm
pull--back by \bs\beta~. Let us note, that in the particular case $p=0$\ we
have for $f\in\mcl F(M)$: $\beta^*f(y)\equiv f\circ\beta(y)$.\hfill\pika
\end{defi}
We can now present a basic tool of the ``machinery'' for such a differential
computation on manifolds which does not need introducing any coordinates on
them; we shall collect also some earlier recognized relations,
cf.\ \cite{bourb;manif,kob&nom,abr&mars,3baby}.
\begin{thm}\label{thm;d-calcul}
For above defined operations on (infinitely) differentiable manifolds
represented by the symbols
$\beta^*,\mL{\boldsymbol{\cdot}}~,\rd,\mip{\cdot}$, as well as by the \emm
commutator~ (if it is defined) of any operations $\tau_j$:
\[ [\tau_1,\tau_2]:=\tau_1\circ\tau_2-\tau_2\circ\tau_1, \]
with \bv, \bw\ any differentiable vector fields on a manifold,
the following identities are valid:
\item{(i)} $[\mL\bv~,\mL\bw~]=\mL{[\bv,\bw]}~$;
\item{(ii)} $[\mL\bv~,\rd]=0$;
\item{(iii)} $[\mL\bv~,\mip\bw]=\mip{[\bv,\bw]}$;
\item{(iv)} $[\beta^*,\rd]=0$,\nl where, for $\beta\in C^\infty(N,M)$,\ \rd\
acts interchangeably on \LM{}, and on \LN{};
\item{(v)} $\rd\circ\rd=0$;
\item{(vi)} $\rd\circ\mip\bv + \mip\bv\circ\rd = \mL\bv~$;
\item{(vii)} $\mip\bv\circ\mip\bw+\mip\bw\circ\mip\bv=0$;\vspace{6pt}\nl
\noidt If $\beta: N\rarw M$\ is a diffeomorphism, and, for any
$\bv\in\mXM$, we define $\beta^*\bv\in\mXN$\ by the identity
\[ (\rd g)_y(\beta^*\bv):=\bigl((\beta^{-1})^*\rd
g\bigr)_{\beta(y)}(\bv),\ \forall g\in\mcl F(N),\ \forall y\in N,\]
then the following two items, (viii), and (ix), also express identities:
\item{(viii)}\ $\beta^*\circ\mL\bv~=\mL{\beta^*\bv}~\circ\beta^*$;
\item{(ix)} $\beta^*\mip\bv=\mip{\beta^*\bv}\beta^*$.\nl\nl
\noidt Moreover, the following elementary properties are identically valid
(with $f\in\mLM0=\mcl F(M)$, and $f\dti$\ means poinwise multiplication,
e.g.\ $f\dti\malp\equiv f\wedge\malp,\ \malp\in\mLM{}$):
\item{(x)} $\mL \bv+\bw~=\mL\bv~+\mL\bw~$;
\item{(xi)} $\mL{f\cdot\bv}~=\rd f\wedge\mip\bv+f\dti\mL\bv~$;
\item{(xii)} $\mip{\bv+\bw}=\mip\bv+\mip\bw$;
\item{(xiii)} $\mip{f\cdot\bv}=f\dti\mip\bv$.\vspace{6pt}\nl
Let us give also the following useful formula for
coordinate--free calculation of the exterior differential:
\item{(xiv)} \bequ\begin{split}
\rd\mome(\bv_0,\bv_1,\dots,\bv_p)&=\sum_{j=0}^p(-1)^j
\mL{\bv_j}~\bigl(\mome(\bv_0,\bv_1,\dots,\hat\bv_j,\dots,\bv_p)\bigr) \\
+&\sum_{0\leq j<k\leq p}(-1)^{j+k}
\mome([\bv_j,\bv_k],\bv_0,\dots,\hat\bv_j,\dots,\hat\bv_k,\dots\bv_p),
\end{split}
\end{equation}
for all $\mome\in\mLM p$, where $\hat\bv_j$\ means {\em skipping of}\ the
vector field $\bv_j$\ in the arguments, so that it is replaced by
$\bv_{j-1}$, and other arguments are also shifted by keeping their original
order unchanged.\hfill\zal
\end{thm}

We shall introduce here also the following standard terminology:
\begin{defi}\label{df;closed}
Let $\mome\in\mLM p$\ be such a $p$--form, that its differential vanishes:
$ \rd\mome=0$, hence it equals to the zero element of \LM{p+1}. In this
situation, \ome\ is a \emm closed \bs p--form~. Clearly, if $\mome=\rd\malp$\
for some
$p-1$--form \alp, then $\rd\mome=0$; for such a closed form we say, that
\ome\ is an \emm exact \bs p--form~. Let us assume
now, that $\dim M<\infty$. Since exact $p$--forms form a linear subspace in
the subspace of all closed $p$--forms, one can form the factorspace of the
later $p$--forms according to its subspace consisting of the former ones. The
resulting linear space is
denoted by $H^p(\mLM{})\equiv H^p(M)$, and it is called the \emm \bs
p--th cohomology group of \bs M~, where the group operation is the vector
addition, cf.\ \cite{kiril,3baby}.
\hfill\pika
\end{defi}
The mentioned cohomology groups are important algebraic--topological
characterizations of manifolds, but we leave it here without giving any
further comments and results, cf.\ \cite{spanier,deRham,kiril,dold,3baby}.

If there is given a bilinear continuous form $\Psi$\ on a vector space \cl E,
it determines a linear mapping $\Psi^\flat$\ from \cl E\ into its topological
dual $\mcl E^*$\ by
\bequ\label{eq;1b-map}
\Psi^\flat:\mcl E\rarw\mcl E^*,\ x\mapsto\Psi^\flat_x,\ \text{with}\
\lb\Psi^\flat_x;y\rb:=\Psi(x,y),\ \forall x,y\in\mcl E.
\end{equation}
The mapping $\Psi^\flat$\ is injective iff
\bequ\label{eq;1nondeg}
x\neq0\imply\Psi^\flat_x\neq0.
\end{equation}
In the case of finite dimensional \cl E, this condition means that \Psib\
is a linear isomorphism (hence also bicontinuous in the natural l.c.
topologies). Otherwise, \Psib\ needn't be even a bijection: it might
injectively map the space \cl E\ onto a proper subspace of $\mcl E^*$.
It is useful to distinguish several cases,~\cite{abr&mars,mars,3baby}:

\begin{defi}[{\bf Nondegenerate 2-tensors}\index{nondegenerate 2-tensors}]\label{df;2nondeg}
\item{(i)} Let the above introduced mapping \Psib\ fulfills the
condition~\rref1nondeg~. Then we say that the bilinear form
\emm \bs{\Psi}\ is weakly nondegenerate~. If \Psib\ is bijective (hence,
\cl E\ is mapped also onto $\mcl E^*$), then $\Psi$\ is called \emm strongly
nondegenerate~.

\item{(ii)} Let now $\Psi\in\mcl T^0_2(M)$\ be a two-covariant tensor field
on a manifold $M$, $\Psi:x(\in M)\mapsto \Psi_x(\in T^0_{2x}M)$.
Let us assume, that  $\Psi$\ is either symmetric (i.e.
$\Psi_x(\bv,\bw)\equiv\Psi_x(\bw,\bv),\ \forall x\in M,\bv,\bw\in\mcl X(M)$),
 or antisymmetric (i.e.
$\Psi_x(\bv,\bw)\equiv-\Psi_x(\bw,\bv),\ \forall x\in M,\bv,\bw\in\mcl X(M)$).
Then $\Psi$\ is
\emm weakly
(resp.\ strongly) nondegenerate~, if all $\Psi_x,\ \forall x\in M$, are weakly
(resp.\ strongly) nondegenerate.

\item{(iii)} Let $\Gamma\in\mcl T^0_2(M)$\ be symmetric. If it is weakly
(strongly) nondegenerate, then it is called \emm weak (strong)
pseudo--Riemannian metric on \bs M~. If $\Gamma$\ is, moreover, positive
definite (i.e.\ $\Gamma_x(\bv,\bv)>0,\forall\bv\neq0,\bv\in T_xM,\forall
x\in M$), then it
is called a \emm weak (resp.\ strong) Riemannian metric~.

\item{(iii)} Let $\Omega\in\mLM2$, and assume, moreover, that it is closed:
$\rd\Omega\equiv0$. If the two--form $\Omega$\ is weakly (strongly)
nondegenerate, it is called \emm weak (strong) symplectic form on \bs M~.
\hfill\pika
\end{defi}

The Riemannian metrics are the basic objects of Riemannian
geometry,~\cite{helgas,3baby,abr&mars}, providing a mathematical formalism
for the relativistic \emn theory of gravitation~ (i.e.\ the \emm general relativity $\equiv$ GR~),\
\cite{einst,pauli2,sachs&wu}, and it is useful also for a description of
classical
``continuous media'' (i.e.\ the phase spaces are infinite--dimensional),\
e.g.~\cite[Appendix 2]{arn1},~\cite{mars,goldin1}.
The symplectic forms are basic for (finite--, or infinite--dimensional)
classical Hamiltonian mechanics (CM), cf., e.g.~\cite{abr&mars,chern&mars,mars}.
In our extension of quantum mechanics (EQM), symplectic forms on manifolds
of density matrices generate dynamics and symmetries with a help of
scalar--valued functions (``Hamiltonians''), and simultaneously canonically
defined Riemannian metrics on that manifolds of density matrices are tools
for determination of specifically quantum probability interpretation of
the theory.

\section{Elementary concepts of Lie groups}\label{A;LieG}

We shall restrict our present brief exposition mainly to finite dimensional Lie
groups; for infinite dimensional Lie groups see,
e.g.~\cite{bourb;Lie,kriegl;michor}. Let us start, however, with some basic
definitions and
relations,~\cite{pontrjag,birk&lane,varad}, concerning general groups.\newpage

\begin{defs}[{\bf Abstract and topological groups}\index{groups}]\label{dg;gen-groups}

\item{(i)} A \emm group~ $G$ is a set with a distinguished element $e\in G$
called the \emm unit element~ of $G$, and with two mappings: (a) a
bijection of $G$ onto itself, $g(\in G)\mapsto g^{-1} (\equiv$\emm the
inverse of \bs g~); and (b) the \emm group multiplication ({\rm equiv.:}
product)~,
\[(g_1;g_2)\ (\in G\times G)\ \mapsto g_1\dti g_2\equiv g_1g_2\ (\in G),\]
 which {\bf is associative} and such, that
 \[ e\dti g=g,\ g^{-1}\dti g=e,\ \forall g\in G. \]
Then it is also $ge\equiv g,\ gg^{-1}\equiv e$. If $g\dti h=h\dti g\
(\forall g,h\in G)$, then $G$ is \emm abelian ({\rm equiv.:} commutative) group~.

\item{(ii)} A subset $H\subset G$\ such, that it is invariant \wrt taking
inverse and also \wrt group multiplication of its elements: $h_1\dti h_2\in
H,\ \forall h_j\in H,\ j=1,2$, is called a \emm subgroup of \bs G~. Any
subgroup $H$ of $G$ is a group with the induced operations from $G$.

\item{(iii)} If the group $G$ is a topological space, and the inverse
operation and group multiplication are in this topology continuous (by
which $G\times G$\ is endowed by the product topology), then G is \emm
topological group~. If $H$ is a subgroup of $G$ {\em and a closed
subspace}, it is a \emm topological subgroup of \bs G~.

\item{(iv)} Let $G$ be a (topological) group, and $H$ its (topological)
subgroup. If $gHg^{-1}:=\{ghg^{-1}: h\in H\}=H,\ \forall g\in G$, then $H$
is \emm normal ({\rm equiv.:} invariant) subgroup of \bs G~.
Since any subgroup contains the unit element $e$ of $G$, the subsets $g\dti
H\subset G, g\in G$\ (called \emm left cosets of \bs G~) cover whole $G$, and
any two of them are either equal, or disjoint: They define an equivalence
relation on $G$. The factor spaces $G/H$\ corresponding to this
decomposition of $G$ to left cosets are important in the theory of actions
of $G$ on some arbitrary spaces. Similarly, another equivalence
relation on $G$ determined by the \emm right cosets~ $\{Hg:g\in G\}$\ of $G$;
for normal subgroups $H$ (and only for them) these two decompositions of $G$
coincide. If $H$ is a normal subgroup, the space
$G/H$\ is again a (topological) group with the group multiplication
\[ (g^{-1}\dti H)\dti(g'\dti H):=(g^{-1}\dti g')\dti H,\ \forall g,g'\in
G.\]

In this case, the factor space $G/H$\ is called the \emm factor group of \bs
G by \bs H~.

\item{(v)} Let $G,G'$ be two (topological) groups and $\phi:G\rarw G'$\ be
such a (continuous) mapping, that
\[ \phi(g_1\dti g_2)\equiv (\phi g_1)\dti(\phi g_2),\ \phi e:=e';\]
the mapping $\phi$\ is a \emm group homomorphism~ of $G$ into $G'$, with
$e'$\ = the identity of $G'$. If $\phi$
is bijective (i.e.\ injective and onto) (resp.\ homeomorphism), it is called
\emm isomorphism~ of (topological) groups $G$, and $G'$.
An isomorphism of $G$ onto itself is an \emm automorphism of \bs G~. The
set of all automorphisms of $G$ forms, \wrt the group multiplication given
by the compositions of mappings, a
group \bs{\mAut G}, called the \emm automorphisms group of \bs G~. Let
any fixed $g\in G$\ be given. Then the mapping
\[ g'(\in G)\mapsto g\dti g'\dti g^{-1},\]
defines an \emm inner automorphism~ of $G$, and all of them form the \emm
group of inner automorphisms \bs{In(G)}~. The group $In(G)$\ is a normal
subgroup of \Aut G, and the factor group\nl
 \bs{\mAut G/In(G)}\ is
called,~\cite{birk&lane}, the \emm group of external automorphisms~ of the
group G.\hfill\pika
\end{defs}

\def\nazov{{
\ref{A;LieG}\quad Elementary concepts of Lie groups}}
The groups defined above are certain abstract sets endowed with their
``inner'' operations. We find usually in applications groups as some sets of
transformations of some other sets of well defined (i.e.\ formalized)
elements, e.g.\ some reversible motions of physical systems. Having defined
a group, on the other hand, we could find some transformations of a set
which act as a homomorphic image of the given group; e.g.\ a group of some
mechanical motions can act on electromagnetic field in some electronic
device. To enforce intuition
about transformations of an arbitrary (in general infinite) set, we can
imagine them as some ``permutations'' of elements of that set: The ``number
of elements'' remains the same (transformation is invertible and onto), but at
least some of elements are ``replaced to places occupied before by some
other replaced elements''.

\begin{defs}[{\bf Actions of groups}\index{actions of groups}]\label{df;G-action}

\item{(i)} The set of (all ``permutations'', i.e.\ of) all
transformations of a set $X$ form a group $\mbG(X)$; it will be called \emm
the transformations group of \bs X~. If the set $X$ is endowed by a
structure (e.g.\ topology, algebra, metrics,\dots), the subgroup of all
transformations of $X$ consisting of the transformations preserving this
structure (e.g.\ homeomorphisms, algebraic automorphisms, isometries,\dots)
will be denoted by \Aut X\ (with a corresponding specification), and called
the \emm automorphism group of \bs
X~. If there is no structure specified on $X$ (i.e.\ the only structure is
the set-structure), then we shall use $\mbG(X)$ and \Aut X\ interchangeably.

\item{(ii)} Let $G$ be a group, and let $X$ be a set. Let $T: g\mapsto
T_g\in \mAut X,\ g\in G$, be a homomorphism of $G$ into \Aut X. The
mapping $T$ is called an \emm action ({\rm or} realization) of
\bs G on \bs X~, and the space
(resp.\ set) $X$ endowed with such an action is called a \emm \bs G
--space~. For any fixed $x\in X$, the set of elements $\{y\in X:\exists
g\in G,\ T_gx=y\}$\ is called the \emm orbit of \bs T ({\rm equiv.:} of \bs G)
through \bs{x\in X}~. The belonging to orbits is an equivalence relation.
If the whole space $X$ coincides with an orbit, it is a \emm
homogeneous ({\rm equiv.:} transitive) \bs G --space~. We shall usually use
notation $g\dti x:= T_g(x):=T_gx,\ \forall x\in X, g\in G$. Each orbit of
any $G$--space is a transitive $G$--space. If $G$ is a topological group
and $X$ a topological space, the mapping $g\mapsto T_g$\ is assumed to be
continuous in a certain topology on \Aut X; usually it is assumed continuity
on the topological product
space: $G\times X\rarw X, (g;x)\mapsto T_gx$ is jointly continuous, cf.
e.g.~\cite[\S 24]{pontrjag}. If $X$ is a linear space, and $T_G\subset\mAut X=
\mcl L(X)$, then \emm\bs{T_G}\ is a representation of \bs G~.

\item{(iii)} Let $X$ be a transitive $G$--space, and let $x\in X$ be fixed.
Clearly, $G\dti x=X$. It is $e\dti x=x$, and the set of all $h\in G$\
such, that $h\dti x=x$\ forms a (closed) subgroup $H\equiv G_x$ of $G$.
The group $G_x$\ is the \emm stability subgroup (of \bs G) at \bs x~. It is
called also the \emm stationary subgroup of \bs x~. Since the left coset
$g\dti H$ consists of all the elements transforming $x$ into $g\dti x$, the
homogeneous space $X$ is isomorphic to the factor space $G/H$.
\hfill\pika
\end{defs}

\begin{defi}\label{df;LR-act}
Let $G$ be any group. Take the space $X:=G$, and define the \emm left
translation \bs{g\mapsto L_g}~ as an action of $G$ on itself by $L_g(g'):=
g\dti g'$. Then
$G$ is a transitive $G$--space. Similarly, another action of $G$ on itself
is defined by the \emm right translations~ $R_g,\ R_g(g'):=g'\dti g$, by taking
the group homomorphism $G\rarw \mbG(G): g\mapsto R_{g^{-1}}$. These two
actions mutually commute: $L_gR_h\equiv R_hL_g$. The mapping $g\mapsto
A(g):= L_g\circ R_{g_{-1}}$ is also an action of $G$ on itself,
$A(g)\in\mAut G$.\hfill\pika
\end{defi}

Let us turn our attention to Lie groups now.

\begin{defi}[\emm Lie groups~]\label{df;LieG}
Let $G$ be a manifold with such a group structure, that the group mapping
$(g_1;g_2)\mapsto g_1\dti g_2^{-1}$ is differentiable (equiv.:
continuous, equiv.: smooth, if $\dim G<\infty$) as a manifold--mapping of
$G\times G\rarw G$. The
group $G$ endowed with such a manifold structure is  a \emm Lie group~.\nl
Equivalently: The Lie group is a topological group with a $C^r$--manifold
structure  consistent with the group topology (for $\dim G<\infty$\ one need
not specify $r$).
\hfill\pika
\end{defi}
\begin{noti}\label{not;5Hilbert}
Let us note, that the mentioned equivalence (i.e.\ sufficiency of mere
topological manifold structure, and continuity of the group operations for
smoothness of these) is contents of positive solution of the fifth Hilbert
problem by Gleason~\cite{gleason1}, and Montgomery with
Zippin~\cite{mont&zip}, for $\dim G<\infty$. A partial solution is given in the
book~\cite{pontrjag}, according to which the original papers are cited
here.\hfill\dovi
\end{noti}
\begin{exmp}
The following groups are simple examples of Lie groups:
\item{(i)} Abelian connected Lie groups are $\mbR^k\times\mbT^n$, where
\bT\
is one dimensional torus (circle), and multiplication is componentwise
addition ($\mod(2\pi)$\ on the torus with a marked element).
\item{(ii)} The group $GL(n,\mbR)$ of all real invertible $n\times n$\
matrices with the matrix multiplication as the group operation, and the
topology given by continuity of all matrix elements. Also all closed
continuous subgroups of this group are Lie groups, e.g.
$O(n),O(p,q),Sp(2p,\mbR)$. Such groups $G$ of matrices $g\in GL(n,\mbR)$\ can
be obtained by specification of a matrix $A$, and by requiring,~\cite[p.
78]{kiril}: $g^TAg=A,\ \forall g\in G$.
\item{(iii)} As an example of infinite-dimensional Lie
group,~\cite[Chap.III.3.10, Proposition 37]{bourb;Lie}, let us take an infinite
dimensional
Hilbert space \H, and let \fk U\ be the group of all unitary operators on
it. Then \fk U\ is a Lie group, if taken in the norm--topology of \LH, as a
submanifold of \LH, what
can be taken, in turn, as a manifold with the single chart with the identity
mapping onto itself, as a B-space \LH.\hfill\dovi
\end{exmp}

Let us consider a Lie group $G$ with unit element $e$, and let $\xi,\eta\in
T_eG$ be arbitrary tangent vectors at $e$ to $G$. We shall construct, to each
$\xi$, a vector field \w\xi\ on the manifold $G$ by a help of left
translations, cf.\dref LR-act~, with a help of their tangent
mappings,\dref2tangent~:
\bequ\label{eq;1L-vfield}
\mw\xi(g) := T_eL_g(\xi),\ g\in G,\ \mw\xi(e):=\xi,\quad\forall\xi\in T_eG.
\end{equation}
These vector fields are \emm left invariant~, i.e.\ for any $g\in G$:
\bequ\label{eq;2L-vfield}
 L_{g*}\mw\xi=\mw\xi,\ \text{i.e.}\ T_hL_g(\mw\xi(h))\equiv \mw\xi(g\dti
 h),\ \forall h\in G,
\end{equation}
what is an immediate consequence of the definition\rref1L-vfield~. The
mapping $\xi\mapsto\mw\xi (\xi\in T_eG)$\ is linear.
Conversely, all left--invariant vector fields on $G$\ are of this form.
These vector fields are complete. Let us form a commutator, cf.\dref1lie~,
of two left--invariant vector fields, $[\mw\xi,\mw\eta]\in\mcl X(G)$. It can be
shown, that the commutator is again left invariant, hence
\bequ\label{eq;3L-vfield}
[\mw\xi,\mw\eta]=:\mw{[\xi,\eta]},\ [\xi,\eta]\equiv\mw{[\xi,\eta]}(e).
\end{equation}
This shows, that the subspace of $\mcl X(G)$\ consisting of all left--invariant
vector fields on $G$\ is also an algebra \wrt commutations.
The mapping $\mw\xi\mapsto\mw\xi(e)\equiv\xi$\ is a linear isomorphism of
the space of left invariant vector fields onto $T_eG$; they are isomorphic
also as algebras with the ``commutation'' $[\cdot,\cdot]$.

\begin{defi}\label{df;1Lie-alg}
 A linear space $X$ is a \emm Lie algebra~, if it is endowed by a
\emm Lie bracket~, i.e.\ by a bilinear mapping $[\cdot,\cdot]:X\times X\rarw X,
(\xi;\eta)\mapsto [\xi,\eta]\in X$, such that it is antisymmetric:
$[\xi,\eta]\equiv-[\eta,\xi]$, and the \emm Jacobi identity~ is
fulfilled:
\bequ\label{eq;Jacobi}
[\xi,[\eta,\zeta]]+[\zeta,[\xi,\eta]]+[\eta,[\zeta,\xi]]=0,\ \forall
\xi,\eta,\zeta\in X.
\end{equation}
The Lie bracket $[\xi,\eta]$\ is called also the \emm commutator~ of the
elements $\xi$\ and $\eta$.
A mapping $\phi$\ between two Lie algebras is a \emm Lie algebra morphism~,
if it is linear, and conserves the Lie brackets: $\phi([\xi,\eta])\equiv
[\phi(\xi),\phi(\eta)]$. If $\phi$ is a bijection, it is a \emm Lie
algebra isomorphism~.\hfill\pika
\end{defi}
 We shall next consider the Lie algebras determined by given Lie
groups.

\begin{defs}\label{df;2Lie-alg}
\item{(i)} Since the commutator of vector fields satisfies the Jacobi identity,
cf.\dref1lie~, the tangent space $T_eG$\ is naturally endowed by the Lie
algebra structure induced by that of vector fields \w\xi. This linear space
with the Lie algebra structure is the \emm Lie algebra of the Lie group \bs G~;
it will be denoted alternatively by \glss$Lie(G)\equiv \mfk g$~. It is
considered also as topological space with the topology of $T_eG$. It is
also a B-space, in this natural way, cf.\ \cite[Chap.III]{bourb;Lie}.
The \emm topological dual of \fk g~\ will be denoted $\mfk g^*=Lie(G)^*$.

\item{(ii)} Let the integral curve through $e$ of the left--invariant field
\w\xi\ be denoted by $t(\in\mbR)\mapsto\exp(t\xi)(\in G)$. This curves form
\emm one--parameter subgroups \bs{\mbR\rarw G} of \bs G~:
\[ t_1+t_2\mapsto\exp((t_1+t_2)\xi)\equiv \exp(t_1\xi)\dti\exp(t_2\xi),
\ t_j\in\mbR,\xi\in\mfk g.\]
The mapping $\xi(\in \mfk g)\mapsto\exp(\xi)(\in G)$\ is called the \emm
exponential mapping~; it is a local
homeomorphism of neighbourhoods of $0\in\mfk g$\ and $e\in G$, hence its
(local) inverse provides a chart of $G$ around $e$.\hfill\pika
\end{defs}
 Let us define now a representation of any Lie group $G$ on its Lie
 algebra \fk g. The action $A: g\mapsto A(g):=L_g\circ R_{g^{-1}}$\ of $G$
 on itself is differentiable, it leaves the unit element $e$ invariant, and
 its tangent at $e,\ T_eA(g)$, is a linear automorphism of the Lie algebra
 (identified with $T_eG$). It is an element of the wanted representation.

\begin{prop}\label{prop;1Ad}
The linear automorphisms $Ad(g):= T_eA(g):\mfk g\rarw\mfk g,\
g\in G$, form a representation of $G$ in linear endomorphisms of \fk g:
\[ Ad(g_1\dti g_2)\equiv Ad(g_1)\circ Ad(g_2),\]
(this is a consequence of the chain rule for the tangent mappings). They are
also Lie algebra automorphisms:
\[ Ad(g)([\xi,\eta])\equiv [Ad(g)\xi,Ad(g)\eta]. \]
The tangent of $Ad(\cdot)$\ in the unit element is a linear mapping denoted
by \glss{\rm ad}~, ${\rm ad}:\xi\mapsto{\rm ad}_\xi$\ of \fk g\ into \cl L(\fk
g,\fk g) such that the identity
\[ T_eAd(\xi)\dti\eta=:{\rm ad}_\xi(\eta)\equiv[\xi,\eta] \]
is satisfied.\hfill\zal
\end{prop}

\begin{defs}\label{df;2Ad}
\item{(i)} The representation $g\mapsto Ad(g)$\ is called the \emm adjoint
representation of \bs G~.

\item{(ii)} Let $F,F'\in\mfk g^*$\ be elements of the dual space of the Lie
algebra \fk g; their values on the elements $\xi\in\mfk g$\ are denoted by
$\lb F;\xi\rb\equiv F(\xi)$, etc. Then the mappings $F\mapsto
\glss Ad^*(g)~F,\ g\in G$, of $\mfk
g^*$ into itself determined by
\[ \lb Ad^*(g)F;\xi\rb := \lb F;Ad(g^{-1})\xi\rb,\quad \xi\in\mfk g,\ g\in
G,\]
form also a (linear) representation of $G$ called the \emm coadjoint
representation~ of the Lie group $G$.
\hfill\pika\end{defs}

Let the tangent spaces $T_F\mfk g^*,\ F\in\mfk g^*$\ are all
identified with $\mfk g^*$\ in the canonical way (as in any linear space).
Their duals $T^*_F\mfk g^*$\ are then canonically identified with the second
dual $\mfk g^{**}$\ of the Lie algebra, and also \fk g\ is canonically
included into $\mfk g^{**}$\ as a $\msg(\mfk g^{**},\mfk g^*)$--dense
subset, but in the norm topology it is identical with a norm-closed
subspace of the (canonically defined) B-space $\mfk g^{**}$. Since the
commutator $(\xi;\eta)\mapsto[\xi,\eta]$\ is continuous in norm (from the
continuity of Fr\'echet derivatives), it is also continuous in $\msg(\mfk
g^{**},\mfk g^*)$ topology, if \fk g\ is considered as a
$\msg(\mfk g^{**},\mfk g^*)$--dense subspace of $\mfk g^{**}$. Hence, the
double dual $\mfk g^{**}$\ is canonically endowed with a Lie bracket -- it is
also a Lie algebra. This is, clearly, trivial for $\dim G<\infty$, in what
case $\mfk g=\mfk g^{**}$, by the canonical identification.

\begin{defi}\label{df;3Ad}
Let $\mcl F(\mfk g^*)$ be the space of infinitely differentiable
functions on $\mfk g^*$. Then differentials $d_Ff\in T_F^*\mfk g^*, f\in\mcl
F(\mfk g^*)$, can be (canonically) considered as elements of the Lie algebra
$\mfk g^{**}$, according to the above written arguments. Let
\[ [d_Ff,d_Fh]\in \mfk g^{**}(\supseteq \mfk g),\ f,h\in\mcl F(\mfk g^*) \]
be the corresponding commutator. Let us define the bilinear mapping
\bequ\label{eq;3Ad-poiss}
\{\cdot,\cdot\}:\mcl F(\mfk g^*)\times\mcl F(\mfk g^*)\rarw\mcl F(\mfk g^*),
\text{with}\ \{f,h\}(F)\equiv -\lb F;[d_Ff,d_Fh]\rb,\ \forall f,h\in \mcl
F(\mfk g^*),
\end{equation}
where the evaluations at $F\in\mfk g^*$ of linear functionals $\gamma\in\mfk
g^{**}$\ are denoted by $\gamma: F\mapsto \lb F;\gamma\rb$. The
mapping\rref3Ad-poiss~ is called the \emm Poisson bracket~, defining the
\emm canonical Poisson structure on \bs{\mfk g^*}~.\hfill\pika
\end{defi}

\begin{lem}\label{lem;poiss-vect}
Let $G$ be a Lie group, and let the canonical Poisson structure
$\{\dti,\dti\}$\ on $Lie(G)^*$\ be given. Let us accept the above mentioned
identifications of $T_F^*Lie(G)^*$\ with the second dual of $Lie(G)$. Then,
for any $f\in\mcl F(Lie(G)^*)$, and for an arbitrary $F\in Lie(G)^*$, the
restriction to $Lie(G)\subset Lie(G)^{**}$\ of the linear map
\bequ\label{eq;2v-f}
\rd_Fh(\in Lie(G)^{**})\mapsto -\lb F;[\rd_Ff,\rd_Fh]\rb,\ h\in\mcl
F(Lie(G)^*),
\end{equation}
to the Lie algebra $Lie(G)$, identified with the set of (differentials of)
the functions
\[h_\xi(F)\equiv \lb F;\xi\rb,\ \xi\in Lie(G),\]
is norm--continuous, cf.\dref2Lie-alg~. Hence,
as an element of $Lie(G)^*$, which in turn is identified with
$T_FLie(G)^*$, the map\rref2v-f~ can be considered as a tangent vector
to $Lie(G)^*$\ at the
point $F$. With $f$ fixed, these tangent vectors (for $F\in Lie(G)^*$)
form a smooth vector field
\vf\ on $Lie(G)^*$.\hfill\zal
\end{lem}
\begin{proof}
The Poisson bracket $\{f,g\}(F)$\ is a norm--continuous bilinear form of the
variables $\rd_Ff,\rd_Fh\in\mfk g^{**}$, hence (with the above mentioned
identification) the linear functionals: $\xi\mapsto\lb F;[\rd_Ff,\xi]\rb$,
are norm continuous on \fk g, representing some vectors $\mvf(F)\in T_F\mfk
g^*$. Let $\tilde\eta\in\mfk g^{**}$\ be an arbitrary element. Then the
(bounded linear) function $h_{\tilde\eta}:\ F(\in\mfk g^*)\mapsto
\lb F;\tilde\eta\rb$\ is smooth, $h_{\tilde\eta}\in\mcl F(\mfk g^*)$, and
its differential $\rd_Fh_{\tilde\eta}$\ (in any point $F$) is identified
with $\tilde\eta$\ itself. Hence, the mapping $f(\in\mcl F(\mfk g^*))\mapsto
\rd_Ff\in\mfk g^{**}$\ is onto.
Since the functions $f,h$ are smooth (in the sense of the underlying
norm--topology), all the functions $F\mapsto \rd_Fh(\mvf(F)),\ h\in\mcl
F(\mfk g^*)$\ are also smooth. This, due the Leibniz property of
derivatives, implies smoothness of \vf.
\end{proof}

Now we can define some of the key structures for the present paper.
\begin{defs}\label{df;1Ad-hamilt}
\item{(i)} Let $f\in\mcl F(\mfk g^*)$\ be given. The vector field \vf\ on
$\mfk g^*$\ determined (according to Lemma~\ref{lem;poiss-vect})
by the canonical Poisson structure:
\[ \rd_Fh(\mvf(F))\equiv\{f,h\}(F), \ \forall h\in\mcl F(\mfk g^*) \]
is called the \emm Hamiltonian vector field~ corresponding to
the \emm Hamiltonian function \bs f~. Let us denote by $\mphi^f$\ the local
flow of \vf, called the \emm Hamiltonian flow of \bs f~.  Let
$\mphi^\xi$\ be the Hamiltonian flow of the linear function $h_\xi$. Then we
have
\[ \mph t,\xi~(F)\equiv Ad^*(\exp(t\xi))F.\]
The \emm stability subgroup~ $G_F$ of the coadjoint action at $F\in\mfk
g^*$\ is a Lie subgroup of $G$ generated by those $\xi\in\mfk g$, for which
\[ \lb F;[\xi,\eta]\rb = 0,\ \forall \eta\in\mfk g,\]
cf.\ Lemma~\ref{lem;2.18}. The Lie algebra generated by these elements is
\emm stability Lie algebra of \bs F~ \wrt the $Ad^*(G)$--representation.

\item{(ii)} The sets $Ad^*(G)F:=\{Ad^*(g)(F):g\in G\}$\ are \emm coadjoint
orbits \bs{\mOGF}\ of \bs G~. They are identical with the  symplectic leaves
(cf.
Section~\ref{I;clmech},\dref P-strct~) of this Poisson structure. They are
conserved by all the Hamiltonian flows: $\mph t,f~F\in Ad^*(G)F:=\mOGF,\
\forall t\in\mbR$. In this sense, all the vectors $\mvf(F'),\ F'\in \mOGF,
f\in\mcl F(\mfk g^*)$, are \emm tangent vectors to the leaf
\bs{\mOGF}~. (These ``tangent vectors'' needn't form a closed tangent space
to a coadjoint orbit $Ad^*(G)F$\ for a general $F\in\mfk g^*$, cf.
Proposition~\ref{prop;2.4}. For $\dim G<\infty$,
all the $Ad^*(G)F$ are smooth submanifolds on $\mfk g^*$, hence the
(``tangent vectors'') $\equiv$\ (tangent vectors) now.)

\item{(iii)} Let us define, on each \OGF, a two form
\glss$F\mapsto\Omega_F$~\
by defining it for all the tangent vectors to \OGF\ by\label{2Sympl}
\[ \Omega_F(\mvf(F),\mvh(F)):=\{f,h\}(F),\ \forall f,h\in\mcl F(\mfk
g^*).\]
This is a well defined (i.e.\ it depends only on the vectors
$\mvf(F),\dots$, and not on the various functions $f,\dots$\ giving the same
vectors), closed (from the Jacobi identity for the commutator in \fk g),
weakly nondegenerate two--form on \OGF called the
\emm canonical symplectic form on the coadjoint orbit \bs{\mOGF}~. Endowed
with this form, \OGF\ is a (weakly) symplectic manifold, called the
\emm symplectic leaf of \bs{\mfk g^*}~.\hfill\pika
\end{defs}
It is clear, that the Hamiltonian flows $\mphi^f$\ of the canonical Poisson
structure on $\mfk g^*$\ are identical on each orbit \OGF\ with the
symplectic flows corresponding to the Hamiltonian functions which are
equal on \OGF\ to the restrictions of $f$'s to that orbit.

\chapter{On Bounded Operators and \Ca s}\label{B;Ca}
\def\autor{{
\ref{B;Ca}\quad On Bounded Operators and \Ca s}}

Conventional nonrelativistic QM is (or can be) formulated with a
help of the algebra \LH\ of all bounded operators on a separable
Hilbert space \H,~\cite{dirac,schiff,pauli1,neum1,peres,messiah}.
This is essentially true also for the conventional (but
mathematically largely heuristic) quantum field theory
(QFT),~\cite{schweber,itz&zub}. That such a formulation is not
satisfactory for systems with infinite number of degrees of
freedom became clear at least since the Haag's paper on
nonexistence of the \emm ``interaction representation''~ in cases
of nontrivially interacting fields,~\cite{haag3}. The problems of
description of ``infinite systems'' (i.e.\ quantum fields, as well
as infinite--particle ``thermodynamic'' systems) were connected
with the mathematical phenomenon of appearance of ``inequivalent
representations'', either of CCR, or CAR, or in some other way
defined sets of observables. This phenomenon was formalized in the
framework of QFT  by Araki, Haag and Kastler (and others, cf. also
Footnote \ref{AQFT} on page \pageref{AQFT}) in terms of \Ca s. It
offered possibilities to describe in mathematically well defined
terms also such physical phenomena as \emm phase
transitions~,~\cite{emch1,bra&rob,sewell,sak2}, or, more
generally, \emm collective phenomena~ in ``large systems'',
including ``macroscopic (classical) variables'' of large quantal
systems.

We shall give here a brief description of several basic concepts of the theory
of \Ca s important for understanding of description of ``the quantum world'',
including our nonlinear extensions of QM: These last mentioned applications
to finite systems with nonlinear ``quantum rules of behaviour'', can also
be included into the
(linear) \Ca ic formalism; in that connection, \Ca s composed of
operator--valued functions on a Hamiltonian (better: Poisson) phase space
consisting of, e.g., density matrices of the traditional QM, were introduced,
cf.\dref2.25b~;
these density
matrices are here, perhaps rather paradoxically, in a r\^ole of (in the
presented proposal of interpretation
of EQM) \emm classical macroscopic parameters~ --- they can be
considered in this place as classical fields describing a ``macroscopic
background'' of the considered microsystem, cf.\ref{sec;IIIB}.

\section{Bounded operators on Hilbert space}\label{B;b-oper}
\def\nazov{{
\ref{B;b-oper}\quad Bounded operators on Hilbert space}}
A \emm linear operator \bs A~ on an infinite--dimensional Hilbert space is a
linear
mapping $A:D(A)\rarw\mH$\ of a linear subset \glss\bs{D(A)\subset\mH}~\
called the \emm domain of \bs A~, into \H. If possible, we shall assume
that $D(A)$ is dense in \H. For bounded operators $A$ it is always either
$D(A) = \mH$, or the domain is a closed subspace of \H. We shall assume
that, for bounded $A$'s, if not explicitly stated a contrary, the
domain is $D(A)=\mH$.

Bounded linear operators $A:\mH\rarw\mH$\ on a complex Hilbert space \H\
form a specific \emm Banach algebra with involution~, i.e.\ they are endowed
with a natural norm $\|A\|:=\sup\{\|A\psi\|:\psi\in\mH,\|\psi\|\leq 1\}$\
(with the Hilbert--space scalar product $(\mphi,\psi)=\overline{(\psi,
\mphi)}$, and
$\|\psi\|:=\sqrt{(\psi,\psi)}$);
their product $AB:=A\circ B$, and the (adjoint--linear, i.e.\ antilinear,
involution, i.e.\ the) ${}^*$--operation $\lb^*\rb: A\mapsto A^*,
(A^*\psi,\mphi)\equiv
(\psi, A\mphi)$, satisfying also (besides the associative linear algebra
and the Banach space properties):
\bequ
\begin{split}
{}&(A^*)^*\equiv A,\ (AB)^*=B^*A^*,\\
{}&\|AB\|\leq\|A\|\dti\|B\|,\ \|A^*\|\equiv\|A\|,\
\|A^*A\|\equiv\|A\|^2,
\end{split}
\end{equation}
and the B-space of all such operators is denoted by \glss \bs{\mLH}~. The
elements $A=A^*$\ are \emm selfadjoint~. The operator $I_{\mH}\equiv
I\equiv\mbI$, for which $IA=AI=A\ (\forall A\in\mLH$), is the \emm identity (or
unit element) of \bs{\mLH}~. If, for a given $A$, there is an $A'\in\mLH$\ such,
that $A'A=AA'=I_{\mH}$, it is called the \emm inverse of \bs A~, denoted
$A'=:A^{-1}$, and $A$ is called an \emm invertible operator~; clearly,
$(A')^{-1}=A$. The subset of all invertible elements of \LH\ will be denoted
\glss \bs{GL(\mH)}~. The operators $U\in GL(\mH):U^*=U^{-1}$\ are called
\emm unitary~, and compose a subset of \LH\ -- the infinite--dimensional
Lie group,~\cite[Chap.III]{bourb;Lie}\ denoted by \bs{\mfk U}\ (:= the \emm
unitary group of \bs{\mH}~).
For any given $A\in\mLH$, the set of complex numbers
$\rho(A):=\{\mlam\in\mbC:(\mlam\mbI-A)\in GL(\mH)\}$\ is called the \emm
resolvent set of \bs A~; it is an open subset of \bC. Its complement
\glss$\msg(A)\equiv sp(A)$~\ is called the \emm spectrum of \bs
A:\ \bs{\msg(A):=\mbC\setminus\rho(A)}~. The spectrum contains also all the
\emm eigenvalues of \bs A~, i.e.\ the numbers $\mlam_j\in\mbC$, for which
there are some (nonzero) vectors $\mphi_j\in\mH$\ such that
\bequ\label{eq;eigenv}
 A\mphi_j=\mlam_j\mphi_j, j\in J\ (:=\text{an index set}).
 \end{equation}
Dimension of the subspace of \H\ spanned by all the vectors $\mphi_j\in\mH$\
satisfying\rref eigenv~ for the same complex value of $\mlam_j$ is called
the \emm degeneracy (i.e. multiplicity) of \bs{\mlam_j} ~,
it will be denoted \deg{\mlam_j}.
Let $A=A^*$. Then $\msg(A)\subset\mbR$.
The set of eigenvalues is denoted by \glss $\msg_{pp}(A)$~.
The closure of the set of all the eigenvalues of $A$:
$\overline{\msg_{pp}(A)}=:\msg_p(A)\subset\msg(A)$\ is called the \emm
pure--point spectrum~.

If the vectors $\mphi_j,j\in J$,\rref eigenv~, form
a basis in \H, the spectrum of the operator $A$ reduces to the pure--point
spectrum: $\msg(A)=\msg_p(A)$. Otherwise,
$A$ has also some \emm continuous spectrum~. As subsets of $\msg(A)$, these
two parts of spectra needn't be disjoint.
The spectrum of any bounded operator $A$
is compact, enclosed in the closed disc centered in $0\in\mbC$\ of radius
$\|A\|$.

The selfadjoint operators form the subspace (a real B-space) \LHs; they have
spectra lying on the real line: $\msg(A)\subset\mbR$. The selfadjoint
operators $A$ with positive (i.e.\ nonnegative) spectra are called \emm positive
operators~, what is denoted by $A\geq0$, or also $A>0$.

The (positive) operators $P\in\mLH$\ such, that $P=P^2=P^*$\ are called
{\bf(orthogonal) projections, or projectors}\index{orthogonal projections}\index{projectors}. There is a natural bijection
between projectors and closed subspaces: $\mcl H_P:=P\mH(\subset\mH)
\leftrightarrow P$. The projection onto the one--dimensional subspace spanned
by a nonzero vector $\psi\in\mH$\ is denoted by $P_\psi$. Projectors
$P_1,P_2$ are \emm mutually orthogonal~ iff $P_1P_2=0$. The projector onto
the subspace of \H\ spanned by all eigenvectors of a selfadjoint operator
$A$ corresponding to the same eigenvalue \lam\ is its \emm eigenprojector
\bs{E_A(\{\mlam\})}~. The dimension of the \emm eigenspace \bs{\mcl
H_\mlam}~$:=E_A(\{\mlam\})\mH$\ is \deg\mlam.

Important objects for analysis of structure and of representations of a
Banach algebra \cl A\ are
its \emm left (resp.\ right, resp.\ two--sided) ideals~, i.e.\ such linear subsets
$\mcl J\subset\mcl A$, $\{0\}\neq\mcl J\neq\mcl A$, that multiplication of
their elements {\em by an
arbitrary element $B$ of \cl A} from left (resp.\ right, resp.\ any) side
gives again elements from \cl J, i.e.\ $\forall B\in \mcl A:B\dti\mcl
J\subset\mcl J, \text{resp.}\ \mcl J\dti B\subset\mcl J, \text{resp.}\
B\dti\mcl J\cup\mcl J\dti B\subset\mcl J$. Two--sided ideals are called just
\emm ideals~. It follows that an (also one--sided) ideal $\mcl J\subset \mcl
A$\ is also a subalgebra of \cl A. For \cl A := \LH, and \H\ separable,
there is only norm--closed ideal \glss\fk C~\ in \LH,~\cite[\S 22]{najm}
consisting of all \emm compact operators~, i.e.\ such linear operators
on \H, that map any norm--bounded subset of \H\ into a norm--compact
subset of \H. There are other important ideals in \LH, which are subsets of
\fk C, e.g.\ the set of all \emm Hilbert--Schmidt operators~:\ \fk H, and its
subset \fk T\ of all \emm trace--class operators~ in \LH; these ideals are
characterized below. All these sets are
(as are all twosided ideals)  \emm symmetric~, i.e.\ they are invariant \wrt
the involution $\lb{}^*\rb$. Hence, they are generated by their selfadjoint
elements: each their element $A$ can be decomposed into the complex--linear
combination of its two selfadjoint elements:
\[ A=\frac{A+A^*}{2}+i\frac{A-A^*}{2i}. \]
The ideal \fk T\ contains exactly those selfadjoint $A$ which have pure point
spectra, and the set of all their
eigenvalues is absolutely summable (by respecting the degeneracy):
\[ A^*=A\in\mfk T\eequiv\text{$A$\ {\it has pure point spectrum and}}
\sum_{\mlam\in\msg_{pp}(A)}\mdeg\mlam|\mlam|=:\mN A,1~<\infty.\]
Then we can define the finite number (for $A=A^*$)
\[ Tr(A):= \sum_{\mlam\in\msg_{pp}(A)}\mdeg\mlam\mlam, \]
called the \emm trace of \bs A~. Its value does not depend on
unitary transformations: $Tr(A) = Tr(UAU^*),$\ $(\forall U\in\mfk
U)$. The trace $A\mapsto Tr(A)$\ can be uniquely extended to the
whole complex space \fk T\ by linearity. Then it is defined for
all products (since \fk T\ is an ideal) $BA:B\in\mLH,\ A\in\mfk
T$, and we have
\[ Tr(AB)\equiv Tr(BA),\forall B\in\mLH, A\in\mfk T.\]
It is also valid:
\bequ\label{eq;11n}
\mN AB,1~\leq\mN A,1~\mN B,{}~,\ \mN BA,1~\leq\mN A,1~\mN B,{}~,\quad\forall
A\in\mfk T, B\in\mLH.
\end{equation}
The Hilbert--Schmidt ideal \fk H\ is defined:
\[ A\in\mfk H\eequiv A\in\mLH\ \&\ A^*A\in\mfk T.\]
Then also $AB\in\mfk T$\ for all $A,B\in\mfk H$; the algebra \fk H\ of
operators in \H\ can be made a Hilbert space in a canonical way, by
defining the scalar product by
\[ (A,B)_2:=Tr(A^*B),\ \forall A,B\in\mfk H.\]
The set \fk H\ is closed \wrt the {\bf Hilbert--Schmidt norm $\|A\|_2
:=\sqrt{Tr(A^*A)}$}\ind{Hilbert--Schmidt norm}. Also \fk T\ is closed \wrt the \emm trace norm
\bs{\mN A,1~:=Tr|A|}~, where the operator $|A|$, the \emm absolute value of
\bs A~ can be defined by a ``functional calculus'' as $|A|:=\sqrt{A^*A}$.
The elements of the  subset $\mfk T_{+1}\subset\mfk T$\ of positive
trace--class operators
with unit norm:
\bequ\label{eq;1densm}
\mrh\in\mfk T_{+1}\eequiv\mrh\in\mfk T\ \&\ \mrh>0\ \&\ \mN\mrh,1~=1,
\end{equation}
are called in physics the \emm density matrices~.

The bilinear form
\[ \lb B;A\rb:= Tr(BA),\ A\in\mfk T,\ B\in\mLH,\]
provides a \emm duality~ between the B-spaces \fk T, and \LH, and similarly
between the B-space of compact operators \fk C, and \fk T, in the sense
that the operators $B$ from the second of a couple of spaces represent
(all) continuous linear functionals $l_B$\ on the first of spaces, by the
evaluations
\[ A\mapsto l_B(A):=Tr(BA)\equiv\lb B;A\rb.\]
In this sense, the following assertions are valid for the topological duals:
\bequ\label{eq;1duals}
\mfk C^*=\mfk T,\quad\mfk T^*=\mLH =\mfk C^{**}.
\end{equation}

Since the mathematical objects $A,B,\dots\in\mLH$\ are not only elements of
a Banach algebra, but they also realize linear transformations of \H, they
are endowed by other natural l.c. topologies. Let us introduce the \emm
weak operator topology \bs{\mcl T_w}~ by the set of seminorms
$\{p^w_\psi:\psi\in\mH\}$:
\bequ\label{eq;0weak}
p^w_\psi:A\mapsto p^w_\psi(A):=|(\psi,A\psi)|,\quad\psi\in\mH, A\in\mLH.
\end{equation}
The \emm strong operator topology \bs{\mcl T_s}~ on \LH\ is determined by
the seminorms $\{p^s_\psi:\psi\in\mH\}$:
\bequ\label{eq;0strong}
p^s_\psi:A\mapsto p^s_\psi(A):=\|A\psi\|,\quad\psi\in\mH, A\in\mLH.
\end{equation}

There are important also other topologies on \LH, namely the
\emm \bs\msg--strong
(equiv.: ultrastrong) topology
\bs{\mcl T_{us}}~, the \emm \bs\msg--weak (equiv.: ultraweak) topology
\bs{\mcl T_{uw}}~, and also the \emm strong\bs{{}^*}\ topology \bs{\mcl
T_{s*}}~,
 and the \emm \bs{\msg}--strong\bs{{}^*} (equiv.: ultrastrong${}^*$) topology
\bs{\mcl T_{us*}}~,~\cite{najm,emch1,bra&rob,takesI}. These
topologies, including also the \emm norm topology \bs{\mcl T_n}~, are ordered in
the following hierarchy (\wrt the ordering $\prec$\
introduced in Definitions~\ref{Adf;top}):
\bequ\label{eq;LH-tops}
\begin{matrix} \mT w&\prec&\mT s&\prec&\mT{s*}&{}&{}\\
                  \cw&{}&\cw&{}&\cw&{}&{}\\
               \mT{uw}&\prec&\mT{us}&\prec&\mT{us*}&\prec&\mT n
\end{matrix}
\end{equation}
We did not consider now the topological dual of \LH\ \wrt the
norm--topology: $\mLH^*$\ contains also ``nonnormal states'' on
\LH, cf.\dref states~. In QM, mainly linear functionals from $\mfk
T \subset\mLH^*$\ are used in the r\^ole of (normal) quantum
states described by density matrices. The space \fk T\ can be
considered as the dual of \LH, if the last space is endowed with
the $\msg(\mLH,\mfk T)$--topology, which is identical with the
\sg--weak topology \T{uw}\ determined by all density matrices \rh\
by seminorms $p^{uw}_\mrh:A(\in\mLH)\mapsto p^{uw}_\mrh(A):=
|Tr(\mrh A)|$. The ``nonnormal'' states from $\mLH^*\setminus\mfk
T$\ include, e.g. {\emm dispersionfree states for observables with
purely continuous spectra~}, e.g.\ the ``eigenstates'' for
position coordinates,~\cite{bon8}.

Theoretical physics is mainly interested in selfadjoint operators, resp.\ in
unitary operators (these all belong to the ``equally nice'' \emm normal
operators~ $A$ characterized by $AA^*=A^*A$).
Normal operators can have their spectrum also in $\mbC\setminus\mbR$.
Selfadjoint operators $A$ (not
only bounded) are generators of one--parameter groups of unitaries:
$t\mapsto\exp(itA)\in\mfk U$, and also are representatives of
``observables'' in QM; the most clear understanding of their interpretation is
expressed, perhaps, via
the \emm spectral theorem~. This theorem shows, that any selfadjoint
operator can be, roughly speaking, expressed as a real linear combination (resp.
integral) of mutually orthogonal ``eigenprojections'', which are multiplied
by the corresponding ``eigenvalues''.

The key concept in this connection is a projection--valued
measure (PM, or PVM). Let us introduce simultaneously its generalizations,
i.e.\ positive operator valued measures (POV, or POVM).

\begin{defs}[{\bf Projection measures, and POV measures}\index{measures (PM and POV)}]\label{df;1p-measure}
\item{(i)} Let $(X;\mcl T)$ be a topological space, and $\mcl B(X)\subset
\mcl P(X)$\ be the set of all subsets obtained from the open {\em and
closed} subsets of $X$\ by countable unions and/or intersections. Elements
$\Lambda\in\mcl B(X)$ are called \emm Borel sets of \bs X~. The class of
Borel sets is topology--dependent; if, on some set $Y$, the topology is standard
(e.g., on $\mbR^n$), then the specification of $\mcl B(Y)$\ is not usually
given.
A function $f$\ from a topological space $(X_1;\mcl T_1)$\ into a
topological space $(X_2;\mcl T_2)$\ is called a \emm Borel function~ iff
for any $V\in\mcl B(X_2)$\ the inverse image $f^{-1}[V]\in\mcl B(X_1)$;
the set of
all such uniformly bounded Borel functions will be denoted by $\mcl
B_b(X_1,X_2)$.

\item{(ii)} Let a mapping $E:\mcl B(X)\rarw\mLH$\ be such that each
$E(\Lambda)$\ is an orthogonal projection, and\nl
\hspace*{.5cm}(a) $E(X)=\mbI,\ E(\emptyset)=0$;\nl
\hspace*{.5cm}(b) for any at most countable collection $\Lambda_j, j\in J,
|J|\leq\aleph_0$\ of mutually disjoint Borel sets $\Lambda_j\in\mcl B(X)$,
$\Lambda_j\cap\Lambda_k=\emptyset\ (\forall j\neq k)$, one has
\[ E(\cup_{j\in J}\Lambda_j)=\sum_{j\in J}E(\Lambda_j),\]
where the sum converges in the strong operator topology of \LH.
\item{{}} The mapping $E$ is a \emm projection (valued) measure
(PM, {\rm equiv.} PVM)~ (on $X$ with values in \LH).

\item{(iii)} Let $S\subset\mLH$ be any set of bounded operators. The \emm
commutant \bs{S'}\ of \bs S~ is the set $S':=\{B\in\mLH:AB-BA=0,\forall
A\in S\}\subset\mLH$. Clearly, $S\subset S'':=(S')'$, and $S''$\ is called
the \emm bicommutant of \bs S~.

\item{(iv)} Let a mapping $F:\mcl B(X)\rarw\mLH$\ be such that each
$F(\Lambda)$\ is a positive operator, and\nl
\hspace*{.5cm}(a) $F(X)=\mbI,\ F(\emptyset)=0$;\nl
\hspace*{.5cm}(b) for any at most countable collection $\Lambda_j, j\in J,
|J|\leq\aleph_0$\ of mutually disjoint Borel sets
$\Lambda_j\in\mcl B(X)$\ one has
\[ F(\cup_{j\in J}\Lambda_j)=\sum_{j\in J}F(\Lambda_j),\]
where the sum converges in the strong operator topology of \LH.
\item{{}} The mapping $F$ is a \emm normalized positive operator
valued (POV) measure~ (on $X$ with values in \LH), called also \emm an
observable on \bs X, {\rm resp.\ also} a POVM,~\cite[Sec.3.1]{davies}. \hfill\pika
\end{defs}
The POV measures are
generalizations of PM; they are useful in description of ``nonideal
measurements'' in QM, cf.\ \cite{davies}. Let's note, that now the operators
$F(\Lambda)$\ for different $\Lambda\in\mcl B(X)$\ needn't mutually commute.
There is an important construction giving also a criterion for distinction
of PM from POVM, cf.\ \cite{davies}:
\begin{subequations}\label{eq;0sp-thm}
\begin{prop}
Let $X$ be a compact Hausdorff space, and let $F:\mcl B(X)\rarw \mLH$\ be a
normalized POV measure (i.e.\ an observable). Let $C(X)$\ be the space of
complex--valued continuous
functions on $X$. Then the strongly convergent integral
\bequ\label{eq;1sp-thm}
 F(f):= \int_X f(x)F(\rd x),\ f\in C(X),
\end{equation}
defines a bijection between observables $F(\cdot)$\ and linear maps $F:
C(X)\rarw \mLH$\ such that $f\geq0\imply F(f)\geq0, F(\mbI)=I_{\mH}$. The
POV measure $F$\ is PM iff the map $F: C(X)\rarw \mLH$\ is a
${}^*$--homomorphism of the algebra $C(X)$\ into the algebra \LH.\hfill\zal
\end{prop}

\begin{thm}[Spectral theorem]\label{thm;b-spectral}
Let $A\in\mLH$\ be a normal operator. Then there is a unique PM, $E_A$, on the
spectrum $\msg(A)$\ such that:\nl
\hspace*{.5cm}(i)\ $E_A(\Lambda)\in\{A\}'',\ \forall
\Lambda\in\mcl B(\msg(A))$;\footnote{$\{A\}''$\ is the bicommutant of the
one--point set $\{A\}\subset\mLH$.}\nl
\hspace*{.5cm}(ii)\ For any $f\in\mcl B_b(\msg(A),\mbC)$, there is
an operator $E_A(f)\equiv f(A)$\ given by the strongly convergent integral
(being the strong-operator limit of any
sequence of expressions $E_A(f_n)$ for ``simple functions''
$f_n(\mlam):=\sum_jc^{(n)}_j\chi_{\Lambda^{(n)}_j}(\lambda)$\ approximating
$f$ by pointwise limits)
\bequ\label{eq;2sp-thm}
 E_A(f)\equiv f(A)=\int_{\msg(A)} f(\mlam)E_A(\rd\mlam) ;
\end{equation}
(iii)\ \ \ The mapping $E_A:\mcl B_b(\msg(A),\mbC)\rarw\mLH,\ f\mapsto
E_A(f)=f(A)$, is a unique continuous ${}^*$--homomorphism (called \emm the
functional calculus~) of commutative
algebras (in $\mcl B_b(\msg(A),\mbC)$, the multiplication and addition
are defined pointwise, and involution is the complex conjugation) determined by
$E_A({\rm id}_{\mbC})=A$. Continuity is here understood so that
$\|f(A)\|\leq\|f\|$, where the norm of $f$ is the supremum norm.\hfill\zal
\end{thm}
\end{subequations}

Important features of the algebras of operators, also from a physical point of
view, are their representations as homomorphic images in some \cl{L(K)},
where \cl K\ is a complex Hilbert space. There are two kinds of nonzero
representations of the algebra \LH, if \H\ is separable~\cite{najm}:
There are orthogonal multiples of the
identical representation, and the representations setting the ideal \fk C\ into
zero, in which case the (simple) factoralgebra \LH/\fk C\ (= the {\em Calkin
algebra}) is isomorphically
represented. Representations of the first mentioned kind are ``trivial'',
and that of the second kind are ``physically irrelevant'' (\wrt the standard
nonrelativistic QM),
since it might
be difficult to interpret states, in which all finite--dimensional
projections in the ``given algebra of observables'' \LH\ are mapped to zero
(probabilities of values of all quantities with pure point spectra with
finite degeneracies would be zero!); cf., however, Note~\ref{note;types}.
More ``colourful'' picture of
``physically interesting'' representations of algebras of observables arise
for some closed symmetric subalgebras of \LH, and, more generally, for
general \Ca s (these might not be faithfully represented on separable
Hilbert spaces).

\section{Elementary properties of \Ca s and \Wa s}\label{B;c-alg}
\def\nazov{{
\ref{B;c-alg}\quad Elementary properties of \Ca s and \Wa s}}

We shall reformulate now algebraic properties of \LH\ to be able to obtain
a more
general framework for quantum theories (QT). All (mathematical) fields of
scalars will be the complex numbers \bC, and in the natural restriction
also the field of reals \bR.

\begin{defs}[{\bf \Ca s and \Wa s}]\label{df;Ca}
\item{(i)} A \emm Banach algebra~ \fk A\ is a B-space endowed with an
associative and distributive multiplication (i.e.\ the \emm algebraic
product~, resp.\ the \emm product~, converting the linear space \fk A\ into
an \emm algebra~):
$(x;y)(\in\mfk A\times\mfk
A)\mapsto x\dti y\equiv xy(\in\mfk A)$,\ $x\dti(y\dti z)=(x\dti y)\dti z,\
(x+\mlam y)\dti z=x\dti z+\mlam y\dti z,\ x\dti(y+\mlam z)=x\dti y+\mlam
x\dti z$; the multiplication is connected with the norm in \fk A by the
requirement: $\|xy\|\leq\|x\|\|y\|; \forall x,y,z\in\mfk A,\mlam\in\mbC$.
If $xy=yx,\forall x,y\in\mfk A$, the algebra \fk A\ is called \emm
abelian~, resp.\ \emm commutative~.

\item{(ii)} If there is an element $e\in\mfk A$\ such that $e\dti x=x\dti
e=x,\forall x\in\mfk A$, the Banach algebra \fk A\ is a \emm unital algebra~
and the element \emm \bs e is the unit element~ (or \emm unit~) of \fk A.
If a unit exists in \fk
A, it is unique. If there is, for an element $x\in\mfk A$, an element
(denoted by) $x^{-1}\in\mfk A$\ such that $x\dti x^{-1}=x^{-1}\dti x=e$,
the element \emm \bs x is invertible~, and \emm \bs{x^{-1}}\ is the inverse of
\bs x~. If $x$ is invertible, the inverse element $x^{-1}$\ for $x$ is unique;
then also $x^{-1}$\ is invertible, and $(x^{-1})^{-1}=x$. The set of all
invertible elements in \fk A\ is its \emm general linear group \bs{G(\mfk
A)}~, denoted also by \glss \bs{\mfk A^{-1}}~.

\item{(iii)}\label{df;symA} An algebra \fk A\ is \emm symmetric~, if
there is defined in
it an (antilinear)\emm involution~, i.e.\ a mapping $\lb{}^*\rb:\mfk
A\rarw\mfk A, x\mapsto x^*: x^{**}:=(x^*)^*\equiv x,\ (x+\mlam y)^*\equiv x^*
+ \overline{\mlam}y^*$; and, moreover, this involution is connected with
the product by: $(xy)^*\equiv y^*x^*$. We shall call the linear
combination, product and involution the \emm algebraic operations~.

\item{(iv)} If, for a symmetric Banach algebra \fk A, it is satisfied the
\emm \bs{C^*}--property~: $\|x^*\dti x\|\equiv \|x\|^2$, then \emm \bs{\mfk
A}\ is a \bs{C^*}--algebra~. If \fk A\ has also the unit element, then it is
a \emm unital \bs{C^*}--algebra~. We
shall usually assume, that \fk A\ is unital, and the converse will be
pointed out.
A Banach subspace \fk B\ of a \Ca\ \fk A, which is invariant \wrt all the
algebraic operations applied to its elements is a \emm \bs{C^*}--subalgebra of
\bs{\mfk A}~. If $\mfk B(\neq \mfk A)$\ is, moreover, invariant \wrt the
multiplication by all
elements of \fk A, it is a \emm closed (two--sided) ideal~; clearly, such a
\Csa\ \fk B\ does not contain the unit element of \fk A. An element $x$ of a
\Ca\ is: \emm selfadjoint~ iff $x^*=x$; \emm normal~ iff $x^*x=xx^*$;
\emm projection~ iff $x=x^*=x^2$; \emm partial isometry~ iff $xx^*$\ is a
projection; \emm unitary~ (in a unital algebra) iff $xx^*=x^*x=e$.

\item{(v)} If, for a \Ca\ \fk A, as a B-space, there is another B-space
(denoted by) $\mfk A_*$\ such that \fk A\ is (isomorphic to) its topological
dual: $(\mfk
A_*)^*=\mfk A$, the \Ca\ \fk A\ is called a \emm \Wa~, and the Banach space
\emm \bs{\mfk A_*}\ {\rm is its} predual~. Any \Ca\ has at most one predual, up
to isomorphisms. The \Wa s (originally: their specific operator
realizations) are called also \emm von Neumann algebras~. Any \Wa\ is a
unital \Ca. Any \Wa\ is generated by its projections (via $\msg(\mfk A,\mfk
A_*)$--closure of their linear combinations). [A general \Ca\ needn't have
any nontrivial projection.]

\item{(vi)} Let \fk A\ be a \Ca, and let $\mfk A^{**}:=(\mfk A^*)^*$\ be its
second topological dual. The \Ca\ \fk A\ is canonically embedded into $\mfk
A^{**}$\ as a $\msg(\mfk A^{**},\mfk A^*)$--weakly dense linear subspace,
cf.\dref top-ls~\
and, in this topology, all the algebraic operations (i.e.\ the linear
combination, addition, multiplication -- with one of the multiplicands fixed,
and the involution) are continuous.
Hence, the algebraic structure of \fk A\ can be unambiguously extended to
the whole $\mfk A^{**}$, endowing this by a (canonical) \Ca ic structure.
The obtained \Wa\ $\mfk A^{**}$\ is denoted also \bs{\mfk A''}, and it is
called the \emm universal enveloping \bs{W^*}--algebra of the
\bs{C^*}--algebra \fk A~.\footnote{The notation $\mfk A''$\ originated in the
realization of this von Neumann algebra as the ultrastrong (hence weak) closure
of a specific faithful representation,~\cite{emch1,pedersen}, called the
\emm universal representation
\bs{\pi_u(\mfk A)} of \bs{\mfk A}~, hence as the bicommutant
$\pi_u(\mfk A)''$, cf.\ Example~\ref{ex;Ca&Wa}(ii).}

\item{(vii)} The \emm centre \bs{\mcl Z(\mfk A)}~ of a \Ca\ \fk A\ is the
commutative \Csa\ of \fk A\ consisting of all elements of \fk A\ commuting
with any element of \fk A: $\mcl Z(\mfk A):=\{z\in\mfk A:z\dti x-x\dti z=0,
\forall x\in\mfk A\}$. A von Neumann algebra \fk B\ with trivial centre:
$\mcl Z(\mfk B)=\{\mlam\dti e:\mlam \in\mbC\}$, is called a \emm factor~.
\hfill\pika
\end{defs}

\begin{noti}[Quotient \Ca]
The factor--space (resp.\ the quotient space) $\mfk A/\mfk B$\ of a \Ca\
\fk A\ over its closed ideal \fk
B\ is canonically endowed with the structure of a \Ca. Let the canonical
projection be $\beta:\mfk A\rarw \mfk A/\mfk B,x\mapsto\beta_x:=\{y\in\mfk
A:y=x-z,z\in\mfk B\}$, and $\beta_x\dti\beta_y:=\beta_{xy},
\beta^*_x:=\beta_{x^*}, \|\beta_x\|:=\inf\{\|x-z\|:z\in\mfk B\}$. Then all
the ``$C^*$--properties'' for $\mfk A/\mfk B:=\{\beta_x:x\in\mfk A\}$\
are valid, cf.\ \cite[1.8.2]{dix2}.\hfill\dovi
\end{noti}
Let us give a list of examples of \Ca s:

\begin{exmp}[Some \Ca s and \Wa s]\label{ex;Ca&Wa}
\item{(i)} Since the dual of the trace class operator--space \fk T\ is $\mfk
T^*=\mLH$, the algebra of all bounded operators on \H\ is a von Neumann algebra.

\item{(ii)} \LH, and all its closed symmetric subalgebras are \Ca s. Any
such \Csa\ \cl B, that is also closed in operator weak (resp.
ultraweak, resp.\ strong, resp.\ ultrastrong) topology is also a \Wa; the
closure  of any \Csa\ \cl A\ of \LH\ (if $I_\mH\in\mcl A$) in any of these
mentioned topologies
equals to its double commutant $\mcl A''$,~\cite{najm,sak1,takesI}, what is a
form of the well known \emm von Neumann bicommutant theorem~. In
separable Hilbert space, the only nontrivial \Csa\ of \LH\ which is also
an ideal of \LH\ is the algebra of all compact operators \fk C.

\item{(iii)} Let $M$ be a compact Hausdorff space, and let $C(M)$ be the
set of all complex valued continuous functions on $M$. Let pointwise linear
combinations, multiplication, and conjugation be defined on $C(M)$\ by:
\begin{equation*}
\begin{split}
f,h\in C(M),\mlam\in\mbC:\quad &(f+\mlam h)(m):=f(m)+\mlam h(m),\\ & (f\dti
h)(m):=f(m)h(m),\ f^*(m):=\overline{f(m)},\quad \forall m\in M,
\end{split}
\end{equation*}
and let the norm be the supremum norm: $\|f\|:=\sup\{|f(m)|:m\in M\}$. Then
$C(M)$, endowed with these structures, is a commutative \Ca. Each unital
commutative \Ca\ is isomorphic to one of this form (the
\emm Gel\acc fand--Najmark theorem~).

\item{(iv)} The factoralgebra \LH/\fk C\ of the algebra of all bounded
operators by the \Csa\ of its compact operators \fk C\ is a unital \Ca, called
(according to~\cite[6.1.2]{pedersen}) the \emm Calkin algebra~. It belongs
to the
class of \emm antiliminary \bs{C^*}--algebras~, playing an important r\^ole
in descriptions of infinite quantum systems.\hfill\dovi
\end{exmp}

An important characterization of elements of a \Ca\ \fk A\ is (as it was in
\LH\ for operators) their \emm spectrum~. Since the definitions and
properties are
identical in this general case with those in the case of bounded operators
in \LH, we shall proceed briefly:

\begin{defi}[{\bf Spectrum}\index{spectrum}]\label{df;sp-x}
Let $x\in\mfk A:=$\ a unital \Ca. The set $\rho(x):=\{\mlam\in\mbC:(\mlam
e-x)\in\mfk A^{-1}\}\subset\mbC$\ is the \emm resolvent set of \bs x~. Its
complement $\msg(x)\equiv {\rm sp}(x):=\mbC\setminus\rho(x)$\ is the \emm
spectrum of \bs x~. The spectrum of any element is closed in \bC. The number
$\|x\|_{\msg}:=\sup\{|\mlam|:\mlam\in\msg(x)\}$\ is called the \emm
spectral radius of \bs x~. Always is $\|x\|_\msg\leq\|x\|$, and
$\|x\|_\msg=\|x\|$\ if $x$ is
normal.
\hfill\pika
\end{defi}

An important property of spectrum of any \Ca ic element $x$\ is
its independence on a choice of  unital \Csa s $\mfk B\subset\mfk
A$\ containing $x$, \wrt which is $\msg(x)$ calculated (instead of
\fk A). Hence, the spectrum of $x$ can be calculated \wrt the
minimal \Csa\ $\mfk A_x\subset\mfk A$\ containing $x$, i.e.\ \wrt
the subalgebra generated by the elements $x,x^*,e$. For a normal
element $x$, the \Ca\ $\mfk A_x$\ is commutative, and it is
isomorphic to $C(\msg(x))$. The algebraic elements in ${\mfk
A}_x$\ corresponding (according to this isomorphism) to some $f\in
C(\msg(x))$\ are denoted by $f(x)$. The association $f\bigl(\in
C(\msg(x))\bigr)\mapsto f(x)\in\mfk A_x\subset\mfk A$ is the
inverse of the \emm Gel\acc fand transform~. It is a
${}^*$--isomorphism of \Ca s, cf.\dref repres~, hence for, e.g.,
$f(\mlam)\equiv \mlam^n$\ one has $f(x)=x^n, n\in\mbZ_+, x^0:=e$.
This mapping of $\mfk A\times C(\mbC)$\ (restricted to normal
elements of \fk A) into \fk A\ is called \emm continuous
functional calculus~ on \fk A. If \fk A\ is a \Wa, then also
complex valued bounded Borel functions $f\in\mcl B_b(\mbC)$ have
their homomorphic images $f(x)$\ in \fk A, for normal elements
$x\in\mfk A$. The ${}^*$--homomorphism determined by an arbitrary
{\em normal} element $x$\ of a \Wa\ \fk A:
\[ f(\in \mcl B_b(\mbC))\mapsto f(x)\in\mfk A \]
is a unique continuous (i.e.\ $\|f(x)\|\leq\|f\|=\|f\|_\infty$) extension of
the continuous functional calculus. These extended mappings are called the
\emm Borel functional calculus~ on a \Wa.

For arbitrary elements (i.e.\ not necessarily normal)
$x$\ of \fk A, we have the \emm analytic functional calculus~: If $f$\ is
holomorphic (i.e.\ analytic) function on an open domain in \bC\ containing
the spectrum
of $x$, and $c$\ is a ``Jordan'' curve (i.e.\ continuous, closed,
nonselfintersecting, of finite length, being a homeomorphic image of a
circle $S^1$) lying in this domain and ``surrounding'' the spectrum
$\msg(x)$, then we can define a Banach space valued integral
\bequ\label{eq;anal-calc}
f(x):=\frac{1}{2\pi i}\oint_c\frac{f(\mlam)\rd\mlam}{\mlam e-x}\in\mfk A,
\end{equation}
what can be defined by a norm--convergent sequence of Riemann sums in \fk A.
Restrictions of above mentioned ``functional calculi'' to analytic
functions give the values expressed by\rref anal-calc~.

If we define \emm positive elements~ $x$\ of a \Ca\ \fk A\ as such
selfadjoint elements of \fk A\ that can be
expressed as $x=y^*y$\ for some $y\in\mfk A$, we can see that these, and only
these elements correspond to positive functions in the mentioned functional
calculi. The positive elements form a cone $\mfk A_+$\ in \fk A, i.e.\ any linear
combination of elements in $\mfk A_+$\ {\em with nonnegative coefficients}
also belongs to $\mfk A_+$.
The isomorphism of commutative \Ca s with spaces of continuous
functions mentioned in Example~\ref{ex;Ca&Wa}(iii) exactly corresponds to
the mentioned functional calculi, but
extended also to such commutative \Ca s, that need--not be generated by a single
normal element. The compact $M$, corresponding to a unital commutative \Ca\
\fk A\ which is
${}^*$--isomorphic to $C(M)$, is called the \emm spectrum of the abelian
\Ca~ \fk A.
If a function $f_x\in C(M)$\ represents the element $x$\ via the Gel\acc fand
transform, the spectrum $\msg(x)$\ is identical with the range of $f_x$:
$\msg(x)=\{f_x(m):m\in M\}$, cf.\ also Examples~\ref{ex;abel-C&W}.

Elements of algebras usually appear in physical theories represented in
forms of linear
operators acting on Hilbert spaces. This is naturally connected, as we
shall also see in the Subsection~\ref{B;state&repr}, with the physical
interpretation of elements of Hilbert spaces as physical states in which
the ``observables'' represented by the elements of algebra are measured
(resp.\ calculated). We shall now turn to an introduction to the
representation theory.

\begin{defs}[{\bf Representations}]\label{df;repres}
\item{(i)} A mapping $\pi:\mfk A\rarw\mfk B$\ between two \Ca s \fk A, and
\fk B, is a \emm \bs{{}^*}--morphism~, iff it satisfies the properties:\nl
\hspace*{.5cm}(l) $\pi$\ is linear: $\pi(x+\mlam y)=\pi(x)+\mlam\pi(y)$,\nl
\hspace*{.5cm}(ll) $\pi(x\dti y)=\pi(x)\dti\pi(y)$,\nl
\hspace*{.5cm}(lll) $ \pi(x^*)=\pi(x)^*$,\nl
\noidt for all $x,y\in\mfk A,\mlam\in \mbC$. The set \bs{{\rm
Ker}(\pi)}\ consisting of such elements $x$ of \fk A, that $\pi(x)=0$, is
called the \emm kernel of \bs\pi~. If ${\rm Ker}(\pi)=0$\ for all nonzero
 morphisms
$\pi$\ of \fk A, then \fk A\ is called a \emm simple \Ca~.

\item{(ii)} If $\mKer\pi=\{0\}$\ then $\pi$\ is a
bijection. If $\mKer\pi=\{0\}$, and also $\pi(\mfk A)=\mfk B$, the morphism
is called a \emm \bs{{}^*}--isomorphism~ (briefly: an isomorphism) of \fk
A\ onto  \fk B, and these two \Ca s
are mutually \emm isomorphic~. Any morphism of \Ca s is continuous:
\[ \|\pi(x)\|\leq \|x\|,\]
and any isomorphism is isometric:
\[ \mKer\pi=0\imply \|\pi(x)\| \equiv\|x\|. \]
\Ker\pi\ is always a closed (twosided) ideal in the \Ca\ \fk A.

\item{(iii)} Let $\pi$\ be a ${}^*$--morphism of a \Ca\ \fk A\ into
\LH, for some Hilbert space \H. It is called a \emm representation of
\bs{\mfk A} in \bs{\mH}~. It will be denoted also \glss $(\pi;\mH)$~.
If there is a nontrivial (i.e.\ $\neq0$, and
$\neq I_\mH$) projection $P$\ in the commutant $\pi(\mfk A)'\in\mLH$, then
the mapping
\[ \pi_P:\mfk A\rarw\mcl L(P\mcl H),\ x\mapsto P\pi(x),\]
is a \emm subrepresentation of \bs{\pi}~. We shall assume, that any
considered
representation is \emm nondegenerate~, i.e.\ that it has {\bf no}\
\emm zero subrepresentations~, i.e.\ that
there is no nonzero projection $P\in\mLH$\ such, that $P\pi(\mfk A)=\{0\}$.
If a representation of a \Ca\ have no nontrivial subrepresentations, it is
called an \emm irreducible representation~.

\item{(iv)} Let us assume that, in the Hilbert space $\mH_\pi$\ of a
representation $\pi(\mfk A)$, there is a vector, say $\psi_\pi\in\mH_\pi$\
such, that the set
\[ \pi(\mfk A)\psi_\pi:=\{\pi(x)\psi_\pi:x\in\mfk A\}\]
is dense in the Hilbert space $\mH_\pi$. Then the representation $\pi$\ is
called a \emm cyclic representation of \bs{\mfk A}~, and the vector
$\psi_\pi$\ is a \emm cyclic vector of the representation $\pi$~. Each
representation of a \Ca\ can be decomposed into an orthogonal sum of cyclic
subrepresentations, i.e.\ there is a system of mutually orthogonal projections
$P_j\in\pi(\mfk A)',\ j\in J$,\ such that $\sum_{j\in J}P_j=I_{\mH_\pi}$\
(strong convergence), and each subrepresentation $x\mapsto P_j\pi(x)\in\mcl
L(P_j\mH)$\ is cyclic. In an irreducible representation space
$\mH_\pi$, any nonzero vector $\psi\in\mH_\pi$\ is cyclic, and $\pi(\mfk A)
\psi=\mH_\pi$.

\item{(v)} Let $(\pi_1;\mH_1),(\pi_2;\mH_2)$\ be two representations of a
\Ca\ \fk A. If there is a linear (unitary) isometry $U:\mH_1\rarw\mH_2$\
such that $\pi_2(x)\equiv U\pi_1(x)U^{-1}$, the representations \emm
\bs{\pi_1}\ and \bs{\pi_2}\ are unitarily (or spatially) equivalent~.\index{unitarily equivalent} We
shall denote this fact by $\pi_1\simeq\pi_2$.
\hfill\pika
\end{defs}
As it was mentioned above, any ${}^*$--representation of a \Ca\ is
continuous in the
norm topology. On each \Wa\ \fk B, moreover, another natural topology,
namely the $\msg(\mfk B,\mfk B_*)$--topology, or briefly the
$w^*$--topology is given. The same is true for the \Wa\ $\mcl L(\mcl
H_\pi)$ for any representation $\pi$. Hence the question on the
$w^*$--continuity
of $\pi$\ arises. As we shall show, this property is also relevant for any
representation of a \Ca\ \fk A\ canonically extended to $\mfk A''$, cf.\
Proposition~\ref{prop;pi-ext}.

\begin{defi}[{\bf \Wrep s}]\label{df;W-rep}
If a representation $\pi$\ of a \Wa\ \fk B\ in the Hilbert space $\mH_\pi$\ is
$\msg(\mfk B,\mfk B_*)-
\msg(\mcl L(\mcl H_\pi),\mfk T(\mcl H_\pi))$\ continuous, it is called a
\emm \Wrep~.
\end{defi}

The image $\pi(\mfk B)\subset\mcl L(\mcl H_\pi)$\ of any \Wrep\ such that
$\pi(e)=I_{\mH_\pi}$\ (nondegeneracy) is again a
\Wsa\ of $\mcl L(\mcl H_\pi)$. Let us mention also that an isomorphism of
two \Wa s is always $w^*$--continuous,~\cite{sak1}.

\begin{prop}\label{prop;pi-ext}
Let \fk A\ be a \Ca, and $(\pi;\mH_\pi)$\ its arbitrary nondegenerate
representation. Let
us consider $\mfk A\subset\mfk A''$, in the canonical way. Then there is
unique extension of $\pi$\ to a \Wrep\ $(\pi'';\mH_\pi)$\ of the univeral
enveloping algebra $\mfk A''$,~\cite[Proposition 2.21.13]{sak1}. The image
of this extension equals to the bicommutant $\pi(\mfk A)''$, if \fk A\ is
unital.\hfill\zal
\end{prop}
It follows from this assertion that to any representation $\pi$\ of a \Ca\
\fk A\
there is a central projection $z(\pi)\in\mcl Z(\mfk A'')$\ such, that its
orthogonal complement $e''-z(\pi)$\ (with $e''$\ the unit of $\mfk A''$)\
supports the kernel of $\pi'': \mKer{\pi''}=(e''-z(\pi))\mfk
A''$, called alternatively the
\emm central support~,~\cite[1.21.14]{sak1}, resp.
\emm central cover~ of $\pi$,~\cite[3.8.1]{pedersen}. The representations
$\pi_1,\pi_2$\ with the same central projection $z(\pi_1)=z(\pi_2)$\ are
called {\bf (quasi--) equivalent}\index{quasi--equivalent}, denoting this by \glss $\pi_1\sim
\pi_2$~. Unitary equivalence implies equivalence, but equivalent
representations are just, roughly speaking, decomposable into various
multiples of the same unitary equivalent
subrepresentations,~\cite[5.3.1]{dix2}. If the central supports are
orthogonal: $z(\pi_1)\dti z(\pi_2)=0$, the two representations are called
\emm disjoint~,\label{disj} we shall denote this by
$\pi_1\mho\pi_2$.\footnote{The
standard symbol for disjoitness,~\cite{dix2,bra&rob}, was not found among the
\LaTeX\ symbols.}
\begin{intpn}[Macro--distinguishability]\label{int;macro-d}
Disjoint representations are interpreted in physics as
\emm macroscopically (or classically) distinguishable representations~:
Since the ``physically most relevant'' seems the $w^*$--topology, it also
seems natural to consider also (some of) elements of the enveloping algebra
 $\mfk
A''$\ of ``the \Ca\ of observables \fk A\,'' which do not belong to \fk A,
as representing
some observable
quantities of the system. The macroscopic (resp.\ classical) quantities of
the considered quantum system are then found in the centre $\mcl Z(\mfk
A'')$. Any two mutually orthogonal projections of the centre then can
represent macroscopically distinguishable values of some observable
quantity. Hence, $\pi_1\mho\pi_2$\ can be interpreted as macroscopic
distinguishability. It would be, perhaps, more intuitive after a discussion
of disjointness of states,~\cite{hp-meas,bon-m,sewell}, cf.\ also
Interpretation~\ref{int;centre}.\hfill\bpika
\end{intpn}

In general theory of \Ca s, and also in physical applications, cyclic
representations arise from given ``states''.

\section{States and representations}\label{B;state&repr}
\def\nazov{{
\ref{B;state&repr}\quad States and representations}}

We shall introduce here the mathematical definition of states on a \Ca, as
well as some connetions with representations, and we shall give some hints
to their physical interpretations.

\begin{defi}[{\bf States}\index{states}]\label{df;states}
\item{(i)} Let \fk A\ be a \Ca, and $\mfk A^*$\ its topological dual.
A continuous linear functional $\mrh\in\mfk A^*$ is \emm symmetric (or real)~
if $\mrh(x^*)\equiv\overline{\mrh(x)}$. It is, moreover, \emm positive~, if
$\mrh(x^*x)\geq 0, \forall x\in\mfk A$.
The set $\mfk A^*_+$\ consists of all positive elements of $\mfk A^*$. The
elements of $\mcl S(\mfk A):=\mfk A^*_{+1}:=\{\mrh\in\mfk
A^*_+:\|\mrh\|=1\}$\ are \emm positive normalized functionals~ on \fk A.
They are called \emm states on the \Ca\ \bs{\mfk A}~. The set $\mS(\mfk
A)\subset \mfk A^*$\ is convex, i.e.\ $\mrh_j\in\mS(\mfk A)\ (j=1,2),\ 0<\mlam<1
\imply \mlam\mrh_1+(1-\mlam)\mrh_2\in\mS(\mfk A)$.

\item{(ii)} An element of $\mome\in\mcl S(\mfk A)$\ is called a \emm pure
state on \bs{\mfk A}~, if it is not an
internal point of any line segment lying in $S(\mfk A)$, i.e.\ if for some
states $\mome_j, j=1,2$, one has $\mome=\frac{1}{2}(\mome_1+\mome_2)$, then
necessarily $\mome_1=\mome_2=\mome$. Such elements \ome\ of a convex set
\cl S\ are called \emm extremal points of \bs{\mcl S}~. The set of all
pure states on \fk A\ will be denoted by \glss $\mcl{ES}(\mfk A)$~.
In the state space $\mcl S=\mcl S(\mfk A)$, the nonextremal elements are
called \emm mixed states, or mixtures~.

\item{(iii)} Let \fk A\ be a \Wa. Then it is canonically
$\mfk A_*\subset\mfk A^*$.
The subset $\mcl S_*(\mfk A):=\mfk A_*\cap\mcl S(\mfk A)$\ of states
consists of all \emm normal states on \fk A~.
\hfill\pika
\end{defi}

The convex set \cl S := $\mcl S(\mfk A)$\ is
contained in the closed unit ball of $\mfk A^*$, what is compact in the
$\msg(\mfk A^*,\mfk A)$--topology, according
to the Banach--Alaoglu theorem~\cite{R&S}. \cl S is also compact iff \fk
A\ is unital. In other cases, it is usually considered~\cite{pedersen}\
the \emm quasi--state space \bs{\mcl Q}~\ defined by $\mcl Q:=\{\mrh\in\mfk
A^*_+:\|\mrh\|\leq1\}$. The quasi--state space \cl Q\ is the convex hull of
the state space \cl S\ and the zero functional. It is always compact in the
$w^*$--topology. We shall use the following
theorem,~\cite{najm,bourb;vect}:
\begin{thm}[Krein--Mil\acc man]\label{thm;kr&mil}
Every compact convex set \cl S\ in a Hausdorff l.c.s. is closure of the set
of all finite convex combinations of the extremal points of \cl S.\hfill\zal
\end{thm}
We see that there is enough pure states on any \Ca\ so that finite convex
combinations of them can approximate any state in the $w^*$--topology.

\begin{intpn}[States in physics]\label{intpn;states}
If the selfadjoint elements of a $C^*$-al\-geb\-ra\ \fk A are
considered as \emm
observables of a physical system~, then the states on \fk A\ are
interpreted as follows: The state $\mome\in\mcl S(\mfk A)$\ represents a
\emm physical situation~ (whatever it means, we do not go to analyze its
meaning here) of the described system and, for any given $x=x^*\in\mfk A$,
the real number $\mome(x)$\ equals to the expectation value (i.e.\ the
arithmetical average -- in ``finite approximations'') of results of
repeated measurements of the observable $x$\ in the (repeatedly prepared)
``situation'' described by \ome.

Hence, the above mentioned possibility of approximation of any state by
convex combinations of pure ones means an approximation by convex
combinations of
(potential) measurement results; this shows in what sense the
$w^*$--topology on the state space $\mfk A^*_s$ := (the set of symmetric
elements of $\mfk A^*$) is ``more physical'', than the topology of norm,
cf.\ \cite{haag&kast}.
\hfill\bpika
\end{intpn}

The following proposition can be proved by the well known Gel\acc
fand--Najmark--Segal (GNS) construction of a canonical representation
$\pi_\mome$\ corresponding to any
given state \ome, with a use of algebraic
concepts,~\cite{najm,dix2,sak1,emch1,bra&rob}:

\begin{prop}\label{prop;GNS}
A cyclic representation $(\pi_\mome;\mH_\mome;\psi_\mome)$\ in $\mH_\mome$\ with
the cyclic vector $\psi_\mome\in\mH_\mome: \overline{\pi_\mome(\mfk
A)\psi_\mome}=\mH_\mome$, $\|\psi_\mome\|=1$, such that
\[ \mome(x)=(\psi_\mome,\pi_\mome(x)\,\psi_\mome),\quad\forall x\in\mfk
A,\]
corresponds to any state $\mome\in\mcl S(\mfk A)$\ on a \Ca\ \fk A.
Such a representation is unique, up to unitary equivalence. Any cyclic
representation of \fk A\ can be obtained in this way.
The cyclic representation $\pi_\mome$ is irreducible iff \ome\ is a pure
state. In that case, $\pi_\mome(\mfk A)\psi_\mome=\mH_\mome$ (without
taking the closure in $\mH_\mome$).
\hfill\zal
\end{prop}
\label{GNS}
The canonical representations satisfying the conditions of
Proposition~\ref{prop;GNS} are also called the \emm GNS
representations~. Two \emm states $\mome_1,\mome_2\in\mcl S(\mfk A)$\ are
mutually disjoint~, cf.\ page~\pageref{disj},
 \bs{\mome_1\mho\mome_2}, iff $\pi_{\mome_1}\mho\pi_{\mome_2}$; such a
two states can be considered
as macroscopically different,~\cite{emch1,hp-meas,bra&rob,sewell}, and any
two macroscopically different states
are mutually disjoint,
cf.\ Interpretations~\ref{int;macro-d}, and~\ref{int;centre}.
\begin{exmp}[Abelian \Ca s and \Wa s]\label{ex;abel-C&W}
\item{(i)} Any unital abelian \Ca\ \fk A\ is isomorphic to the space of all
continuous complex valued functions $C(M)$\ on a Hausdorff compact $M$. The
set $M$\ can be constructed as the set of all irreducible representations
$\pi_\chi$\ (which are all one--dimensional), and their kernels are \emm
maximal ideals~ of \fk A. The corresponding pure states $\chi\in\mcl S(\mfk
A)$\ are \emm characters on \fk A~, i.e.\ they satisfy also: $\chi(x\dti
y)\equiv\chi(x)\chi(y)$. Since the three sets: the set of irreducible
representations, the set of maximal ideals, and the set of pure states are
in bijective correspondence, they can be endowed, together with the
\emm spectrum space~ $M$, with the induced topology from the
$w^*$--topology of
$\mfk A^*$. Let us denote by $\chi_m$\ the pure state on $C(M)$\
corresponding to the maximal ideal $m\in M$, i.e.\ to the irreducible
representation with the kernel $m$. Let $f\in C(M)$\ be any element of the
commutative \Ca. Then
\[ \chi_m(f)=f(m),\quad\forall m\in M,\]
hence the pure states $\chi_m$\ correspond to the Dirac measures $\delta_m$\
on $M$. Arbitrary states \ome\ are then realized as probability ``regular Borel
measures'' $\mu_\mome$\ on $M$, symbolically
\[ \mome(f)=\int_Mf(m)\mu_\mome(\rd m).\]
The correspondence between states on $C(M)$\ and probability measures on
$M$\ is a bijection, according to Riesz-Markov theorem,~\cite[Theorems
IV.14, and IV.18]{R&S}. Hence, a decomposition of an arbitrary state on an
abelian \Ca\ into a convex combination (here: integral) of pure states, so
called extremal decomposition, is unique.
\item{(ii)} Let us assume, that the abelian \Ca\ $C(M)$\ is a \Wa. We know,
that it is generated by its projections. But a projection in $C(M)$\ is
just a characteristic function $\chi_B$\ of a subset $B\subset M$, which is
also continuous: Hence the set $B=\chi_B^{-1}(\{1\})$ should be a clopen
subset in $M$. It can be shown,~\cite{gamelin,takesI}, that the topology of
$M$ is now generated by its clopen sets. The chracteristic functions of
one--point sets corresponding to {\em normal pure states} $\chi_m$\ can
be considered also as elements of the \Wa: $\delta_{m,m'}=:f_m(m'), f_m\in
C(M)$, because these $m$'s are just the isolated points of $M$.
\hfill\dovi
\end{exmp}

Let $\mome\in\mcl S(\mfk A)$ be a mixed state on a unital \Ca\ \fk A.
Then it can
be \emm decomposed~ into a (generally ``continuous'') convex combination of
other states. Looking for such convex decompositions of a given
$\mome\in\mS(\mfk A)$, we are interested in such probability
measures $\mu_\mome$\ on the compact $\mcl S:=\mcl S(\mfk A)$, that for all
affine continuous functions $f\in C(\mcl S):f(\mlam
\mome_1+(1-\mlam)\mome_1)\equiv \mlam f(\mome_1)+(1-\mlam)f(\mome_2)$, we
have
 \[ f(\mome)=\int_{\mcl S} f(\nu)\mu_\mome(\rd\nu).\]
 The state $\mome$\ is a \emm barycentre of the measure \bs{\mu_\mome}~.
 Specific examples $\hat x$\ of the affine functions are given by arbitrary
 elements  $x\in\mfk A:\ \hat x(\mome):=\mome(x),\forall \mome\in\mS$.
 The measures $\mu_\mome$\ can be ``concentrated'' on various subsets of
 \cS. We have seen that if one assumes that $\mu_\mome$\ is concentrated on pure
 states $\mcl E\mS$, then $\mu_\mome$\ is uniquely determined in the commutative
 case. For general \Ca s, this uniqueness is absent (it is in a sense
equivalent to
 commutativity of the \Ca). Hence, we have to choose some of the measures
 with the barycentre \ome\ to obtain some (barycentric) decomposition.
 These decompositions might be chosen in different ways. We cannot give
 here details; let us see at least some of techniques for construction
 of such decompositions.

 If $\mome=\sum_j\mlam_j\mome_j$\ is a (discrete, convex) decomposition of
 a state \ome\ to states $\mome_j,\mlam_j\neq0$, then the states $\mome_j$\
 are {\em majorized by} \ome.
 Important might be the following lemma stating
conditions for $f\in\mfk A^*_+$\ to be \emm majorized by
\bs\mome~, i.e.
stating conditions when there is a number \lam\ for which
$(\mlam\mome-f)\in\mfk A^*_+$.
 \begin{lem}
Let, with the introduced notation, $\mome\in\mS$, and
$(\pi_\mome;\mH_\mome;\psi_\mome)$\ is the corresponding cyclic
representation. Then there is a bijection between the set of all
$f\in\mfk A^*_+$\ majorized by \ome\nolinebreak, and the set of all
positive elements:
$B\geq0, B\in \pi_\mome(\mfk A)'$,\ of the commutant of $\pi_\mome(\mfk A)$.
The correspondence $f\mapsto B_f\in \pi_\mome(\mfk A)'\cap\mcl L(\mcl
H_\mome)_+$\ is by the relation
\[ f(x)=(\psi_\mome,\pi_\mome(x)\dti B_f\psi_\mome),\quad \forall x\in \mfk
A,\]
 determined uniquely.\hfill\zal
 \end{lem}

Most
simple and useful decompositions are such that are derived from some abelian
\Wsa s \fk B\ of the von Neumann algebra $\pi_\mome(\mfk A)'$.
Let us mention here just a very simple example when the algebra \fk B\ is
generated by a (``discrete'') projection measure $E_d$\ defined on $\mbZ_+$:\
$E_d: j\ (\in\mbZ_+)\mapsto E_d(j)\in\pi_\mome(\mfk A)',
\sum_{j=0}^\infty E_d(j)=\mbI$. Then we
define the states $\mome_j\in\mS$, for those values of the indices $j$ for
which $(\psi_\mome,E_d(j)\psi_\mome)=:\mlam_j\neq0$, by the relation
\[ \mome_j(x):=\mlam_j^{-1}(\psi_\mome,\pi_\mome(x)E_d(j)\psi_\mome)\ (\forall
x\in\mfk A).\]
It is trivially clear that now we can write a (``orthogonal'') decomposition
of the \ome\ by:
\[ \mome(x)\equiv\sum_j\mlam_j\mome_j(x).\]
If the \Wa\ \fk B\ is contained in the centre $\mcl Z(\pi_\mome(\mfk A)') :=
\pi_\mome(\mfk A)''\cap \pi_\mome(\mfk A)'$, the decomposition is called a
{\bf (sub--) central decomposition}\index{decomposition (central, subcentral)}.
\begin{intpn}\label{int;centre}
By the extended representation
$\pi_\mome'':\mfk A''\rarw \mLHo$, cf.\ Proposition~\ref{prop;pi-ext}, the
centre $\mcl Z(\mfk A'')$\ is mapped onto the centre $\mcl Z(\pi_\mome(\mfk
A)'')$. Hence, the subcentral decompositions might be interpreted
physically as decompositions according to values of macroscopic observables.
\hfill\bpika
\end{intpn}
 General theories of decompositions can be found
 in~\cite{sak1,bra&rob,pedersen}.



\section{Symmetries and automorphisms}\label{B;sym&repr}
\def\nazov{{
\ref{B;sym&repr}\quad Symmetries and automorphisms}}

Symmetries appear in quantum theory (QT) either in a form of transformations of
``states'' (the {\em Schr\"odinger picture}), or as transformations
of ``observables'' (the {\em Heisenberg picture}). Although, in the ``standard''
QM, these two forms of symmetry transformations are usually considered as
equivalent, for more general formulations of QT it needn't be so. Some
relations between these two descriptions of symmetry operations in QT are
described in~\cite[Chap. 3.2]{bra&rob}. We shall restrict here our attention
mainly to the
formulation in the ``Heisenberg form'', what is the most usual form of the
description of time evolution in quantum
theories of large systems.

Theory of symmetries, resp.\ automorphism groups of \Ca s is a rather
extensive field, cf.\ \cite{ruelle1,sak1,bra&rob,pedersen,sak2}. There
are known many related and mutually connected fields
 like the ergodic theory, decomposition theory, various kinds of
``spectra'' connected with analysis of structure of algebras and of their
automorphism
groups $\malp_G$\ etc. which we shall not consider in this paper.
 We shall present
here just some notes for a first orientation in approaches to
formulations and investigation of several problems concerning symmetries
of physical systems described by \Ca ic theories, and such which are
connected with techniques used in this work.

Let \fk A\ be a \Ca. The \autm s of \fk A\ (i.e.\ the ${}^*$--isomorphisms
of \fk A\ onto itself) form a group \Aut{\mfk A}\ \wrt composition as the
group multiplication. Each $\malp\in\mAut{\mfk A} $\ is a continuous linear
transformation of the B--space \fk A, hence $\mAut{\mfk A}\subset\mcl
L(\mfk
A)$\ (=the space of bounded linear mappings of \fk A\ into itself), where
$\mcl L(\mfk A)$\ is again, canonically, a B--space. With the
induced topology, \Aut{\mfk A}\ is a topological group; it is also a closed
subset of $\mcl L(\mfk A)$,~\cite[Proposition 4.1.13]{sak1}.
There are also several other useful ``natural''
(weaker--than--norm) topologies introduced on \Aut{\mfk A}, namely the \emm
strong topology~ given by the seminorms
\[\mfk p_x(\malp):=\|\malp(x)\|,\quad \forall\malp\in\mAut{\mfk
A},\quad x\in\mfk A,\]
in which \Aut{\mfk A}\ is also a topological group,~\cite{sak2},
and also some of the \emm \bs{\msg(\mAut{\mfk A},\mcl F^*)}--weak topologies~,
where $\mcl F^*$\ is a ``conveniently chosen'' subset of linear functionals
on the B--space $\mcl L(\mfk A)$, to make \Aut{\mfk A}\ a Hausdorff space,
cf.\ also\dref top-ls~(v). The subset $\mcl F^*$\ is often given by
the requirement of continuity of the mappings
\[ \malp(\in \mAut{\mfk A}) \mapsto \mome(\malp(x)),\quad\forall x\in\mfk
A,\ \mome\in\mcl F',\]
where we have different useful possibilities,~\cite[Definition
2.5.17]{bra&rob}, for a choice of the set $\mcl F'\subset\mfk A^*$.
If \fk A\ is a \Wa, then its automorphisms are continuous mappings of \fk
A\ onto itself not only in the norm--topology, but also in the \sg(\fk
A,$\mfk A_*$)--topology determined by its normal states, cf.\dref states~.
These states are ``often'' chosen in the r\^ole of the set $\mcl F'$\ above,
in the case of a \Wa s \fk A.
The automorphism \alp\ of \fk A\ is called \emm inner~ if there is a
unitary element $u_\malp\in\mfk A$\ such that $\malp(x)\equiv u_\malp
xu^*_\malp$. For $\mfk A:=\mLH$ is each automorphism inner ,~\cite[Corollary
2.9.32]{sak1}.

Any $\malp\in\mAut{\mfk A}$\ determines a unique affine isometry $\malp^*:
\mS(\mfk A)\rarw\mS(\mfk A)$\ of the state space of \fk A\ by the
transposing:
\[ \malp^*(\mome)(x)\equiv \mome(\malp(x)),\quad\mome\in\mS(\mfk A),\
x\in\mfk A.\]
If \fk A\ is a \Wa, the transposed map leaves its normal states invariant:
\bequ\label{eq;1trans}
\malp^*:\mSs(\mfk A)\rarw\mSs(\mfk A).
\end{equation}
This is the transition to the ``Schr\"odinger picture''. The converse needn't
be so immediate:

If it is given an affine mapping $\malp^*: \mS(\mfk A)\rarw\mS(\mfk A)$,
its transpose $\malp^{**}$ determines a linear map of the double dual $\mfk
A^{**}$\ into itself, that leaves its (in a canonical way defined) subset
\fk A\ invariant only in specific cases: some ``sufficient continuity''
conditions should be satisfied, cf.\ \cite[Theorem 3.2.11]{bra&rob}; only
then one can consider the restriction \alp\ of $\malp^{**}$\ to the
subspace \fk A\ of $\mfk A^{**}$\ and ask, whether is $\malp\in\mAut{\mfk
A}$, hence, whether there exists the corresponding ``Heisenberg picture''.
In the case of a \Wa\ \fk A, if the
condition\rref1trans~ is fulfilled, there is a unique Jordan\footnote{Jordan
automorphisms \alp\ of a \Ca\ are a certain ``combinations'' of morphisms,
cf.\dref repres~(i)\
(satisfying: $\malp(xy)\equiv\malp(x)\malp(y)$), and \emm antimorphisms~
(satisfying: $\malp(xy)\equiv\malp(y)\malp(x)$, with other morphism
properties unchanged); hence, by definition, instead of satisfying the
property $(ll):\malp(xy)\equiv\malp(x)\malp(y)$ of the\dref repres~ of
${}^*$--isomorphisms, Jordan automorphisms satisfy the following property:
$\malp(xy+yx)\equiv\malp(x)\malp(y)+\malp(y)\malp(x)$.}
automorphism \alp\ of
\fk A\ obtained by the above mentioned ``transposing'' of
$\malp^{*}$\ and by the subsequent restriction.

A physical meaning is usually given to an automorphism \alp\ according to its
belonging to some subgroup of \Aut{\mfk A}\ which is a homomorphic image
of a topological (usually Lie) group $G$: $\malp\in\malp_G$, where
\[ \malp_{g_1\dti g_2}=\malp_{g_1}\dti\malp_{g_2},\quad\forall g_j\in G,\
j=1,2.\]
The homomorphisms $g(\in G)\mapsto \malp_g$, i.e.\ \emm representations of \bs G~
represent groups of ``physical motions'', or transformations. If the group
is $G=\mbR$, we have a one parameter transformations group $\malp_\mbR$;
such groups describe also time evolutions of the physical systems with \fk
A\ as the ``algebra of observables''.

There are traditional reasons in QM (e.g.\ the spectra of generators of
$U^\pi_G$\
represent measurable values) for interest in such representations
$\{\pi,\mH_\pi\}$\ of the \Ca\ \fk
A\ with a given \emm symmetry \bs{\malp_G}~, in which the automorphisms
$\malp_g,\ g\in G$, are expressed by a unitary strongly continuous
representation $U_G^\pi\in\mcl U(\mH_\pi)$\ of $G$ ``in the usual way'', i.e.
\bequ\label{eq;1cov}
\pi\bigl(\malp_g(x)\bigr)\equiv U_g^\pi \pi(x)U_{g^{-1}}^\pi,\quad\forall
g\in G.
\end{equation}
Such representations $\pi(\mfk A)$\ are called \emm covariant
representations~. A simple important case of a covariant representation is
obtained (we omit here specification of necessary continuity conditions
imposed to $\malp_G$), if there is an \emm \bs{\malp^*_G}--invariant state
\bs{\mome}~ given,~\cite{seg,ruelle1}; the corresponding cyclic representation
$(\pi_\mome;\mH_\mome;\psi_\mome)$\ ensures existence of a unique (continuous)
unitary representation $U^\mome_G$\ satisfying\rref1cov~\ (with
$\pi_\mome\mapsto\pi,\ U^\mome\mapsto U^\pi$), and such that the cyclic
vector corresponding to the state \ome\ is $U^\mome_G$--invariant:
\[ U^\mome_g\psi_\mome=\psi_\mome,\quad\forall g\in G.\]
In more general situations (e.g.\ of cyclic representations with
noninvariant cyclic vectors), all  covariant representations of a
{\em dynamical system} $\{\mfk A,\malp_G\}$ are in a bijective correspondence
with representations of another \Ca\ \bs{\mfk A\otimes_\malp G}\ constructed
from
functions on the group $G$ with values in \fk A\ with a help of the action
of $\malp_G$, and called the \emm crossed product of the dynamical system~
 \bs{\{\mfk A,\malp_G\}}, cf.\ \cite{pedersen}, or also~\cite{naudts}.

Let us consider now $G:=\mbR$, i.e.\ one--parameter automorphism groups.
For \fk A\ = \LH, all one parameter automorphism groups $t\mapsto \malp_t$\
are ``covariant'', i.e.\ they are representable in the form\rref1cov~, i.e.
$\malp_t(x)\equiv u_tx
u^*_t$\ for a one--parameter group of unitary operators $u_t,\ t\in\mbR$.
If the group $\malp_\mbR$\ is ``sufficiently continuous'', e.g.\ if the
functions $t\mapsto Tr(\mrh\malp_t(x)),\ \mrh\in\mfk T_s, x\in\mLH$\ are
all continuous, then $t\mapsto u_t$\ is strongly continuous and, according
to Stone's theorem, cf.\ Theorem~\ref{thm;Stone}, there is a selfadjoint
operator $A$ on (a dense domain of) \H\ such, that
\bequ
u_t\equiv\exp(itA).
\end{equation}
The operator $A$ is determined by the automorphism group $\malp_\mbR$\ up
to an additive real constant.
The operator $A$ is called a {\bf (selfadjoint) generator of \bs{u_\mbR}}\index{generator (selfadjoint) of \bs{u_\mbR}}.
The \emm generator of \bs{\malp_\mbR}~ is obtained as a linear operator
$\delta_\malp(x):=i[A,x]$\ on (a dense subset of) \LH($\ni x$). In a general
case:
\bequ\label{eq;deriv}
\mome\bigl(\delta_\malp(x)\bigr):= \left.\frac{\rd}{\rd
t}\right|_{t=0}\mome\bigl(\malp_t(x)\bigr)
\end{equation}
for all $x\in D(\delta_\malp)\subset\mfk A$. The generator \emm
\bs{\delta_\malp} is called the derivation of \bs{\malp_\mbR}~.
Some details of a theory of (unbounded) derivations can be found
in~\cite{bra&rob,sak2}.

If the group $\malp_\mbR$\ is not ``sufficiently continuous'', the generator
needn't exist. Moreover, some of the covariant representations $\pi$\ of the
same $\{\mfk A,\malp_\mbR\}$\ might be continuous with well defined selfadjoint
generators $A_\pi$, and in other ``covariant'' representations
the unitary groups
$t\mapsto U^\pi_t$\ might be discontinuous (i.e.\ there is no
``Hamiltonian'' there). For different
continuous covariant representations $\pi$ the ``Hamiltonians $A_\pi$'' are
generally mutually
different (e.g., their spectra might be mutually ``very different''). In
the examples of states describing
thermodynamic equilibria for different temperatures the
selfadjoint generators describing time evolution of local perturbations
are mutually different
in known solvable examples, e.g.\ for simple versions of the BCS model of
superconductivity, cf.\ \cite{haag1,thir&wehrl,bon2}. In this last mentioned
example, the representations of the (``quasilocal'') algebra of observables
corresponding to different equilibrium states are all faithful, they mutually
differ,
however, in representing  ``macroscopic quantities'' of the
described infinite quantal system by different operators (resp.\ numbers).
Also in more
general cases, mutually
disjoint representations are distinguished by values of some ``macroscopic
quantities''.

Thermodynamic equilibrium states (also of infinite systems, corresponding
to the ``thermodynamic limit'',~\cite{ruelle1}) can be
defined for any ``sufficiently continuous'' one--parameter automorphism
group $\malp_\mbR$ of a \Ca. This fact is interesting as such, from the
point of view of traditional techniques for statistical--mechanical
description of thermodynamic equilibria by Gibbs statistical ensembles,
because for time evolutions $(t;x)\mapsto \malp_t(x)$ of an infinite
system there is no global Hamiltonian operator $H$ to be inserted into the
expression of a ``statistical sum'', e.g.\footnote{Even in some
``traditional'' cases, when the Hamiltonian $H$ is a well defined
selfadjoint operator, the trace in the following formula does not exist.
Take, e.g.\ a hydrogen atom in a box.}\ into
\[ Z(T,H):=Tr\exp\left(-\frac{1}{kT}H\right).\]
The definition of the thermodynamic equilibrium states $\mome=\mome_\beta$\
for a temperature $T=:(k\beta)^{-1}$ ($k$ is here the Boltzmann constant) of
infinite (and other) systems is expressed by the \emm KMS
condition~ for states \ome,~\cite{HHW,bra&rob,pedersen}:\footnote{KMS is for
Kubo, Martin, and Schwinger.}
\bequ\label{eq;KMS}
\mome(\malp_{\mlam}(y)x)\equiv\mome(x\malp_{\mlam+i\beta}(y)),\
\forall\mlam\in\mbC, x\in\mfk A, y\in\mfk A_a,
\end{equation}
\label{KMS}
where $\mfk A_a\subset \mfk A$\ is the set of analytic elements \wrt \alp\
(i.e.\ $x\in\mfk A_a\eequiv \mlam\mapsto \malp_\mlam(x)$\ is an
entire--analytic \fk A--valued function, cf.\ \cite[Chap.VI]{R&S}).
States $\mome_\beta$\ satisfying the condition\rref KMS~ are the \emm
\bs{\beta}--KMS states for \alp~, with a given $0<\beta<+\infty$.

It is an interesting result of the \emm Tomita--Takesaki theory~ of modular
Hilbert algebras,~\cite{tomita,pedersen,bra&rob}, that for a class of states
\ome\ of any \Ca\ \fk A\ one can find a canonical one--parameter automorphism
group (called the \emm modular group for \bs{\{\mfk A,\mome\}}~)
of the weak closure $\pi_\mome(\mfk A)''$ of the GNS--representation of \fk
A\ such, that the chosen state \ome\ is a KMS--state of that automorphism
group at $\beta=1$ (this finite nonzero value of $\beta$\ is chosen
arbitrarily). The
condition for the class of states \ome\ allowing this ``creation of
dynamics from states'' is, that \ome\ is faithful for $\pi_\mome(\mfk
A)''$, i.e.\ that for any positive (nonzero) operator $B\geq0$ in this \Wa,
its diagonal
matrix element with the cyclic vector $\psi_\mome$\ is strictly positive:
$(\psi_\mome,B\psi_\mome)\neq 0$ (hence, the \Wa\ $\pi_\mome(\mfk A)''$
does not contain any ``annihilation operators'' \wrt the ``vacuum vector''
$\psi_\mome$).

Let us mention also the phenomenon
of ``spontaneous symmetry breaking'' in a
stationary state \ome\ \wrt ``dynamical evolution group''
$\tau_\mbR\subset\mAut{\mfk A}$. Assume that there is another automorphism
group $\malp_G$\ commuting with the time evolution $\tau$:
\[ \tau_t\circ\malp_g=\malp_g\circ\tau_t,\quad\forall t\in\mbR,\ g\in G.\]
This situation ``corresponds'', e.g.\ to commutativity of the Hamiltonian
as the selfadjoint generator of the unitary group implementing the time
evolution $\tau_\mbR$\
with generators of  the transformation group $\malp_G$ for a Lie
group $G$.

The notion of states with ``broken'' symmetry comes from expectations that
a certain states will have larger symmetry than they really
have,~\cite{curie}. Let us assume that, e.g., in the usual formulation of QM,
the Hamiltonian $H$ is invariant \wrt a unitary representation $U_G$\ of a
finite--dimensional Lie group $G$ in \H: $H\equiv U_gHU^*_g$.
If there is an eigenvector $\psi_\mveps\in\mH$\ of $H$:
$H\psi_\mveps=\mveps\psi_\mveps$, then also all the vectors
$\{U_g\psi_\mveps:g\in G\}$\ are eigenvectors of $H$ with the same
eigenvalue \veps.
Then a nondegenerate eigenvector $\psi_\mveps\in\mH$\ is
proportional to all the vectors $U_G\psi_\mveps\in\{\mlam
\psi_\mveps:\mlam\in\mbC\}$, hence the state
\[x\mapsto
(\psi_\mveps,x\psi_\mveps)\equiv
(U_g\psi_\mveps,xU_g\psi_\mveps),\ x\in\mLH\] is also
``$G$--invariant''. If the eigenvalue \veps is of higher
multiplicity, the $G$--invariance of $\psi_\mveps$\ might be
``broken''. Similar considerations apply to equilibrium states at
fixed temperature: If, in the above situation, there is only one
KMS--state for a given $\beta$, then it is invariant also \wrt
$\malp_G$. The phenomenon of \emm phase transitions~ is usually
considered as equivalent to existence of several KMS states for
any temperature of ``phase coexistence'', e.g.\ below the critical
temperature of a ferromagnet. In the last mentioned case, e.g.,
the group $G$ might be the Euclidean group in $\mbR^3$\ (or only
its rotation subgroup $O(3)$) \wrt  which the Hamiltonian of the
ferromagnet is invariant. Different (extremal) KMS states
correspond to different directions of the magnetization of the
ferromagnet, hence the rotation symmetry $\malp_G$ is broken;
translation symmetry is broken in states of any crystal state of
many--particle systems (with translation invariant Hamiltonians).
The stationary {\em and $G$--invariant} states always exist, but
in the mentioned ``degenerate'' situations they are not
``extremal'': they have nontrivial convex decompositions to states
(e.g.\ equilibrium) with lower, hence ``broken'', symmetry. These
situations are considered in the above mentioned decomposition
theory, resp.\ in a part of the ergodic
theory,~\cite{ruelle1,bra&rob,emch1,walters,K&S&F,sak2,connes}.

\begin{noti}\label{note;types}
Let us add several words on possible structures of physically relevant \Ca
s, resp.\ \Wa s. It is useful to classify \Ca s according to the sets of
projections contained in (the \Wa s obtained by) the weak closures of their
(e.g.\ GNS) representations. Let us concentrate on a von Neumann
classification of
\Wa s. Let \fk M\ be a \Wa. Let us denote by \fk{P(M)} the set of all
projections
in \fk M. Two projections $p_j\in\mfk{P(M)}, j=1,2$\ are \emm
equivalent~: \bs{p_1\sim p_2}, if $\exists u\in\mfk M: p_1=uu^*, p_2=u^*u$.
This allows us to introduce an ordering between projections in \fk M: $p\prec
q\eequiv \{\exists p'\leq q\ \&\ p\sim p'\}$. If $\{p\sim q\leq p\imply
p=q\}$,
then $p$ is \emm finite~. If $\exists q<p\ (q\neq p)\ \&\ q\sim p$, then $p$
is
\emm infinite~. If $0\neq q\prec p\imply q$\ is infinite, then $p$ is \emm
purely infinite~. A projection $p\in\mfk{P(M)}$ is \emm abelian~,
if $p\mfk M p:=\{p\dti x\dti p:\ x\in\mfk M\}\subset\mfk M$\ is an abelian
algebra; $p$ is \emm minimal~, if $p\mfk M p\sim\mbC$. We call \fk M\ \emm
finite~ (resp.\ \emm infinite~, resp.\ \emm purely infinite~), if its identity
$e:={\rm id}_{\mfk M}$, as a projection, is finite (resp.\ infinite, resp.
purely infinite). \fk M\ is \emm continuous~, if $\forall p\in\mfk{P(M)}$\
there
are $q,q'\in\mfk{P(M)}:\ p=q+q',\ q\dti q'=0,\ q\sim q'$. Now we can introduce
the types of \Wa s:
\bs{\mfk M}\ \emm is of type I~\quad $\eequiv\quad\forall p\in\mfk{P(M)}\
\exists$\  abelian
$q\leq p\quad\eequiv\quad\mfk M$ is isomorphic to a \Wa\ with abelian
commutant.
\bs{\mfk M}\ \emm is of type II~, iff it is continuous and its center \cl
Z(\fk M)\ does not contain purely infinite projection.
\bs{\mfk M}\ \emm is of type III~, iff it is purely infinite ($\imply$\ \fk
M\ is continuous, and each nonzero $p\in\mfk{P(M)}$\ is purely infinite).
Any \fk M\ can be written as $\mfk M_I\oplus\mfk M_{II}\oplus\mfk M_{III}$,
with $\mfk M_\malp$\ of type $\malp\in\{I,II,III\}$,
cf.\ \cite{connes,sak1,takesI}.

Let $\mbI_\mH\in\mfk M\subset\mLH$. Then the type (I, II, or III) of the
commutant $\mfk M'=$\ the type of \fk M. For \fk M\ of type III, no pure
state is normal (hence no vector--state given by $\psi\in\mH$\ is pure).
Von Neumann even doubted existence of type III algebras,~\cite{neum2}.
Now we know, that perhaps ``most'' of \Wa s occuring in QT are of type III:
Many KMS states lead to type III representations, and also many algebras of
observables ``localized'' in restricted domains of Minkowski space in
relativistic QFT
are of type III, cf.\ \cite{bra&rob2,sewell,horuz,haag2}.
Such a ``wild'' structure of the physical \Ca s is (also) a consequence of
imposed symmetries.\hfill\dovi
\end{noti}


\chapter{Notes on Unbounded Operators in Hilbert Space}\label{C;oper}
\def\autor{{
\ref{C;oper}\quad Notes on Unbounded Operators in Hilbert Space}}

Unbounded operators usually appear in QM as selfadjoint generators $A$ of
one--parameter unitary
groups $t\mapsto u_t\equiv\exp(-itA)$ which are not continuous in norm topology
of \LH, but they are operator--weakly continuous. Such generators seem to
be unavoidable in the present--day formalism of QM, since their presence
is a consequence of usage of ``nontrivial'' unitary representations of
noncompact Lie groups $G$ ``of motions'', such as Galileo, or Poincar\'e
groups.
Hence, necessary unboundedness of some operators in QM can be connected,
e.g.\ with our common models of noncompact space--times.

Unbounded linear operators $A$ are also characterized by their domains of
definition $D(A)$ which are, as a rule, dense but not equal to the B-space,
 on which the operators $A$ act. This is especially a property of
unbounded symmetric operators on an infinite--dimensional Hilbert space \H,
and these will be the object of our interest in this Section. The reason
for a necessity of dealing with unbounded symmetric operators in some details
in framework of papers on physical
applications is that ignorance of several basic facts can lead to serious
ambiguities in obtained results. Several methods and results presented in
the following subsections can be generalized to other spaces and operators
 than Hilbert spaces and operators acting on them.

\section{Unbounded operators, their domains and adjoints}\label{C;adj}
\def\nazov{{
\ref{C;adj}\quad Unbounded operators, their domains and adjoints}}

Let \H\ be an infinite--dimensional Hilbert space with scalar product
$(x,y)=\overline{(y,x)},\ x,y\in\mH$, and let $A$ be a linear
mapping from a linear subset $D(A)\subset\mH$\ into \H. The linear set
$D(A)$\ is called the \emm domain of \bs A~, and the mapping $A$ is a \emm
linear operator on (a domain \bs{D(A)} in) \bs{\mH}~. We shall usually
assume (if it will be possible) that \DA\ is dense in \H, i.e.\ the norm closure
$\overline{D(A)}=\mH$. The operator $A$ is \emm symmetric~ if
\bequ\label{eq;1symm}
(x,Ay) = (Ax,y),\quad \forall x,y\in D(A),\quad \overline{D(A)}=\mH.
\end{equation}
If $D(A)=\mH$\ for a symmetric $A$ (now symmetry means $(x,Ay)\equiv(Ax,y)$),
then $A$ is
bounded (Hellinger--Toeplitz).
We shall introduce now a useful description of operators on \H. Let us consider
the Hilbert space $\mH\oplus\mH$\ consisting of ordered couples $(x;y),\
x,y\in\mH$, with pointwise linear combinations
$(x_1;y_1)+\mlam(x_2;y_2)\equiv (x_1+\mlam x_2;y_1+\mlam y_2)$, and with
scalar product $((x_1;y_1),(x_2;y_2))\equiv(x_1,x_2)+(y_1,y_2)$.
For any operator $A:\ D(A)\mapsto \mH$, let us define the \emm graph
\bs{\Gamma(A)}\ of \bs A~ as a subset of $\mH\oplus\mH$:
\bequ\label{eq;1graph}
\Gamma(A):=\{(x;Ax):x\in D(A)\}.
\end{equation}
If $\Gamma(A)$ is closed in the norm of $\mH\oplus\mH$, the operator $A$ is
\emm closed~. If the closure of the graph of an operator $A$ is again a
graph of a (uniquely defined) operator, we denote this operator $\overline
A$, it is called the \emm closure of \bs A~, and that operator $A$ (with
$\overline{\Gamma(A)}=\Gamma(\overline A)$) is called a \emm closable
operator~. The closure of an (closable) operator is a closed operator.

Let $A$ be now a densely defined linear operator on \H\ (such are, e.g.\ all
bounded operators $A\in\mLH$). Let us define, for any $x\in\mH$, the linear
functional
\[ f_x^A:=(x,A\cdot):y(\in D(A))\mapsto f_x^A(y):=(x,Ay)(\in\mbC) \]
on the dense domain of $A$. If this linear functional is continuous (in
the induced topology from the norm--topology of \H), hence bounded, it can
be uniquely
extended by linearity and continuity to the whole Hilbert space \H. We
shall denote these extensions by the unchanged symbols. In that
case $f_x^A\in\mH^*$. The dual $\mH^*$\ of \H\ is antilinearly isomorphic
to \H\ itself; hence, each its element $f\in\mH^*$ is uniquely represented by
an element $y_f\in\mH$\ by the identification $f(x)\equiv(y_f,x)$ (this is
the \emm Riesz' lemma~,~\cite{R&S}). Let us denote, with $A$ fixed,
by $\tilde x\in \mH$\ the vector
corresponding by the Riesz lemma to (the continuous extension of) $f_x^A
\in\mH^*$. The \emm adjoint \bs{A^*}~ of $A$ is a linear operator on \H\ with
the domain
\bequ\label{eq;1dom*}
D(A^*):=\{x\in\mH:\ f_x^A \in\mH^*,\text{i.e.\ there is}\
\tilde x\in\mH,(\tilde x,y) \equiv(x,Ay)\},
\end{equation}
and with the values
\[ A^*x:=\tilde x,\quad\forall x\in D(A^*).\]
It is seen that the density of \DA\ in \H:\ $\overline{D(A)}=\mH$, is essential
for possibility of definition of the adjoint operator $A^*$.

For $D(A)=\mH$, this definition of adjointness is the ``usual one'', valid
also for the bounded $A$'s. It is easily seen that $A^*$\ is a linear
operator (hence $D(A^*)$\ is a linear subset of \H), but it needn't be
densely defined.

The reader can check that this definition of \As\ can be expressed in terms
of graphs as follows: Let $V$ be the unitary operator on \HH\ defined by
$V:(x;y)\mapsto(-y;x)$. Then the graph of \As\ is expressed as an
orthogonal complement
\bequ\label{eq;2graph}
\Gamma(A^*)=[V\Gamma(A)]^\perp,
\end{equation}
hence it is closed. It follows that the adjoint operator is always closed.

For two operators $A,B$ on \H, we write $A\subset B$\ iff $D(A)\subset
D(B)$, and $Ax=Bx,\forall x\in D(A)$, i.e.\ $A\subset
B\eequiv\Gamma(A)\subset\Gamma(B)$. In this case, \emm \bs B is an
extension of \bs A~, or \emm \bs A is a restriction of \bs B~.
It is clear from this that a restriction of any closed operator is closable.

\section{Symmetric operators and their (selfadjoint ?)
extensions}\label{C;symm}
\def\nazov{{
\ref{C;symm}\quad Symmetric operators and their (selfadjoint ?)
extensions}}

A symmetric operator $A$ is \emm selfadjoint~ if $A=A^*$, i.e.\ if for the above
defined domain\rref1dom*~ we have \DAs=\DA. If ``$\ i\,\cdot\ $'' is the
multiplication by the imaginary unit $i\in\mbC$ in \H, and an operator
$A$\ on \H\ is selfadjoint, its multiple \bs{i\dti A}\ is called \emm
antiselfadjoint~. Only (anti-)selfadjoint operators
can determine one parameter weakly--continuous unitary groups uniquely.
e.g.\ generators of time
evolution (Hamiltonians) in QM should be selfadjoint, and not just symmetric.

It is seen
from the definition\rref1symm~ of a symmetric operator $A$ that the
definition is equivalent to the condition $A\subset A^*$.
The \emm Hellinger--Toeplitz
theorem~ states,~\cite{R&S}, that if a symmetric operator $A$ is everywhere
defined : \DA=\H, then it is continuous: $A\in\mLH$. This shows, that an
unbounded
symmetric operator cannot be defined on the whole Hilbert space \H.
Most of the Hamiltonians $H$ of particle systems in models of QM are unbounded
symmetric operators, e.g.\ formally defined second order differential
operators $\sum a_{jk}(q)\partial_j\partial_k+v(q)$\ on
$\mH:=L^2(\mbR^n,\rd^nq)$,
where  an ``initial domain'' can be chosen such that $H$
is symmetric, e.g.\ $D(H):= C^\infty_0(\mbR^n)$,  but it is not there
self\-ad\-joint.  The natural question arises, whether
there is a selfadjoint extension of such a $H$. The answer needn't be, in
a general case, positive: Besides an ``ideal possibility'' of existence of
a unique selfadjoint extension, one can have, for some $H$'s, infinitely
many (physically distinct) possibilities, or also there could be no
selfadjoint extension of some $H$'s. The theory analyzing this situation
was formulated by J. von Neumann, known sometimes as \emm deficiency indices
theory~. Let us describe briefly its results.

Let $A$ be symmetric, hence densely defined with densely defined adjoint
\As. Then there is defined the second adjoint $A^{**}$\ of $A$, and from
the graph formulation\rref2graph~ of definition of the adjoint operator one
can see that
\bequ
A\subset A^{**}\subset A^*,
\end{equation}
and that $\overline A=A^{**}$. If $A^*=A^{**}$, the operator $A$ is called
\emm essentially selfadjoint~, and this is the only case, when $A$ has a
unique selfadjoint extension which is then equal to \As=$\overline A$.
Since any symmetric operator $A$ is closable, we can assume, that we have a
closed $A=A^{**}\subset A^*$\ from the beginning. Our present problem is
about classification of conditions for existence of possible selfadjoint
extensions of a (generally not essentially selfadjoint) closed symmetric
operator $A$.

Let us introduce, for a given $A=A^{**}\subset A^*$\ two linear subsets
$\mcl K_\pm^A:=\mKer{A^*\pm iI_\mH}$\ of $D(A^*)\subset\mH$.\footnote{Remember
that $\mKer F$\
for a linear operator $F$ is the subset of its domain on which its values
vanish.} Their dimensions $n_\pm(A)$
(finite, or not) are called the \emm deficiency indices of \bs A~. A closed
symmetric operator $A$ is selfadjoint iff both of its deficiency indices are
equal to zero: $n_+(A)=n_-(A)=0$, i.e.\ if the adjoint operator \As\ has no
eigenvalues equal to $\mp i$.
The domain \DAs\ can be endowed with the scalar product
\bequ\label{eq;1GrA}
 (x,y)_A:=(x,y)+(A^*x,A^*y),\ \forall x,y\in D(A^*),
\end{equation}
and it becomes a new Hilbert space $\mcl H_A=D(A^*)$\ in this way.
The three linear subspaces \DA, $\mK_\mp^A$\ are closed, mutually
orthogonal subspaces of $\mH_A$ providing its orthogonal
decomposition. This ``reorganization'' of the dense subspace \DAs\
of \H\ allows us to find an elegant expression for all closed
symmetric extensions of $A$; this is done with a help of the
bilinear form $\msg_A$\ on $\mH_A$\ defined by:
\[\msg_A[x,y]:= (A^*x,y)-(x,A^*y),\quad\forall x,y\in D(A^*).\]
Closed symmetric extensions $A_W$\ of $A$ are exactly all the restrictions
of \As\
onto arbitrary closed linear subspaces $D_W$ of $\mH_A$ that contain \DA,
and annihilate the form $\msg_A$:
\bequ
\msg_A[x,y]=0,\quad\forall x,y\in D_W.
\end{equation}

From these results, one is able to construct domains $D_W$\ of the symmetric
extensions $A_W$\ with a help of linear isometries $W$\ (in the original
Hilbert
space \H) from closed linear subspaces $S_W$\ of $\mK_-^A$\ into $\mK_+^A$,
$\dim S_W\leq \min\{n_-(A),n_+(A)\}$. The domain $D_W$\ is
\bequ\label{eq;2dom}
D_W:=\{y+x+Wx: y\in D(A), x\in S_W\},
\end{equation}
and the wanted symmetric closed extension $A_W$\ of $A$ is:
\bequ\label{eq;1ext}
 A_W(y+x+Wx):= Ay+ix-iWx,\quad \forall y\in D(A),\ x\in S_W.
\end{equation}
The deficiency indices of this $A_W$ are $n_\pm(A_W)=n_\pm(A)-\dim S_W$, if
$\dim S_W<\infty$. We see that selfadjoint extensions of $A$ exist iff
it is $n_-(A)=n_+(A)$. In that case, all the selfadjoint
extensions are in the easily definable bijective correspondence with all
linear isometries $W$ of $\mK^A_-$\ onto $\mK^A_+$. Hence, the selfadjoint
extensions $A_W$\ of a symmetric operator $A$ with equal deficiency indices
$n_\mp(A)=:n$ are in bijective correspondence with the elements of the Lie
group $U(n)$ of all unitary operators of an $n$--dimensional complex
Hilbert space onto itself. The action of $A_W$'s on the corresponding domains
is given by\rref1ext~, where $S_W:=\mK^A_-$.

\section{The spectral theorem. Stone's theorem}\label{C;spect}

The resolvent set and spectrum of a selfadjoint unbounded operator $A$ is
defined essentially in the same way
as it was done for bounded operators in Subsection~\ref{B;b-oper}:
The resolvent set $\rho(A):=\{\mlam\in\mbC:(\mlam\mbI-A)^{-1}\in\mLH\}$,
but the spectrum $\msg(A):=\mbC\setminus \rho(A)\subset\mbR$\ is not
compact now. Also in this case, however, it is possible to associate unique
projection measure $E_A$\ on the real line (supported by the spectrum
$\msg(A)$) to any selfadjoint $A$, and to formulate the corresponding spectral
theorem expressed by the same formula, as it was done in the ``bounded
case'', cf.\ Theorem~\ref{thm;b-spectral}. This
projection measure provides a transparent representation of the \emm
functional calculus~ also for unbounded $A$,
cf.\ Subsection~\ref{B;b-oper}, and
Subsection~\ref{B;c-alg}. It is now natural, however, to use
also unbounded real Borel functions $f$ on \bR\ for construction of other
unbounded operators $f(A)$ from the given one, cf.
Theorem~\ref{thm;b-spectral}. In the case of unbounded functions $f\in\mcl
F(\mbR,E_A):=${\em the set of measurable, $E_A$--almost everywhere finite
(i.e.\ $E_A\bigl(f^{-1}(\{\infty\})\bigr)=0$)  real functions on}\ \bR, the
domain questions arise. One has (cf.\ \cite{bir&sol})
\begin{prop}\label{prop;fA-dom}
Let $A$ be a selfadjoint (generally unbounded) operator, and let $E_A$\ be
its canonical spectral (projection valued) measure. Let $f\in\mcl
F(\mbR,E_A)$, and let\footnote{We skip here details on exact meaning of
the integral in the spectral representation of $f(A)$.}
\[ f(A)=\int_\mbR f(\mlam) E_A(\rd\mlam).\]
The operator $f(A)$ is selfadjoint, with the (dense) domain
\[ D(f(A)):=\{x\in\mH: \int_\mbR |f(\mlam)|^2 (x,E_A(\rd
\mlam)x)<\infty\}.\]
For any two functions $f,h\in\mcl F(\mbR,E_A)$, and for
$0\neq\mlam\in\mbR$, one
has \item{(i)} $D(f(A)+\mlam h(A))=D(f(A))\cap D(h(A))\subset D((f+\mlam
h)(A))$;
\item{(ii)} $D(f(A)h(A))=D((f\dti h)(A))\cap D(h(A))$.
All these operators $\{f(A):f\in\mcl F(\mbR,E_A)\}$\ mutually commute, i.e.
their projection measures commute.\hfill\zal
\end{prop}
Clearly, the special choice $f(\mlam)\equiv\mlam$\ gives $f(A)=A$.
Another (bounded, but complex) choice $f(\mlam)\equiv \exp(it\mlam)$\ gives
a one--parameter unitary group $U(t)$:
\[ t\mapsto U(t):=\exp(itA)=\int_\mbR \exp(it\mlam) E_A(\rd \mlam).\]
This group is strongly continuous, and it is also norm--continuous iff $A$
is bounded. Different operators $A$ give different groups $U(t)$.

The converse statement is the celebrated \emm Stone's
theorem~,~\cite{riesz&nagy,R&S}:
\begin{thm}[Stone]\label{thm;Stone}
Let $t(\in\mbR)\mapsto U(t)$\ be a weakly continuous one--parameter unitary
group on
a Hilbert space \H, i.e.\ $U(t_1+t_2)\equiv U(t_1)U(t_2)\in\mcl U(\mH)\
(\forall t_1,t_2\in\mbR)$, and
all the complex--valued functions $t\mapsto (x,U(t)y),\forall x,y\in\mH$\
are continuous. Then there is a unique selfadjoint operator $A$ such, that
\[ U(t)\equiv\exp(itA).\]
(Let us note, that strong and weak continuity of the unitaries $U(t)$ are
equivalent.) \hfill\zal
\end{thm}

This theorem has a natural generalization to many--dimensional commutative
locally compact groups of continuous unitary transformations of \H\ known
as the \emm
SNAG theorem~ (by Stone--Najmark--Ambrose--Godement), cf.\ \cite[Chap.
X.140]{riesz&nagy},~\cite[Chap. IV]{GRS},~\cite[Theorem VIII. 12]{R&S}. The
SNAG theorem can be used naturally also for construction of ``macroscopic
(classical) subalgebras'' of large quantal systems determined by a group
action,~\cite{bon1}. Let us formulate it in some details here.

Let us remember first here that any locally compact commutative ($\equiv$ abelian)
group $G$ has a unique \emm dual group $\hat{G}$~ consisting of \emm continuous
characters of $G$~, i.e. of continuous one dimensional unitary representations
$\chi: G\rightarrow\mathbb{C}, \gamma\mapsto\chi(\gamma)\in\mathbb{C}, |\chi(\gamma)|\equiv 1.$
The group composition law in $\hat{G}$ is just pointwise multiplication:
$(\chi_1\cdot\chi_2)(\gamma) \equiv \chi_1(\gamma)\chi_2(\gamma).$ The relevant topology on $\hat{G}$ is now
the $w^*-$ topology determined by (continuity of) the functions $\chi\mapsto\chi(\gamma), \forall\gamma\in\hat{G}.$
The group $\hat{G}$ is then also a \emm locally compact group~. Let us formulate now the \emn SNAG theorem~.

\begin{thm}[SNAG]\label{snag}
Let G be a locally compact \emm abelian group~ and $U(G)$ be its \emn weakly continuous~
\emn unitary representation~ in a Hilbert space \H, $\gamma(\in G)\mapsto U(\gamma)\in\mcl U(\mH),
U(\gamma_1\gamma_2)=U(\gamma_1)U(\gamma_2), \forall \gamma_j\in G.$ Let $\hat{G}$ be the dual group of $G$.
Then there is a unique orthogonal \emn projection valued measure~ (defined on Borel subsets of $\hat G$), $E_U:
B(\subset\hat{G})\mapsto E_U(B)=E_U(B)^*=E_U(B)^2 (\in\mLH)$ contained in the commutant of $U(G)$ such, that
$$ U(\gamma)\equiv \int_{\hat G}\chi(\gamma)E_U({\rd \gamma}).$$
The integral representing the unitary group $U(G)$ converges in the sense, that for any $\varepsilon>0$ there is
a finite Borel decomposition $\hat{G}=\cup_jB_j$ such, that
$$\|U(\gamma)-\sum_j\chi_j(\gamma)E_U(B_j)\|<\varepsilon,$$
for all $\gamma\in G$ and any $\chi_j\in B_j.$ \hfill\zal
\end{thm}
Let us note that the SNAG theorem was used also in our work \cite[Eqs. (2.8)-(2.16) and Section IV]{bon1} to express connections
of classical \emn macroscopic variables~ \index{classical observables}(resp. observables) of  \emn macroscopic subsystems~ of large quantal systems and their dynamics with local
quantal observables and their dynamics, cf. also Section \ref{sec;IIIB} of this book.

\vspace{8cm}

\centerline{\Huge $\clubsuit\ \ \diamondsuit\ \ \spadesuit\ \ \heartsuit\ \ \clubsuit$}

\newpage
\def\autor{Pavel B\'ona: Extended Quantum Mechanics}

\label{index}\hypertarget{index}{}
\printindex


\twocolumn[{\Huge\bf List of Symbols}\vspace{40pt}]\label{symbols}\hypertarget{symbols}{}
\markboth{LIST OF SYMBOLS}{LIST OF SYMBOLS}


\nl{$P({\cal H})$\ }{{ }\ \bf\bf\pageref{mot;modif}, \bf\pageref{PH}}

\nl{$U(G)$}{{ }\ \bf\pageref{rem;G-unb},\ \bf\pageref{eq;2.25}}

\nl{$Ad^*(G)$}{{ }\ \bf\pageref{AdGo},\ \bf\pageref{AdG}}

\nl{$F(\xi) \equiv <F;\xi>$}{{ }\ \bf\pageref{Fxi},\ \bf\pageref{df;2Ad}}

\nl{PM=PVM, POV=POVM}{{ }~\bf\pageref{PVM},\ \bf\pageref{df;1p-measure}}

\nl{$\Delta(B)$}{{ }~\bf\pageref{PVM}}

\nl{$\mfk{F},\mfk{T},\mfk{H},\mfk{C},\mLH$}{{}~\bf\pageref{q-phsp;basic}}

\nl{$\|a\|$},\ {\Ss},\ {\cS},\ {\fk U},\ {$\tilde{\mfk A}$},\ {${\cal U(H)}$},\ {$\mgam_{\ru}$}{{ }~\bf\pageref{PH}}

\nl{$Ad(\ru)$},\ {$\mcl O_{\mrh}$},\ {$\nu(\ry)$},\ {$\mrh({\rm b})$},\ {$E_j$}{{ }\bf\pgr2.1~}

\nl{$\mfk U_{\mrh}$},\ {$\mfk M_{\mrh}$},\ {$\mfk N_{\mrh}$},\ {${\rm p}_{\mrh}$},\ {\qr}{{ }~\bf\pageref{lem;2.2}}

\nl{$\ad^*$},\ {\adrs},\ {\ber},\ {$\mTr\mOU$},\ {\N\cdot,\mrh~}{{ }\bf\pgr2.3c~}

\nl{${\cal F}$(${\cal B}$)},\ {$\{f,h\}$},\ {$h_{\rm  y}(\nu )$}{{ }~\bf\pageref{df;2.5}}

\nl{${\cal D}({h}_{X})$},\ {$D(X)$},\ {$U^{X}$}{{ }\bf\pgr2.18~}

\nl{$D_a(X):=D^{\omega }(X),\ D_d(X),\ D(X)$}{{ }~\bf\pageref{df;domains}}

\nl{${\cal D}(\delta_{X})$},\ {${\cal D}_r(\delta_X) :={\cal D}(\delta_X)\cap {\mathfrak F}_s\cap {\cal S}_*$}{{ }~\bf\pageref{df;domains}}

\nl{${\cal D}_r(X)$},\ {${\cal D}_{ra}(X)$},\ {${\cal D}_{rd}(X)$}{{ }~\bf\pageref{df;domains}}

\nl{${\cal D}_{ra}(\delta_X):= {\cal D}_r(\delta_X)\cap {\cal D}_{ra}(X)$}{{ }~\bf\pageref{df;domains}}

\nl{${d_{\varrho }{{ }  {h}_{X}}}\in T_{\varrho }^*{\cal O}_{\nu}({\mathfrak U})$}{{ }~\bf\pageref{df;dhX}}

\nl{${\mathbf{\mathaccent "7014 {v}}}_f(\nu )$},\ {${\mathbf v}_f^{(n)}(\nu )$}{{ }\bf\pgr1vf~}

\nl{\bcD,\ {\bf D}}{~\bf\pageref{notat;domains}}

\nl{{${\rm q}_{\varrho }({\boldsymbol{{\cal D}}}_rh)$},\ {$d_{\varrho}h(i[\varrho,{\rm  b}]),\ \boldsymbol{{\cal D}_{r+}^1}$}}{\bf\pgd g-dif~}

\nl{${\mathbf v}_{h}(\varrho ),\ \mcl V_\nu$}{\bf\pgr3g-dif~}

\nl{$D^{\omega }(G)$,\ {$Lie(G)\equiv {\mathfrak g}$},\ {${\cal D}^{\omega }_r(G)$}}{\bf\pgr2.26~}

\nl{$Lie(G)^*\equiv {\mathfrak g}^*$}{\bf\pgd2.17~,\bf\pgd2Lie-alg~}

\nl{${\boldsymbol{{\cal D}}_r}({\mathbb F}):={\cal D}^{\omega}_r(G)\subset{\boldsymbol{{\cal D}}}({\mathbb F})$}{\bf\pgd2.17~}

\nl{${\mathbb F}:\ {\boldsymbol{{\cal D}}}({\mathbb F})\rightarrow Lie(G)^*$}{\bf\pgr2.28~}

\nl{$\varrho \DOTSB \mapstochar \rightarrow{\mathbb F}(\varrho ):=F_{\varrho }$}{\bf\pgr2.28~}

\nl{$f_{\xi
}:Lie(G)^*\rightarrow {\mathbb R}$}{\bf\pgr2.28~}

\nl{$\boldsymbol{G_{{\mathbb F}(\varrho
)}}$}{~\bf\pageref{lem;stabG}}

\nl{$\{\mbF^*h,\rf\}(\nu)$}{\bf\pgr2.b33~}

\nl{${\rm  f}^{\nu
}={\mathbb F}^*f^{\nu },\ \boldsymbol{{\cal E}_{{\mathbb
F}}}$}{\bf\pgd2.19~}

\nl{{\rm  Ran}($\mathbb F),\ {\cal E}_{{\mathbb
F}}^0$}{~\bf\pageref{prop;2.24}}

\nl{conv$_0(B)$},\
{conv$(B)$}{~\bf\pageref{prop;2.24}}

\nl{${\cal G}^G_{cl},\
\tilde\mphi^\rf_t,\ \mphi^\rf_t$}{\bf\pgr2.13~,\ \bf\pgd2.25a~}

\nl{${{\rm  u}_{\rm  f}(\cdot ,\cdot )}:{\mathbb R}\times {\cal
S}_*\DOTSB \mapstochar \rightarrow {\mathfrak
U}$}{\bf\pgr2.11~,\bf\pgd2.25a~}

\nl{${{\rm  u}_{\rm  f}(t,F):={\rm
u}_{\rm  f}(t,\nu ')}$},\ {$\tau^{{\rm f}}_t$}{\bf\pgd2.25a~}

\nl{$s^*({\cal L(H)},{\boldsymbol{{\cal D}}}({\mathbb F})),\ \mfk
h_{\cdot}(F)$}{\bf\pgd2.25b~}

\nl{$p_\nu , p^*_\nu \ (\nu \in
{\boldsymbol{{\cal D}}}({\mathbb F}))$}{ \bf\pgd2.25b~}

\nl{${\cal
C}_{bs}$},\ {${\cal C}^G$},\ {${\cal C}^G_U$},\ {${\cal
C}^G_q$}{\bf\pgd2.25b~}

\nl{${\cal C}^G_{cl}:= {\mathbb I}\tmspace
-\thinmuskip {.1667em}\cdot \tmspace  -\thinmuskip
{.1667em}C({\cal E}_{{\mathbb F}},{\mathbb C})\subset {\cal
C}^G_U$}{\bf\pgd2.25b~}

\nl{$Y(F)$\ },\ {$E_{Y(F)}$},\ {$\mathaccent
"705E h_{{\mathfrak f}}(\varrho ,\nu )$},\ {$\mathaccent "705E
{\cal C}^G$}{\bf\pgd2.25c~}

\nl{${\cal M}^G$},\ {${\cal S}_G$},\
{${\cal S}^{cl}_G$\ },\ {$\omega_{\mu ,\mathaccent "705E \varrho
}\in {\cal S}({\cal C}^G)$}{ \bf\pgr2.42d~}

\nl{$g_Q(t,F)$},\ $R_g$,\
{$\tilde\varphi_t^{\rQ}$}{\bf\pgr2.39a~}

\nl{${\mathbb F}_\xi (\nu
):={\mathbb F}(\nu )(\xi )$},\ {$\mu_\xi (B)$}{\bf\pgr2.42a~}

\nl{$\tau ^{\rm  Q}_t$},\ {$\sigma (G)$},\ {$\mathaccent "705E
\tau_t^{\rm  Q}$}{\bf\pgr2.42a~}

\nl{$\msg(g),\
\hat{\rf}_t(\mrh,\nu)$}{\bf\pgr2.43~}

\nl{${\rm  f}^\psi_\xi,\
\Phi^\psi_\xi,\ \tau_t^{\xi,\psi} $}{~\bf\pageref{prop;2.31}}

\nl{$\mome_{\mfk A},\ F_\mome$}{{ } \bf\pageref{lem;AxC}}

\nl{$\gamma,\ \hat\gamma_F,\ \mphi_\gamma$}{{ }
\bf\pageref{prop;AxC-symm}}

\nl{$\mcl V_\nu$}{{ } \bf\pageref{cl
V}},{$\quad{\cal V}_{\bf  y}$}{{ } \bf\pgr3.1~}

\nl{$G_{WH}$\ }{{ }
\bf\pgr GWH-m~,\ \bf\pgr1GWH~}

\nl{$x\tmspace  -\thinmuskip
{.1667em}\cdot \tmspace -\thinmuskip {.1667em}\nu,\ W(x)\nu
W(x)^*,\ Ad^*(W(x))\nu $}{{ } \bf\pageref{notat;W(x)}}

\nl{$\Omega
^{{\cal V}}_\varrho $},\ {$\varrho_0$},\ {$h_X^\varrho $}{{ }
 \bf\pageref{rem;h-bar}}

\nl{$V_\varrho (q)$},\ {$\mathaccent "707E z(q):=z(-q)$}{{ }
\bf\pageref{eq;hHcl}}

\nl{$p_{\leftrightarrow }$}{{ } \bf\pgr H-F-hamilt~}

\nl{${\cal P(X)}$}{{ } \bf\pageref{Adf;top}}

\nl{$\sigma $(${\cal L}$,${\cal M}$)}{{ }  \bf\pageref{sLM}}

\nl{\D f,{}~},\ {\D f,{\nu}~}{{ }  \bf\pageref{def;frechet}}

\nl{$c:=(U;{\varphi };{\cal L})$}{{ } \bf\pageref{df;1manif}}

\nl{$n=\dim(M)$}{{ } \bf\pageref{dimM}}

\nl{${\cal F}(M)$}{{ }  \bf\pageref{df;1difmap}}

\nl{$T_xM\equiv T_x(M),\ [c]_x$}{{ } \bf\pgd1tangent~}

\nl{$T_xf,\ TM$}{{ }  \bf\pgd1tangent~}

\nl{$T^p_qM\equiv T^p_q(M)$}{{ } \bf\pgr2TrM~}

\nl{$f_*=Tf,\
\mL{\bv}~f(x)=\bv_x(f)$}{{ } \bf\pgr1lie~}

\nl{$\mL\bv~\bw=[\mL\bv~,\mL\bw~]$}{{ } \bf\pgr2lie~}

\nl{$ d_xf$}{{ }  \bf\pgr1df~}

\nl{$\mome_1\wedge\mome_2,\ \mome\wedge df$}{{ } \bf\pg1wedg~}

\nl{$\beta^*\mome,\ \beta^*f$}{{ } \bf\pgd1pullb~}

\nl{$Lie(G)\equiv {\mathfrak g}$}{{ }  \bf\pgd2Lie-alg~}

\nl{\rm  ad},\ $Ad(g)$,\ {$Ad^*(g)$}{{ } \bf\pgd2Ad~}

\nl{$F\DOTSB \mapstochar \rightarrow \Omega_F$}{{ } \bf\pg2Sympl~}

\nl{${D(A)}$},\ $\cal L(H)$,\
{$GL({\cal H})$}{{ } \bf\pg B;b-oper~}

\nl{$\sigma (A)=sp(A),\ \sigma_{pp}(A)$}{{ }
\bf\pg B;b-oper~}

\nl{$E(\Lambda)$,\ $F(\Lambda)$}{{ }~\bf\pageref{df;1p-measure}}

\nl{$\mfk A,\ G(\mfk A),\ {\mathfrak A}^{-1}$}{{ } \bf\pgd symA~}

\nl{$\pi(\mfk A),\ \mH_\pi,\ \psi_\pi$}{{ } \bf\pgd repres~}

\nl{$\mfk A'',\ \pi'',\ \pi_1\sim \pi_2,\ z(\pi),\ \pi_1\mho\pi_2$}{{
} \bf\pg prop;pi-ext~}

\nl{${\cal ES}({\mathfrak A})$}{{ } \bf\pgd states~}

\nl{$(\pi_\mome;\mH_\mome;\psi_\mome)$,\ GNS,\ $\mome_1\mho\mome_2$}{{ }
 \bf\pageref{GNS}}

\nl{$\mome_\beta,\ \beta$--KMS}{{ } \bf\pageref{KMS}}

\nl{$\mfk{P(M)}$,\ \fk P(\LH)}{{ } \bf\pageref{note;types}}

\nl{$\Gamma(A),\ D(A),\ D(A^*),\ A^*$}{{ }\bf\pageref{eq;1dom*}}

\nl{$n_{\pm}(A):=\dim\mcl K_{\pm},\ (x,y)_A$}{{ }\bf\pageref{eq;1GrA}}
\label{end}
\onecolumn
\vspace*{8cm}
\nopagebreak
\centerline{\bf\Large $\star$ The End of this Book $\star$}


\begin{thebibliography}{999}
\addcontentsline{toc}{chapter}{\protect\numberline{}{\noidt\bf Bibliography}}

\bibitem{abr&mars} R. Abraham, J. E. Marsden: {\sl Foundations of Mechanics},
(second edition), Benjamin/Cummings, Reading, Mass., 1978.

\bibitem{ali}\refer{S. T. Ali, E. Prugove\v cki}{Physica}{89A}{1977}{501-521}
 \refer{S.T. Ali,  G.G. Emch}{J. Math. Phys.}{15}{1974}{176}  \refer{S.T.
Ali}{J. Math. Phys.}{21}{1980}{818} \refer{F.E.
Schroeck}{Found. Phys.}{12}{1982}{825}
 \refer{S.T. Ali,
E. Prugove\v cki}{Acta Applic. Math.}{6}{1986}{
1, 19, 47}

\bibitem{ali&doeb}\refer{S. T. Ali, H.-D. Doebner}{Phys. Rev.
A}{41}{1990}{1199-1210}

\bibitem{ali2}\refer{S. T. Ali, J.-P. Antoine, J.-P. Gazeau, U. A.
Mueller}{Rev. Math. Phys.}{7}{1995}{1013-1104}

\bibitem{alonzo}\refer{L. M. Alonzo}{J. Math. Phys.}{18}{1977}{1577-1581}
\refer{L. M. Alonzo}{J. Math. Phys.}{20}{1979}{219-230}


\bibitem{araki}\refer{H. Araki}{J. Math. Phys.}{4}{1963}{1343}
\refer{H. Araki}{J. Math. Phys.}{5}{1964}{1}



\bibitem{arn1} V. I. Arnol\acc d: {\sl Matemati\v ceskije osnovy klassi\v
 ceskoj mechaniki},\ (third edition), Nauka, Moscow, 1989.
\bibitem{arn2} V. I. Arnol\acc d, V. V. Kozlov, A. I. Neustadt: {\sl Matemati\v
ceskije aspekty klassi\v ceskoj i nebesnoj mechaniki}, in Sovremennyje
Problemy Matematiki - Fundamental\acc nyje napravlenija, Vol. 3, VINITI, Moscow
1985.

\bibitem{arn3} V. I. Arnol\acc d: {\sl Dopol\acc nitel\acc nyje glavy teorii
obyknovennych differencial\acc nych uravnenij}; Nauka, Moscow, 1978.

\bibitem{arn&avez} V. I. Arnol\acc d, A. Avez: {\sl Ergodic Problems of
Classical Mechanics}, W. A. Benjamin, Inc,\., New York - Amsterdam, 1968.

\bibitem{ashtekar} A. Ashtekar, T. A. Schilling: {\sl Geometrical Formulation
of Quantum Mechanics}, 1997,\ \url{http://arxiv.org/abs/gr-qc/9706069}.

\bibitem{lieb2}\refer{V. Bach, E. H. Lieb, M. Loss, J. P. Solovej}{Phys.
Rev. Lett.}{72}{1994}{2981-2983}

\bibitem{bar&racz} A. O. Barut, R. Raczka: {\sl Theory of Group
Representations and Applications}, PWN, Warsaw, 1977; Russ. edition:
Mir,  Moscow, 1980.
\bibitem{bell}J. S. Bell: {\sl Speakable and unspeakable in quantum
mechanics}, Cambridge University Press, Cambridge, 1988.
\bibitem{beltram}\refer{E. G. Beltrametti, S. Bugajski}{J. Phys. A: Math.
Gen.}{28}{1995}{3329-3343}
\bibitem{berez1}\refer{F. A. Berezin}{Commun. Math. Phys.}{63}{1978}{131}
\bibitem{berez2}\refer{F. A. Berezin}{Izvestija Akad.Nauk SSSR, Ser.Matem.}{36
}{1972}{1134}

\bibitem{bial&myc}\refer{J. Bialynicki-Birula, J. Mycielski}{Ann.
Phys.}{100}{1976}{62}.
\bibitem{birk&lane} G. Birkhoff, S. Mac Lane: {\sl Algebra}, Macmillan Co.,
New York, 1968.

\bibitem{bir&sol} M. Sh. Birman, M. Z. Solomyak: {\sl Spektral\acc naya
Teoriya Samosopryazhonnykh Operatorov v Gil\acc bertovom Prostranstve}, in
Russ., {\sl (Spectral Theory of Selfadjoint Operators in Hilbert
Space)}, Leningrad University Press, Leningrad (=Sankt Petersburg), 1980.

\bibitem{bohr1} N. Bohr: {\sl Discussion with Einstein on epistemological
problems in atomic physics}, pp. 200-241 in P. A. Schlipp (Ed.): {\sl
Albert Einstein: Philosopher - Scientist}, The Library of Living
Philosophers, Evanston, 1949.
\bibitem{bohr2}\refer{N. Bohr}{Phys. Rev.}{48}{1935}{696-702}

\bibitem{b-yc}\refer{Bon-Yao Chu}{Trans. Am. Math. Soc.}{197}{1974}{145}

\bibitem{bon10} P. B\'ona: {\sl Quantum Mechanics with Mean - Field
Back\-grounds}, Preprint No. {\bf Ph10-91}, Comenius University,
Faculty of Mathematics and Physics, Bratislava, October 1991;\url{http://sophia.dtp.fmph.uniba.sk/~bona/preprint%20Ph10-91.html}.
\bibitem{bon11} P. B\'ona: {\sl On Nonlinear Quantum Mechanics}, pp.
185-192 in {\bf
Differential Geometry and Its Applications}, Proc. Conf. Opava
(Czechoslovakia), August 24-28, 1992, Silesian University, Opava, 1993;\url{http://sophia.dtp.fmph.uniba.sk/~bona/NLQM-Opava1992.pdf}.
\bibitem{bon4}\refer{P. B\'ona}{Czech. J. Phys.}{B 33}{1983}{837}\url{http://sophia.dtp.fmph.uniba.sk/~bona/QC.html}.\footnote{This
 paper essentially coincides with the talk given by the author
at VIIth International Congress on Mathematical Physics, Boulder, Colorado,
USA, August 1-10, 1983.}
\bibitem{bon8} P. B\'ona: {\sl Classical Projections and Macroscopic Limits of
Quantum Mechanical Systems}, unpublished monograph, Bratislava, 1984,
revised version 1986, \url{http://sophia.dtp.fmph.uniba.sk/~bona/monograph.html}.
\bibitem{bon-m}\refer{P. B\'ona}{acta phys. slov.}{23}{1973}{149}\url{http://www.physics.sk/aps/pubs/1973/aps_1973_23_3_149.pdf}.
P. B\'ona: {\sl Interakcia makrosyst\'emu s mikroobjektom v kvantovej te\'orii.}
($\equiv$ {\sl ``Interaction of Macrosystem with
Microobject in Quantum Theory'', in Slovak}), dissertation,
Comenius University, Bratislava, 1974;\url{http://sophia.dtp.fmph.uniba.sk/~bona/dissertation.html}.
\refer{P. B\'ona}{acta phys.slov.}{27}{1977}{101}\url{http://www.physics.sk/aps/pubs/1977/aps_1977_27_2_101.pdf};
\refer{P. B\'ona}{ACTA  F.R.N.Univ.Comen.-PHYSICA}{XX}{1980}{65}
P. B\'ona: Selfconsistency and objectification, in {\sl Quantum
Measurement, Irreversibility and the Physics of Information}, Proceedings
of the Symposium on the Foundations of Modern Physics, Eds: P.
Busch, P. Lahti, P. Mittelstaedt, World Scientific 1993, \url{http://sophia.dtp.fmph.uniba.sk/~bona/selfconsist+object.pdf}.

\bibitem{bon6}\refer{P. B\'ona}{acta phys.slov.}{27}{1977}{101}\url{http://sophia.dtp.fmph.uniba.sk/~bona/aps/Quantum%20Domino%20(APS).pdf};

\bibitem{bon7}\refer{P. B\'ona}{ACTA  F. R. N. Univ. Comen. -
PHYSICA}{XX}{1980}{65}\url{http://sophia.dtp.fmph.uniba.sk/~bona/Model-Q-Detect/AFRN-1980-QD.pdf}

\bibitem{bon-sir} P. B\'ona, M. \v Sira\v n: {\sl A Radiating Spin Chain as a Model of Irreversible Dynamics}, Bratislava 2012,
\url{http://arxiv.org/abs/1211.6783}.

\bibitem{bon-d} P. B\'ona: {\sl Unpublished proof of the
formula~\eqref{eq;3.9}}, Dubna, 1987.

\bibitem{bon1}\refer{P. B\'ona}{J. Math. Phys.}{29}{1988}{2223}\url{http://sophia.dtp.fmph.uniba.sk/~bona/JMP-1-1988.pdf}
\bibitem{bon2}\refer{P. B\'ona}{J. Math. Phys.}{30}{1989}{2994}\url{http://sophia.dtp.fmph.uniba.sk/~bona/JMP-2-1989.pdf}
\bibitem{bon3}\refer{P. B\'ona}{Czech. J. Phys.}
{B 37}{1987}{482}\url{http://sophia.dtp.fmph.uniba.sk/~bona/CJP87/CJP-1987.pdf}

\bibitem{bon-dens} P. B\'ona: {\sl Geometric formulation of nonlinear
quantum mechanics for density matrices}, in {\bf Trends in Quantum
Mechanics}, Proceedings of International Symposium, Goslar,
Germany, 31 Aug.-- 3 September 1998, (Eds. H.-D. Doebner, S. T.
Ali, M. Keyl and R. F. Werner) World Scientific, Singapore-New
Jersey-London-Hong Kong, 2000; \url{arxiv:quant-ph/9910011}.

\bibitem{bon-tr} P. B\'ona: {\sl On Symmetries in Nonlinear Quantum
Mechanics},\url{arxiv:quant-ph/9910012}.

\bibitem{bon-sym} P. B\'ona: {\sl On Symmetries in Mean-Field Theories}, in
{\bf Selected Topics in Quantum Field Theory and Mathematical Physics}\
(Eds. J. Niederle, J. Fischer), World Scientific, Singapore, 1990;\url{http://sophia.dtp.fmph.uniba.sk/~bona/Sym-in-MFT-1989.pdf}.

\bibitem{bon-orbit}\refer{P. B\'ona}{Journal of Geometry and Physics}{51}{2004}{256-268}\url{http://sophia.dtp.fmph.uniba.sk/~bona/JGP-2004.pdf}.

\bibitem{borchers1}\refer{H. J. Borchers, R. Haag, B. Schroer}{Nuovo
Cimento}{24}{1963}{214} \refer{H. J. Borchers}{Commun. Math.
Phys.}{1}{1965}{281}

\bibitem{borchers} H.-J. Borchers: {\sl Translation Group and Particle
Representations in Quantum Field Theory}, Springer, Berlin, 1996.

\bibitem{bourb;Lie} N. Bourbaki: {\sl Groupes  et  Alg\`ebres  de  Lie},
Hermann,  Paris, 1972.
\bibitem{bourb;manif} N. Bourbaki: {\sl Varietes differentielles et
analytiques. Fascicu\-le de resultats}, Hermann, Paris, 1967 and 1971;
Russ.  ed.  Mir, Moscow, 1975.
\bibitem{bourb;vect} N. Bourbaki: {\sl Espaces vectoriels topologiques},
Hermann,  Paris; Russ. translation: Izdatelstvo  Inostrannoj  Literatury,
 Moscow, 1959;


\bibitem{bra&rob} O. Bratteli, D. W. Robinson: {\sl Operator Algebras  and
Quantum Statistical Mechanics}, Vol.I, Springer, New York -  Heidelberg  -
Berlin, 1979.
\bibitem{bra&rob2} O. Bratteli, D.W. Robinson: {\sl Operator Algebras  and
Quantum Statistical Mechanics}, Vol.II, Springer, New York -  Heidelberg  -
Berlin, 1980.

\bibitem{breuer}\refer{T. Breuer}{Phil. Sci.}{62}{1995}{197-214}

\bibitem{deBrog1}\refer{L. de Broglie}{Phil. Mag. and Journ. of
Science}{XLVII/No. CCLXXVIII}{1924}{446} (Cited according to~\cite{polak}.)
\bibitem{deBrog2}\refer{L. de Broglie}{Ann. d. Phys.}{III, dixi\`eme
s\'erie}{1925}{22-128}  (Cited according to~\cite{polak}.)
\bibitem{deBrog} L. de Broglie: {\sl Non-Linear Wave Mechanics - A Causal
Interpretation}, Elsevier, Amsterdam,  1950; \refer{D. Bohm, J. Bub}
{Rev. Mod. Phys.}{38}{1966}{453} \refer{P. Pearle}{Phys. Rev.}{D
13}{1969}{857}

\bibitem{buch&yng}\refer{D. Buchholz, J. Yngvason}{Phys. Rev.
Lett.}{73}{1994}{613-616}

\bibitem{bug}\refer{S. Bugajski}{Physics Letters A}{190}{1994}{5-8}
\bibitem{QTM} P. Busch, P. J. Lahti, P. Mittelstaedt: {\sl The quantum theory
  of measurement}, Springer Verlag, Berlin, 1992.

\bibitem{busch}\refer{P. Busch}{Int. J. Theor. Phys.}{24}{1985}{63-92}

\bibitem{QTM1} P. Busch, P. J. Lahti, P. Mittelstaedt (Eds.):
{\sl Quantum
Measurement, Irreversibility and the Physics of Information}, Proceedings
of the Symposium on the Foundations of Modern Physics, World Scientific 1993.

\bibitem{buz1}\refer{V. Bu\v zek, R. Derka, G. Adam, P. L. Knight}{Ann.
Phys.}{266}{1998}{454-496}
\bibitem{buz3}\refer{V. Bu\v zek}{Phys. Rev. A}{58}{1998}{1723-1727}
\bibitem{buz4} V. Bu\v zek, M. Hillery, R. F. Werner: quant-ph/99 01 053,
to appear in {\sl Phys. Rev. Lett.}


\bibitem{cant}\refer{V. Cantoni}{Acad. Naz. dei Lincei}{Ser.VIII, LXII/5}
{1977}{628}\ \refer{V. Cantoni}{Commun. Math. Phys.}{44}{1975}{125}
\bibitem{cant2}\refer{V. Cantoni}{Helvetica Phys. Acta}{58}{1985}{956-968}

\bibitem{h-cartan} H. Cartan: {\sl Calcul diff\`erentiel. Formes
diff\`erentielles}, Hermann, Paris, 1967.

\bibitem{chern&mars} P. R. Chernoff, J. E. Marsden: {\sl Properties of
Infinite Dimensional Hamiltonian Systems}, LNM 425, Springer,
Berlin-Heidelberg-New York, 1974.

\bibitem{choquet} G. Choquet: {\sl Lectures on Analysis}, Benjamin, Reading
- Mass., 1969.

\bibitem{3baby} Y. Choquet-Bruhat,  C. DeWitt-Morette, M. Dillard-Bleick:
{\sl Analysis, Manifolds, and Physics}, Revised edition, North-Holland,
Amsterdam - New York - Oxford, 1982.

\bibitem{PH}\refer{R. Cirelli, P. Lanzavecchia, A. Mani\'a}{J. Phys.
A}{16}{1983}{3829}
\refer{R. Cirelli, P. Lanzavecchia}{Nuovo Cimento}{B 79}{1984}{271-283}
 \refer{M. C. Abbati, R. Cirelli, P. Lanzavecchia, A. Mani\'a}{Nuovo
Cimento}{B 83}{1984}{43}  \refer{A. Heslot}{Phys. Rev. D}{31}{1985}{1341}

\bibitem{cir}\refer{R. Cirelli, A. Mani\'a, L.
Pizzocchero}{J. Math. Phys.}{31}{1990}{2891-2897, 2898-2903}

\bibitem{cir5}\refer{R. Cirelli, A. Mani\'a, L. Pizzocchero}{Int. J. Mod.
Phys. A}{6}{1991}{2133-2146}

\bibitem{cir1}\refer{R. Cirelli}{J. Math. Phys.}{32}{1991}{1235-1281}

\bibitem{cir3}\refer{R. Cirelli, A. Mani\'a, L.
Pizzocchero}{Rev. Math. Phys.}{6}{1994}{675-697}
\bibitem{cir4}\refer{R. Cirelli, M. Gatti, A. Mani\'a}{J. Geom.
Phys.}{29}{1999}{64-86}


\bibitem{connes} A. Connes: {\sl Noncommutative Geometry}, Academic Press,
San Diego, 1994.

\bibitem{curie} P. Curie: {\sl Oeuvres}, Gauthier-Villars, Paris, 1908;
esp. J.de phys.(III){\bf 3}(1894)393, according to Ch. IX-X of russ.
ed. Nauka, Moscow 1966;

\bibitem{czachor} M. Czachor: {\sl ``Entropic'' framework for nonlinear
quantum mechanics}, preprint, Warsaw, 1993;
\refer{M. Czachor}{acta phys. slov.}{48}{1998}{1-6}
\refer{M. Czachor, M. Marciniak}{Physics Letters}{A 239}{1998}{353-358}
\refer{M. Czachor}{Phys. Rev. A}{57}{1998}{4122-4129}
\refer{M. Czachor, M. Kuna}{Phys. Rev. A}{58}{1998}{128-134}
\refer{M. Czachor}{Int. J. Theor. Phys.}{}{1999}{}
\refer{M. Czachor}{Phys. Rev. A}{57}{1998}{R2263-R2266}

\bibitem{davies} E. B. Davies: {\sl Quantum Theory of  Open  Systems},
Academic,  New York, 1976.
\bibitem{davies1}\refer{E. B. Davies}{Commun. Math. Phys.}{55}{1977}{231}
\bibitem{davies2} E. B. Davies: {\sl Non-Linear Functionals in Quantum Mechanics},
Troisieme Cycle de la Physique, En Suisse Romande (Lausanne)
Semestre d'\'et\'e 1980.

\bibitem{buz2}\refer{R. Derka, V. Bu\v zek, A. K. Ekert}{Phys. Rev.
Lett.}{80}{1998}{1571-1575}

\bibitem{dirac} P. A. M. Dirac: {\sl The principles of Quantum mechanics}
(4-th ed.), Claredon, Oxford, 1958.

\bibitem{divak}\refer{P. P. Divakaran}{Phys. Rev. Lett.}{79}{1997}{2159}

\bibitem{dix1} J. Dixmier: {\sl Les alg\`ebres d'op\'erateurs  dans  l'espace
Hilbertien}, Second edition - revised and  completed, Gauthier-Villars,
Paris, 1969.
\bibitem{dix2} J. Dixmier: {\sl Les $C^*$-alg\`ebres et leurs
repr\'esentations}, Gauthier-Villars, Paris, 1969; Russ. transl. Nauka,
Moscow, 1974.

\bibitem{doebner}\refer{H.-D. Doebner, G. A. Goldin}{Phys. Rev.
A}{54}{1996}{3764-3771}
\bibitem{doeb&tol}\refer{H. D. Doebner, J. Tolar}{J. Math.
Phys.}{16}{1975}{975-984} \refer{J. Niederle, J. Tolar}{Czech. J. Phys.}{B
29}{1979}{1358-1368}
\refer{P. \v St\acc ov\'\i\v cek, J. Tolar}{Acta
Polytechnica - Pr\'ace \v CVUT, Prague}{No. 6}{1984}{37-75}

\bibitem{dold} A. Dold: {\sl Lectures on Algebraic Topology}, Springer,
Berlin - Heidelberg - new York, 1980.

\bibitem{dopl1}\refer{S. Doplicher, R. Haag, J. E. Roberts}{Commun.
Math. Phys.}{13}{1969}{1-23} \refer{S. Doplicher, R. Haag, J. E. Roberts}{
Commun. Math. Phys.}{15}{1969}{173-200}
\bibitem{dopl2}\refer{S. Doplicher, R. Haag, J. E. Roberts}{Commun.
Math. Phys. }{23}{1971}{199-230} \refer{S. Doplicher, J. E. Roberts}{Commun.
Math. Phys.}{28}{1972}{331-348}
\refer{S. Doplicher, R. Haag, J. E. Roberts}{Commun. Math.
Phys.}{35}{1974}{49-85}
\bibitem{dopl3}\refer{S. Doplicher}{Commun. Math. Phys.}{85}{1982}{73-86}
\refer{S. Doplicher, R. Longo}{Commun. Math. Phys.}{88}{1983}{399-409}

\bibitem{dowk&kent}\refer{F. Dowker, A. Kent}{Phys. Rev. Lett.}{75}{1995}{3038}


\bibitem{DNF} B. A. Dubrovin, S. P. Novikov, A. T. Fomenko: {\sl
Sovremennaja Geometrija - Metody i prilo\v zenia (Modern Geometry - Methods
and applications)}, Nauka, Moscow, 1979.

\bibitem{d+wer1}\refer{N. G. Duffield, R. F. Werner}{Helv. Phys.
Acta}{65}{1992}{1016 - 1054}\url{http://sophia.dtp.fmph.uniba.sk/~bona/Duffield-Werner.pdf};
\refer{N. G. Duffield, R. F. Werner}{Rev. Math. Phys.}{4}{1992}{383-424}

\bibitem{duf&rieck}\refer{E. Duffner, A. Rieckers}{Z.
Naturforsch.}{43a}{1988}{321}

\bibitem{einst} A. Einstein: {\sl The Meaning of Relativity}, Princeton UP,
Princeton, NJ, 1955.
\bibitem{einst1}\refer{A. Einstein}{Ann. d. Phys.}{17}{1905}{132}
\refer{A. Einstein}{Ann. d. Phys.}{20}{1906}{199}

\bibitem{EPR}\refer{A. Einstein,  B. Podolsky, N. Rosen}{Phys. Rev. D}{47}{
1935}{777}
\bibitem{emch1} G. G. Emch: {\sl Algebraic  Methods  in  Statistical
Mechanics  and Quantum Field Theory}, Wiley, New York, 1972.

\bibitem{eva&lewi} D. E. Evans and J. T. Lewis: {\sl Dilations of Irreversible
Evolutions in  Algebraic Quantum  Theory},
Commun. of DIAS, Ser.A (Theoretical Physics), No.24, IAS-Dublin, 1977.

\bibitem{ever&whee}\refer{H. Everett, III}{Rev. Mod.
Phys.}{29}{1957}{454-462}\refer{J. A. Wheeler}{\sl ibid.}{}{1957}{463-465}

\bibitem{fecko}\refer{M. Fecko}{acta phys. slov.}{44}{1994}{445}
\refer{M. Fecko}{J. Math. Phys.}{36}{1995}{6709}\ \refer{{\sl
ibid}}{}{36}{1995}{1198-1207}\
\refer{{\sl ibid}}{}{38}{1997}{4542-4560}

\bibitem{feller} W. Feller: {\sl An Introduction to Probability Theory and
its  Applications}, Vol.I, and Vol.II, Russ. edition Mir, Moscow, 1984.

\bibitem{fermi}\refer{E. Fermi}{Rev. Mod. Phys.}{4}{1932}{87}

\bibitem{flato} M. Flato, D. Sternheimer: {\sl Star Products,
Quantum Groups, Cyclic Cohomology and Pseudodifferential calculus}, lectures
at the 10-th annual Joint Summer Research Conference (AMS -- IMS -- SIAM) on
Conformal Field Theory, Topological Field Theory and Quantum Groups (Mount
Holyoke College, South Hadley, MA; June 13-19, 1992), in {\bf Contemporary
Mathematics}, AMS-series (Eds. M. Flato, J. Lepowsky, N. Reshetikin, P.
Sally); D. Arnal, J. C. Cortet, M. Flato, D. Sternheimer: {\sl Star
Products and Representations Without Operators}, in {\bf Field Theory,
Quantization and Statistical Physics} (Ed. E. Tirapegui), D. Reidel Pub.
Co., Dordrecht-Holland, 1981.

\bibitem{f&lewis&c}\refer{G. W. Ford, J. T. Lewis, R. F. O'Connell}{Phys.
Rev. Lett.}{55}{1985}{2273}\refer{G. W. Ford, J. T. Lewis, R. F.
O'Connell}{Ann. Phys. (NY)}{185}{1988}{270-283}\refer{G. W. Ford, J. T. Lewis, R. F.
O'Connell}{Phys. Rev. A}{37}{1988}{4419}\refer{G. W. Ford, J. T. Lewis, R. F.
O'Connell}{J. Stat. Phys}{53}{1988}{439-455}

\bibitem{fredenh} K. Fredenhagen: {\sl Global Observables in Local Quantum
Physics}, Preprint DESY 93-009, Hamburg 1993.

\bibitem{fronsdal}\refer{C. Fronsdal}{Rep. Math. Phys.}{15}{1978}{111-145}

\bibitem{gamelin} T. W. Gamelin: {\sl Uniform Algebras}, Prentice-Hall,
Englewood Cliffs, N. J., 1969.

\bibitem{gatti} Mauro Gatti: {\sl Private communications}, June 1992, and
June 1993; M. Gatti: {\sl From Quantum Phase Space to Classical Phase Space,
a draft of PhD Dissertation -- a concise summary of the first
6 chapters}, Universit\`a degli Studi di Milano, June 1993.

\bibitem{GRS} I. M. Gel\acc fand, D. A. Rajkov, G. E. \v Silov: {\sl
Kommutativnyje normirovannyje kol\acc ca (Commutative Normed Rings)},
GIFML, Moscow, 1960.


\bibitem{gilmore} R. Gilmore: in {\sl Symmetries in Science}, Eds. B. Gruber,
 and R. S. Millman, Plenum, New York, 1980.

\bibitem{gisin}\refer{N. Gisin}{Helv. Phys. Acta}{62}{1989}{363}

\bibitem{noncausal}\refer{N. Gisin}{Phys. Lett. A}{143}{1990}{1}
  \refer{ J. Polchinski}{Phys. Rev. Lett.}{66}{1991}{397}
\refer{M. Czachor}{Found. Phys. Lett.}{4}{1991}{351}
\refer{H. Scherer, P. Busch}{Phys. Rev. A}{47}{1993}{1647}

\bibitem{glaub} R. Glauber: {\sl Optical coherence and statistics of photons},
 in Quantum Optics and Elec\-tro\-nics, Gordon and Breach, New York, 1965;
\bibitem{gleason1}\refer{A. M. Gleason}{Ann. Math.}{56}{1952}{193-212}


\bibitem{goldin}\refer{G. A. Goldin}{Nonlin. Math. Phys.}{4}{1997}{6-11}
\bibitem{goldin1}\refer{G. A. Goldin, R. Menikoff, D. H. Sharp}{Phys. Rev.
Lett.}{51}{1983}{2246}

\bibitem{zeil1}\refer{D. M. Greenberger, M. A. Horne, A. Zeilinger}{Physics
Today}{August}{1993}{22}
\refer{D. Bouwmeester, J.-W. Pan, M. Daniell, H. Weinfurter,
A. Zeilinger}{Phys. Rev. Lett.}{82}{1999}{1345}
\refer{A. Zeilinger, M. A. Horne, H. Weinfurter, M. \.Zukowski}{Phys.Rev.
Lett.}{78}{1997}{3031}

\bibitem{gresak} Eduard Gre\v s\'ak: {\sl Communications in the framework of
PhD--seminar on mathematical physics}, Comenius
University, Bratislava 1992.

\bibitem{gr&pr}\refer{H. Grosse, P. Pre\v snajder}{Lett. Math.
Phys.}{28}{1993}{239}

\bibitem{gudder}\refer{S. Gudder}{Commun. Math. Phys.}{29}{1973}{249-264}

\bibitem{gugg} E. A. Guggenheim: {\sl Thermodynamics, Classical and
Statistical}, in Handbuch der Physik, Band III/2, Springer, Berlin -
G\"ottingen - Heidelberg, 1959.

\bibitem{gutzw} M. C. Gutzwiller: {\sl Chaos in Classical and Quantum
Mechanics}, Springer, New York, 1991;

\bibitem{HHW}\refer{R. Haag, N. M. Hugenholtz, M. Winnink}{Commun. Math.
Phys.}{5}{1967}{215}
\bibitem{haag&kast}\refer{R. Haag, D. Kastler}{J. Math. Phys.}{5}{1964}{848-861}
\bibitem{haag1}\refer{R. Haag}{Nuovo Cimento}{25}{1962}{287-299}
\bibitem{haag2} R. Haag: {\sl Local Quantum Physics}, Springer, New York,
1992;
\bibitem{haag3}\refer{R. Haag}{Dan. Mat. Fys. Med.}{29}{1955}{No.12}
\bibitem{haag4}\refer{R. Haag, B. Schroer}{J. Math. Phys.}{3}{1962}{248}

\bibitem{hagedorn}\refer{G. A. Hagedorn}{Commun. Math.
Phys.}{71}{1980}{77-93}

\bibitem{halmos} P. R. Halmos: {\sl Introduction to the Theory of Hilbert
space and Spectral Multiplicity}, Chelsea Pub. Co., New York, 1957;

\bibitem{hamel}\refer{G. Hamel}{Zs. Math. Phys.}{50}{1904}{1}

\bibitem{hartree} D. R. Hartree: {\sl The Calculation of Atomic
Srtuctures}, John Wiley \& Sons, Inc., New York, 1957;

\bibitem{heisenberg} W. Heisenberg: {\sl Physik und Philosophie}, Ullstein
B\"ucher, West-Berlin, 1961; {\sl Physics and beyond: Encounters and
Conversations}, Harper\& Row Pub., 1972, New York.

\bibitem{hegerfeldt}\refer{G. C. Hegerfeldt}{Phys. Rev.
Lett.}{72}{1994}{596-599}

\bibitem{helgas} S. Helgason: {\sl Differential Geometry and Symmetric
Spaces}, Academic, New York, 1962.

\bibitem{hp+lie1}\refer{K. Hepp, E. H. Lieb}{Helvetica Phys.Acta}{46}{1973}{573}
\bibitem{hp-meas}\refer{K. Hepp}{Helvetica Phys. Acta}{45}{1972}{237}
H.Primas: {\sl preprint ETH}, Z\"urich, 1972;

\bibitem{hp3}\refer{K. Hepp}{Commun. Math. Phys.}{35}{1974}{265-277}


\bibitem{herman}\refer{R. Hermann}{J. Math. Phys.}{6}{1965}{1768}

\bibitem{hill} T. L. Hill: {\sl Statistical Mechanics - Principles and
Selected Applications}, McGraw-Hill, New York - Toronto - London, 1956.

\bibitem{hirsch} M. W. Hirsch: {\sl Differential Topology}, Springer, New
York - Heidelberg - Berlin, 1976;

\bibitem{holevo} A. S. Holevo: {\sl Probabilistic and Statistical Aspects
of Quantum Theory}, Russ. Edition, Nauka, Moscow, 1980; \refer{A. S.
Holevo}{Rep. Math. Phys.}{22}{1985}{385-407} A. S. Holevo: {\sl Statistical
Structure of Quantum Mechanics and Hidden Variables}, in Russian, Matematika
i Kibernetika, 6/1985, Znanije, Moscow, 1985.

\bibitem{3horod}\refer{M. Horodecki, P. Horodecki, R. Horodecki}{Phys. Rev.
Lett.}{80}{1998}{5239}

\bibitem{HVW}\refer{R. M. F. Houtappel, H. Van Dam, E. P. Wigner}{Rev. Mod.
Phys.}{37}{1965}{595-632}

\bibitem{hughston}\refer{L. P. Hughston}{Proc. Roy. Soc.
Lond.}{A452}{1996}{953-979}


\bibitem{horuz} S. S. Horuzhy: {\sl Vvedenie v algebrai\v ceskuiyu kvantovuiyu
teoriyu polia (= Introduction to Algebraic Quantum Field Theory)}, Nauka,
Moscow, 1986.

\bibitem{isham} C. J. Isham: {\sl Topological and Global Aspects of Quantum
Theory}, in the proceedings of Les Houches XL (1983): {\bf Relativity,
Groups and Topology II} (Eds. B. S. DeWitt, R. Stora), pp. 1059-1290,
North-Holland, Amsterdam - New York - Tokyo, 1984.

\bibitem{itz&zub} C. Itzykson, J.-B. Zuber: {\sl Quantum Field Theory},
McGraw-Hill, New York - London - Toronto, 1980.

\bibitem{jauch} J. M. Jauch: {\sl Foundations of Quantum Mechanics},
Reading, Mass., 1968.

\bibitem{jones}\refer{K. R. W. Jones}{Annals of Physics}{233}{1994}{295-316}
\refer{K. R. W. Jones}{Phys. Rev. A}{48}{1993}{822-825}
\refer{K. R. W. Jones}{Phys. Rev. A}{50}{1994}{1062-1070}

\bibitem{jordan}\refer{T. F. Jordan}{Ann. Phys. (N. Y.)}{225}{1993}{83}

\bibitem{kay}\refer{K. G. Kay}{Phys. Rev. A}{42}{1990}{3718-3725}

\bibitem{kibble}\refer{T. W. B. Kibble}{Commun. Math. Phys.}{65}{1979}{189-201}

\bibitem{kiril} A. A. Kirillov: {\sl Elementy teorii predstavleniyi
(Elements of Representations Theory)}, Nauka,
Moscow, 1978, Second edition.

\bibitem{klaud1}\refer{J. R. Klauder}{J. Math. Phys.}{4}{1963}{1058}

\bibitem{klein}\refer{F. Klein}{Math. Ann.}{43}{1893}{63}


\bibitem{kob&nom} S. Kobayashi, K. Nomizu: {\sl Foundations of Differential
Geome\-try}, Vols. 1 and 2, Interscience, New York, 1963 and 1969;  Russ.
ed. Nauka, Moscow, 1981.
\bibitem{koopman}\refer{B. O. Koopman}{Proc. Natl. Acad.
Sci.}{17}{1931}{315-318}

\bibitem{K&S&F} I. P. Kornfel\acc d, Ja. G. Sinaj, C. V. Fomin: {\sl
Ergodi\v ceskaja Teorija (Ergodic Theory)}, Nauka, Moscow, 1980.

\bibitem{kra&sar} P. Kramer, M. Saraceno: {\sl Geometry of the Time-Dependent
Varia\-ti\-onal Principle in Quantum Mechanics}, LNP 140, Springer, New York,
1981.
\bibitem{kriegl;michor} A. Kriegl, P. W. Michor: {\sl Foundations of Global
Analysis}, manuscript of monograph, Institut f\"ur Mathematik der
Universit\"at Wien, Vienna, 1995.

\bibitem{kuhn} T. S. Kuhn: {\sl The Structure of Scientific Revolutions},
Second Edition, Chicago, 1970; Slovak translation: Pravda, Bratislava,
1982.

 \bibitem{landau1} L. D. Landau, E. M. Lif\v sic: {\sl Mechanika (Mechanics)},
 Theoretical Physics, vol. I, FM, Moscow, 1958.
\bibitem{landau5} L. D. Landau, E. M. Lif\v sic: {\sl Statisti\v ceskaya
Fizika (Statistical Physics)}, Theoretical Physics, vol. V, 2nd Edition,
Nauka, Moscow, 1964.
\bibitem{landau3} L. D. Landau, E. M. Lif\v sic: {\sl Kvantovaya Mechanika
- nerelativistskaya teoriya
(Nonrel. Quantum Mechanics)}, Theoretical Physics, vol.
III, 2nd Edition, GIFML, Moscow, 1963.
\bibitem{landau2} L. D. Landau, E. M. Lif\v sic: {\sl Teoriya Polya
(Field Theory)}, Theoretical Physics, vol.
II, 3rd Edition, GIFML, Moscow, 1960.

\bibitem{landsman} N. P. Landsman: {\sl Mathematical Topics Between
Classical and Quantum Mechanics}, Springer Monographs in Mathematics,
Springer, New York, 1998.
\bibitem{lands1}\refer{N. P. Landsman}{Rev. Math. Phys.}{2}{1990}{45}\
 \refer{{\sl ibid}}{}{9}{1997}{29-57} \refer{N. P.
Landsman}{Int. J. Mod. Phys.}{A30}{1991}{5349-5371}\ \refer{{\sl
ibid}}{}{B10}{1996}{1545-1554}\
\refer{N. P. Landsman}{J. Geom. Phys.}{15}{1995}{285-319}

\bibitem{lassner}\refer{G. Lassner}{Rep. Math. Phys.}{3}{1972}{279}
G. Lassner: {\sl Algebras of Unbounded Operators and Quantum Dynamics},
pp. 471--480 in Proceedings of the VIIth International Congress on Mathematical
Physics, Boulder, Colorado, USA, August 1--10, 1983 (Eds. W. E. Brittin, K.
E. Gustafson, W. Wyss), North--Holland, Amsterdam, 1984.

\bibitem{leifer} P. Leifer: {\sl Nonlinear modification of quantum
mechanics}, hep-th/9702160.

\bibitem{lew&samp}\refer{M. Lewenstein, A. Sampera}{Phys. Rev.
Lett.}{80}{1998}{2261}

\bibitem{lieb1}\refer{E. H. Lieb}{Commun. Math. Phys.}{31}{1973}{327}

\bibitem{ludwig}\refer{G. Ludwig}{Z. Phys.}{135}{1953}{483}
\refer{G. Ludwig}{Phys. B1}{11}{1955}{489}\refer{G. Ludwig}{Z.
Phys.}{152}{1958}{98}\refer{G. Ludwig}{Commun. Math. Phys.}{4}{1967}{331}
\refer{G. Ludwig}{Commun. Math. Phys.}{9}{1968}{}
\bibitem{ludw1}G. Ludwig: {\sl An Axiomatic Basis of Quantum Mechanics,
Vols.I, II}, Springer, New-York, 1985 \& 1987.

\bibitem{luecke} W. L\"ucke: {\sl Nonlinear Schr\"odinger Dynamics and
Nonlinear Observables}, in ``Nonlinear, Deformed, and Irreversible Quantum
Systems'', Ed. H.-D. Doebner, V.K. Dobrev, P. Nattermann, World Scientific,
Singapore, 1995; W. L\"ucke: {\sl Gisin Nonlocality of the Doebner-Goldin
2-Particle Equation}, quant-ph/9710033, 1997; W. L\"ucke, R.F. Werner: {\sl
Inconsistency of Bialynicki-Birula and Mycielski's Nonlinear Quantum
Mechanics}, a talk in the Doppler Institute, Czech Technical University,
Prague, April 1997.

\bibitem{mach} E. Mach: {\sl Die Mechanik in ihrer Entwicklung}, Leipzig,
1933.
\bibitem{mach1} E. Mach: {\sl Popul\"ar-Wissenschaftliche Vorlesungen},
Johann Ambrosius Barth, Leipzig, 1910.

\bibitem{mack1} G. W. Mackey: {\sl Mathematical Foundations of Quantum
Mechanics}, Benjamin, Reading, Mass., 1963.
\bibitem{mack2} G. W. Mackey: {\sl Induced Representations},
Benjamin, Reading, Mass., 1968.
\bibitem{mack3} G. W. Mackey: {\sl Unitary Representations in Physics,
Probability, and Number Theory}, Benjamin, Reading, Mass., 1978.

\bibitem{march} N. H. March, W. H. Young, S. Sampanthar: {\sl The
Many-Body Problem in Quantum Mechanics}, Cambridge University Press,
Cambridge, 1967;


\bibitem{markus&meyer} L. Markus, K. R. Meyer: {\sl Generic Hamiltonian
Dynamical Systems are neither Integrable nor Ergodic}, Memoirs of AMS,
{\bf Number 144}, Providence, Rhode Island, 1974;

\bibitem{marle} C. M. Marle, in {\sl Bifurcation Theory, Mechanics and
Physics}, (Eds. C. P. Bruter, A. Aragnol, A. Lichnerowicz), D. Reidel,
Dordrecht - Boston - Lancaster, 1983;

\bibitem{mars} J. E. Marsden: {\sl Geometric Methods in Mathematical
Physics}, LNM 775, Springer, New York, 1980.
\bibitem{mars&rati} J. E. Marsden, T. S. Ratiu: {\sl Introduction to Mechanics
and Symmetry}, Springer, New York, 1999.

\bibitem{mermin}\refer{N. D. Mermin}{Rev. Mod. Phys.}{65}{1993}{803-815}

\bibitem{messiah} A. Messiah: {\sl Quantum Mechanics} (Russ. transl. from
French), Nauka, Moscow, 1978.


\bibitem{meyer} P. A. Meyer: {\sl Probability and Potentials}, Blaisdell
Pub. Co., Waltham, Mass. - Toronto - London, 1966.

\bibitem{mielnik}\refer{B. Mielnik}{Commun. Math. Phys.}{37}{1974}{221-256}

\bibitem{mont&zip}\refer{D. Montgomery, L. Zippin}{Ann.
Math.}{56}{1952}{213-241}

\bibitem{morch&stroc}\refer{G. Morchio, F. Strocchi}{Commun. Math.
Phys.}{99}{1985}{153}

\bibitem{stroc}\refer{G. Morchio, F. Strocchi}{J. Math. Phys.}{28}{1987}{622}
F. Strocchi: {\sl Long-range dynamics and spontaneous symmetry breaking in
many-body systems}, preprint ISAS-Trieste, 1987.

\bibitem{najm} M. A. Najmark: {\sl Normirovannye kol\acc ca (Normed Rings)},
Nauka, Moscow, 1968;

\bibitem{naudts} J. Naudts: {\sl $C^*$--Multipliers, crossed product
algebras, and canonical commutation relations}, preprint, Dept. Natuurkunde,
 Universiteit Antverpen UIA, July 1999.

\bibitem{neum1}J. von Neumann: {\sl Mathematische Grundlagen der
Quantenmechanik}, Springer, Berlin, 1932.
\bibitem{neum2}\refer{J. von Neumann}{Compos. Math.}{6}{1938}{1-77}

\bibitem{nielsen}\refer{M. A. Nielsen}{Phys. Rev. Lett.}{79}{1997}{2915}

\bibitem{noether}\refer{E. Noether}{Nachr. K\"onigl. Ges. Wiss.
G\"ottingen,\ }{math.-phys. K1}{1918}{258}

\bibitem{qm-hist}\refer{R. Omn\`es}{Rev. Mod. Phys.}{64}{1992}{339-382}
\refer{R. B. Griffiths}{Phys. Rev. Lett.}{70}{1993}{2201} \refer{M.
Gell-Mann, J. B. Hartle}{Phys. Rev. D}{47}{1993}{3345-3382} \refer{L.
Di\'osi, N. Gisin, J. Halliwell, I. C. Percival}{Phys. Rev.
Lett.}{74}{1995}{203} \refer{C. J. Isham}{J. Math.
Phys.}{35}{1994}{2157-2185} \refer{C. J. Isham, N. Linden}{J. Math.
Phys.}{35}{1994}{5452-5476} \refer{F. Dowker, A. Kent}{Phys. Rev.
Lett.}{75}{1995}{3038}


\bibitem{pauli1} W. Pauli: {\sl General Principles of Quantum Mechanics},
Springer, New York, 1980.
\bibitem{pauli2} W. Pauli: {\sl Teorija Otnositel\acc nosti (Theory of
Relativity)} (translated from an English edition), Nauka, Moscow, 1983.

\bibitem{pedersen} G. K. Pedersen: {\sl \Ca s and their Automorphism Groups},
Academic Press, London - New York -San Francisco, 1979.

\bibitem{penrose1} R. Penrose: {\sl The Emperor's New Mind: Concerning
Computers, Minds and the Laws of Physics}, Oxford University Press,
Oxford, 1989.

\bibitem{penrose2} R. Penrose: {\sl Shadows of the Mind: An Approach to the
Missing Science of Consciousness}, Oxford University
Press, Oxford, 1994; R. Penrose: {\sl The Large, the Small, and the Human
Mind}, The
Press Syndicate of the University of Cambridge, Cambridge, 1997 (Czech
translation, Prague, 1999).

\bibitem{perel1}\refer{A. M. Perelomov}{Commun. Math. Phys.}{26}{1972}{222}
\bibitem{koh-st}
A. M. Perelomov: {\sl Generalized coherent states and their applica\-tions},
Springer, Berlin, 1986;
J. R. Klauder and B. S. Skagerstam:
{\sl Coherent States. Applications in Physics  and  Mathematical
Physics}, World  Scientific,  Singapore, 1984;
\refer{S. T. Ali, G. G. Emch}
{J. Math. Phys.}{27}{1986}{2936-2943}

\bibitem{peres} A. Peres: {\sl Quantum Theory: Concepts and Methods},
Kluver Academic Publishers, Dordrecht-Boston-London, 1994.

\bibitem{peres2}\refer{A. Peres}{Phys. Rev. Lett.}{74}{1995}{4571}

\bibitem{piron}\refer{C. Piron}{Helv. Phys. Acta}{42}{1969}{330}\
C. Piron: {\sl Foundations of Quantum
Physics}, Benjamin,  Reading,  Mass., 1976; \refer{A. Amann, U. M\"uller-Herold}
{Helv. Phys. Acta}{59}{1986}{1311} \refer{A. Amann}{Helv.
Phys. Acta}{60}{1987}{384} \refer{A. Amann}{Fortschr. Phys.}{34}{1986}{167}

\bibitem{poincare}\refer{H. Poincar\'e}{Compt. Rend.}{140}{1905}{1504}

\bibitem{polak} L. S. Polak (Ed.): {\sl Variacionnyie Principy Mechaniki},
G.I. Fiz.-Mat. Lit., Moscow, 1959.

\bibitem{polakovic}\refer{M. Polakovi\v c}{Int. J. Theor.
Phys.}{37}{1998}{2923-2934}

\bibitem{pontrjag} L. S. Pontrjagin: {\sl Nepreryvnyie Gruppy (Continuous
Groups)}, Nauka, Moscow, 1973.

\bibitem{pulmann}\refer{S. Pulmannov\'a}{Commun. Math. Phys.}{49}{1976}{47-51}

\bibitem{popper} K. R. Popper: {\sl The Logic of Scientific Discovery},
Routledge, London - New York, 1994; Czech translation: {\sl Logika
v\v edeck\'eho b\'ad\'an\'\i}, OIKOIMENH, Prague, 1997, completed
from 10th German edition: {\sl Logik der Forschung}, J. C. B. Mohr
(P. Siebeck), T\"ubingen, 1994.
\bibitem{popper1} K. R. Popper: {\sl Objective Knowledge (An Evolutionary
Approach)}, Revised edition, Oxford UP, at the Claredon Press, Oxford, 1979.

\bibitem{povzner}\refer{A. Povzner}{Trans. Am. Math. Soc.}{51}{1966}{189}\
cited from~\cite{abr&mars}.

\bibitem{pr+v}\refer{J. P. Provost, G. Vallee}{Commun. Math. Phys.
}{76}{1980}{289}

\bibitem{pru1}\refer{E. Prugove\v cki}{J. Math. Phys.}{7}{1966}{1680}
\bibitem{pru2}\refer{E. Prugove\v cki}{Canad. J. Phys.}{45}{1967}{2173}
\bibitem{pru3}\refer{E. Prugove\v cki}{Physica}{91A}{1978}{202}
\bibitem{prugovecki} E. Prugove\v cki: {\sl Stochastic  Quantum  Mechanics
and Quantum Spacetime}, Reidel, Dordrecht,
1984;\
\refer{J.A. Brooke, E. Prugove\v cki}{Nuovo Cimento}{89 A}{1985}{126}

\bibitem{zeil2}\refer{M. Reck, A. Zeilinger, H. J. Bernstein, P.
Bertani}{Phys. Rev. Lett.}{73}{1994}{58}

\bibitem{R&S} M. Reed, B. Simon: {\sl Methods of Modern Mathematical
Physics}, Vols.I and II, Academic Press, New York - London, 1972 and 1975;

\bibitem{deRham} G. de Rham: {\sl Vari\'et\'es Diff\'erentiables}, Hermann,
Paris, 1955.

\bibitem{riesz&nagy} F. Riesz, B. Sz.- Nagy: {\sl Vorlesungen \"uber
Functionalanalysis}, DVW, Berlin, 1956.

\bibitem{rowe1}\refer{D. J. Rowe, A. Ryman, G. Rosensteel}{Phys. Rev.
A}{22}{1980}{2362}
\bibitem{rowe2}\refer{D. J. Rowe}{Nucl. Phys. A}{391}{1982}{307}

\bibitem{ruelle}\refer{D. Ruelle}{Commun. Math. Phys.}{3}{1966}{133}
\bibitem{ruelle1} D. Ruelle: {\sl Statistical Mechanics - Rigorous
Results}, W. A. Benjamin, Inc., New York - Amsterdam, 1969.
\bibitem{ruelle2} D. Ruelle: {\sl Thermodynamic Formalism: The Mathematical
Structures of Classical Equilibrium Mechanics}, Addison-Wesley,
Reading-Mass. - London, 1978.

\bibitem{sachs&wu} R. K. Sachs, H. Wu: {\sl General Relativity for
Mathematicians}, Springer, New York - Heidelberg - Berlin, 1977.

\bibitem{sak1} S. Sakai: {\sl \Ca s and $W^*$-algebras}, Springer, New
York, 1971.
\bibitem{sak2} S. Sakai: {\sl Operator algebras in dynamical systems},
Cambridge University Press, Cambridge - New York - Sydney, 1991.
\bibitem{sch&busch}\refer{H. Scherer, P. Busch}{Phys. Rev.
A}{47}{1993}{1647-1651}

\bibitem{schiff} L. I. Schiff: {\sl Quantum Mechanics}, McGraw-Hill, New
York - Toronto - London, 1955.

\bibitem{schlieder}\refer{S. Schlieder}{Commun. Math.
Phys.}{13}{1969}{216-225}

\bibitem{schroed1}\refer{E. Schr\"odinger}{Ann. d. Phys., vierte
Volge}{79/No.6}{1926}{489-527} (Cited from~\cite{polak}.)

\bibitem{schulman}\refer{L. S. Schulman}{Ann. Phys.
(NY)}{212}{1991}{315-370}

\bibitem{jt-schw} J. T. Schwartz: {\sl Nonlinear Functional Analysis}, Gordon
and Breach, New York, 1969.
\bibitem{l-schw} L. Schwartz: {\sl Analyse  math\`ematique}, Hermann,
Paris,  1967.
\bibitem{schweber} S. S. Schweber: {\sl An Introduction to Relativistic
Quantum Field Theory}, Row-Peterson, New York, 1961.

\bibitem{seg}\refer{I. E. Segal}{Duke Math J.}{18}{1951}{221-265}

\bibitem{sewell} G. L. Sewell : {\sl Quantum theory of collective
phenomena}, Oxford Science Publ., Claredon Press, Oxford, 1989.
\bibitem{sewell1}\refer{G. L. Sewell}{J. Math. Phys.}{26}{1985}{2324-2334}
\bibitem{sewell2} G. L. Sewell: {\sl Macroscopic Quantum Electrodynamics of
a Plasma Model: Derivation of the Vlasov Kinetics}, Texas Math.-Phys.
Database:  Preprint No. 96-179, Queen Mary and Westfield College, London 1996.
\bibitem{sewell3} G. L. Sewell : {\sl Quantum Mechanics and Its Emergent Macrophysics}, Princeton University Press, Princeton and Oxford, 2002.

\bibitem{sikela} R. Sikela, Thesis (in Slovak), Comenius University,
Bratislava, 1991.

\bibitem{sim1}\refer{B. Simon}{Commun. Math. Phys.}{71}{1980}{247}
\bibitem{str-atr} J. G. Sinaj, L. P. \v Silnikov (Eds.): {\sl Strannye
Attraktory (Strange Attractors)}, ser. Mathematics in Science Abroad No.
22, Mir. Moscow, 1981.

\bibitem{sladecek} L. Sl\'ade\v cek, Thesis (in Slovak), Comenius University,
Bratislava, 1991.

\bibitem{sniatycki} J. \'Sniatycki: {\sl Geometric Quantization and Quantum
Mechanics}, Springer, New York, 1980.

\bibitem{spanier} E. H. Spanier: {\sl Algebraic Topology}, McGraw-Hill,
New York, 1966.
\bibitem{spohn}\refer{H. Spohn}{Rev. Mod. Phys.}{53}{1980}{569-615}

\bibitem{storm}\refer{E. St\"ormer}{J. Funct. Analysis}{3}{1969}{48}

\bibitem{streater}\refer{R. F. Streater}{Rep. Math. Phys.}{33}{1993}{203-219}

\bibitem{str&wight} R. F. Streater, A. S. Wightman: {\sl PCT, Spin, and
Statistics}, Pergamon, London, 1964.

\bibitem{strocchi}\refer{F. Strocchi}{Rev. Mod. Phys.}{38}{1966}{36}
\bibitem{strocchi1} F. Strocchi (Lecture notes by F.S.): {\sl An Introduction to the Mathematical Structure of Quantum Mechanics} -- {{\small  A Short Course for Mathematicians}}\ (2nd Edition), World Scientific, Singapore 2008.
\bibitem{gisin1}\refer{W. T. Strunz, L. Di\'osi, N. Gisin}{Phys. Rev.
Lett.}{82}{1998}{1801}

\bibitem{tomita} M. Takesaki: {\sl Tomita's Theory of Modular Hilbert
Algebras and its Applications}, {\bf LNM 128}, Springer, Berlin - Heidelberg -
New York, 1970.
\bibitem{takesI} M. Takesaki: {\sl Theory of Operator Algebras I},
 Springer, New-York, 1979.


\bibitem{thirr1} W. Thirring: {\sl Lehrbuch der Mathematischen Physik, Band
1}, Springer, Wien, 1977.
\bibitem{thirr2} W. Thirring: {\sl Lehrbuch der Mathematischen Physik, Band
2}, Springer, Wien, 1978.
\bibitem{thirr4} W. Thirring: {\sl Lehrbuch  der  Mathematischen  Physik},
Band  4, Springer, Wien, 1980.
\bibitem{thir&wehrl}\refer{W. Thirring, A. Wehrl}{Commun. Math.
Phys.}{4}{1967}{303-314}

\bibitem{tol}\refer{J. Tolar}{\v Cs. \v Cas. Fyz. A}{25}{1975}{576-588
(in Czech)}

\bibitem{tolar} J. Tolar: {\sl Obecn\' e Metody Kvantov\'an\'\i\ (General
Quantization Methods)}, DrSc Dissertation, FJFI \v CVUT, Prague, 1985.

\bibitem{tol&chadzi}\refer{J. Tolar, G. Chadzitaskos}{J. Phys. A: Math.
Gen.}{30}{1997}{2509-2517}


\bibitem{uhl&ford} G. E. Uhlenbeck, G. W. Ford: {\sl Lectures in
Statistical Mechanics}, AMS, Providence - Rhode Island, 1963.

\bibitem{unner0} T. Unnerstall: {\sl Dynamische Beschreibungen und
extensive  physi\-kalische Gr\"o\ss en makroskopischer Quantensysteme mit
Anwendung auf
den Josephson-Kontakt},  Dissertation  (Eberhard-Karls-Universit\"at
zu T\"ubingen), T\"ubingen, 1990.
\bibitem{unner1}\refer{T. Unnerstall}{J. Math. Phys.}{31}{1990}{680}
\bibitem{unner2}\refer{T. Unnerstall}{Commun. Math. Phys.}
{130}{1990}{237}
\bibitem{unner3}\refer{T. Unnerstall}{Lett. Math. Phys.}{20}{1990}{183}

\bibitem{varad} V. S. Varadarajan: {\sl Geometry of  quantum  theory},
Vols.I and II, Van  Nostrand Reinhold, New York, 1970;

\bibitem{knight}\refer{V. Vedral, M. B. Plenio, M. A. Rippin, P. L.
Knight}{Phys. Rev. Lett.}{78}{1997}{2275}

\bibitem{verh} F. Verhulst: {\sl Nonlinear Differential Equations and
Dynamical Systems}, Springer, Berlin - Heidelberg, 1990;


\bibitem{votruba} V. Votruba: {\sl Z\'aklady Speci\'aln\'\i\ Teorie
Relativity (Foundations of Special  Relativity Theory)}, Academia, Prague,
1962.

\bibitem{walters} P. Walters: {\sl An Introduction to Ergodic Theory},
Springer, New York, 1982.

\bibitem{mobil}(a) \refer{R. L. Walsworth, I. F. Silvera}{Phys. Rev.
A}{42}{1990}{63}
(b) \refer{M. Czachor}{Found. Phys. Lett.}{4}{1991}{351}
\bibitem{weinb}(a) \refer{S. Weinberg}{Phys. Rev. Lett.}{62}{1989}{485}  (b)
\refer{S. Weinberg}{Ann. Phys.}{194}{1989}{336}

\bibitem{weinst}\refer{A. Weinstein: {\sl The Local Structure of Poisson
Manifolds}}{J. Differential Geometry}{18}{1983}{523-557}

\bibitem{weyl} H. Weyl: {\sl The Theory of Groups and  Quantum  Mechanics},
 Dover, New York, 1931.

\bibitem{w+z}J. A. Wheeler, W. H. \.Zurek: {\sl Quantum Theory and Measurement},
Princeton University Press, 1983.
\bibitem{whitt}E. T. Whittaker: {\sl A Treatise on the Analytical Dynamics
of Particles and Rigid Bodies with an Introduction to the Problem of Three
bodies}, Cambridge University Press, Cambridge, 1952.

\bibitem{whit+emch}\refer{B. Whitten-Wolfe, G. G. Emch}{Helvetica Phys.
Acta}{49}{1976}{45-55}

\bibitem{WWW}\refer{G. C. Wick, A. S. Wightman, E. P. Wigner}{Phys.
Rev.}{88}{1952}{101-105}
\bibitem{wigner1} E. P. Wigner: {\sl Gruppentheorie und ihre Anwendung auf
die  Quantenme\-chanik der Atomspektren}, Vieweg, Braunschweig, 1931.
\bibitem{wigner2} E. P. Wigner: {\sl Symmetries and Reflections}, Indiana Univ.
Press, Bloomington-London, 1970.
\bibitem{wigner3} E. P. Wigner {\sl in Symmetries in  Science},
Ed. B. Gruber, and R. S. Millman;  Plenum, New York, 1980.

\bibitem{woron} S. Woronowicz: {\sl Podstawy Aksiomatycznej Kwantowej
Teorii  Pola, I. Mechanika Kwantowa}, Preprint No.82, N. Copernicus
 University, Toru\v n, 1969.

\bibitem{yaffe}\refer{L. G. Yaffe}{Rev. Mod. Phys.}{54}{1982}{407}

\bibitem{zeh1}\refer{H. Zeh}{Found.Phys}{1}{1970}{1}

\bibitem{zeil3}\refer{M. \.Zukowski, A. Zeilinger, M. A. Horne, A. K.
Ekert}{Phys. Rev. Lett.}{71}{1993}{4287}
\refer{A. Zeilinger, M. A. Horne, H. Weinfurter, M.
\.Zukowski}{Phys. Rev. Lett.}{78}{1997}{3031}

\bibitem{zelob&stern} D. P. \v Zelobenko, A. I. Stern: {\sl Predstavlenija
Grupp Li}, Nauka, Moscow, 1983.


\bibitem{zurek}\refer{W. \.Zurek}{Phys. Rev. D}{24}{1981}{1516} \refer{W.
\.Zurek}{Phys. Rev. D}{26}{1981}{1862} \refer{W. \.Zurek}{Phys.
Today}{44}{1991}{36}\refer{S. Habib, K. Shizume, W. H. Zurek}{Phys. Rev.
Lett.}{80}{1998}{4361}
\vfill\pagebreak
\end{thebibliography}
\end{document}